\documentclass[%
    a4paper,               
    pagesize,
    twoside=true,           
    openright,              
    parskip=false,           
    chapterprefix=true,     
    10pt,                   
    headings=normal,        
    bibliography=totoc,     
    listof=totoc,           
    titlepage=on,           
    captions=tablebelow,    
    chapterprefix=false,    
    appendixprefix=false,    
    draft=false,            
]{scrreprt}%

\PassOptionsToPackage{utf8}{inputenc}
\usepackage{inputenc}

\newcommand{\thesisTitleCut}{Effective-theory description \\ of heavy-flavored hadrons \\and their properties in a hot medium}
\newcommand{\thesisTitle}{Effective-theory description of heavy-flavored hadrons and their properties in a hot medium}
\newcommand{\thesisName}{Glòria Montaña Faiget}
\newcommand{\thesisSubject}{Documentation}
\newcommand{\thesisDate}{May 2022}
\newcommand{\tesisData}{Maig 2022}

\newcommand{\thesisFirstSupervisor}{Dra. Àngels Ramos Gómez}
\newcommand{\thesisSecondSupervisor}{Dra. Laura Tolós Rigueiro}
\newcommand{\thesisTutor}{Dr. Joan Soto Riera}

\newcommand{\thesisFirstSupervisorEng}{Dr. Àngels Ramos Gómez}
\newcommand{\thesisSecondSupervisorEng}{Dr. Laura Tolós Rigueiro}

\newcommand{\thesisUniversity}{Universitat de Barcelona}
\newcommand{\thesisUniversityDepartment}{Departament de Física Quàntica i Astrofísica}
\newcommand{\thesisUniversityInstitute}{Institut de Ciències del Cosmos}

\newcommand{\thesisUniversityCity}{Barcelona}
\newcommand{\thesisUniversityStreetAddress}{Martí i Franquès, 1}
\newcommand{\thesisUniversityPostalCode}{08028}

\usepackage[english]{babel} 

\PassOptionsToPackage{
    figuresep=colon,%
    hangfigurecaption=false,%
    hangsection=true,%
    hangsubsection=true,%
    sansserif=false,%
    configurelistings=true,%
    colorize=bw,          
    colortheme=bluegreen,%
    configurebiblatex=true,%
    bibsys=biber,%
    bibfile=bib-refs,%
    bibstyle=alphabetic,%
    bibsorting=nty,%
}{cleanthesis}
\usepackage{cleanthesis}

\hypersetup{
    pdftitle={\thesisTitle},    
    pdfsubject={\thesisSubject},
    pdfauthor={\thesisName},    
    plainpages=false,           
    colorlinks=false,           
    pdfborder={1 1 1},          
    breaklinks=true,            
    bookmarksnumbered=true,     %
    bookmarksopen=true          %
    ,linkbordercolor=ctcoloraccessory
    ,citebordercolor=ctcolormain
    ,urlbordercolor=SkyBlue
}

\newacronym{bs}{BS}{Bethe-Salpeter}

\newacronym{be}{BE}{Bose-Einstein}

\newacronym[first={Brookhaven National Laboratory (BNL, New York, NY, USA)}]{bnl}{BNL}{Brookhaven National Laboratory}

\newacronym[first={CERN (\textit{Conseil Européen pour la Recherche Nucléaire}, Geneva, Switzerland)}]{cern}{CERN}{in French \textit{Conseil Européen pour la Recherche Nucléaire}}

\newacronym[first={\textit{chiral perturbation theory} ($\chi$PT)}]{chpt}{$\chi$PT}{chiral perturbation theory}

\newacronym{dr}{DR}{dimensional regularization}

\newacronym{ds}{D-S}{Dyson-Schwinger}

\newacronym[plural={EFTs},firstplural={\textit{effective field theories} (EFTs)}]{eft}{EFT}{effective field theory}

\newacronym[plural={HICs},firstplural={heavy-ion collisions (HICs)}]{hic}{HIC}{heavy-ion collision}

\newacronym{hgs}{HGS}{hidden-gauge symmetry}

\newacronym{hmet}{HMET}{heavy-meson effective theory}

\newacronym[first={\textit{heavy-quark effective theory} (HQET)}]{hqet}{HQET}{heavy-quark effective theory}

\newacronym{hqfs}{HQFS}{heavy-quark flavor symmetry}

\newacronym{hqsfs}{HQSFS}{heavy-quark spin-flavor symmetry}

\newacronym{hqss}{HQSS}{heavy-quark spin symmetry}

\newacronym{itf}{ITF}{imaginary-time formalism}

\newacronym{kms}{KMS}{Kubo-Martin-Schwinger}

\newacronym{ksfr}{KSFR}{Kawarabayashi-Suzuki-Fayyazuddin-Riazuddin}

\newacronym[plural={LECs},firstplural={\textit{low-energy constants} (LECs)}]{lec}{LEC}{low-energy constant}

\newacronym{lhc}{LHC}{Large Hadron Collider}

\newacronym{lhs}{l.h.s.}{left-hand side}

\newacronym{lh}{LH}{left-handed}

\newacronym{lqcd}{LQCD}{lattice QCD}

\newacronym[first={leading-order (LO)}]{lo}{LO}{leading order}

\newacronym{nlo}{NLO}{next-to-leading order}

\newacronym{nrqcd}{NRQCD}{nonrelativistic QCD}

\newacronym{qcd}{QCD}{quantum chromodynamics}

\newacronym{qed}{QED}{quantum electrodynamics}

\newacronym[first={\textit{quark-gluon plasma} (QGP)}]{qgp}{QGP}{quark-gluon plasma}

\newacronym{rhic}{RHIC}{Relativistic Heavy Ion Collider}

\newacronym{rhs}{r.h.s.}{right-hand side}

\newacronym{rh}{RH}{right-handed}

\newacronym{rs}{RS}{Riemann sheet}

\newacronym{tvme}{TVME}{$t$-channel Vector-Meson-Exchange}

\newacronym[first={\textit{unitarized chiral perturbation theory} (U$\chi$PT)}]{uchpt}{U$\chi$PT}{unitarized chiral perturbation theory}

\newacronym{bottom}{$B$}{beauty}

\newacronym{isospin}{$I$}{isospin}

\newacronym{strangeness}{$S$}{strangeness}

\newacronym{sps}{SPS}{Super Proton Synchrotron}

\newacronym{charm}{$C$}{charm}

\newacronym[plural={$\mathcal{B}$},firstplural={baryons ($\mathcal{B}$)}]{baryon}{$\mathcal{B}$}{baryon}

\newacronym[plural={$\mathcal{M}$},firstplural={mesons ($\mathcal{M}$)}]{meson}{$\mathcal{M}$}{meson}

\newacronym{meson-baryon}{$\mathcal{MB}$}{meson--baryon}

\newacronym{meson-meson}{$\mathcal{MM}$}{meson--meson}

\newacronym{pseudoscalar-baryon}{$\mathcal{PB}$}{pseudoscalar--baryon}

\newacronym{vector-baryon}{$\mathcal{VB}$}{vector--baryon}

\newacronym[plural={$\mathcal{P}$},firstplural={pseudoscalar mesons ($\mathcal{P}$)}]{pseudoscalar}{$\mathcal{P}$}{pseudoscalar meson}

\newacronym[plural={$\mathcal{V}$},firstplural={vector mesons ($\mathcal{V}$)}]{vector}{$\mathcal{V}$}{vector meson}

\newacronym{vmd}{VMD}{vector-meson dominance}

\newacronym{ckm}{CKM}{Cabibbo-Kobayashi-Maskawa}

\newacronym{pdg}{PDG}{Particle Data Group}

\newacronym{rpp}{RPP}{Review of Particle Physics}

\newcommand{\ii}{\textrm{i}\,}

\begin{document}



\pagenumbering{roman}			
\pagestyle{empty}				
%


\begin{titlepage}
	\tgherosfont
	\centering
    \hfill
	\vfill
	{\large\textcolor{gray}{ -- Tesi doctoral -- }\par} \vspace{5mm}
	{\fontsize{25}{30}\selectfont \thesisTitleCut \\[15mm]}
	{\fontsize{11.5}{13}\selectfont Memòria presentada per optar al grau de doctor per la \thesisUniversity} \\[5mm]
	{\large Programa de Doctorat en Física } \\[12mm]

		\includegraphics[width=3.5cm]{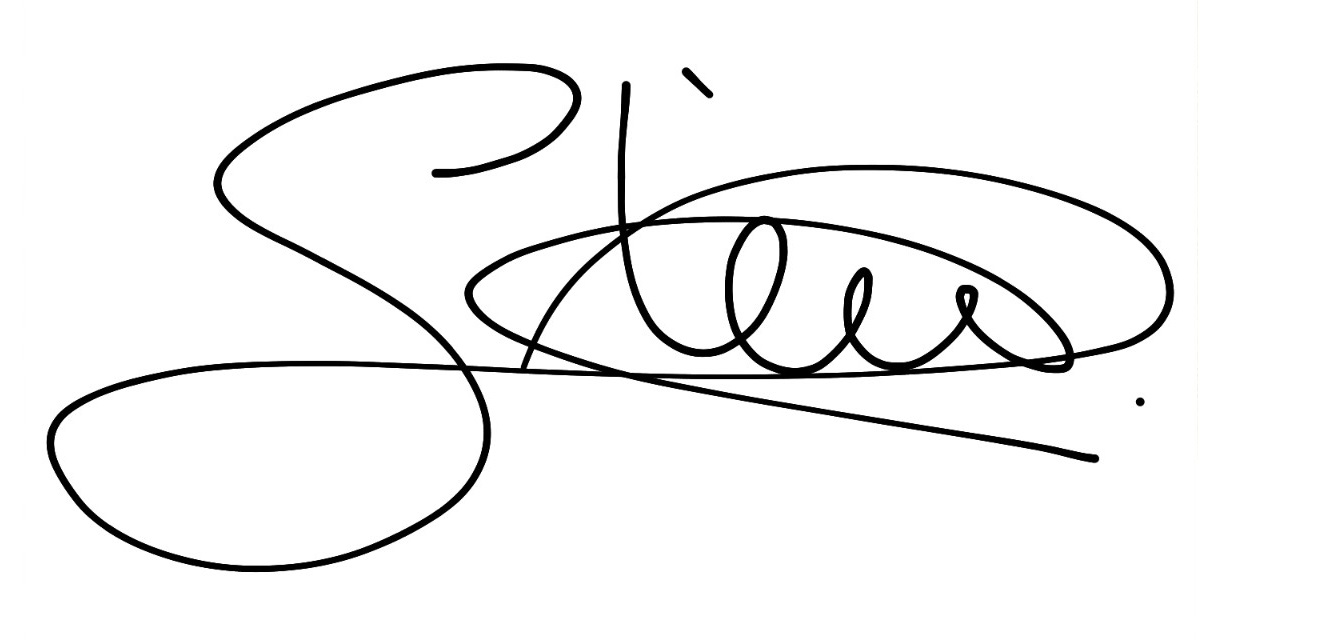}\\[2mm] 

	\begin{minipage}[t]{.3\textwidth}
	    \large
		\raggedleft
		\textbf{Autora:}
	\end{minipage}
	\hspace*{15pt}
	\begin{minipage}[t]{.62\textwidth}
	    \Large
		\thesisName
	\end{minipage} \\[8mm]
	\begin{minipage}[t]{.3\textwidth}
	    \large
		\raggedleft
		\textbf{Directores de tesi:}
	\end{minipage}
	\hspace*{15pt}
	\begin{minipage}[t]{.62\textwidth}
	    \Large
		\thesisFirstSupervisor \\ \thesisSecondSupervisor
	\end{minipage} \\[8mm]
	\begin{minipage}[t]{.3\textwidth}
	    \large
		\raggedleft
		\textbf{Tutor:}
	\end{minipage}
	\hspace*{15pt}
	\begin{minipage}[t]{.62\textwidth}
	    \Large
		\thesisTutor
	\end{minipage} \\[15mm]

	\vfill
	{\large \thesisUniversityDepartment \\ \thesisUniversityInstitute \\[1mm] \thesisUniversity} \\[5mm]
    {\fontsize{11.5}{13}\selectfont \tesisData} \\[10mm]

	\includegraphics[width=8cm]{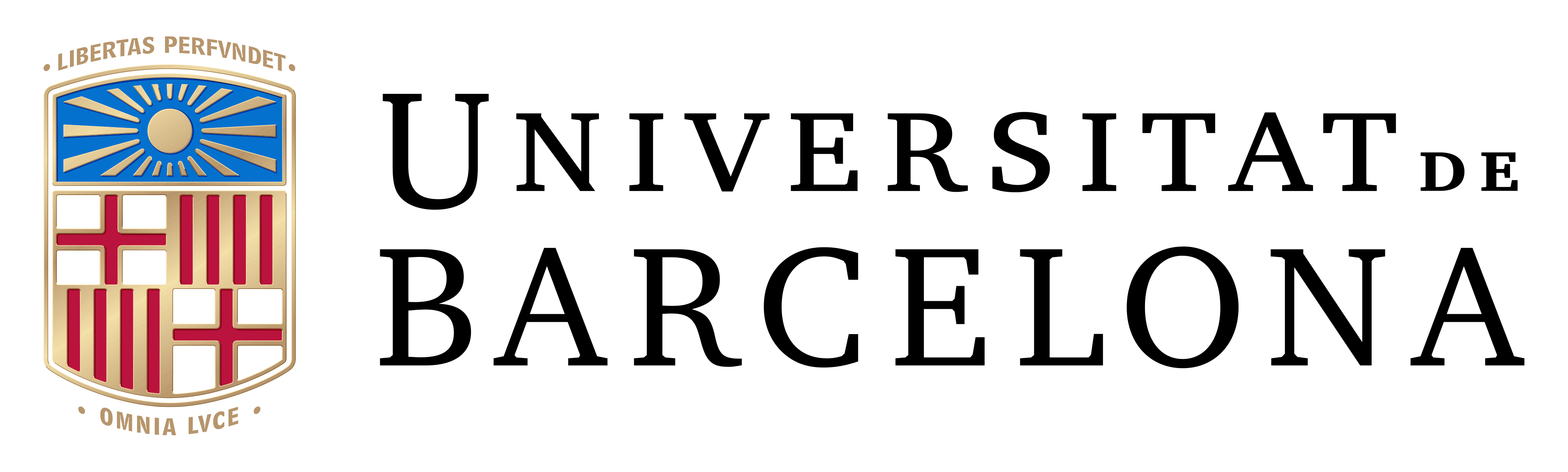} 

\end{titlepage}

\hfill
\vfill
{
	\small\noindent
	\textbf{\thesisName} \\
	\textit{\thesisTitle} \\
	\thesisDate \\
	Ph.D. supervisors: \thesisFirstSupervisorEng\ and \thesisSecondSupervisorEng \\[1.5em]
	\textbf{\thesisUniversity} \\
	\thesisUniversityDepartment \\
	\thesisUniversityInstitute \\
	\thesisUniversityStreetAddress \\
	\thesisUniversityPostalCode\ - \thesisUniversityCity
}

\cleardoublepage
\null\vfill


\hspace{6.5cm}
\parbox{7.8cm}{%
  \raggedright{\normalsize\textit{%
   ``Fall in love with some activity, and do it! \\
   Nobody ever figures out what life is all about,\\
   and it doesn't matter. Explore the world. \\
   Nearly everything is really interesting if you go\\
   into it deeply enough. Work as hard and as much \\
   as you want to on the things you like to do the best. \\
   Don't think about what you want to be, but what\\
   you want to do. Keep up some kind of a minimum \\
   with other things so that society doesn't stop you\\
   from doing anything at all.''\\ \vspace{5mm}
  }   
  }
  \raggedleft\normalsize\MakeUppercase{Richard P. Feynman}\par%
}

\vfill\vfill

\clearpage
\cleardoublepage

\pagestyle{plain}				
%

\addchap{Abstract}
\label{sec:abstract}
For many decades after the conception of the quark model in 1964, and the development of quantum chromodynamics (QCD) a few years later as the theory governing the strong interaction between quarks and gluons, there was no experimental evidence of the existence of hadronic states beyond the quark-antiquark mesons and the three-quark baryons. In the last two decades, however, with the explosion of data in electron--positron and hadron colliders, many states have been observed that do not fit in this picture, especially in the heavy-flavor sector.
Evidence of the existence of the so-called exotic hadrons has recently prompted a lot of activity in the field of hadron physics, with experimental programs in ongoing and upcoming facilities dedicated to the search for new exotic mesons and baryons, and many theoretical efforts trying to disentangle, for instance, compact multiquark structures from hadronic molecules. 

In this dissertation, we focus on recently seen exotic hadrons with heavy-quark content that may be understood as being generated dynamically from the hadron--hadron interaction. This interaction is derived from a suitable effective Lagrangian and properly unitarized in a full coupled-channel basis. In particular, we discuss the possible interpretation of some of the $\Omega_c^*$ excited states recently discovered at LHCb as being meson--baryon molecular states. We also discuss the dynamical generation of excited open-charm mesons from the scattering of pseudoscalar and vector charmed mesons off light mesons. We show that a double-pole structure is predicted for the $D_0^*(2300)$ state, as well as for the $D_1(2430)$, within the molecular picture, while the $D_{s0}^*(2317)$ and the $D_{s1}(2460)$ may be interpreted as molecular bound states. Extensions of these calculations to the bottom sector are also presented.

Moreover, charmed hadrons are a promising probe of the quark-gluon plasma (QGP) phase that is expected to be created in heavy-ion collision experimental facilities. Charm and anticharm quarks are produced in the early stages of the collision and experience the whole evolution of the QGP, before hadronizing predominantly into open-charm mesons. To describe the experimental data, it is necessary to understand, from the theoretical side, the propagation of the $D$ mesons in the hadronic phase and their interaction with the surrounding medium of light mesons. The approach that we employ in this thesis to study the thermal modification of open heavy-flavor mesons in a hot medium is based on the use of effective theories. By means of an extension to finite temperature of the unitarized effective interactions with the light mesons, we obtain the in-medium spectral properties of the $D$, $D^*$, $D_s$, and $D_s^*$ ground-state mesons. We also analyze the temperature dependence of the masses and the decay widths of the dynamically generated $D_0^*(2300)$, $D_1(2430)$, $D_{s0}^*(2317)$, and $D_{s1}(2460)$ states. Additionally, we provide results for the bottomed mesons by exploiting the heavy-quark flavor symmetry of the Lagrangian.

In order to test the results of the thermal effective theory against lattice QCD calculations, we further employ the temperature-dependent scattering amplitudes and spectral functions to compute charm Euclidean correlators.
The spectral properties of charmed mesons at finite temperature can be extracted from lattice QCD data of meson Euclidean correlators, yet relying on a priori assumptions about the shape of the spectral function. Hence we compare both approaches at the level of Euclidean correlators
and find that they compare reasonably well at temperatures below the QCD phase transition temperature. 
We also present calculations of off-shell transport coefficients in the hadronic phase, such as the drag force and the diffusion coefficients. Contrary to previous approaches in the literature, we implement in-medium scattering amplitudes and the thermal dependence of the heavy-meson spectral properties. The transport coefficients in the QGP phase have been recently computed with lattice QCD and extracted from Bayesian analyses of heavy-ion collision data. We observe a smooth matching with our results at the QCD phase transition temperature.
\cleardoublepage

%


%


\addchap*{Acknowledgement}

\label{sec:acknowledgement}

\textit{I would like to dedicate some lines to the people that have accompanied me during the last years in the journey that culminates with this thesis manuscript. It has been an exciting and challenging journey, not exempt from difficulties and stress that have nevertheless been overcome by fulfillment, gratitude, and joy. I would like to express my gratitude to the people that have helped, guided, advised, supported, and encouraged me throughout the Ph.D. adventure.
But before starting, I would like to extend my thanks and apologize to all those that I will forget to mention on this page. }
\\[-3mm]

Voldria començar donant les gràcies a les que han estat les millors directores de tesi que podria haver tingut. Àngels i Laura, gràcies per introduir-me en el món de la física hadrònica, per ensenyar-me, guiar-me i ajudar-me en tot el que ha sigut necessari, però també per inculcar-me la intuïció física, el pensament crític i les bones pràctiques científiques que m'acompanyaran al llarg de la meva carrera com a investigadora. Durant aquests gairebé sis anys, ja des del màster, us heu involucrat en aquesta aventura més enllà del pla professional, pensant sempre en el que era millor per a mi. I per això us estic profundament agraïda. 
\\[-3mm]

\textsl{Si hay alguien a quien tengo que dar las gracias en segundo lugar, este es Juan Torres-Rincón. Sin ti gran parte de esta tesis, tal como aparece en esta memoria, no habría sido posible.} Gràcies Juan per ensenyar-me sobre transport i per ajudar-me sempre que m'ha fet falta. Ha estat un plaer col·laborar amb tu aquests darrers anys.
\\[-3mm]

\textit{I would also like to thank those people with whom I have had the fortune to collaborate abroad. I would like to thank Prof. Luciano Rezzolla and Matthias Hanauske for welcoming me to Frankfurt and introducing me to the field of neutron-star mergers. I would also like to express my gratitude to Prof. Olaf Kaczmarek for his hospitality during my visit to Bielefeld and for providing me with knowledge about lattice QCD. I am also very grateful to Prof. Elena Santopinto for kindly hosting me in Genova during the last stages of my Ph.D.}
\\[-3mm]

També voldria expressar la meva gratitud a tot el Departament de Física Quàntica i Astrofísica i l'Institut de Ciències del Cosmos, en especial a l'Assumpta Parreño, el Volodymyr Magas, el Mario Centelles, l'Arnau Ríos, el Vincent Mathieu, el Javier Menéndez, el Federico Mescia, el Bruno Julià, la Carme Jordi i el Francesc Salvat, així com també a l'Isaac Vidaña. Igualment, voldria tenir un record especial per l'Artur Polls. El buit que la teva marxa prematura va deixar en el grup serà difícilment reemplaçable.
\\[-3mm]

A l'Albert, en Jordi i en Pere voldria agrair que m'acollissin al seu despatx de ``doctorands'' durant el meu últim any del grau i durant el màster, així com també la seva predisposició a aconsellar-me i ajudar-me en tot el que fes falta durant els primers anys del doctorat. Així mateix, voldria dedicar una menció especial al que ha estat el meu company de despatx més longeu, tot i la interrupció imprevista i desafortunada del covid. Gràcies Marc per fer que els ``dies de feina'' fossin més entretinguts després d'una batalla de buscar diferències, i també per compartir l'aventura de creuar l'Atlàntic per assistir a HUGS, amb viatge a NY inclòs.
\\[-3mm]

No voldria oblidar-me d'agrair també al ``grup del ñam ñam'': l'Adrià, l'Albert, l'Alejandro, l'Andreu, el Chiranjib, la Clàudia, l'Iván, el Javi, els Joseps i els Marcs. Gràcies per fer dels dinars un moment de trobada per xerrar i riure plegats, i per estendre l'amistat més enllà de les quatre parets del menjador de la planta 5. Com tampoc voldria oblidar-me dels amics del grau de física. En especial del Pol, del Pau i de la Clara, amb qui vam compartir tantes tardes a la biblio i dinars al Palau Reial. Espero que malgrat la distància ens puguem retrobar al xiringuito per fer una excursió amb caiac. Però tampoc de l'Adrià, l'Aleix, el Borja, el Bruno, l'Elena, l'Eloy, la Gemma, el Guillem, la Júlia, les Núries, el Pere i la Xènia. Gràcies per les trobades regulars, que espero que puguem mantenir al llarg de molt de temps.
\\[-3mm]

\textsf{C'est à mon tour de remercier celui qui a été mon complice et mon compagnon de route. Merci Pierre de m'avoir accompagnée dans ce chemin, de m'avoir compris et aidée ``presque à chaque instant''. Merci pour ta patience ces derniers mois. Je me sens très chanceuse de t'avoir rencontré à ce stade de ma vie et de pouvoir partager avec toi des projets de vie futures. Et n'oublies pas ton gâteau au chocolat! Je t'aime.}
\\[-3mm]

Només em queda agrair a la meva família tot el suport i comprensió que he rebut al llarg d'aquest viatge. Gràcies papa i mama per sempre creure en mi i motivar-me a perseguir els meus somnis, en aquesta etapa i sempre al llarg de la meva vida. Gràcies Àngels per ser al meu costat des del moment que vam néixer, per ser la meva germana i amiga, i per fer-me sentir que sempre podré confiar amb tu sigui on sigui. També a l'Adrià, gràcies per formar part de les nostres vides aquests últims anys. En definitiva, gràcies pel vostre amor. Us estimo!
\\[-3mm]

Aquesta tesi s'ha realizat amb un ajut per a la Contractació de Personal Investigador Novell FI-DGR (referència 2018 FI\_B 00234) finançat per l'Agència de Gestió d'Ajuts Universitaris i de Recerca de la Generalitat de Catalunya i, a partir de l'octubre del 2018, amb un ajut del Programa de Formació de Professorat Universitari FPU (referència FPU2017/04910) finançat primer pel Ministeri de Ciència, Innovació i Universitats i posteriorment pel Ministeri d'Universitats d'Espanya. A més, s'ha rebut una beca STSM (referència ECOST-STSM-Request-CA15213-44423) de l'acció THOR COST Action CA15213. També s'ha rebut suport a través dels projectes No. MDM-2014-0369 i No. CEX2019-000918-M de l'ICCUB (Unitat d'Excel·lència María de Maeztu) finançats, respectivament, pel Ministeri d'Economia i Competitivitat (MINECO) i pel Ministeri de Ciència i Innovació (MCIN), del projecte No. FPA2016-81114-P del MINECO, dels projectes No. PID2019-110165GB-I00 i No. PID2020-118758GB-I00 del MCIN, així com també dels fons europeus FEDER, sota el contracte FIS2017-87534-P, i el projecte No. 411563442 finançat per la \textit{Deutsche Forschungsgemeinschaft} (DFG, Fundació Alemanya d'Investigació). 
\cleardoublepage
%
%

\addchap{List of publications}

The following articles have been published as a result of the work presented in this dissertation:
\begin{itemize}
 \item G. Montaña, A. Feijoo, and A. Ramos,  \textbf{``A meson--baryon molecular interpretation for some $\Omega_c$ excited states''}. Eur. Phys. J. A \textbf{54}.4 (2018) 64 
 \item G. Montaña, A. Ramos, L. Tolos, and J.~M. Torres-Rincon, \textbf{``Impact of a thermal medium on $D$ mesons and their chiral partners''}. Phys. Lett. B \textbf{806} (2020) 135464 
 \item G. Montaña, A. Ramos, L. Tolos, and J.~M. Torres-Rincon, \textbf{``Pseudoscalar and vector open-charm mesons at finite temperature''}. Phys. Rev. D \textbf{102}.9 (2020) 096020 
 \item G. Montaña, O. Kaczmarek, L. Tolos, and A. Ramos, \textbf{``Open-charm Euclidean correlators within heavy-meson EFT interactions''}. Eur. Phys. J. A \textbf{56}.11 (2020) 294 
 \item A. Ramos, A. Feijoo, Q. Llorens, and G. Montaña, \textbf{``The molecular nature of some exotic hadrons''}. Few Body Syst. \textbf{61}.4 (2020) 34 
 \item J.~M. Torres-Rincon, G. Montaña, A. Ramos, and L. Tolos, \textbf{``In-medium kinetic theory of $D$ mesons and heavy-flavor transport coefficients''}. Phys. Rev. C \textbf{105}.2 (2022) 025203
\end{itemize}

The author has also contributed to the following conference proceedings:
\begin{itemize}
 \item \underline{G. Montaña}, A. Ramos, and A. Feijoo, \textbf{``Exotic $\Omega_c^0$ baryons from meson--baryon scattering''}. J. Phys. Conf. Ser. \textbf{1137}.1 (2019) 012040. Contribution to the 13th International Conference on Beauty, Charm and Hyperons (BEACH2018). 17--23 June 2018, Peniche, Portugal
 \item \underline{G. Montaña}, A. Ramos, and A. Feijoo, \textbf{``The molecular nature of some $\Omega_c^0$ states''}. Springer Proc. Phys. \textbf{238} (2020) 729. Contribution to the 22nd International Conference on Few-Body Problems in Physics (FB22). 9--13 July 2018, Caen, France
 \item \underline{G. Montaña}, A. Ramos, and L. Tolos, \textbf{``Properties of heavy mesons at finite temperature''}. SciPost Phys. Proc. \textbf{3} (2020) 038. Contribution to the 24th European Conference on Few-Body Problems in Physics (EFB24). 2--6 September 2019, Surrey, United Kingdom
 \item \underline{G. Montaña}, A. Ramos, L. Tolos, and J.~M. Torres-Rincon, \textbf{``Temperature dependence of the properties of open heavy-flavor mesons''}. EPJ Web Conf. \textbf{259} (2022) 12008. Contribution to the 19th International Conference on Strangeness in Quark Matter (SQM2021). 17--22 May 2021, online, United States
 \item \underline{J.~M. Torres-Rincon}, G. Montaña, A. Ramos and L. Tolos, \textbf{``Finite-temperature effects on $D$-meson properties''}. PoS \textbf{CHARM2020} (2021) 040. Contribution to the 10th International Workshop on Charm Physics (CHARM 2020). 31 May--4 June 2021, online, Mexico
 \item \underline{G. Montaña}, \textbf{``Thermal modification of open heavy-flavor mesons from an effective hadronic theory''}. EPJ Web Conf. \textbf{258} (2022) 04004. Contribution to the virtual tribute to Quark Confinement and the Hadron Spectrum (vConf21). 2--6 August 2021, online, Norway
\end{itemize}

In parallel to the work of this thesis, the author has participated in the development of the following work, although it has not been included in this dissertation:
\begin{itemize}
 \item G. Montaña, L. Tolos, M. Hanauske, and L. Rezzolla, \textbf{``Constraining twin stars with GW170817''}. Phys. Rev. D \textbf{99}.10 (2019) 103009
\end{itemize}
\cleardoublepage
{
\setstretch{1.6}
\printglossary[type=\acronymtype,title=List of abbreviations, toctitle=List of abbreviations]
\cleardoublepage
}

\currentpdfbookmark{\contentsname}{toc}
\setcounter{tocdepth}{2}		
\tableofcontents				
\cleardoublepage

\pagenumbering{arabic}			
\setcounter{page}{1}			
\pagestyle{scrheadings}			

%
\chapter{Introduction}
\label{ch:intro}
With the work that is presented in this dissertation, we pursue two distinct yet related goals: the study of heavy flavored hadrons in the vacuum, paying special attention to those that can be interpreted as molecular states, and the effects that a hot medium may have in their properties.
Given this double purpose, in Section~\ref{sec:intro-exotics} of this introductory chapter, we present a brief description of the quark model and the basic aspects of the theory of quantum chromodynamics. We then motivate the importance of studying the nature of exotic hadrons that cannot be explained by the constituent quark model, in particular those containing heavy quarks. Later, in Section~\ref{sec:intro-extreme}, we summarize the current knowledge on the phases of strongly interacting matter when subject to extreme conditions of temperature and/or density, and finally, we describe the role of heavy-flavor mesons in probing the hot deconfined phase, which constitutes ultimately our motivation to study the modification of heavy mesons in a hot medium.

\section{A brief overview of the quark model and the theory of quantum chromodynamics}
\label{sec:intro-exotics}
\subsection{The quark model}
\label{subsec:intro-quarkmodel}

The modern understanding of hadron physics was conceived with the notion of \textit{quarks} that was independently proposed in 1964 by Gell-Mann \cite{Gell-Mann:1964ewy} and Zweig \cite{Zweig:1964ruk}. The quark model originated as an attempt to classify and understand the properties (mass, spin, charge, isospin, strangeness) of the large number of particles that were being discovered since the 1950s in cosmic rays and particle experiments. It was built on top of the eightfold way organizational scheme, conceived by Gell-Mann himself \cite{Gell-Mann:1961omu,Gell-Mann:1962yej} and Ne'eman \cite{Neeman:1961jhl} a few years before. The key idea behind it is flavor $\textrm{SU}(3)$ symmetry, since at that time only three quarks were known: the up quark (or $u$), with isospin $(I,I_z)=\big(\frac12,+\frac12\big)$; the down quark (or $d$), with isospin $(I,I_z)=(\frac12,-\frac12)$; and the strange quark (or $s$), which carries strangeness $S=-1$.
Nevertheless, it was not until the formalization of quantum chromodynamics by Fritzsch, Leutwyler and Gell-Mann \cite{Fritzsch:1973pi} in the early 1970s that the strong force was better understood in terms of quarks and gluons. 

According to the quark model, hadrons ($q\bar{q}$ mesons and $qqq$ baryons) are classified into multiplets with definite quantum numbers: orbital angular momentum $\ell$, parity $P=(-1)^{\ell+1}$, spin $J$ as the sum of the intrinsic spin $\mathcal{S}$ and $\ell$ given by the relation $|\ell-\mathcal{S}|\leq J\leq|\ell+\mathcal{S}|$, charge conjugation $\mathcal{C}=(-1)^{\ell+\mathcal{S}}$ (for flavor neutral mesons), and its generalization to the $G$-parity, $G=(-1)^{I+\ell+\mathcal{S}}$ (for mesons with isospin $I$ in multiplets with neutral average charge).

Since quarks have spin $\frac12$, the meson and baryon ground-state configurations can carry spin $J=0,\,1$, and $J=\frac12,\,\frac32$, respectively. Besides, the quark model predicts excited states corresponding to orbital and radial excitations.

Based on their $J^{P\mathcal{C}}$ numbers, the mesons with $\ell=0$ are classified into pseudoscalars ($0^{-+}$) and vectors ($1^{--}$), while the orbital excitations with $\ell=1$ are divided into scalars ($0^{++}$), axial vectors or pseudovectors ($1^{+-},1^{++}$), and tensors ($2^{++}$). 

In group theory language, and neglecting the mass difference between the $s$ and the $\{u,d\}$ quarks, the nine $q\bar{q}$ combinations in $\textrm{SU}(3)$ are grouped into an octet and a singlet:
\begin{equation}
 \mathbf{3}\otimes\bar{\mathbf{3}}=\mathbf{8}\oplus\mathbf{1}\ ;\quad (\textrm{SU}(3)) \ .
\end{equation}

While the original quark model was built on top of $\textrm{SU}(3)$, the existence of a fourth quark, designated $c$ and carrying a property called \textit{charm}, was soon proposed \cite{Bjorken:1964gz}. This introduced a new quantum number $C$, representing charm ($C=+1$ for the $c$ quark), and evidence of its existence came in 1974 with the discovery of the $J/\psi$ meson, which is a $c\bar{c}$ bound state \cite{E598:1974sol,SLAC-SP-017:1974ind}.
The addition of a fourth quark flavor such as charm requires the extension from $\textrm{SU}(3)$ to $\textrm{SU}(4)$, although the $\textrm{SU}(4)$ symmetry is clearly broken by the large mass of the $c$ quark. Nevertheless, using $\textrm{SU}(4)$ turns out to be very useful to classify the heavy hadrons. The meson combinations can be grouped in a $15$-plet and a singlet:
\begin{equation}
 \mathbf{4}\otimes\bar{\mathbf{4}}=\mathbf{15}\oplus\mathbf{1}\ ;\quad (\textrm{SU}(4)) \ .
\end{equation}
The weight diagrams for the ground-state pseudoscalar ($J^P=0^-$) and vector ($J^P=1^-$) mesons are depicted in Figs.~\ref{fig:intro-mesonsSU4a} and \ref{fig:intro-mesonsSU4b}, respectively, where the vertical direction is the charm $C$, the horizontal direction is the $z$ component of the isospin $I_z$, and the depth corresponds to the hypercharge $Y$, defined in terms of the baryon number, the strangeness, and the charm as $Y=\mathcal{B}+S-C/3$. The $\textrm{SU}(3)$ nonets are contained in the middle plane, together with the singlet of $\textrm{SU}(4)$.

\begin{figure}[b!]
 \tikzset{isometricYXZ/.style={x={(1cm,0cm)}, z={(-0.1cm,-0.5cm)}, y={(0cm,1cm)}}}
\begin{subfigure}[b]{0.1\textwidth}\centering 
\captionsetup{skip=0pt}
 \centering
\begin{tikzpicture}[scale=1.5,isometricYXZ,rotate around x=-5,
                    grid/.style={very thin,gray}]
    \draw[grid] (0,0,0) -- (0.6,0,0);
    \draw[grid] (0,0,0) -- (0,0.6,0);
    \draw[grid] (0,0,0) -- (0,0,-1);
            \node at (0.8,0,0) {\(I_z\)};
            \node at (0,0.8,0) {\(C\)};
            \node at (0.2,0,-1) {\(Y\)};
\end{tikzpicture}
\end{subfigure}
\begin{subfigure}[b]{0.35\textwidth}\centering 
\captionsetup{skip=0pt}
 \centering
\begin{tikzpicture}[scale=1.5,isometricYXZ,rotate around x=-5,
                    grid/.style={very thin,gray},
                    grid2/.style={very thin,gray!50}]
    \draw[grid2] (-0.5,-0.8,0) -- (-0.5,0,-1) -- (0,0.8,-1) -- (0.5,0,-1);
    \draw[grid,fill=ctcolormagenta!15] (-0.5,0,-1) -- (0.5,0,-1) -- (1,0,0) -- (0.5,0,1) -- (-0.5,0,1) -- (-1,0,0) -- cycle;
    \draw[grid,fill=ctcolorblue!15] (-0.5,0.8,0) -- (0,0.8,-1) -- (0.5,0.8,0) -- cycle;
    \draw[grid,fill=ctcolorgreen!15] (-0.5,-0.8,0) -- (0,-0.8,1) -- (0.5,-0.8,0) -- cycle;
    \draw[grid2]  (0.5,-0.8,0) -- (1,0,0) -- (0.5,0.8,0) -- (0.5,0,1) -- (0,-0.8,1) -- (-0.5,0,1) -- (-0.5,0.8,0) -- (-1,0,0) -- (-0.5,-0.8,0) ;
    \draw[ctcolormagenta,fill=ctcolormagenta] (0.08,0,0.13) circle(0.5mm);
    \draw[ctcolormagenta,fill=ctcolormagenta] (0.08,0,-0.13) circle(0.5mm);
    \draw[ctcolormagenta,fill=ctcolormagenta] (-0.08,0,0.13) circle(0.5mm);
    \draw[ctcolormagenta,fill=ctcolormagenta] (-0.08,0,-0.13) circle(0.5mm);
    \draw[ctcolormagenta,fill=ctcolormagenta] (-0.5,0,-1) circle(0.5mm);
    \draw[ctcolormagenta,fill=ctcolormagenta] (0.5,0,-1) circle(0.5mm);
    \draw[ctcolormagenta,fill=ctcolormagenta] (1,0,0) circle(0.5mm);
    \draw[ctcolormagenta,fill=ctcolormagenta] (0.5,0,1) circle(0.5mm);
    \draw[ctcolormagenta,fill=ctcolormagenta] (-0.5,0,1) circle(0.5mm);
    \draw[ctcolormagenta,fill=ctcolormagenta] (-1,0,0) circle(0.5mm);
    \draw[ctcolorblue,fill=ctcolorblue] (-0.5,0.8,0) circle(0.5mm);
    \draw[ctcolorblue,fill=ctcolorblue] (0,0.8,-1) circle(0.5mm);
    \draw[ctcolorblue,fill=ctcolorblue] (0.5,0.8,0) circle(0.5mm);
    \draw[ctcolorgreen,fill=ctcolorgreen] (-0.5,-0.8,0) circle(0.5mm);
    \draw[ctcolorgreen,fill=ctcolorgreen] (0,-0.8,1) circle(0.5mm);
    \draw[ctcolorgreen,fill=ctcolorgreen] (0.5,-0.8,0) circle(0.5mm);
            \node at (-0.2,0.2,0) {\(\pi^0\)};
            \node at (0.2,0.2,0) {\(\eta\)};
            \node at (-0.2,-0.2,0) {\(\eta_c\)};
            \node at (0.2,-0.2,0) {\(\eta'\)};
            \node at (-0.75,0.8,0) {\(D^0\)};
            \node at (0.75,0.8,0) {\(D^+\)};
            \node at (0.,1,-1) {\(D_s^+\)};
            \node at (-0.75,-0.8,0) {\(D^-\)};
            \node at (0.75,-0.8,0) {\(\bar{D}^0\)};
            \node at (0.,-1,1) {\(D_s^-\)};
            \node at (-0.5,0.17,-1) {\(K^0\)};
            \node at (0.6,0.17,-1) {\(K^+\)};
            \node at (-0.5,-0.17,1) {\(K^-\)};
            \node at (0.6,-0.17,1) {\(\bar{K}^0\)};
            \node at (-1.25,0,0) {\(\pi^-\)};
            \node at (1.25,0,0) {\(\pi^+\)};
\end{tikzpicture}
\caption{}
\label{fig:intro-mesonsSU4a}
\end{subfigure}
\begin{subfigure}[b]{0.45\textwidth}\centering 
\captionsetup{skip=0pt}
 \centering
\begin{tikzpicture}[scale=1.5,isometricYXZ,rotate around x=-5,
                    grid/.style={very thin,gray},
                    grid2/.style={very thin,gray!50}]
    \draw[grid2] (-0.5,-0.8,0) -- (-0.5,0,-1) -- (0,0.8,-1) -- (0.5,0,-1);
    \draw[grid,fill=ctcolormagenta!15] (-0.5,0,-1) -- (0.5,0,-1) -- (1,0,0) -- (0.5,0,1) -- (-0.5,0,1) -- (-1,0,0) -- cycle;
    \draw[grid,fill=ctcolorblue!15] (-0.5,0.8,0) -- (0,0.8,-1) -- (0.5,0.8,0) -- cycle;
    \draw[grid,fill=ctcolorgreen!15] (-0.5,-0.8,0) -- (0,-0.8,1) -- (0.5,-0.8,0) -- cycle;
    \draw[grid2]  (0.5,-0.8,0) -- (1,0,0) -- (0.5,0.8,0) -- (0.5,0,1) -- (0,-0.8,1) -- (-0.5,0,1) -- (-0.5,0.8,0) -- (-1,0,0) -- (-0.5,-0.8,0) ;
    \draw[ctcolormagenta,fill=ctcolormagenta] (0.08,0,0.13) circle(0.5mm);
    \draw[ctcolormagenta,fill=ctcolormagenta] (0.08,0,-0.13) circle(0.5mm);
    \draw[ctcolormagenta,fill=ctcolormagenta] (-0.08,0,0.13) circle(0.5mm);
    \draw[ctcolormagenta,fill=ctcolormagenta] (-0.08,0,-0.13) circle(0.5mm);
    \draw[ctcolormagenta,fill=ctcolormagenta] (-0.5,0,-1) circle(0.5mm);
    \draw[ctcolormagenta,fill=ctcolormagenta] (0.5,0,-1) circle(0.5mm);
    \draw[ctcolormagenta,fill=ctcolormagenta] (1,0,0) circle(0.5mm);
    \draw[ctcolormagenta,fill=ctcolormagenta] (0.5,0,1) circle(0.5mm);
    \draw[ctcolormagenta,fill=ctcolormagenta] (-0.5,0,1) circle(0.5mm);
    \draw[ctcolormagenta,fill=ctcolormagenta] (-1,0,0) circle(0.5mm);
    \draw[ctcolorblue,fill=ctcolorblue] (-0.5,0.8,0) circle(0.5mm);
    \draw[ctcolorblue,fill=ctcolorblue] (0,0.8,-1) circle(0.5mm);
    \draw[ctcolorblue,fill=ctcolorblue] (0.5,0.8,0) circle(0.5mm);
    \draw[ctcolorgreen,fill=ctcolorgreen] (-0.5,-0.8,0) circle(0.5mm);
    \draw[ctcolorgreen,fill=ctcolorgreen] (0,-0.8,1) circle(0.5mm);
    \draw[ctcolorgreen,fill=ctcolorgreen] (0.5,-0.8,0) circle(0.5mm);
            \node at (-0.2,0.2,0) {\(\rho^0\)};
            \node at (0.2,0.2,0) {\(\omega\)};
            \node at (-0.2,-0.2,0) {\(J/\psi\)};
            \node at (0.2,-0.2,0) {\(\phi\)};
            \node at (-0.8,0.8,0) {\(D^{*0}\)};
            \node at (0.8,0.8,0) {\(D^{*+}\)};
            \node at (0.,1,-1) {\(D_s^{*+}\)};
            \node at (-0.8,-0.8,0) {\(D^{*-}\)};
            \node at (0.8,-0.8,0) {\(\bar{D}^{*0}\)};
            \node at (0.,-1,1) {\(D_s^{*-}\)};
            \node at (-0.5,0.17,-1) {\(K^{*0}\)};
            \node at (0.6,0.17,-1) {\(K^{*+}\)};
            \node at (-0.5,-0.17,1) {\(K^{*-}\)};
            \node at (0.6,-0.17,1) {\(\bar{K}^{*0}\)};
            \node at (-1.25,0,0) {\(\rho^-\)};
            \node at (1.25,0,0) {\(\rho^+\)};
\end{tikzpicture}
\caption{}
\label{fig:intro-mesonsSU4b}
\end{subfigure}
\caption{The $\textrm{SU}(4)$ weight diagrams of the $16$-plets of the ground-state pseudoscalar (a) and vector mesons (b), containing the $\textrm{SU}(3)$ nonets.}
\label{fig:intro-mesonsSU4}
\end{figure}
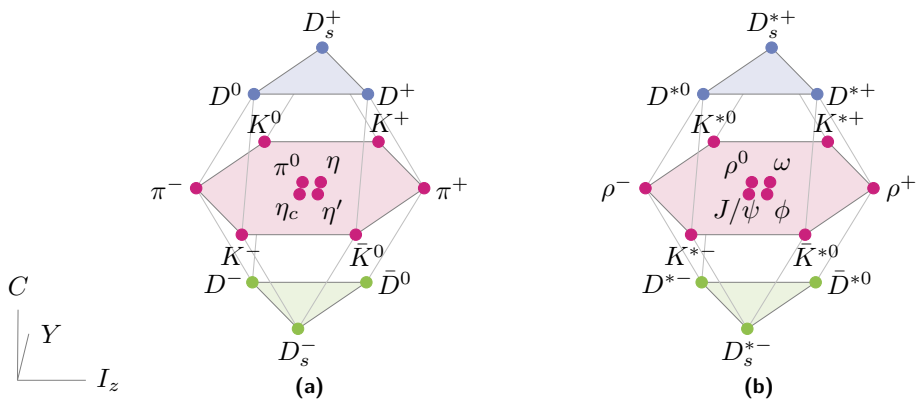

\begin{figure}[t!]
 \tikzset{isometricYXZ/.style={x={(1cm,0cm)}, z={(-0.1cm,-0.5cm)}, y={(0cm,1cm)}}}
\begin{subfigure}[b]{0.1\textwidth}\centering 
\captionsetup{skip=0pt}
 \centering
\begin{tikzpicture}[scale=1.5,isometricYXZ,rotate around x=-5,
                    grid/.style={very thin,gray}]
    \draw[grid] (0,0,0) -- (0.6,0,0);
    \draw[grid] (0,0,0) -- (0,0.6,0);
    \draw[grid] (0,0,0) -- (0,0,-1);
            \node at (0.8,0,0) {\(I_z\)};
            \node at (0,0.8,0) {\(C\)};
            \node at (0.2,0,-1) {\(Y\)};
\end{tikzpicture}
\end{subfigure}
\begin{subfigure}[b]{0.35\textwidth}\centering 
\captionsetup{skip=0pt}
 \centering
\begin{tikzpicture}[scale=1.5,isometricYXZ,rotate around x=-5,
                    grid/.style={very thin,gray},
                    grid2/.style={very thin,gray!50}]
    \draw[grid2] (-0.5,0,0) -- (-1,0.8,0) -- (-0.5,1.6,0);
    \draw[grid2] (0.5,0,0) -- (1,0.8,0) -- (0.5,1.6,0);
    \draw[grid2] (-1,0,1) -- (-1,0.8,0);
    \draw[grid2] (1,0,1) -- (1,0.8,0);
    \draw[grid,fill=ctcolormagenta!15] (-0.5,0,0) -- (0.5,0,0) -- (1,0,1) -- (0.5,0,2) -- (-0.5,0,2) -- (-1,0,1) -- cycle;
    \draw[grid,fill=ctcolorblue!15] (-1,0.8,0) -- (0,0.8,0) -- (1,0.8,0) -- (0.5,0.8,1) -- (0,0.8,2) -- (-0.5,0.8,1) -- cycle;
    \draw[grid,fill=orange!15] (-0.5,1.6,0) -- (0,1.6,1) -- (0.5,1.6,0) -- cycle;
    \draw[grid2] (0,1.6,1) -- (0,0.8,2);
    \draw[grid2] (0.5,0,2) -- (0,0.8,2) -- (-0.5,0,2);
    \draw[ctcolormagenta,fill=ctcolormagenta] (-0.06,0,1) circle(0.5mm);
    \draw[ctcolormagenta,fill=ctcolormagenta] (0.06,0,1) circle(0.5mm);
    \draw[ctcolormagenta,fill=ctcolormagenta] (-0.5,0,0) circle(0.5mm);
    \draw[ctcolormagenta,fill=ctcolormagenta] (0.5,0,0) circle(0.5mm);
    \draw[ctcolormagenta,fill=ctcolormagenta] (1,0,1) circle(0.5mm);
    \draw[ctcolormagenta,fill=ctcolormagenta] (0.5,0,2) circle(0.5mm);
    \draw[ctcolormagenta,fill=ctcolormagenta] (-0.5,0,2) circle(0.5mm);
    \draw[ctcolormagenta,fill=ctcolormagenta] (-1,0,1) circle(0.5mm);
    \draw[ctcolorblue,fill=ctcolorblue] (-1,0.8,0) circle(0.5mm);
    \draw[ctcolorblue,fill=ctcolorblue] (-0.05,0.8,0) circle(0.5mm);
    \draw[ctcolorblue,fill=ctcolorblue] (0.05,0.8,0) circle(0.5mm);
    \draw[ctcolorblue,fill=ctcolorblue] (1,0.8,0) circle(0.5mm);
    \draw[ctcolorblue,fill=ctcolorblue] (0.55,0.8,0.9) circle(0.5mm);
    \draw[ctcolorblue,fill=ctcolorblue] (0.45,0.8,1.1) circle(0.5mm);
    \draw[ctcolorblue,fill=ctcolorblue] (0,0.8,2) circle(0.5mm);
    \draw[ctcolorblue,fill=ctcolorblue] (-0.55,0.8,0.9) circle(0.5mm);
    \draw[ctcolorblue,fill=ctcolorblue] (-0.45,0.8,1.1) circle(0.5mm);
    \draw[orange,fill=orange] (-0.5,1.6,0) circle(0.5mm);
    \draw[orange,fill=orange] (0,1.6,1) circle(0.5mm);
    \draw[orange,fill=orange] (0.5,1.6,0) circle(0.5mm);
            \node at (-0.21,0.,1) {\(\Lambda\)};
            \node at (0.24,0.,1) {\(\Sigma^0\)};
            \node at (-0.75,0.,0) {\(n\)};
            \node at (0.75,0.,0) {\(p\)};
            \node at (-0.5,-0.17,2) {\(\Xi^-\)};
            \node at (0.6,-0.17,2) {\(\Xi^0\)};
            \node at (-1.25,0,1) {\(\Sigma^-\)};
            \node at (1.25,0,1) {\(\Sigma^+\)};
            \node at (0.0,0.85,1.5) {\(\Omega_c^0\)};
            \node at (1.1,0.8,1) {\(\Xi_c^+,\Xi_c'^+\)};
            \node at (-1,0.8,1) {\(\Xi_c^0,\Xi_c'^0\)};
            \node at (-1.25,0.8,0) {\(\Sigma_c^0\)};
            \node at (1.35,0.8,0) {\(\Sigma_c^{++}\)};
            \node at (0.2,1,0) {\(\Lambda_c^+,\Sigma_c^+\)};
            \node at (-0.25,1.6,1) {\(\Omega_{cc}^+\)};
            \node at (-0.75,1.6,0) {\(\Xi_{cc}^+\)};
            \node at (0.85,1.6,0) {\(\Xi_{cc}^{++}\)};           
\end{tikzpicture}
\caption{}
\label{fig:intro-baryonsSU4a}
\end{subfigure}
\begin{subfigure}[b]{0.45\textwidth}\centering 
\captionsetup{skip=0pt}
 \centering
\begin{tikzpicture}[scale=1.5,isometricYXZ,rotate around x=-5,
                    grid/.style={very thin,gray},
                    grid2/.style={very thin,gray!50}]
    \draw[grid2] (-1.5,0,0) -- (-1,0.8,0) -- (-0.5,1.6,0) -- (0,2.4,0);
    \draw[grid2] (1.5,0,0) -- (1,0.8,0) -- (0.5,1.6,0) -- (0,2.4,0);
    \draw[grid,fill=ctcolormagenta!15] (-1.5,0,0) -- (-0.5,0,0) -- (0.5,0,0) -- (1.5,0,0) -- (1,0,1) -- (0.5,0,2) -- (0,0,3) -- (-0.5,0,2) -- (-1,0,1) -- cycle;
    \draw[grid,fill=ctcolorblue!15] (-1,0.8,0) -- (0,0.8,0) -- (1,0.8,0) -- (0.5,0.8,1) -- (0,0.8,2) -- (-0.5,0.8,1) -- cycle;
    \draw[grid,fill=orange!15] (-0.5,1.6,0) -- (0,1.6,1) -- (0.5,1.6,0) -- cycle;
    \draw[grid2] (0,1.6,1) -- (0,0.8,2) -- (0,0,3);
    \draw[ctcolormagenta,fill=ctcolormagenta] (0,0,1) circle(0.5mm);
    \draw[ctcolormagenta,fill=ctcolormagenta] (-0.5,0,0) circle(0.5mm);
    \draw[ctcolormagenta,fill=ctcolormagenta] (0.5,0,0) circle(0.5mm);
    \draw[ctcolormagenta,fill=ctcolormagenta] (1.5,0,0) circle(0.5mm);
    \draw[ctcolormagenta,fill=ctcolormagenta] (-1.5,0,0) circle(0.5mm);
    \draw[ctcolormagenta,fill=ctcolormagenta] (1,0,1) circle(0.5mm);
    \draw[ctcolormagenta,fill=ctcolormagenta] (0.5,0,2) circle(0.5mm);
    \draw[ctcolormagenta,fill=ctcolormagenta] (-0.5,0,2) circle(0.5mm);
    \draw[ctcolormagenta,fill=ctcolormagenta] (-1,0,1) circle(0.5mm);
    \draw[ctcolormagenta,fill=ctcolormagenta] (0,0,3) circle(0.5mm);
    \draw[ctcolorblue,fill=ctcolorblue] (-1,0.8,0) circle(0.5mm);
    \draw[ctcolorblue,fill=ctcolorblue] (0,0.8,0) circle(0.5mm);
    \draw[ctcolorblue,fill=ctcolorblue] (1,0.8,0) circle(0.5mm);
    \draw[ctcolorblue,fill=ctcolorblue] (0.5,0.8,1) circle(0.5mm);
    \draw[ctcolorblue,fill=ctcolorblue] (0,0.8,2) circle(0.5mm);
    \draw[ctcolorblue,fill=ctcolorblue] (-0.5,0.8,1) circle(0.5mm);
    \draw[orange,fill=orange] (-0.5,1.6,0) circle(0.5mm);
    \draw[orange,fill=orange] (0,1.6,1) circle(0.5mm);
    \draw[orange,fill=orange] (0.5,1.6,0) circle(0.5mm);
    \draw[ctcolordarkgreen,fill=ctcolordarkgreen] (0,2.4,0) circle(0.5mm);
            \node at (0.2,0.,1) {\(\Sigma^0\)};
            \node at (-1.76,0,0) {\(\Delta^-\)};
            \node at (-0.6,0.15,0) {\(\Delta^0\)};
            \node at (0.7,0.15,0) {\(\Delta^+\)};
            \node at (1.8,0,0) {\(\Delta^{++}\)};
            \node at (-0.75,0,2) {\(\Xi^-\)};
            \node at (0.8,0,2) {\(\Xi^0\)};
            \node at (-1.25,0,1) {\(\Sigma^-\)};
            \node at (1.3,0,1) {\(\Sigma^+\)};
            \node at (0,-0.15,3) {\(\Omega^-\)};
            \node at (0.0,0.85,1.5) {\(\Omega_c^0\)};
            \node at (0.85,0.8,1) {\(\Xi_c^+\)};
            \node at (-0.8,0.8,1) {\(\Xi_c^0\)};
            \node at (-1.25,0.8,0) {\(\Sigma_c^0\)};
            \node at (1.35,0.8,0) {\(\Sigma_c^{++}\)};
            \node at (0.1,1,0) {\(\Sigma_c^+\)};
            \node at (-0.25,1.6,1) {\(\Omega_{cc}'^+\)};
            \node at (-0.75,1.6,0) {\(\Xi_{cc}'^+\)};
            \node at (0.85,1.6,0) {\(\Xi_{cc}'^{++}\)};  
            \node at (0,2.6,0) {\(\Omega_{ccc}^{++}\)};
\end{tikzpicture}
\caption{}
\label{fig:intro-baryonsSU4b}
\end{subfigure}
\caption{The $\textrm{SU}(4)$ weight diagrams of the ground-state baryons. (a) The spin-$1/2$ $20$-plet containing the $\textrm{SU}(3)$ octet; (b) the spin-$3/2$ $20$-plet containing the $\textrm{SU}(3)$ decuplet.}
\label{fig:intro-baryonsSU4}
\end{figure}

The construction of the baryons in $\textrm{SU}(3)$ involves the following multiplets:
\begin{equation}
 \mathbf{3}\otimes\mathbf{3}\otimes\mathbf{3}=\mathbf{10}_S\oplus\mathbf{8}_M\oplus\mathbf{8}_M\oplus\mathbf{1}_A\ ;\quad (\textrm{SU}(3)) \ ,
\end{equation}
where the states in the decuplet are symmetric under the exchange of two quarks, the singlet is antisymmetric, and the states in the octets have mixed symmetry. When considering $\textrm{SU}(4)$ we have a totally symmetric $20$-plet, two $20$-plets with mixed symmetry, and a totally antisymmetric quadruplet,
\begin{equation}
 \mathbf{4}\otimes\mathbf{4}\otimes\mathbf{4}=\mathbf{20}_S\oplus\mathbf{20}_M\oplus\mathbf{20}_M\oplus\mathbf{4}_A\ ;\quad (\textrm{SU}(4)) \ .
\end{equation}
The symmetric $\mathbf{20}_S$ contains the $\textrm{SU}(3)$ decuplet as a subset, forming the ``ground floor'' of the weight diagram shown in Fig.~\ref{fig:intro-baryonsSU4b}, and the baryons in this multiplet have $J^P=\frac32^+$. The mixed-symmetric $\mathbf{20}_M$'s correspond to the $J^P=\frac12^+$ baryons shown in Fig.~\ref{fig:intro-baryonsSU4a}, with the $\textrm{SU}(3)$ octets on the lowest level. The $\textrm{SU}(3)$ antisymmetric singlet $\Lambda_1$ is contained in the antisymmetric $\mathbf{4}_A$, in which the baryons have $J^P=\frac12^-$, but these states are forbidden in the ground-state multiplets by Fermi statistics.

One can also construct $\textrm{SU}(4)$ multiplets in which charm is replaced by beauty and classify the corresponding hadrons with $b$ quarks, or even combine the two sets of $\textrm{SU}(4)$ structures into larger $\textrm{SU}(5)$ multiplets that contain all possible ground-state mesons.

\begin{figure}[t!]
\centering
\begin{tikzpicture}[x=1.15cm, y=1.15cm,scale=1.8]
  \draw[rounded corners=0.3cm] (-0.5,0.5) rectangle (4.4,-1.5);
  \draw[rounded corners=0.3cm] (-0.6,0.6) rectangle (5.0,-2.5);
  \draw[rounded corners=0.3cm] (-0.7,0.7) rectangle (5.6,-3.5);

  \node at(0, 0)   {\particle
                   {$u$}        {up}       {$2.2$ MeV}{\large\sfrac{1}{2}}{\;\large\sfrac{2}{3}}{$r$, $g$, $b$}};
  \node at(0,-1)   {\particle
                   {$d$}        {down}    {$4.7$ MeV}{\large\sfrac{1}{2}}{\large\sfrac{--1}{3}}{$r$, $g$, $b$}};
  \node at(0,-2)   {\particle[gray!20!white]
                   {$e$}        {electron}       {$511$ keV}{\large\sfrac{1}{2}}{$-1$}{}};
  \node at(0,-3)   {\particle[gray!20!white]
                   {$\nu_e$}    {$e$ neutrino}         {$<1.0$ eV}{\large\sfrac{1}{2}}{}{}};
  \node at(1, 0)   {\particle
                   {$c$}        {charm}   {$1.28$ GeV}{\large\sfrac{1}{2}}{\;\large\sfrac{2}{3}}{$r$, $g$, $b$}};
  \node at(1,-1)   {\particle 
                   {$s$}        {strange}  {$96$ MeV}{\large\sfrac{1}{2}}{\large\sfrac{--1}{3}}{$r$, $g$, $b$}};
  \node at(1,-2)   {\particle[gray!20!white]
                   {$\mu$}      {muon}         {$105.66$ MeV}{\large\sfrac{1}{2}}{$-1$}{}};
  \node at(1,-3)   {\particle[gray!20!white]
                   {$\nu_\mu$}  {$\mu$ neutrino}    {$<170$ keV}{\large\sfrac{1}{2}}{}{}};
  \node at(2, 0)   {\particle
                   {$t$}        {top}    {$173.1$ GeV}{\large\sfrac{1}{2}}{\;\large\sfrac{2}{3}}{$r$, $g$, $b$}};
  \node at(2,-1)   {\particle
                   {$b$}        {bottom}  {$4.18$ GeV}{\large\sfrac{1}{2}}{\large\sfrac{--1}{3}}{$r$, $g$, $b$}};
  \node at(2,-2)   {\particle[gray!20!white]
                   {$\tau$}     {tau}          {$1.7768$ GeV}{\large\sfrac{1}{2}}{$-1$}{}};
  \node at(2,-3)   {\particle[gray!20!white]
                   {$\nu_\tau$} {$\tau$ neutrino}  {$<18.2$ MeV}{\large\sfrac{1}{2}}{}{}};
  \node at(3,-3)   {\particle[orange!20!white]
                   {$W^{\hspace{-.3ex}\scalebox{.5}{$\pm$}}$}
                                {}              {$80.39$ GeV}{$1$}{$\pm1$}{}};
  \node at(4,-3)   {\particle[orange!20!white]
                   {$Z$}        {}                    {$91.19$ GeV}{$1$}{}{}};
  \node at(3.5,-2) {\particle[ctcoloraccessory!35!white]
                   {$\gamma$}   {photon}                        {}{$1$}{}{}};
  \node at(3.5,-1) {\particle[ctcolormain!20!white]
                   {$g$}        {gluon}                    {}{$1$}{}{8 colors}};
  \node at(5,0)    {\particle[gray!50!white]
                   {$H$}        {Higgs}              {$124.97$ GeV}{0}{}{}};
%
  \node at(4.25,-0.5) [force]      {strong nuclear force (color)};
  \node at(4.85,-1.5) [force]    {electromagnetic force (charge)};
  \node at(5.45,-2.4) [force] {weak nuclear force (weak isospin)};
%
  \draw [<-] (2.5,0.3)   -- (2.7,0.3)                node [legend] {spin};
  \draw [<-] (2.15,0.25) -- (2.3,0.15) -- (2.7,0.15) node [legend] {mass};
  \draw [<-] (2.5,0)  -- (2.7,0)                     node [legend] {color};
  \draw [<-] (2.5,-0.3)  -- (2.7,-0.3)               node [legend] {charge};
  \draw [mbrace] (-0.8,0.5)  -- (-0.8,-1.5)
                 node[leftlabel] {QUARKS}; 
  \draw [mbrace] (-0.8,-1.5) -- (-0.8,-3.5)
                 node[leftlabel] {LEPTONS}; 
%
  \draw [brace] (-0.5,1.) -- (2.5,1.) node[toplabel]         {Generations of matter (fermions)};
  \draw [brace] (2.5,.8)  -- (4.5,.8) node[toplabel]          {Force carriers (bosons)};
%
  \node at (0,0.8)   [generation] {1\tiny st};
  \node at (1,0.8)   [generation] {2\tiny nd};
  \node at (2,0.8)   [generation] {3\tiny rd};
\end{tikzpicture}
\caption{Particle content of the Standard Model.}
 \label{fig:intro-StandardModel}
\end{figure}
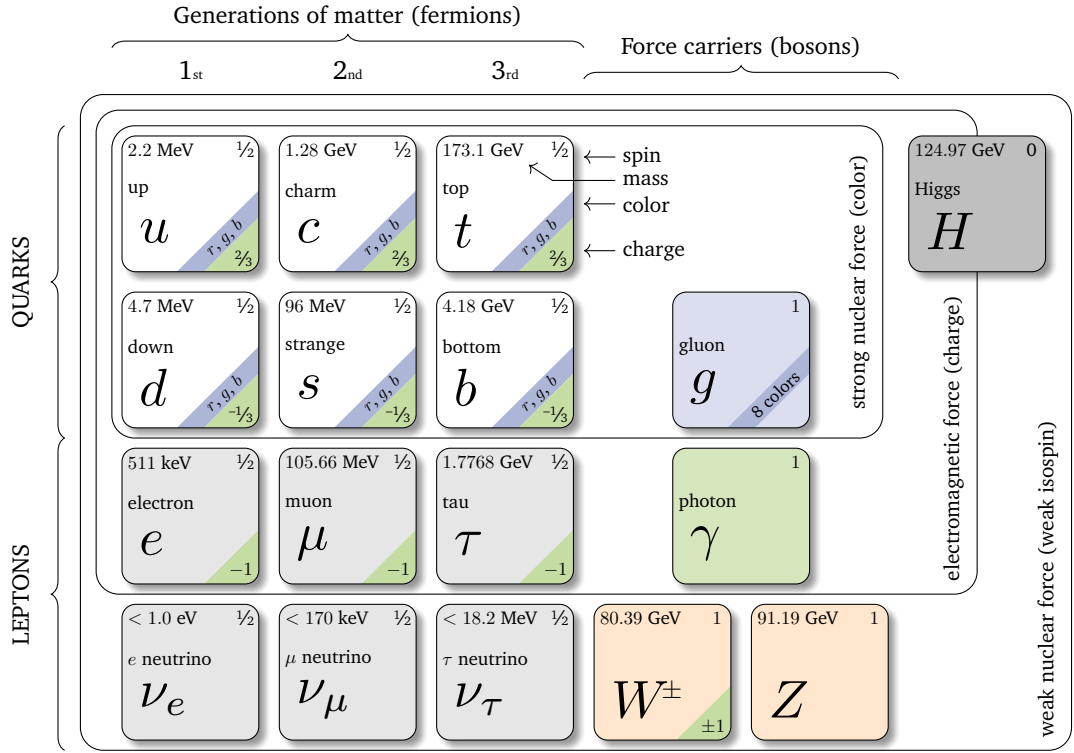

\subsection{Quantum chromodynamics}
\label{subsec:intro-qcd}

Shortly after the concept of quarks was proposed, it became obvious that certain hadrons had quark compositions that violated the exclusion principle. For instance, the $\Omega^-$ ($sss$) baryon contains three strange quarks with spin $\frac12$. In an attempt to solve this problem, it was suggested that quarks possess an additional property, the so-called \textit{color charge}, which laid the basis of the theory of \gls{qcd}.

\begin{figure}[t!]
\centering 
\begin{subfigure}[b]{0.32\textwidth}\centering 
\captionsetup{skip=0pt}
 \begin{tikzpicture}[baseline=(i.base)]
    \begin{feynman}[small]
      \vertex (i);
      \vertex [above = of i] (a);
      \vertex [below left = of i] (b);
      \vertex [below right = of i] (c);
      \diagram* {
        (a) -- [gluon] (i) , 
        (b) -- [fermion] (i) -- [fermion] (c),
       };
    \end{feynman}
  \end{tikzpicture}
\caption{}
\label{fig:intro-verticesqcd-a}
\end{subfigure}
\begin{subfigure}[b]{0.32\textwidth}\centering 
\captionsetup{skip=0pt}
 \begin{tikzpicture}[baseline=(i.base)]
    \begin{feynman}[small]
      \vertex (i);
      \vertex [above = of i] (a);
      \vertex [below left = of i] (b);
      \vertex [below right = of i] (c);
      \diagram* {
        (a) -- [gluon] (i) , 
        (b) -- [gluon] (i) -- [gluon] (c),
       };
    \end{feynman}
  \end{tikzpicture}
\caption{}
\label{fig:intro-verticesqcd-b}
\end{subfigure}
\begin{subfigure}[b]{0.32\textwidth}\centering 
\captionsetup{skip=0pt}
 \begin{tikzpicture}[baseline=(i.base)]
    \begin{feynman}[small]
      \vertex (i);
      \vertex [above left = of i] (a);
      \vertex [above right = of i] (d);
      \vertex [below left = of i] (b);
      \vertex [below right = of i] (c);
      \diagram* {
        (a) -- [gluon] (i) -- [gluon] (d), 
        (b) -- [gluon] (i) -- [gluon] (c),
       };
    \end{feynman}
  \end{tikzpicture}
\caption{}
\label{fig:intro-verticesqcd-c}
\end{subfigure}
 \caption{Interaction vertices of \gls{qcd}: (a) quark-gluon vertex, (b) three-gluon vertex and (c) four-gluon vertex. Straight lines represent quarks and coiled lines represent gluons.}
 \label{fig:intro-verticesqcd}
\end{figure}
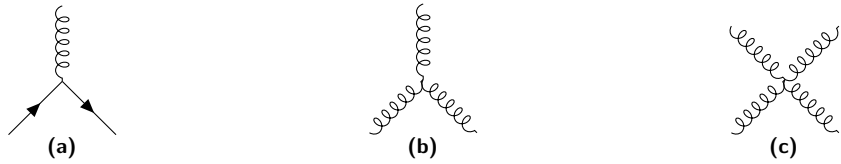

\Gls{qcd} is the fundamental quantum field theory that describes the physics of strongly interacting particles and it is by now well established. It is a non-Abelian gauge theory with color $\textrm{SU}(3)_c$ as the underlying gauge group, in which color is hypothesized to be the equivalent of the electric charge in \gls{qed}. Its fundamental degrees of freedom are \textit{quarks}, which are spin-$\frac12$ matter fields, and \textit{gluons}, which are the massless spin-$1$ fields mediating the strong force. In the Standard Model (see Fig.~\ref{fig:intro-StandardModel}) there are six types, or flavors, of quarks: up ($u$), down ($d$), strange ($s$), charm ($c$), bottom ($b$), and top ($t$), each of which comes in three colors (red, green, blue) and transforms as a triplet under the fundamental representation of $\textrm{SU}(3)_c$. Moreover, there are eight flavorless and necessarily colored gluons carrying color ($r$, $g$, $b$) and anticolor ($\bar{r}$, $\bar{g}$, $\bar{b}$), which transform under the adjoint representation of $\textrm{SU}(3)_c$. The interaction between quarks and gluons does not depend on the flavor, and gluons not only couple to quark fields (Fig.~\ref{fig:intro-verticesqcd-a}), but they also interact among themselves via three-gluon (Fig.~\ref{fig:intro-verticesqcd-b}) and four-gluon vertices (Fig.~\ref{fig:intro-verticesqcd-c}). This is a fundamental difference between \gls{qed} and \gls{qcd} that makes the latter richer but mathematically more complex.

The Lagrangian density of \gls{qcd} for $N_f$ quark flavors reads\footnote{The so-called $\theta$-term responsible for the violation of CP symmetry has been omitted here, as it is not relevant for the discussions in this thesis.}
\begin{equation}\label{eq:intro-lagrangianQCD}
 \mathcal{L}_\textrm{QCD}=\sum_f\bar{\psi}_{f,a}(\ii\slashed{D}_{ab}-m_f\delta_{ab})\psi_{f,b}-\frac{1}{4}F_{\alpha\beta}^AF_A^{\alpha\beta} \ ,
\end{equation}
where repeated indices are summed over. The spinor $\psi_{f,a}$ denotes a quark field with flavor $f$, mass $m_f$ and color charge $a=1,...,N_c$ ($N_c=3$). The spinor indices have been suppressed for simplicity. The purely gluonic term in Eq.~(\ref{eq:intro-lagrangianQCD}) is given in terms of the gluonic field-strength tensor
\begin{equation}
 F_{\alpha\beta}^A=\partial_\alpha\mathcal{A}^A_\beta-\partial_\beta\mathcal{A}^A_\alpha+g_sf^{ABC}\mathcal{A}_\alpha^B\mathcal{A}_\beta^C \ ,
\end{equation}
where $\mathcal{A}_\alpha^A$ are the gluon fields, the indices $A$, $B$, $C$ run from $1$ to $N_c^2-1$ (the eight color degrees of freedom of the gluon field in the case of $N_c=3$), and the factors $f^{ABC}$ in the non-Abelian term are the structure constants of \gls{qcd}.
The notation $\slashed{D}=\gamma^\alpha D_\alpha$ has been used for the covariant derivative, $D_\alpha$, which in \gls{qcd} is a matrix in color space with matrix elements
\begin{equation}
 (D_\alpha)_{ab}=\partial_\alpha\delta_{ab}-\ii g_st^C_{ab}\mathcal{A}_\alpha^C \ ,
\end{equation}
with $t^C$ proportional to the Gell-Mann matrices, $t^C=\frac12\lambda^C$.
The \gls{qcd} Lagrangian is symmetric under the local color gauge symmetry, under global Lorenz transformations, and under the discrete parity, charge conjugation and time-reversal symmetries. Other important symmetries of \gls{qcd} are the so-called \textit{chiral} and \textit{heavy-quark symmetries}, which are discussed in Section~\ref{subsec:free-th-symmetries}.
The quark masses $m_f$ and the strong coupling constant $g_s$, which determines the strength of the interaction between colored objects, are free parameters of the Lagrangian. To keep a notation equivalent to that in \gls{qed}, it is common to use  $\alpha_s$ ($\alpha_s=g_s^2/(4\pi)$) to denote the strength of the strong interaction.

The coupling constant $\alpha_s$ becomes asymptotically weaker at increasing energies (or equivalently, at short distances) in a phenomenon which is known as \textit{asymptotic freedom}. This feature of \gls{qcd} allows perturbative calculations in terms of $\alpha_s$ in the high-energy regime. However, its behavior changes at smaller energies (of the order of the hadron masses, below a few GeV), where the coupling constant becomes too large for perturbation theory to be applied. In this regime, a perturbative description of \gls{qcd} in terms of quarks and gluons as degrees of freedom is not valid anymore.

Another important feature of \gls{qcd} is the so-called \textit{color confinement}: neither quarks nor gluons have hitherto been observed isolated in experiments, but rather confined inside colorless hadrons. These include the \textit{mesons}, which are made up of a valence $q\bar{q}$ pair with the quark carrying color and the antiquark carrying the corresponding anticolor; and the \textit{baryons}, which are made up of three valence quarks, $qqq$, carrying three different colors that add up to color neutral. These states are explained within \gls{qcd} as color-singlet clusters.

\subsection{Exotics: nonconventional hadrons}
\label{subsec:intro-exotics}

In conventional quark models, the interpretation of ordinary mesons as composed by a quark--antiquark pair and ordinary baryons as bound states of three quarks gives rise to a rather successful description of a wealth of data \cite{Capstick:1986bm}. 
Still, there is more to the picture than this because \gls{qcd} does not exclude hadrons having a valence composition different from the ordinary hadrons. Allowed combinations include pairs of valence quarks and antiquarks (\textit{tetraquarks}: $qq\bar{q}\bar{q}$), four quarks and an antiquark (\textit{pentaquarks}: $qqqq\bar{q}$) or six quarks (\textit{hexaquarks}: $qqqqqq$), as far as they are color singlets. 
Actually, the general concept of multiquark states arose at the same time as the birth of the quark model, as Gell-Mann and Zweig themselves speculated about the possibility of having such more complex assemblies of quarks~\cite{Gell-Mann:1964ewy,Zweig:1964ruk}.
In the literature, these structures are encompassed in the so-called \textit{exotic hadron} terminology. The attribute ``exotic'' is sometimes reserved for hadrons having quantum numbers not encountered in the conventional quark model, for example, exotic $J^{P\mathcal{C}}=0^{--},0^{+-},1^{-+},2^{+-}$,..., exotic flavor, etc. Then, the name \textit{cryptoexotic} is used to designate hadrons that do not possess explicit exotic values of the quantum numbers but are incompatible with standard mesons and baryons because of their properties, such as their mass and decay width. However, in practice, any hadronic state beyond $q\bar{q}$ or $qqq$ is usually referred to as exotic in the literature, and this is also the terminology adopted in this thesis. 

In addition to multiquark hadrons, which are the object of research in this dissertation, exotic states comprised exclusively of gluonic fields (\textit{glueballs}) and exotic combinations of excited gluons together with quarks and antiquarks (\textit{hybrids}) are also theoretically possible within \gls{qcd}.

Multiquark states can take the form of genuine compact four-, five- or six-quark states bound together by gluons, or, alternatively, they can form \textit{hadronic molecules}, namely bound or quasi-bound states of two or more hadrons. In this latter framework, tetraquarks are interpreted as meson--meson states, pentaquarks are organized in the form of meson--baryon molecules, and hexaquarks correspond to two baryons bound together (dibaryons). The deuteron, which is composed of a proton and a neutron barely bound by a few MeV per nucleon, can indeed be considered as the first bound state of two hadrons discovered in 1932 \cite{PhysRev.39.164}.

All of the exotic candidates mentioned above have been the subject of thorough theoretical and experimental searches for almost six decades with still no unambiguous evidence of such exotic configurations. Admittedly, a great number of states that do not fit in the original quark model have been reported. In the early years, the experimental efforts for searching exotic states focused on the light hadron sector.

Regarding the nonet of the lowest scalar mesons that includes the $f_0(500)$ (also called $\sigma$), the $\kappa(800)$, the $a_0(980)$, and the $f_0(980)$, their interpretation as the $\ell=1$ orbital excitations within the $q\bar{q}$ picture fails to explain, for instance, the mass degeneracy of the $a_0(980)$ and the $f_0(980)$, as it rather expects the $a_0$ to be close to the $f_0(500)$. If they are considered to be tetraquarks instead, either in the form of diquark--antidiquark~\cite{Jaffe:1976ig,Close:2002zu,Maiani:2004uc} or meson--meson molecules~\cite{Janssen:1994wn,Pelaez:2004xp,RuizdeElvira:2010cs}, the correct mass ordering is reproduced. 

In the baryon sector, a paradigmatic example is that of the $\Lambda(1405)$ resonance, the mass of which is systematically overestimated by quark models. Predicted by Dalitz and Tuan in 1959~\cite{Dalitz:1959dn} as a meson--baryon molecule before the quark model was proposed, many experimental and theoretical efforts have been dedicated to revealing its nature. In particular, studies of the meson--baryon interaction employing chiral effective Lagrangians and implementing unitarization predicted the $\Lambda(1405)$ as being the superposition of two poles in the meson--baryon scattering matrix \cite{pdg,Oller:2000fj,Jido:2003cb,Hyodo:2011ur}, a particularity that has been recently acknowledged by the \gls{pdg}~\cite{pdg} in favor of its molecular structure.

A substantial activity escalation in the heavy sector took place after 2003 when the $X(3872)$ (also known as $\chi_{c1}(3872)$) was observed by the Belle collaboration~\cite{Belle:2003nnu}, and confirmed and extensively studied later by other experiments at electron--positron~\cite{BaBar:2004oro,BESIII:2013fnz} and hadron colliders~\cite{CDF:2003cab,D0:2004zmu,LHCb:2011zzp,CMS:2013fpt,ATLAS:2016kwu}. Despite its $c\bar{c}$ content, the fact that its quantum numbers and mass do not fit those of an ordinary quarkonium state turned the $X(3872)$ into the first clear exotic candidate. A decade after its discovery, its nature is still under intense debate. Due to its extreme proximity to the $D^0\bar{D}^{*0}$ threshold, a natural interpretation is that of a loosely bound $D\bar{D}^*$ molecule. The possible existence of a charmed meson--anticharmed meson molecule was already proposed in Refs.~\cite{Voloshin:1976ap,Tornqvist:1993ng} some decades before the experimental observation, and since then the $D\bar{D}^*$ molecular picture for the $X(3872)$ has received a lot of support \cite{Close:2003sg,Gamermann:2009fv}. Another popular scenario for the $X(3872)$ is that of a compact tetraquark state, structured into a diquark--antidiquark~\cite{Maiani:2004vq}, as well as the admixture of conventional charmonium with molecular and tetraquark components~\cite{Hanhart:2011jz}.

Since then, many other exotic candidates have been found in the heavy quarkonium sector, the so-called $XYZ$ mesons. In the past, isospin $I=0$ states with $J^{P\mathcal{C}}=1^{--}$ have been traditionally designated as $Y$, those with isospin $I=0$ and quantum numbers other than $1^{--}$ have been called $X$, and $Z$ has been used to refer to quarkonium-like states with isospin different than $0$. In particular, the charged $Z$ states are manifestly exotic. More recently, the \gls{pdg} has developed a new naming scheme for these states~\cite{pdg} based on their quantum numbers, in an attempt to extend the nomenclature used for ordinary quarkonia to the newly discovered states. Despite being well established for conventional $q\bar{q}$ states, the new names used by the~\gls{pdg} may well designate states with a dominant molecular, tetraquark, etc. component. As a matter of fact, the $X(3872)$ is called ``$\chi_{c1}(3872)$ (aka $X(3872)$)'' in the latest version of the \gls{pdg} compilation~\cite{pdg}.

Also, since 2003, new open heavy-flavor mesons have been observed experimentally, some of them not fully consistent with the excited states predicted by the conventional quark model. Particularly interesting are the positive-parity charm-strange $D_{s0}^*(2317)$ and $D_{s1}(2460)$ observed in 2003 by the BaBar~\cite{BaBar:2003oey} and CLEO~\cite{CLEO:2003ggt} collaborations, respectively. The masses of these two states stay lie just below the $DK$ and $DK^*$ thresholds, respectively, a fact which turns them into natural candidates for hadronic molecules~\cite{Barnes:2003dj,Szczepaniak:2003vy,Kolomeitsev:2003ac,Hofmann:2003je,Guo:2006fu,Gamermann:2006nm,Faessler:2007gv,Flynn:2007ki}, although other explanations, such as a conventional $c\bar{s}$ meson \cite{Dai:2003yg,Narison:2003td,Bardeen:2003kt}, a compact tetraquark \cite{Cheng:2003kg,Terasaki:2003qa,Chen:2004dy,Maiani:2004vq,Bracco:2005kt,Wang:2006bs}, and a mixture of $c\bar{s}$ with tetraquark \cite{Browder:2003fk} and $D^{(*)}K$ molecular \cite{vanBeveren:2003kd} components also exist in the literature. We will come back to the properties of these states in this dissertation and study them within the molecular picture.


%
%
The existence of pentaquark baryons has been made evident from the recent discovery by the LHCb collaboration \cite{Aaij:2015tga} of the excited nucleon resonances $P_c(4380)^+$ and $P_c(4450)^+$, seen in the invariant mass distribution of $J/\psi \, p$ pairs from the decay of the $\Lambda_b$. More recently, the new $P_c(4312)^+$ has been reported by the LHCb collaboration from the same decay~\cite{LHCb:2019kea}, where they also observed that the formerly reported  $P_c(4450)^+$ consists of two overlapping peaks, $P_c(4440)^+$ and $P_c(4457)^+$. 
The high mass of these excited nucleons inevitably demands the presence of an additional $c\bar{c}$ pair. Hidden-charm baryons having a meson--baryon structure had already been predicted previously \cite{Wu:2010jy,Wu:2010vk,Yang:2011wz,Xiao:2013yca,Karliner:2015ina}, and later studies confirmed that the narrow pentaquark seen from the $\Lambda_b \to J/\Psi~K^- p$ decay at \gls{cern} could receive a molecular interpretation \cite{Chen:2015loa,Roca:2015dva,He:2015cea,Meissner:2015mza,Ortega:2016syt}. For recent reviews on multiquark states and hadronic molecules see Refs.~\cite{Esposito:2016noz,Chen:2016qju,Guo:2017jvc,Olsen:2017bmm,Brambilla:2019esw}.

In Chapter~\ref{ch:exoticsinfreespace} we will show some examples of hadronic states that may be understood within the molecular picture as dynamically generated from the hadron--hadron interaction, using effective theories.




\section{Strongly interacting matter in extreme conditions}
\label{sec:intro-extreme}
\subsection{Phases of QCD matter}
\label{subsec:intro-phases}

At extremely high temperatures and/or baryon densities, hadrons are expected to lose their identity and dissolve into a ``soup'' of their constituents, quarks and gluons, the so-called \gls{qgp}. The existence of this deconfined phase of strongly interacting matter was proposed in the seventies~\cite{Collins:1974ky,Cabibbo:1975ig}, just a couple of years after the formulation of the theory of \gls{qcd}.
Indeed, even though color confinement prevents the direct observation of isolated quarks and gluons in experiments, asymptotic freedom predicts the weakening of inter-quark forces at high energies, thus expecting a transition from a phase in which quarks and gluons are confined inside hadrons to the \gls{qgp} at extreme conditions of temperature and/or baryon density.  
Such conditions can be obtained experimentally by colliding heavy ions at relativistic energies at \gls{cern} and at \gls{bnl}. While in the year 2000, at the end of the main part of the \gls{hic} program at the \gls{sps}, \gls{cern} announced possible evidence of the formation of the \gls{qgp} in $\textrm{Pb+Pb}$ collisions~\cite{Heinz:2000bk}, the actual discovery took place in 2005 with $\textrm{Au+Au}$ collisions at the \gls{rhic} at \gls{bnl}~\cite{BRAHMS:2004adc,PHOBOS:2004zne,STAR:2005gfr,PHENIX:2004vcz}.

From the cosmological point of view, a very hot \gls{qgp} filled the universe during its very early existence, and a transition to hadron matter took place as the universe cooled down to $T\lesssim 150$~MeV. 
The nature of this phase transition affects considerably our understanding of the evolution of the early universe. Considering that the typical baryon densities $\rho_{\textrm{B}}$ of the early-universe \gls{qgp} are negligible, in this regime, numerical simulations of \gls{qcd} on the lattice have been very helpful. In particular, nonperturbative \gls{lqcd} calculations not only confirmed the existence of the deconfined phase of quarks and gluons at high temperatures~\cite{Susskind:1979up} but also provided strong evidence that the \gls{qcd} phase transition at $\rho_{\textrm{B}}=0$ is a crossover~\cite{Aoki:2006we,Bhattacharya:2014ara}. Furthermore, \gls{lqcd} results indicate that the two kinds of phase transition that are possible in \gls{qcd}, that is, the deconfining transition and the restoration of chiral symmetry, occur essentially at the same temperature~\cite{Cheng:2007jq}. 

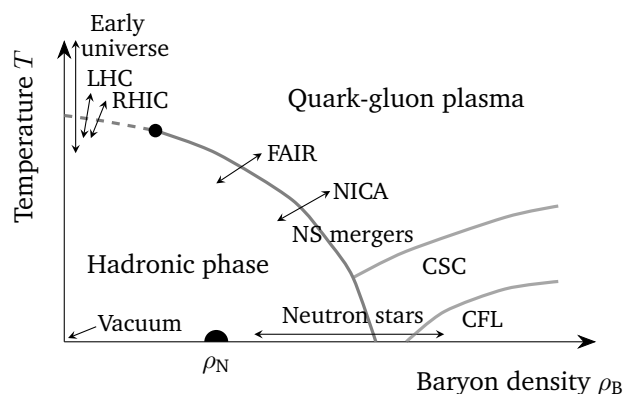
\begin{figure}[b!]\centering 
\captionsetup{skip=0pt}
\begin{tikzpicture}[baseline=0,
  ]
  \draw [->,-{Stealth[scale=1.5]}] (0,0) -- (7,0) coordinate (xaxis);
  \draw [->,-{Stealth[scale=1.5]}] (0,0) -- (0,4) coordinate (yaxis);
  \node [below] at (6,-0.25) {Baryon density $\rho_{\textrm{B}}$};
  \node [left,rotate=90] at (-0.5,4) {Temperature $T$};
  \draw[dashed,very thick,gray] plot [smooth] coordinates {(0,3) (0.5,2.93) (1.2,2.8)};
  \draw[very thick,gray] plot [smooth] coordinates {(1.2,2.8) (2,2.5) (3,1.88) (3.5,1.3) (3.8,0.85) (4.1,0)};
  \node at (1.2,2.8) [circle,fill,inner sep=1.8pt] {};
  \node at (2,0.07) [semicircle,fill,inner sep=2pt] {};
  \node at (2,-0.3) {$\rho_{\textrm{N}}$};
  \node at (4.5,3.2) {Quark-gluon plasma};
  \node at (1.5,1) {Hadronic phase};
  \node at (3.8,0.35) {\small Neutron stars};
  \draw [<->,>=stealth] (2.5,0.1) -- (5,0.1);
  \node at (3.8,1.4) {\small NS mergers};
  \node at (0.7,4.2) {\small Early};
  \node at (0.8,3.9) {\small universe};
  \draw [<->,>=stealth] (0.15,4) -- (0.15,2.5);
  \node at (0.6,3.5) {\small LHC};
  \draw [<->,>=stealth] (0.35,3.3) -- (0.25,2.7);
  \node at (1,3.2) {\small RHIC};
  \draw [<->,>=stealth] (0.55,3.2) -- (0.35,2.7);  
  \node at (3,2.5) {\small FAIR};
  \draw [<->,>=stealth] (2.6,2.5) -- (2,2.1);  
  \node at (3.9,2) {\small NICA};
  \draw [<->,>=stealth] (3.5,2) -- (2.8,1.6);  
  \draw[very thick,gray,opacity=0.7] plot [smooth] coordinates {(3.8,0.85) (4.5,1.2) (5.8,1.65) (6.5,1.8)};
  \draw[very thick,gray,opacity=0.7] plot [smooth] coordinates {(4.5,0) (5,0.4) (5.8,0.7) (6.5,0.8)};
  \node at (5,1) {\small CSC};
  \node at (5.5,0.3) {\small CFL};
  \node at (1,0.25) {\small Vacuum};
  \draw [->,>=stealth] (0.42,0.2) -- (0.05,0.05);
\end{tikzpicture}
\caption{Schematic phase diagram of \gls{qcd}. The solid line between the hadronic phase and the \gls{qgp} corresponds to a first-order transition, while the dashed line represents the crossover transition. The black circle denotes the critical endpoint of the deconfining and chiral transitions. LHC: Large Hadron Collider; RHIC: Relativistic Heavy Ion Collider; FAIR: Facility for Antiproton and Ion Research; NICA: Nuclotron-based Ion Collider fAcility; $\rho_{\textrm{N}}$: nuclear saturation density; NS: neutron star; CSC: color superconducting phase; CFL: color-flavor locked phase. }
\label{fig:intro-phasediagram}
\end{figure}

The current knowledge of the \gls{qcd} phase diagram is schematically summarized in Fig.~\ref{fig:intro-phasediagram}, in the plane of temperature, $T$, and baryon density, $\rho_{\textrm{B}}$. The dashed line represents the crossover transition at low densities, separated from the first-order transition (solid line) at higher densities by the \gls{qcd} critical point (black circle). In this diagram, the matter created shortly after the Big Bang is located in the upper-left corner, which is the region that is accessible with \glspl{hic} at the \gls{rhic} at \gls{bnl} and the \gls{lhc} at \gls{cern}. 

On the other hand, when the energy per nucleon of the colliding nuclei is of the order of a few tens of GeV, such as in the future Facility for Antiproton and Ion Research (FAIR) at GSI (\textit{Gesellschaft f\"ur Schwerionenforschung}, Darmstadt, Germany) and in the Nuclotron-based Ion Collider fAcility (NICA) at the Joint Institute for Nuclear Research (JINR, Dubna, Russia), not only high temperature but also high baryon density can be achieved.

A different scenario for the realization of deconfined \gls{qcd} matter is encountered in the interior of neutron stars. In the core of these compact astrophysical objects, baryon densities as high as several times nuclear saturation density ($\rho_{\textrm{N}}$) are reached. It is thus plausible that neutrons (and other hadrons present in the composition of the neutron star) melt into a deconfined phase of cold quark matter. During the merger of two neutron stars, temperatures reach several tens of MeV and central densities can increase further. Therefore, a lot of effort is currently devoted to unraveling the composition of neutron stars and the equation of state of cold and hot neutron-star matter, from both the nuclear-physics and the general-relativity communities (see \cite{Ozel:2016oaf,Oertel:2016bki,Baiotti:2016qnr,Baym:2017whm} for some recent reviews).

At ultra-high baryon densities, several additional phases of \gls{qcd} matter are predicted. In particular, a color-superconducting (CSC) phase and a color-flavor locked (CFL) phase of deconfined \gls{qcd} matter are expected~\cite{Bailin:1983bm,Alford:1998mk,Buballa:2002wy,Braun-Munzinger:2008szb,Fukushima:2010bq}.

Thermal effective field theories provide a nonperturbative tool, complementary to \gls{lqcd}, to study strongly interacting matter at finite temperature that permits one to approach the \gls{qcd} phase transition from the hadronic chirally-broken phase.
Let us imagine that starting from the vacuum ($\mu_{\textrm{B}}=T=0$, bottom left corner of Fig.~\ref{fig:intro-phasediagram}) we heat the system. That is to say, we move vertically upwards along the temperature axis. A hadron gas is formed, in which the abundances of each species in thermal equilibrium are dictated by the appropriate thermal distribution functions (Bose-Einstein distribution for mesons, Fermi-Dirac distribution for baryons). At temperatures below $T\sim 150$~MeV, this medium is dominated by light mesons (pions, mostly). Eventually, the hadron gas reaches the critical temperature $T_c\sim 150$~MeV and the thermal fluctuations can break up the pions and there is a crossover transition to the \gls{qgp}.
This is the picture that will be exploited in this dissertation and hadronic effective theories will be employed to analyze the effects of a thermal medium of light mesons on the properties and the propagation of heavy hadrons, as will be described in Chapters~\ref{ch:hot-medium} and \ref{ch:lattice}.

\subsection{Heavy-flavor mesons as a probe of the quark-gluon plasma}
\label{sec:intro-heavy}

A \gls{hic} is a dynamic many-body process and, even if the \gls{qgp} is expected to be created in the initial stage of the collision, it cools rapidly as the system expands and emits various forms of radiation, and eventually reaches the transition temperature and undergoes hadronization. It may experience further expansion and cooling before the chemical and kinetic freeze-outs take place. Therefore, what is detected in the experiments are the hadronic and leptonic residues, and the very early formation of the \gls{qgp} needs to be retraced by using the observed data. Examples of such detected particles that are used to probe the formation of the \gls{qgp} are electromagnetic probes, that is, photons and dileptons ($e^+e^-$ or $\mu^+\mu^-$ pairs), and also quarkonia and jets. Quarkonia and jets (as well as open heavy-flavor mesons and very energetic dileptons and photons) constitute the so-called hard probes, whose production takes place at the initial stages of the collision, through the hard scattering of the colliding nucleons.

Heavy-flavor mesons, both heavy quarkonia and open heavy-flavor mesons, are particularly good probes. Due to the fact that the masses of the charm ($m_c\approx 1.3$~GeV) and the bottom ($m_b\approx 4.2$~GeV) quarks are significantly larger than the typical temperatures attained in \glspl{hic}, the thermal creation of heavy quark--antiquark pairs during the evolution of the \gls{qgp} is extremely suppressed, and their production is essentially restricted to the early stages of the collision. Moreover, heavy quarks have large relaxation times in contrast to the light partons of the medium ($\sim m_Q/T$), and hence they are not expected to fully thermalize within the timescale of the \gls{qgp} lifetime in \glspl{hic}. Therefore, sensitive information about the interaction history of the heavy quarks (or the heavy mesons, after hadronization) in the hot medium may be preserved in the heavy-flavor observables.

The phenomenon of quarkonia suppression is expected to be a particularly important signature of the \gls{qgp}. It was first discussed in the context of the $J/\psi$ meson in 1986 when Matsui and Satz suggested that the binding of a $c\bar{c}$ pair to form a charmonium system may be precluded by the effect of the deconfined quarks and gluons in the \gls{qgp} in a process known as \textit{color screening}~\cite{Matsui:1986dk}, analogous to the Debye screening in an electromagnetic plasma.
This effect can be probed experimentally by comparing the quarkonia yields (not only of the ground states $J/\psi$ and $\Upsilon$, but also of excited charmonia and bottomonia) in collisions of heavy nuclei ($AA$) with the properly normalized yields in proton--proton ($pp$) collisions, as the \gls{qgp} is more likely to be produced in the former case~\cite{Karsch:1987pv,Digal:2001ue}. The suppression is quantified through the so-called nuclear modification factor $R_{AA}$. The ``melting'' (disappearance of the spectral peak) of quarkonia states at some temperature above $T_c$ has been confirmed by \gls{lqcd} studies~\cite{Datta:2003ww,Aarts:2007pk}, which also corroborated the fact that higher excited states may dissociate at lower temperatures, as they are less tightly bound.

There are however additional processes that may affect the yields of heavy quarkonia and that make the interpretation of the experimental data rather complicated~\cite{Braun-Munzinger:2007edi}. For instance, the absorption of the quarkonia states in the nuclear medium~\cite{Gerschel:1988wn}, or their break-up due to collisions with secondary hadrons produced in the collision and comoving with quarkonium in the medium (comovers)~\cite{Gerschel:1998zi,Vogt:1999cu,Cassing:1999es,Capella:2000zp}, are examples of mechanisms that can lead to quarkonia suppression even in the absence of the \gls{qgp} formation. Furthermore, regeneration (also called recombination or coalescence) of quarkonia from uncorrelated heavy quark--antiquark pairs generated in different hard collisions may lead to an enhancement of the quarkonia yields in $AA$ collisions compared to those in $pp$ collisions. This process is especially important if the abundance of heavy quarks and antiquarks in the heavy-ion environment is large, and it thus increases with the centrality and the energy of the collision~\cite{Thews:2001hy,Rapp:2008tf}.

While the melting of quarkonia states has long been considered an important signature of the \gls{qgp} formation, the evolution of open heavy-flavor mesons in \glspl{hic} is also of great interest for the understanding of the strongly interacting regime close to the \gls{qcd} phase transition. The existence of a connection between observables in the heavy-quarkonium and the open heavy-flavor sectors has become evident in recent years. As a matter of fact, the yield and spectra of the regenerated quarkonia are sensitive to the abundance and momentum distributions of the open heavy-flavor mesons in the system. Also, in the comover scattering scenario, quarkonium dissociation (and recombination) are explained by the inelastic interactions with the comoving hadrons. Processes such as $J/\psi+\Phi\rightarrow D+\bar{D}$ and $D+\bar{D}\rightarrow J/\psi+\Phi$, where $\Phi$ is a comoving light meson, lead to the production of new open-charm mesons ($J/\psi$ suppression) or charmonia states ($J/\psi$ regeneration). If the properties (masses and decay widths) of the open-charm mesons are modified in a hot medium, the thresholds for the processes above would vary accordingly, thus providing a complementary explanation for the $J/\psi$ yields. This is a key point that motivates the study of the properties of open heavy-flavor mesons in a hot medium, and a substantial portion of the work described in this dissertation is devoted to this aim. 

Due to their large mass, the propagation of heavy quarks (and early produced quarkonia) in the \gls{qgp}, as well as the propagation of the heavy mesons in the hot mesonic medium after hadronization, can be treated as ``Brownian motion'', with momentum transfers being relatively small. Then, the motion of the heavy particles can be characterized by transport coefficients, such as the heavy flavor spatial diffusion coefficient, $D_s$. The study of charm and bottom transport coefficients provides valuable access to the nonperturbative regime of \gls{qcd}, and universal information about the transport properties of the \gls{qgp} is carried, in particular, by the normalized diffusion coefficient, $2\pi TD_s$. This coefficient can be determined from \gls{lqcd} calculations\footnote{In fact, the momentum diffusion coefficient, $\kappa=4\pi T^3/(2\pi T D_s)$, is the object that is computed in the lattice.} or, conversely, with effective hadronic models below the transition temperature, as we will show in Chapter~\ref{ch:transport} of this thesis. Knowledge of this coefficient is necessary to model the diffusion of heavy quarks and heavy mesons in a hot medium, which will in turn help interpret the data from \glspl{hic}. For recent reviews on the extraction of heavy-flavor transport coefficients we refer the reader to Refs.~\cite{Rapp:2018qla,Dong:2019byy}.

\chapter{Exotics in free space within unitarized effective theories}
\label{ch:exoticsinfreespace}

This chapter is devoted to the study of the vacuum properties of exotic hadrons with open heavy flavor within the hadronic molecular description.
The structure of the chapter is the following. Section~\ref{sec:free-theoryremarks} begins with an overview of the symmetries of \gls{qcd} before introducing the basic ideas of the effective theories that will be at the core of the study of baryonic and mesonic molecular systems. Some remarks on unitarization in coupled channels and the analytic structure of the $T$ matrix are also given. The next two sections are devoted to the study of some particular hadronic sectors. In Section~\ref{sec:free-mb} we investigate the dynamical generation of $\Omega_c^{*0}$ (and $\Omega_b^{*-}$) excited states from the interaction of the lowest-lying pseudoscalar and vector mesons with the ground-state baryons in the charm $+1$ (bottom $-1$), strangeness $-2$ and isospin $0$ sector. This work was published in Ref.~\cite{Montana:2017kjw}. In Section~\ref{sec:free-mm} we analyze the interaction of the open heavy-flavor ground-state mesons ($D^{(*)}$ and $D_s^{(*)}$, also $\bar{B}^{(*)}$ and $\bar{B}_s^{(*)}$) with the Goldstone bosons, as well as the excited states that are dynamically generated in these sectors. This analysis, together with the extension to finite temperature presented in Chapter~\ref{ch:hot-medium}, was published in Refs~\cite{Montana:2020lfi,Montana:2020vjg}.

\section{Theory remarks}
\label{sec:free-theoryremarks}

\subsection{Symmetries of QCD}
\label{subsec:free-th-symmetries}

It is well known that symmetries and the breaking of symmetries play a crucial role in physics and in particular in modern particle physics. We may distinguish between discrete and continuous symmetries. According to Noether's theorem, every continuous symmetry under which the Lagrangian (or the Hamiltonian) of a physical system is invariant leads to a conservation law, particularly to a conserved current. The conservation of energy, momentum, and electrical charge are some well-known examples. Hence symmetries introduce restrictions in the choice of the interactions that are allowed for the description of a given physical problem and they often entirely fix the Lagrangian of the theory. In addition, symmetries may be broadly classified into two categories: \textit{global} symmetries involve spacetime independent transformations, while \textit{local} symmetries allow for an arbitrary dependence of the parameters of the transformation on the spacetime location.
Local symmetries are usually called \textit{gauge} symmetries because they characterize gauge theories, such as the local gauge color $\textrm{SU}(3)_c$ symmetry in the case of \gls{qcd}, while global symmetries allow one to classify particles according to quantum numbers. 

The phenomenon of symmetry breaking in physics is as ubiquitous as symmetry itself. Depending on the mechanism breaking the symmetry, one may distinguish three kinds of symmetry breaking: explicit, spontaneous, and anomalous breaking. \textit{Explicit} symmetry breaking happens when the symmetry is not respected by any of the terms of the Lagrangian.  If the explicit symmetry breaking is small, one often says that the would-be symmetry is an approximate one. \textit{Spontaneous} symmetry breaking appears when the Lagrangian obeys the symmetry but the ground state does not. And \textit{anomalous} symmetry breaking or anomalies arise when the symmetry of a classical theory does not remain at the quantum-mechanical level.
The spontaneous breakdown of global symmetries leads to the appearance of massless modes that are known as \textit{Goldstone} (or Nambu-Goldstone) \textit{bosons} \cite{Nambu:1960tm,Goldstone:1961eq}. The $\textrm{SU}(2)$ isospin symmetry and its extension to the $\textrm{SU}(3)$ flavor symmetry are examples of global symmetries of \gls{qcd} that are spontaneously broken, and the pions and the octet of pseudoscalar mesons, consisting of the pions and the kaons, are the corresponding Goldstone bosons, respectively. 

Besides the evident symmetries like Lorentz invariance, $\textrm{SU}(3)_c$ gauge invariance and the discrete $P$ (parity), $\mathcal{C}$ (charge conjugation), and $\mathcal{T}$ (time reversal) symmetries, the Lagrangian $\mathcal{L}_{\textrm{QCD}}$ of Eq.~(\ref{eq:intro-lagrangianQCD}) exhibits symmetries associated to the wide range of the quark masses.

The six quark flavors are usually divided into two groups according to their mass: the light quarks ($u$, $d$, $s$) and the heavy quarks ($c$, $b$, $t$), in comparison with the scale $\Lambda_{\textrm{QCD}}$ separating the perturbative and the nonperturbative regimes of \gls{qcd}:
\begin{equation}
 m_u,\,m_d,\,m_s\ll \Lambda_{\textrm{QCD}} \ll m_c,\,m_b,\,m_t \ .
\end{equation}
The value of $\Lambda_{\textrm{QCD}}$ has been determined empirically to be of the order of a few hundred MeV~\cite{Deur:2016tte} and the values of the current quark masses are shown in Fig.~\ref{fig:intro-StandardModel}.
Furthermore, we recall that the current quark masses are fundamental parameters of the \gls{qcd} Lagrangian of Eq.~(\ref{eq:intro-lagrangianQCD}) and thus one may be tempted to formulate the theory for any value of the quark mass. These considerations motivate the interest in investigating the symmetries of \gls{qcd} in two opposite scenarios: the massless quark limit (chiral limit) and the infinite quark mass limit (heavy-quark limit). These symmetries become approximate in the real world, where quarks have a finite mass, yet they remain very useful to simplify \gls{qcd} interactions in situations where they are well justified. In this section, we briefly discuss some approximate symmetries of \gls{qcd} that are relevant for this thesis.
 
\subsubsection{Light-quark symmetries}
We may assume 
massless quarks. When dealing with hadrons composed of $N_f=3$ light quarks, $q=\{u,d,s\}$, for which $m_q\ll\Lambda_{\textrm{QCD}}$, it is a good approximation to consider the limit of the quark masses going to zero, $m_q\rightarrow 0$.   
Fermionic fields have a \gls{rh} helicity component, with the spin parallel to the momentum, and a \gls{lh} helicity component, for which the spin is antiparallel to the momentum. In the case of massless quarks, the \gls{rh} and \gls{lh} components, defined as $\psi_{\textrm{R},q}=P_{\textrm{R}}\psi_q$ and $\psi_{\textrm{L},q}=P_{\textrm{L}}\psi_q$ in terms of the projection operators $P_{\textrm{R,L}}=\frac12 (1\pm \gamma_5)$, respectively, are completely decoupled from each other.
This is referred to as the \textit{chiral limit} of \gls{qcd}. In this limit the kinetic term in the \gls{qcd} Lagrangian of Eq.~(\ref{eq:intro-lagrangianQCD}) can be written in terms of \gls{rh} and \gls{lh} quark fields,
\begin{equation}\label{eq:free-th-lagrangianChiral}
 \mathcal{L}_{\textrm{QCD}}^0=\sum_q\big(\bar{\psi}_{\textrm{L},q}\ii\slashed{D}\psi_{\textrm{L},q}+\bar{\psi}_{\textrm{R},q}\ii\slashed{D}\psi_{\textrm{R},q}\big) -\frac14 F_{\alpha\beta}F^{\alpha\beta}\ ,
\end{equation}
where we have dropped the color indices for simplicity. 

The Lagrangian of Eq.~(\ref{eq:free-th-lagrangianChiral}) is invariant under separate unitary global transformations of the \gls{rh} and \gls{lh} quarks, known as chiral rotations, which means that it has a $\textrm{U}(3)_\textrm{L}\times\textrm{U}(3)_\textrm{R}$ global symmetry. This symmetry decomposes into $\textrm{SU}(3)_\textrm{L}\times\textrm{SU}(3)_\textrm{R}\times\textrm{U}(1)_\textrm{V}\times\textrm{U}(1)_\textrm{A}$. The axial current associated with the $\textrm{U}(1)_\textrm{A}$ is anomalous and thus it is not conserved, leaving us with a $\textrm{SU}(3)_\textrm{L}\times\textrm{SU}(3)_\textrm{R}\times\textrm{U}(1)_\textrm{V}$ symmetry. The singlet vector current associated with the $\textrm{U}(1)_\textrm{V}$ subgroup is responsible for baryon number conservation, and the remaining $G\equiv\textrm{SU}(3)_\textrm{L}\times\textrm{SU}(3)_\textrm{R}$ is the so-called chiral $\textrm{SU}(3)$ symmetry group. The invariance of the Lagrangian of Eq.~(\ref{eq:free-th-lagrangianChiral}) under independent global $G$ transformations of the \gls{lh} and \gls{rh} components in flavor space:
\begin{equation}
 \psi_{\textrm{L},q}\xrightarrow[]{G}L\psi_{\textrm{L},q} \ ,\qquad \psi_{\textrm{R},q}\xrightarrow[]{G}R\psi_{\textrm{R},q} \ , \qquad L,R\in \textrm{SU}(3)_\textrm{L,R}
\end{equation}
allows one to define the following conserved Noether currents:
\begin{equation}
 J_{\textrm{L}}^{\mu,A}=\bar{\psi}_{\textrm{L}}\gamma^\mu\frac{\lambda^A}{2}\psi_{\textrm{L}} \ , \qquad J_{\textrm{R}}^{\mu,A}=\bar{\psi}_{\textrm{R}}\gamma^\mu\frac{\lambda^A}{2}\psi_{\textrm{R}} \ , \qquad (A=1,...,8) \ .\end{equation}
 
It is common to define linear combinations of the left- and right-handed currents, 
\begin{align}
 J_{\textrm{V}}^{\mu,A}&\ =J_{\textrm{R}}^{\mu,A}+J_{\textrm{L}}^{\mu,A}=\bar{\psi}\gamma^\mu\frac{\lambda^A}{2}\psi \ , \\
 J_{\textrm{A}}^{\mu,A}&\ =J_{\textrm{R}}^{\mu,A}-J_{\textrm{L}}^{\mu,A}=\bar{\psi}\gamma^\mu\gamma_5\frac{\lambda^A}{2}\psi \ ,
\end{align}
which, under a parity transformation, transform as vector and axial-vector currents,
\begin{align}
 J_{\textrm{V}}^{\mu,A}(t,\vec{x})&\ \xrightarrow[]{P}J_{\textrm{V}}^{\mu,A}(t,-\vec{x}) \ , \\
 J_{\textrm{A}}^{\mu,A}(t,\vec{x})&\ \xrightarrow[]{P}-J_{\textrm{A}}^{\mu,A}(t,-\vec{x}) \ ,
 \end{align}
respectively.

Note that one can also consider the case of two massless quarks $u$ and $d$, for which one speaks of chiral $\textrm{SU}(2)$ symmetry.

Of course, in \gls{qcd} the light-quark masses are small but not exactly zero. The quark mass term in Eq.~(\ref{eq:intro-lagrangianQCD}),
\begin{equation}
 \sum_q m_q\bar{\psi}_q\psi_q=\sum_{i,j}\bar{\psi}_{\textrm{R},i}M_{ij}\psi_{\textrm{L},j}+\textrm{h.c.}  \ , \quad M=\textrm{diag}\,(m_u,\, m_d,\,m_s) \ ,
\end{equation}
couples \gls{lh} and \gls{rh} quarks, leading to the explicit breaking of chiral symmetry.
Since $m_u$ and $m_d$ are much smaller than $m_s$, chiral $\textrm{SU}(2)$ symmetry is less badly broken than chiral $\textrm{SU}(3)$ symmetry. 

Furthermore, we might take
the limit in which the $N_f$ quarks have the same mass. For three equal-mass quarks, that is, for $N_f=3$, the \gls{qcd} Lagrangian respects an exact $\textrm{SU}(3)_\textrm{V}\subset\textrm{SU}(3)_\textrm{L}\times\textrm{SU}(3)_\textrm{R}$ symmetry, which is nothing else than the $\textrm{SU}(3)$ flavor symmetry that led Gell-Mann and Ne'eman to the eightfold way \cite{Gell-Mann:1961omu,Gell-Mann:1962yej,Neeman:1961jhl}. The $\textrm{SU}(3)$ flavor symmetry is also approximate since $m_s>m_d\gtrsim m_u$. For $m_u=m_d$ the recovered symmetry in the two-flavor sector is $\textrm{SU}(2)_\textrm{V}\subset\textrm{SU}(3)_\textrm{V}$, which is the known isospin symmetry responsible for hadrons appearing in isospin multiplets. The small splittings within members of the same multiplet are due to the breaking of isospin symmetry caused by the small difference between the up and down quark masses, $m_d-m_u$, and additional electromagnetic effects that are of the same order.

Another crucial aspect of \gls{qcd} is the so-called \textit{spontaneous chiral symmetry breaking}. The chiral symmetry of the Lagrangian is not a symmetry of the ground state of the system, the \gls{qcd} vacuum. This is expected from the fact that hadrons do not appear in degenerate parity doublets. When the Lagrangian of a theory is invariant under the group of transformations $G$ but the vacuum is not, then the particle spectrum does not manifest the symmetry of $G$ but that of a certain subgroup $H\subset G$. 
In the case of \gls{qcd}, the spontaneous breakdown of chiral symmetry results in the $H\equiv\textrm{SU}(3)_\textrm{V}$ subgroup. The nonvanishing quark condensate $\langle0|\bar{\psi}_f\psi_f|0\rangle=\langle0|\bar{\psi}_{\textrm{R},q}\psi_{\textrm{L},q}+\bar{\psi}_{\textrm{L},q}\psi_{\textrm{R},q}|0\rangle\neq 0$ plays the role of an order parameter of the spontaneous symmetry violation.

The appearance of massless bosons is the consequence of the spontaneous symmetry breakdown of $G\rightarrow H$. According to the Goldstone's theorem \cite{Goldstone:1961eq,Goldstone:1962es}, the number of massless bosons is given by the difference between the number of generators of the full symmetry group $G$ and that of the subgroup $H$ that remains unbroken. Hence, from the spontaneous breaking of chiral symmetry one expects $(N_f^2-1)$ massless Goldstone bosons. The members of the isotriplet of pions are good candidates for such bosons for $N_f=2$ as they are considerably lighter than the rest of the hadronic spectra and they have the expected quantum numbers. However, in the limit of massless quarks the mass of the pions would have to be exactly zero.

The massless Goldstone bosons acquire nonzero masses if the symmetry to which they are associated is also explicitly broken in the theory. This is precisely the case of chiral symmetry in \gls{qcd}, which is explicitly broken by the nonzero quark masses as explained above. Then the Goldstone bosons are rather called \textit{pseudo-Goldstone bosons}, although we may simply refer to them as \textit{Goldstone bosons} throughout this dissertation for simplicity.
The explicit breaking of chiral symmetry in \gls{qcd} accounts for the nonzero mass of the pions, as well as for the corresponding eight Goldstone bosons that appear for $N_f=3$ and that are associated with the pseudoscalar octet of mesons containing the pions, the kaons, the antikaons, and the eta meson ($\pi^0,\,\pi^\pm,\,K^0,\,\bar{K}^0,\,K^\pm,\,\eta$). While all their masses are still small in comparison to the proton mass, the mass of the mesons containing strange quarks being larger than that of the pions shows that chiral $\textrm{SU}(3)$ symmetry is less reliable than chiral $\textrm{SU}(2)$ symmetry. 

The chiral $\textrm{SU}(3)_\textrm{L}\times\textrm{SU}(3)_\textrm{R}$ group can be generalized on a theoretical ground to the larger $\textrm{SU}(4)_\textrm{L}\times\textrm{SU}(4)_\textrm{R}$ group. In other words, one can imagine that, in addition to the $u$, $d$ and $s$ quarks, either the charm or the bottom quarks are also massless.
Chiral symmetry is explicitly broken for the $\textrm{SU}(4)$ case due to the large mass of the heavy quarks ($c$ and $b$). Despite this, it is common in hadron physics to rely on this $\textrm{SU}(4)$ symmetry to obtain the interactions involving charmed and bottomed hadrons, yet using the observed heavy-hadron masses to account for the symmetry breakdown. 

Besides, in the same way that $\textrm{SU}(2)$ flavor symmetry can be enhanced to the higher $\textrm{SU}(3)$ flavor symmetry by considering the $s$ quark being degenerate in mass with the $u$ and $d$ quarks, one might think of four mass-degenerate quark flavors, for instance $\{u,d,s,c\}$ or $\{u,d,s,b\}$. Then one has instead an $\textrm{SU}(4)$ flavor symmetry. Indeed $\textrm{SU}(4)$ flavor symmetry can explain the classification of charmed and bottomed hadrons in $\textrm{SU}(4)$ multiplets, yet the comparatively large masses of the $c$- and $b$-quarks strongly break $\textrm{SU}(4)$ flavor symmetry, giving rise to considerably large mass splittings within multiplets.


\subsubsection{Heavy-quark symmetries}
Heavy-quark symmetries appear in the limit of \gls{qcd} where the masses of the heavy quarks, $Q=\{c,b,t\}$, are taken to infinity, $m_Q\rightarrow\infty$. In practice, the quarks that can potentially be treated as heavy are the charm and the bottom, since the top quark decays through the weak interaction too rapidly to form hadronic bound states.

In a hadron containing a single infinitely heavy quark, for instance in a $Q\bar{q}$ meson or a $Qqq$ baryon, the heavy quark can be seen as a static source of a color field. The reason for this is that the momenta of the light degrees of freedom are of the order of $\Lambda_{\textrm{QCD}}\ll m_Q$, and so is the momentum transfer to the heavy quark, $\Delta p$. Therefore, the interactions of the light partons with the heavy quark do not change the kinematics of the heavy quark; that is, the variation in the four-velocity of the heavy hadron, 
\begin{equation}
 \Delta v=\frac{\Delta p}{m_Q}\sim\frac{\Lambda_{\textrm{QCD}}}{m_Q} \ , 
\end{equation}
is negligible.
In the $m_Q\rightarrow\infty$ limit the actual value of the heavy-quark mass becomes irrelevant. For $N_h$ heavy flavors, this leads to a $\textrm{U}(N_h)$ \textit{\gls{hqfs}}:
the dynamics of the hadron are independent of the heavy-quark flavor. 

In addition, the chromomagnetic moment associated with the spin of the quarks is suppressed in the heavy-quark limit, so the interaction of the static heavy quark with gluons does not depend on its spin. This brings us to an $\textrm{SU}(2)$ \textit{\gls{hqss}}:
the dynamics of the hadron do not change if the heavy quark is replaced by another one with a different spin. 

These two heavy-quark symmetries can be combined in a larger $\textrm{SU}(2N_h)$ \gls{hqsfs}.  For two heavy-quark flavors ($c$ and $b$) the corresponding symmetry group is $\textrm{SU}(4)$. In this case, the $c$ and $b$ quarks with spin up, $\uparrow$, and spin down, $\downarrow$, belong to the same multiplet, $\{c\uparrow,\,c\downarrow,\,b\uparrow,\,b\downarrow\}$.

\Gls{hqsfs} is only an approximate symmetry to the extent that $m_c$ and $m_b$ are large but corrections arise from the fact that the $c$ and $b$ quarks are not infinitely heavy. These corrections are of the order of $\Lambda_{\textrm{QCD}}/m_Q$ and hence \gls{hqsfs} is totally justified in the bottom sector ($\Lambda_{\textrm{QCD}}/m_b\sim 0.05$) and good enough in the charm sector ($\Lambda_{\textrm{QCD}}/m_c\sim 0.2$). It is also important to note that, whereas chiral symmetry is a symmetry of the \gls{qcd} Lagrangian of Eq.~(\ref{eq:intro-lagrangianQCD}) in the limit $m_q\rightarrow 0$, the \gls{qcd} Lagrangian does not exhibit heavy-quark symmetries in the $m_Q\rightarrow\infty$, not even in an approximate way. Instead, \gls{hqsfs} is a symmetry of an effective theory of \gls{qcd} that will be described in the section below. The idea is that hadrons with a heavy quark are characterized by a large separation of mass scales, as the heavy-quark mass is much larger than the mass of the light degrees of freedom. After separating the physics associated with these two scales, the long-distance (low-energy) physics may simplify due to the realization of an approximate spin-flavor symmetry.

For reviews on heavy-quark symmetries, see Refs.~\cite{Neubert:1993mb,Shifman:1995dn,Manohar:2000dt}.

\subsection{Effective theories of QCD}
\label{subsec:free-th-effective theories}
We have seen in Section~\ref{subsec:intro-qcd} that, as a result of the asymptotic freedom of \gls{qcd}, the coupling constant $\alpha_s$ at high momenta, or equivalently, at short distances, is small enough for perturbation theory to be applied. Conversely, at long distances, that is, of the order of $\sim 1~\textrm{fm}$, $\alpha_s$ becomes large and perturbative methods are inapplicable. In order to perform quantitative calculations in the latter regime we need other methods like, for instance, \gls{lqcd} or \glspl{eft}.

\Gls{lqcd} has proved to be powerful a tool to numerically solve \gls{qcd} in terms of quarks and gluons in a discretized Euclidean spacetime lattice. However, calculations of \gls{qcd} on the lattice can be extremely computationally demanding and larger-than-physical quark masses are used to speed up the computations, resulting in  pion masses in the lattice ($\sim 200-400~\textrm{MeV}$) that are larger than the physical pion mass ($135$~MeV). Later extrapolations to the physical pion mass are required. In addition, the numerical sign problem makes the extension of the computations to finite density very challenging. In any case, \gls{lqcd} has provided precise results and predictions in many sectors and it has a promising future (see Ref.~\cite{FlavourLatticeAveragingGroup:2019iem} for a review of recent lattice results). A more extensive description of \gls{lqcd} methods is given in Chapter~\ref{ch:lattice}.

\Glspl{eft} are a valuable theoretical technique to describe physics in a limited range of scales. \Glspl{eft} rely on the fact that the relevant degrees of freedom depend on the typical energy scale of the problem and they are practical phenomenological tools when the characteristic energy scales are well separated \cite{Weinberg:1978kz,Polchinski:1983gv,Georgi:1993mps,Manohar:1996cq}. This is the case in \gls{qcd}, where the matter fields are separated because the mass of the light quarks ($u$, $d$, $s$) is much smaller than that of the typical hadron scale set by the mass of the proton ($938$~MeV) or the $\rho$ meson ($776$~MeV).

Although far apart in conception and origin, there is a close link between \gls{lqcd} and \glspl{eft}, and expert communities in each of these methods typically work with one another. On the one hand, \glspl{eft} are useful for lattice calculations to help make extrapolations to extract physical results, as well as to understand the physics of the simulations and artifacts caused by the discrete lattice. On the other hand, \glspl{eft} require \gls{lqcd} to compute the values of the parameters directly from the underlying dynamics.
In this thesis, we adopt the point of view of \glspl{eft} and we employ those that are the most suitable to deal with the heavy hadrons that are of our interest, as will be explained in the next sections.

The key motivation behind \glspl{eft} is to mimic the theory governing a physical system that would be otherwise intractable at low energies; in some cases, the underlying theory is not even known.  For instance, although we know that quantum mechanics is a more fundamental theory, we use classical mechanics to describe the motion of the Earth around the Sun.
Similarly, strong interactions are formulated in terms of quark and gluon fields through the \gls{qcd} Lagrangian of Eq.~(\ref{eq:intro-lagrangianQCD}), but the corresponding low-energy \gls{eft} provides a description in terms of hadronic asymptotic states. In \glspl{eft}, the low-energy physics, with ``low'' defined in comparison to some energy scale $\Lambda$, is described using only the degrees of freedom that are relevant at the energy scale of interest, that is, those states with $m\ll \Lambda$, while integrating those with $M\gg \Lambda$. Thus, the states that are heavier than a certain energy scale do not appear explicitly in the theory. This is possible because the heavier degrees of freedom decouple at energies lower than their mass and the effect of their inclusion is suppressed by powers of the inverse of their mass.

The procedure was formulated in 1979 by Weinberg in the form of a conjecture that is considered to be the guiding theorem in quantum \glspl{eft} \cite{Weinberg:1978kz}. It states that, for a given set of asymptotic states, a perturbative description in terms of the most general effective Lagrangian, containing all terms compatible with the assumed symmetries, yields the most general possible $S$ matrix, consistent with all fundamental principles of quantum field theory, that is, analyticity\footnote{The $S$ matrix can be analytically continued to complex values of the momenta.}, perturbative unitarity\footnote{The optical theorem is fulfilled at each order in perturbation theory.}, and cluster decomposition\footnote{Two scattering experiments distant in spacetime should not interfere with each other.}, as well as the assumed symmetry principles. 
Such a ``most general'' Lagrangian contains an infinite number of terms, each of them multiplied by a coupling constant. The so-called \glspl{lec} encode information about the heavy states that have been integrated out. While the symmetries of the underlying theory impose restrictions on the structure of the operators, they do not determine the values of the \glspl{lec}, which are a priori unknown. Ideally, these should be calculable from the underlying theory, but this is not yet the case in \gls{qcd} and they have to be fitted to experimental data, when available, or \gls{lqcd} calculations. 

What makes \glspl{eft} a particularly valuable tool is that these terms can be organized according to their relevance in a systematic and consistent way. Then, the importance of the diagrams generated by the interaction terms of this effective Lagrangian can be assessed through a power-counting scheme giving rise to an expansion in powers of energy$/\Lambda$. Since $m/\Lambda\ll 1$, the importance of each consecutive term is less than the previous one and the expansion may be cut at a given order. The first omitted term gives an estimate of the error. Moreover, since low-energy \glspl{eft} are specified by a finite number of \glspl{lec} at a given order in the energy expansion, renormalization has to be performed order by order.


Effective theories are useful in many fields of physics, particularly in theoretical particle and hadron physics.
In the previous section, we have highlighted the importance of symmetries in \gls{qcd}. These come into play in the context of effective theories because one has to be extremely careful with the symmetries when building an appropriate effective theory of \gls{qcd}.
The \glspl{eft} treated in this thesis include \gls{chpt}, as it is the \gls{eft} of \gls{qcd} consistent with the chiral symmetry, and \gls{hqet}, which is a useful effective theory to deal with hadrons that contain one heavy quark. The basics of these two \glspl{eft} are described in some detail below, following essentially Refs.~\cite{Pich:1995bw,Manohar:1996cq,Scherer:2005ri,Petschauer:2020urh}.

\subsubsection{Chiral perturbation theory}
\gls{chpt} is the effective theory of light-quark \gls{qcd} in the low-energy regime and it gets its name from the approximate chiral symmetry of \gls{qcd}, upon which it is built.

The Lagrangian of \gls{chpt} contains as degrees of freedom the octet of Goldstone bosons of the subgroup $\textrm{SU}(3)_\textrm{V}$. These eight fields, $\phi_A$ ($A=1,...,8$), are collected in a unitary $3\times3$ matrix, $U(\phi)$, which under the chiral transformations $(L,R)\in G=\textrm{SU}(3)_{\textrm{L}}\times\textrm{SU}(3)_{\textrm{R}}$ transforms as $U(\phi)\xrightarrow[]{G} LU(\phi)R^\dagger$. Among the different parametrizations that one can take for $U(\phi)$, it is customary to choose
\begin{equation}\label{eq:free-th-paramFields}
 U(\phi)=u(\phi)^2=\exp\Big\{\ii\frac{\sqrt{2}}{f}\Phi\Big\} \ ,
\end{equation}
where $f$ is the pion decay constant in the chiral limit, $f=92.4\,\textrm{MeV}$, and\footnote{The physical $\eta$ and $\eta'$ mesons are mixtures of the $\textrm{SU}(3)$ octet $\eta_8$ and singlet $\eta_1$ states.}
\begin{equation}\label{eq:free-th-matrixPseudoscalars}
 \Phi(x)\equiv\frac{\lambda^A}{\sqrt{2}}\phi_A=
 \begin{pmatrix}
\frac{1}{\sqrt{2}}\pi^0+\frac{1}{\sqrt{6}}\eta_8 & \pi^+ & K^{+} \\
\pi^- & -\frac{1}{\sqrt{2}}\pi^0+\frac{1}{\sqrt{6}}\eta_8 & K^{0} \\
K^{-} & \bar{K}^{0} & -\frac{2}{\sqrt{6}}\eta_8\\
\end{pmatrix} \ .
\end{equation}

Then the effective chiral Lagrangian involving $U(\phi)$ is given by the most general Lagrangian that is consistent with chiral symmetry. This Lagrangian is organized in separated terms with increasing powers of momentum or, equivalently, with an increasing number of derivatives acting on the fields,
\begin{equation}\label{eq:free-th-expansionChiralLag}
 \mathcal{L}_{\chi}(U)=\sum_{n=0}^\infty\mathcal{L}_{\chi}^{(2n)}(U) \ ,
\end{equation}
where the number in superindex brackets refers to the number of derivatives, which has to be even as required by parity conservation. This Lagrangian consists of an infinite number of operators, so one has to apply a power counting to establish a hierarchy among them. The soft scale of the low-energy expansion is given by the small external momenta and the small masses of the pseudo-Goldstone bosons, while the large scale is associated with a certain hadronic scale $\Lambda_\chi$. Since the \gls{chpt} Lagrangian is expanded in powers of momenta, at low momenta compared to $\Lambda_\chi$, the contribution of higher-order operators involving higher powers of momenta in Eq.~(\ref{eq:free-th-expansionChiralLag}) is suppressed. The power counting that follows from this is
\begin{equation}
 U(\phi)\sim \mathcal{O}(1) \ , \qquad \partial_\mu\sim\mathcal{O}\Bigg(\frac{p}{\Lambda_\chi}\Bigg) \ ,
\end{equation}
where the value of $\Lambda_\chi$ is associated with the mass of lightest \gls{qcd} resonance that is not included in \gls{chpt}, that is, the $\rho$ vector meson,
\begin{equation}
 \Lambda_\chi\sim m_\rho\approx 1\, \textrm{GeV} \ .
\end{equation}

Due to the unitary nature of $U$, $UU^\dagger=\mathbb{1}$, the term with no derivatives gives rise to a trivial interaction that is independent of the Goldstone fields, so one needs at least two derivatives. This means that the Goldstone bosons couple through derivatives, and hence their scattering amplitudes vanish at zero momentum. The lowest-order effective chiral Lagrangian is therefore
\begin{equation}\label{eq:free-th-LOChiralLag}
 \mathcal{L}_\chi^{(2)}=\frac{f^2}{4}\langle\partial_\mu U^\dagger\partial^\mu U\rangle \ ,
\end{equation}
where $\langle \cdots\rangle$ denotes the trace in flavor space. This Lagrangian contains an infinite number of interactions. Indeed, expanding $U$ in terms of $\Phi$ in Eq.~(\ref{eq:free-th-LOChiralLag}),
\begin{equation}
 U=\mathbb{1}+\frac{\sqrt{2}\ii}{f}\Phi-\frac{1}{f^2}\Phi^2-\frac{\sqrt{2}\ii}{3f^3}\Phi^3+\mathcal{O}\Bigg(\frac{\Phi^4}{f^4}\Bigg) \ ,
\end{equation}
 and noting that the $\Phi$ matrix is Hermitian, $\Phi=\Phi^\dagger$, one obtains the kinetic term for the Goldstone bosons plus an infinite number of interacting terms with increasing number of fields,
\begin{equation}\label{eq:free-th-LOChiralLagExpanded}
 \mathcal{L}_\chi^{(2)}=\frac12\langle\partial_\mu\Phi\partial^\mu\Phi\rangle+\frac{1}{12f^2}\langle [\Phi,\partial_\mu\Phi][\Phi,\partial^\mu\Phi]\rangle+\mathcal{O}\Bigg(\frac{\Phi^6}{f^4}\Bigg) \ ,
\end{equation}
where $[\cdot ,\cdot]$ is the commutator. All the interactions among the Goldstone bosons are given in terms of the same coupling $f$. This follows from the common prefactor $f^2/4$ in Eq.~(\ref{eq:free-th-LOChiralLag}), where it is fixed to recover the standard normalized form of the kinetic term, $\frac12\partial_\mu\phi_A\partial^\mu\phi_A$, for a meson $\phi_A$ that belongs to the octet. The invariance of the Lagrangian in Eq.~(\ref{eq:free-th-LOChiralLag}) under $G=\textrm{SU}(3)_{\textrm{L}}\times\textrm{SU}(3)_{\textrm{R}}$ is easily verified:
\begin{equation}
 \langle\partial_\mu U^\dagger\partial^\mu U\rangle\xrightarrow[]{G}\langle R\partial_\mu U^\dagger L^\dagger L\partial^\mu UR^\dagger\rangle=\langle\partial_\mu U^\dagger\partial^\mu U\rangle \ .
\end{equation}

In order to ascertain the physical meaning of the chiral coupling $f$, we need to calculate the chiral Noether currents associated to the invariance of $\mathcal{L}^{(2)}_\chi$ under $\textrm{SU}(3)_{\textrm{L}}\times\textrm{SU}(3)_{\textrm{R}}$. The \gls{lh} and \gls{rh} currents are given by:
\begin{align}
 J_{\textrm{L}}^{\mu}&\ =\ii\frac{f^2}{2}\partial_\mu U^\dagger U \ , \\
 J_{\textrm{R}}^{\mu}&\ =\ii\frac{f^2}{2}\partial_\mu UU^\dagger \ .
\end{align}
Using $\partial_\mu U U^\dagger=-U\partial_\mu U^\dagger$, which follows from differentiating $U U^\dagger=\mathbb{1}$, the vector and axial-vector currents follow,
\begin{align}
 J_{\textrm{V}}^{\mu}&\ =J_{\textrm{R}}^{\mu}+J_{\textrm{L}}^{\mu}=-\ii\frac{f^2}{2}[U,\partial^\mu U^\dagger] \ , \\
  J_{\textrm{A}}^{\mu}&\ =J_{\textrm{R}}^{\mu}-J_{\textrm{L}}^{\mu}=-\ii\frac{f^2}{2}\{U,\partial^\mu U^\dagger\} \ , 
\end{align}
where $\{\cdot,\cdot\}$ is the anti-commutator. After expanding these expressions in terms of the fields, it is easy to see that the matrix element of the axial-vector current evaluated between a one-Goldstone boson state and the vacuum is different from zero:
\begin{equation}
 \langle0|J^{\mu,A}_{\textrm{A}}|\phi^B(p)\rangle=-\sqrt{2}f\langle0|\partial^\mu\phi^A(x)|\phi^B(p)\rangle=-\sqrt{2}f\partial^\mu \textrm{e}^{-\ii p\cdot x}\delta^{AB}\approx \ii\sqrt{2}f p^\mu \ .
\end{equation}
Hence we can conclude that $f$ can be identified with the pion (meson) decay constant in the chiral limit, which has been measured through the charged pion weak decay process, $\pi^+\rightarrow l^+ +\nu_l$, giving $f=f_\pi=92.4\,\textrm{MeV}$ \cite{pdg}. This result was anticipated when introducing a parametrization for the fields in Eq.~(\ref{eq:free-th-paramFields}).

As discussed in Section~\ref{subsec:free-th-symmetries}, chiral symmetry is explicitly broken by the nonzero quark masses. The \gls{chpt} Lagrangian can be extended so as to incorporate spontaneous chiral symmetry breaking. The most general extension of the \gls{qcd} Lagrangian in the chiral limit of Eq.~(\ref{eq:free-th-lagrangianChiral}) is achieved through the so-called \textit{external-field method}. Following the procedure of Gasser and Leutwyler \cite{Gasser:1983yg,Gasser:1984gg}, we consider the coupling of the scalar, pseudoscalar, vector, and axial-vector currents of quarks to the color-neutral external fields given by the Hermitian $3\times 3$ matrices $s(x)$, $p(x)$, $v_\mu(x)$, and $a_\mu(x)$,
\begin{equation}
 \mathcal{L}_{\textrm{QCD}}=\mathcal{L}_{\textrm{QCD}}^0+\bar{\psi}\gamma^\mu(v_\mu+\gamma_5a_\mu)\psi-\bar{\psi}(s-\ii\gamma_5p)\psi \ .
\end{equation}
These external fields fulfill the following chiral transformation relations
\begin{align}\nonumber
 r_\mu\equiv v_\mu+a_\mu&\ \xrightarrow[]{G}R(v_\mu+a_\mu)R^\dagger+\ii R\partial_\mu R^\dagger \ , \\ \nonumber
 l_\mu\equiv v_\mu-a_\mu&\ \xrightarrow[]{G}L(v_\mu+a_\mu)L^\dagger+\ii L\partial_\mu L^\dagger \ , \\ \nonumber
 s+\ii p&\ \xrightarrow[]{G}R(s+\ii p)L^\dagger \ , \\
 s-\ii p&\ \xrightarrow[]{G}L(s-\ii p)R^\dagger \ .
\end{align}
In order to incorporate them to the chiral Lagrangian one has to define the following building blocks:
\begin{align}\label{eq:free-th-buildingbloc1}
 u_\mu&\ =\ii [u^\dagger(\partial_\mu-\ii r_\mu)u-u(\partial_\mu-\ii l_\mu)u^\dagger] \ , \\ \label{eq:free-th-buildingbloc2}
 \chi_\pm&\ =u^\dagger\chi u^\dagger\pm u\chi^\dagger u \ , \\
 \label{eq:free-th-buildingbloc3}
 f_{\mu\nu}^\pm&\ =uf_{\mu\nu}^\textrm{L}u^\dagger\pm u^\dagger f_{\mu\nu}^\textrm{R}u \ ,
\end{align}
where
\begin{equation}
 \chi=2B_0(s+\ii p) 
\end{equation}
is given in terms of the external scalar and pseudoscalar fields and a constant $B_0$ which, like $f$, is not fixed by the symmetry. The external field strength tensors, $f^\pm_{\mu\nu}$, are defined in terms of the \gls{rh} and \gls{lh} external fields,
\begin{equation}
 f_{\mu\nu}^\textrm{R}=\partial_\mu r_\nu-\partial_\nu r_\mu -\ii[r_\mu,r_\nu] \ , \quad
 f_{\mu\nu}^\textrm{L}=\partial_\mu l_\nu-\partial_\nu l_\mu -\ii[l_\mu,l_\nu] \ .
\end{equation}
The covariant derivative of a building block $A$ is given by
\begin{equation}\label{eq:free-th-covDerivative}
 D_\mu A=\partial_\mu A+[\Gamma_\mu,A] \ ,
\end{equation}
with the chiral connection defined as
\begin{equation}\label{eq:free-th-chiralConnection}
 \Gamma_\mu=\frac12\big[u^\dagger(\partial_\mu-\ii r_\mu)u+u(\partial_\mu-\ii l_\mu)u^\dagger] \ .
\end{equation}
In particular, for $U$ and $U^\dagger$ the minimal coupling introduced in the covariant derivative reads
\begin{equation}\label{eq:free-th-covDerivativeU}
 D_\mu U=\partial_\mu U-\ii r_\mu U+\ii U l_\mu \ , \quad D_\mu U^\dagger=\partial_\mu U^\dagger-\ii U^\dagger r_\mu+\ii l_\mu U^\dagger \ .
\end{equation}

In the presence of external fields, the lowest-order effective chiral Lagrangian in Eq.~(\ref{eq:free-th-LOChiralLag}) turns into \cite{Gasser:1984gg}
\begin{equation}\label{eq:free-th-LOChiralLagExternalFields}
 \mathcal{L}_\chi^{(2)}=\frac{f^2}{4}\langle D_\mu U^\dagger D^\mu U+U^\dagger \chi+\chi^\dagger U\rangle ,
\end{equation}
which is invariant under chiral symmetry, Lorentz transformations, $\mathcal{C}$ and $P$. We recall that our goal was to use the external-field method to introduce the explicit chiral symmetry breaking, for which one needs to give particular values to the external fields. To reach the expression of the leading chiral Lagrangian for Goldstone with nonvanishing quark masses in the isospin limit with $m=(m_u+m_d)/2$, we take $s(x)=M=\textrm{diag}\,(m,m,m_s)$ and switch the other external fields off, that is, $v^\mu(x)=a^\mu(x)=p(x)=0$: 
\begin{equation}
  \mathcal{L}_\chi^{(2)}=\frac{f^2}{4}\langle \partial_\mu U^\dagger \partial^\mu U\rangle+\frac12B_0f^2\langle U^\dagger M+M^\dagger U\rangle .
\end{equation}
The mass term in this Lagrangian follows from the $\chi$ terms in Eq.~(\ref{eq:free-th-LOChiralLagExternalFields}), and the meson decay constants $f_\pi\ne f_K\ne f_\eta$ contain chiral symmetry breaking effects: $f_\Phi=f\{1+\mathcal{O}(m,m_s)\}$. Upon expanding in powers of $\Phi$,
\begin{equation}\label{eq:free-th-ChiralMassTerm}
 \frac12f^2B_0\langle  M(U+U^\dagger)\rangle=B_0\Bigg\{(2m+m_s)f^2-\langle M\Phi^2\rangle+\frac{1}{6f^2}\langle M\Phi^4\rangle+\mathcal{O}\Bigg(\frac{\Phi^6}{f^4}\Bigg)\Bigg\} \ ,
\end{equation}
and focusing only on the term quadratic in the fields, the Gell-Mann--Oakes--Renner relations arise \cite{Gell-Mann:1968hlm},
\begin{align}\label{eq:free-th-GOR}\nonumber
 m_\pi^2&\ =2mB_0 \ , \\
 m_K^2&\ =(m+m_s)B_0 \ , \\
 m_\eta^2&\ =\frac{2}{3}(m+2m_s)B_0 \ . \nonumber
\end{align}
Using these relations, the mass matrix $\chi$ can be written as
\begin{equation}\label{eq:free-th-massmatrix}
 \chi=\begin{pmatrix}
      m_\pi^2 & 0 & 0 \\ 
      0 & m_\pi^2 & 0 \\
      0 & 0 & 2m_K^2-m_\pi^2 
      \end{pmatrix} \ .
\end{equation}

One can easily see that \gls{lec} $B_0$ is related to the chiral quark condensate through \cite{Scherer:2005ri}
\begin{equation}
 f^2B_0=-\frac13\langle\bar{\psi}_q\psi_q\rangle \ .
\end{equation}
Furthermore, the Gell-Mann--Okubo \cite{Gell-Mann:1962yej,Okubo:1961jc} mass relation,
\begin{equation}
 4m_K^2=3m_\eta^2+2m_\pi^2 \ .
\end{equation}
which is independent of the value of $B_0$, follows directly from Eqs.~(\ref{eq:free-th-GOR}). Without a numerical value of $B_0$, the quark masses $m$ and $m_s$ cannot be extracted from the pseudoscalar meson masses, only their ratios.

It is important to note that the external-field method can also be used to systematically incorporate the electromagnetic and semileptonic weak interactions to the chiral effective Lagrangian through appropriate external vector and axial-vector fields, upon the identification of the \gls{lh} and \gls{rh} external fields with the external photon $A_\mu$ and $W$-boson fields,
\begin{equation}\label{eq:free-th-extFieldsPhotonW}
 r_\mu=e\mathcal{Q}A_\mu + \cdots \ , \quad l_\mu=e\mathcal{Q}A_\mu+\frac{e}{\sqrt{2}\theta_W}(W_\mu^\dagger T_++\textrm{h.c.})+\cdots \ .
\end{equation}
In these definitions $\mathcal{Q}$ denotes the quark-charge matrix, $\mathcal{Q}=\frac13\textrm{diag}(2,-1,-1)$, $\theta_W$ is the electroweak angle, and 
\begin{equation}
 T_+=\begin{pmatrix}
      0 & V_{ud} & V_{us} \\
      0 & 0 & 0 \\
      0 & 0 & 0 \\
     \end{pmatrix} \ ,
\end{equation}
where $V_{ij}$ are the Cabibbo-Kobayashi-Maskawa (or CKM) matrix elements.
For further details, see for example Refs.~\cite{Pich:1995bw,Scherer:2005ri}.


\subsubsection{Chiral perturbation theory with baryons}
So far we have considered an \gls{eft} for the interaction of Goldstone bosons among themselves and with external fields. One may be interested in extending this theory to also describe the dynamics of the ground-state baryons at low energies. 

The baryons of the octet ($J^P=\frac12^+$) are described by four-component Dirac spinor fields that can be arranged in a traceless $3\times3$ matrix in flavor space,
\begin{equation}
 B(x)\equiv\frac{\lambda^A}{\sqrt{2}}B_A=
 \begin{pmatrix}
\frac{1}{\sqrt{2}}\Sigma^0+\frac{1}{\sqrt{6}}\Lambda & \Sigma^+ & p \\
\Sigma^- & -\frac{1}{\sqrt{2}}\Sigma^0+\frac{1}{\sqrt{6}}\Lambda & n \\
\Xi^{-} & \Xi^{0} & -\frac{2}{\sqrt{6}}\Lambda \\
\end{pmatrix} \ .
\end{equation}
In contrast to the mesonic case of Eq.~(\ref{eq:free-th-matrixPseudoscalars}), the matrix of baryon fields $B(x)$ is not Hermitian ($B\ne B^\dagger$). It transforms under the chiral symmetry group $G=\textrm{SU}(3)_{\textrm{L}}\times\textrm{SU}(3)_{\textrm{R}}$ as
\begin{equation}
 B\xrightarrow[]{G}KBK^\dagger \ ,
\end{equation}
with the compensator field $K(L,R,U)$ that depends on $(L,R)\in G$ and the meson matrix $U$. Its definition follows from the transformation of $\sqrt{U}$, $u=\sqrt{U}\rightarrow\sqrt{RUL^\dagger}=: RuK^{-1}=KuL^\dagger$, giving $K(L,R,U) =\sqrt{LU^\dagger R^\dagger}R\sqrt{U}$.

The chiral effective Lagrangian involving baryons can be organized in terms according to the number of baryon fields:
\begin{equation}\label{eq:free-th-ChiralLagBaryons}
 \mathcal{L}_{\chi}(U,B)=\mathcal{L}_{\chi}(U)+\mathcal{L}_{\chi,B}(U,B)+\mathcal{L}_{\chi,BB}(U,B)+\mathcal{L}_{\chi,BBB}(U,B)+\cdots \ ,
\end{equation}
where $\mathcal{L}_{\chi}(U)$ is the purely mesonic Lagrangian of Eq.~(\ref{eq:free-th-expansionChiralLag}). Each contribution to the Lagrangian of Eq.~(\ref{eq:free-th-ChiralLagBaryons}) is organized in separated terms with an increasing number of derivatives, but contrary to the mesonic sector, where only even powers of the momentum where allowed, in the baryonic sector odd powers are possible:
\begin{equation}
 \mathcal{L}_{\chi}(U,B)=\sum_{n=0}^{\infty}\mathcal{L}_{\chi}^{(n)}(U,B) \ .
\end{equation}
In order to give an expression of the above Lagrangian up to a given order in the chiral expansion, one has to consider the power counting rules for the building blocks defined in Eqs.~(\ref{eq:free-th-buildingbloc1})--(\ref{eq:free-th-buildingbloc3})~\cite{Krause:1990xc},
\begin{equation}
 u_\mu \sim\mathcal{O}(p) \ , \quad f_{\mu\nu}^\pm,\,\chi_\pm\sim\mathcal{O}(p^2) \ ,
\end{equation}
for the baryon fields,
\begin{equation}\label{eq:free-th-transbaryon}
 B,\,\bar{B},\,D_\mu B\sim\mathcal{O}(1) \ , \quad (-\ii\slashed{D}-M_B)B\sim\mathcal{O}(p) \ ,
\end{equation}
and for the baryon bilinears $\bar{B}\Gamma B$,
\begin{equation}
 \bar{B}B,\, \bar{B}\gamma_\mu B,\,\bar{B}\gamma_5\gamma_\mu B,\,\bar{B}\sigma_{\mu\nu}B\sim\mathcal{O}(1) \ , \quad \bar{B}\gamma_5 B\sim\mathcal{O}(p) \ .
\end{equation}
The chiral covariant derivative of the baryon field has the same form as that defined for the building blocks $A$ in Eq.~(\ref{eq:free-th-covDerivative}) and transforms in the same way as $B$ under chiral transformations, $D_\mu B\xrightarrow[]{G}K(D_\mu A)K^\dagger$. The constant $M_B$ is the mass of the baryons of the octet in the chiral limit, and its large value is responsible for the time derivative of the field $B$ not being counted as small in Eq.~(\ref{eq:free-th-transbaryon}), where only the baryon three-momenta are small. With this, the explicit form of the leading-order chiral effective Lagrangian with at most two baryons is
\begin{equation}\label{eq:free-th-LOChiralLagBaryons}
 \mathcal{L}_{\chi,BB}=\langle\bar{B}(\ii\slashed{D}-M_B)B\rangle+\frac{D}{2}\langle\bar{B}\gamma^\mu\gamma_5\{u_\mu,B\}\rangle+\frac{F}{2}\langle\bar{B}\gamma^\mu\gamma_5[u_\mu,B]\rangle \ ,
\end{equation}
with
\begin{align}
 D_\mu B&\ =\partial_\mu B+[\Gamma_\mu,B] \ , \\ \label{eq:free-th-LOChiralLagBaryonsGammamu}
 \Gamma_\mu&\ =\frac12(u^\dagger\partial_\mu u+u\partial_\mu u^\dagger) \ , \\ \label{eq:free-th-LOChiralLagBaryonsumu}
 u_\mu&\ =\ii(u^\dagger\partial_\mu u-u\partial_\mu u^\dagger) \ .
\end{align}
The \glspl{lec} $D$ and $F$ are the axial-vector coupling constants and their sum is related to the axial-vector coupling constant of nucleons, $g_A=D+F=1.27$.

In particular, the $\mathcal{B}\mathcal{B}\mathcal{P}\mathcal{P}$ vertex for the interaction between \glspl{pseudoscalar} and \glspl{baryon} arises from the term with the covariant derivative. After expanding the meson fields, one gets
\begin{equation}
 \mathcal{L}_{\mathcal{B}\mathcal{B}\mathcal{P}\mathcal{P}}=\frac{\ii}{4f^2}\langle\bar{B}\gamma_\mu\big[[\Phi,\partial_\mu\Phi],B\big]\rangle \ .
\end{equation}
Moreover, the $D$ and $F$ terms provide the Lagrangian for the $\mathcal{B}\mathcal{B}\mathcal{P}$ vertex:
\begin{equation}
 \mathcal{L}_{\mathcal{B}\mathcal{B}\mathcal{P}}=-\frac{D}{\sqrt{2}f}\langle\bar{B}\gamma^\mu\gamma_5\{\partial_\mu\Phi,B\}\rangle-\frac{F}{\sqrt{2}f}\langle\bar{B}\gamma^\mu\gamma_5[\partial_\mu\Phi,B]\rangle \ . 
\end{equation}

\subsubsection{Vector-meson fields within the hidden-gauge formalism}
Different alternative schemes have been proposed to incorporate vector mesons in chiral effective Lagrangians. These approaches differ in the kind of fields used (vector fields $V_\mu$ or antisymmetric tensor fields $V_{\mu\nu}$), their transformation properties under chiral symmetry (linear or nonlinear), and whether they are gauge bosons of a certain symmetry or not (see \cite{Birse:1996hd} for a review). In general, they are based on the phenomenological ideas of \gls{vmd} and universal coupling \cite{1969CurrentsAM,Alfaro1975CurrentsIH}. The \gls{vmd} model was developed by Sakurai in the 1960s \cite{Sakurai:1960ju} to describe the interactions between energetic photons and hadrons through the coupling of the first to neutral vector mesons ($\rho^0,\,\omega,\,\phi$), which have the same quantum numbers as the photon ($J^{PC}=1^{--}$). 

``Massive Yang-Mills'' \cite{Gasiorowicz:1969kn,Kaymakcalan:1983qq,Meissner:1987ge} and ``hidden gauge'' theories \cite{Bando:1984ej,Bando:1987br} are among the more used approaches. Let us introduce in some more detail the hidden-gauge formalism, as it will be used later on in this dissertation to describe the interaction between vector mesons and pseudoscalar mesons, between vector mesons and baryons, and among vector mesons themselves.

In the hidden-gauge, formalism vector mesons are introduced as the gauge bosons of a hidden local symmetry, the so-called \gls{hgs}. This artificial symmetry has no physics associated with it and it can be removed by fixing the gauge \cite{Georgi:1989xy}. Taking the unitary gauge, this symmetry reduces to a nonlinear realization of chiral symmetry \cite{Weinberg:1968de}, under which vector mesons transform inhomogeneously. Following the procedure in Ref.~\cite{Ecker:1988te}, the Lagrangian involving pseudoscalar mesons, photons, and vector mesons can be written as \cite{Nagahiro:2008cv}
\begin{equation}\label{eq:free-th-LagHiddenGauge}
 \mathcal{L}_{\textrm{HGS}}=\mathcal{L}_\chi^{(2)}+\mathcal{L}_{III} \ ,
\end{equation}
where $\mathcal{L}_\chi^{(2)}$ is the lowest-order chiral Lagrangian in the presence of external fields of Eq.~(\ref{eq:free-th-LOChiralLagExternalFields}) and 
\begin{equation}
 \mathcal{L}_{III}=-\frac14\langle V_{\mu\nu}V^{\mu\nu}\rangle+\frac12 m_V^2\left\langle\left(V_\mu-\frac{\ii}{g}\Gamma_\mu\right)^2\right\rangle \ .
\end{equation}
Here, $V_\mu$ is the $\textrm{SU}(3)$ matrix containing the vector meson fields,
\begin{equation}\label{eq:free-th-matrixVectors}
V_\mu=
 \begin{pmatrix}
  \frac{1}{\sqrt{2}}\rho^0+\frac{1}{\sqrt{2}}\omega & \rho^+ & K^{*+} \\
  \rho^- & -\frac{1}{\sqrt{2}}\rho^0+\frac{1}{\sqrt{2}}\omega & K^{*0} \\
  K^{*-} & \bar{K}^{*0} & \phi \\
 \end{pmatrix}_\mu \ ,
\end{equation}
which have a polarization vector $\epsilon_\mu(p)$. 
The tensor $V_{\mu\nu}$ is defined as
\begin{equation}
 V_{\mu\nu}=\partial_\mu V_\nu-\partial_\nu V_\mu-\ii g[V_\mu,V_\nu] \ .
\end{equation}
The covariant derivative in $\mathcal{L}_\chi^{(2)}$ provides the coupling of the pseudoscalars to the photon field through the definitions in Eqs.~(\ref{eq:free-th-covDerivativeU}) and (\ref{eq:free-th-extFieldsPhotonW}),
\begin{equation}
 D_\mu U=\partial_\mu U-\ii e\mathcal{Q}A_\mu U+\ii eU\mathcal{Q}A_\mu \ ,
\end{equation}
while the chiral connection in $\mathcal{L}_{III}$ reads
\begin{equation}\label{eq:free-th-chiralConnectionPhoton}
 \Gamma_\mu=\frac12[u^\dagger(\partial_\mu-\ii e \mathcal{Q}A_\mu)u+u(\partial_\mu-\ii e\mathcal{Q}A_\mu)u^\dagger] \ ,
\end{equation}
where $U=u^2$ is the usual matrix containing the Goldstone bosons of Eq.~(\ref{eq:free-th-matrixPseudoscalars}). The hidden-gauge coupling constant $g$ is related to the constant $f$ and the vector-meson mass $m_V$ through the relation
\begin{equation}\label{eq:free-th-KSFRrelation}
 g=\frac{m_V}{2f} \ ,
\end{equation}
which fulfills the Kawarabayashi-Suzuki-Fayyazuddin-Riazuddin (or KSFR) rule \cite{Kawarabayashi:1966kd,Riazuddin:1966sw} of \gls{vmd}.

After combining the expansions in powers of $\Phi$ in Eqs.~(\ref{eq:free-th-LOChiralLagExpanded}) and (\ref{eq:free-th-ChiralMassTerm}), it is straightforward to see that $\mathcal{L}_\chi^{(2)}$ gives rise to the lowest-order chiral Lagrangian involving four pseudoscalar mesons,
\begin{equation}\label{eq:free-th-LagHG_PPPP}
 \mathcal{L}_{\mathcal{P}\mathcal{P}\mathcal{P}\mathcal{P}}=\frac{1}{12f^2}\langle[\Phi,\partial_\mu\Phi][\Phi,\partial^\mu\Phi]+\chi\Phi^4\rangle \ .
\end{equation}
The Lagrangian $\mathcal{L}_\chi^{(2)}$ also contains the coupling between two pseudoscalars and a photon ($\gamma$),
\begin{equation}\label{eq:free-th-LagHG_gPP}
 \mathcal{L}_{\gamma\mathcal{P}\mathcal{P}}=-\ii eA^\mu\langle\mathcal{Q}[\Phi,\partial_\mu\Phi]\rangle \ .
\end{equation}
Similarly, one can derive the following Lagrangians involving pseudoscalar and \glspl{vector}, as well as photons, from $\mathcal{L}_{III}$ \cite{Nagahiro:2008cv}:
\begin{align}\label{eq:free-th-LagHG_Vg}
\mathcal{L}_{\mathcal{V}\gamma}&\ =-m_V^2\frac{e}{g}A_\mu\langle V^\mu\mathcal{Q}\rangle \ , \\ \label{eq:free-th-LagHG_VgPP}
\mathcal{L}_{\mathcal{V}\gamma\mathcal{P}\mathcal{P}}&\ =e\frac{m_V^2}{4g f^2}A_\mu\langle V^\mu (\mathcal{Q}\Phi^2+\Phi^2\mathcal{Q}-2\Phi\mathcal{Q}\Phi)\rangle \ , \\ \label{eq:free-th-LagHG_VPP}
 \mathcal{L}_{\mathcal{V}\mathcal{P}\mathcal{P}}&\ =-\ii \frac{m_V^2}{4gf^2}\langle V^\mu[\Phi,\partial_\mu\Phi]\rangle \ , \\ \label{eq:free-th-LagHG_gPP2}
\tilde{\mathcal{L}}_{\gamma\mathcal{P}\mathcal{P}}&\ =\ii eA^\mu\langle\mathcal{Q}[\Phi,\partial_\mu\Phi]\rangle \ , \\ \label{eq:free-th-LagHG_PPPP2}
 \tilde{\mathcal{L}}_{\mathcal{P}\mathcal{P}\mathcal{P}\mathcal{P}}&\ =-\frac{1}{8f^2}\langle[\Phi,\partial_\mu\Phi][\Phi,\partial^\mu\Phi]\rangle \ , \\ 
 \mathcal{L}_{\mathcal{V}\mathcal{V}\mathcal{V}\mathcal{V}}&\ =\frac{g^2}{2}\langle V_\mu V_\nu V^\mu V^\nu-V_\nu V_\mu V^\mu V^\nu\rangle \ , \\
 \mathcal{L}_{\mathcal{V}\mathcal{V}\mathcal{V}}&\ =\ii g\langle(\partial_\mu V_\nu-\partial_\nu V_\mu)V^\mu V^\nu\rangle \ .
\end{align}
The Lagrangians in Eqs.~(\ref{eq:free-th-LagHG_gPP}) and (\ref{eq:free-th-LagHG_gPP2}) cancel each other. Hence, pseudoscalars do not couple to photons directly but through vector-meson exchange, as a consequence of \gls{vmd}. In addition, the Lagrangian in Eq.~(\ref{eq:free-th-LagHG_PPPP2}) has the same structure as the first term of the chiral Lagrangian in Eq.~(\ref{eq:free-th-LagHG_PPPP}). Nevertheless, the contact term between four pseudoscalars in $\mathcal{L}_{III}$ is canceled by the term resulting from the exchange of a vector meson between two pseudoscalars that is built using $\mathcal{L}_{\mathcal{V}\mathcal{P}\mathcal{P}}$, in the limit $q^2/m_V^2\rightarrow 0$, where $q$ is the momentum carried by the exchanged vector meson. This way chiral symmetry is preserved \cite{Weinberg:1968de}. Besides the Lagrangians listed above, one could expect also a term for the $\mathcal{V}\mathcal{V}\mathcal{P}$ vertex. However this vertex is anomalous \cite{Wess:1967jq,Meissner:1987ge,Pallante:1992qe}, as it violates parity, and its contribution is usually small.

The hidden-gauge approach can be extended to incorporate baryons.
We have seen that the second and third terms of $\mathcal{L}_{\chi,BB}$ in Eq.~(\ref{eq:free-th-LOChiralLagBaryons}) give rise to the Lagrangian coupling the pseudoscalar mesons with baryons through the $\mathcal{B}\mathcal{B}\mathcal{P}$ vertex. By considering the electromagnetic part of the building block
\begin{equation}\label{eq:free-th-umu}
 u_\mu=-\frac{\sqrt{2}}{f}(\partial_\mu\Phi-\ii e[\mathcal{Q},\Phi]A_\mu) \ ,
\end{equation}
one can see that Eq.~(\ref{eq:free-th-LOChiralLagBaryons}) also provides the Lagrangian coupling the photon with the baryons and a pseudoscalar meson, i.e. the so-called Kroll-Ruderman term,
\begin{equation}
 \mathcal{L}_{\mathcal{B}\mathcal{B}\gamma\mathcal{P}}=\frac{\ii e}{\sqrt{2}f}\bigg(D\langle\bar{B}\gamma^\mu\gamma_5\big\{[\mathcal{Q},\Phi],B\big\}\rangle+F\langle\bar{B}\gamma^\mu\gamma_5\big[[\mathcal{Q},\Phi],B\big]\rangle\bigg)A_\mu \ .
\end{equation}

In Refs.~\cite{Klingl:1996by,Klingl:1997kf}\footnote{The normalizations that we use for the $\Phi$, $V_\mu$ and $u_\mu$ matrices differ from those in Refs. \cite{Klingl:1996by,Klingl:1997kf}.} the interaction between vector mesons and baryons was introduced through the minimal coupling scheme\footnote{Also the prescription taken for the minimal coupling is different from that in \cite{Klingl:1997kf}.}, in which we substitute $e\mathcal{Q}A^\mu$ by the vector field $gV^\mu$. With this, Eq.~(\ref{eq:free-th-umu}) becomes
\begin{equation}\label{eq:free-th-umu2}
 u_\mu=-\frac{\sqrt{2}}{f}\bigg(\partial_\mu\Phi+\ii {g}[V_\mu,\Phi]\bigg) \ ,
\end{equation}
and the Kroll-Ruderman term leads to the $\mathcal{B}\mathcal{B}\mathcal{V}\mathcal{P}$ Lagrangian,
\begin{equation}
 \mathcal{L}_{\mathcal{B}\mathcal{B}\mathcal{V}\mathcal{P}}=\frac{\ii}{2f}\bigg(D\langle\bar{B}\gamma^\mu\gamma_5\big\{[V_\mu,\Phi],B\big\}\rangle+F\langle\bar{B}\gamma^\mu\gamma_5\big[[V_\mu,\Phi],B\big]\rangle\bigg) \ .
\end{equation}

The authors of Ref.~\cite{Klingl:1997kf} also derived the direct coupling of the photon to the baryons:
\begin{equation}\label{eq:free-th-LagBBg}
 \mathcal{L}_{\mathcal{B}\mathcal{B}\gamma}=e\big(\langle\bar{B}\gamma_\mu[\mathcal{Q},B]\rangle +\langle\bar{B}\gamma_\mu B\rangle\langle \mathcal{Q}\rangle \big)A^\mu \ .
\end{equation}
The first term of this Lagrangian follows from the term with the covariant derivative in Eq.~(\ref{eq:free-th-LOChiralLagBaryons}) and the electromagnetic part of $\Gamma_\mu$ (see Eq.~(\ref{eq:free-th-chiralConnectionPhoton})).
The second term in the Lagrangian of Eq.~(\ref{eq:free-th-LagBBg}) involves the $\textrm{SU}(3)$ singlet part of the nonet of vector mesons. 

In the limit in which the number of colors $N_c$ in \gls{qcd} is taken to be large, the $U(1)_A$ anomaly is absent. In this situation, there are nine Goldstone bosons associated with the spontaneous chiral symmetry breaking of $\textrm{U}(3)_{\textrm{L}}\times\textrm{U}(3)_{\textrm{R}}$ to $\textrm{U}(3)_{\textrm{V}}$. They are collected in a $3\times 3$ unitary matrix $\tilde{U}(\phi)=\langle 0|\tilde{U}|0\rangle\exp\{\ii\sqrt{2}\tilde{\Phi}/f\}$, with $\tilde{\Phi}=\Phi+\eta_1\mathbb{1}/\sqrt{3}$. See Refs.~\cite{Pich:1991fq,Pich:1995bw} for details. As a result, an  additional singlet piece has to be added to the lowest-order Lagrangian of Eq.~(\ref{eq:free-th-LOChiralLagBaryons}) for the interaction of the baryons with the Goldstone bosons,
\begin{equation}
 \mathcal{L}_{\chi,BB}+\mathcal{L}_{\chi,BB}^{(1)}=\mathcal{L}_{\chi,BB}+g_S\langle\tilde{\xi}_\mu\rangle\langle\bar{B}\gamma^\mu\gamma_5B\rangle \ ,
\end{equation}
where $\xi_\mu$ replaces $u_\mu$ for large $N_c$. 

The singlet term in Eq.~(\ref{eq:free-th-LagBBg}) vanishes because the trace of the quark-charge matrix is exactly zero. However, in the case of the Lagrangian coupling the vector meson to the baryons obtained from the replacement $e\mathcal{Q}A^\mu\rightarrow gV^\mu$, the contribution from the singlet of the vector fields survives and we have
\begin{equation}\label{eq:free-th-LagBBV}
 \mathcal{L}_{\mathcal{B}\mathcal{B}\mathcal{V}}=g\big(\langle\bar{B}\gamma_\mu[V^\mu,B]\rangle +\langle\bar{B}\gamma_\mu B\rangle\langle V^\mu\rangle \big) \ .
\end{equation}

Similarly, one can also incorporate the interactions with the baryons of the decuplet to the effective chiral Lagrangians \cite{Jenkins:1991es}.

\subsubsection{Chiral perturbation theory with heavy hadrons}
The next step that one shall take is to incorporate hadrons that contain a single heavy quark $Q$. This step is especially relevant because in the next sections we will be concerned with the interaction of heavy mesons and heavy baryons with light mesons. Because of the large separation of mass scales between the heavy quark and the light constituents (quarks, antiquarks, and gluons), one can take advantage of heavy-quark symmetries. In this context, \gls{hqet} provides us with a convenient tool to separate the physics associated with the two scales \cite{Eichten:1989zv,Georgi:1990um}. 
As we noted before, the typical momentum exchange between the heavy quark and the light degrees of freedom is of the order of $\Lambda_{\textrm{QCD}}$. Since $m_Q\gg \Lambda_{\textrm{QCD}}$, the heavy quark is close to its mass shell ($p_Q^2=m_Q^2$) and its momentum can be decomposed into an on-shell part $m_Qv$ and a small deviation from the mass shell $|k|\ll m_Q$,
\begin{equation}
 p_Q^\mu=m_Qv^\mu+k^\mu \ ,
\end{equation}
where $v^\mu$ is the four-velocity of the heavy quark and satisfies $v^2=1$. The heavy-quark propagator reads
\begin{equation}
 \frac{\ii}{\slashed{p}-m_Q+\ii\varepsilon}=\frac{\ii(\slashed{p}+m_Q)}{p^2-m_Q^2+\ii\varepsilon}=\frac{\ii[m_Q(1+\slashed{v})+\slashed{k}]}{k^2+2m_Qv\cdot k+\ii\varepsilon}\xrightarrow[]{m_Q\gg k}\frac{1+\slashed{v}}{2}\frac{\ii}{v\cdot k+\ii\varepsilon} \ ,
\end{equation}
and it is independent of the heavy-quark mass. In addition, the heavy-quark field $\psi_Q$ can be decomposed into a ``large'' component $h_v$ and a ``small'' component $H_v$,
\begin{equation}
 \psi_Q(x)=e^{-\ii m_Qv\cdot x}(h_v(x)+H_v(x)) \ ,
\end{equation}
with
\begin{equation}
 h_v=e^{\ii m_Qv\cdot x} \mathcal{P}^v_+\psi_Q \ , \quad H_v= e^{\ii m_Qv\cdot x} \mathcal{P}^v_-\psi_Q \ ,
\end{equation}
where the action of the projectors $\mathcal{P}^v_\pm=(1\pm\slashed{v})/2$ on $\psi_Q(x)$ is given by
\begin{equation}
 \mathcal{P}^v_+\psi_Q(x)=\psi_Q(x)+\mathcal{O}\bigg(\frac{1}{m_Q}\bigg) \ , \quad \mathcal{P}^v_-\psi_Q(x)=0+\mathcal{O}\bigg(\frac{1}{m_Q}\bigg) \ .
\end{equation}
Introducing these expressions into the \gls{qcd} Lagrangian for a single heavy quark one gets
\begin{align}
 \mathcal{L}_{\textrm{QCD}}^Q &\ =\bar{\psi}_Q(\ii\slashed{D}-m_Q)\psi_Q \\ \nonumber &\ =\bar{h}_v\ii v\cdot Dh_v-\bar{H}_v[\ii v\cdot D-2m_Q]H_v+\bar{h}_v\ii\slashed{D}_\perp H_v+\bar{H}_v\ii\slashed{D}_\perp h_v \ ,
\end{align}
where we have introduced the perpendicular component of the covariant derivative $D^\mu_\perp=D^\mu-(v\cdot D) v^\mu$. We can see that this Lagrangian contains a massless field $h_v$ and a heavy field $H_v$ with a mass $2m_Q$.  The latter can be integrated out of the theory in the limit $m_Q\rightarrow\infty$. Using
\begin{equation}
 (\ii v\cdot D+2m_Q)H_v=\ii\slashed{D}_\perp h_v  \quad \rightarrow\quad H_v=\frac{1}{\ii v\cdot D+2m_Q}\ii\slashed{D}_\perp h_v \ ,
\end{equation}
we find the \gls{hqet} Lagrangian \cite{Eichten:1989zv,Georgi:1990um},
\begin{equation}
 \mathcal{L}_{\textrm{HQET}}=\bar{h}_v\ii v\cdot Dh_v-\frac{1}{2m_Q}\bar{h}_v\slashed{D}_\perp^2h_v+\mathcal{O}\bigg(\frac{1}{m_Q^2}\bigg) \ .
\end{equation}
It consists of an expansion in powers of the inverse of the heavy-quark mass.

While the \gls{hqet} approach deals with quark degrees of freedom, one may be interested in building an effective Lagrangian in terms of heavy mesons and heavy baryons.
Let us concentrate on heavy-light mesons and briefly introduce the formalism of the \gls{hmet} \cite{Wise:1992hn,Burdman:1992gh,Casalbuoni:1996pg}.
Heavy-light mesons contain one heavy quark $Q$ ($c$ or $b$) and one light antiquark $\bar{q}_a$ ($\bar{u}$, $\bar{d}$ or $\bar{s}$)\footnote{This means that we are dealing with $D$ (or $c\bar{q}$) and $\bar{B}$ (or $b\bar{q}$) states.}. In the heavy-quark limit, the lowest-lying pseudoscalar ($J^P=0^-$) and vector ($J^P=1^-$) heavy-light mesons form degenerate multiplets:
\begin{equation}
 H_{a}=\begin{cases}
     (D^0,  D^+,  D_s^+ ) \ , \\
     (\bar{B}^-,  \bar{B}^0,  \bar{B}_s^0 ) \ ,
   \end{cases}  \quad 
 H_{a}^{*\mu}=\begin{cases}
     (D^{*0},  D^{*+},  D_s^{*+} )^\mu \ , \\
     (\bar{B}^{*-},  \bar{B}^{*0},  \bar{B}_s^{*0})^\mu      \ .
     \end{cases} 
\end{equation}
 It is customary to combine these fields into a supermultiplet state in the form of a $4\times 4$ matrix $\mathcal{H}_a$, given by
\begin{equation}
 \mathcal{H}_a(v)=\frac{1+\slashed{v}}{2}(H_a^{*\mu}\gamma_\mu-H_a\gamma_5) \ ,
\end{equation}
where $v$ is the four-velocity of the meson, and the polarization vector of the vector-meson field satisfies the condition $v_\mu \epsilon^{\mu}=0$. Under $G\equiv \textrm{SU}(3)_\textrm{L}\times \textrm{SU}(3)_\textrm{R}$ chiral symmetry, $\mathcal{H}_a$ transforms as
\begin{equation}
 \mathcal{H}_a\xrightarrow[]{G} \mathcal{H}_bU_{ba}^\dagger \ ,
\end{equation}
where $U$ is the $3\times 3$ matrix introduced in Eq.~(\ref{eq:free-th-paramFields}) for the Goldstone bosons and the repeated light-flavor index $b$ is summed over $\{1,2,3\}\rightarrow \{u,d,s\}$. Under the heavy-quark spin rotation, it transforms as
\begin{equation}
  \mathcal{H}_a\xrightarrow[]{\textrm{SU}(2)_v} S_v\,\mathcal{H}_a \ ,
\end{equation}
with $S_v\in\textrm{SU}(2)_v$, that is, the \gls{hqss} group boosted by the velocity $v$.

Taking this into account, the interactions of the heavy mesons with the Goldstone bosons are described by a \gls{lo} effective Lagrangian, in both chiral and $1/m_Q$ expansion, satisfying both light-quark chiral symmetry and heavy-quark spin symmetry. This Lagrangian reads \cite{Wise:1992hn,Burdman:1992gh,Yan:1992gz,Goity:1992tp,Boyd:1994pa}
\begin{equation}\label{eq:free-th-LagHMET-LO}
 \mathcal{L}_{\textrm{HMET}}=-\ii v_\mu\langle\bar{\mathcal{H}}_a \nabla^{\mu}_{ba} \mathcal{H}_b\rangle_{\textrm{D}}+\frac{g}{2}\langle \bar{\mathcal{H}_a}\mathcal{H}_b\gamma_\mu\gamma_5u^\mu_{ba}  \rangle_{\textrm{D}} \ ,
\end{equation}
where $\langle\cdots\rangle_{\textrm{D}}$ denotes the trace over Dirac indices, and flavor indices $a$, $b$ are summed over. The covariant derivative of the heavy fields reads
\begin{align}
 \nabla^\mu \mathcal{H}_{a}^{\dagger}&\ =(\partial^\mu+\Gamma^\mu)\mathcal{H}_{a}^{\dagger} \ , \\
 \nabla^\mu \mathcal{H}_{a}&\ =\mathcal{H}_{a} (\overleftarrow{\partial}^\mu+\Gamma^\mu)=(\partial^\mu+\Gamma^{\mu \dagger})\mathcal{H}_a\ , 
\end{align}
and $\Gamma_\mu$ ($\Gamma_\mu^{\dagger}=-\Gamma_\mu$ is the Hermitian conjugate matrix) and $u_\mu$ are given in Eqs.~(\ref{eq:free-th-LOChiralLagBaryonsGammamu}) and (\ref{eq:free-th-LOChiralLagBaryonsumu}), respectively. The Hermitian conjugate field is defined as $\bar{\mathcal{H}}_b=\gamma_0\mathcal{H}_b^\dagger\gamma_0$. In the Lagrangian of Eq.~(\ref{eq:free-th-LagHMET-LO}), the parameter $g$ is a universal coupling constant for the $\Phi HH^*$ and $\Phi H^*H^*$ interactions \cite{Casalbuoni:1996pg,Scoccola:2009au}.

It is important to note that the heavy fields in Eq.~(\ref{eq:free-th-LagHMET-LO}) have dimension $3/2$ because a factor $\sqrt{m_H}$ has been absorbed in their definition.

A more explicit form of the \gls{lo} Lagrangian coupling the heavy pseudoscalar and vector mesons to light mesons is obtained by expanding the superfield $\mathcal{H}_a$ in Eq.~(\ref{eq:free-th-LagHMET-LO}) in terms of $H_a$ and $H_a^{*\mu}$ and taking the Dirac traces. The first term in Eq.~(\ref{eq:free-th-LagHMET-LO}) contains the kinetic terms for the heavy mesons $H_a$ and $H_a^{*\mu}$ as well as the interactions between the heavy mesons and an even number of Goldstone bosons, obtained from the expansion of the chiral connection $\Gamma^\mu$. The interactions with an odd number of Goldstone bosons originate from the second term. Taking $v_\mu=(1,\vec{0}\,)+\mathcal{O}(\vec{p}\,/m_H)$, one can write \cite{Lutz:2007sk,Guo:2009ct,Geng:2010vw}
\begin{align}\nonumber\label{eq:free-th-LagHMET-LO2}
 \mathcal{L}_{\Phi HH^*}^{\textrm{LO}}&\ =\langle\nabla^\mu H\nabla_\mu H^\dagger\rangle-m_H^2\langle HH^\dagger\rangle-\langle\nabla^\mu H^{*\nu}\nabla_\mu H_\nu^{*\dagger}\rangle+m_H^2\langle H^{*\nu}H_\nu^{*\dagger}\rangle \\ &\ +\ii g\langle H^{*\mu} u_\mu H^\dagger-Hu^\mu H_\mu^{*\dagger}\rangle+\frac{g}{2m_H}\langle H_\mu^*u_\alpha\nabla_\beta H_\nu^{*\dagger}-\nabla_\beta H_\mu^* u_\alpha H_\nu^{*\dagger}\rangle\varepsilon^{\mu\nu\alpha\beta} \ ,
\end{align}
where $m_H$ is the mass of the heavy mesons in the chiral limit and $\langle\cdots\rangle$ denotes the trace in flavor space.

The \gls{lo} Lagrangian of the \gls{hmet} of Eq.~(\ref{eq:free-th-LagHMET-LO}) describes the interactions of heavy and light mesons in the limit of combined chiral and heavy-quark symmetries. Indeed, these symmetries are approximate and corrections of order $m_q/\Lambda_{\textrm{QCD}}$ and $\Lambda_{\textrm{QCD}}/m_Q$ can be considered.

The correction terms that take into account chiral symmetry breaking due to nonvanishing light-quark masses read
\begin{align} \nonumber
 \delta\mathcal{L}_{\textrm{HMET}}^{m_q}&\ =-h_0\langle\bar{\mathcal{H}}_a\chi_{+,bb}\mathcal{H}_a\rangle_{\textrm{D}}+h_1\langle\bar{\mathcal{H}}_a\tilde\chi_{+,ba}\mathcal{H}_b\rangle_{\textrm{D}} \\ \nonumber
 &\ +h_2\langle\bar{\mathcal{H}}_a (u_\mu u^\mu)_{bb} \mathcal{H}_a\rangle_{\textrm{D}} +h_3\langle\bar{\mathcal{H}}_a(u_\mu u^\mu)_{ba}\mathcal{H}_b \rangle_{\textrm{D}} \\
 &\ +h_4\langle\bar{\mathcal{H}}_a(-\ii v_\mu)(u^\mu u^\nu)_{bb}(\ii v_\nu)\mathcal{H}_a\rangle_{\textrm{D}}  +h_5\langle\bar{\mathcal{H}}_a(-\ii v_\mu)\{u^\mu, u^\nu\}_{ba}(\ii v_\nu)\mathcal{H}_b \rangle_{\textrm{D}} \ , 
\end{align}
where 
\begin{align}
 \chi_\pm=u^\dagger\chi u^\dagger\pm u\chi^\dagger u \ , \\
 \tilde{\chi}_\pm=\chi_\pm-\frac{1}{3}\chi_{\pm,aa} \ .
\end{align}
The mass matrix $\chi$ is given in Eq.~(\ref{eq:free-th-massmatrix}) and $\chi_{+,aa}$ is the trace of $\chi_+$ in flavor space. The explicit form of the Lagrangian for the interaction of heavy-light mesons with Goldstone bosons at \gls{nlo} in the chiral expansion is given by \cite{Guo:2009ct,Geng:2010vw,Abreu:2011ic,Liu:2012zya,Tolos:2013kva}
\begin{align}\nonumber\label{eq:free-th-LagHMET-NLO}
 \delta\mathcal{L}_{\Phi HH^*}^\textrm{NLO}=&\ -h_0\langle HH^\dagger\rangle\langle\chi_+\rangle+h_1\langle H\chi_+H^\dagger\rangle+h_2\langle HH^\dagger\rangle\langle u^\mu u_\mu\rangle  +h_3\langle Hu^\mu u_\mu H^\dagger\rangle \\ \nonumber
 &\ +h_4\langle\nabla_\mu H\nabla_\nu H^\dagger\rangle\langle u^\mu u^\nu\rangle+h_5\langle\nabla_\mu H\{u^\mu,u^\nu\}\nabla_\nu H^\dagger \rangle \\ \nonumber
 &\ +\tilde{h}_0\langle H^{*\mu}H^{*\dagger}_\mu\rangle\langle\chi_+\rangle-\tilde{h}_1\langle H^{*\mu}\chi_+H^{*\dagger}_\mu\rangle-\tilde{h}_2\langle H^{*\mu}H^{*\dagger}_\mu\rangle\langle u^\nu u_\nu\rangle  -\tilde{h}_3\langle H^{*\mu}u^\nu u_\nu H^{*\dagger}_\mu\rangle\\ 
 &\ -\tilde{h}_4\langle\nabla_\mu H^{*\alpha}\nabla_\nu H^{*\dagger}_\alpha\rangle\langle u^\mu u^\nu\rangle-\tilde{h}_5\langle\nabla_\mu H^{*\alpha}\{u^\mu,u^\nu\}\nabla_\nu H^{*\dagger}_\alpha\rangle \ .
\end{align}

The violation of \gls{hqss} at order $1/m_H$ is introduced by 
\begin{equation}\label{eq:free-th-LagMQ}
 \delta\mathcal{L}_{\textrm{HMET}}^{1/m_Q}=\frac{\lambda}{m_Q}\langle\bar{\mathcal{H}}_a\sigma^{\mu\nu}\mathcal{H}_a\sigma_{\mu\nu}\rangle_{\textrm{D}}-\frac{g_1}{2m_Q}\langle \bar{\mathcal{H}}_a \mathcal{H}_b\gamma_\mu\gamma_5u^\mu_{ba}\rangle_{\textrm{D}}-\frac{g_2}{2m_Q}\langle\bar{\mathcal{H}}_a \gamma_\mu\gamma_5 u^\mu_{ba}\mathcal{H}_b\rangle_{\textrm{D}} \ ,
\end{equation}
where $\sigma^{\mu\nu}$ follows the usual definition in terms of the gamma matrices, $\sigma^{\mu\nu}\equiv(\gamma^\mu\gamma^\nu-\gamma^\nu\gamma^\mu)$. The first term introduces the splitting between the members of the heavy-meson supermultiplet,
\begin{equation}
 m_{H^*}-m_H=\frac{-2\lambda}{m_Q} \ ,
\end{equation}
while the couplings $g_1$ and $g_2$ renormalize the coupling $g$ introduced in Eq.~(\ref{eq:free-th-LagHMET-LO}),
\begin{align}
 g_{\Phi HH^*}=g+\frac{1}{m_Q}(g_1-g_2) \ , \\
 g_{\Phi H^*H^*}=g+\frac{1}{m_Q}(g_1+g_2) \ ,
\end{align}
for the $\Phi HH^*$ and $\Phi H^*H^*$ couplings, respectively.
Since the difference between the heavy-quark and the heavy-meson masses appears at higher orders in the $1/m_Q$ expansion, writing $1/m_Q$ or $1/m_H$ in Eq.~(\ref{eq:free-th-LagMQ}) is equivalent.

In practice, in the calculations described in Section~\ref{sec:free-mm}, the breaking of \gls{hqss} is taken into account by using the physical masses for the pseudoscalar and vector heavy-light mesons.

\subsection{Unitarization in coupled channels and the $T$-matrix formalism}
\label{subsec:free-th-unitarization}
The techniques based on effective Lagrangians such as \gls{chpt} have become a practical tool to address the study of the low-energy interactions between hadrons. The Lagrangians presented in Section~\ref{subsec:free-th-effective theories} consist of a controlled expansion in powers of the external momenta of the hadrons over the hadronic scale $\Lambda_\chi$. With these Lagrangians one can obtain the hadron--hadron interaction amplitude at a given order in the expansion and, thus, the results of the effective theory can be improved by going to higher orders, with the drawback of a large number of free parameters, that is, the \glspl{lec}, that appear in higher-order Lagrangians, and the fact that the amplitudes obtained from these Lagrangians do not satisfy the exact unitarity condition.

Furthermore, up to now, we have not considered the possibility of resonant states, which are very rich phenomena of strong interactions. Resonances appear as poles of the so-called $S$ matrix, and chiral Lagrangians by themselves are not able to reproduce them. Hence, the range of applicability of chiral Lagrangians is limited by the masses of the lowest resonances in each hadron--hadron scattering sector. For instance, in the case of $s$-wave meson--meson scattering, ordinary \gls{chpt} breaks down at the energies where the pole of the $\sigma$ meson appears, that is, around $500$~MeV. Thus, the question that arises is whether the range of applicability of chiral Lagrangians can be extended to higher energies by using some suitable unitarization (or resummation) technique.

Before introducing unitary extensions of \gls{chpt}, let us review some basic ideas on unitarity in scattering theory \cite{Peskin:257493}. 

\begin{figure}[b!]
\centering 
 \begin{tikzpicture}[baseline=(i.base)]
    \begin{feynman}[small]
      \vertex (i);
      \vertex [above left = 2cm of i] (a) {\(A\)};
      \node [below = 5mm of a] (pa) {\(p_1\)};
      \vertex [above right = 2cm of i] (c) {\(C\)};
      \node [below = 5mm of c] (pc) {\(p_3\)};
      \vertex [below right = 2cm of i] (d) {\(D\)};
      \node [above = 5mm of d] (pd) {\(p_4\)};
      \vertex [below left= 2cm of i] (b) {\(B\)};
      \node [above = 5mm of b] (pb) {\(p_2\)};
      \diagram* {
        (a) -- [fermion] (i), 
        (i) -- [fermion] (c),
        (b) -- [fermion] (i),
        (i) -- [fermion] (d),
       };
     \draw[dot,minimum size=10mm,thick,fill=gray] (i) circle(5mm);
    \end{feynman}
  \end{tikzpicture}
\caption{Diagram of the $A(p_1)+B(p_2)\rightarrow C(p_3)+D(p_4)$ scattering.}
\label{fig:free-th-scattering}
\end{figure}
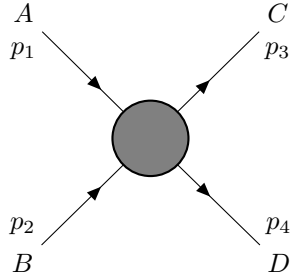

For a  $2\rightarrow 2$ multi-channel scattering process, $A(p_1)+B(p_2)\rightarrow C(p_3)+D(p_4)$, where $p_i$ are the momenta of the particles involved, as illustrated in Fig.~\ref{fig:free-th-scattering}, the usual Mandelstam variables are defined as $s=(p_1+p_2)^2=(p_3+p_4)^2$, $t=(p_1-p_3)^2=(p_2-p_4)^2$ and $u=(p_1-p_4)^2=(p_2-p_3)^2$. We recall that only two of these variables are independent, since $s+t+u=\sum_i m_i^2$. The scattering operator or $S$ matrix is the unitary operator that connects the asymptotic \textit{in} and \textit{out} states. The transition scattering amplitude or $T$ matrix is defined as the interacting part of the $S$ matrix
\begin{equation}
 S=\mathbb{1}+\ii T \ ,
\end{equation}
and the unitarity condition for the $S$ matrix, $S^\dagger S=\mathbb{1}$, implies
\begin{equation}\label{eq:free-th-unitarycondTmatrix}
 -\ii(T-T^\dagger)=T^\dagger T \ .
\end{equation}
If we consider the case of two interacting particles, the scattering amplitude is defined in terms of the invariant matrix $\mathcal{M}(\textrm{in};\textrm{out})$ as
\begin{equation}\label{eq:free-th-scatteringamplitude}
 _{\textrm{out}}\langle p_3p_4,\beta|\ii T|p_1p_2,\alpha\rangle_{\textrm{in}}=(2\pi)^4\delta^{(4)}(p_1+p_2-p_3-p_4)\ii\mathcal{M}(p_1,p_2;p_3,p_4)_{\beta\alpha} \ ,
\end{equation}
where $|p_1p_2,\alpha\rangle$ and $|p_3p_4,\beta\rangle$ are asymptotic states of two noninteracting particles with momentum $p_1$, $p_2$, and $p_3$, $p_4$, and the labels $\alpha$ and $\beta$ refer to all the additional properties of the channel.
In general, $\mathcal{M}(\textrm{in};\textrm{out})$ is a matrix in channel space and it is an analytic function of the Mandelstam variables $s$, $t$, and $u$ up to poles and kinematic singularities.

The optical theorem, which gives a nonperturbative relation between the imaginary part of the scattering amplitudes and the total cross sections, follows from the unitarity of the $S$-matrix. Indeed, using the definition in Eq.~(\ref{eq:free-th-scatteringamplitude}), the \gls{lhs} of Eq.~(\ref{eq:free-th-unitarycondTmatrix}) becomes
\begin{align}\nonumber
 -\ii\langle \Phi_\beta|(T-T^\dagger)|\Phi_\alpha\rangle&=-\ii(\langle\Phi_\beta|T|\Phi_\alpha\rangle-\langle\Phi_\alpha|T|\Phi_\beta\rangle^*)\\
 & =-\ii(2\pi)^4\delta^{(4)}(p_\alpha-p_\beta)(\mathcal{M}_{\beta\alpha}-\mathcal{M}_{\alpha\beta}^*) \ ,
\end{align}
where we have defined $|\Phi_\alpha\rangle\equiv|p_1p_2,\alpha\rangle$, $|\Phi_\beta\rangle\equiv|p_3p_4,\beta\rangle$, $p_\alpha=p_1+p_2$, and $p_\beta=p_3+p_4$ to simplify the notation.
After the insertion of a complete set of states, the \gls{rhs} of Eq.~(\ref{eq:free-th-unitarycondTmatrix})can be written as
\begin{align}\nonumber
 \langle\Phi_\beta|T^\dagger T|\Phi_\alpha\rangle&=\sum_\gamma\int d\Pi_\gamma \langle\Phi_\beta|T^\dagger|\Phi_\gamma\rangle\langle\Phi_\gamma|T|\Phi_\alpha\rangle \\ 
 &=\sum_\gamma\int d\Pi_\gamma (2\pi)^4\delta^{(4)}(p_\beta-p_\gamma)(2\pi)^4\delta^{(4)}(p_\alpha-p_\gamma)\mathcal{M}_{\gamma\beta}^*\mathcal{M}_{\gamma\alpha} \ ,
\end{align}
with $\Pi_\gamma$ being the invariant phase space for a channel $\gamma$, and the generalized optical theorem in the coupled-channel case is obtained:
\begin{equation}
-\ii( \mathcal{M}_{\beta\alpha}-\mathcal{M}_{\alpha\beta}^*)=\sum_\gamma\int d\Pi_\gamma(2\pi)^4\delta^{(4)}(p_\alpha-p_\gamma)\mathcal{M}_{\gamma\beta}^*\mathcal{M}_{\gamma\alpha} \ .
\end{equation}
For a diagonal matrix element ($\alpha=\beta$), and using $\textrm{Disc\,}\mathcal{M}=\mathcal{M}-\mathcal{M}^*=2\ii \textrm{Im\,}\mathcal{M}$, we get the optical theorem for elastic forward scattering,
\begin{equation}\label{eq:free-th-opticaltheorem}
 2\textrm{Im\,}\mathcal{M}_{\alpha\alpha}=\sum_\gamma\int d\Pi_\gamma(2\pi)^4\delta^{(4)}(p_\alpha-p_\gamma)|\mathcal{M}_{\gamma\alpha}|^2 \ ,
\end{equation}
which tells that the imaginary part of the forward scattering amplitude is a sum over squared matrix elements of the transition $\alpha\rightarrow\gamma$ for all kinematically allowed intermediate channels $\gamma$. 

Equation (\ref{eq:free-th-opticaltheorem}) can be written in the standard form in terms of the total cross section,
\begin{equation}
 \textrm{Im\,}\mathcal{M}_{\alpha\alpha}=2E_{\textrm{cm}}p_{\textrm{cm}}\sigma_{\textrm{tot}}(\alpha\rightarrow\textrm{anything}) \ ,
\end{equation}
where $E_{\textrm{cm}}$ and $p_{\textrm{cm}}$ are the energy and momentum in the center-of-mass frame.

The fact that resonances cannot be described by an expansion in \gls{chpt} is closely related to the particularity that the resulting amplitudes are polynomials in the masses and momenta, and, thus, the elastic unitarity condition of Eq.~(\ref{eq:free-th-opticaltheorem}) cannot be fulfilled. In a \gls{chpt} calculation at a given order, this condition is only satisfied perturbatively,
\begin{equation}
 2\textrm{Im\,}\mathcal{M}_{\alpha\alpha}^{(1)}=\sum_\gamma\int d\Pi_\gamma(2\pi)^4\delta^{(4)}(p_\alpha-p_\gamma)|\mathcal{M}_{\gamma\alpha}^{(0)}|^2 \ .
\end{equation}

For an $s$-wave projected amplitude such as the ones that will be derived in the next chapter for meson--baryon and meson--meson scattering, this condition would read (for a 1-channel case) $\textrm{Im } V(s) = \rho_{\textrm{scatt}}(s) |V(s)|^2$, where $\rho_{\textrm{scatt}}(s)$ is the 2-body phase space. This relation is not satisfied because the potential, $V(s)$, is real. A unitarization method aims at constructing from $V^{ij}(s)$ a new amplitude $T^{ij}(s)$ which satisfies (for the 1-channel case)
$\textrm{Im }T(s) = \rho_{\textrm{scatt}}(s) |T(s)|^2$.

There are different procedures in the literature to implement the ideas of unitarity in chiral amplitudes. Among the most popular ones, there is the Inverse Amplitude Method (IAM) \cite{Dobado:1996ps,Oller:1997ng,Oller:1998hw,}, the N/D Method \cite{Oller:1998zr,Oller:2000fj}, and the \gls{bs} approach \cite{Kaiser:1996js,Oller:1997ti,Oset:1997it}, all of which reach the same results for the unitarized scattering amplitudes. Due to its nonperturbative nature, these methods allow one to extend the energy range of applicability of the chiral Lagrangians because the growth of the amplitude at larger energies is tamed, and the cross sections saturate. Being a rational function of $s$, the unitarized amplitudes also allow for the dynamical generation of resonances from the interaction between the hadrons that are considered as degrees of freedom. 

The use of the so-called \gls{uchpt} was well established already two decades ago after it was able to provide good reproduction of the experimental data on meson--meson scattering below $1.2$~GeV. By imposing unitarity constraints on the amplitudes obtained from the lowest-order meson chiral Lagrangian, \gls{uchpt} essentially led to the appearance of the lowest-lying scalar mesons (for example, the $\sigma(500)$, the $\rho$, the $f_0(980)$, the $a_0(980)$, and the $K^*$ resonances \cite{Oller:1997ng,Oller:1997ti,Oller:1998zr}) without introducing them explicitly in the formalism. The analytical properties of the unitarized scattering amplitudes and the dynamically generated states are described in the next section.

In summary, the goal of unitarization is to solve a scattering equation respecting the unitarity of the scattering matrix and using the chiral amplitudes as a kernel. Let us illustrate the fundamental ideas with the \gls{bs} approach in coupled channels, as this is the unitarization technique used for the calculations presented in this dissertation. 

\subsubsection{The Bethe-Salpeter equation}
The so-called \textit{Bethe-Salpeter equation}, or \gls{bs} equation, was introduced by Bethe and Salpeter in 1951~\cite{Salpeter:1951sz} to describe the generation of bound states in the two-body scattering of two interacting relativistic particles. Within the \gls{bs} formalism, the $T$ matrix is obtained by solving an integral equation in momentum space,
\begin{equation}\label{eq:free-th-BSintegral}
 T_{ij}(k_i,k_j;P)=V_{ij}(k_i,k_j;P)+\ii\sum_l\int\frac{d^4q}{(2\pi)^4}V_{il}(k_i,q;P)G_{l}(q;P)T_{lj}(q,k_j;P) \ ,
\end{equation}
where $k_i$ and $k_j$ are the relative momenta of the particles in the initial and final states, with a total four-momentum $P$, whereas $q$ is the momentum of the species propagating in the intermediate loop. The sum over $l$ runs over the different channels involved in the sector of interest. The kernel $V_{ij}$ describes the interaction between channels $i$ and $j$, and $G_l$ is the two-body propagator of the two intermediate particles, also called the \textit{loop function}:
\begin{equation}
 G_l(q;P)=\mathcal{D}_l(q;P)\tilde{\mathcal{D}}_l(q;P) \ .
\end{equation}
In our case, the functions $\mathcal{D}_l$ and $\tilde{\mathcal{D}}_l$ can correspond either to a pseudoscalar meson propagator,
\begin{equation}\label{eq:free-th-pprop}
 \mathcal{D}_l^{\mathcal{P}}=\frac{1}{{q}^2-m_l^2+\ii\varepsilon} \ ,
\end{equation}
with an additional factor in the case of a vector meson,
\begin{equation}\label{eq:free-th-vprop}
 \mathcal{D}_l^{\mathcal{V}}=\frac{1}{{q}^2-m_l^2+\ii\varepsilon}\left(-g^{\mu\nu}+\frac{q^\mu q^\nu}{m_l^2}\right) \ ,
\end{equation}
or a baryon propagator,
\begin{equation}\label{eq:free-th-bprop}
 \mathcal{D}_l^{\mathcal{B}}=\frac{1}{\slashed{q}-M_l+\ii\varepsilon} \ ,
\end{equation}
where $m_l$ and $M_l$ denote the masses of the meson and the baryon, respectively.

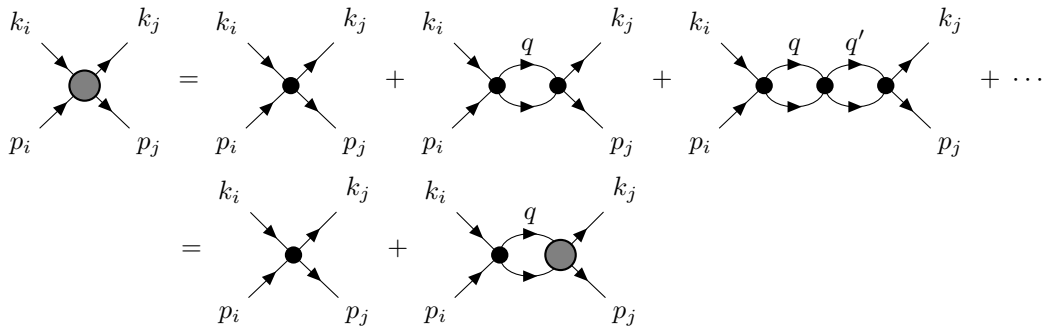
\begin{figure}[b!]%
 \centering
\begin{tikzpicture}[baseline=(i.base)]
    \begin{feynman}[small]
      \vertex (i);
      \vertex [above left = 0.8cm of i] (a) {\(k_i\)};
      \vertex [above right = 0.8cm of i] (c) {\(k_j\)};
      \vertex [below left = 0.8cm of i] (b) {\(p_i\)};
      \vertex [below right = 0.8cm of i] (d) {\(p_j\)};
      \diagram* {
        (a) -- [fermion] (i) -- [fermion] (c), 
        (b) -- [fermion] (i) -- [fermion] (d),
       };
     \draw[dot,minimum size=1mm,thick,fill=gray] (i) circle(2mm);
    \end{feynman}
  \end{tikzpicture}
  $=$
  \begin{tikzpicture}[baseline=(i.base)]
    \begin{feynman}[small]
      \vertex (i);
      \vertex [above left = 0.8cm of i] (a) {\(k_i\)};
      \vertex [above right = 0.8cm of i] (c) {\(k_j\)};
      \vertex [below left = 0.8cm of i] (b) {\(p_i\)};
      \vertex [below right = 0.8cm of i] (d) {\(p_j\)};
      \diagram* {
        (a) -- [fermion] (i) -- [fermion] (c), 
        (b) -- [fermion] (i) -- [fermion] (d),
       };
     \draw[dot,minimum size=1mm,thick,fill=black] (i) circle(1mm);
    \end{feynman}
  \end{tikzpicture}
  $+$
  \begin{tikzpicture}[baseline=(i.base)]
    \begin{feynman}[small]
      \vertex (i);
      \vertex [right = 0.8cm of i] (j);
      \vertex [above left = 0.8cm of i] (a) {\(k_i\)};
      \vertex [above right = 0.8cm of j] (c) {\(k_j\)};
      \vertex [below left = 0.8cm of i] (b) {\(p_i\)};
      \vertex [below right = 0.8cm of j] (d) {\(p_j\)};
      \diagram* {
        (a) -- [fermion] (i) -- [fermion, half left, looseness=1.2,edge label={\(q\)}] (j) -- [fermion] (c), 
        (b) -- [fermion] (i) -- [fermion, half right, looseness=1.2] (j) -- [fermion] (d),
       };
     \draw[dot,minimum size=1mm,thick,fill=black] (i) circle(1mm);
     \draw[dot,minimum size=1mm,thick,fill=black] (j) circle(1mm);
    \end{feynman}
  \end{tikzpicture}
  $+$
  \begin{tikzpicture}[baseline=(i.base)]
    \begin{feynman}[small]
      \vertex (i);
      \vertex [right = 0.8cm of i] (j);
      \vertex [right = 0.8cm of j] (k);     
      \vertex [above left = 0.8cm of i] (a) {\(k_i\)};
      \vertex [above right = 0.8cm of k] (c) {\(k_j\)};
      \vertex [below left = 0.8cm of i] (b) {\(p_i\)};
      \vertex [below right = 0.8cm of k] (d) {\(p_j\)};
      \diagram* {
        (a) -- [fermion] (i) -- [fermion, half left, looseness=1.2,edge label={\(q\)}] (j) -- [fermion, half left, looseness=1.2,edge label={\(q'\)}] (k) -- [fermion] (c), 
        (b) -- [fermion] (i) -- [fermion, half right, looseness=1.2] (j) -- [fermion, half right, looseness=1.2] (k) -- [fermion] (d),
       };
     \draw[dot,minimum size=1mm,thick,fill=black] (i) circle(1mm);
     \draw[dot,minimum size=1mm,thick,fill=black] (j) circle(1mm);
     \draw[dot,minimum size=1mm,thick,fill=black] (k) circle(1mm);
    \end{feynman}
  \end{tikzpicture}
  $+ \; \cdots$ \\
  $=$
  \begin{tikzpicture}[baseline=(i.base)]
    \begin{feynman}[small]
      \vertex (i);
      \vertex [above left = 0.8cm of i] (a) {\(k_i\)};
      \vertex [above right = 0.8cm of i] (c) {\(k_j\)};
      \vertex [below left = 0.8cm of i] (b) {\(p_i\)};
      \vertex [below right = 0.8cm of i] (d) {\(p_j\)};
      \diagram* {
        (a) -- [fermion] (i) -- [fermion] (c), 
        (b) -- [fermion] (i) -- [fermion] (d),
       };
     \draw[dot,minimum size=1mm,thick,fill=black] (i) circle(1mm);
    \end{feynman}
  \end{tikzpicture}
    $+$
  \begin{tikzpicture}[baseline=(i.base)]
    \begin{feynman}[small]
      \vertex (i);
      \vertex [right = 0.8cm of i] (j);
      \vertex [above left = 0.8cm of i] (a) {\(k_i\)};
      \vertex [above right = 0.8cm of j] (c) {\(k_j\)};
      \vertex [below left = 0.8cm of i] (b) {\(p_i\)};
      \vertex [below right = 0.8cm of j] (d) {\(p_j\)};
      \diagram* {
        (a) -- [fermion] (i) -- [fermion, half left, looseness=1.2,edge label={\(q\)}] (j) -- [fermion] (c), 
        (b) -- [fermion] (i) -- [fermion, half right, looseness=1.2] (j) -- [fermion] (d),
       };
     \draw[dot,minimum size=1mm,thick,fill=black] (i) circle(1mm);
     \draw[dot,minimum size=1mm,thick,fill=gray] (j) circle(2mm);
    \end{feynman}
  \end{tikzpicture}
  \hspace{2.8cm}
\caption{Diagrams representing the \gls{bs} equation of a two-body scattering. The big grey circle corresponds to the $T_{ij}$ matrix element, the black small circles correspond to the potential $V_{ij}$ and the loops represent the propagator $G_{l}$ function. The $i, j, l$ indices stand for the channels of the coupled-channel theory.}
\label{fig:free-mb-BS}
\end{figure}

The \gls{bs} equation performs a resummation of an infinite series of ladder-type diagrams with an increasing number of loops, as depicted diagrammatically in Fig.~\ref{fig:free-mb-BS}.

Due to its generality, this covariant scattering equation has been extensively applied in many branches of physics.
Solving the \gls{bs} equation entails some difficulties related to its off-shell nature, as well as to the presence of the poles of the propagators along the integration contour. It was shown by the authors of Refs.~\cite{Oller:1997ti,Oller:2000ma} that in the on-shell approximation the integral equation reduces to an algebraic one for $s$-wave scattering. Indeed, it was shown in \cite{Oller:1997ti} that only the on-shell information of $V$ and $T$ is needed to solve Eq.~(\ref{eq:free-th-BSintegral}) and that the off-shell part goes into the renormalization of the couplings and the masses. When taken on shell, $V$ and $T$ can be factorized outside the integral, so that the on-shell \gls{bs} equation can be written as
\begin{equation}\label{eq:free-th-BSeq}
T_{ij}=V_{ij}+V_{il}G_{l}T_{lj} \ ,
\end{equation}
with
\begin{equation}\label{eq:free-th-2bodyprop}
 G_l=\ii\int\frac{d^4q}{(2\pi)^4}\mathcal{D}_l(q;P)\tilde{\mathcal{D}}_l(q;P) \ ,
\end{equation}
and the integration over $dq^0$ can be done analytically by choosing the contour in the lower half of the complex plane.
Equation~(\ref{eq:free-th-BSeq}) represents a set of coupled equations which can be written in matrix form as
\begin{equation}\label{eq:free-th-BSmatrix}
 T=V+VGT \ ,
\end{equation}
with a purely algebraic solution,
\begin{equation}\label{eq:free-th-BSmatrixinv}
 T=(\mathbb{1}-VG)^{-1}V \ .
\end{equation}
This is the expression of the \gls{bs} equation that will be solved to unitarize the scattering amplitudes in coupled channels throughout this thesis.

The problem of solving the on-shell \gls{bs} equation reduces to the calculation of the two-body propagator of Eq.~(\ref{eq:free-th-2bodyprop}), which is a divergent integral and therefore has to be regularized with a proper scheme. Two regularization techniques are extensively used in the literature to achieve this: 
\begin{itemize}
 \item \textbf{Cut-off regularization}: it consists in replacing the infinite upper limit of the three-momentum integral with a large enough cut-off momentum, $\Lambda$, for the UV divergence
 \begin{equation}
  G_l^{\textrm{cut}}=\int_{|\vec{q}\,|<\Lambda}\frac{d^3q}{(2\pi)^3}\mathcal{D}_l(q;P)\bar{\mathcal{D}}_l(q;P) \ .
 \end{equation}
 This upper bound is put over the modulus of the momenta only, as the integral over angles is not divergent.
 The value of $\Lambda$ determines the maximum on-shell momentum of the particle in the loop and, thus, it has to be large enough for all channels (typically, several hundreds of MeV).
\item \textbf{\Gls{dr}}: it consists in lowering the dimensionality of the integral to $D=4-2\eta$, with $\eta>0$ and not necessarily integer, and taking the finite part of the two-body propagator in the limit $\eta\rightarrow 0$. This method is far more efficient than the use of a hard cut-off, and it also has the advantage of preserving translational invariance and gauge invariance. The disadvantage of \gls{dr} is that it is less intuitive. We are analytically continuing in $D$, the dimension of spacetime, the results calculated for an arbitrary $D$, and the physical meaning of a real-valued dimension might not be clear. It gives rise to a subtraction constant $a(\mu)$ at a given regularization scale $\mu$. Typical values of these parameters are $\mu\approx 630$~MeV and $a(\mu)\sim-2$ in the case of chiral effective theories in the light sector \cite{Oller:2000fj}, and $\mu\approx 1$~GeV and $a(\mu)\sim -2.3$ in the charm sector \cite{Wu:2010jy,Wu:2010vk}.
\end{itemize}

Both regularization schemes for the loop function can lead to similar results of a particular \gls{uchpt} model upon reasonably varying their parameters. In general, one demands that both regularization procedures give the same value of the loop function at the two-hadron energy threshold, and so an expression for the subtraction constants as a function of the cut-off and the regularization scale is obtained.

These two methods to regularize the two-body propagator will be particularized for the meson--baryon and meson--meson cases in the next sections.

\subsection{Analytic continuation of the $T$ matrix and dynamically generated states}
\label{subsec:free-th-dynamicallygeneratedstates}
The unitarization process leads to the potential emergence of poles (singularities) in the resummed amplitude $T(s)$ at the zeros of the denominator of Eq.~(\ref{eq:free-th-BSmatrixinv}). These poles correspond to states that are dynamically generated by the attractive coupled-channel hadron--hadron interactions. The characterization of these states requires analytically continuing the $T$ matrix to the complex-energy plane, by allowing the Mandelstam $s$ to be a complex variable, that is, $s\rightarrow z\in \mathbb{C}$. Besides, the search for poles should be performed in the correct \gls{rs}. 

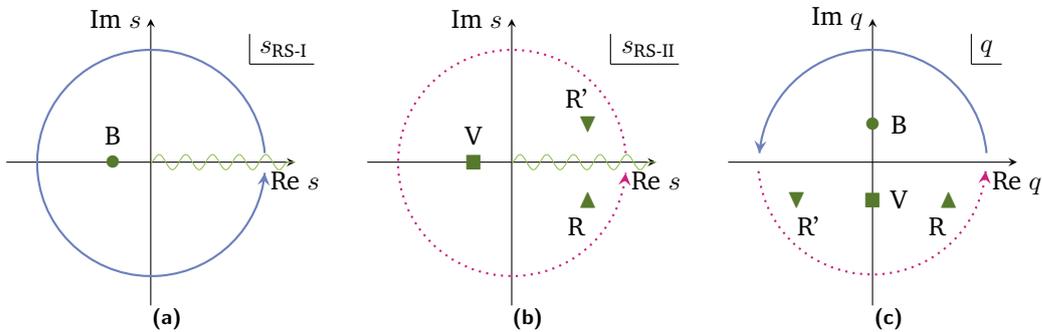
\begin{figure}[b!]\centering 
\begin{subfigure}[b]{0.32\textwidth}\centering 
\captionsetup{skip=0pt}
\begin{tikzpicture}[baseline=-2,
    decoration={%
      markings,
      mark=at position 9.16cm with {\arrow[line width=1pt,>=stealth]{>}},
    }
  ]
  \draw [>=stealth,->] (-1.9,0) -- (1.9,0) coordinate (xaxis);
  \draw [>=stealth,->] (0,-1.9) -- (0,1.9) coordinate (yaxis);
  \node [color=ctcolordarkgreen] at (-0.5,0) {\scalebox{1.2}{$\bullet$}};
  \path [draw, line width=0.8pt, postaction=decorate, color=ctcolorblue] (1.5,0.12)  arc (5:355:1.5) ;
  \node [below] at (xaxis) {Re $s$};
  \node [left] at (yaxis) {Im $s$};
  \node at (1.75,1.5) {$s_{\textrm{\gls{rs}-I}}$};
  \draw (1.3,1.3) -- (1.3,1.7);
  \draw (1.3,1.3) -- (2.15,1.3);
  \path [draw,decorate,decoration={snake,amplitude=0.1cm},color=ctcolorgreen] (0,0) -- (1.9,0);
  \node at (-0.5,0.35) {B};
\end{tikzpicture}
\caption{}
\label{fig:free-th-polesRS-a}
\end{subfigure}
\begin{subfigure}[b]{0.32\textwidth}\centering 
\captionsetup{skip=0pt}
\begin{tikzpicture}[baseline=-2,
    decoration={%
      markings,
      mark=at position 9.16cm with {\arrow[line width=1pt,>=stealth]{>}},
    }
  ]
  \draw [>=stealth,->] (-1.9,0) -- (1.9,0) coordinate (xaxis);
  \draw [>=stealth,->] (0,-1.9) -- (0,1.9) coordinate (yaxis);
  \node [color=ctcolordarkgreen] at (-0.5,0.) {\scalebox{0.8}{$\blacksquare$}};
  \node [color=ctcolordarkgreen] at ( 1,0.5) {$\blacktriangledown$};
  \node [color=ctcolordarkgreen] at ( 1,-0.5) {$\blacktriangle$};
  \path [draw, dotted, line width=0.8pt, postaction=decorate, color=ctcolormagenta] (1.5,0.12)  arc (5:355:1.5) ;
  \node [below] at (xaxis) {Re $s$};
  \node [left] at (yaxis) {Im $s$};
  \node at (1.8,1.5) {$s_{\textrm{\gls{rs}-II}}$};
  \draw (1.3,1.3) -- (1.3,1.7);
  \draw (1.3,1.3) -- (2.2,1.3);
  \path [draw,decorate,decoration={snake,amplitude=0.1cm},color=ctcolorgreen] (0,0) -- (1.9,0);
  \node at (-0.5,0.35) {V};
  \node at (0.85,0.85) {R'};
  \node at (0.85,-0.85) {R};
\end{tikzpicture}
\caption{}
\label{fig:free-th-polesRS-b}
\end{subfigure}
\begin{subfigure}[b]{0.32\textwidth}\centering 
\captionsetup{skip=0pt}
\begin{tikzpicture}[baseline=-2,
    decoration={%
      markings,
      mark=at position 4.45cm with {\arrow[line width=1pt,>=stealth]{>}},
    }
  ]
  \draw [>=stealth,->] (-1.9,0) -- (1.9,0) coordinate (xaxis);
  \draw [>=stealth,->] (0,-1.9) -- (0,1.9) coordinate (yaxis);
  \node [color=ctcolordarkgreen] at (0,0.5) {\scalebox{1.2}{$\bullet$}};
  \node [color=ctcolordarkgreen] at (0,-0.5) {\scalebox{0.8}{$\blacksquare$}};
  \node [color=ctcolordarkgreen] at (-1,-0.5) {$\blacktriangledown$};
  \node [color=ctcolordarkgreen] at ( 1,-0.5) {$\blacktriangle$};
  \path [draw, line width=0.8pt, postaction=decorate, color=ctcolorblue] (1.5,0.12)  arc (5:175:1.5) ;
  \path [draw, dotted, line width=0.8pt, postaction=decorate, color=ctcolormagenta] (-1.5,-0.12)  arc (185:355:1.5) ;
  \node [below] at (xaxis) {Re $q$};
  \node [left] at (yaxis) {Im $q$};
  \node at (1.5,1.5) {$q$};
  \draw (1.3,1.3) -- (1.3,1.7);
  \draw (1.3,1.3) -- (1.7,1.3);
  \node at (0.35,0.5) {B};
  \node at (0.35,-0.5) {V};
  \node at (-0.85,-0.85) {R'};
  \node at (0.85,-0.85) {R};
\end{tikzpicture}
\caption{}
\label{fig:free-th-polesRS-c}
\end{subfigure}
\caption{Location of the poles in the complex plane: (a) $s$-plane (\gls{rs}-I), (b) $s$-plane (\gls{rs}-II), (c) $q$-plane. They are labeled as B for bound state, V for virtual state, and R/R' for conjugate resonances.}
\label{fig:free-th-polesRS}
\end{figure}

Let us consider first the simple case where we only have one channel with two particles of masses $m_1$ and $m_2$. The scattering amplitude $T(z)$ is an analytical function of $z$ up to a right-hand cut along the real axis starting at the threshold energy $z=s_{\textrm{thr}}=(m_1+m_2)^2$, the so-called \textit{unitary cut}, and the poles generated by the dynamics of the underlying theory. As a consequence, the $T$ matrix can be defined in two \glspl{rs} that are usually called \textit{physical sheet} or \textit{first \gls{rs}} (\gls{rs}-I) and \textit{unphysical sheet} or \textit{second \gls{rs}} (\gls{rs}-II). Depending on their location on the Riemann surface, poles can be classified into different categories, as shown schematically in Fig.\ref{fig:free-th-polesRS}: poles that appear in the real axis of the physical sheet below the threshold correspond to \textit{bound states} (B) (see Fig.~\ref{fig:free-th-polesRS-a}), poles in the real axis of the unphysical sheet are called \textit{virtual states} (V), and \textit{resonances} (R/R') are identified with poles located outside the real axis in the unphysical sheet (see Fig.~\ref{fig:free-th-polesRS-b}). It follows from analyticity that, if there is a pole at some complex energy $z=z_p$, there must be another pole at its complex energy, $z=z_p^*$. That is, poles outside the real axis (resonant poles) appear in conjugate pairs.

An equivalent way to discuss the analytical structure of the $T$ matrix is to use the relative three-momentum $q$ instead of the Mandelstam $s$, which are related through
\begin{equation} \label{eq:free-th-relmom}
 q=\frac{\sqrt{\left(s-(m_1+m_2)^2\right)\left(s-(m_1-m_2)^2\right)}}{2\sqrt{s}} 
\end{equation}
As there is no right-hand cut, in the complex $q$-plane there is only one sheet (Fig.~\ref{fig:free-th-polesRS-c}). The relative momentum $q$ itself is a two-valued function of $s$, with two solutions $q_+$ (with $\textrm{Im }q_+>0$) and $q_-=q_+e^{\ii\pi}$ (with $\textrm{Im }q_-<0$) for a given value of $s$. Therefore, the upper (lower) half of the complex $q$-plane, corresponding to positive (negative) values of the imaginary part of $q$, has a mapping onto the physical (unphysical) \gls{rs} of the complex $s$-plane. The denomination physical/unphysical sheet is more conventional than meaningful. Amplitudes for complex values of the kinematic parameters have to be connected with the physical amplitudes for the real values of the kinematic parameters, and the usual convention is to adopt $T(q+\ii\beta)\rightarrow T_{\textrm{phys}}(q)$, with $q\in\mathbb{R}$ and $\beta\rightarrow 0^+$.

In the $q$-plane bound states (B) and virtual states (V) are located on the imaginary axis on the upper and lower half-planes, respectively, while resonances (R/R') appear symmetrically with respect to the imaginary axis in the lower half-plane.

The multi-channel case is somewhat more complicated, as there are two sheets for each channel. To select the correct \gls{rs} of the $T$ matrix, we observe that the loop function in Eq.~(\ref{eq:free-th-2bodyprop}) is a multivalued function with two \glspl{rs} above the two-hadron energy threshold. Taking the principal value of the argument, that is, $\textrm{Arg}(G_l)\in (0,2\pi]$, defines the first \gls{rs}. The expression of $G_l$ in the second \gls{rs} is obtained by adding a contribution to the imaginary part~\cite{Roca:2005nm},
\begin{equation}\label{eq:free-th-RS2}
G_l^\textrm{II}(\sqrt{s}+\ii\varepsilon)=G_l^\textrm{I}(\sqrt{s}-\ii\varepsilon)=[G_l^\textrm{I}(\sqrt{s}+\ii\varepsilon)]^*=\ G_l^\textrm{I}(\sqrt{s}+\ii\varepsilon)-\ii 2\,\textrm{Im}\,G_l^\textrm{I}(\sqrt{s}+\ii\varepsilon) \ .
\end{equation}
The superindices I and II denote the first and second \glspl{rs}, respectively.
In the case of a two-meson loop function, one gets
\begin{equation}\label{eq:free-th-MMRS2}
G_{l,\mathcal{MM}}^\textrm{II}(\sqrt{s}+\ii\varepsilon)=G_{l,\mathcal{MM}}^\textrm{I}(\sqrt{s}+\ii\varepsilon)+\ii\frac{q}{4\pi\sqrt{s}} \ ,
\end{equation}
while for meson--baryon scattering there is an additional factor $2M_l$,
\begin{equation}\label{eq:free-th-MBRS2}
G_{l,\mathcal{MB}}^\textrm{II}(\sqrt{s}+\ii\varepsilon)=G_{l,\mathcal{MB}}^\textrm{I}(\sqrt{s}+\ii\varepsilon)+\ii 2M_l\frac{q}{4\pi\sqrt{s}} \ .
\end{equation}
The same results follow from changing the sign of the momentum $q$ in the regularized expressions of the two-body propagator, explicitly given in the sections below for the cases of meson--baryon and meson--meson scattering, and taking the phase prescription of the logarithms $\ln z=\ln|z|+\ii\theta$ as $0\le\theta<2\pi$.

Thus, the analytic structure of $G_l$ provides the unitarized amplitude $T_{ij}(\sqrt{z})$ in a set of $2^n$ \glspl{rs}, where $n$ is the number of coupled channels. The \gls{rs}-I or physical sheet of the $T$ matrix follows from solving the coupled-channel \gls{bs} equation with $G_l^\textrm{I}$ for all channels.
The \gls{rs}-II is defined as the unphysical sheet that is connected to the real-energy axis from below. It is obtained by taking $G_l^\textrm{I}$ for $\textrm{Re } \sqrt{z}< \sqrt{s_l^{\textrm{thr}}}$ and $G_l^\textrm{II}$ for $\textrm{Re } \sqrt{z}> \sqrt{s_l^{\textrm{thr}}}$, with $\sqrt{s_l^{\textrm{thr}}}$ the threshold energy for channel $l$ in the center-of-mass frame. 
The different \glspl{rs} are connected to each other in the regions between thresholds in a nontrivial way due to the presence of branch cuts.

With this prescription, the scattering amplitude close to a pole, and if not very close to a threshold, can be parametrized with a Breit-Wigner distribution,
\begin{equation}
 T(\sqrt{z})=\frac{1}{\sqrt{z}-M_R+\ii\Gamma_R/2} \ .
\end{equation}
The real and imaginary parts of the pole positions $\sqrt{z_p}$ in the complex plane give the mass and the half width of the dynamically generated states, respectively:
\begin{equation}
  M_R=\textrm{Re\,}\sqrt{z_p} \ ,\quad \Gamma_R/2=-\textrm{Im\,} \sqrt{z_p} \ .
\end{equation}

In the general case of one physical \gls{rs} and multiple unphysical sheets, the definitions of the pole categories are the following:
\begin{itemize}
 \item \textit{Bound states} are identified with poles located on the real axis of the physical \gls{rs}, below the lowest threshold energy. As a result, bound states cannot decay. That is, they are stable and have no width.
 \item \textit{Resonances} are those poles located on an unphysical \gls{rs} at necessarily complex energy. Due to their nonzero width, resonances are associated with unstable states that can decay to open channel states. \\
 The meaningful physics lies in the reflection of the resonances on the real axis. From all the unphysical sheets, the resonant poles that are physically more relevant are those located in the lower half-plane of the \gls{rs}-II, as they are the ones closest to the physical axis.
 \item \textit{Virtual poles} are poles that lie on the real axis of an unphysical sheet, below the lowest threshold.
\end{itemize}
Resonance poles that are located on the \gls{rs}-II are the ones that, together with bound states, are more likely to generate structures in the scattering amplitude in the real axis. Therefore, it is common to call resonances only the resonant poles in the second sheet and generalize the term ``virtual state'' to resonant poles in any other unphysical sheet, which can still yield to structures and cusps near the thresholds.

\subsubsection{Couplings and compositeness}
The scattering amplitude can be expanded in a Laurent series around the pole position
\begin{equation}\label{eq:free-th-couplings}
T^{ij}(z)=\frac{g_ig_j}{z-z_p}+\sum_{n=0}^\infty T_{ij}^{(n)}(z-z_p)^n \ ,
\end{equation}
where $g_i$ is the coupling of the resonance or bound state to the channel $i$ and $g_ig_j$ is the residue around the pole. Therefore, from the residue of the different components of the $T$ matrix around the pole, one can extract the coupling constants to each of the channels. The residue of a simple pole (of order 1) can be calculated with different methods, all of them giving similar numerical values. One can apply the limit formula,
\begin{equation}\label{eq:free-th-couplim}
 g_ig_j=\lim_{z\rightarrow z_p}(z-z_p)T_{ij}(z) \ ,
\end{equation}
and numerically extrapolate to the pole position. An equivalent method is to perform a contour integral along a path of radius $r$ around the pole,
\begin{equation}\label{eq:free-th-coupcauchy}
 g_ig_j=\frac{1}{2\pi\ii}\oint dz\, T_{ij}(z)=\frac{r}{2\pi\ii}\int_0^{2\pi}d\theta\, T_{ij}(z_p+re^{\ii\theta}) \ .
\end{equation}
And it can also be calculated from the numerical derivative of the inverse $T$ matrix,
\begin{equation}\label{eq:free-th-coupder}
g_ig_j=\Big[\frac{\partial}{\partial (z)^2}\Big(\frac{1}{T_{ij}(z)}\Big)\Big|_{z_p}\Big]^{-1} \ .
\end{equation}

The concept of compositeness of shallow bound states was formulated by Weinberg in Refs.~\cite{Weinberg:1962hj,Weinberg:1965zz}, applied to narrow resonances close to the threshold in~\cite{Baru:2003qq,Hanhart:2011jz}, and subsequently extended to the complex pole position of an unstable resonant state by analytical continuation in~\cite{Gamermann:2009uq,Hyodo:2011qc,Aceti:2012dd,Aceti:2014ala,Sekihara:2014kya}. An appropriate unitary transformation~\cite{Guo:2015daa} permits assigning real values to the compositeness of complex poles lying in the second Riemann sheet as
\begin{equation}\label{eq:free-th-compositeness}
\chi_i= \left|g_i^2\frac{\partial G_i(z_p)}{\partial z}\right| \ ,
\end{equation}
which effectively measures the amount of $i^\textrm{th}$ channel component in the dynamically generated state. The authors of Ref.~\cite{Aceti:2014ala} noted that the real part of $-g_i^2\partial G_i(z_p)/\partial z$ can be rigorously interpreted as a probability only for bound states, but that it can still be regarded as the relevance of a given channel in the wave function in the case of resonant states.

\section{Meson-baryon interaction in the open heavy-flavor sector}
\label{sec:free-mb}

In Section \ref{sec:free-theoryremarks} we have given some remarks on effective theories and the chiral Lagrangians have been reviewed, including the formal extensions of \gls{chpt} to incorporate baryons and vector mesons into the theory, as well as hadrons with heavy flavor. Preservation of the symmetries of \gls{qcd} has been the guiding principle to achieve this. In addition, a unitarization technique based on the \gls{bs} equation has been introduced to preserve the unitary and analytic structure of the scattering amplitudes, and hadronic molecules have been described as dynamically generated states in the complex-energy plane.
Now we make use of these methods to describe the scattering between mesons (pseudoscalar and vector mesons) with baryons within the formalism of the hidden gauge applied to sectors with open heavy flavor. For this, the Lagrangians of the local hidden-gauge formalism will be extended to $\textrm{SU}(4)$.
The aim is to obtain heavy excited baryons, in particular those with the quantum numbers of the $\Omega_c$ and $\Omega_b$ states, that is strangeness $S=-2$ and either charm $C=1$ or beauty $B=-1$, respectively. 

\subsection{Introduction}

\begin{figure}[htbp!]
 \centering
   \includegraphics[width=0.405\textwidth]{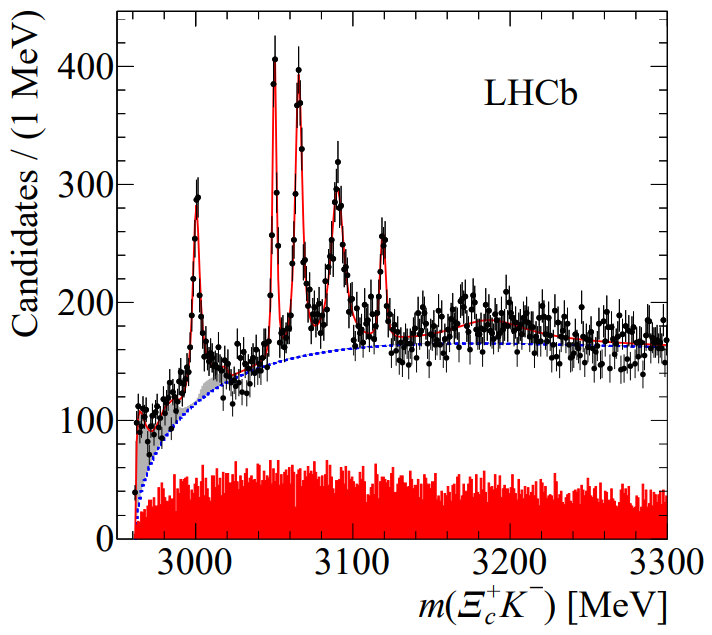}
   \includegraphics[width=0.495\textwidth]{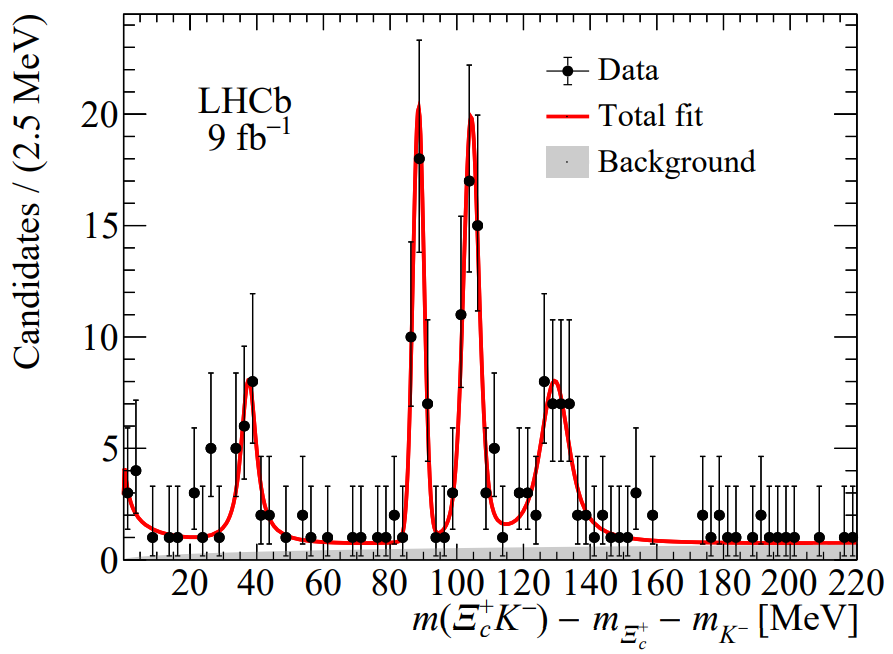}
 \caption{Reconstructed invariant mass distributions from the analyses in Ref.~\cite{Aaij:2017nav} (left panel) and Ref.~\cite{LHCb:2021ptx} (right panel), showing the peaks corresponding to the observed $\Omega_c(3000)^0$, $\Omega_c(3050)^0$, $\Omega_c(3066)^0$, $\Omega_c(3090)^0$ and $\Omega_c(3119)^0$ excited states (the highest state is not observed in the analysis of Ref.~\cite{LHCb:2021ptx}).}
 \label{fig:free-mb-omegacLHCb}
\end{figure}

In recent years, the LHCb collaboration has made an important contribution to the spectroscopy of heavy baryons by observing several new excited states \cite{Aaij:2017nav,LHCb:2018vuc,LHCb:2018haf,LHCb:2019soc,LHCb:2020tqd,LHCb:2020lzx,LHCb:2020iby,LHCb:2020xpu,LHCb:2021ptx}. Among these, the $\Omega_c^0$ has drawn a lot of attention. 
Five narrow $\Omega_c^0$ excited resonances, denoted $\Omega_c^{*0}$ from now on in this dissertation, have been observed in proton-proton ($pp$) collisions decaying into $\Xi_c^+ K^-$ states \cite{Aaij:2017nav,Belle:2017ext,LHCb:2021ptx}. The reconstructed invariant mass distributions of the experimental analysis of the LHCb collaboration are shown in Fig.~\ref{fig:free-mb-omegacLHCb}, and the properties of the states as of Ref.~\cite{Aaij:2017nav} are listed in Table~\ref{tab:free-mb-explhc}. Their first observation in 2017 triggered a lot of activity in the immediate time in the field of baryon spectroscopy aiming at understanding their inner structure and possibly establishing their unknown values of spin-parity \cite{Montana:2017kjw,Karliner:2017kfm,Wang:2017vnc,Wang:2017zjw,Chen:2017gnu,Padmanath:2017lng,Chen:2017sci,Agaev:2017jyt,Agaev:2017lip,Cheng:2017ove,Wang:2017hej,Huang:2017dwn,Yang:2017rpg,An:2017lwg,Kim:2017jpx}.

\begin{table}[htbp!]
\setlength{\tabcolsep}{10pt}
\renewcommand{\arraystretch}{1.2}
\centering
\begin{tabular}{l c c }
\hline
Resonance & Mass (MeV) & Width (MeV) \\
\hline
$\Omega_c(3000)^0$ & $3000.4\pm0.2\pm0.1^{+0.3}_{-0.5}$ & $4.5\pm0.6\pm0.3$ \\
$\Omega_c(3050)^0$ & $3050.2\pm0.1\pm0.1^{+0.3}_{-0.5}$ & $0.8\pm0.2\pm0.1$ \\
$\Omega_c(3066)^0$ & $3065.6\pm0.1\pm0.3^{+0.3}_{-0.5}$ & $3.5\pm0.4\pm0.2$ \\
$\Omega_c(3090)^0$ & $3090.2\pm0.3\pm0.5^{+0.3}_{-0.5}$ & $8.7\pm1.0\pm0.8$ \\
$\Omega_c(3119)^0$ & $3119.2\pm0.3\pm0.9^{+0.3}_{-0.5}$ & $1.1\pm0.8\pm0.4$ \\
\hline
\end{tabular}
\caption{Properties of the experimental $\Omega_c^{*0}$ states from the analysis of the LHCb collaboration reported in Ref.~\cite{Aaij:2017nav}.}
\label{tab:free-mb-explhc}
\end{table}

The conventional quark model can naturally explain the presumed spin-parity of the $1/2^+$ and $3/2^+$ $\Omega_c^0$ ground states \cite{pdg} within a $css$ quark content picture and it also predicts the lowest-lying excited states \citep{Maltman:1980er,Migura:2006ep,Roberts:2007ni,Valcarce:2008dr,Ebert:2011kk,Vijande:2013yxa,SilvestreBrac:1996bg,Yoshida:2015tia}. 
The discovery of excited $\Omega_c^{*0}$ states provided new information against which revisited quark models can be tested. Different theoretical approaches including \gls{lqcd}, \gls{qcd} sum rules, and potential models have been used, and several interpretations have been proposed for these states. In general, they benefit from the symmetries that arise from the fact that the charm quark has a much larger mass than its strange companions. Some works interpret all the observed states as $P$-wave orbital excitations of the $ss$ diquark with respect to the $c$ quark \cite{Karliner:2017kfm,Wang:2017vnc,Wang:2017zjw,Chen:2017gnu}, a result which finds support from an \gls{lqcd} simulation reporting the energy spectra of $\Omega_c^{*0}$ baryons with spin up to $7/2$ for both positive and negative parity \cite{Padmanath:2017lng}. In other works, some of the states would be associated with $1P$ orbital excitations \cite{Chen:2017sci} and others to radial $2S$ ones \cite{Agaev:2017jyt,Agaev:2017lip,Cheng:2017ove,Wang:2017hej}.  
Some earlier quark-model predictions, before the LHCb observations, give only two negative-parity states in the energy region of interest, all having spin $1/2$ \citep{Valcarce:2008dr,SilvestreBrac:1996bg}.
A pentaquark structure within a quark picture has also been advocated for the excited $\Omega_c^{*0}$ baryons, either from a model that includes the exchange of Goldstone bosons in the interaction between the constituent quarks \citep{Huang:2017dwn,Yang:2017rpg,An:2017lwg}, by employing the quark-soliton model \citep{Kim:2017jpx} or within a harmonic oscillator based model \cite{Santopinto:2018ljf}.
We refer to the recent review on charmed baryons of Ref.~\cite{Cheng:2021qpd} for a comprehensive discussion.

A complementary scenario is provided by models that can interpret some resonances as being composed by five quarks, but structured
in the form of a quasi-bound state of an interacting meson--baryon pair. Different approaches have been explored for the effective interaction between the two hadrons, including an $\textrm{SU}(4)$ extension of the hidden-gauge formalism \cite{Montana:2017kjw}, a hidden-gauge approach in $\textrm{SU}(3)$ exploiting the dominance of the exchange of light vectors \cite{Debastiani:2017ewu}, and a scheme based on \gls{hqss} \cite{Nieves:2017jjx}. All of these works succeed in interpreting some of the observed excited $\Omega_c^{*0}$ resonances as meson--baryon molecules upon unitarization of the interaction kernel in coupled channels.

In this section, we describe the formalism that we used in \cite{Montana:2017kjw} and present the results obtained in this work for $\Omega_c^{*0}$ and $\Omega_b^{*-}$ states, prior to the experimental observation of the bottomed states. The motivation to investigate these heavy baryons with a molecular model arises from the fact that, similarly to the $P_c$ pentaquarks, which find more natural having a $c\bar{c}$ pair in their composition rather than being an extremely high energy excitation of the $3q$ system, it is also plausible to expect that some of the excitations in the $C=1$ (or $B=-1$), $S=-2$ sector can be obtained by adding a $u\bar{u}$ pair to the natural $ssc$ content of the $\Omega_c^{*0}$ (or $\Omega_b^{*-}$). The hadronization of the five quarks could then lead to bound states, generated by the meson--baryon interaction in coupled channels. This possibility is supported by the fact that the $\bar K\Xi_c$ and $\bar K\Xi_c^\prime$ thresholds are in the energy range of interest, and that the excited $\Omega_c^{*0}$ baryons under study have been observed from the invariant mass spectrum of $K^-\Xi^+_c$ pairs.

Predictions of dynamically generated meson--baryon molecules in the open charm sector with strangeness $S=-2$ before the first experimental observation of the excited $\Omega_c^{*0}$ states were already made \cite{Hofmann:2005sw,Garcia-Recio:2008rjt,JimenezTejero:2009vq,Romanets:2012hm}. The authors of Ref.~\cite{Hofmann:2005sw} found a rich spectrum of molecules in this sector employing a zero-range exchange of vector mesons as the driving force for the $s$-wave scattering of pseudoscalar mesons off the baryon ground states. Similar qualitative findings were obtained in the work of Ref.~\cite{JimenezTejero:2009vq}, where finite-range effects were explored. \gls{hqss} is explicitly considered in the model of Refs.~\cite{Garcia-Recio:2008rjt,Romanets:2012hm}, thus treating the pseudoscalar and vector mesons, as well as the ground-state $1/2^+$ and $3/2^+$ baryons, on the same footing. Despite their qualitative differences, the three works predicted the existence of $\Omega_c^{*0}$ resonances as poles of the coupled-channel meson--baryon scattering amplitude in the complex plane. However, most of the predicted quasibound states were below $3$~GeV, too low to explain the observed states. Only the model of Ref.~\cite{JimenezTejero:2009vq} predicts a state at $3117$~MeV, but its width turns out to be one order of magnitude larger than that of the closer observed state. The new experimental results \cite{Aaij:2017nav,Belle:2017ext,LHCb:2021ptx} justify a revision of the original model of Ref.~\cite{Hofmann:2005sw}, which we published in Ref.~\cite{Montana:2017kjw} and describe in the following.

\subsection{Formalism}
\label{subsec:free-mb-formalism}

Assuming perfect isospin symmetry, the mesonic and baryonic fields are written in terms of isospin multiplets and the scattering of mesons off baryons decouples into isospin-defined channels of the so-called isospin basis, as opposed to the physical or charge basis. 

We consider the sector with charm $C=1$, strangeness $S=-2$, and isospin $I=0$ quantum numbers, where the $\Omega_c^{*0}$ excited states are located. Our building blocks are the pseudoscalar ($\pi,\, K,\,\bar{K},\,\eta,\,\eta',\,D,\,D_s,\,\bar{D},\,\bar{D}_s$) and vector ($\rho,\, K^*,\,\bar{K}^*,\,\omega,\,\phi,\,D,^*\,D_s^*,\,\bar{D}^*,\,\bar{D}_s^*$) mesons and the spin-$1/2$ baryons ($N,\,\Lambda,\,\Sigma,\,\Xi,\,\Lambda_c,\,\Sigma_c,\,\Xi_c,\,\Xi_c',\Omega_c,\,\Xi_{cc},\,\Omega_{cc}$). The formalism presented in this section can be also applied to the bottom sector upon the direct substitution of the $c$ quark in the content of the heavy hadrons by a $b$ quark. This is possible thanks to \gls{hqfs} and it leads to the straightforward prediction of $\Omega_b^{*-}$ excited states.

In $\textrm{SU}(4)$, the pseudoscalar and vector meson fields are organized in $16$-plets ($\textbf{15}\oplus\textbf{1}$, see Section~\ref{subsec:intro-quarkmodel}) that can be written in the form of $4\times 4$ matrices containing the $15$-plet and the singlet pieces, 
$ \Phi_{[16]}=\Phi_{[15]}+\eta_1\mathbb{1}_4/\sqrt{3}$ and 
$ V^\mu_{[16]}=V^\mu_{[15]}+\omega_1\mathbb{1}_4/\sqrt{3}$, respectively. 

For the pseudoscalars, the $16$-plet containing the $\pi$ is given by 
\begin{equation}\label{eq:free-mb-matrixphi}
\Phi_{[16]} \! = \!
\begin{pmatrix}
\frac{1}{\sqrt{2}}\pi^0+\frac{1}{\sqrt{6}}\eta+\frac{1}{\sqrt{3}}\eta^\prime & \pi^+ & K^{+} &\ \bar{D}^{0} \\
\pi^- & \!\!\!\!\!\!\! -\frac{1}{\sqrt{2}}\pi^0+\frac{1}{\sqrt{6}}\eta+\frac{1}{\sqrt{3}}\eta^\prime & K^{0} & D^{-} \\
K^{-} & \bar{K}^{0} & \!\!\!\!\!\!\!-\sqrt{\frac{2}{3}}\eta+\frac{1}{\sqrt{3}}\eta^\prime & D_s^{-} \\
D^{0} & D^{+} & D_s^{+} & \eta_c \\
\end{pmatrix} \ ,
\end{equation}
where the $\eta$ and $\eta'$ mesons have been identified with the mathematical $\eta_8$ state and the singlet $\eta_1$, respectively.
Isoscalar states with the same $J^{P\mathcal{C}}$ numbers mix. This is the case of the $\eta$, $\eta'$, and $\eta_c$ mesons, with $J^{P\mathcal{C}}=0^{-+}$. While the mixing of the light-quark isoscalar mesons with the heavier charmonium state is negligible, the $\eta$ and $\eta'$ are admixtures of the $\textrm{SU}(3)$ octet $\eta_8$ and singlet $\eta_1$ states,
\begin{align}
 \eta=\eta_8\cos\theta_{\mathcal{P}}-\eta_1\sin\theta_{\mathcal{P}} \ , \\
 \eta'=\eta_8\sin\theta_{\mathcal{P}}+\eta_1\cos\theta_{\mathcal{P}} \ , 
\end{align}
where $\theta_{\mathcal{P}}$ is the pseudoscalar mixing angle. The $\eta-\eta'$ mixing is small, with $\theta_{\textrm{P}}\approx-19^{\circ}$ \cite{Bramon:1994cb}, and hence the octet is the largest component of the $\eta$, while the largest component the $\eta'$ is the singlet\footnote{It is also common in the literature to introduce the mixing between the singlet and octet $\textrm{SU}(3)$ states for the $\eta$ and $\eta'$ as \cite{Bramon:1994cb,Roca:2015tea}
\begin{equation*}
 \eta=\frac{1}{3}\eta_1+\frac{2\sqrt{2}}{3}\omega_8 \ ,\quad \eta'=\frac{2\sqrt{2}}{3}\eta_1-\frac{1}{3}\eta_8 \ .
\end{equation*}
}.

In the case of the $\phi$ and $\omega$ vector mesons, the mixture of the $\textrm{SU}(3)$ singlet $\omega_1$ and octet $\omega_8$ states,
\begin{align}
 \phi=\omega_8\cos\theta_{\mathcal{V}}-\omega_1\sin\theta_{\mathcal{V}} \ , \\
 \omega=\omega_8\sin\theta_{\mathcal{V}}+\omega_1\cos\theta_{\mathcal{V}} \ , 
\end{align}
is close to the ideal mixing, for which $\theta_{\mathcal{V}}\approx 35.3^{\circ}$\footnote{The ideal mixing between the singlet and octet $\textrm{SU}(3)$ states for the $\omega$ and $\phi$ gives \cite{Kucukarslan:2006wk}
\begin{equation*}
 \omega=\sqrt{\frac{2}{3}}\omega_1+\frac{1}{\sqrt{3}}\eta_8 \ ,\quad \phi=\frac{1}{\sqrt{3}}\omega_1-\sqrt{\frac{2}{3}}\omega_8 \ .
\end{equation*}
}. With this, the $16$-plet of the vector fields reads 
\begin{equation}\label{eq:free-mb-matrixVmu}
 V_{[16]}^\mu =
\begin{pmatrix}
\frac{1}{\sqrt{2}}(\rho^0+\omega) & \rho^+ & K^{\ast +} & \bar{D}^{\ast 0} \\
\rho^- & \frac{1}{\sqrt{2}}(-\rho^0+\omega) & K^{\ast 0} & D^{\ast -} \\
K^{\ast -} & \bar{K}^{\ast 0} & \phi & D_s^{\ast -} \\
D^{\ast 0} & D^{\ast +} & D_s^{\ast +} & J/\psi \\
\end{pmatrix}^\mu \ .
\end{equation}

As for the baryons, the $20$-plet of the proton in $\textrm{SU}(4)$ is given by a tensor $B^{ijk}$, which is antisymmetric under the exchange of the first two indices,
\begin{equation}
\begin{tabular}{lll}
$B^{121}=p$, & ~~~~$B^{122}=n$, & ~~~~$B^{132}=\frac{1}{\sqrt{2}}\Sigma^0-\frac{1}{\sqrt{6}}\Lambda$, \\
$B^{213}=\sqrt{\frac{2}{3}}\Lambda$,  & ~~~~$B^{231}=\frac{1}{\sqrt{2}}\Sigma^0+\frac{1}{\sqrt{6}}\Lambda$, & ~~~~$B^{232}=\Sigma^-$, \\
$B^{233}=\Xi^-$, & ~~~~$B^{311}=\Sigma^+$, & ~~~~$B^{313}=\Xi^0$, \\
$B^{141}=-\Sigma_c^{++}$, & ~~~~$B^{142}=\frac{1}{\sqrt{2}}\Sigma_c^++\frac{1}{\sqrt{6}}\Lambda_c$, & ~~~~$B^{143}=\frac{1}{\sqrt{2}}\Xi_c^{'+}-\frac{1}{\sqrt{6}}\Xi_c^+$,\\
$B^{241}=\frac{1}{\sqrt{2}}\Sigma_c^+-\frac{1}{\sqrt{6}}\Lambda_c$, & ~~~~$B^{242}=\Sigma_c^0$, & ~~~~$B^{243}=\frac{1}{\sqrt{2}}\Xi_c^{'0}+\frac{1}{\sqrt{6}}\Xi_c^0$,\\
$B^{341}=\frac{1}{\sqrt{2}}\Xi_c^{'+}+\frac{1}{\sqrt{6}}\Xi_c^+$, & ~~~~$B^{342}=\frac{1}{\sqrt{2}}\Xi_c^{'0}-\frac{1}{\sqrt{6}}\Xi_c^0$, & ~~~~$B^{343}=\Omega_c$, 
\\
$B^{124}=\sqrt{\frac{2}{3}}\Lambda_c$, & ~~~~$B^{234}=\sqrt{\frac{2}{3}}\Xi_c^0$, 
& ~~~~$B^{314}=\sqrt{\frac{2}{3}}\Xi_c^+$, \\
$B^{144}=\Xi_{cc}^{++}$, & ~~~~$B^{244}=-\Xi_{cc}^+$, & ~~~~$B^{344}=\Omega_{cc}$, 
\\ 
\end{tabular}
\label{eq:free-mb-baryons}
\end{equation}
where the indices $i,j,k$ denote the quark content of the baryon fields with the identification $1\leftrightarrow u$, $2\leftrightarrow d$, $3\leftrightarrow s$ and $4\leftrightarrow c$. 

The phase convention for the isospin states $|I,I_3\rangle$, consistent with the structure of the $\Phi_{[16]}$, $V^\mu_{[16]}$, and $B^{ijk}$ fields, is $|\pi^+\rangle=-|1,1\rangle$, $|K^{ -}\rangle=-|1/2,-1/2\rangle$, and $|D^{ 0}\rangle=-|1/2,-1/2\rangle$ for the pseudoscalar mesons and, analogously, $|\rho^+\rangle=-|1,1\rangle$, $|K^{*\, -}\rangle=-|1/2,-1/2\rangle$, and $|D^{* 0}\rangle=-|1/2,-1/2\rangle$ for the vector mesons. For the baryons, we take $|\Sigma^+\rangle=-|1,1\rangle$ and $|\Xi^-\rangle=-|1/2,-1/2\rangle$. It is the one followed in Refs. \cite{Wu:2010jy,Gamermann:2006nm} and it differs from that in Ref.~\cite{Hofmann:2005sw} in the sign of the $D^+(D^{* +})$ and $D^-(D^{* -})$ mesons.

The \gls{pseudoscalar-baryon} and \gls{vector-baryon} channels involved in the $(I,S,C)=(0,-2,1)$ sector are listed in Table~\ref{tab:free-mb-channels} in order of increasing value of their threshold. The corresponding ones in the $(I,S,B)=(0,-2,-1)$ sector are given in Table~\ref{tab:free-mb-channelsbottom}. The doubly heavy $\Omega_{cc}$ and $\Omega_{bb}$ states have not been reported experimentally yet and the values of their masses used for the calculation of the threshold mass of the channels in which they are involved have been taken from Ref.~\cite{Albertus:2009ww}. Nevertheless, the triply heavy $\bar{D}_s \Omega_{cc}$ and $\eta_c \Omega_c $ channels will be neglected, as their energy is much larger than that of the other channels. Also their counterparts in the \gls{vector-baryon} scattering problem and in the bottom sector. We have checked that their inclusion barely influences the results presented here.

\begin{table}[hb!]
\setlength{\tabcolsep}{10pt}
\renewcommand{\arraystretch}{1.2}
\centering
\begin{tabular}{l c | l c }
\hline
$\mathcal{PB}$ & Threshold (MeV) & $\mathcal{VB}$ & Threshold (MeV) \\
\hline  
{${\bar K}\Xi_c$}    &  $2964.7$   &  {$D^*\Xi$}      &      $3326.5$     \\
{${\bar K}\Xi_c^\prime$}  &  $3072.2$   & {${\bar K}^*\Xi_c$}    &  $3362.9$  \\
{$D\Xi$}        &     $3185.0$     &  {${\bar K}^*\Xi_c^\prime$}  &  $3470.4$  \\
{$\eta\Omega_c$}   &   $3245.0$  &  {$\omega\Omega_c$}  &   $3480.1$   \\
{$\eta^\prime\Omega_c$} &  $3655.3$ &  {$\phi\Omega_c$} & $3717.0$  \\          
\hdashline
$\eta_c\Omega_c$ & $5677.2$ & $J/\psi\Omega_c$ & $5794.4$ \\
{$\bar{D}_s\Omega_{cc}$} & $5680.5$ & $\bar{D}_s^*\Omega_{cc}$ & $5824.4$ \\
\hline
\end{tabular}
\caption{The \gls{pseudoscalar-baryon} ($J^P=1/2^-$) and \gls{vector-baryon} ($J^P=1/2^-,\,3/2^-$) channels for the sector with quantum numbers $(I,S,C)=(0,-2,1)$, and the threshold mass in MeV.}
\label{tab:free-mb-channels}
\end{table}

\begin{table}[hb!]
\setlength{\tabcolsep}{10pt}
\renewcommand{\arraystretch}{1.2}
\centering
\begin{tabular}{l c | l c }
\hline
$\mathcal{PB}$ & Threshold (MeV) & $\mathcal{VB}$ & Threshold (MeV)  \\
\hline  
{${\bar K}\Xi_b$}  &   $6288.9$    &  {$\bar{B}^*\Xi$}    & $6642.8$\\
{${\bar K}\Xi_b^\prime$}  & $6430.7$ & {${\bar K}^*\Xi_b$}   &   $6687.1$ \\
{$\eta\Omega_b$}  & $6593.6$       &  {$\omega\Omega_b$}   &  $6828.7$ \\
{$\bar{B}\Xi$}    &      $6597.6$      &  {${\bar K}^*\Xi_b^\prime$}  & $6828.9$   \\
{$\eta^\prime\Omega_b$} &  $7003.9$ &  {$\phi\Omega_b$} & $7065.6$ \\          
\hdashline
 $\eta_b\Omega_b$ & $15444.8$ & $\Upsilon\Omega_b$ & $15506.4$ \\
$B_s\Omega_{bb}$ & $15635.9$ & $B_s^*\Omega_{bb}$ & $15684.4$ \\
\hline
\end{tabular}
\caption{The \gls{pseudoscalar-baryon} ($J^P=1/2^-$) and \gls{vector-baryon} ($J^P=1/2^-,\,3/2^-$) channels for the sector with quantum numbers $(I,S,B)=(0,-2,-1)$, and the threshold mass in MeV.}
\label{tab:free-mb-channelsbottom}
\end{table}

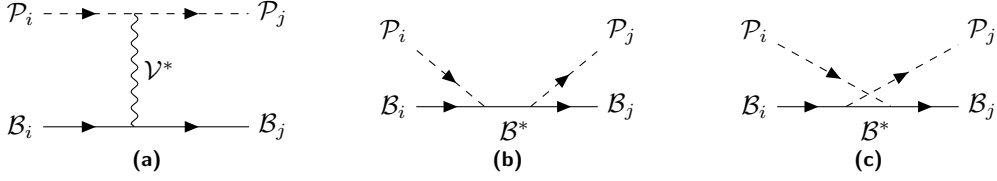
\begin{figure}[t!]
\centering 
\begin{subfigure}[b]{0.32\textwidth}\centering 
\captionsetup{skip=0pt}
 \begin{tikzpicture}[baseline=(i.base)]
    \begin{feynman}[small]
      \vertex (a) {\(\mathcal{P}_i\)};
      \vertex [right = 1.5cm of a] (i);
      \vertex [right = 1.5cm of i] (c) {\(\mathcal{P}_j\)};
      \vertex [below = 1.5cm of i] (j);
      \vertex [below = 1.5cm of a] (b) {\(\mathcal{B}_i\)};
      \vertex [below = 1.5cm of c] (d) {\(\mathcal{B}_j\)};
      \diagram* {
        (a) -- [charged scalar] (i) -- [charged scalar] (c), 
        (i) -- [boson,edge label={\(\mathcal{V}^*\)}] (j),
        (b) -- [fermion] (j) -- [fermion] (d),
       };
    \end{feynman}
  \end{tikzpicture}
\caption{}
\label{fig:free-mb-feynmandiagram_ps_a}
\end{subfigure}
\begin{subfigure}[b]{0.32\textwidth}\centering 
\captionsetup{skip=0pt}
 \begin{tikzpicture}[baseline=(i.base)]
    \begin{feynman}[small]
      \vertex (a) {\(\mathcal{P}_i\)};
      \vertex [right = 3cm of a] (c) {\(\mathcal{P}_j\)};
      \vertex [below = 1cm of a] (b) {\(\mathcal{B}_i\)};
      \vertex [below = 1cm of c] (d) {\(\mathcal{B}_j\)};
      \vertex [right = 1.2cm of b] (i);
      \vertex [right = 0.6cm of i] (j);
      \node [below right = 0.1cm of i] (bb) {\(\mathcal{B}^*\)};
      \diagram* {
        (a) -- [charged scalar] (i),
        (j) -- [charged scalar] (c), 
        (b) -- [fermion] (i) --  (j) -- [fermion] (d),
       };
    \end{feynman}
  \end{tikzpicture}
\caption{}
\label{fig:free-mb-feynmandiagram_ps_b}
\end{subfigure}
\begin{subfigure}[b]{0.32\textwidth}\centering 
\captionsetup{skip=0pt}
 \begin{tikzpicture}[baseline=(i.base)]
    \begin{feynman}[small]
      \vertex (a) {\(\mathcal{P}_i\)};
      \vertex [right = 3cm of a] (c) {\(\mathcal{P}_j\)};
      \vertex [below = 1cm of a] (b) {\(\mathcal{B}_i\)};
      \vertex [below = 1cm of c] (d) {\(\mathcal{B}_j\)};
      \vertex [right = 1.2cm of b] (i);
      \vertex [right = 0.6cm of i] (j);
      \node [below right = 0.1cm of i] (bb) {\(\mathcal{B}^*\)};
      \diagram* {
        (a) -- [charged scalar] (j),
        (i) -- [charged scalar] (c), 
        (b) -- [fermion] (i) --  (j) -- [fermion] (d),
       };
    \end{feynman}
  \end{tikzpicture}
\caption{}
\label{fig:free-mb-feynmandiagram_ps_c}
\end{subfigure}
 \caption{The leading-order tree-level diagrams contributing to the \gls{pseudoscalar-baryon} interaction in the (a) $t$-channel, (b) $s$-channel, and (c) $u$-channel. Baryons, pseudoscalar, and vector mesons are depicted by solid, dashed, and wiggly lines, respectively.}
 \label{fig:free-mb-feynmandiagram_ps}
\end{figure}

Next, we discuss the \gls{meson-baryon} scattering process, which we will consider to be  in $s$-wave, as the lowest term in the partial wave decomposition is the main contribution at low energies. The diagrams contributing to the interaction between a meson and a baryon at tree level are depicted in Fig.~\ref{fig:free-mb-feynmandiagram_ps}. In the case of the $s$-wave amplitude, the $t$-channel vector-meson exchange term of Fig.~\ref{fig:free-mb-feynmandiagram_ps_a} is the most important one due to \gls{vmd} on top of which the local hidden-gauge approach is built. The $s$- and $u$-channel terms of Figs.~\ref{fig:free-mb-feynmandiagram_ps_b} and \ref{fig:free-mb-feynmandiagram_ps_c}, respectively, contribute mainly to $p$-wave, and they may begin to take relevance as the energy increases. In the calculation of Ref.~\cite{Oller:2000fj} for the light $S=-1$, $C=0$ sector appropriate for the ${\bar K}N$ interaction generating the $\Lambda(1405)$, it is pointed out that these terms can reach around $20\%$ of the dominant $t$-channel contribution around $200$~MeV above the threshold. In the heavy $S=-2$, $C=1$ sector addressed here, one may expect a similar behavior, even reduced by the fact that the mass of the intermediate baryon in the $s$- or $u$-channel diagrams is more than twice that in the $S=-1$, $C=0$ sector. Therefore, one may safely conclude that the $s$- or $u$-channel contributions will be less than $10\%$ in the energy range $3000-3200$~MeV of interest here. 

Let us start with \gls{pseudoscalar-baryon}. The vertices that appear in the diagram of Fig.~\ref{fig:free-mb-feynmandiagram_ps_a}, coupling the vector meson to pseudoscalar mesons ($\mathcal{VPP}$) and baryons ($\mathcal{VBB}$), 
are described by the chiral effective Lagrangians within the hidden-gauge formalism introduced in Section~\ref{subsec:free-th-effective theories}. The first attempt to describe the scattering of Goldstone bosons off $1/2^+$ charmed baryons was carried out by the authors of Ref.~\cite{Lutz:2003jw}, but a substantial improvement was achieved with the generalization of the hidden-gauge approach to the charm sector by assuming an approximate $\textrm{SU}(4)$ symmetry, broken by the universal vector-meson coupling hypothesis \cite{Hofmann:2005sw}. The resulting $\textrm{SU}(4)$ Lagrangians read
\begin{equation}\label{eq:free-mb-vertexVPPsu4}
\mathcal{L}_{\mathcal{VPP}}^{\textrm{SU}(4)}=\ii g\langle\left[\partial_\mu\Phi_{[16]}, \Phi_{[16]}\right] V_{[16]}^\mu\rangle \ ,
\end{equation}
\begin{equation}\label{eq:free-mb-vertexBBVsu4}
\mathcal{L}_{\mathcal{VBB}}^{\textrm{SU}(4)}=\frac{g}{2}\sum_{i,j,k,l=1}^4\bar{B}_{ijk}\gamma^\mu\left(V_{\mu,l}^{{[16]},k}B^{ijl}+2V_{\mu,l}^{{[16]},j}B^{ilk}\right) \ ,
\end{equation}
where the symbol $\langle\cdots\rangle$ denotes the trace of the $\textrm{SU}(4)$ matrices in flavor space. The universal coupling constant $g$ is related to the pion decay constant $f_\pi=93~\textrm{MeV}$ by Eq.~(\ref{eq:free-th-KSFRrelation}), that we extend here to $\textrm{SU}(4)$:
\begin{equation}\label{eq:free-mb-g_coup}
g=\frac{m_V}{2f_\pi} \ ,
\end{equation}
with $m_V$ being a representative mass of the light (uncharmed) vector mesons. The authors of Ref.~\cite{Hofmann:2005sw} found that this expression in the limit of a light charm-quark mass is in accordance with the Kawarabayashi-Suzuki-Fayyazuddin-Riazuddin (or KSFR) relation \cite{Kawarabayashi:1966kd,Riazuddin:1966sw}.

Equations~(\ref{eq:free-mb-vertexVPPsu4}) and (\ref{eq:free-mb-vertexBBVsu4}) are the $\textrm{SU}(4)$-symmetric generalizations of Eqs.~(\ref{eq:free-th-LagHG_VPP}) and (\ref{eq:free-th-LagBBV}), respectively, and thus the $\Phi$ and $V_\mu$ matrices are those of Eqs.~(\ref{eq:free-mb-matrixphi}) and (\ref{eq:free-mb-matrixVmu}) and $B_{ijk}$ is the tensor given in Eq.~(\ref{eq:free-mb-baryons}).


Using the $\mathcal{VPP}$ and $\mathcal{VBB}$ vertices above one obtains the \gls{tvme} potential \cite{Hofmann:2005sw} between pseudoscalar mesons and baryons:
\begin{equation}\label{eq:free-mb-Vij_1}
 V_{ij}=g^2 \sum_v C_{ij}^v \bar{u}\left(p_j\right)\gamma^\mu u\left(p_i\right)\frac{1}{t-m_v^2} \left[\left(k_i+k_j\right)_\mu -\frac{k_i^2-k_j^2}{m_v^2}\left(k_i-k_j\right)_\mu\right] \ ,
\end{equation}
where $p_i$, $p_j$ ($k_i$, $k_j$) are the four-momenta of the baryons (mesons) in the $i$, $j$ channels, $t=(k_i-k_j)^2=(p_i-p_j)^2$ is the usual Mandelstam variable, and $m_v$ is the mass of the vector meson exchanged. The coefficients $C_{ij}^v$ are symmetric with respect to the indices and are related to the strength of the interaction between two \gls{meson-baryon} channels $i$ and $j$ mediated by the exchange of a vector meson $v$. The sign convention is such that a positive value of $C_{ij}^v$ implies an attractive potential, while a negative value implies repulsion. 
Adopting the same mass $m_v=m_V$ for the light vector mesons ($\rho$, $\omega$, $\phi$, $K^*$) and accounting for the higher mass of the charmed mesons ($D^*$, $D_s^*$) with a common multiplying factor $\kappa_c$,
\begin{equation}
 \kappa_c=\left(\frac{m_V}{m_V^c} \right)^2\approx \frac14 \ ,
\end{equation}
 as in \cite{Mizutani:2006vq}, Eq.~(\ref{eq:free-mb-Vij_1}) simplifies to
\begin{equation}\label{eq:free-mb-Vij_2}
 V_{ij}=-C_{ij}\frac{1}{4f^2}\bar{u}\left(p_j\right)\gamma^\mu u\left(p_i\right)\left(k_i+k_j\right)_\mu \ .
\end{equation}
The limit $t\ll m_V$ has been taken to reduce the $t$-channel diagram to a contact term as that depicted in Fig.~\ref{fig:free-mb-feynmandiagram_ps_wt}. The new coefficients $C_{ij}$  are obtained by summing the various vector-meson exchange contributions, $ C_{ij}=\sum_v C_{ij}^v$  \cite{Hofmann:2005sw}, including the factor $\kappa_c$ in the case of charmed mesons that breaks that $\textrm{SU}(4)$ symmetry. We note that the exchange of the $J/\psi$ meson should be suppressed further than the exchange of charmed vector mesons due to its larger mass, that is, with $\kappa_{c\bar{c}}=(m_V/m_{J/\psi})^2$, but we do not consider here the channels for which such transition is allowed.

\begin{figure}[t!]
\centering 
\begin{subfigure}[b]{0.32\textwidth}\centering 
\captionsetup{skip=0pt}
  \begin{tikzpicture}[baseline=(i.base)]
    \begin{feynman}[small]
      \vertex (a) {\(\mathcal{P}_i\)};
      \vertex [right = 1.5cm of a] (i);
      \vertex [right = 1.5cm of i] (c) {\(\mathcal{P}_j\)};
      \vertex [below = 1.5cm of i] (j);
      \vertex [below = 1.5cm of a] (b) {\(\mathcal{B}_i\)};
      \vertex [below = 1.5cm of c] (d) {\(\mathcal{B}_j\)};
      \diagram* {
        (a) -- [charged scalar] (j) -- [charged scalar] (c), 
        (b) -- [fermion] (j) -- [fermion] (d),
       };
     \draw[dot,minimum size=1mm,thick,fill=black] (j) circle(1mm);
    \end{feynman}
  \end{tikzpicture}
\caption{}
\label{fig:free-mb-feynmandiagram_ps_wt}
\end{subfigure}
\begin{subfigure}[b]{0.32\textwidth}\centering 
\captionsetup{skip=0pt}
   \begin{tikzpicture}[baseline=(i.base)]
    \begin{feynman}[small]
      \vertex (a) {\(\mathcal{V}_i\)};
      \vertex [right = 1.5cm of a] (i);
      \vertex [right = 1.5cm of i] (c) {\(\mathcal{V}_j\)};
      \vertex [below = 1.5cm of i] (j);
      \vertex [below = 1.5cm of a] (b) {\(\mathcal{B}_i\)};
      \vertex [below = 1.5cm of c] (d) {\(\mathcal{B}_j\)};
      \diagram* {
        (a) -- [charged boson] (j) -- [charged boson] (c), 
        (b) -- [fermion] (j) -- [fermion] (d),
       };
     \draw[dot,minimum size=1mm,thick,fill=black] (j) circle(1mm);
    \end{feynman}
  \end{tikzpicture}
\caption{}
\label{fig:free-mb-feynmandiagram_v_wt}
\end{subfigure}
 \caption{Contact terms of the (a) \gls{pseudoscalar-baryon} and (b) \gls{vector-baryon}  interactions.}
 \label{fig:free-mb-feynmandiagram_wt}
\end{figure}

We also note that the expression derived for the contact \gls{pseudoscalar-baryon} interaction of Eq.~(\ref{eq:free-mb-Vij_1}) is the same as that usually obtained from the lowest-order chiral Lagrangian of $\textrm{SU}(3)$ coupling the octet of light pseudoscalar mesons and the octet of $1/2^+$ baryons presented in Section~\ref{subsec:free-th-effective theories}. This kind of contact interaction, the so-called Weinberg-Tomozawa term, follows from the term with the covariant derivative in Eq.~(\ref{eq:free-th-LOChiralLagBaryons}).

In Eq.~(\ref{eq:free-mb-Vij_2}), $u(p_i)$ denotes the Dirac spinor of the incoming baryon with momentum $p_i$, while $\bar{u}(p_j)=u^\dagger(p_j)\gamma^0$ is the corresponding one for the outgoing baryon with momentum $p_j$. Next, one has to work out the Dirac algebra up to order ${\cal O}(p^2/M^2)$ and compute the $s$-wave component of these tree-level scattering amplitudes. The $s$-wave projected amplitude is computed as
\begin{equation}
  V^{ij}(s)= \frac{1}{2} \int_{-1}^{+1} d(\cos\theta) \ V^{ij}(s,t(s,\cos \theta)) P_{L=0} (\cos \theta) \ , \label{eq:free-mb-proj}
\end{equation}
where $t(s,\cos \theta)$ is given in terms of $s$, $\cos \theta$, and the hadron masses $m_k$ through the relation $u=\sum_{k=1}^4 m_k^2-s-t$. The function $P_L (\cos \theta)$ is the Legendre polynomial of order $L$ normalized to $\int_{-1}^{+1}  dx P_L (x) P_{L'} (x)=2\delta_{LL'}/(2L+1)$. In particular, for the $s$-wave projection we need $P_{L=0}=1$. The analysis of higher partial waves is not carried out here as the interaction in $p$-wave is expected to be weak compared to the $s$-wave interaction, especially for energies close to the thresholds, at which the molecular states can be potentially generated.

The resulting expression for the scattering amplitude reads:
\begin{equation}\label{eq:free-mb-Vij}
 V_{ij}(\sqrt{s})=-C_{ij}\frac{1}{4f^2}\left(2\sqrt{s}-M_i-M_j\right) N_i N_j \ 
\end{equation}
where $M_i$, $M_j$ and $E_i$, $E_j$ are the masses and the energies of the baryons and the normalization factors $N_i$, $N_j$ read $N=\sqrt{(E+M)/2M}$. Note that, while $\textrm{SU}(4)$ symmetry is encoded in the values of the coefficients $C_{ij}$, the interaction potential is not $\textrm{SU}(4)$ symmetric due to the use of physical masses for the mesons and baryons involved, as well as to the factor $\kappa_c$.  Actually, the transitions mediated by the exchange of light vector mesons do not make explicit use of the $\textrm{SU}(4)$ symmetry of the interaction \cite{Sakai:2017avl}. This is for instance the case of the dominant diagonal transitions, which are effectively projected into their $\textrm{SU}(3)$ content. Therefore, a moderate breaking of the $\textrm{SU}(4)$ symmetry in the transitions mediated by heavy vector mesons will have a limited effect on the results, as it will be shown in the next section.

The matrix of $C_{ij}$ coefficients for the resulting 5-channel \gls{pseudoscalar-baryon} interaction in isospin basis is given in Table~\ref{tab:free-mb-coeff}. The values of the coefficients for the exchange of a single vector meson $v$, $C_{ij}^v$, can be found in Appendix~\ref{sec:appendix:mb_coeff}. We note that the nondiagonal coefficients in which the $c$ quark is transferred from the meson to the baryon, or vice versa, that is, those that involve the exchange of a charmed vector meson, are suppressed by the factor $\kappa_c$.

\begin{table}[t!]
\setlength{\tabcolsep}{10pt}
\renewcommand{\arraystretch}{1.2}
\centering
\begin{tabular}{l | c c c c c}
\hline
  &{${\bar K}\Xi_c$}  & {${\bar K}\Xi_c^\prime$}  & { $D\Xi$}  & { $\eta\Omega_c$} &{$\eta^\prime\Omega_c$}  \\
\hline
 & & & & &  \\[-4.5mm]
{${\bar K}\Xi_c$}         & $1$ & $0$ & $\sqrt{\frac{3}{2}}\kappa_c$  & $0$ & $0$     \\
{${\bar K}\Xi_c^\prime$}   &     & $1$ & $\frac{1}{\sqrt{2}}\kappa_c$ & $-\sqrt{6}$ & $0$   \\
{ $D\Xi$} &     &     &    $2$        &  $-\frac{1}{\sqrt{3}}\kappa_c$  & $-\sqrt{\frac{2}{3}}\kappa_c$ \\
{ $\eta\Omega_c$} &     &     &     &  $0$  &  $0$\\
{ $\eta^\prime\Omega_c$}  &     &     &    &     &  $0$   \\             
\hline
\end{tabular}
\caption{The $C_{ij}$ coefficients of the \gls{pseudoscalar-baryon} interaction for the $(I,S,C)=(0,-2,1)$ sector.}
\label{tab:free-mb-coeff}
\end{table}

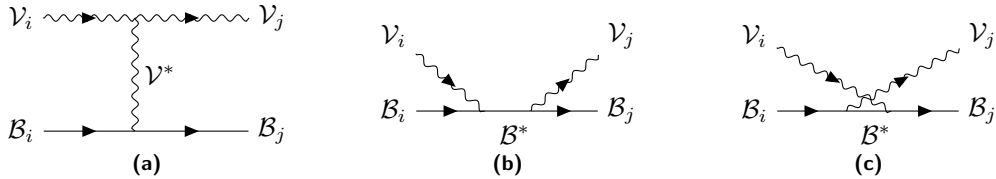
\begin{figure}[b!]
\centering 
\begin{subfigure}[b]{0.32\textwidth}\centering 
\captionsetup{skip=0pt}
 \begin{tikzpicture}[baseline=(i.base)]
    \begin{feynman}[small]
      \vertex (a) {\(\mathcal{V}_i\)};
      \vertex [right = 1.5cm of a] (i);
      \vertex [right = 1.5cm of i] (c) {\(\mathcal{V}_j\)};
      \vertex [below = 1.5cm of i] (j);
      \vertex [below = 1.5cm of a] (b) {\(\mathcal{B}_i\)};
      \vertex [below = 1.5cm of c] (d) {\(\mathcal{B}_j\)};
      \diagram* {
        (a) -- [charged boson] (i) -- [charged boson] (c), 
        (i) -- [boson,edge label={\(\mathcal{V}^*\)}] (j),
        (b) -- [fermion] (j) -- [fermion] (d),
       };
    \end{feynman}
  \end{tikzpicture}
\caption{}
\label{fig:free-mb-feynmandiagram_v_a}
\end{subfigure}
\begin{subfigure}[b]{0.32\textwidth}\centering 
\captionsetup{skip=0pt}
 \begin{tikzpicture}[baseline=(i.base)]
    \begin{feynman}[small]
      \vertex (a) {\(\mathcal{V}_i\)};
      \vertex [right = 3cm of a] (c) {\(\mathcal{V}_j\)};
      \vertex [below = 1cm of a] (b) {\(\mathcal{B}_i\)};
      \vertex [below = 1cm of c] (d) {\(\mathcal{B}_j\)};
      \vertex [right = 1.2cm of b] (i);
      \vertex [right = 0.6cm of i] (j);
      \node [below right = 0.1cm of i] (bb) {\(\mathcal{B}^*\)};
      \diagram* {
        (a) -- [charged boson] (i),
        (j) -- [charged boson] (c), 
        (b) -- [fermion] (i) --  (j) -- [fermion] (d),
       };
    \end{feynman}
  \end{tikzpicture}
\caption{}
\label{fig:free-mb-feynmandiagram_v_b}
\end{subfigure}
\begin{subfigure}[b]{0.32\textwidth}\centering 
\captionsetup{skip=0pt}
 \begin{tikzpicture}[baseline=(i.base)]
    \begin{feynman}[small]
      \vertex (a) {\(\mathcal{V}_i\)};
      \vertex [right = 3cm of a] (c) {\(\mathcal{V}_j\)};
      \vertex [below = 1cm of a] (b) {\(\mathcal{B}_i\)};
      \vertex [below = 1cm of c] (d) {\(\mathcal{B}_j\)};
      \vertex [right = 1.2cm of b] (i);
      \vertex [right = 0.6cm of i] (j);
      \node [below right = 0.1cm of i] (bb) {\(\mathcal{B}^*\)};
      \diagram* {
        (a) -- [charged boson] (j),
        (i) -- [charged boson] (c), 
        (b) -- [fermion] (i) --  (j) -- [fermion] (d),
       };
    \end{feynman}
  \end{tikzpicture}
\caption{}
\label{fig:free-mb-feynmandiagram_v_c}
\end{subfigure}
 \caption{The leading-order tree-level diagrams contributing to the \gls{vector-baryon} interaction in the (a) $t$-channel, (b) $s$-channel, and (c) $u$-channel. Baryons and vector mesons are depicted by solid and wiggly lines, respectively.}
 \label{fig:free-mb-feynmandiagram_v}
\end{figure}

The interaction of vector mesons with baryons is obtained following the formalism presented in Ref.~\cite{Oset:2009vf}, which is extended to $\textrm{SU}(4)$ here. Similarly as for pseudoscalar mesons, from the diagrams contributing to the tree-level \gls{vector-baryon} interaction (Fig.~\ref{fig:free-mb-feynmandiagram_v}), we only retain the $t$-channel vector-exchange term (Fig.~\ref{fig:free-mb-feynmandiagram_v_a}).  Employing the effective Lagrangian
\begin{align}\label{eq:free-mb-vertexVVVsu4}\nonumber
\mathcal{L}^{\textrm{SU}(4)}_{\mathcal{VVV}}&=\ii g \langle (\partial_\mu V_\nu^{[16]}-\partial_\nu V_\mu^{[16]})V_{[16]}^\mu V_{[16]}^\nu\rangle \\
&=\ii g\langle {[V_{[16]}^\mu,\partial_\nu V^{[16]}_\mu] V_{[16]}^\nu}\rangle 
\end{align}
for the three-vector $\mathcal{VVV}$ vertex and that of Eq.~(\ref{eq:free-mb-vertexBBVsu4}) for the $\mathcal{VBB}$ one, and under the approximation of small momentum transfer $q=t$ in which we neglect the $q^2/m_V^2$ term in the propagator of the exchanged vector, the resulting interaction kernel for the contact \gls{vector-baryon} interaction of Fig.~\ref{fig:free-mb-feynmandiagram_v_wt} is identical to that obtained for the \gls{pseudoscalar-baryon} one (see Eq.~(\ref{eq:free-mb-Vij})), multiplied by the product of polarization vectors, $\vec{\epsilon}_i\cdot\vec{\epsilon}_j$. Contrary to the \gls{pseudoscalar-baryon} case, in which there is only one vector meson in the $\mathcal{PPV}$ vertex and this must necessarily correspond to the exchanged vector meson in the $t$-channel diagram, in the \gls{vector-baryon} case any of the three vectors of the $\mathcal{VVV}$ vertex could be in principle the vector exchanged in the diagram of Fig.~\ref{fig:free-mb-feynmandiagram_v_a}. Nevertheless, the assumption of neglecting the three-momenta of the external mesons in comparison with the vector-meson mass implies that the polarization vectors of the external vector mesons have only spatial components, $\epsilon_i\rightarrow\vec{\epsilon}_i$. Then, by rewriting the Lagrangian of Eq.~(\ref{eq:free-mb-vertexVVVsu4}) as
\begin{equation}\label{eq:free-mb-vertexVVVsu4_2}
\mathcal{L}^{\textrm{SU}(4)}_{\mathcal{VVV}}=\ii g\langle {(V_{[16]}^\mu\partial_\nu V^{[16]}_\mu -\partial_\nu V^{[16]}_\mu V_{[16]}^\mu )V_{[16]}^\nu}\rangle \ ,
\end{equation}
one can see that the field $V^\nu_{[16]}$ cannot correspond to an external vector meson because the $\nu$ index would be spatial and the derivative $\partial_\nu$ would be replaced by the three-momentum of the vector meson, which is neglected in this approach. 
 
The allowed \gls{vector-baryon} states are listed in the right columns of Table~\ref{tab:free-mb-channels}, where, again, we neglect the states with three heavy quarks. The corresponding coefficients $C_{ij}$ are listed in Table~\ref{tab:free-mb-coeff-v} and can be straightforwardly obtained from those for the \gls{pseudoscalar-baryon} interaction in Table~\ref{tab:free-mb-coeff}, by considering the following correspondences: 
\begin{equation}
 \begin{matrix}
  \pi \rightarrow \rho, & K\rightarrow K^\ast, & \bar{K}\rightarrow\bar{K}^\ast, & D\rightarrow D^\ast, & \bar{D}\rightarrow\bar{D}^\ast, 
 \end{matrix}
\end{equation}
\begin{equation}
 \begin{matrix}
  \frac{1}{\sqrt{3}}\eta+\sqrt{\frac{2}{3}}\eta^\prime\rightarrow\omega {\qquad\textrm{and}\;} & -\sqrt{\frac{2}{3}}\eta+\frac{1}{\sqrt{3}}\eta^\prime\rightarrow\phi\ .
 \end{matrix}
 \label{eq:free-mb-eta_phi}
\end{equation}

\begin{table}[t!]
\setlength{\tabcolsep}{10pt}
\renewcommand{\arraystretch}{1.2}
\centering
\begin{tabular}{l | c c c c c}
\hline
  &{ $D^*\Xi$}  & {${\bar K}^*\Xi_c$}  & {${\bar K}^*\Xi_c^\prime$}  & { $\omega\Omega_c$} &{$\phi\Omega_c$}  \\
\hline
 & & & & &  \\[-4.5mm]
{ $D^*\Xi$}        & $2$ & $\sqrt{\frac{3}{2}}\kappa_c$ & $\frac{1}{\sqrt{2}}\kappa_c$  & $-\kappa_c$ & $0$     \\
{${\bar K}^*\Xi_c$}   &     & $1$ & $0$ & $0$ & $0$   \\
{${\bar K}^*\Xi_c^\prime$}&     &     &    $1$        &  $-\sqrt{2}$  & $2$ \\ 
{ $\omega\Omega_c$} &     &     &     &  $0$  &  $0$\\
{ $\phi\Omega_c$}  &     &     &    &     &  $0$   \\             
\hline
\end{tabular}
\caption{The $C_{ij}$ coefficients of the \gls{vector-baryon} interaction for the $(I,S,C)=(0,-2,1)$ sector.}
\label{tab:free-mb-coeff-v}
\end{table}

The scattering amplitude of Eq.(\ref{eq:free-mb-Vij_2}) is unitarized via the coupled-channel \gls{bs} equation given in Eq.~(\ref{eq:free-th-BSeq}) and reproduced here, 
\begin{equation}\label{eq:free-mb-BSeq}
T_{ij}=V_{ij}+V_{il}G_{l}T_{lj} \ ,
\end{equation}
which implements the resummation of loop diagrams to infinite order as described in Section~\ref{subsec:free-th-unitarization}. 
The sought resonances are generated as poles of the unitarized scattering amplitude.
In the \gls{vector-baryon} case, in the iteration of diagrams implicit in the \gls{bs} equation, the $\vec{\epsilon}_i\cdot\vec{\epsilon}_j$ factor of the interaction forces the vector meson in the loops to propagate with spatial components. Then, the sum 
over the polarizations of the internal vector mesons gives \cite{Sarkar:2010saz,Oset:2009vf}
\begin{equation}
 \sum_{\textrm{pol}}\epsilon_i\epsilon_j=\delta_{ij}+\frac{q_iq_j}{m_V^2} \ .
\end{equation}
The factor $\vec{\epsilon}_i\cdot\vec{\epsilon}_j$ can be factorized out in all the terms of the \gls{bs} equation in the on-shell factorization approach \cite{Oset:1997it}, leading to \cite{Roca:2005nm}
\begin{equation}
 T=(\mathbb{1}-V\hat{G})^{-1}V\vec{\epsilon}\cdot\vec{\epsilon}\,' \ ,
\end{equation}
where
\begin{equation}
 \hat{G}=G\left(1+\frac13\frac{\vec{q}\,^2}{m_V^2}\right) \ ,
\end{equation}
and the correction $\sim q^2/(3m_V^2)$ can be neglected considering the approximations done so far.

The meson--baryon loop function is constructed from the meson and baryon propagators as
 \begin{equation}\label{eq:free-mb-Gmatrix}
G_{l}=\ii\int \frac{d^4q}{(2\pi)^4}\frac{2M_l}{(P-q)^2-M_l^2+\ii\varepsilon}\frac{1}{q^2-m_l^2+\ii\varepsilon} \ ,
\end{equation}
where $M_l$ corresponds to the mass of the intermediate baryon, $m_l$ is the mass of the intermediate meson, $P=(\sqrt{s},\vec{0}\,)$ is the total four-momentum of the system in the center-of-mass frame, and $q$ denotes the four-momentum of the meson propagating in the intermediate loop. We note that we have taken the nonrelativistic approximation of Eq.~(\ref{eq:free-th-bprop}) for the baryon propagator.

This function diverges for $|\vec{q}\,|\rightarrow\infty$ and it must be regularized with a proper scheme. As discussed in Section~\ref{subsec:free-th-unitarization}, one may employ the cut-off regularization method, which consists in replacing the infinite upper limit of the three-momentum integral with a large enough cut-off momentum $\Lambda$. After the $q^0$ contour integration, the two-meson propagator regularized with a momentum cut-off reads
\begin{equation}\label{eq:free-mb-Gmatrixcut}
 G_{l}^{\textrm{cut}}=\int_{0}^{\Lambda}\frac{d^3q}{(2\pi)^3}\frac{1}{2\omega_l(\vec{q}\,)}\frac{M_l}{E_l(\vec{q}\,)}\frac{1}{\sqrt{s}-\omega_l(\vec{q}\,)-E_l(\vec{q}\,)+\ii\varepsilon} \ ,
\end{equation}
where $E_l=\sqrt{\vec{q}\,^2+M_l^2}$ and $\omega_l=\sqrt{\vec{q}\,^2+m_l^2}$ are the energies of the intermediate baryon and meson, respectively. 

The loop function of Eq.~(\ref{eq:free-mb-Gmatrixcut}) admits an analytical expression \cite{Oller:1998hw}
\begin{align}\label{eq:free-mb-Gmatrixcutanalytic}\nonumber
  G_{l}^{\textrm{cut}}(s)&=\frac{2M_l}{32\pi^2}\Bigg\{\ln\left(\frac{m_l^2M_l^2}{\Lambda^4}\right)-\frac{M_l^2-m_l^2}{s}\ln\left(\frac{m_l^2}{M_l^2}\right)\\ \nonumber
  +&2\frac{M_l^2-m_l^2}{s}\ln\left(\frac{1+\sqrt{1+m_l^2\Lambda^{-2}}}{1+\sqrt{1+M_l^2\Lambda^{-2}}}\right)-2\ln\left[\left(1+\sqrt{1+\frac{m_l^2}{\Lambda^2}}\right)\left(1+\sqrt{1+\frac{M_l^2}{\Lambda^2}}\right)\right] \\  \nonumber
  +&\frac{2q_l}{\sqrt{s}}\left[\ln\left(s-(M_l^2-m_l^2)+2\sqrt{s}\,q_l\sqrt{1+m_l^2\Lambda^{-2}}\right)\right. \\ \nonumber 
  &\quad+\ln\left(s+(M_l^2-m_l^2)+2\sqrt{s}\,q_l\sqrt{1+M_l^2\Lambda^{-2}}\right) \\ \nonumber
  &\quad-\ln\left(-s+(M_l^2-m_l^2)+2\sqrt{s}\,q_l\sqrt{1+m_l^2\Lambda^{-2}}\right)\\ 
  &\left.\quad-\ln\left(-s-(M_l^2-m_l^2)+2\sqrt{s}\,q_l\sqrt{1+M_l^2\Lambda^{-2}}\right)\right] \Bigg\} \ ,
\end{align}
where we have designated as $q_l\equiv|\vec{q}\,|$ the relative momentum between the hadrons of channel $l$ propagating in the loop and which defined in Eq.~(\ref{eq:free-th-relmom}).

In the alternative \gls{dr} approach, which is the one adopted in this section, the \gls{meson-baryon} loop function reads:
\begin{align}\label{eq:free-mb-GmatrixDR} \nonumber
 G_{l}^{\textrm{DR}}=&\frac{2M_l}{16\pi^2}\Big\{ a_l(\mu)+\ln\frac{M_l^2}{\mu^2}+\frac{m_l^2-M_l^2+s}{2s}\ln\frac{m_l^2}{M_l^2} \\  \nonumber
  &+\frac{q_l}{\sqrt{s}}\left[ \ln\left(s-(M_l^2-m_l^2)+2q_l\sqrt{s}\right)+\ln\left(s+(M_l^2-m_l^2)+2q_l\sqrt{s}\right) \right.\\
  &\left. -\ln\left(-s+(M_l^2-m_l^2)+2q_l\sqrt{s}\right)-\ln\left(-s-(M_l^2-m_l^2)+2q_l\sqrt{s}\right) \right] \Big\} \ ,
\end{align}
where $a_l(\mu)$ is the channel-dependent subtraction constant at the regularization scale $\mu$.
The choice of the regularization scale $\mu$ and the corresponding subtraction constants $a_l(\mu)$ can be obtained by demanding that, at an energy close to the channel threshold, $G_l^{\textrm{DR}}$ is similar to $G_l^\textrm{cut}$ for a certain cut-off $\Lambda$, namely
\begin{equation}\label{eq:free-mb-a(mu)}
 a_l({\mu})= \frac{16\pi^2}{2M_l}\left(G_{l}^{\textrm{cut}}(s_l^{\textrm{thr}},\Lambda)-G_{l}^{\textrm{DR}}(s_l^{\textrm{thr}},\mu,a_l=0)\right) \ ,
\end{equation}
for a given $\mu$. Notice that the running of $a_l(\mu)$ cancels the explicit $\mu$ dependence in Eq.~(\ref{eq:free-mb-GmatrixDR}), so the loop function does not depend on the regularization scale.

The coupling $g$ of Eq.~(\ref{eq:free-mb-g_coup}) could have been considered dependent on the energy through the incorporation of additional form factors at the vertices of the $t$-channel diagram of Fig.~\ref{fig:free-mb-feynmandiagram_ps}. This is not usually done in the works that employ the on-shell factorization approach to solve the \gls{bs} equation, as done in this thesis, since this poses additional technical difficulties. However, practically speaking, the lack of the form factor in our formalism can be compensated by an appropriate value of the equivalent cut-off employed.

The expression for the loop function $G_l^{\textrm{DR}}$ in Eq.~(\ref{eq:free-mb-GmatrixDR}) assumes that the baryon and the meson have fixed masses and no width.
When the \gls{bs} equation involves channels that include particles with a large width, which is the case of the $\rho$ ($\Gamma_\rho=149.4~\textrm{MeV}$) and $K^\ast$ ($\Gamma_{K^\ast}~=~50.5~\textrm{MeV}$) mesons, this function has to be convoluted with the mass distribution of the particle. Following the method described in \cite{Oset:2009vf}, the loop function in these cases is replaced by
\begin{equation}\label{eq:free-mb-loop_conv}
 \tilde{G}_{l}^{\textrm{DR}}(s)=-\frac{1}{N}\int_{(m_l-2\Gamma_l)^2}^{(m_l+2\Gamma_l)^2} \frac{d\tilde{m}_l^2}{\pi}{\, \textrm{Im}\,}\frac{1}{\tilde{m}_l^2-m_l^2+\ii m_l\Gamma(\tilde{m}_l)}
 G_{l}^{\textrm{DR}}\left(s,\tilde{m}_l^2,M_l^2\right),
\end{equation}
where the limits of the integral have been taken to extend over a couple of times the width of the meson, and the normalization factor $N$ reads
\begin{equation}
 N=\int_{(m_l-2\Gamma_l)^2}^{(m_l+2\Gamma_l)^2}d\tilde{m}_l^2\left(-\frac{1}{\pi}\right){\, \textrm{Im}\,}\frac{1}{\tilde{m}_l^2-m_l^2+\ii m_l\Gamma(\tilde{m}_l)} \ .
\end{equation}
The energy-dependent width $\Gamma(\tilde{m}_l)$ is given by
\begin{equation}\label{eq:G_kallen}
 \Gamma(\tilde{m}_l)=\Gamma_l\frac{m_l^5}{\tilde{m}_l^5}\frac{\lambda^{3/2}(\tilde{m}_l^2,m_1^2,m_2^2)}{\lambda^{3/2}(m_l^2,m_1^2,m_2^2)}\,\theta_{\textrm{H}}(\tilde{m}_l-m_1-m_2),
\end{equation}
where $\lambda(x,y,z)=(x-(\sqrt{y}+\sqrt{z})^2)(x-(\sqrt{y}-\sqrt{z})^2)$ is the K\"{a}ll\'{e}n function, $m_1$ and $m_2$ are the masses of the lighter mesons to which the vector meson in the loop decays, that is, $m_1=m_2=m_\pi$ for the $\rho$ and $m_1=m_\pi$, $m_2=m_K$ for the $K^\ast$, and $\theta_ {\textrm{H}}$ is the usual Heaviside step function. 

The bound states and resonances generated dynamically from the coupled-channel \gls{meson-baryon} interaction appear as poles of the scattering amplitude in the \gls{rs}-I and the \gls{rs}-II of the complex-energy plane, respectively. See the discussion in Section~\ref{subsec:free-th-dynamicallygeneratedstates}, where definitions are also given for the coupling constants $g_i$ of the resonance to the various channels $i$ (Eq.~(\ref{eq:free-th-couplings})), and for the amount of $i^{\textrm{th}}$-channel meson--baryon component in a given resonance as given by the compositeness (Eq.~(\ref{eq:free-th-compositeness})).


%
%

\subsection{Results in the open-charm sector: $\Omega_c^{*0}$ states}
\label{subsec:free-mb-omegac}

Now we proceed to present the results obtained employing the unitarized model for \gls{meson-baryon} scattering in coupled channels described above and which we published in Ref.~\cite{Montana:2017kjw}. This section is divided into two parts: first, we describe the results obtained with the \gls{pseudoscalar-baryon} interaction kernel of  Eq.~(\ref{eq:free-mb-Vij}) corresponding to $J^P=\frac12^-$ dynamically generated states, and later we present the corresponding results for the \gls{vector-baryon} scattering case leading to the dynamical generation of degenerate $J^P=\frac12^-,\,\frac32^-$ states.

\subsubsection{$0^-\oplus\frac12^+$ states}
A specific effective model for the scattering of pseudoscalar mesons with baryons in the sector with quantum numbers $(I,S,C)=(0,-2,1)$ is built by fixing a regularization scheme for the loop function $G_l$ in Eq.~(\ref{eq:free-mb-BSeq}) and determining the values of the parameters, that is, $\Lambda$ in the case of using a hard cut-off, or the subtraction constants $a_l(\mu)$ when using the \gls{dr} scheme.
We use the latter one here and determine the subtraction constants of ``Model~1'' for a regularization scale of $\mu=1$~GeV by imposing the loop function of each \gls{pseudoscalar-baryon} channel to coincide, at the corresponding threshold, with the cut-off loop function evaluated for $\Lambda=800$~MeV, using Eq.~(\ref{eq:free-mb-a(mu)}). 
This value roughly corresponds to the mass of the exchanged vector mesons in the $t$-channel diagram that has been eliminated in favor of the contact interaction employed in this model and therefore it can be regarded to be a natural choice.
The values obtained with this prescription are listed in the upper part of Table~\ref{tab:free-mb-a_pseudo}.

\begin{table}[b!]
\setlength{\tabcolsep}{10pt}
\renewcommand{\arraystretch}{1.2}
\centering
\begin{tabular}{lccccc}
\hline 
    &  $a_{ \bar{K}\Xi_c}$  &  $a_{\bar{K}\Xi'_c}$  &  $a_{D\Xi }$  &  $a_{\eta \Omega_c}$  & $a_{\eta' \Omega_c }$ \\
\hline                           
   Model~1     &  $-2.19$  &  $-2.26$  &  $-1.90$  & $-2.31$  &  $-2.26$  \\
   $\Lambda$ (MeV) & $800$ & $800$ & $800$ & $800$ & $800$ \\
   \hline
    Model~2     &  $-1.69$  &  $-2.09$  &  $-1.93$ &  $-2.46$  &  $-2.42$  \\
       $\Lambda$ (MeV) & $320$ & $620$ & $830$ & $980$ & $980$ \\
\hline
\end{tabular}
\caption{Values of the subtraction constants at a regularization scale $\mu=1$ GeV and the equivalent cut-off $\Lambda$ for the two models discussed in the text.} 
\label{tab:free-mb-a_pseudo}
\end{table}

The scattering amplitude $T$ resulting from solving the \gls{bs} equation shows two poles on the \gls{rs}-II that have the following properties:
\begin{align}
M_1 &=  {\textrm{Re}\,}z_1= 3051.6~\textrm{MeV}\ , \quad\Gamma_1 = -2\, {\textrm{Im}\,}z_1= 0.45~\textrm{MeV}\ , \nonumber \\
M_2 &=  {\textrm{Re}\,}z_2 = 3103.3~\textrm{MeV}\ , \quad\Gamma_2 = -2\, {\textrm{Im}\,}z_2= 17~\textrm{MeV} \ .
\label{eq:free-mb-model1}
\end{align}
These resonances have spin-parity $J^P=1/2^-$, as they are obtained from the scattering amplitude of pseudoscalar mesons with baryons of the ground-state octet in $s$-wave. 
The couplings of each resonance to the various \gls{pseudoscalar-baryon} channels are displayed in Table~\ref{tab:free-mb-pseudo} under the label ``Model~1'', where one can also find the corresponding values of the compositeness. We observe that the lowest-energy state at $3052$~MeV couples appreciably to the channels $\bar{K}\Xi'_c$, $D\Xi $ and  $\eta \Omega_c$. Although the coupling to  $\eta \Omega_c$ states is the strongest, the compositeness is larger in the $\bar{K}\Xi'_c$ channel, to which the resonance also couples strongly, and its threshold lies closer to the pole. The higher-energy resonance at $3103$~MeV, with a strong coupling to $D\Xi$ and a compositeness in this channel of $0.90$, clearly qualifies as being a $D\Xi $ quasi-bound state.

\begin{table}[t!]
\setlength{\tabcolsep}{10pt}
\renewcommand{\arraystretch}{1.2}
\centering
\begin{tabular}{l|cc|cc}
\hline
  &  \multicolumn{4}{c}{Model~1}   \\ 
\hline
$M$ (MeV)     &    \multicolumn{2}{c|}{$3051.6$}    &   \multicolumn{2}{c}{$3103.3$}  \\
$\Gamma$ (MeV)   &     \multicolumn{2}{c|}{$0.45$}  &     \multicolumn{2}{c}{$17$}   \\ 
\hline
&   $| g_i|$   & $-g_i^2 dG/dE$  &   $| g_i|$   & $-g_i^2 dG/dE$     \\
\hline
$\bar{K}\Xi_c\, (2964)$   &  $0.11$  & $0.00+\ii 0.00$  &   $0.58$  & $0.01+\ii 0.03$      \\
$\bar{K}\Xi'_c\, (3072)$  &  $1.67$  & $0.54+\ii 0.01$   &   $0.30$  & $0.01-\ii 0.01$      \\
$D\Xi\, (3185)$           &  $1.10$  & $0.05-\ii 0.01$  &   $4.08$  & $0.90-\ii 0.05$      \\
$\eta \Omega_c\, (3245)$  &  $2.08$  & $0.23+\ii 0.00$    &   $0.44$  & $0.01+\ii 0.01$      \\
$\eta' \Omega_c\, (3655)$ &  $0.04$  & $0.00+\ii 0.00$    &   $0.28$  & $0.00+\ii 0.00$      \\
\hline
 &  \multicolumn{4}{c}{Model~2} \\ 
\hline 
$M$ (MeV)    &    \multicolumn{2}{c|}{$3050.3$}    &    \multicolumn{2}{c}{$3090.8$}  \\
$\Gamma$ (MeV)   &  \multicolumn{2}{c|}{$0.44$}     &    \multicolumn{2}{c}{$12$}   \\ 
\hline
&      $| g_i|$    & $-g_i^2 dG/dE$    &   $| g_i|$  & $-g_i^2 dG/dE$ \\
\hline
$\bar{K}\Xi_c\, (2964)$   &  $0.11$  & $0.00+\ii 0.00$     &    $0.49$  & $-0.02+\ii 0.01$  \\
$\bar{K}\Xi'_c\, (3072)$  &  $1.80$  & $0.61+\ii 0.01$     &    $0.35$  & $0.02-\ii 0.02$  \\
$D\Xi\, (3185)$           &  $1.36$  & $0.07-\ii 0.01$     &    $4.28$  & $0.91-\ii 0.01$ \\
$\eta \Omega_c\, (3245)$  &  $1.63$  & $0.14+\ii 0.00$     &    $0.39$  & $0.01+\ii 0.01$ \\
$\eta' \Omega_c\, (3655)$ &  $0.06$  & $0.00+\ii 0.00$     &    $0.28$  & $0.00+\ii 0.00$ \\
\hline
\end{tabular}
\caption{The $\Omega^{*0}_c$ ($1/2^-$) states dynamically generated employing zero-range interactions (Model~1, Model~2) between a pseudoscalar meson ($0^-$) and a ground-state baryon ($1/2^+$), within a coupled-channel approach, for the two models described in the text.}
\label{tab:free-mb-pseudo}
\end{table}

The energies at which these resonances appear are very similar to the 
second and fourth $\Omega_c^{*0}$ states discovered by LHCb \cite{Aaij:2017nav} and which we have listed in Table~\ref{tab:free-mb-explhc}.
Even if the mass of our heavier state is larger by $10$ MeV and its width is about twice the experimental one, our results clearly show the ability of the \gls{meson-baryon} dynamical models for generating states in the energy range of interest. 

Let us comment on the differences between our results and those of previous works in the literature. The approach developed in Ref.~\cite{Hofmann:2005sw}, on which our model is based, imposes approximate crossing symmetry by demanding the loop functions to vanish at a subtraction scale $\sqrt{m_{\textrm{th}}^2+M_{\textrm{th}}^2}$, where $m_{\textrm{th}}$ and $M_{\textrm{th}}$ are, respectively, the mass of the meson and the baryon for the lightest \gls{meson-baryon} state of the sector considered. Translating the values of the subtraction constants resulting from this prescription to equivalent cut-off values, we find that they turn out to be generally larger than $1000$~MeV, even reaching a value of $1650$~MeV for the highest-mass channel. This is the reason why all the $\Omega_c^{*0}$ states generated in Ref.~\cite{Hofmann:2005sw} appear below $3000$~MeV. The comparison with the work of Ref.~\cite{JimenezTejero:2009vq} is less straightforward, as the Lipmann-Schwinger equation is solved there with a nonlocal kernel and momentum-dependent form factors are employed, while here we solve the \gls{bs} equation in its on-shell factorized form with a local interaction. In any case, the work of Ref.~\cite{JimenezTejero:2009vq} employs a common cut-off value, taken from the analysis of their model in the $I=0$, $S=0$, $C=1$ sector, finding one $\Omega_c^{*0}$ state in the region of interest, at $3117$~MeV, in addition to another state at lower energy, at $2966$~MeV, and hence located outside the range of states found by the LHCb collaboration.

\begin{figure}[b!]
\centering
  \includegraphics[width=0.7\textwidth]{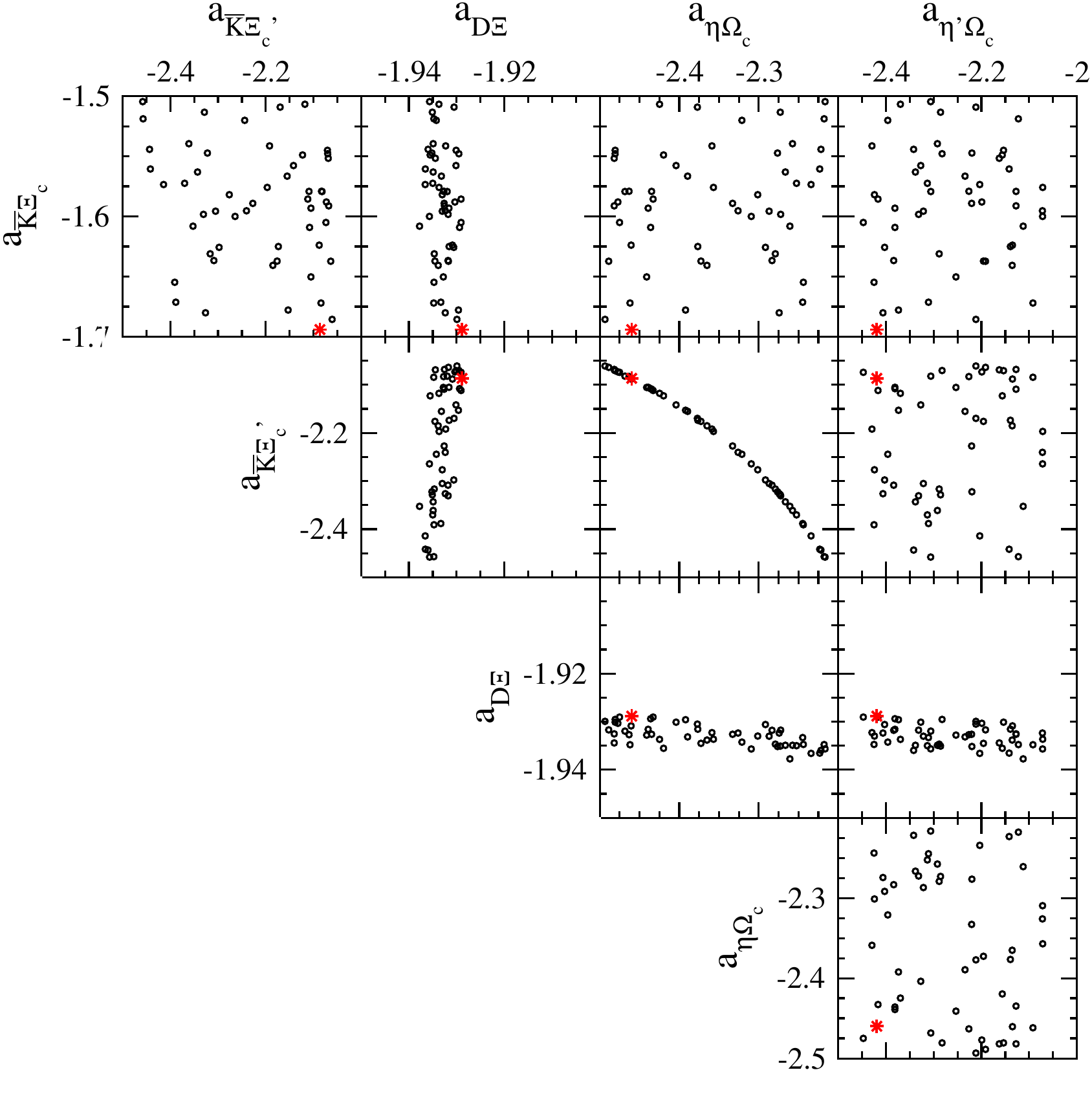}
\caption{Correlations between the various subtraction constants. The circles represent different configurations of subtraction constants that reproduce the experimental resonances $\Omega_c(3050)^0$ and $\Omega_c(3090)^0$, within $2\sigma$ of the experimental errors. The red asterisks denote one particular representative set.  } 
  \label{fig:free-mb-a_corr}
\end{figure}

In an attempt to accommodate better to the experimental data, we relax the condition of forcing that each loop function calculated with \gls{dr} matches, at the corresponding threshold, the cut-off loop function evaluated for $\Lambda=800$~MeV.
To this end, we let the values of the five subtraction constants vary freely within a reasonably constrained range and look for sets that reproduce the characteristics of the two observed states, $\Omega_c(3050)^0$ and $\Omega_c(3090)^0$, within $2\sigma$ of the experimental errors (see Table~\ref{tab:free-mb-explhc}). In order to analyze the correlations, we represent in Fig.~\ref{fig:free-mb-a_corr} the values of each subtraction constant against all the others in the sets that comply with the experimental constraints. We observe an anti-correlation between the subtraction constants $a_{\bar{K}\Xi'_c}$ and $a_{\eta \Omega_c}$. This can be simply understood by noting that the resonance at $3050$~MeV couples mostly to these two meson--baryon states, as can be seen from the results in Table~\ref{tab:free-mb-pseudo}, implying that, if one subtraction constant becomes more negative, favoring a stronger attraction for the pole, the other subtraction constant needs to compensate this effect by being less negative. We also find the subtraction constant $a_{D\Xi}$ to acquire a rather stable value between -1.94 and -1.93. This is a reflection of the resonance at $3090$~MeV being essentially a ${D\Xi}$ bound state, which requires a particular value of the subtraction constant $a_{D\Xi}$ to generate the pole at the appropriate experimental energy. 

Among all the possible configurations of subtraction constants producing the experimental $\Omega_c^{*0}$ states at $3050$~MeV and $3090$~MeV represented in Fig.~\ref{fig:free-mb-a_corr}, we select a representative set (``Model~2''), denoted by red asterisks in the figure, the values of which are listed in the bottom part of Table~\ref{tab:free-mb-a_pseudo}. 
The properties of the two poles obtained with this model are:
\begin{align}
M_1 &=  {\textrm{Re}\,}z_1= 3050.3~\textrm{MeV}\ , \quad\Gamma_1 = -2\, {\textrm{Im}\,}z_1= 0.44~\textrm{MeV}\ , \nonumber \\
M_2 &=  {\textrm{Re}\,}z_2 = 3090.8~\textrm{MeV}\ , \quad\Gamma_2 = -2\, {\textrm{Im}\,}z_2= 12~\textrm{MeV} \ .
\label{eq:free-mb-model2}
\end{align}
which are similar for any of the sets of subtracting constants represented in Fig.~\ref{fig:free-mb-a_corr}. The stronger changes with respect to the properties of the poles generated with ``Model~1'' are found in the higher resonance. Apart from having been lowered to the experimental energy, its width has been substantially decreased to agree with the experiment at the $2\sigma$ level.
We see from Table~\ref{tab:free-mb-a_pseudo} that the equivalent values of the cut-off for this new set of subtracting constants now lie
in the range $[320-950]$~MeV. The strongest change corresponds to the subtraction constant $a_{\bar{K}\Xi_c}$, needed to decrease the width of the $\Omega_c(3090)^0$ towards its experimental value. The equivalent cut-off value of $320$~MeV is on the low side of the usually employed values, but it is still naturally sized.

The five $\Omega_c^{*0}$ states were observed from the $K^-\Xi^+_c$ invariant mass spectrum obtained from a sample of $pp$ collision data at center-of-mass energies of $7$, $8$, and $13$~TeV, recorded by the LHCb experiment \cite{Aaij:2017nav}. To model such a spectrum from the elementary $pp$ collision reaction is a tremendously difficult task, but we can give a taste of the spectrum that our models would predict by representing, as a function of the $\bar{K}\Xi_c$ center-of-mass energy, the sum of the amplitudes for the $i \to \bar{K}\Xi_c$ transition, with $i$ being any of the five coupled channels involved in this sector, multiplied by the momentum of the $\bar{K}$ in the $\bar{K}\Xi_c$ center-of-mass frame,
\begin{equation}
q_{\bar{K}} \left| \sum_{i} T_{i\to \bar{K}\Xi_c} \right| ^2 \ .
\label{eq:free-mb-t2}
\end{equation}

\begin{figure}[b!]
\centering
  \includegraphics[width=0.5\textwidth]{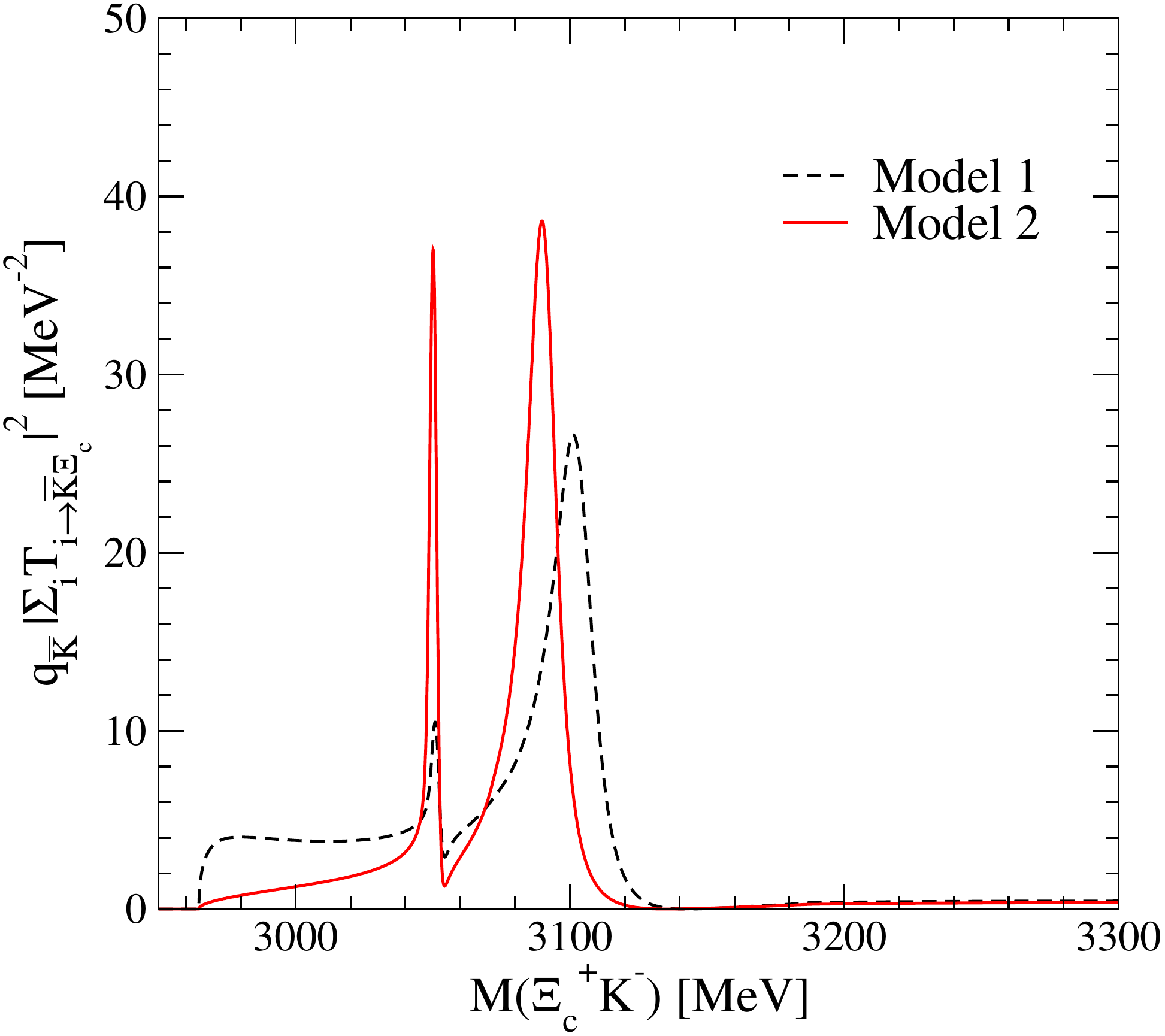}
\caption{Spectrum obtained from the sum of amplitudes squared times a phase space factor, for the two models discussed in the text.} 
  \label{fig:free-mb-t2}
\end{figure}

This is shown in Fig.~\ref{fig:free-mb-t2} for ``Model~1'' (black dashed line) and ``Model~2'' (red solid line). The momentum $q_{\bar{K}}$ acts as a phase-space modulator. The calculated spectrum has been convoluted with the energy-dependent resolution of the experiment, which runs linearly from $0.75$~MeV at $3000$~MeV to $1.74$~MeV at $3119$~MeV, employing a Gaussian function. 
In front of each amplitude $T_{i\to \bar{K}\Xi_c}$ in Eq.~(\ref{eq:free-mb-t2}), one should include in principle a coefficient gauging the strength with which the production mechanism excites the particular \gls{pseudoscalar-baryon} channel $i$. Nevertheless, given the limited understanding of the production dynamics, we have assumed all these coefficients to be equal. Therefore, the spectrum displayed in Fig.~\ref{fig:free-mb-t2} is merely illustrative as it also lacks the background contributions. However, one can still see certain similarities with the experimental spectrum shown in the left panel of Fig.~\ref{fig:free-mb-omegacLHCb}, extracted from Ref.~\cite{Aaij:2017nav}, in the energy regions of the $3050$~MeV and $3090$~MeV states.

\begin{figure}[b!]
\centering
  \includegraphics[width=0.75\textwidth]{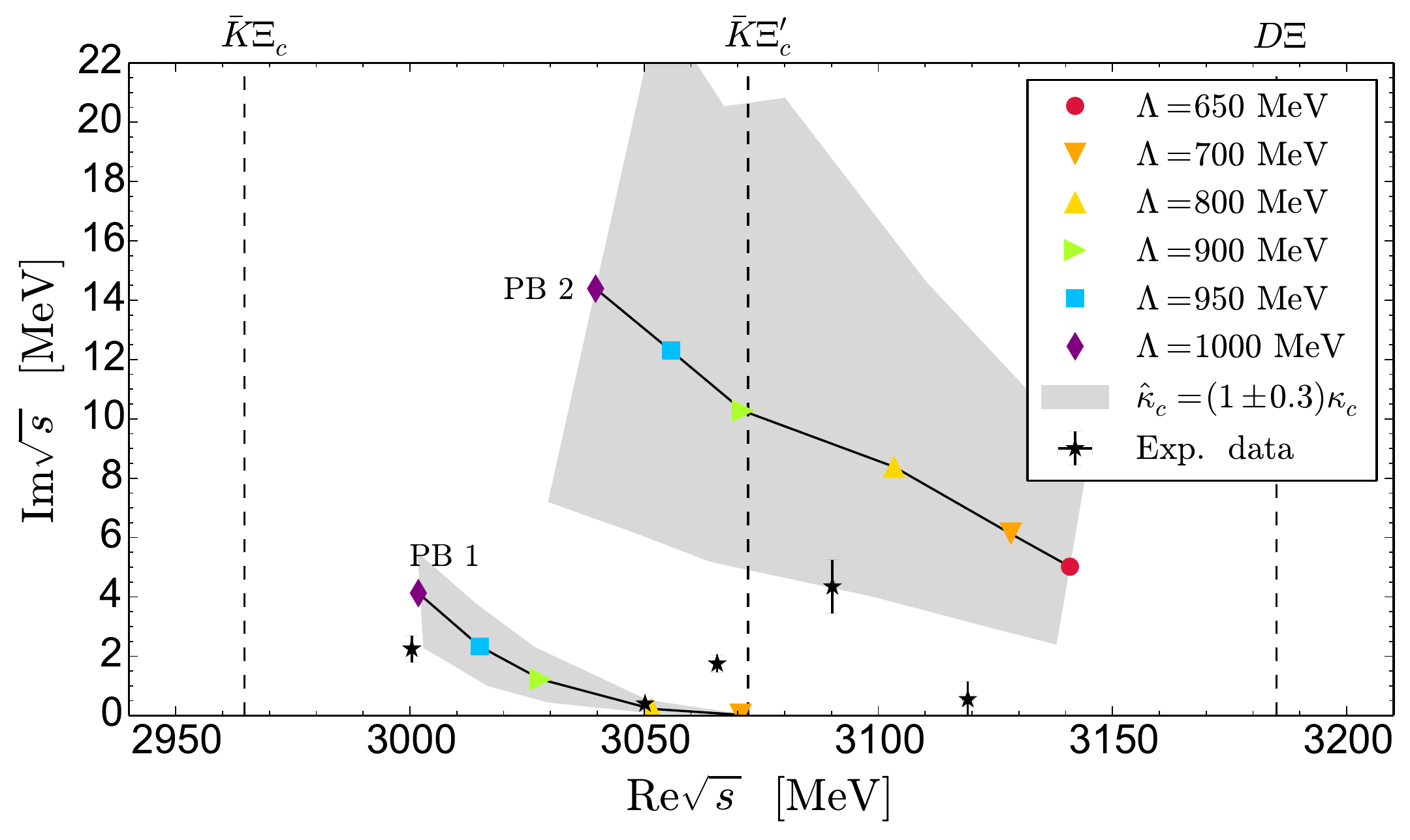}
  \includegraphics[width=0.8\textwidth]{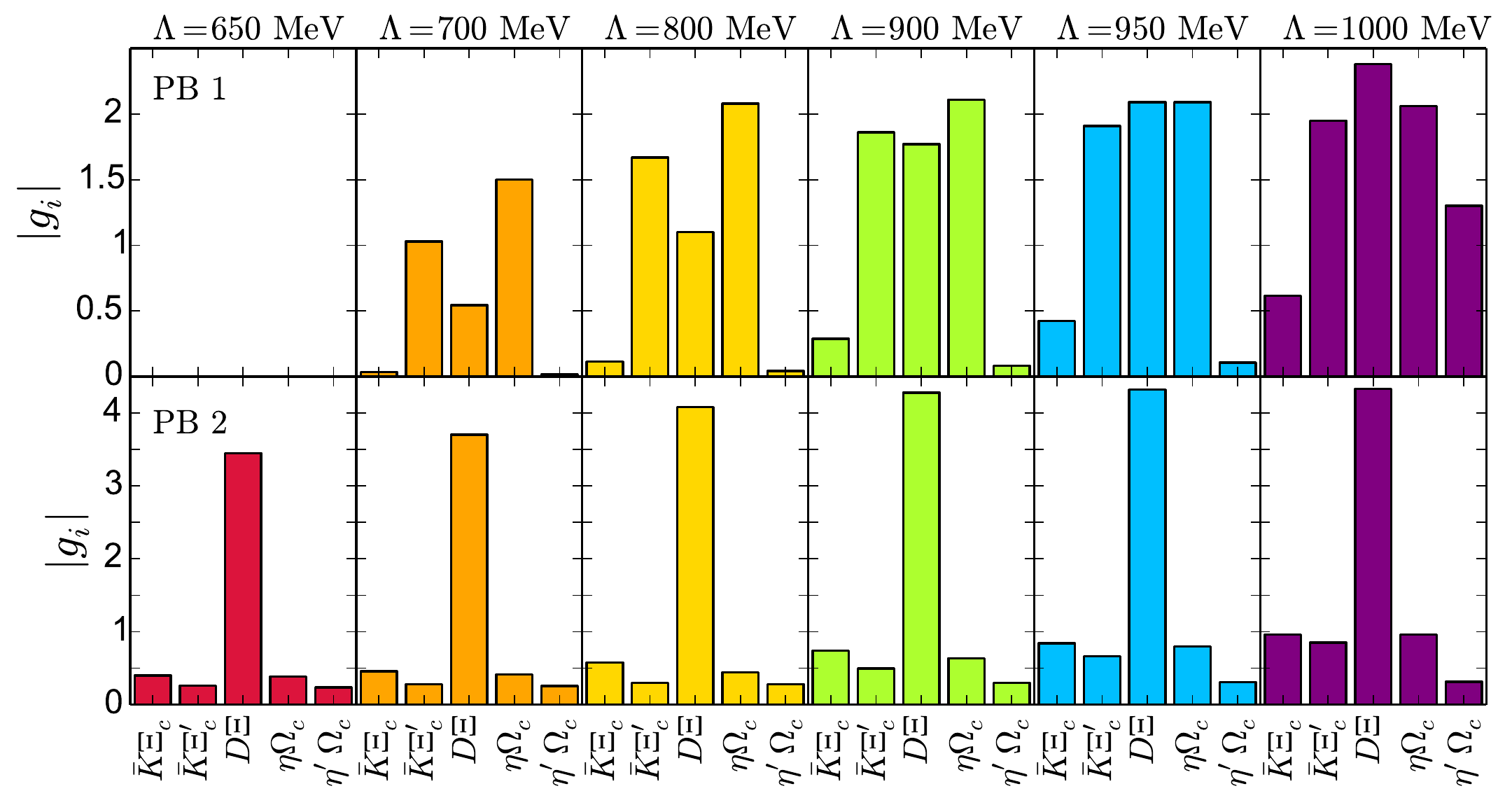}
\caption{Upper panel: evolution of the position of the pole resonances for various cut-off values. The shaded area indicates the region of results covered when an additional $30\%$ of $\textrm{SU}(4)$ symmetry breaking is assumed in the transitions mediated by a heavy-meson exchange. Lower panel: schematic representation of the couplings of each resonance to the \gls{pseudoscalar-baryon} channels for various cut-off values.} 
  \label{fig:free-mb-cut-off}
\end{figure}

Finally, we discuss the dependence of the results in the \gls{pseudoscalar-baryon} sector on the assumed value of the cut-off, as well as the influence of a certain amount of $\textrm{SU}(4)$ symmetry violation associated with the fact that the charm-quark mass is substantially heavier than that of the light quarks. The solid lines in the upper panel of Fig.~\ref{fig:free-mb-cut-off} show the evolution of the resonance poles as the value of the cut-off is varied between $650$~MeV and $1000$~MeV. One can see that, when the cut-off value is increased, the resonances become more bound because the number of intermediate states included in the unitarized amplitudes also increases. When a resonance moves to lower energies, one usually expects it to be narrower since there is less phase space to decay to open channels. We observe, however, an increase in the width, a behavior that can be understood upon examining, in the lower panel of Fig.~\ref{fig:free-mb-cut-off}, the evolution of the coupling constants to the various channels as the value of the cut-off is varied. For larger cut-off values we observe a larger absolute value of the coupling to ${\bar K}\Xi_c$, and this is either the only possible decay channel or the main decay one for these resonances, thereby explaining their increased width.

The implementation of $\textrm{SU}(4)$ symmetry violation is already partly done by the use of the physical meson and baryon masses in the interaction kernel, as well as by the introduction of the factor $\kappa_c=m_V/m_V^c$ to suppress the matrix elements that connect states via the $t$-channel exchange of a vector meson with charm (see Table~\ref{tab:free-mb-coeff}). Therefore, we leave the transitions mediated by light vector mesons untouched, as only $\textrm{SU}(3)$ is effectively acting there, and implement up to an additional $30\%$ of $\textrm{SU}(4)$ symmetry breaking by allowing the factor $\kappa_c\rightarrow\hat{\kappa}_c$ to vary in the range $(0.7-1.3)\kappa_c$. The shaded area in the upper panel of Fig.~\ref{fig:free-mb-cut-off} denotes the region in the complex plane where the resonances can be found by varying the cut-off value in the range $[650-1000]$~MeV and also by admitting up to $30\%$ $\textrm{SU}(4)$ symmetry breaking in certain transitions, as described above. We see that the band of uncertainties includes the $3050$~MeV and the $3090$~MeV resonances measured at LHCb, a fact that gives strength to their interpretation as \gls{pseudoscalar-baryon} molecules.

\subsubsection{$1^-\oplus \frac12^+$ states}
We next construct the unitarized interaction between vector mesons and baryons in the sector with $(I,S,C)=(0,-2,1)$, employing the set of subtraction constants of Table~\ref{tab:free-mb-a_vector}. In the same way a for ``Model~1'' in the \gls{pseudoscalar-baryon} case, the values have been obtained for a regularization scale of $\mu=1$~GeV and imposing the loop function of each \gls{vector-baryon} channel to coincide, at the corresponding threshold, with the cut-off loop function evaluated for $\Lambda=800$~MeV. The mass and other properties of the resonances found from the \gls{vector-baryon} interaction in this sector are listed in Table~\ref{tab:free-mb-vector}. Each of these states corresponds to a degenerate $J^P=\frac12^-,\frac32^-$ doublet.

\begin{table}[b!]
\setlength{\tabcolsep}{10pt}
\renewcommand{\arraystretch}{1.2}
\centering
\begin{tabular}{ccccc}
\hline 
      $a_{ D^*\Xi}$  &  $a_{\bar{K}^*\Xi_c}$  &  $a_{\bar{K}^*\Xi'_c}$  &  $a_{\omega \Omega_c}$  & $a_{\phi \Omega_c}$ \\
\hline                            
   $-1.97$  &  $-2.15$  &  $-2.20$  & $-2.27$  &  $-2.26$   \\
\hline
\end{tabular}
\caption{Values of the subtraction constants at a regularization scale $\mu=1$ GeV and an equivalent cut-off of $\Lambda=800$~MeV.} 
\label{tab:free-mb-a_vector}
\end{table}

\begin{table}[b!]
\setlength{\tabcolsep}{10pt}
\renewcommand{\arraystretch}{1.2}
\centering
\begin{tabular}{l|cc|cc|cc}
\hline
$M$ (MeV)             &    \multicolumn{2}{c|}{$3231.2$}    &  \multicolumn{2}{c|}{$3352.4$ 
}   
 & \multicolumn{2}{c}{$3419.3$}  \\
$\Gamma$ (MeV)  &     \multicolumn{2}{c|}{$0.0$}         &   \multicolumn{2}{c|}{$1.3$}    &  \multicolumn{2}{c}{$4.8$}  \\
\hline
&   $| g_i|$   & $-g_i^2 dG/dE$    &   $| g_i|$   & $-g_i^2 dG/dE$        &   $| g_i|$   & $-g_i^2 dG/dE$     \\
$D^*\Xi\, (3327)$          &  $4.30$  & $0.90+i\,0.00$   &  $0.31$  & $0.01-i\,0.01$   &   $0.24$  & $0.00+i\,0.00$      \\
$\bar{K}^*\Xi_c\, (3363)$  &  $0.64$  & $0.03+i\,0.00$   &  $1.74$  & $0.91+i\,0.01$   &   $0.13$  & $0.00+i\,0.00$      \\
$\bar{K}^*\Xi'_c\, (3470)$ &  $0.26$  & $0.00+i\,0.00$   &  $0.15$  & $0.00+i\,0.00$   &   $1.83$  & $0.42+i\,0.02$      \\
$\omega \Omega_c\, (3480)$ &  $0.34$  & $0.01+i\,0.00$   &  $0.16$  & $0.00+i\,0.00$   &   $1.56$  & $0.28+i\,0.00$      \\
$\phi \Omega_c\, (3717)$   &  $0.00$  & $0.00+i\,0.00$   &  $0.09$  & $0.00+i\,0.00$   &   $2.31$  & $0.22+i\,0.00$      \\
\hline
\end{tabular}
\caption{The $\Omega^{*0}_c$ ($1/2^-,\,3/2^-$) states dynamically generated employing a zero-range interaction between a vector meson ($1^-$) and a ground-state baryon ($1/2^+$), within a coupled-channel approach.}
\label{tab:free-mb-vector}
\end{table}

A similar pattern to that found for the \gls{pseudoscalar-baryon} case is seen for the \gls{vector-baryon} scattering: one resonance coupling strongly to $D^*\Xi$ and another coupling strongly to $\bar{K}^*\Xi'_c $ and to $\phi \Omega_c$, which mainly takes the role of the $\eta \Omega_c$ state of the pseudoscalar case according to the tranformation of Eq.~(\ref{eq:free-mb-eta_phi}). However, the ordering in energies of these resonances appears interchanged compared to that found in \gls{pseudoscalar-baryon} scattering. This is simply related to the fact that the energy thresholds of the various \gls{vector-baryon} channels have also changed in comparison with their \gls{pseudoscalar-baryon} counterparts. The lower energy resonance at $3231$~MeV is mainly a $D^*\Xi$ bound state, while the resonance at $3419$~MeV, is mainly a $\bar{K}^*\Xi'_c $ composite state with some admixture of  $\omega \Omega_c$ and
$\phi \Omega_c$ components. An additional resonance is found between these two, coupling strongly to $\bar{K}^*\Xi_c$ states. We note that the proximity of this resonance to the channel threshold to which it couples more strongly makes it impossible to be found employing the convoluted \gls{meson-baryon} loop of Eq.~(\ref{eq:free-mb-loop_conv}) that incorporates the width of the $K^*$ meson. This is due to the fuzzy transition between Riemann sheets in the energy region of the resonance and, in this case, the pole has been found assuming a zero width in the vector meson propagators.  The three resonances are located at energy values well above the states found by the LHCb collaboration, in a region where no narrow structures have been seen \cite{Aaij:2017nav}. Nevertheless, the states found here from the \gls{vector-baryon} interaction are artificially narrow as they do not couple to, and hence cannot decay into, the \gls{pseudoscalar-baryon} states that lie at lower energy. In order to account for this possibility in our model, one should incorporate the coupling of \gls{vector-baryon} states to the \gls{pseudoscalar-baryon} ones, via for example box diagrams \cite{Garzon:2012np,Liang:2014kra}, or employing the methodology of Refs.~\cite{Romanets:2012hm,GarciaRecio:2008dp} where, based on \gls{hqss}, the pseudoscalar and vector mesons, as well as the baryons of the octet and those of the decuplet, are treated on the same footing. It would be interesting to perform such calculations to see if these structures remain narrow or widen up sufficiently to accommodate the apparently featureless spectrum (within experimental errors) in this higher energy range, as seen in Fig.~\ref{fig:free-mb-omegacLHCb}. 

It would also be interesting to explore how the \gls{pseudoscalar-baryon} resonances with $J^P=1/2^-$ studied in the present thesis would be affected by considering their coupling to the \gls{vector-baryon} states, a task that goes beyond the scope of the exploratory study done here. Note, however, that the energy threshold of the lighter $D^*\Xi$ \gls{vector-baryon} channel lies above those of the \gls{pseudoscalar-baryon} channels, except for the $\eta^\prime \Omega_c$ one, which plays a quite irrelevant role in the dynamical generation of the states discussed above. We, therefore, expect limited changes in their energy positions and widths, which could anyway be compensated by appropriate changes in the subtraction constants.

The results presented in this thesis correspond to those that we published in \cite{Montana:2017kjw}. Several posterior works also addressed the possible interpretation of some of the $\Omega_c^{*0}$ states seen at \gls{cern} as being quasi-bound meson--baryon systems \cite{Debastiani:2017ewu,Wang:2017smo,Chen:2017xat,Nieves:2017jjx}.
In particular, a very similar approach to the one presented here is followed in \cite{Debastiani:2017ewu}, although the interaction is slightly different.
While we employ $\textrm{SU}(4)$ symmetry at the vertices of the vector-meson exchange diagram, the model of \cite{Debastiani:2017ewu} breaks explicitly this symmetry 
by using, at the $\mathcal{BBV}$ vertex, baryon wave functions that incorporate flavor-spin symmetry in the light-quark sector but keep the charm quark 
factorized. In this way, \gls{hqss} is respected in the diagonal transitions, which are mediated by the exchange of light vector mesons and have the heavy quark as a spectator. Nevertheless, the work of \cite{Debastiani:2017ewu} finds the diagonal transitions to be the same as those in this thesis, as the exchanged vector meson is light, and this effectively projects the $\textrm{SU}(4)$ interaction into its $\textrm{SU}(3)$ content, which is identical in both models. The subleading nondiagonal components mediated by the exchange of a heavy vector meson break \gls{hqss} somewhat differently in each model. This explains why the results found in \cite{Debastiani:2017ewu} are qualitatively very similar to those presented in this dissertation, the most notable difference being the narrower width obtained for the $\Omega_c(3090)^0$. Moreover, the channel space covering the states composed of pseudoscalar mesons and $J=3/2^+$ baryons was also considered in \cite{Debastiani:2017ewu}, finding an additional $J^P=3/2^-$ state which could be identified with the observed $\Omega_c(3119)^0$.

Relatedly, the work of \cite{Chen:2017xat} addresses a three-channel problem, $\bar{K} \Xi_c^*/ \bar{K}^*\Xi_c / \bar{K}^*\Xi_c^\prime$, permitting the coupling of a state with a pseudoscalar meson and a $3/2^+$ baryon with states with a vector meson and a $1/2^+$ baryon via a one-boson-exchange potential. A loosely bound molecular state of mainly $ \bar{K} \Xi_c^*$ nature is found around a mass of $3140$~MeV, being several MeV wide and decaying mostly to $\bar{K} \Xi_c^\prime$.        
The authors of Ref.~\cite{Nieves:2017jjx} revisit the renormalization scheme adopted in their earlier work \cite{Romanets:2012hm}, where an extension of the Weinberg-Tomozawa $\pi N$ interaction that incorporates light-quark spin-flavor symmetry plus explicit \gls{hqss} was employed. This model permits transitions between pseudoscalar--$1/2^+$ baryon and vector--$3/2^+$ baryon channels in $J^P=1/2^-$, and between pseudoscalar--$3/2^+$ baryon, vector--$1/2^+$ baryon, vector--$3/2^+$ baryon channels in $J^P=3/2^-$. In \cite{Nieves:2017jjx} it is found that, by modifying moderately the subtraction point in the renormalization scheme of \cite{Romanets:2012hm}, two of the $\Omega_c^{*0}$ states would move above $3000$~MeV. Moreover, adopting an alternative cut-off regularization scheme, with $\Lambda=1090$~MeV, there appear three states, with a significant contribution of $\{\bar{K}\Xi_c^\prime,\bar{K}\Xi_c,\eta\Omega_c\}$, $\{\bar{K}\Xi_c^*,\bar{K}^*\Xi_c,\eta\Omega_c^*\}$, and $\{D \Xi,\bar{K}^*\Xi_c,D^*\Xi,\bar{K}\Xi_c\}$ channels, that can be related to the states found in the present work and in \cite{Debastiani:2017ewu}, as well as be identified with some of the LHCb resonances. They find a different assignment of experimental masses, that is, $J=1/2$ $\Omega_c(3000)^{0}$,  $J=3/2$ $\Omega_c(3050)^{0}$, and $J=1/2$ for either $\Omega_c(3119)^{0}$ or $\Omega_c(3090)^{0}$, due to their proximity in mass, respectively. 

Although subject to inevitable uncertainties, the results of this thesis, as well as those of the models quoted here, indicate that one cannot ignore the possibility that some of the $\Omega^{*0}_c$ resonances observed by the LHCb collaboration could be interpreted as quasi-bound \gls{meson-baryon} states. Additional experiments, establishing the spin-parity of the $\Omega^{*0}_c$ states or observing them in the invariant masses of different decay channels, will certainly provide valuable information to help clarify their nature.

\subsection{Results in the open-beauty sector: $\Omega_b^{*-}$ states}
\label{subsec:free-mb-omegab}
To conclude this section on meson--baryon scattering, we complement the results on the charm sector by presenting predictions for the analog bottom resonances $\Omega_b^{*-}$. These states are generated from \gls{meson-baryon} interaction kernels obtained from the Lagrangians of Eqs.~(\ref{eq:free-mb-vertexVPPsu4}),(\ref{eq:free-mb-vertexBBVsu4}) and (\ref{eq:free-mb-vertexVVVsu4}), where the pseudoscalar- and vector-meson matrices are those of Eqs.~(\ref{eq:free-mb-matrixphi}) and (\ref{eq:free-mb-matrixVmu}), respectively, but changing the charm mesons by their bottom counterparts, and, analogously, the charm baryons in Eq.~(\ref{eq:free-mb-baryons}) are replaced by the bottomed ones. The matrices of coefficients are then the same as that in Tables~\ref{tab:free-mb-coeff} and ~\ref{tab:free-mb-coeff-v} but involve the \gls{pseudoscalar-baryon} and \gls{vector-baryon} channels in the sector with quantum numbers $(I,S,B)=(0,-2,-1)$ displayed in Table~\ref{tab:free-mb-channelsbottom}. In addition, the coefficient $\kappa_c$ in the nondiagonal transitions involving the exchange of a charmed vector meson is replaced by
\begin{equation}
\kappa_b = -\frac{m_V^2}{t - m_V^{b\, 2}} \simeq  -\frac{m_V^2}{(m_K-m_B)^2 - m_{B_s^*}^2} \sim 0.1 \ ,
\end{equation}
to account for the much larger mass of the exchanged bottomed vector mesons compared to the light ones, as done in the work of Ref.~\cite{Liang:2017ejq}. The \gls{meson-baryon} loops are regularized employing the subtraction constants displayed in Table~\ref{tab:free-mb-a_bottom}, which are obtained assuming a common cut-off value of $\Lambda=800$~MeV in Eq.~(\ref{eq:free-mb-a(mu)}), in the same way that we proceeded in the case of ``Model~1'' for charm.

The results obtained are presented in Tables~\ref{tab:free-mb-pseudo_bottom} and \ref{tab:free-mb-vector_bottom} for the interaction of baryons with pseudoscalar and vector mesons, respectively.
 \begin{table}[t!]
\setlength{\tabcolsep}{10pt}
\renewcommand{\arraystretch}{1.2}
\centering
\begin{tabular}{ccccc}
\hline 
$a_{ \bar{K}\Xi_b}$  &  $a_{\bar{K}\Xi'_b}$  &  $a_{\eta \Omega_b}$  & $a_{\bar{B}\Xi }$  &  $a_{\eta' \Omega_b }$ \\
\hline                          
 $-3.57$  &  $-3.62$  &  $-3.63$  & $-3.24$  &  $-3.53$  \\
\hline 
  $a_{ \bar{B}^*\Xi}$  &  $a_{\bar{K}^*\Xi_b}$  &  $a_{\omega \Omega_b}$  & $a_{\bar{K}^*\Xi'_b}$  &  $a_{\phi \Omega_b}$ \\
\hline                             
  $-3.26$  &  $-3.46$  &  $-3.57$  & $-3.51$  &  $-3.52$   \\
\hline
\end{tabular}
\caption{Values of the subtraction constants for \gls{meson-baryon} channels in the $(I,S,B)=(0,-2,-1)$  sector, corresponding to a cut-off $\Lambda=800\;\textrm{MeV}$ at a regularization scale $\mu=1$ GeV.} 
\label{tab:free-mb-a_bottom}
\end{table}

These results for the $\Omega_b^{*-}$ resonances are completely analog to those found in the charm sector. The interaction of pseudoscalar mesons with baryons generates two states with spin $J^P=1/2^-$, one at $6418$~MeV, coupling strongly to ${\bar K}\Xi^\prime_c$ and $\eta\Omega_b$ states, and another at $6519$~MeV, being essentially a $\bar{B}\Xi$ bound state. 
The interaction of vector mesons with baryons generates $J=1/2^-,3/2^-$ spin-degenerate $\Omega^{*-}_b$ resonances at energies $6560$~MeV, coupling strongly to ${\bar B}^*\Xi$, $6665$~MeV, coupling strongly to $K^*\Xi_b$, and $6797$~MeV, being a mixture of $\omega \Omega_b$, ${\bar K}^* \Xi'_b$ and $\phi \Omega_b$. These results, which were published in Ref.~\cite{Montana:2017kjw} are very much in line with what is found in Ref.~\cite{Liang:2017ejq}, which is an extension to the bottom sector of their previous work on $\Omega^{*0}_c$ resonances \cite{Debastiani:2017ewu}. 

\begin{table}[b!]
\setlength{\tabcolsep}{10pt}
\renewcommand{\arraystretch}{1.2}
\centering
\begin{tabular}{c|cc|cc}
\hline
$M$ (MeV)   &       \multicolumn{2}{c|}{$6418.2$}  &   \multicolumn{2}{c}{$6518.8$}  \\
$\Gamma$ (MeV)   &            \multicolumn{2}{c|}{$0.00$}  &     \multicolumn{2}{c}{$1.24$}   \\ 
\hline
&      $| g_i|$   & $-g_i^2 dG/dE$     &   $| g_i|$   & $-g_i^2 dG/dE$   \\
$\bar{K}\Xi_b\, (6289)$   &  $0.01$  & $0.00+i\,0.00$  &   $0.13$  & $0.00+i\,0.00$      \\
$\bar{K}\Xi'_b\, (6431)$  &  $1.32$  & $0.55+i\,0.00$  &   $0.01$  & $0.00+i\,0.00$      \\
$\eta \Omega_b\, (6594)$  &  $2.02$  & $0.33+i\,0.00$  &   $0.01$  & $0.00+i\,0.00$      \\
$\bar{B}\Xi\, (6598)$     &  $0.28$  & $0.00+i\,0.00$  &   $5.23$  & $0.97+i\,0.00$      \\
$\eta' \Omega_b\, (7004)$ &  $0.00$  & $0.00+i\,0.00$  &   $0.12$  & $0.00+i\,0.00$      \\
\hline
\end{tabular}
\caption{The $\Omega^{*-}_b$ ($1/2^-$) states dynamically generated employing zero-range interactions between a pseudoscalar meson ($0^-$) and a ground-state baryon ($1/2^+$), within a coupled-channel approach.}
\label{tab:free-mb-pseudo_bottom}
\end{table}

\begin{table}[b!]
\setlength{\tabcolsep}{10pt}
\renewcommand{\arraystretch}{1.2}
\centering
\begin{tabular}{c|cc|cc|cc}
\hline
$M$ (MeV)   &    \multicolumn{2}{c|}{$6559.9$}  &  \multicolumn{2}{c|}{$6662.5$}  &   \multicolumn{2}{c}{$6797.1$}\\
$\Gamma$ (MeV)     &       \multicolumn{2}{c|}{$0.0$}   &   \multicolumn{2}{c|}{$19.1$}  &     \multicolumn{2}{c}{$1.39$}\\ 
\hline
 &      $| g_i|$   & $-g_i^2 dG/dE$  &      $| g_i|$   & $-g_i^2 dG/dE$     &   $| g_i|$   & $-g_i^2 dG/dE$ \\
$\bar{B}^*\Xi\, (6643)$    &  $5.31$  & $0.97+i\,0.00$  &  $0.22$  & $0.00+i\,0.00$  &   $0.11$  & $0.00+i\,0.00$    \\
$\bar{K}^*\Xi_b\, (6687)$  &  $0.23$  & $0.01+i\,0.00$  &  $2.32$  & $1.32+i\,0.28$  &   $0.02$  & $0.00+i\,0.00$    \\
$\omega \Omega_b\, (6829)$ &  $0.10$  & $0.00+i\,0.00$  &  $0.04$  & $0.00+i\,0.00$  &   $1.36$  & $0.41+i\,0.01$    \\
$\bar{K}^*\Xi'_b\, (6829)$ &  $0.14$  & $0.00+i\,0.00$  &  $0.03$  & $0.00+i\,0.00$  &   $1.13$  & $0.29+i\,0.00$    \\
$\phi \Omega_b\, (7066)$   &  $0.07$  & $0.00+i\,0.00$  &  $0.00$  & $0.00+i\,0.00$  &   $1.95$  & $0.27+i\,0.00$    \\
\hline
\end{tabular}
\caption{The $\Omega^{*-}_b$ ($1/2^-,\,3/2^-$) states dynamically generated employing a zero-range interaction between a vector meson ($1^-$) and a ground-state baryon ($1/2^+$), within a coupled-channel approach.}
\label{tab:free-mb-vector_bottom}
\end{table}

The narrow $\Omega_b^{*-}$ states listed in Tables~\ref{tab:free-mb-pseudo_bottom} and \ref{tab:free-mb-vector_bottom} lie quite above the energy region $6300-6350$~MeV of the invariant $K^-\Xi_b^0$ mass spectrum, where four $\Omega_b^{*-}$ excited states were reported recently by the LHCb collaboration~\cite{LHCb:2020tqd}. The authors of Ref.~\cite{Liang:2020dxr}, in relation to their results in \cite{Liang:2017ejq}, argued that it is quite unlikely that the states reported in~\cite{LHCb:2020tqd} correspond to molecular states dynamically generated from the \gls{meson-baryon} interaction. Some works have suggested that these observed $\Omega_b^{*-}$ states could correspond to ordinary $1P$ excitations~\cite{Chen:2020mpy,Liang:2020hbo}. An exception could be the molecular $\Omega_b^{*-}$ state predicted in \cite{Nieves:2019jhp} at $6360$~MeV, coupling strongly to $\bar{K}\Xi_b$ and belonging to a $J^P=1/2^-$ sextet of excited bottomed baryons, using an \gls{hqss}-extended approach similar to that in \cite{Nieves:2017jjx}.
The $\Omega_b^{*-}$ reported in this section and in Refs.~\cite{Montana:2017kjw,Liang:2017ejq} could be rather identified with some structures seen in the experimental spectrum at higher energy, although they are statistically not significant. If the properties of these structures are close to those of the states predicted here, upon the adjustment of the subtraction constants as in the charm sector, they could be interpreted as having a molecular origin.

\section{Interaction of heavy mesons with light mesons}
\label{sec:free-mm}
In this section, we describe the scattering of open heavy-flavor mesons with light mesons using the \gls{hmet} Lagrangian at \gls{nlo} in the chiral expansion and keeping the \gls{lo} in the heavy mass expansion. We analyze the properties of the dynamically generated states that appear in the unitarized amplitudes in the charm and bottom sectors.

\subsection{Introduction}

\begin{figure}[b!]
\centering
   \includegraphics[width=0.9\textwidth]{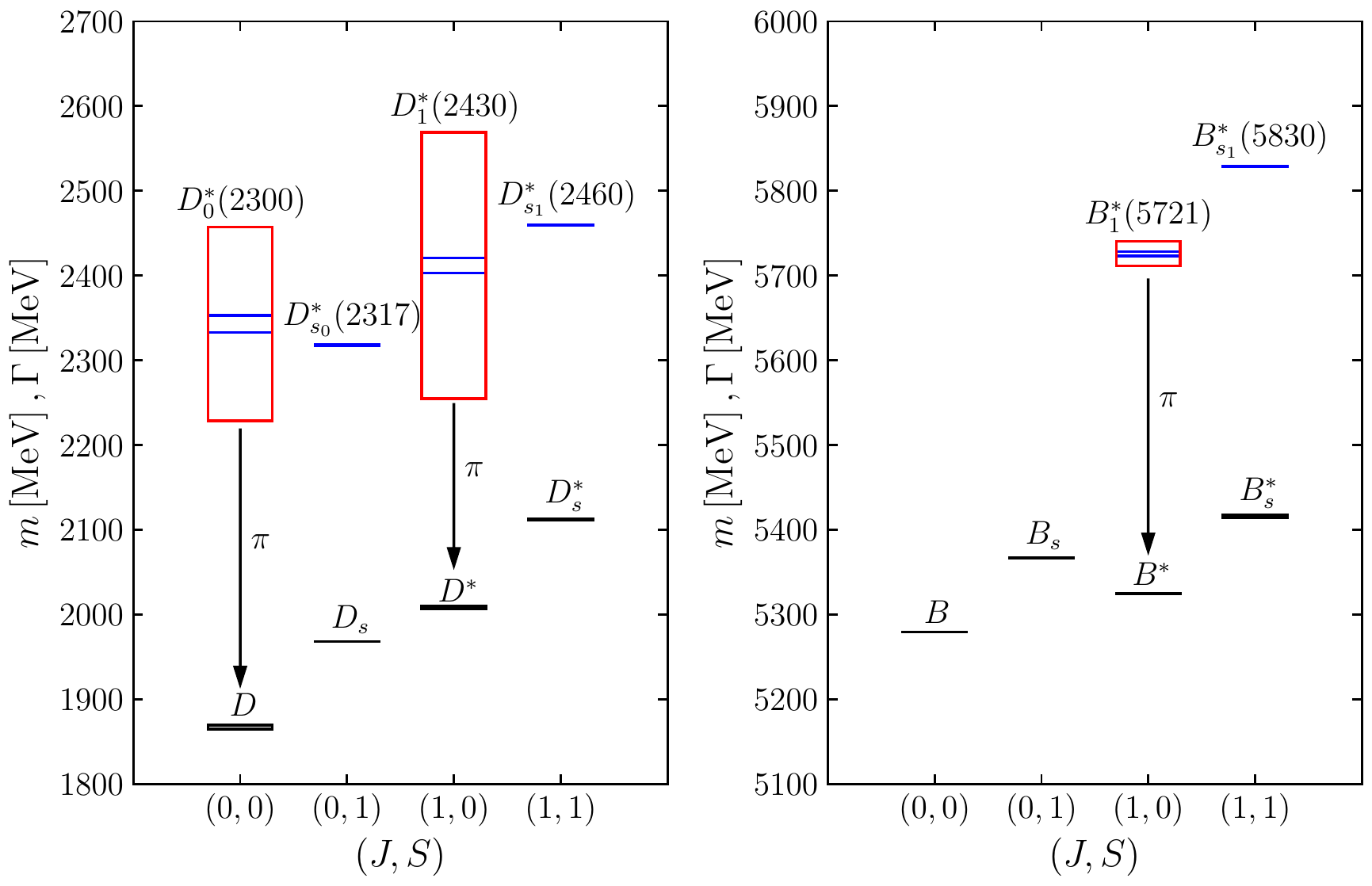}
   \caption{Spectrum of the lowest-lying $D$ mesons (left) and $B$ mesons (right) according to the \gls{rpp}~\cite{pdg}. The height of the black and blue boxes shows the uncertainty in the values of the mass of the ground states and the excited states, respectively, while the height of the red boxes represents the decay width.}
   \label{fig:free-mm-spectrumpdg}
\end{figure}

The interest in the spectrum of heavy mesons containing a heavy quark $Q$ and a light antiquark $\bar{q}=\{\bar{u},\bar{d},\bar{s}\}$ has been recently renewed in view of the vast new experimental data collected, in particular of exotic states that cannot be accommodated within quark-model predictions. The pseudoscalar $D$ (with $J^P=0^-$) and vector $D^*$ (with $J^P=1^-$) isospin doublets and the strange $D_s$ and $D_s^*$ isospin singlets, being the $s$-wave predictions of the quark model, are well established by the \gls{pdg}~\cite{pdg}, as well as their bottomed counterparts. However, the spectrum of excited states is less well understood. In particular, the scalar ($J^P=0^+$) $D_{s0}^*(2317)^\pm$ \cite{BaBar:2003oey} and the pseudovector ($J^P=1^+$) $D_{s1}^*(2460)^\pm$ \cite{CLEO:2003ggt} mesons have attracted a lot of attention, as they are significantly lighter than the quark-model expectations for the $1P$ excitations \cite{Godfrey:1985xj,Ebert:2009ua,Godfrey:2015dva}. Moreover, the mass difference between these two excited states is similar to that between the $D$ and $D^*$ ground states ($\sim 140$~MeV) which, together with the fact that their masses lie very close to the $DK$ and $D^*K$ thresholds, respectively, suggested the possible interpretation of the $D_{s0}^*(2317)^\pm$ and the $D_{s1}(2460)^\pm$ mesons as hadronic molecules \cite{Barnes:2003dj,Szczepaniak:2003vy,Kolomeitsev:2003ac,Hofmann:2003je,Guo:2006fu,Gamermann:2006nm,Faessler:2007gv,Flynn:2007ki} soon after their discovery. Descriptions such as a conventional $c\bar{s}$ meson \cite{Dai:2003yg,Narison:2003td,Bardeen:2003kt}, a compact tetraquark structure \cite{Cheng:2003kg,Terasaki:2003qa,Chen:2004dy,Maiani:2004vq,Bracco:2005kt,Wang:2006bs}, and a mixture of $c\bar{s}$ with tetraquark \cite{Browder:2003fk} and $D^{(*)}K$ molecular \cite{vanBeveren:2003kd} components have also been advocated for these states. The case of the nonstrange scalar $D_0^*(2300)$ and pseudovector $D_1(2430)^0$ states \cite{Belle:2003nsh,FOCUS:2003gru,BaBar:2009pnd}, which have large natural widths ($\sim 300$~MeV), is equally interesting. 
The properties of these states and their bottomed counterparts according to the \gls{rpp} compilation of the \gls{pdg}~\cite{pdg}, which are pictorially summarized in Fig.~\ref{fig:free-mm-spectrumpdg}, are the following:
\begin{equation}\label{eq:free-mm-proppdg}
\begin{aligned}[b]
 & D_0^*(2300):~~ && M=2343\pm 10~\textrm{MeV} ~~ & & \Gamma=229\pm16~\textrm{MeV} ~~ && I(J^P)=\frac12(0^+) \ , \\
 & D_{s0}^*(2317)^\pm:~~ &&  M=2317.8\pm0.5~\textrm{MeV} ~~ & & \Gamma<3.8~\textrm{MeV} ~~ && I(J^P)=0(0^+) \ , \\ 
 & D_1(2430)^0:~~ && M=2412\pm 9~\textrm{MeV} ~~ & & \Gamma=314\pm29~\textrm{MeV} ~~  && I(J^P)=\frac12(1^+) \ , \\ 
 & D_{s1}(2460)^\pm:~~ && M=2459.6\pm0.6~\textrm{MeV} ~~ & & \Gamma<3.5~\textrm{MeV} ~~ && I(J^P)=0(1^+) \ , \\ 
 & B_1(5721)^+:~~ && M=5725.9^{+2.5}_{-2.7}~\textrm{MeV} ~~ & & \Gamma=31\pm6~\textrm{MeV} ~~ && I(J^P)=\frac12(1^+) \ , \\ 
 & B_1(5721)^0:~~ && M=5726.1\pm1.3~\textrm{MeV} ~~ & & \Gamma=27.5\pm3.4~\textrm{MeV} ~~ && I(J^P)=\frac12(1^+) \ , \\ 
 & B_{s1}(5830)^0:~~ && M=5828.70\pm0.20~\textrm{MeV} ~~ & & \Gamma=0.5\pm0.4~\textrm{MeV} ~~ && I(J^P)=0(1^+) \ .
\end{aligned}
\end{equation}

The value reported for the mass of the $D_0^*(2300)^0$ from $\gamma\,A$ reactions, $2407\pm21\pm35$~MeV (FOCUS \cite{FOCUS:2003gru}), differs considerably from that obtained from $B$-factories, $2308\pm17\pm36$~MeV (Belle, \cite{Belle:2003nsh}) and $2297\pm8\pm20$~MeV (BaBar, \cite{BaBar:2009pnd}), while the values reported by the LHCb collaboration for the charged partner \cite{LHCb:2015klp,LHCb:2015klp} lie in the middle. The experimental value of the mass of the $D_0^*(2300)^0$ being close, or even larger, to that of the $D_{s0}^*(2317)$, which is the opposite of what one expects from the constituent quark model, poses a puzzle in the excited open heavy-flavor spectrum, with two-meson thresholds definitely playing an important role. 

Analysis combining \gls{eft} methods with \gls{lqcd} can be actually used to shed light on this puzzle.
\Gls{lqcd} calculations in the charmed scalar sector started to develop soon after the experimental observation of these states. The first studies obtained masses for the $D_{s0}^*(2317)$ larger than the experimental ones as they included only $c\bar{s}$ interpolators \cite{Bali:2003jv,Dougall:2003hv}. After incorporating also meson--meson interpolators, the authors of Refs.\cite{Mohler:2012na,Mohler:2013rwa} were able to obtain values consistent with the experimental masses of the $D_{s0}^*(2317)^\pm$ and the $D_{0}^*(2300)$. The Hadron Spectrum collaboration investigated the coupled-channel $D\pi$, $D\eta$, and $D_s\bar{K}$ scattering at a value of the pion mass $m_\pi=391$~MeV and found one pole coupling largely to $D\pi$ that was assigned to the $D_{0}^*(2300)$ state \cite{Moir:2016srx}. Given these results, a unitarized effective model in coupled channels was revisited by the authors of Ref.~\cite{Albaladejo:2016lbb}, where strong support for the two-pole structure of the $D_{0}^*(2300)$, previously claimed in \cite{Kolomeitsev:2003ac,Guo:2006fu,Guo:2009ct}, was found. The evidence came from a remarkably good agreement with \gls{lqcd} results of the lowest-lying energy levels.

This is not the first evidence in hadron physics where two poles appear together and are dynamically related to each other in some scattering process. The most famous example is the case of the $\Lambda(1405)$, for which the interplay between two poles dynamically generated from the meson--baryon interaction in the sector with strangeness $S=-1$ and isospin $I=0$ but representing the same state was proposed to explain the experimental data in the $\bar{K}N$, $\pi\Sigma$, and $\pi\Lambda$ coupled-channel system \cite{Oller:2000fj,Jido:2003cb,Magas:2005vu,Hyodo:2007jq} in the neighborhood of the $\Lambda(1405)$. The origin of the two-pole structure of the $\Lambda(1405)$ is now well understood from $\textrm{SU}(3)$ symmetry considerations and group theory, each of the poles being generated by the two attractive channels of the leading order interaction in the $\textrm{SU}(3)$ basis (singlet and octet) \cite{Jido:2003cb} and isospin basis ($\bar{K}N$ and $\pi\Sigma$) \cite{Hyodo:2007jq}. More details about the double-pole nature of the $\Lambda(1405)$ can be found in the last issue of the \gls{rpp}~\cite{pdg}, as well as in some recent reviews \cite{Yao:2020bxx,Mai:2020ltx,Hyodo:2020czb}.

Similarly, the double-pole structure that is found in the $D^{(*)}\pi$, $D^{(*)}\eta$, and $D^{(*)}_s\bar{K}$ coupled-channel meson--meson scattering problem  can be traced back to the $\textrm{SU}(3)$ attractive interactions in the $\mathbf{6}$ and $\bar{\textbf{3}}$ irreps \cite{Albaladejo:2016lbb}. In the bottom sector, similar resonance patterns follow from \gls{hqfs}.

In this section, we revisit the effective field theory describing the dynamics of open heavy-flavor ground-state mesons ($D$, $D_s$, $D^*$, $D_s^*$, $\bar{B}$, $\bar{B}_s$, $\bar{B}^*$, $\bar{B}_s^*$), that we will generally denote as $D$ mesons and $\bar{B}$ mesons in the following for simplicity, and their interaction with the light Goldstone bosons ($\pi,\,K,\,\bar{K}$, $\eta$). Particular attention is paid to the dynamical generation of heavy meson-light meson molecular states in the sectors with $(S,I)=(1/2,0)$ and $(S,I)=(0,1)$, where the scalars $D_0^*(2300)$ and $D_{s0}^*(2317)$ are found, respectively, as well as the pseudovectors $D_1(2430)$ and $D_{s1}(2460)$, and their bottomed counterparts. The goal is to set a clear description of the model in the vacuum, as temperature corrections to the scattering of open-heavy flavor mesons off light mesons will be introduced in Chapter~\ref{ch:hot-medium} and the thermal modification of the properties of both ground states and excited states will be investigated. The results presented here were published in Refs.~\cite{Montana:2020lfi,Montana:2020vjg}.

The heavier $\eta'$ mesons are ignored, as well as the light vector mesons ($\rho,\, \omega$, $\phi$), as we only consider the interaction with Goldstone bosons and neglect the $\eta-\eta'$ mixing.

In the same way as in the description of meson--baryon systems in Section~\ref{sec:free-mb}, isospin breaking is not considered, and therefore the isospin basis is adopted for the mesonic fields. The quantum numbers and isospin-averaged masses of the light and heavy mesons grouped in the corresponding isospin multiplets are summarized in Table~\ref{tab:free-mm-isospin_multiplets}.

\begin{table}[t!]
\setlength{\tabcolsep}{10pt}
\renewcommand{\arraystretch}{1.2}
\begin{tabular}{ccccccc}
\hline
$J^P=0^-$ & Multiplet & $I$ & $S$ & $C$ & $B$ & $m\,\textrm{(MeV)}$ \\ 
\hline
$\pi$ & $(\pi^+,\pi^0,\pi^-)$ & $1$ & $0$ & $0$ & $0$ & $138.04$ \\
$\eta$ & $(\eta^0)$ & $0$ & $0$ & $0$ & $0$ & $547.86$ \\
$K$ & $(K^{ +},K^{ 0})$ & $1/2$ & $+1$ & $0$ & $0$ & $495.64$ \\
$\bar{K}$ & $(\bar{K}^{ 0},K^{ -})$ & $1/2$ & $-1$ & $0$ & $0$ & $495.64$  \\
$D$ & $(D^{ +},D^{ 0})$ & $1/2$ & $0$ & $+1$ & $0$ & $1867.24$ \\
$\bar{D}$ & $(\bar{D}^{ 0},D^{ -})$ & $1/2$ & $0$ & $-1$ & $0$ & $1867.24$ \\
$D_s$ & $(D_s^{+})$ & $0$ & $+1$ & $+1$ & $0$ & $1968.34$ \\
$\bar{D}_s$ & $(D_s^{-})$ & $0$ & $-1$ & $-1$ & $0$ & $1968.34$ \\
$B$ & $(B^{ +},B^{0})$ & $1/2$ & $0$ &  0 & $+1$ & $5279.48$ \\
$\bar{B}$ & $(\bar{B}^{0},B^{-})$ & $1/2$ & $0$ & $0$ & $-1$ & $5279.48$ \\
$B_s$ & $(B_s^{0})$ & $0$ & $-1$ & $0$ & $+1$ & $5366.89$ \\
$\bar{B}_s$ & $(\bar{B}_s^{0})$ & $0$ & $+1$ & $0$ & $-1$ & $5366.89$ \\
\hline
$J^P=1^-$ & Multiplet & $I$ & $S$ & $C$ & $B$ & $m\,\textrm{(MeV)}$  \\
\hline
$D^*$ & $(D^{* +},D^{* 0})$ & $1/2$ & $0$ & $+1$ & $0$ & $2008.56$ \\
$\bar{D}^*$ & $(\bar{D}^{* 0},D^{* -})$ & $1/2$ & $0$ & $-1$ & $0$ & $2008.56$ \\
$D_s^{*}$ & $(D_s^{*+})$ & $0$ & $+1$ & $+1$ & $0$ & $2112.20$ \\
$\bar{D}_s^{*}$ & $(D_s^{*-})$ & $0$ & $-1$ & $-1$ & $0$ & $2112.20$ \\
$B^*$ & $(B^{* +},B^{* 0})$ & $1/2$ & $0$ & $0$ & $+1$ & $5324.65$ \\
$\bar{B}^*$ & $(\bar{B}^{* 0},B^{* -})$ & $1/2$ & $0$ & $0$ & $-1$ & $5324.65$ \\
$B_s^{*}$ & $(B_s^{*0})$ & $0$ & $-1$ & $0$ & $+1$ & $5415.40$ \\
$\bar{B}_s^{*}$ & $(\bar{B}_s^{*0})$ & $0$ & $+1$ & $0$ & $-1$ & $5415.40$ \\
\hline
\end{tabular}
\centering
\caption{Isospin multiplets of the light and heavy pseudoscalar mesons ($J^P=0^-$) and the heavy vector mesons ($J^P=1^-$), together with the isospin, strangeness, charm and bottom quantum numbers and the isospin-averaged values of the masses in the \gls{rpp}~\cite{pdg}.}
\label{tab:free-mm-isospin_multiplets}
\end{table}

\subsection{Formalism}
\label{subsec:free-mm-formalism}

The interactions between open-heavy flavor mesons and light mesons are described in this section in terms of an effective Lagrangian for the degrees of freedom contained in Table~\ref{tab:free-mm-isospin_multiplets}. It is based on chiral symmetry, involving the physics of the Goldstone bosons at low energies, as well as \gls{hqsfs} for the singly heavy mesons, both pseudoscalar $D$/$\bar{B}$ and vector $D^*$/$\bar{B}^*$ mesons. In the heavy-quark mass counting only the \gls{lo} terms of the effective Lagrangian are considered, whereas in the chiral power counting both \gls{lo} and \gls{nlo} orders are taken into account. From now on, the terminology \gls{lo} and \gls{nlo} will exclusively refer to the chiral power counting in the effective Lagrangian,
\begin{equation}
  {\cal L}= {\cal L}_\textrm{LO}+{\cal L}_\textrm{NLO} \ . 
\end{equation}
    
The \gls{lo} Lagrangian contains the kinetic terms and the interactions of the heavy mesons with the Goldstone bosons (see Eq.~(\ref{eq:free-th-LagHMET-LO2})), as well as self-interactions of the Goldstone bosons. The pure light-meson sector is described by the standard \gls{chpt} \cite{Gasser:1983yg}, presented in some detail in Section~\ref{subsec:free-th-effective theories}. 
    
In the following, we particularize the description of the scattering formalism off light mesons for charmed mesons. The formalism for the bottomed mesons can be obtained straightforwardly upon the replacement $D\rightarrow \bar{B}$. The \gls{lo} Lagrangian of Eq.~(\ref{eq:free-th-LagHMET-LO2}) for $D$ mesons reads 
\begin{align}\label{eq:free-mm-LagDpiLO}
\mathcal{L}_\textrm{LO}&\ =\mathcal{L}^\textrm{ChPT}_\textrm{LO}+\langle\nabla^\mu D\nabla_\mu D^\dagger\rangle-m_D^2\langle DD^\dagger\rangle-\langle\nabla^\mu D^{*\nu}\nabla_\mu D^{*\dagger}_{\nu}\rangle+m_D^2\langle D^{*\nu}D^{*\dagger}_{\nu}\rangle \nonumber \\
&\ +ig\langle D^{*\mu}u_\mu D^\dagger-Du^\mu D^{*\dagger}_\mu\rangle+\frac{g}{2m_D}\langle D^*_\mu u_\alpha\nabla_\beta D^{*\dagger}_\nu-\nabla_\beta D^*_\mu u_\alpha D^{*\dagger}_\nu\rangle\epsilon^{\mu\nu\alpha\beta} \ ,
\end{align}
where $D$ denotes the antitriplet of $0^-$ $D$ mesons, $D=\left(D^0 \; D^+ \; D^+_s\right)$, and similarly for the vector $1^-$ states, $D^*_\mu=\left(D^{*0} \; D^{*+} \; D^{*+}_s\right)_\mu$. When studying the bottom sector the fields of the $D$ mesons in Eq.~(\ref{eq:free-mm-LagDpiLO}) will be replaced by the bottomed ones, namely $\bar{B}=\left( B^- \; \bar{B}^0 \; \bar{B}^0_s\right)$ and $\bar{B}^*_\mu=\left( B^{*-} \; \bar{B}^{*0} \; \bar{B}^{*0}_s\right)_\mu$. The light mesons are encoded into $u_\mu=\ii(u^\dagger\partial_\mu u-u\partial_\mu u^\dagger)$, where $u=\exp (\ii\Phi/\sqrt{2} f_\pi)$ is the unitary matrix containing the Goldstone bosons in the exponential representation presented in Eq.~(\ref{eq:free-th-matrixPseudoscalars}) and that we reproduce here for completeness,    
\begin{equation}
    \Phi = \left(
    \begin{array}{ccc}
    \frac{1}{\sqrt{2}} \pi^0 +\frac{1}{\sqrt{6}} \eta & \pi^+ & K^+ \\
    \pi^- & -\frac{1}{\sqrt{2}} \pi^0+\frac{1}{\sqrt{6}} \eta & K^0 \\
    K^- & \bar{K}^0 & -\sqrt{\frac23} \eta \\
    \end{array}
    \right)\ , 
\end{equation}
where we have identified the mathematical $\eta_8$ state with the physical $\eta$ meson by neglecting the $\eta-\eta'$ mixing, and $f_\pi$ is the pion decay constant, $f_\pi=92.4$ MeV. A brief discussion on $\eta-\eta'$ mixing is given in Section~\ref{subsec:free-mb-formalism}.

We remind the reader that angle brackets in the Lagrangian denote the trace in flavor space and the connection of the covariant derivative $\nabla_\mu D^{(*)}=\partial_\mu D^{(*)} -D^{(*)}\Gamma^\mu$ reads $\Gamma_\mu=\frac{1}{2}(u^\dagger\partial_\mu u+u\partial_\mu u^\dagger)$.

The \gls{nlo} part of the Lagrangian is given in Eq.~(\ref{eq:free-th-LagHMET-NLO}) and for $D$ mesons can be written as
\begin{align}\label{eq:free-mm-lagrangianNLO}\nonumber
 \mathcal{L}_\textrm{NLO}=&\ \mathcal{L}^{\chi\textrm{PT}}_\textrm{NLO} -h_0\langle DD^\dagger\rangle\langle\chi_+\rangle+h_1\langle D\chi_+D^\dagger\rangle+h_2\langle DD^\dagger\rangle\langle u^\mu u_\mu\rangle  +h_3\langle Du^\mu u_\mu D^\dagger\rangle \\ \nonumber
 &\ +h_4\langle\nabla_\mu D\nabla_\nu D^\dagger\rangle\langle u^\mu u^\nu\rangle+h_5\langle\nabla_\mu D\{u^\mu,u^\nu\}\nabla_\nu D^\dagger \rangle \\ \nonumber
 &\ +\tilde{h}_0\langle D^{*\mu}D^{*\dagger}_\mu\rangle\langle\chi_+\rangle-\tilde{h}_1\langle D^{*\mu}\chi_+D^{*\dagger}_\mu\rangle-\tilde{h}_2\langle D^{*\mu}D^{*\dagger}_\mu\rangle\langle u^\nu u_\nu\rangle -\tilde{h}_3\langle D^{*\mu}u^\nu u_\nu D^{*\dagger}_\mu\rangle  \\ 
 &\ -\tilde{h}_4\langle\nabla_\mu D^{*\alpha}\nabla_\nu D^{*\dagger}_\alpha\rangle\langle u^\mu u^\nu\rangle-\tilde{h}_5\langle\nabla_\mu D^{*\alpha}\{u^\mu,u^\nu\}\nabla_\nu D^{*\dagger}_\alpha\rangle,
\end{align}
where ${\mathcal L}^{\chi\textrm{PT}}_\textrm{NLO}$ represents the \gls{nlo} \gls{chpt} Lagrangian involving only $\Phi$, and $\chi_+=u^\dagger\chi u^\dagger+u\chi u$, with the quark mass matrix $\chi=\textrm{diag}(m_\pi^2,m_\pi^2,2m_K^2-m_\pi^2)$. Further details  regarding these Lagrangians are given in Section~\ref{subsec:free-th-effective theories} as well as in Refs.~\cite{Kolomeitsev:2003ac,Lutz:2007sk,Guo:2009ct,Geng:2010vw,Abreu:2011ic,Liu:2012zya,Tolos:2013kva,Albaladejo:2016lbb,Guo:2018tjx}.
  
\begin{table}[t!]
\setlength{\tabcolsep}{5pt}
\renewcommand{\arraystretch}{1.2}
 \begin{tabular}{c|cc|cc}
 \hline
$(S,I)$ & Channels ($J^P=0^+$) & Threshold (MeV) & Channels ($J^P=1^+$) & Threshold (MeV) \\
\hline
$(-1,0)$  & $D\bar{K}$ & $2364.88$ & $D^* \bar{K}$ & $2504.20$ \\
$(-1,1)$  & $D\bar{K}$ & $2364.88$ & $D^* \bar{K}$ & $2504.20$ \\
$(0,\frac12)$ & $D\pi$ & $2005.28$ & $D^* \pi$ & $2146.59$ \\
        & $D\eta$ & $2415.10$ & $D^* \eta$ & $2556.42$ \\
        & $D_s\bar{K}$ & $2463.98$ & $D_s^* \bar{K}$ & $2607.84$ \\
$(0,\frac32)$ & $D\pi$ & $2005.28$ & $D^* \pi$ & $2146.59$ \\
$(1,0)$   & $DK$ & $2364.88$ & $D^*K$ & $2504.20$ \\
        & $D_s\eta$ & $2516.20$ & $D_s^* \eta$ & $2660.06$ \\
$(1,1)$   & $D_s\pi$ & $2106.38$ & $D_s^* \pi$ & $2250.24$ \\
        & $DK$ & $2364.88$ & $D^* K$ & $2504.20$ \\
$(2,\frac12)$ & $D_sK$ & $2463.98$ & $D_s^* K$ & $2607.84$ \\
\hline
$(-1,0)$  & $\bar{B}\bar{K}$ & $5775.12$ & $\bar{B}^* \bar{K}$ & $5820.29$ \\
$(-1,1)$  & $\bar{B}\bar{K}$ & $5775.12$ & $\bar{B}^* \bar{K}$ & $5820.29$ \\
$(0,\frac12)$ & $\bar{B}\pi$ & $5417.51$ & $\bar{B}^* \pi$ & $5462.69$ \\
        & $\bar{B}\eta$ & $5827.34$ & $\bar{B}^* \eta$ & $5872.51$ \\
        & $\bar{B}_s\bar{K}$ & $5862.53$ & $\bar{B}_s^* \bar{K}$ & $5911.04$ \\
$(0,\frac32)$ & $\bar{B}\pi$ & $5417.51$ & $\bar{B}^* \pi$ & $5462.29$ \\
$(1,0)$   & $\bar{B}K$ & $5775.12$ & $\bar{B}^*K$ & $5820.29$ \\
        & $\bar{B}_s\eta$ & $5914.75$ & $\bar{B}_s^* \eta$ & $5963.26$ \\
$(1,1)$   & $\bar{B}_s\pi$ & $5504.93$ & $\bar{B}_s^* \pi$ & $5553.44$ \\
        & $\bar{B}K$ & $5775.12$ & $\bar{B}^* K$ & $5820.29$ \\
$(2,\frac12)$ & $\bar{B}_sK$ & $5862.53$ & $\bar{B}_s^* K$ & $5911.04$ \\
\hline
 \end{tabular}
 \centering
 \caption{Meson-meson channels in the charm sector (upper part) and in the bottom sector (lower part) together with their threshold energy, their total spin-parity $J^P$, isospin $I$ and strangeness $S$ quantum numbers, for channels involving the pseudoscalar ($J^P=0^-$) $D$/$\bar{B}$ meson (left) and the vector ($J^P=1^-$) $D^*$/$\bar{B}^*$ meson (right).}
 \label{tab:free-mm-channels}
 \end{table}

 The tree-level amplitudes are extracted from the \gls{lo}+\gls{nlo} Lagrangian and they are kept at strictly lowest order in the heavy-quark mass expansion, that is, only the amplitudes at order ${\cal O}(1/m_D^{0},1/m_{D^*}^{0})$ are considered. 
At the lowest order, there are no tree-level diagrams converting $D$ mesons into $D^*$ mesons~\cite{Abreu:2011ic}, and the two sectors are independent but related by \gls{hqss}. Furthermore, at \gls{lo} in the heavy-quark expansion, one has $h_i=\tilde{h}_i$ for the \glspl{lec}. 

For a scattering from an incoming channel $i$ to an outgoing channel $j$, each one involving a charmed meson and a light meson, the amplitudes read
\begin{align} \nonumber\label{eq:free-mm-potential}
 V^{ij}(s,t,u)=&\ \frac{1}{f_\pi^2}\Big[\frac{C_\textrm{LO}^{ij}}{4}(s-u)-4C_0^{ij}h_0+2C_1^{ij}h_1\\ \nonumber
 &\ -2C_{24}^{ij}\Big(2h_2(p_2\cdot p_4)+h_4\big((p_1\cdot p_2)(p_3\cdot p_4)+(p_1\cdot p_4)(p_2\cdot p_3)\big)\Big)\\ 
 &\ +2C_{35}^{ij}\Big(h_3(p_2\cdot p_4)+h_5\big((p_1\cdot p_2)(p_3\cdot p_4)+(p_1\cdot p_4)(p_2\cdot p_3)\big)\Big)
 \Big],
\end{align}
where $s=(p_1+p_2)^2$, $t=(p_1-p_3)^2$, and $u=(p_1-p_4)^2$ are the Mandelstam variables. The different coupled channels $i,j$ in the sectors considered in this dissertation are listed in Table~\ref{tab:free-mm-channels}. The isospin coefficients $C^{ij}_k$, which give the strength of the $k$-term ($k=\{\textrm{LO},0,1,24,35\}$) between channels $i$ and $j$ in the isospin basis, are given in Table~\ref{tab:free-mm-coeffiso} for the charm sector, whereas the corresponding ones in charge basis are given in Appendix \ref{sec:appendix:mm_coeff}, as well as the relations to transform from the charge basis to the isospin basis. The corresponding coefficients in the bottom sector are obtained straightforwardly upon the replacement $D\rightarrow \bar{B}$.
 
\begin{table}[t!]
\setlength{\tabcolsep}{10pt}
\renewcommand{\arraystretch}{1.2}
\begin{tabular}{c l c c c c c}
 \hline
 $(S,I)$  &  Channel $i\rightarrow j$ & $C_\textrm{LO}^{ij}$ & $C_0^{ij}$ & $C_1^{ij}$ & $C_{24}^{ij}$ &  $C_{35}^{ij}$  \\  
\hline
 $(-1,0)$  & $D\bar{K}\rightarrow D\bar{K}$ & $-1$ & $m_K^2$ & $m_K^2$ & $1$ & $-1$ \\
 $(-1,1)$ & $D\bar{K}\rightarrow D\bar{K}$ & $1$ & $m_K^2$ & $-m_K^2$ & $1$ & $1$ \\
 $(0,\frac{1}{2})$  & $D\pi\rightarrow D\pi$ & $-2$ & $m_\pi^2$ & $-m_\pi^2$ & $1$ & $1$ \\
   & $D\pi\rightarrow D\eta$ & $0$ & $0$ & $-m_\pi^2$ & $0$ & $1$ \\
   & $D\pi\rightarrow D_s\bar{K}$ & $-\sqrt{\frac{3}{2}}$ & $0$ & $-\frac{\sqrt{3}}{2\sqrt{2}}(m_K^2+m_\pi^2)$ & $0$ & $\sqrt{\frac{3}{2}}$ \\
   & $D\eta\rightarrow D\eta$ & $0$ & $m_\eta^2$ & $-\frac{1}{3}m_\pi^2$ & $1$ & $\frac{1}{3}$ \\
   & $D\eta\rightarrow D_s\bar{K}$ &$-\sqrt{\frac{3}{2}}$ & $0$ & $\frac{1}{2\sqrt{6}}(5m_K^2-3m_\pi^2)$ & $0$ & $-\frac{1}{\sqrt{6}}$ \\
   & $D_s\bar{K}\rightarrow D_s\bar{K}$ & $-1$ & $m_K^2$ & $-m_K^2$ & $1$ & $1$ \\
 $(0,\frac{3}{2})$  & $D\pi\rightarrow D\pi$ & $1$ & $m_\pi^2$ & $-m_\pi^2$ & $1$ & $1$ \\
 $(1,0)$   & $DK\rightarrow DK$ & $-2$ & $m_K^2$ & $-2m_K^2$ & $1$ & $2$ \\
   & $DK\rightarrow D_s\eta$ & $-\sqrt{3}$ & $0$ & $-\frac{1}{2\sqrt{3}}(5m_K^2-3m_\pi^2)$ & $0$ & $\frac{1}{\sqrt{3}}$ \\
   & $D_s\eta\rightarrow D_s\eta$ & $0$ & $m_\eta^2$ & $-\frac{4}{3}(2m_K^2-m_\pi^2)$ & $1$ & $\frac{4}{3}$ \\
 $(1,1)$  & $D_s\pi\rightarrow D_s\pi$ & $0$ & $m_\pi^2$ & $0$ & $1$ & $0$ \\
   & $D_s\pi\rightarrow DK$ & $1$ & $0$ & $-\frac{1}{2}(m_K^2+m_\pi^2)$ & $0$ & $1$ \\
   & $DK\rightarrow DK$ & $0$ & $m_K^2$ & $0$ & $1$ & $0$ \\
 $(2,\frac{1}{2})$  & $D_sK\rightarrow D_sK$ & $1$ & $m_K^2$ & $-m_K^2$ & $1$ & $1$ \\
\hline
\end{tabular}
\centering
\caption{Isospin coefficients $C_k^{ij}$ for the sectors with isospin $I$ and strangeness $S$.}
\label{tab:free-mm-coeffiso}
\end{table}

The values of the \glspl{lec} $h_i$, with $i=0,...,5$, appearing in the \gls{nlo} part of the interaction kernel of Eq.~(\ref{eq:free-mm-potential}) have to be fixed from fits to \gls{lqcd} data. In the first works describing the interaction of heavy mesons with light mesons with unitarized models at \gls{nlo} \cite{Guo:2008gp,Guo:2009ct}, the terms with the \glspl{lec} $h_{0,2,4}$ were dropped to reduce the number of parameters, since they are suppressed in the large $N_c$ limit. The full \gls{nlo} Lagrangian could be kept in more recent studies~\cite{Liu:2012zya,Guo:2018tjx} thanks to the availability of lattice data at several unphysical quark masses.

The \gls{lec} $h_0$ is determined in the latest two references through the fit of \gls{lqcd} data for the masses of the $D$ and $D_s$ at different pion masses, while the value of $h_1$ can be fixed from the physical mass splitting between the $D$ and $D_s$,
\begin{equation}\label{eq:free-mm-h1}
 h_1=\frac{m_{D_s}^2-m_D^2}{4(m_K^2-m_\pi^2)} \ .
\end{equation}

Dimensionless linear combinations of the remaining \glspl{lec}, $h_{24}=h_2+h_4\hat{M}_D^2$ and $h_{35}=h_3+2h_5\hat{M}_D^2$, with $\hat{M}_D^2=(m_D+m_{D_s})/2$, are determined in~\cite{Liu:2012zya} by fits to the scattering lengths calculated on the lattice, simultaneously also to lattice finite-volume energy levels in~\cite{Guo:2018tjx}, for the sector with charmed pseudoscalar mesons. Here we take the values of the \glspl{lec} from the Fit-2B in~\cite{Guo:2018tjx}, for which the full amount of lattice data available and the physical value of $f_\pi$ are considered, and which is the preferred fit of the authors according to large $N_c$ arguments. The values are shown in the upper rows of Table~\ref{tab:free-mm-LECs}, for both pseudoscalar- and vector-charmed mesons, where we have used the different physical values of the average masses $\hat{M}_D=(m_D+m_{D_s})/2$ and $\hat{M}_{D^*}=(m_{D^*}+m_{D_s^*})/2$ in the determination of the dimensionful \glspl{lec} $h_4$ and $h_5$. 
The difference between our unitarized amplitudes and those in~\cite{Guo:2018tjx} lies in the regularization procedure, as explained below.

\begin{table}[t!]
\setlength{\tabcolsep}{10pt}
\renewcommand{\arraystretch}{1.2}
\begin{tabular}{l | c c c c c c}
 \hline
 & $h_0$ & $h_1$ & $h_2$ & $h_3$ & $h_4\,(\textrm{MeV}^{-2})$ & $h_5\,(\textrm{MeV}^{-2})$ \\
\hline
$D\Phi$ & $0.033$ & $0.45$ & $-0.12$ & $1.67$ & $-0.0054\cdot10^{-6}$ & $-0.22\cdot10^{-6}$ \\
$D^*\Phi$ &  $0.033$ & $0.45$ & $-0.12$ & $1.67$ & $-0.0047\cdot10^{-6}$ & $-0.19\cdot10^{-6}$ \\ 
\hline
$\bar{B}\Phi$ & $0.089$ & $1.05$ & $-0.32$ & $4.49$ & $-0.0019\cdot10^{-6}$ & $-0.077\cdot10^{-6}$ \\
$\bar{B}^*\Phi$ &  $0.089$ & $1.05$ & $-0.32$ & $4.49$ & $-0.0019\cdot10^{-6}$ & $-0.075\cdot10^{-6}$ \\ 
\hline
\end{tabular}
\centering
\caption{Values of the \glspl{lec} for the interaction of charmed pseudoscalar (first row) and vector (second row) mesons with the light mesons. They have been taken from Fit-2B in~\cite{Guo:2018tjx} for the $D\Phi$ case. The corresponding values for the interaction of bottomed pseudoscalar (third row) and vector (fourth row) mesons with light mesons are obtained from the scaling of the \glspl{lec} in the charm sector, as explained in the text.}
\label{tab:free-mm-LECs}
\end{table}

The extension of  the interaction potential of Eq.~(\ref{eq:free-mm-potential}) to the open-bottom sector requires the rescaling of the \glspl{lec} with the heavy-meson masses. Relying on \gls{hqfs}, the heavy-flavor
scaling relating the constants $h_i^B$ in the bottomed sector and corresponding $h_i^D$ in the charmed sector can be written as
\begin{equation}\label{eq:free-mm-scalinglecs}
 \frac{h_i^B}{\hat{M}_B}=\frac{h_i^D}{\hat{M}_D} \ , \quad \textrm{for} \ i=\{0,1,2,3\}\ , \quad \textrm{and} \quad h_i^B\hat{M}_B=h_i^D\hat{M}_D \ , \quad \textrm{for} \ i=\{4,5\} \ ,
\end{equation}
up to corrections of order $\mathcal{O}(1/\hat{M}_D,1/\hat{M}_B)$ \cite{Altenbuchinger:2013vwa}. Alternatively, the value of $h_1^B$ can be calculated from the mass splitting between the $B^{(*)}$ and $B_s^{(*)}$ mesons. Employing this latter method is justified since the values of the \gls{lec} $h_1$ in the different sectors can indeed be used to estimate the size of the \gls{hqsfs} breaking. This symmetry demands $h_1^D/\hat{M}_D=\tilde{h}_1^{D^*}/\hat{M}_{D^*}=h_1^B/\hat{M}_B=\tilde{h}_1^{B^*}/\hat{M}_{B^*}$. Applying Eq.~(\ref{eq:free-mm-h1}) to the different heavy sectors and using $\textrm{SU}(3)$ averaged values of the masses, we find $h_1^D/\hat{M}_D=2.24\cdot10^{-4}$~MeV$^{-1}$, $h_1^{D^*}/\hat{M}_{D^*}=2.28\cdot10^{-4}$~MeV$^{-1}$, $h_1^B/\hat{M}_B=1.93\cdot10^{-4}$~MeV$^{-1}$ and $h_1^{B^*}/\hat{M}_{B^*}=2.00\cdot10^{-4}$~MeV$^{-1}$, giving a breaking $\lesssim 4\%$ for \gls{hqss}, while the \gls{hqfs} breaking is $\sim 15\%$. Considering that the violation of \gls{hqss} is rather small, we take a common value of $h_1^B$ for $B$ and $B^*$ mesons, as well as for the rest of the dimensionless \glspl{lec}, as we have done for $D$ and $D^*$ mesons. With this criterion in mind, we use spin-averaged masses, $\bar{M}_D=(\hat{M}_D+\hat{M}_{D^*})/2$ and $\bar{M}_B=(\hat{M}_B+\hat{M}_{B^*})/2$, in Eq.~(\ref{eq:free-mm-scalinglecs}), which gives the values of the \glspl{lec} in the bottom sector that are displayed in the lower part of Table~\ref{tab:free-mm-LECs}.

We focus on $s$-wave scattering and compute the $s$-wave component of the tree-level scattering amplitudes by using the expression for partial-wave projection presented in Eq.~(\ref{eq:free-mb-proj}) in the previous section for the \gls{meson-baryon} interaction amplitudes.
The analysis of higher partial waves is out of the scope of the work of the present dissertation, as we focus on dynamically generated resonances that can correspond to some observed states listed in the \gls{pdg} compilation~\cite{pdg}. In the energy region explored here, no potential candidates are found that can correspond to states generated dynamically from the meson--meson interaction in partial waves higher than $L=0$. In addition, we note that, for a particular channel, the $p$-wave interaction kernel is of reduced strength compared to the $s$-wave one, especially close to the thresholds, around which the molecular states appear. 

As we have described in Section~\ref{subsec:free-th-unitarization}, the amplitudes found in Eq.~(\ref{eq:free-mm-potential}) have to be unitarized so as to satisfy the exact unitarity condition and generate resonances, which are signaled by the presence of poles in the scattering amplitude. To achieve this, we use the unitarization method based on the \gls{bs} equation in coupled channels because, within a straightforward extension of field theory, it can be simply applied to finite temperature, as will be explained in Chapter~\ref{ch:hot-medium}.

In the on-shell factorization approach~\cite{Oller:1997ti,Oset:1997it}, the \gls{bs} equation for the unitarized amplitude was given in Eq.~(\ref{eq:free-th-BSeq}) and reproduced here for completeness:
\begin{equation} \label{eq:free-mm-BetheSalpeter}
  T_{ij} (s)= V_{ij} (s) + V_{il}(s) G_l(s) T_{lj} (s) \ . 
\end{equation}
The subindices $i,j$ represent the incoming and outgoing channels (see Table~\ref{tab:free-mm-channels}), and we sum over the possible intermediate channels $l$. The two-meson propagator function gives the loop function
\begin{equation}\label{eq:free-mm-loop}
  G_l(s)=\ii\int\frac{d^4q}{(2\pi)^4} \frac{1}{q^2-m_D^2+i\varepsilon} \frac{1}{(p-q)^2-m_\Phi^2+i\varepsilon} \ ,
\end{equation}
with $p^\mu=(E,\vec{p}\,)$. We make explicit that at $T=0$ the loop function is given as a function of the Mandelstam variable $s=p^2$. This function should be regularized, for which we use a hard momentum cut-off $\Lambda$, and the corresponding expression of the loop reads the same as that given in Eq.~(\ref{eq:free-mb-Gmatrixcutanalytic}) for meson--baryon channels, except for the factor $2M_l$ that is missing in the case of meson--meson channels, and with the replacements $M_l\rightarrow m_D$ and $m_l\rightarrow m_\Phi$. Alternatively, the two-meson loop function can be calculated in \gls{dr} using Eq.~(\ref{eq:free-mb-GmatrixDR}), with the same modifications as for the cut-off approach.

In Ref.~\cite{Guo:2018tjx} the loop function is regularized with \gls{dr} and the subtraction constants are considered as fit parameters together with the \glspl{lec}. In this dissertation, we prefer to use the cut-off regularization scheme to follow the same convention that we will use in the next chapter for $T > 0$. The cut-off value is adjusted to a representative scale of the degrees of freedom that are implicitly integrated out in the construction of the \gls{meson-meson} interaction amplitude from the effective Lagrangian. Indeed, the contact interaction of Eq.~(\ref{eq:free-mm-potential}) could have also been obtained from a $t$-channel diagram similar to that of Fig.~\ref{fig:free-mb-feynmandiagram_ps_a} for the \gls{meson-baryon} interaction, involving two three-meson vertices and the propagator of a vector meson of mass $m_V\sim m_\rho$, in the limit $m_V^2\gg t $, with $t$ being the four-momentum exchanged in the process. 
Equation~(\ref{eq:free-mb-a(mu)}) above can be used to compute the equivalent cut-off values for which the loop function coincides at each channel threshold with the corresponding one calculated in \gls{dr} with the subtraction constants in Ref.~\cite{Guo:2018tjx}. We find that the values of $\Lambda$ are rather small in some channels and we, therefore, find it more convenient to use a cut-off of the order of the $\rho$-meson mass, namely $\Lambda=800$~MeV, for all the channels. 

We have checked that the values of the subtraction constants obtained using \gls{dr} while demanding the same value of the loop function at the channel threshold as that of the loop regularized with a cut-off of $800$~MeV are compatible with those employed in~\cite{Guo:2018tjx}, with $\mu= 1$~GeV. Furthermore, the reproduction of scattering observables with our prescription is of comparable quality. In particular the $s$-wave scattering lengths obtained from the values of the unitarized amplitudes at the channel threshold\footnote{Not to be confused with the subtraction constants $a_l(\mu)$.},
\begin{equation}\label{eq:free-mm-scattlengths}
 {a}_{0,i}=-\frac{1}{8\pi(m_D+m_\Phi)}T_{ii}(s_{\textrm{thr}}) \ , \quad s_{\textrm{thr}}=(m_D+m_\Phi)^2 \ ,
\end{equation}
 are shown in Table~\ref{tab:free-mm-scattlengths} together with the results on the lattice \cite{Liu:2012zya} and the predictions of the fits in Refs.~\cite{Liu:2012zya,Guo:2018tjx}. We note that the diagonal elements of the $T$ matrix, $T_{ii}(s)$, are evaluated for real $s$. In order to search for poles, in the sections below the amplitudes will be analytically continued to complex energies.

\begin{table}[t!]
\setlength{\tabcolsep}{5pt}
\renewcommand{\arraystretch}{1.2}
\begin{tabular}{c c l l l}
 \hline
 $(S,I)$ & $i$ & Eq.~(\ref{eq:free-mm-scattlengths}) & Fit-2B \cite{Guo:2018tjx} & 4-parameter fit \cite{Liu:2012zya}  \\
\hline
 $(-1,0)$   & $D\bar{K}$ & $\phantom{-}0.63$ & $\phantom{-}0.68^{+0.17}_{-0.16}$ & $\phantom{-}0.84(15)^*$  \\
 $(-1,1)$   & $D\bar{K}$ & $-0.19$ & $-0.19^{+0.02}_{-0.02}$ & $-0.21(1)^*$\\
 $(0,\frac12)$ & $D\pi$  & $\phantom{-}0.45$ & $\phantom{-}0.34^{+0.00}_{-0.03}$ & $\phantom{-}0.37\pm0.01$ \\
               & $D\eta$ & $\phantom{-}0.14+\ii 0.10$ & $\phantom{-}0.16^{+0.11}_{-0.06}+\ii 0.13^{+0.07}_{-0.03}$ & \\
               & $D_s\bar{K}$ & $-0.18+\ii 0.53$ & $-0.26^{+0.05}_{-0.10}+\ii 0.52^{+0.06}_{-0.03}$ & $-0.06^{+0.01}_{-0.05}+\ii(0.45\pm 0.05)$ \\
 $(0,\frac32)$ & $D\pi$ & $-0.10$ & $-0.099^{+0.003}_{-0.004}$ & $-0.100(2)^*$  \\
 $(1,0)$    & $DK$ & $-0.67$ & $-1.87^{+0.85}_{-1.98}$ & $-0.86\pm0.04$ \\
            & $D_s\eta$ & $-0.27+\ii 0.05$ & $-0.33^{+0.03}_{-0.05}+\ii 0.07^{+0.02}_{-0.02}$ & \\
 $(1,1)$    & $D_s\pi$ & $\phantom{-}0.01$ & $\phantom{-}0.003^{+0.002}_{-0.002}$  & $-0.002(1)^*$\\
            & $DK$ & $\phantom{-}0.04+\ii 0.16$ & $\phantom{-}0.05^{+0.04}_{-0.03}+\ii 0.17^{+0.03}_{-0.03}$ & $\phantom{-}0.04^{+0.05}_{-0.01}+\ii 0.16^{+0.02}_{-0.01}$ \\
 $(2,\frac12)$ & $D_sK$ &  $-0.18$ &  $-0.19^{+0.01}_{-0.01}$  & $-0.18(1)^*$\\
 \hline
\end{tabular}
\centering
\caption{Values of the $s$-wave scattering lengths for various channels obtained with the cut-off unitarization prescription as explained in the text, together with the results on the lattice \cite{Liu:2012zya} (marked with asterisks) and the predictions of the fits in Refs.~\cite{Liu:2012zya,Guo:2018tjx}.}
\label{tab:free-mm-scattlengths}
\end{table}

This agreement reinforces the choice of $\Lambda=800$~MeV as an appropriate selection, a value which, in turn, determines the position of the dynamically generated resonances in the Riemann surface. We note that the largest differences, which are found for the scattering lengths in the $D\pi$ $(0,1/2)$ and $DK$ $(1,0)$ channels, are responsible for the variations in the pole positions that we show in the next sections when compared to those in \cite{Guo:2018tjx}. 

%
%

\subsection{Results for $D$ mesons}
\label{subsec:free-mm-Dmesons}
Let us first discuss the findings for both $J=0$ and $J=1$ charmed sectors. That is, we analyze the interaction of the scalar $D$ mesons and the vector $D^*$ mesons with light mesons separately. Notice that these two sectors are not mixed when keeping the scattering diagrams at \gls{lo} in the heavy-mass expansion, as argued above. After the analytical continuation of the energy to the complex-energy plane, we look for poles in the appropriate \gls{rs} of the $T$ matrix to find bound, resonant, and virtual states (see Section~\ref{subsec:free-th-dynamicallygeneratedstates} for the definitions). The pole position $\sqrt{s_R}$ provides the pole mass, $M_R=\textrm{Re\,}\sqrt{s_R}$, and the width, $\Gamma_R=2\,\textrm{Im\,}\sqrt{s_R}$. We also report the couplings $|g_i|$ of each pole to each of the channels $i$ to which the pole can couple, as well as the compositeness $\chi_i$.

\begin{figure}[b!]
 \centering
 \includegraphics[width=1.02\textwidth]{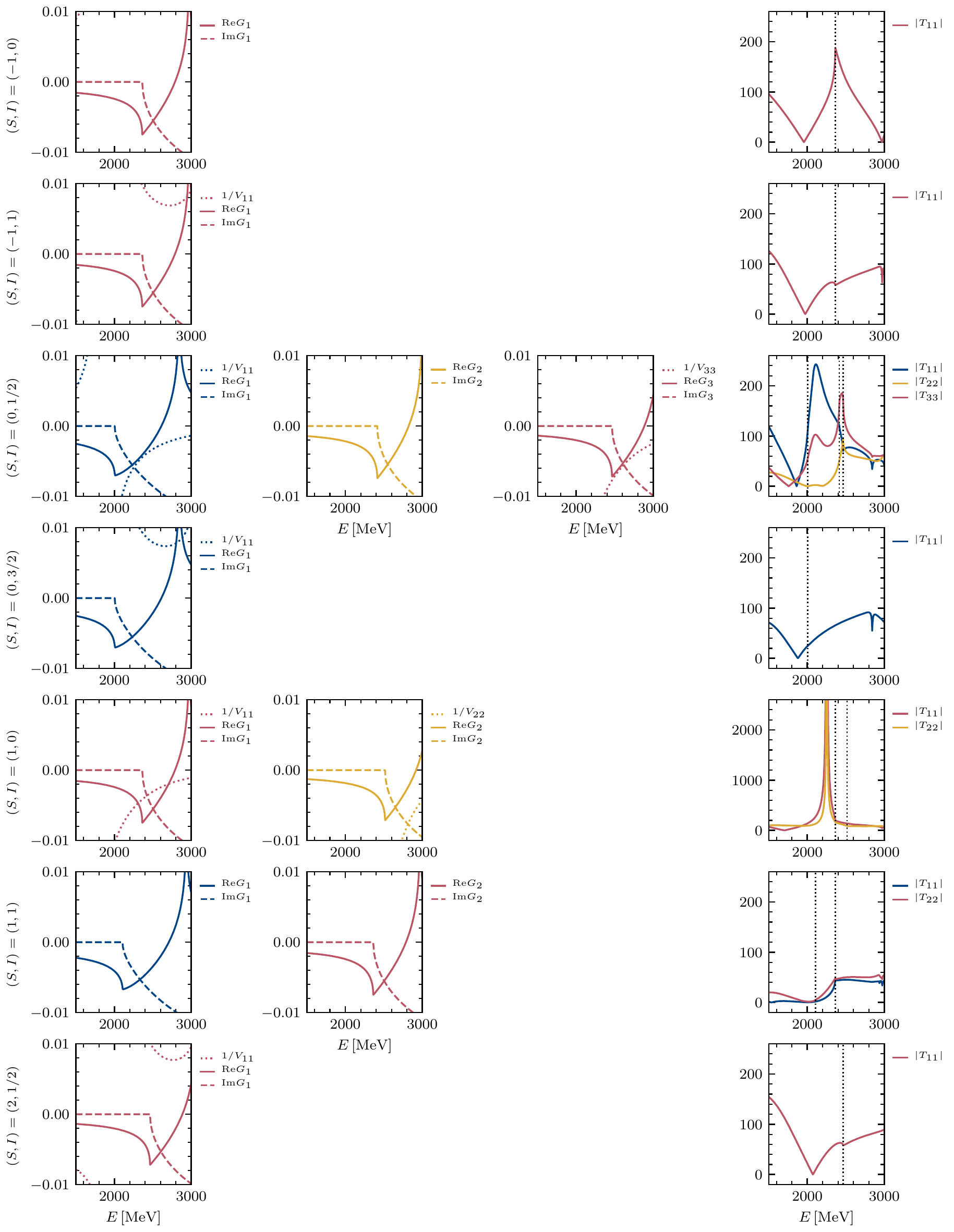}
 \caption{The real and imaginary parts of the loop function, $G_i$, and the inverse of the interaction kernel, $1/V_{ij}$ (left and middle panels), and the absolute value of the diagonal components of the $T$ matrix, $T_{ii}$ (right panels), in the sectors with strangeness and isospin $(S,I)$. The subindices $1$, $2$, $3$ refer to the channels $D\Phi$ in the order in which they are listed in Table~\ref{tab:free-mm-channels}. All the quantities are dimensionless, as they have units of MeV$^0$.}
 \label{fig:free-mm-VGT_allsectors}
\end{figure}

\begin{figure}[b!]
 \centering
 \includegraphics[width=1.02\textwidth]{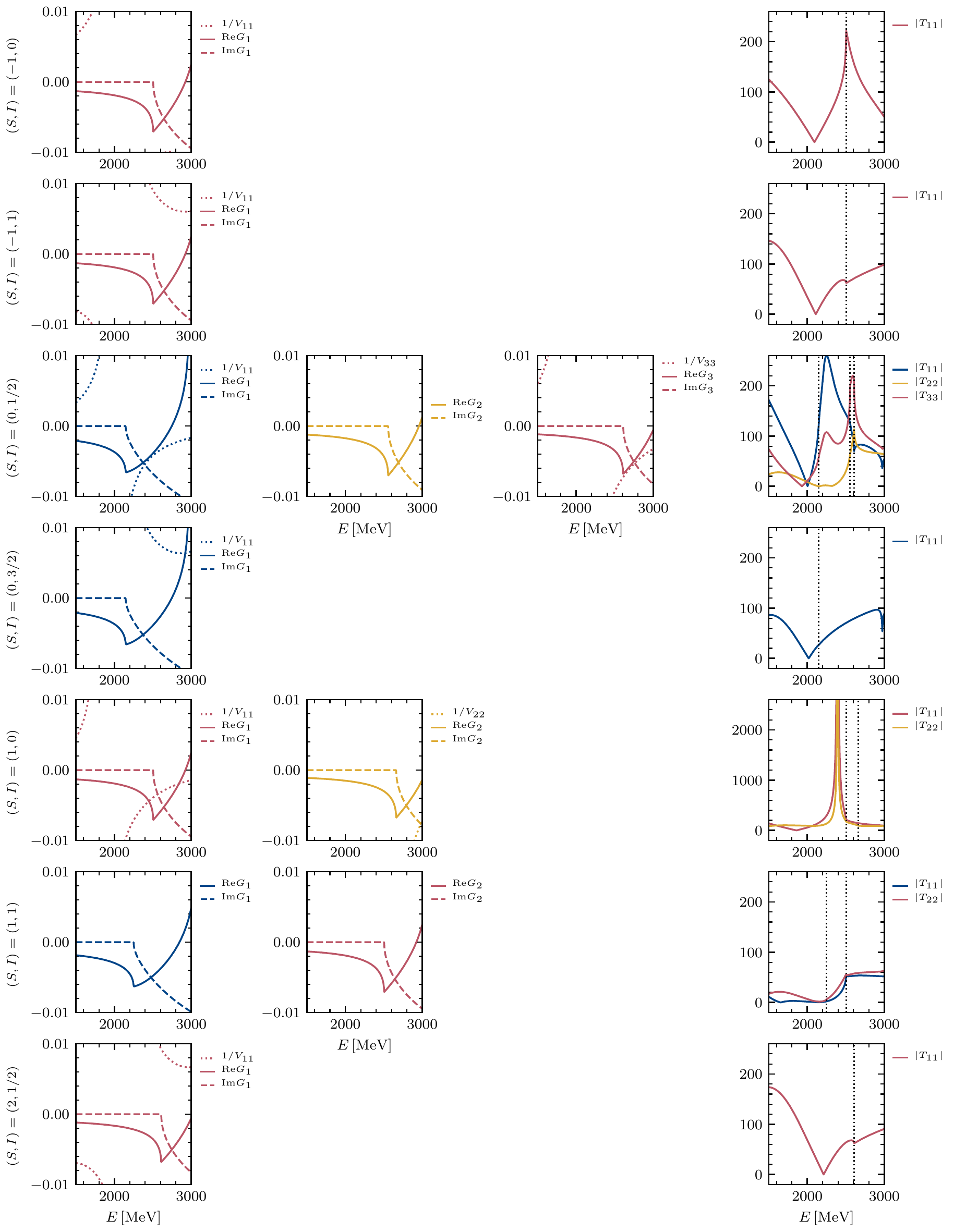}
 \caption{The same as in Fig.~\ref{fig:free-mm-VGT_allsectors} for the case of $D^*\Phi$ scattering.}
 \label{fig:free-mm-VGT_allsectors_v}
\end{figure}

The reflection of the poles in the real-energy axis can be seen from the analysis of the unitarized amplitudes for real $s$.
The diagonal amplitudes in the different sectors with charm $C=1$ and all possible values of strangeness and isospin $(S,I)$ considered in this thesis are shown in the right panels of Figs.~\ref{fig:free-mm-VGT_allsectors} and \ref{fig:free-mm-VGT_allsectors_v} for $D\Phi$ and $D^*\Phi$ channels, respectively, as functions of the total energy and for a center-of-mass momentum $\vec{P}=0$. From the comparison of both figures, one can see that the results are very similar in all the sectors involving either the $D$ or $D^*$ mesons, as we expect from \gls{hqss}, and the small differences are associated with the larger mass of the charmed vector mesons in comparison to that of the pseudoscalars.


To understand the structures appearing in these amplitudes it is convenient to analyze first the energy dependence of the loop function $G$, which is displayed in the left and middle panels of the same figure. The imaginary part of the loop function (dashed lines) starts to have a significant strength from the value of $m_{D^{(*)}}+m_\Phi$ onwards, which is the energy at which the right-hand unitarity cut starts. 
The inverse of the $s$-wave interaction kernel obtained from Eq.~(\ref{eq:free-mm-potential}), $1/V_{ij}$ (dotted lines), is displayed together with the loop function if it falls within the vertical scale employed for each channel. In an uncoupled-channel calculation, one should expect an enhancement in the corresponding unitarized amplitude when this quantity equals or becomes very close to the real part of the loop function (solid lines). This is just the reflection on the real-energy axis of the pole generated by the solution of Eq.~(\ref{eq:free-mm-BetheSalpeter}). The consideration of coupled channels, apart from modifying slightly the energy position of the structures, makes them present in all the amplitudes, with more or less intensity depending on the coupling strength of the pole to each particular channel. As can be seen in the panels on the right, in the $(S,I)=(0,1/2)$ sectors, besides cusps related to thresholds, we see two clear enhancements that are connected to poles of the amplitude in the complex plane, as we will discuss next. Even more clear is the narrow structure appearing in the sectors with $(S,I)=(1,0)$, tied to the position of the crossing of the inverse of the $D^{(*)} K$ potential with the real part of the loop function, which occurs below the $m_{D^{(*)}}+m_K$ threshold and, therefore, leads to a bound state in the real axis, as we will see below.

\subsubsection{$J=0$ case: Interactions and $D^*(2300)$ and $D_{s0}^*(2317)$ }
\label{subsubsec:Dmesons}
   
\begin{figure}[b!]
    \centering
   \includegraphics[width=0.7\textwidth]{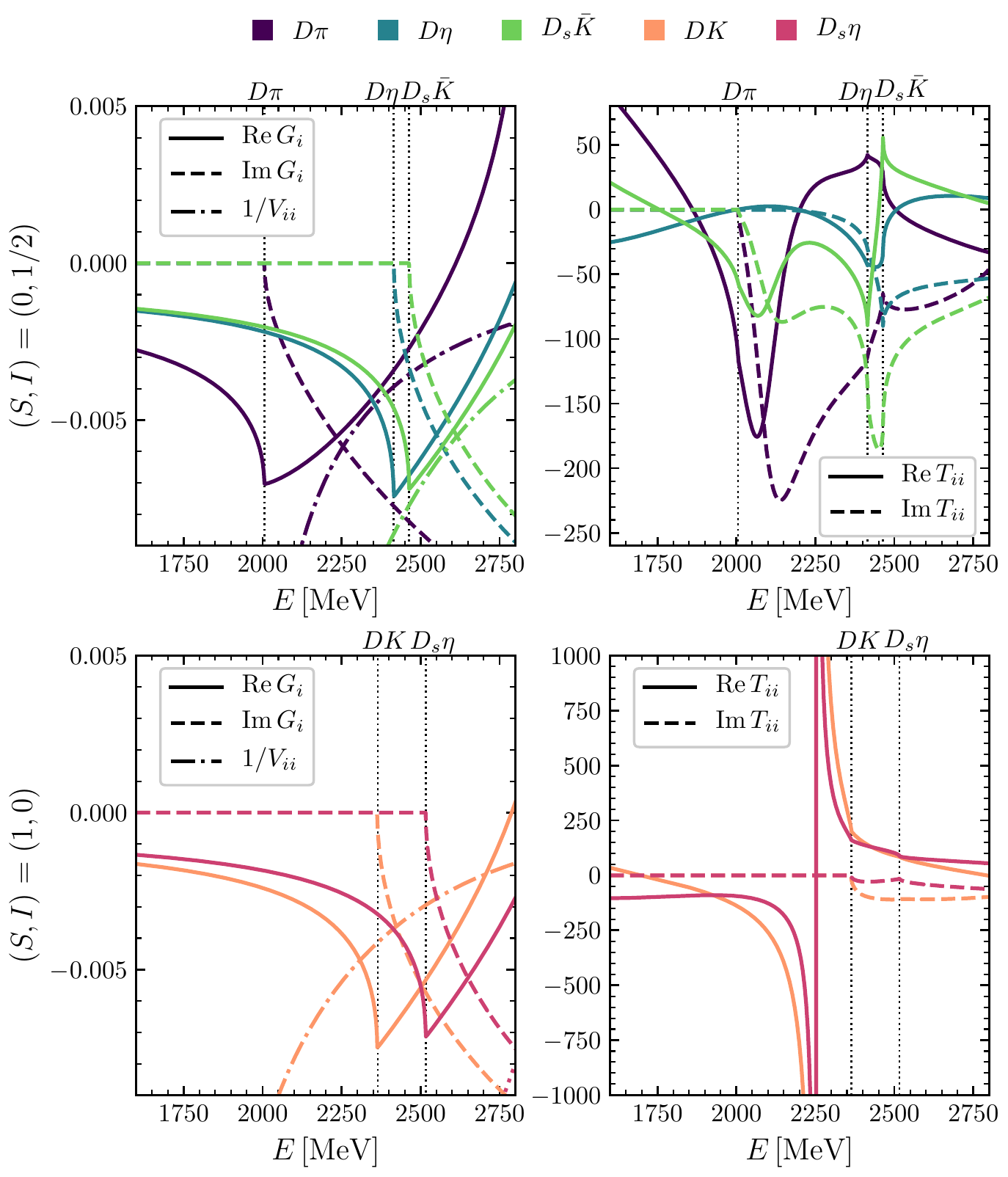}
   \caption{The inverse of the interaction kernel, $1/V_{ii}$, the real and imaginary parts of the loop function, $G_i$, and the real and imaginary parts of the diagonal components of the $T$ matrix, $T_{ii}$, in units of MeV$^0$. Top panels correspond to the sector with strangeness and isospin $(S,I)=(0,1/2)$ (top panels), where the subindices $1$, $2$, $3$ refer to the channels $D\pi$, $D\eta$, and $D_s\bar{K}$, respectively, and the bottom panels to the sector with $(S,I)=(1,0)$, where $1$ and $2$ refer to $DK$ and $D_s\eta$. The vertical dotted lines indicate the corresponding threshold energies. }
   \label{fig:free-mm-VGT_temp0}
   \end{figure}

We start analyzing the $D$ and $D_s$ interactions with light pseudoscalar mesons for the sectors with total strangeness and isospin $(S,I)=(0,1/2)$, corresponding to the $D \pi$, $D \eta$, and $D_s \bar K$ coupled-channels calculation, and $(S,I)=(1,0)$, built from the $D K$ and $D_s \eta$ channels.
A magnification of the plots of the loops, the inverse of the interaction kernels, and the unitarized amplitudes in Fig.~\ref{fig:free-mm-VGT_allsectors} is displayed in Fig.~\ref{fig:free-mm-VGT_temp0}.
We focus on these two sectors since there appear several resonant states, among them two that can be identified with the experimental $D_0^*(2300)$ and $D_{s0}^*(2317)$. 

For the $(S,I)=(0,1/2)$ sector shown in the top panels of Fig.~\ref{fig:free-mm-VGT_temp0}, one can see two clear structures appearing in the imaginary part of the scattering amplitudes (dashed lines in the panel on the right), at around $2100$~MeV and $2450$~MeV. These structures are related to poles of the $T$ matrix in the complex-energy plane.


Our results for the strangeness $S=1$ and isospin $I=0$ sector are shown in the bottom panels of the same figure. In this case, we only observe a narrow structure around $2250$~MeV, which is very close to the position of the crossing of the inverse of the potential in the $DK$ channel (dash-dotted line in the left panel) with the real part of the loop function (orange solid line in the same panel). Since this crossing takes place below the $m_D+m_K$ threshold, a bound state appears in the real axis on the physical \gls{rs} of the unitarized amplitudes in coupled channels (solid lines in the right panel).

As discussed in the previous plots, apart from threshold effects, the different structures that are present in the scattering amplitudes correspond to poles or dynamically generated states that appear due to the attractive coupled-channel meson--meson interactions. 
The poles that we find in the $J^P=0^+$ sectors are summarized in Table~\ref{tab:free-mm-poles0}. 
The first column of this table indicates the possible experimental assignment of the poles according to the \gls{pdg}~\cite{pdg}, whereas the second column shows the strangeness and isospin content of the state. The third column indicates the \gls{rs} where the pole is found, with the convention that the \gls{rs} of the loop function for each of the coupled channels is indicated as ``$+$'' for the first (\gls{rs}-I) and ``$-$'' for the second (\gls{rs}-II). In the fourth and fifth columns, we give the mass $M_R$ and the half width $\Gamma_R/2$ of the state, while in the sixth column  $g_i$ denotes the effective coupling to the different channels, and in the seventh column $\chi_i$ is the compositeness of the state. 

\begin{table}[b!]\centering
 \setlength{\tabcolsep}{9pt}
\renewcommand{\arraystretch}{1.2}
\begin{tabular}{cccccr@{\,=\,}lr@{\,=\,}l}
\hline
& $(S,I)$ & RS & $M_R$ & $\Gamma_R/2$ & \multicolumn{2}{c}{$|g_i|$} & \multicolumn{2}{c}{$\chi_i$} \\
& & & (MeV) & (MeV) & \multicolumn{2}{c}{(GeV)} & \multicolumn{2}{c}{ } \\
\hline
$D_0^*(2300)$ & $(0,\frac12)$ & $(-,+,+)$ & $2081.9$ & $86.0$  & $|g_{D\pi}|$&$8.9$  & $\chi_{D\pi}$&$0.45$ \\
 &    &           &        &       & $|g_{D\eta}|$&$0.4$  & $\chi_{D\eta}$&$0.00$ \\
 &   &           &        &       & $|g_{D_s\bar{K}}|$&$5.4$  & $\chi_{D_s\bar{K}}$&$0.02$ \\
 &   & $(-,-,+)$ & $2529.3$ & $145.4$ & $|g_{D\pi}|$&$6.7$  & $\chi_{D\pi}$&$0.20$ \\
 &   &           &        &       & $|g_{D\eta}|$&$9.9$  & $\chi_{D\eta}$&$0.55$ \\
 &   &           &        &       & $|g_{D_s\bar{K}}|$&$19.4$ & $\chi_{D_s\bar{K}}$&$0.95$ \\
    \hline
$D_{s0}^*(2317)$ &   $(1,0)$ & $(+,+)$ & $2252.5$ & $0.0$ & $|g_{DK}|$&$13.3$ & $\chi_{DK}$&$0.44$ \\
 &    &           &        &       & $|g_{D_s\eta}|$&$9.2$ & $\chi_{D_s\eta}$&$0.08$ \\
\hline
 \end{tabular}
 \centering
 \caption{Properties of the dynamically generated poles in the $J^P=0^+$ sectors with $(S,I)=(0,1/2)$ and $(S,I)=(1,0)$. The first column is reserved for the state listed in the \gls{rpp}~\cite{pdg} to which the pole can be assigned. The following columns display, in this order, the \gls{rs} with the convention given in the main text, the real and the imaginary parts of the pole location in the complex-energy plane, the effective coupling to different channels, and the compositeness of the state. }
 \label{tab:free-mm-poles0}
 \end{table}
 

In the $(S,I)=(0,1/2)$ sector we find two poles that correspond to the $D_0^* (2300)$ state. This double pole structure of the $D_0^* (2300)$ is well documented~\cite{pdg}, being our results compatible with those given in Refs.~\cite{Guo:2018tjx,Albaladejo:2016lbb}. For the position of the lower pole ($2082-\ii 86$~MeV), we find that the real part lies between the $D\pi$ and $D\eta$ thresholds, at $2005$~MeV and $2415$~MeV respectively, and
it has a sizable imaginary part, which is a consequence of the large value of the coupling of the generated resonance to the $D\pi$ channel, to which it can decay.
As for the higher pole ($2529-\ii 145$~MeV), the mass is above the last threshold, that is, the $D_s \bar{K}$ one at $2464$~MeV, and also has a large decay width, as it couples sizably to the channels opened for its decay. However, this pole appears in the $(-,-,+)$ \gls{rs}, with this \gls{rs} being only connected to the real axis between the $D\eta$ and the $D_s\bar{K}$ thresholds. We recall that the \gls{rs} of the $T$ matrix closest to the physical region above the largest threshold corresponds to that for which the loop functions of the three coupled channels have been rotated to the \gls{rs}-II, that is, the $(-,-,-)$ \gls{rs}. In fact, for different values of the parameters~\cite{Guo:2018tjx,Albaladejo:2016lbb}, this pole appears between the $D\eta$ and $D_s\bar{K}$ thresholds or even below the $D\eta$ threshold. 
On the other hand, the reflection of this pole in the real axis is somewhat complicated due to its proximity to the $D\eta$ and the $D_s\bar{K}$ thresholds. Indeed, by inspecting the upper-right panel of Fig.~\ref{fig:free-mm-VGT_temp0} one can see that the width of the structure at higher energies in the $D_s\bar{K}$ amplitude is substantially smaller than the imaginary part of the pole and that the shape is highly dominated by the presence of the thresholds.
Moreover, it is worth noticing that the lower pole qualifies mainly as a $D\pi$ state, as indicated by the large value of the compositeness, whereas the higher one is essentially a $D_s\bar{K}$ system, although we should note that this case does not correspond to a canonical resonance in the sense that the associated pole does not reside in the \gls{rs} that is directly accessible from the physical one. Therefore, as discussed in Ref.~\cite{Guo:2015daa}, the physical interpretation of Eq.~(\ref{eq:free-th-compositeness}) as a probabilistic compositeness is not valid for this resonance, a fact that is corroborated by the sum over the different channels being larger than one in this case. However, it can still be regarded as the strength of the different channels in the wave function of the state~\cite{Aceti:2014ala}.

In the $(S,I)=(1,0)$ sector we find only one pole at $2252-\ii 0$~MeV. It lies on the real axis below the $DK$ threshold (at $2365$~MeV), that is in the $(+,+)$ \gls{rs}. It is identified with the $D_{s0}^*(2317)$ resonance, and has sizable couplings $DK$, as given by the compositeness. With the present model, the pole mass turns out to be smaller than that of the experimental resonance, but a small variation of the parameters can easily accommodate this state to the observed position, in line with similar models in the literature that have advocated this resonance to be mostly a $DK$ hadronic molecule (see Ref.~\cite{Guo:2017jvc} and references therein). Indeed this can be easily understood by looking at the bottom-left panel of Fig.~\ref{fig:free-mm-VGT_temp0}. In the case of uncoupled channels, the pole position is simply given by the crossing of $1/V_{DK\rightarrow DK}$ with the real part of the loop function $G_{DK}$, as it corresponds to the zero of the denominator of the \gls{bs} equation of Eq.~(\ref{eq:free-th-BSmatrixinv}). Then, a smaller value of the cut-off in the regularization of the loop (or a less negative subtraction constant in \gls{dr}) will move $\textrm{Re\,}G_{DK}$ upwards and the crossing with $1/V_{DK\rightarrow DK}$ will take place closer to the $DK$ threshold. That is, the $D_{s0}^*(2317)$ will be less bound. The dynamical generation of the pole in the coupled-channel case is more complicated, but the conclusion remains the same.  

\begin{figure}[t!]\centering 
\begin{subfigure}[b]{0.49\textwidth}\centering 
\captionsetup{skip=0pt}
 \includegraphics[width=\textwidth]{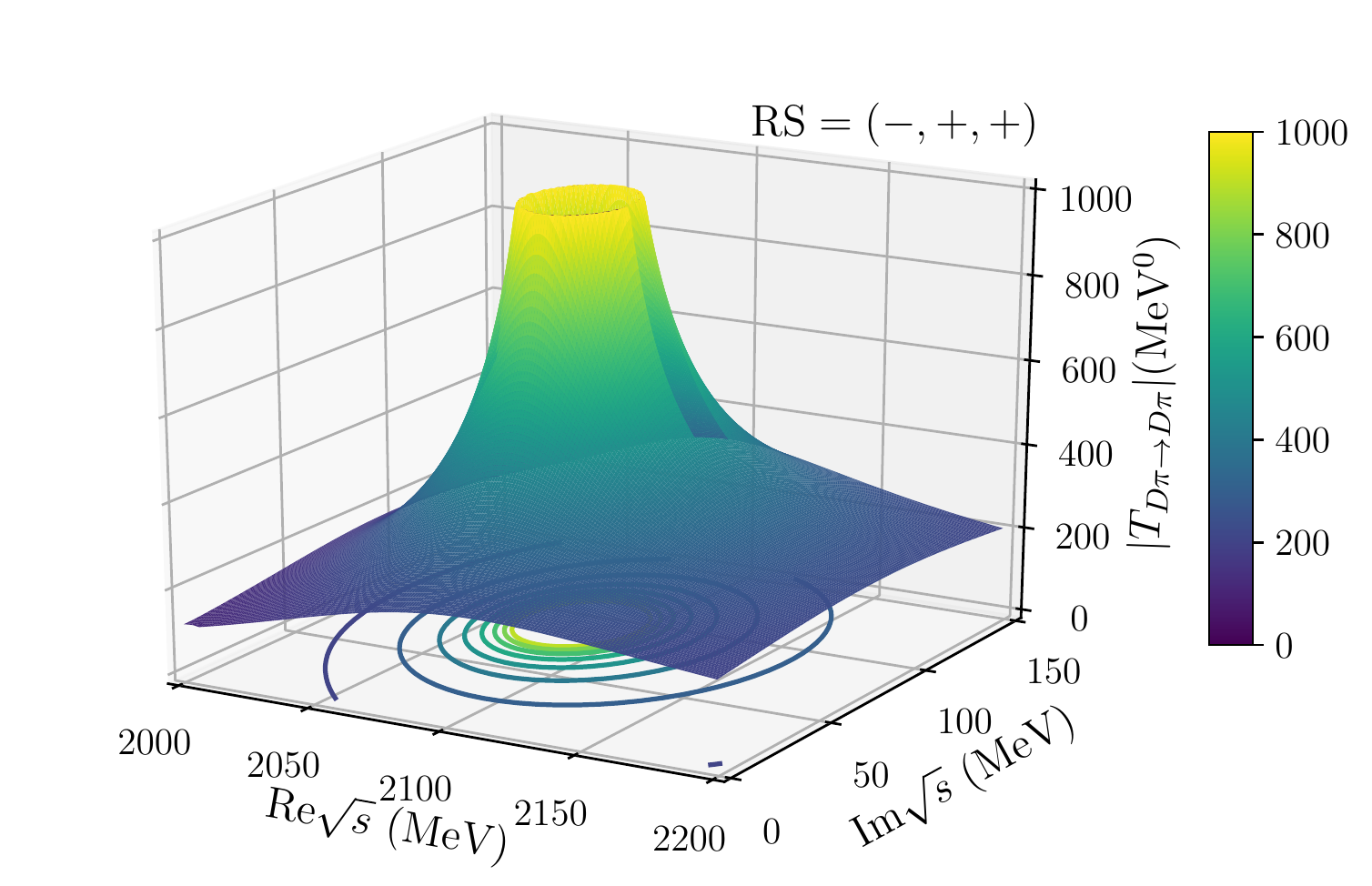}
\caption{}
\label{fig:free-mm-RS2300-a}
\end{subfigure}
\begin{subfigure}[b]{0.49\textwidth}\centering 
\captionsetup{skip=0pt}
 \includegraphics[width=\textwidth]{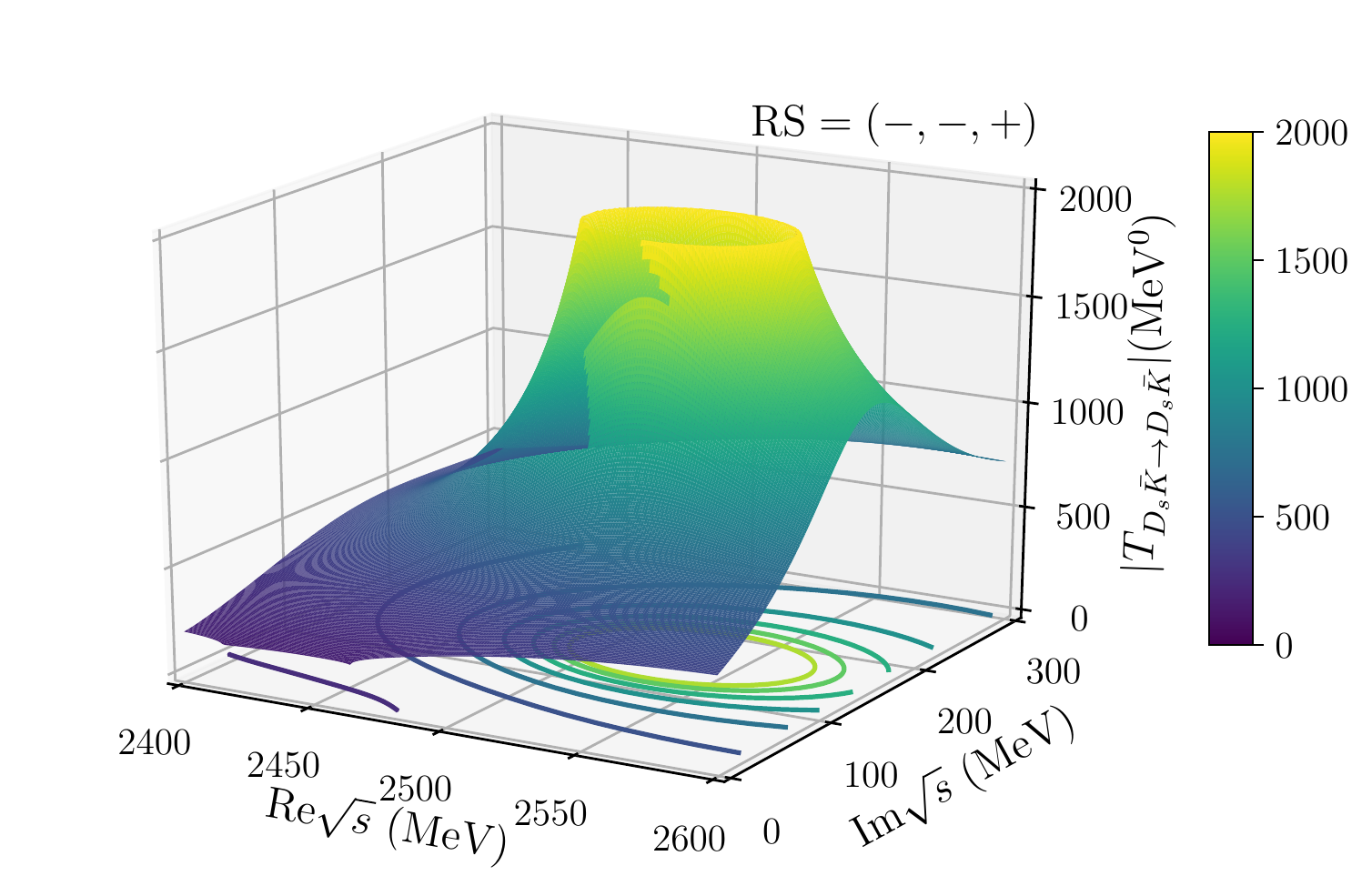}
\caption{}
\label{fig:free-mm-RS2300-b}
\end{subfigure}
\caption{Riemann surfaces of the $T$ matrix around the complex energies of the two poles of the $D_0^*(2300)$. (a) Diagonal $D\pi$ element in the $\textrm{RS}=(-,+,+)$. (b) Diagonal $D_s\bar{K}$ element in the $\textrm{RS}=(-,-,+)$. }
\label{fig:free-mm-RS2300}
\end{figure}
\begin{figure}[t!]\centering 
 \includegraphics[width=0.49\textwidth]{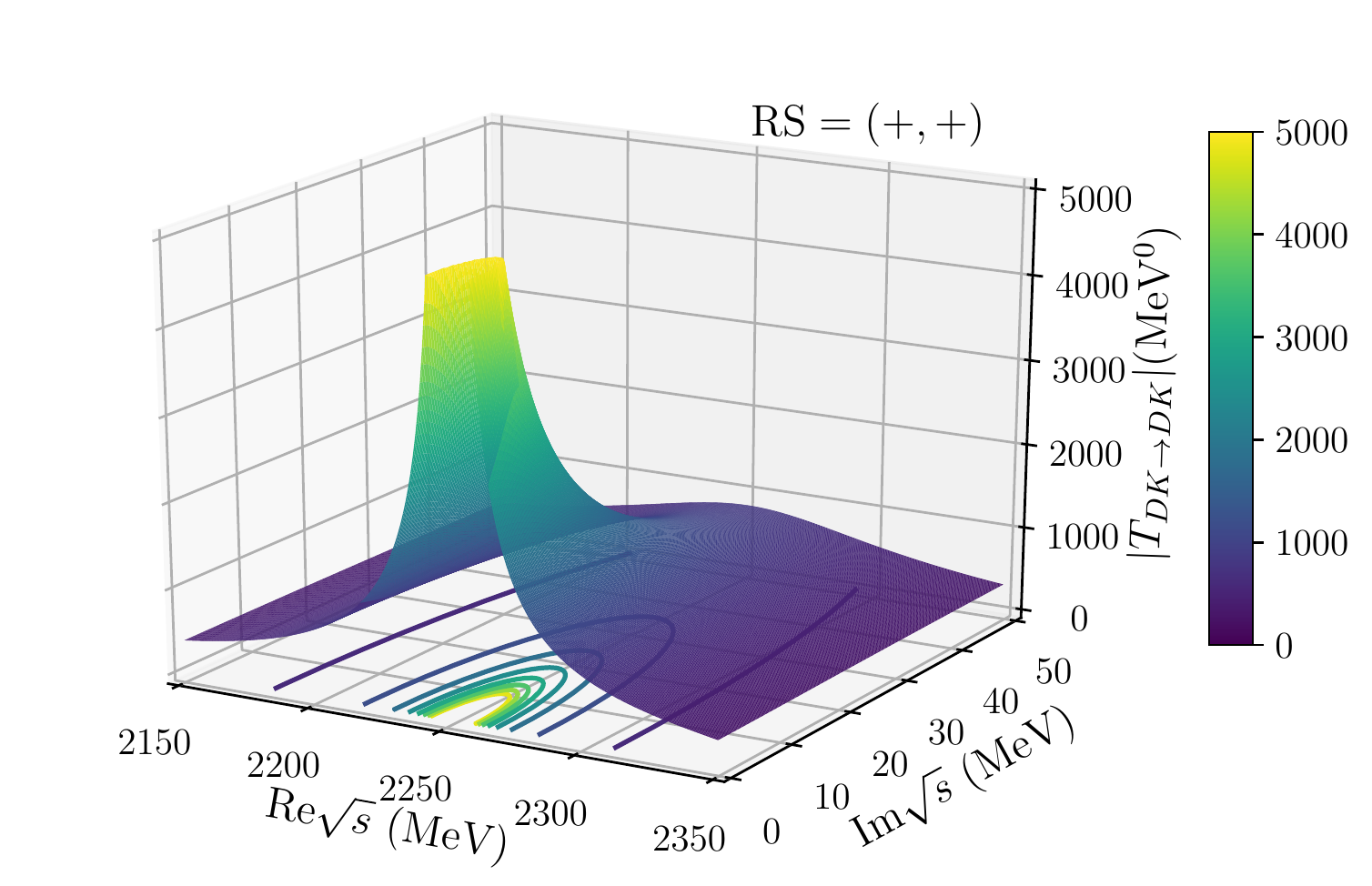}
\caption{Riemann surface of the diagonal $DK$ element of the $T$ matrix in the $\textrm{RS}=(+,+)$ around the complex energy of the $D_{s0}^*(2317)$ pole. }
\label{fig:free-mm-RS2317}
\end{figure}

The Riemann surfaces of the modulus of the $T$ matrix, $|T|$, for complex energies around the location of the poles associated with the $D_0^* (2300)$ and the $D_{s0}^*(2317)$ states are plotted in Figs.~\ref{fig:free-mm-RS2300} and \ref{fig:free-mm-RS2317}, respectively. From these plots, one can see the enormous increase of $|T|$ as the pole position. In the case of the lower pole of the $D_0^*(2300)$ (Fig.~\ref{fig:free-mm-RS2300-a}), the $(-,+,+)$ \gls{rs} is the \gls{rs}-II and thus the slice of the Riemann surface along the real-energy axis corresponds with the structure of the $T$ matrix shown in the upper-left panel of Fig.~\ref{fig:free-mm-VGT_temp0} between the $D\pi$ and $D\eta$ thresholds, after taking the modulus. As for the higher pole of the $D_0^*(2300)$ (Fig.~\ref{fig:free-mm-RS2300-b}), what we see in the same panel between the $D\eta$ and $D_s\bar{K}$ thresholds is the slice along the real axis of the tail on the right of the pole, which is then connected at the thresholds with the corresponding \glspl{rs}-II. Therefore, the mass and width that one can extract from the pole position differ from the location and the width of the real-energy structure.
Regarding the pole of the $D_{s0}^*(2317)$ (Fig.\ref{fig:free-mm-RS2317}), the divergence that appears in the physical sheet of the $T$ matrix on the real-energy axis corresponds to the singularity seen in the bottom-left panel of Fig.~\ref{fig:free-mm-VGT_temp0}.
 
\subsubsection{$J=1$ case: Interactions and $D_1(2430)$ and $D_{s1}(2460)$}
\label{subsubsec:D*mesons}

\begin{figure}[t!]
    \centering
   \includegraphics[width=0.7\textwidth]{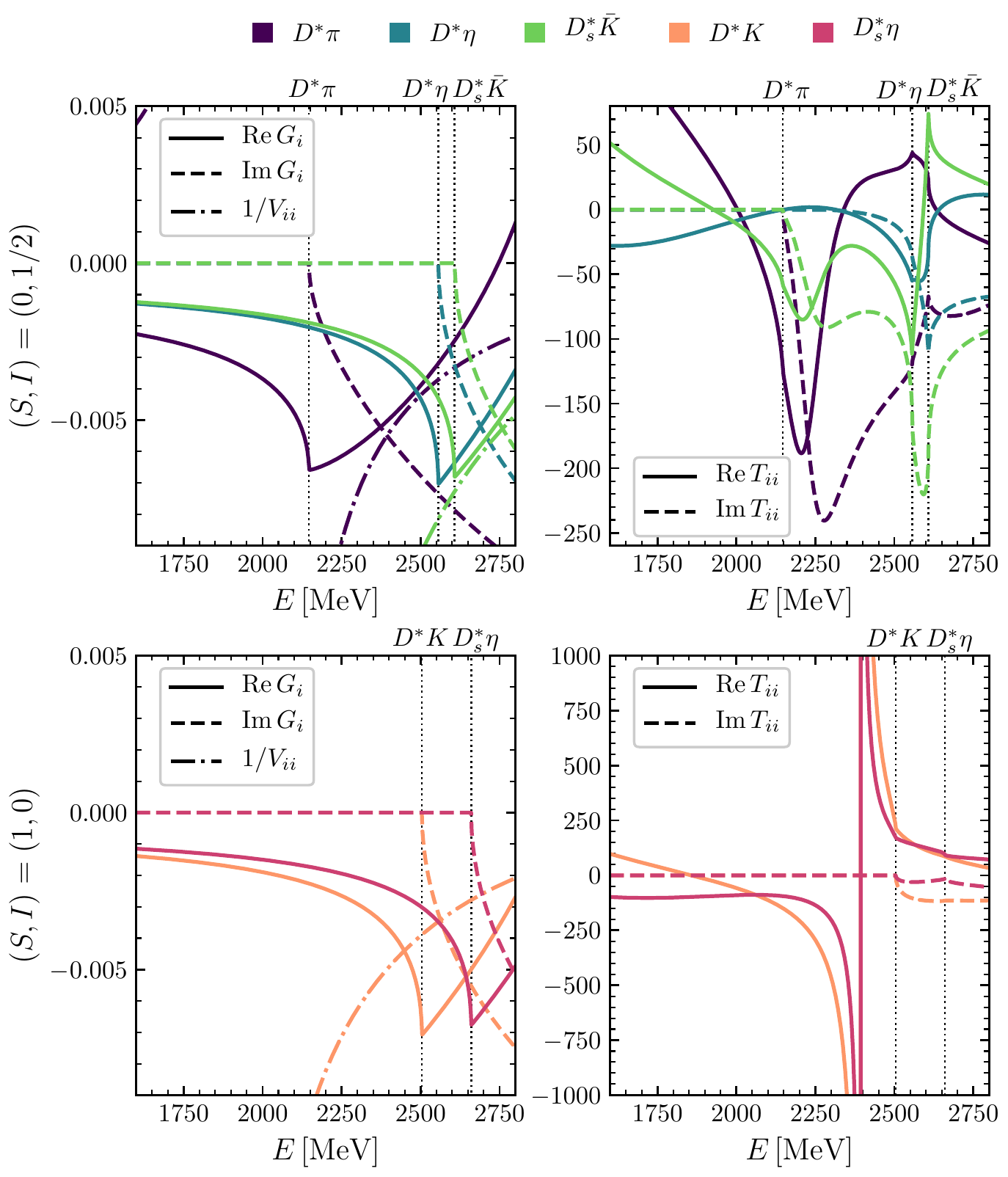}
   \caption{The same as in Fig.~\ref{fig:free-mm-VGT_temp0} for the $J^P=1^+$ states. The top panels correspond to the sector with $(S,I)=(0,1/2)$, where the subindices $1$, $2$, $3$ refer to the channels $D^*\pi$, $D^*\eta$, and $D_s^*\bar{K}$, respectively, and the bottom panels to the sector with $(S,I)=(1,0)$, where $1$ and $2$ refer to $D^*K$ and $D_s^*\eta$. }
   \label{fig:free-mm-VGT_temp0_vec}
\end{figure}

The coupled-channel interaction of the pseudoscalar meson octet with the heavy vector mesons gives rise to a very similar phenomenology to that found for the interaction with the heavy pseudoscalars, only displaced towards higher energies by the increase of the thresholds due to the mass difference between the vector and pseudoscalar heavy mesons. This is seen in Fig.~\ref{fig:free-mm-VGT_temp0_vec}, where the loop functions, the inverse of the diagonal potentials, and the diagonal amplitudes of the $J^P=1^+$ interaction in the $(S,I)=(0,1/2)$ and $(1,0)$ sectors are shown, respectively, as functions of the total energy for a total momentum $\vec{P}=0$.

\begin{figure}[htbp!]\centering 
\begin{subfigure}[b]{0.49\textwidth}\centering 
\captionsetup{skip=0pt}
 \includegraphics[width=\textwidth]{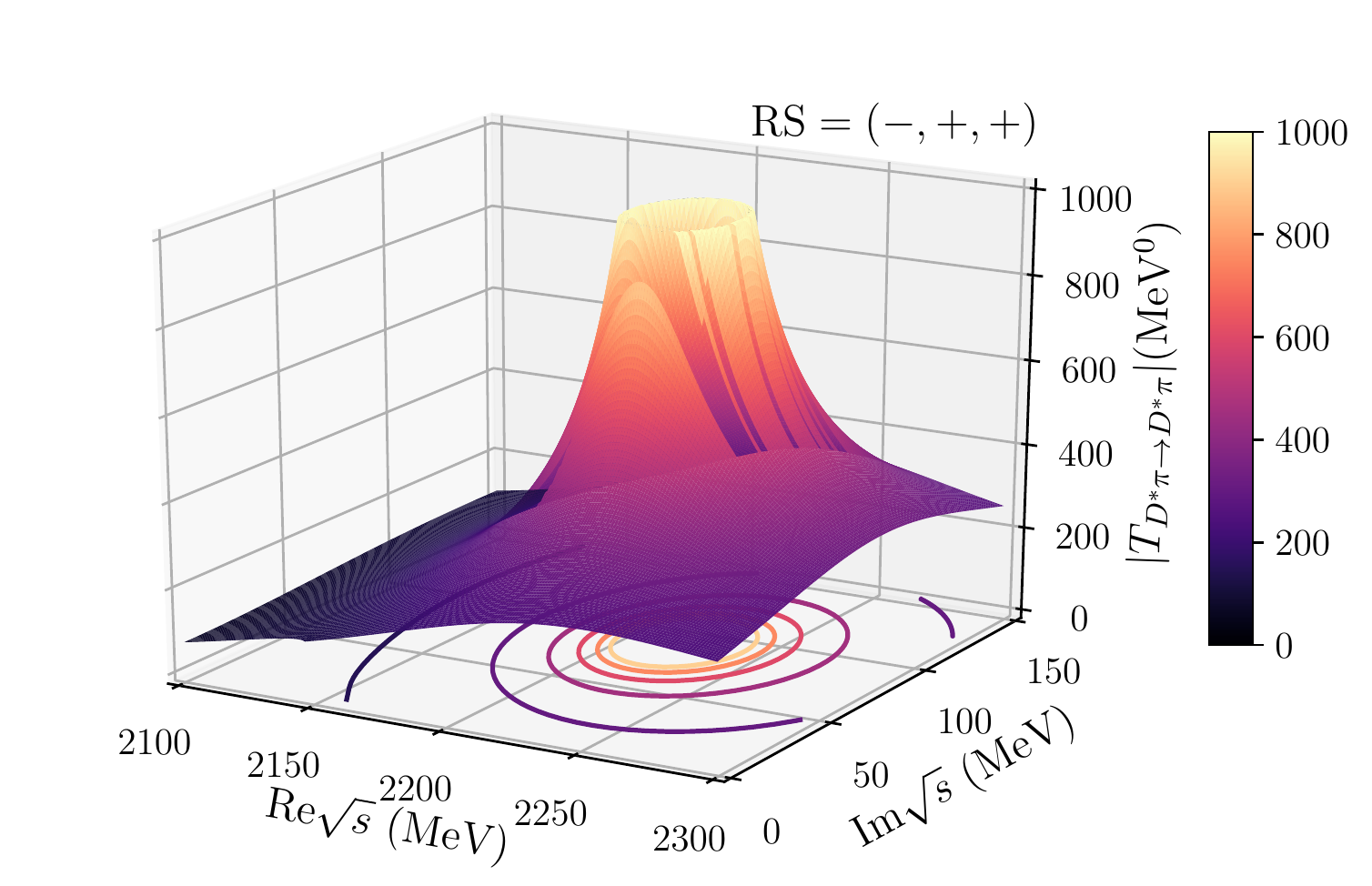}
\caption{}
\label{fig:free-mm-RS2430-a}
\end{subfigure}
\begin{subfigure}[b]{0.49\textwidth}\centering 
\captionsetup{skip=0pt}
 \includegraphics[width=\textwidth]{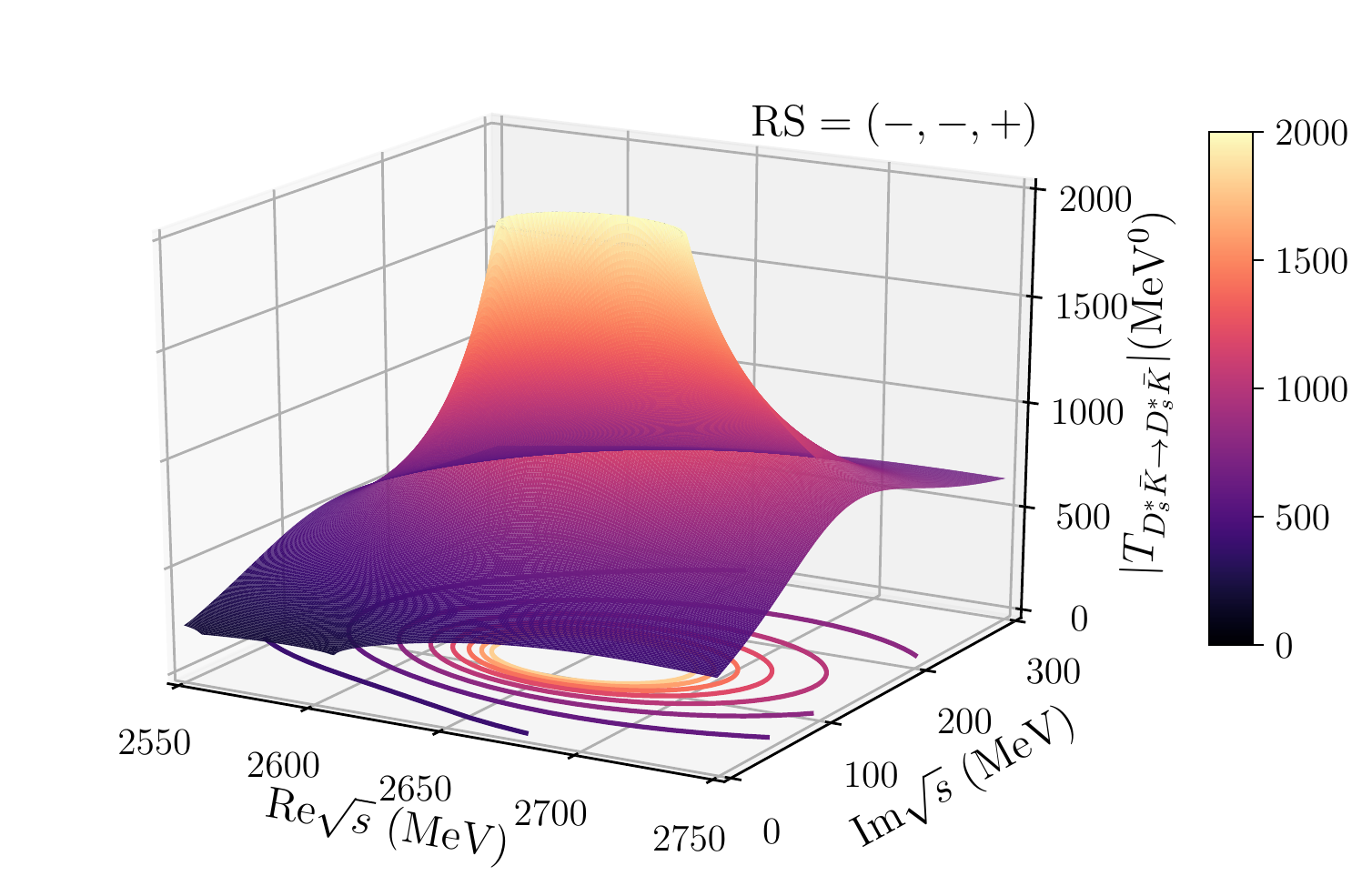}
\caption{}
\label{fig:free-mm-RS2430-b}
\end{subfigure}
\caption{Riemann surfaces of the $T$ matrix around the complex energies of the two poles of the $D_1(2430)$. (a) Diagonal $D^*\pi$ element in the $\textrm{RS}=(-,+,+)$. (b) Diagonal $D_s^*\bar{K}$ element in the $\textrm{RS}=(-,-,+)$. }
\label{fig:free-mm-RS2430}
\end{figure}

\begin{figure}[htbp!]\centering 
 \includegraphics[width=0.49\textwidth]{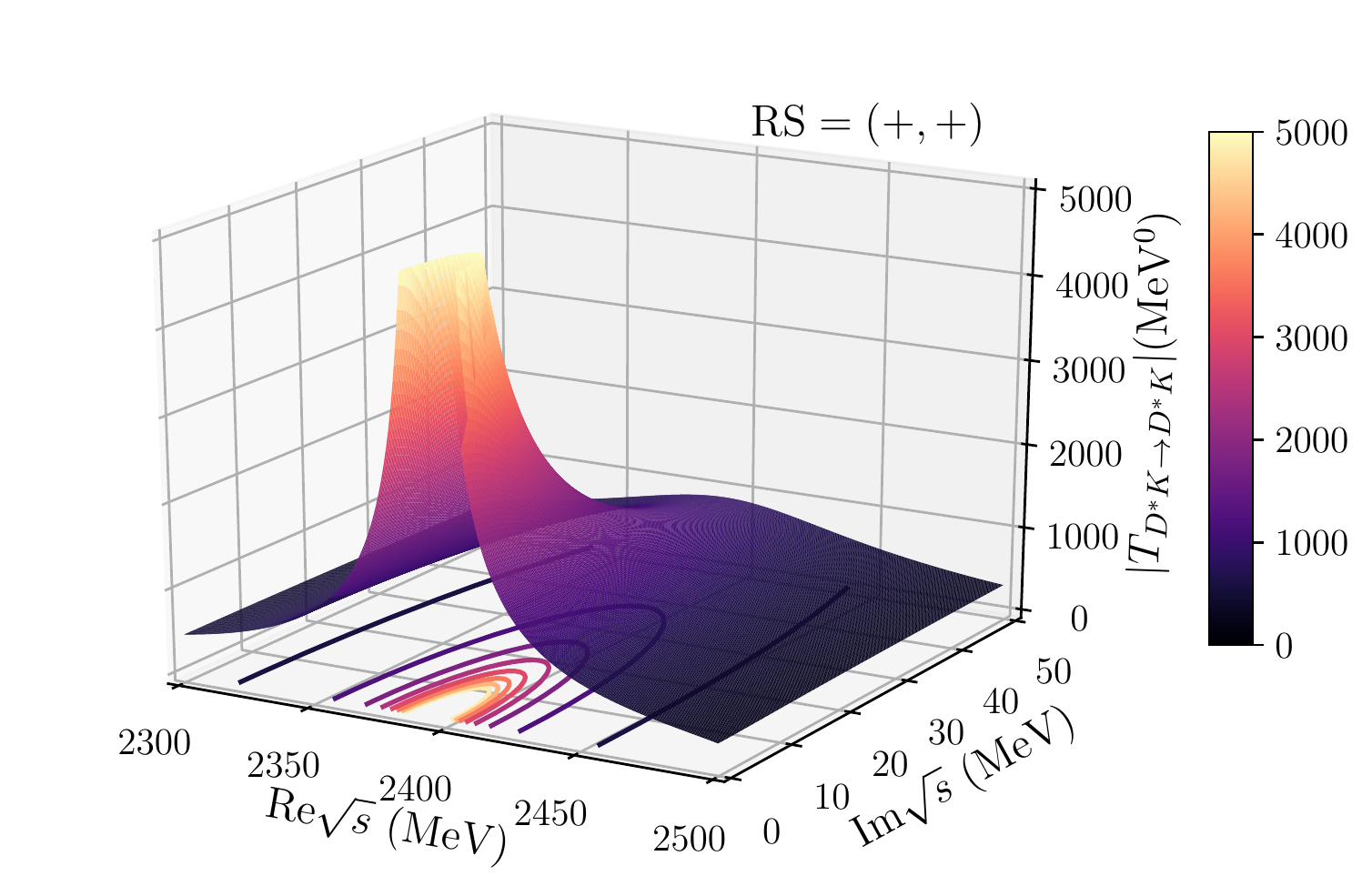}
\caption{Riemann surface of the diagonal $D^*K$ element of the $T$ matrix in the $\textrm{RS}=(+,+)$ around the complex energy of the $D_{s1}(2460)$ pole. }
\label{fig:free-mm-RS2460}
\end{figure}

The poles that we find in the $J^P=1^+$ amplitudes are summarized in Table~\ref{tab:free-mm-poles1}. As seen, a double-pole structure, which can be identified with the $D_1(2430)$ resonance listed in the \gls{rpp}~\cite{pdg}, is obtained in the $(S,I)=(0,1/2)$ sector. The pole at $2222-\ii 85$~MeV qualifies as a mostly ${D^*\pi}$ state. The $(-,+,+)$ \gls{rs} of the $|T_{D^*\pi\rightarrow D^*\pi}|$ amplitude around the complex-energy location of this pole is plotted in Fig.~\ref{fig:free-mm-RS2430-a}. The pole at $2655-\ii 117$~MeV is mostly a ${D_s^*\bar{K}}$ molecule and it is located above this channel threshold in the $(-,-,+)$ \gls{rs}, as it is shown in Fig.~\ref{fig:free-mm-RS2430-b}. A bound state coupling mostly to ${D^*K}$ states is obtained in the $(S,I)=(1,0)$ sector at $2393-\ii 0$~MeV, although the present model locates it at a somewhat lower energy than the mass of the $D_{s1}(2460)$ reported by the \gls{pdg}~\cite{pdg} due to the choice of the cut-off, as discussed for the $0^+$ $D_{s0}(2317)$. Figure~\ref{fig:free-mm-RS2460} shows the $(+,+)$ \gls{rs} of the $|T_{D^*K\rightarrow D^*K}|$ amplitude around the location of the pole.

\begin{table}[htbp!]\centering
 \setlength{\tabcolsep}{9pt}
\renewcommand{\arraystretch}{1.2}
\begin{tabular}{cccccr@{\,=\,}lr@{\,=\,}l}
\hline
& $(S,I)$ & RS & $M_R$ & $\Gamma_R/2$ & \multicolumn{2}{c}{$|g_i|$} & \multicolumn{2}{c}{$\chi_i$} \\
& & & (MeV) & (MeV) & \multicolumn{2}{c}{(GeV)} & \multicolumn{2}{c}{}  \\
\hline
$D_1(2430)$ & $(0,\frac12)$ & $(-,+,+)$ & $2222.3$ & $84.7$  & $|g_{D^*\pi}|$&$9.5$  & $\chi_{D^*\pi}$&$0.45$ \\
 &    &           &        &       & $|g_{D^*\eta}|$&$0.4$  & $\chi_{D^*\eta}$&$0.00$ \\
 &   &           &        &       & $|g_{D_s^*\bar{K}}|$&$5.7$  & $\chi_{D_s^*\bar{K}}$&$0.02$ \\
 &  & $(-,-,+)$ & $2654.6$ & $117.3$ & $|g_{D^*\pi}|$&$6.5$  & $\chi_{D^*\pi}$&$0.17$ \\
 &   &           &        &       & $|g_{D^*\eta}|$&$10.0$  & $\chi_{D^*\eta}$&$0.54$ \\
 &   &           &        &       & $|g_{D_s^*\bar{K}}|$&$18.5$ & $\chi_{D_s^*\bar{K}}$&$0.90$ \\
    \hline
$D_{s1}(2460)$ &   $(1,0)$ & $(+,+)$ & $2393.3$ & $0.0$ & $|g_{D^*K}|$&$14.2$ & $\chi_{D^*K}$&$0.45$ \\
 &    &           &        &       & $|g_{D_s^*\eta}|$&$9.7$ & $\chi_{D_s^*\eta}$&$0.08$ \\
\hline
 \end{tabular}
 \centering
 \caption{Dynamically generated poles in the $J^P=1^+$ sectors with $(S,I)=(0,1/2)$ and $(S,I)=(1,0)$. The structure of the table is the same as that of Table~\ref{tab:free-mm-poles0}.}
 \label{tab:free-mm-poles1}
 \end{table}

We note that the cut-off dependence analyses performed in Section~\ref{subsec:free-mb-omegac} when studying the $\Omega_c^{*0}$ states dynamically generated from the interaction between mesons and baryons (see also Refs.~\cite{Montana:2017kjw,Ramos:2020bgs}) indicate that employing higher (lower) values of the cut-off lowers (increases) the energies of the dynamically generated states, due to the larger (smaller) amount of phase-space included in the unitarized amplitudes. In the case that we are considering here, when the cut-off value is varied between $600$ and $1000$ MeV, the mass of the resonances in the $(S, I) = (0, 1/2)$ sectors, that is, the mass of the $D^*_0(2300)$ and $D_1(2430)$, changes moderately by $^{+5}_{-15}$ MeV. A larger change, of $\pm 70$ MeV, is observed for both bound states in the $(S, I) = (1, 0)$ sector where the $D^*_{s0}(2317)$ and $D_{s1}(2460)$ are generated, indicating that these latter states are rather sensitive to the strength of the interaction. A similar modification of the dynamically generated states is attained by taking different values for the channel-dependent subtraction constants in the \gls{dr} scheme for the regularization of the two-meson propagator (see for example Refs.~\cite{Guo:2018tjx,Albaladejo:2016lbb}).
In relation to this, it is worth commenting on the discrepancy of $\sim 70$~MeV between the position of the poles of the dynamically generated states that we have identified with the $D^*_{s0}(2317)$ and $D_{s1}(2460)$ and the experimental value for their mass. Indeed, taking a smaller cut-off of $\sim 600$~MeV for the coupled channels in the $(S,I)=(1,0)$ sector, which is still physically sized, allows us to dynamically generate these states in agreement with the values reported by the \gls{pdg} (see Eq.~(\ref{eq:free-mm-proppdg})), as well as with previous works \cite{Lutz:2007sk,Liu:2012zya,Albaladejo:2016lbb,Guo:2018tjx}. Although the binding energy that we get for these states is somewhat larger, we consider that it is desirable to fix a simple regularization scheme, with a channel-independent cut-off of the order of the $\rho$-meson mass ($\sim 800$~MeV), as our aim is to analyze in Chapter~\ref{ch:hot-medium} the modification of the heavy-flavor mesons (ground states and dynamically generated states) from their interactions with the light mesons in a hot medium, where the exact location of the $D^{(*)}K$ bound-state may play a minor role.

\subsection{Results for $\bar{B}$ mesons}
\label{subsec:free-mm-Bmesons}
Finally, we discuss the results of the unitarized effective theory for the interaction of $J=0$ $\bar{B}$ and $\bar{B}_s$ mesons and $J=1$ $\bar{B}^*$ and $\bar{B}^*_s$ mesons with light mesons. Motivated by the success of the model in describing the narrow $D_{s0}^*(2317)$ and $D_{s1}(2460)$ states as $DK$ and $D^*K$ bound states, as well as the double-pole structure of the $D_0^*(2300)$ and the $D_1(2430)$ in the open-charm sector, in this section we give quantitative predictions for the open-beauty partners that are foreseen due to \gls{hqfs}.
While there are no experimental states reported with $J^P=0^+$ in the bottomed sector, there are two states that have been observed with $J^P=1^+$, the $B_1(5721)$ and $B_{s1}(5830)$ (see the experimental properties in Eq.~(\ref{eq:free-mm-proppdg})), and which could presumably be the bottom-flavor partners of the $D_1(2430)$ and $D_{s1}(2460)$.

\subsubsection{$J=0$ case: Interactions and predicted states}
\label{subsubsec:Bmesons}

\begin{figure}[b!]
    \centering
   \includegraphics[width=0.7\textwidth]{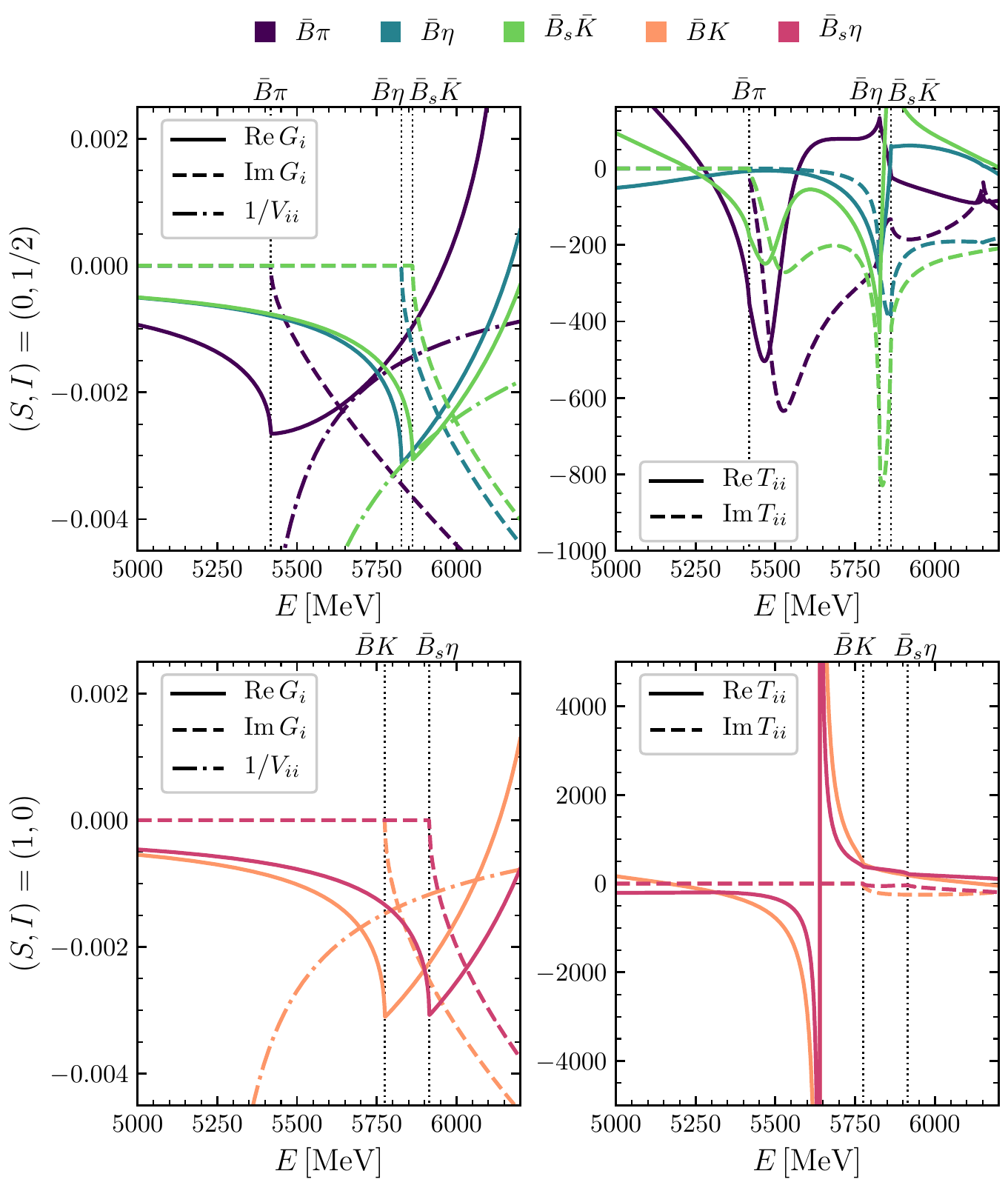}
   \caption{The same as in Fig.~\ref{fig:free-mm-VGT_temp0} for the bottomed $J^P=0^+$ states. The top panels correspond to the sector with $(S,I)=(0,1/2)$, where the subindices $1$, $2$, $3$ refer to the channels $\bar{B}\pi$, $\bar{B}\eta$, and $\bar{B}_s\bar{K}$, respectively, and the bottom panels to the sector with $(S,I)=(1,0)$, where $1$ and $2$ refer to $\bar{B}K$ and $\bar{B}_s\eta$. }
   \label{fig:free-mm-VGT_temp0_B}
\end{figure}

We are interested in the sectors with $S=0$, $I=1/2$ and $S=1$, $I=0$. The loop functions and the inverse of the diagonal elements of the potential for the $0^+$ channels in these sectors are shown in the left panels of Fig.~\ref{fig:free-mm-VGT_temp0_B}, and the corresponding unitarized amplitudes of the coupled-channel scattering are plotted in the right panels of the same figure. We see a similar pattern to that found for $D\Phi$ states, with a shift of the structures to the higher energy region where the thresholds of the $\bar{B}\Phi$ channels are located.

In Table~\ref{tab:free-mm-polesb0} we display the properties of the poles of the unitarized amplitudes in the complex-energy plane. In the $(0,1/2)$ sector there is also a two-pole structure, with a lower resonant pole at $5483-\ii 72$~MeV coupling largely to $\bar{B}\pi$ and a higher resonance at $5848-\ii 66$~MeV mostly coupling to the $\bar{B}_s\bar{K}$ channel. Regarding this latter pole, it is interesting to note that it is located in the $(-,-,+)$ \gls{rs} as its counterpart in the open-charm sector. However, it lies between the thresholds of the $\bar{B}\eta$ and the $\bar{B}_s\bar{K}$ channels, at $5827$~MeV and $5863$~MeV, respectively, and therefore it is located in the unphysical \gls{rs} closest to the physical region, in what we refer to as the \gls{rs}-II. 
The extrapolation of the definition of the compositeness in Eq.~(\ref{eq:free-th-compositeness}) to complex energies for resonant states prevents its strict interpretation as a probability. Still, this quantity provides the weight of each of the coupled channels in the wave function of the states~\cite{Aceti:2014ala}. Thus the lowest pole contains a sizable amount of $\bar{B}\pi$ channel, while the highest one has a large $\bar{B}_s\bar{K}$ molecular component.

\begin{table}[t!]\centering
 \setlength{\tabcolsep}{9pt}
\renewcommand{\arraystretch}{1.2}
\begin{tabular}{ccccr@{\,=\,}lr@{\,=\,}l}
\hline
$(S,I)$ & RS & $M_R$ & $\Gamma_R/2$ & \multicolumn{2}{c}{$|g_i|$} & \multicolumn{2}{c}{$\chi_i$} \\
 & & (MeV) & (MeV) & \multicolumn{2}{c}{(GeV)} & \multicolumn{2}{c}{}  \\
\hline
 $(0,\frac12)$ & $(-,+,+)$ & $5483.1$ & $71.8$  & $|g_{\bar{B}\pi}|$&$22.4$  & $\chi_{\bar{B}\pi}$&$0.44$ \\
     &           &        &       & $|g_{\bar{B}\eta}|$&$0.8$  & $\chi_{\bar{B}\eta}$&$0.00$ \\
    &           &        &       & $|g_{\bar{B}_s\bar{K}}|$&$14.4$  & $\chi_{\bar{B}_s\bar{K}}$&$0.02$ \\
   & $(-,-,+)$ & $5848.0$ & $65.9$ & $|g_{\bar{B}\pi}|$&$11.0$  & $\chi_{\bar{B}\pi}$&$0.11$ \\
    &           &        &       & $|g_{\bar{B}\eta}|$&$18.0$  & $\chi_{\bar{B}\eta}$&$0.49$ \\
    &           &        &       & $|g_{\bar{B}_s\bar{K}}|$&$32.0$ & $\chi_{\bar{B}_s\bar{K}}$&$0.76$ \\
    \hline
$(1,0)$ & $(+,+)$ & $5639.3$ & $0.0$ & $|g_{\bar{B}K}|$&$35.6$ & $\chi_{\bar{B}K}$&$0.68$ \\
    &           &        &       & $|g_{\bar{B}_s\eta}|$&$23.8$ & $\chi_{\bar{B}_s\eta}$&$0.17$ \\
\hline
\end{tabular}
 \centering
 \caption{Dynamically generated poles in the $J^P=0^+$ bottomed sectors with $(S,I)=(0,1/2)$ and $(S,I)=(1,0)$. The structure of the table is similar to that of Table~\ref{tab:free-mm-poles0}. In this case there are no experimental states reported.}
 \label{tab:free-mm-polesb0}
 \end{table}
 
For $(S,I)=(1,0)$, we find a $\bar{B}K$ bound state at $5639$~MeV, with a binding energy of about $130$~MeV, larger than what is found in other works \cite{Guo:2006fu,Albaladejo:2016lbb}. This is mainly due to the prescription that we have taken for the regularization of the loops, as we have explained in the case of the $D_{s0}^*(2317)$ above.

Despite the lack of experimentally observed states with the quantum numbers and the properties of those dynamically generated by the heavy chiral unitary approach in the $0^+$ open-bottom sector, the molecular model predicts two broad resonant poles in the nonstrange sector and a narrow bound $\bar{B}K$ state in the strange sector. These are the bottomed $B_0^*$ and $B_{s0}^*$ partners of the $D_0^*(2300)$ and the $D_{s0}^*(2317)$ and are expected to exist from \gls{hqfs}. 

\subsubsection{$J=1$ case: Interactions and $B_1(5721)$ and $B_{s1}(5830)$ }
\label{subsubsec:B*mesons}

The dynamically generated $B$ states with $J^P=1^+$ follow again the same pattern of a nonstrange two-pole structure and a strange bound state that has been described above for the $0^+$ and $1^+$ charmed states, as well as for the $0^+$ bottomed states. This is a clear confirmation of the small breaking of \gls{hqsfs} of the model.

\begin{figure}[t]
    \centering
   \includegraphics[width=0.7\textwidth]{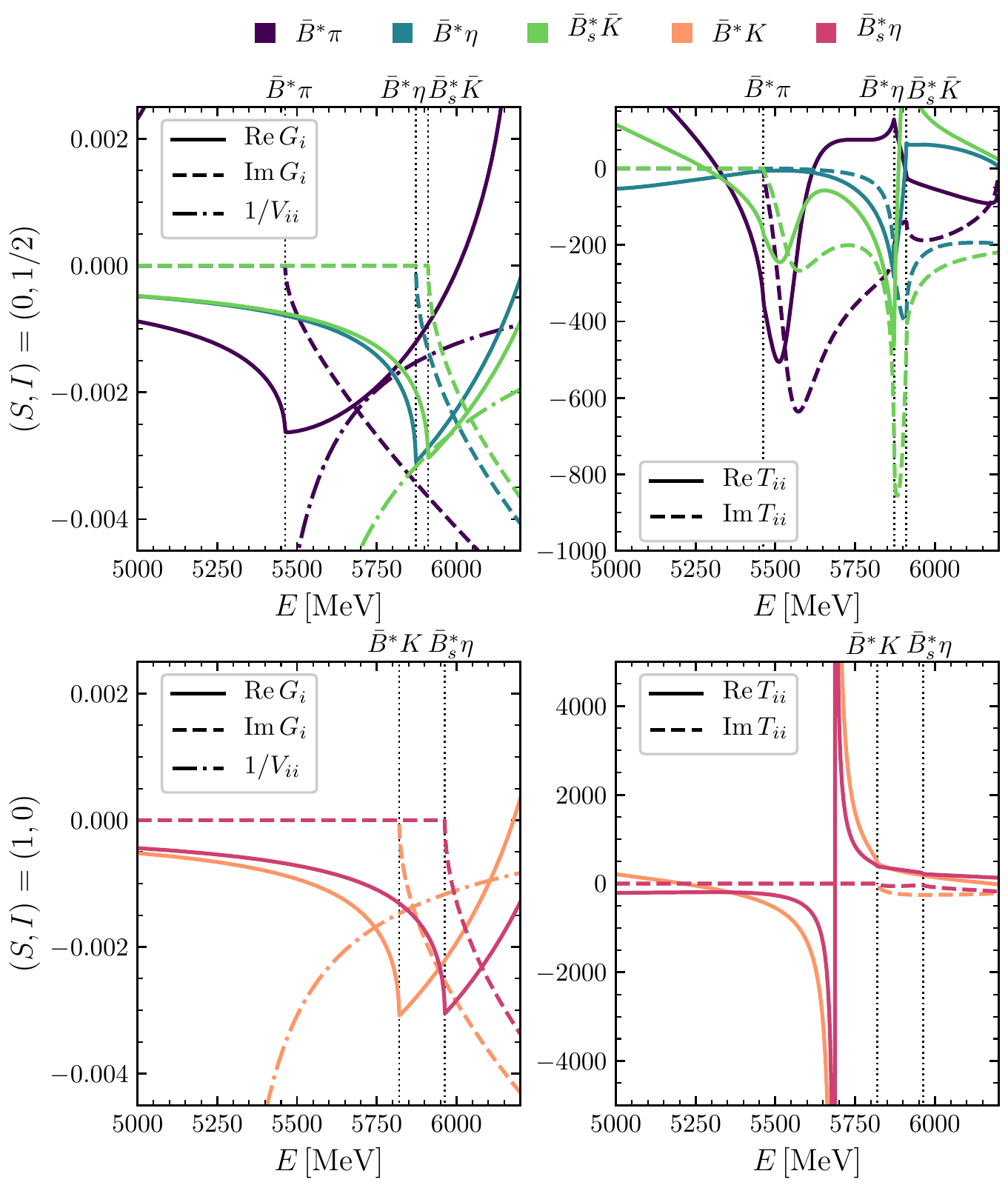}
   \caption{The same as in Fig.~\ref{fig:free-mm-VGT_temp0} for the bottomed $J^P=1^+$ states. The top panels correspond to the sector with $(S,I)=(0,1/2)$, where the subindices $1$, $2$, $3$ refer to the channels $\bar{B}^*\pi$, $\bar{B}^*\eta$, and $\bar{B}^*_s\bar{K}$, respectively, and the bottom panels to the sector with $(S,I)=(1,0)$, where $1$ and $2$ refer to $\bar{B}^*K$ and $\bar{B}^*_s\eta$. }
   \label{fig:free-mm-VGT_temp0_Bvec}
\end{figure}
 \begin{table}[h!]\centering
 \setlength{\tabcolsep}{9pt}
\renewcommand{\arraystretch}{1.2}
\begin{tabular}{cccccr@{\,=\,}lr@{\,=\,}l}
\hline
& $(S,I)$ & RS & $M_R$ & $\Gamma_R/2$ & \multicolumn{2}{c}{$|g_i|$} & \multicolumn{2}{c}{$\chi_i$} \\
& & & (MeV) & (MeV) & \multicolumn{2}{c}{(GeV)} & \multicolumn{2}{c}{}  \\
\hline
$B_1(5721)$ & $(0,\frac12)$ & $(-,+,+)$ & $5528.6$ & $72.3$  & $|g_{\bar{B}^*\pi}|$&$22.6$  & $\chi_{\bar{B}^*\pi}$&$0.44$ \\
 &    &           &        &       & $|g_{\bar{B}^*\eta}|$&$0.8$  & $\chi_{\bar{B}^*\eta}$&$0.00$ \\
 &   &           &        &       & $|g_{\bar{B}^*_s\bar{K}}|$&$14.4$  & $\chi_{\bar{B}^*_s\bar{K}}$&$0.02$ \\
 &  & $(-,-,+)$ & $5893.3$ & $64.9$ & $|g_{\bar{B}^*\pi}|$&$10.7$  & $\chi_{\bar{B}^*\pi}$&$0.10$ \\
 &   &           &        &       & $|g_{\bar{B}^*\eta}|$&$18.0$  & $\chi_{\bar{B}^*\eta}$&$0.49$ \\
 &   &           &        &       & $|g_{\bar{B}^*_s\bar{K}}|$&$32.1$ & $\chi_{\bar{B}^*_s\bar{K}}$&$0.74$ \\
    \hline
$B_{s1}(5830)$ &   $(1,0)$ & $(+,+)$ & $5686.0$ & $0.0$ & $|g_{\bar{B}^*K}|$&$35.8$ & $\chi_{\bar{B}^*K}$&$0.46$ \\
 &    &           &        &       & $|g_{\bar{B}^*_s\eta}|$&$23.9$ & $\chi_{\bar{B}^*_s\eta}$&$0.09$ \\
\hline
 \end{tabular}
 \centering
 \caption{Dynamically generated poles in the $J^P=1^+$ bottomed sectors with $(S,I)=(0,1/2)$ and $(S,I)=(1,0)$. The structure of the table is the same as that of Table~\ref{tab:free-mm-poles0}.}
 \label{tab:free-mm-polesb1}
 \end{table}
 
In Fig.~\ref{fig:free-mm-VGT_temp0_Bvec} we show the loop functions, the inverse of the diagonal interaction kernels, and the diagonal elements of the unitarized scattering amplitudes in the $(S,I)=(0,1/2)$ sector (upper panels) and the $(S,I)=(1,0)$ sector (lower panels). The properties of the corresponding poles are listed in Table~\ref{tab:free-mm-polesb1}. In the $(0,1/2)$ sector, the $T$ matrix presents a lower-energy pole at $5529-\ii 72$~MeV, coupling largely to $\bar{B}^*\pi$, and a higher-energy pole at $5893.3-\ii 65$~MeV, with a large coupling to $\bar{B}_s\bar{K}$. These two poles lie in the energy region where the experimental $B_1(5721)$ state is observed, although the identification as the bottomed partner of the $D_0^*(2300)$ with two poles dynamically generated by the molecular model is not clear, as the $B_1(5721)$ is substantially narrower than predicted by the unitarized heavy chiral approach. The experimental value of the width of this state is $\sim 30$~MeV, which is similar to the width of the structure observed in the $\bar{B}_s^*\bar{K}$ amplitude in the upper-right panel of Fig.~\ref{fig:free-mm-VGT_temp0_Bvec} due to the presence of the $\bar{B}^*\eta$ and $\bar{B}_s^*\bar{K}$ thresholds which, however, lie at energies about $150$~MeV higher than the experimental mass of the $B_1(5721)$ state.

The $\bar{B}^*K$ bound state generated in the $(1,0)$ sector is located at $5686$~MeV, and thus our model predicts a substantially larger binding than that of the experimental $B_{s1}(5830)$. However, as explained above for the charm sector, a modification of the parameters of the regularization scheme might permit the model to generate the state at energies closer to the experimental one, although the value of the cut-off may become unphysically small ($<300$~MeV). Furthermore, in this bottomed case, the experimental $B_{s1}(5830)$ state lies above the $\bar{B}^*K$ threshold, located at $5820$~MeV, which poses an additional difficulty. By decreasing the value of the cut-off, the pole of the bound state generated from the $\bar{B}^*K$ interaction moves to a \gls{rs} different from the \gls{rs}-II, as it crosses the channel threshold. Therefore, the identification of the $B_{s1}(5830)$ with the bottomed partner of the $D_{s1}(2460)$, for which with a unitarized heavy-light molecular model we predict a mass in the range $[5600-5800]$~MeV for a cut-off in the range $[500-1000]$~MeV, is not clear.

Furthermore, it is interesting to notice that the breaking of \gls{hqss} is smaller in the open-bottom sector compared to the open-charm sector, as can be seen by comparing the properties of the $0^+$ and the $1^+$ dynamically generated states in Tables~\ref{tab:free-mm-poles0} and \ref{tab:free-mm-poles1} with those in Tables~\ref{tab:free-mm-polesb0} and \ref{tab:free-mm-polesb1}, respectively. Indeed, the $\mathcal{O}(1/m_H)$ corrections are expected to be smaller for $\bar{B}$ mesons due to their heavier mass in comparison to the mass of the $D$ mesons. For the bottomed states, the difference between the $0^+$ and the $1^+$ is observed in the masses found for these states, since their widths and couplings to the various channels are very similar. The mass shift is related to the mass difference between the pseudoscalar $\bar{B}$ mesons and the vector $\bar{B}^*$ mesons. 
%
\chapter{Heavy mesons in a hot medium}
\label{ch:hot-medium}
In this chapter, we present a detailed and comprehensive discussion on a few basic thermal field theory concepts, essential to incorporate finite temperature effects in the unitarized heavy effective theory that we have employed in the second part of Chapter~\ref{ch:hot-medium} to describe the scattering of heavy mesons off Goldstone bosons in the vacuum. This is necessary because it is known that \gls{hic} experiments, for instance in the \gls{rhic} at \gls{bnl} and the \gls{lhc} at \gls{cern}, produce a hot plasma, the so-called \gls{qgp} that cools down into a large number of light mesons, mostly pions, and other particles that behave thermally, while interacting obeying the laws of \gls{qcd}. 

The chapter is structured as follows. After the introduction given in Section~\ref{sec:hot-intro}, we describe in Section~\ref{sec:hot-formalism} the theoretical framework used to obtain the unitarized amplitudes and the spectral properties of the heavy mesons at finite temperature, including the calculation of the thermal two-meson propagator and the heavy-meson self-energy, while the results for the thermal modification of open-heavy flavor mesons ($D$ and $\bar{B}$ mesons) when immersed in a hot medium of light Goldstone bosons are analyzed in Section~\ref{sec:hot-results}.
The work presented in this chapter lead to the publication of Refs.~\cite{Montana:2020lfi,Montana:2020vjg}.  


\section{Introduction}
\label{sec:hot-intro}

The in-medium properties of mesons with charm content have been a matter of high interest over the years (see \cite{Rapp:2011zz,Tolos:2013gta,Hosaka:2016ypm,Aarts:2016hap} for reviews). This interest was triggered because of the phenomenon of $J/\Psi$ suppression in heavy-ion collisions \cite{Gonin:1996wn}, which was predicted in Ref.~\cite{Matsui:1986dk} as a signature of the existence of the \gls{qgp} due to color screening. The $J/\Psi$ absorption in hot dense matter could be also modified due to the change of the properties of open-charm mesons in matter in the comover scattering scenario (see, for example, the initial works of Refs.~\cite{Capella:2000zp,Cassing:1999es,Vogt:1999cu,Gerschel:1998zi}), thus providing a complementary explanation for $J/\Psi$ suppression.

Several studies have been devoted to examining the properties of charmed mesons in a meson-rich environment. Most of these works, though, have been concentrated on the determination of the hidden-charm $J/\Psi$ dissociation cross sections in heavy-ion collisions (see \cite{Rapp:2008tf} for a review). There are several studies on the $J/\Psi$-hadron interaction at finite temperature based on chiral Lagrangians
\cite{Haglin:2000ar,Bourque:2008es,Blaschke:2008mu}, quark-model calculations \cite{Zhou:2012vv,Maiani:2004py,Maiani:2004qj,Bourque:2008es}, schemes using \gls{qcd} sum rules \cite{Duraes:2002px,Duraes:2002ux}, \gls{lqcd} (see \cite{Rothkopf:2019ipj} and references therein), and $\textrm{SU}(4)$ effective Lagrangians \cite{Mitra:2014ipa,Abreu:2017cof}. With regards to open-charm mesons, the studies on open-charm thermal relaxation in heavy-ion collisions \cite{Laine:2011is,He:2011yi,Ghosh:2011bw,Abreu:2011ic,Tolos:2013kva,Ozvenchuk:2014rpa,Song:2015sfa} have prompted the interest in open charm at finite temperatures. Investigations of open-charm mesons at finite temperature have been performed using \gls{qcd} sum-rule-based approaches \cite{Hilger:2011cq,Buchheim:2018kss,Gubler:2020hft} and calculations on the lattice \cite{Bazavov:2014yba,Bazavov:2014cta,Kelly:2018hsi}. Also, effective models in a hot hadronic bath have been developed in Refs.~\cite{Mishra:2003se,Fuchs:2004fh,He:2011yi,Blaschke:2011yv,Ghosh:2013xea,Sasaki:2014asa}.

Recently, a finite-temperature unitarized approach based on an $\textrm{SU}(4)$ effective Lagrangian has been put forward \cite{Cleven:2017fun}, where the implications of pionic matter at finite temperatures on the properties of open- and hidden-charm mesons have been studied. Whereas the $J/\Psi$ stays narrow even at $T= 150$ MeV,  the  $D$ and $D^*$ mesons acquire a substantial width in the pionic bath, reaching $30-40$~MeV at $T= 150$ MeV. 

In the present chapter, we address the properties of open heavy-flavor mesons in a hot mesonic bath (mainly formed by pions), within a finite-temperature unitarized approach~\cite{Montana:2020lfi,Montana:2020vjg}, following the path of Ref.~\cite{Cleven:2017fun}. The dynamics of the heavy mesons with the light mesons are based on the effective theory presented in Section~\ref{sec:free-mm}, which is built on chiral symmetry at \gls{nlo} and implements \gls{hqsfs} at \gls{lo}. 
We discuss not only the thermal modification of the heavy ground states but also that of the dynamically generated states, paying special attention to the fact the latter ones are the positive parity chiral partners of the former.

The idea that chiral partners become degenerate above the chiral restoration temperature $T_\chi$~\cite{Hatsuda:1985eb,Rapp:1999ej} has motivated a large number of works in which low-lying hadronic states of opposite parities have been studied in a thermal medium and their masses have been seen to merge at large temperatures $T>T_\chi$.

The canonical example resides in the light-meson sector, where the pseudoscalar isotriplet ($\pi$) and the scalar isoscalar ($\sigma$ meson) acquire similar masses above $T_\chi$. This system has been studied in the linear sigma model~\cite{Bochkarev:1995gi}, the (P)NJL model~\cite{Klevansky:1992qe,Florkowski:1993br,Hansen:2006ee}, the quark-meson model~\cite{Tripolt:2013jra}, and others. On the other hand, vector and axial-vector interactions, which have been studied in the (P)NJL model~\cite{Sintes:2014lka} and the gauge linear-sigma model~\cite{Pisarski:1995xu}, for example, allow one to study the chiral-symmetry restoration of the $\rho$ and the $a_1$ states~\cite{Rapp:1999ej}. Opposite-parity diquarks also present such degeneracy in the (P)NJL model~\cite{Torres-Rincon:2015rma}, whereas there exist also indications from \gls{lqcd} calculations of the chiral restoration of opposite-parity baryons~\cite{Aarts:2017rrl,Aarts:2018glk}.

In many of the theoretical models, the parity partners are fundamental degrees of freedom, for instance, the $\pi$ and the $\sigma$ in the linear sigma model~\cite{Bochkarev:1995gi}, and interactions in a thermal/dense medium dress them producing in-medium mass modifications. In another set of models, for example, the NJL and PNJL model, the parity partners (either $0^+/0^-$ or $1^+/1^-$) are not part of the degrees of freedom of the Lagrangian but are instead generated from the few-body dynamics, like those implemented by the \gls{bs} equation for a quark-antiquark pair. In this case, masses and decay widths seem to converge in the chirally-restored phase~\cite{Hansen:2006ee}.

All these models provide insights into the effects of chiral restoration, both below and above $T_\chi$. However, one should keep in mind that, although well-motivated by the \gls{qcd} symmetries and dynamics, they are not usually the correct \gls{eft} of \gls{qcd}. In the light-meson sector, for instance, we know that the low-energy effective theory is \gls{chpt}~\cite{Gasser:1983yg}, which can lead to model-independent results, also at finite temperatures. However, this approach is valid at low energies and temperatures, always below $T_\chi$, and only timid indications of a chiral-symmetry restoration can be expected from it. 

Even if limited to $T<T_\chi$, the chiral approach is quite interesting because a combined picture of the chiral partners comes into play. The negative parity partner ($\pi$) is a degree of freedom of the Lagrangian~\cite{Gasser:1983yg}, whose vacuum mass is dressed by interactions with the whole set of Goldstone bosons. However, the positive parity partner ($\sigma$) is not part of the Lagrangian. In unitarized versions of \gls{chpt}~\cite{Dobado:1989qm,Dobado:1996ps}, it can be associated with the $J^\pi=0^+$ resonant state, appearing in the scalar isoscalar channel of the meson-meson scattering amplitude. This state, experimentally identified with the scalar $f_0(500)$ of the \gls{pdg} compilation~\cite{Tanabashi:2018oca}, can be generated at finite temperature as well~\cite{Dobado:2002xf,Rapp:1995fv}. This scenario, where one of the chiral companions is a degree of freedom of the theory and the other a dynamically-generated state, is the one we consider in this dissertation in the open heavy-flavor sector.

\section{Formalism}
\label{sec:hot-formalism}

The standard theoretical treatment of strongly interacting fields within the theory of \gls{qcd} uses the language of quantum field theory and it has been historically very successful in describing the behavior of hadrons in free space, or at conditions of zero temperature and zero baryon density. However, this theoretical framework does not naturally incorporate the effects of the medium that are relevant when one wants to study \gls{qcd} matter at high temperatures and/or densities.

\glsunset{itf}
In the last decades, there has been an increasing interest in understanding particle interactions in hot and dense systems and an enormous work has been done towards developing a formalism to treat in-medium phenomena, both in and out of equilibrium. The pioneering work in thermal quantum field theory dates back to Matsubara~\cite{Matsubara:1955ws} who developed the \textit{imaginary-time}, or \textit{Matsubara}, \textit{formalism} (\gls{itf}) to describe systems in equilibrium. This approach has many similarities with what we know at zero temperature, in particular the form of the propagators and the diagrammatic structure of the perturbative expansion, but it differs in the way in which time is treated, as it is considered a purely imaginary quantity. Soon after the foundations of the Matsubara formalism, Kubo~\cite{Kubo:1957mj} and Martin and Schwinger~\cite{Martin:1959jp} provided an important relation between thermal propagators in equilibrium, the so-called Kubo-Martin-Schwinger condition. Further relevant developments of equilibrium thermal theories were carried out by Keldysh~\cite{Keldysh:1964ud}, who considered explicitly the evolution in real time.

\begin{figure}[b!]\centering 
\begin{tikzpicture}[baseline=-2
  ]
  \draw [>=stealth,->] (-3,0) -- (4,0) coordinate (xaxis);
  \draw [>=stealth,->] (0,-1.9) -- (0,1.9) coordinate (yaxis);
  \path [draw, |-, line width=0.8pt, postaction={decoration={markings,mark=at position 2cm with {\arrow[line width=1pt,>=stealth]{>}}},decorate}, color=ctcolorblue] (-2.2,0.2) -- (3,0.2)  arc (90:-90:0.2) ;
  \path [draw, line width=0.8pt, postaction={decoration={markings,mark=at position 2cm with {\arrow[line width=1pt,>=stealth]{>}}},decorate}, color=ctcolorblue] (3,-0.2) -- (-2.,-0.2)  arc (90:180:0.2) ;
  \path [draw, -|,  line width=0.8pt, postaction={decoration={markings,mark=at position 0.5cm with {\arrow[line width=1pt,>=stealth]{>}}}, decorate}, color=ctcolorblue] (-2.2,-0.4) -- (-2.2,-1.3) ;
  \node [below] at (xaxis) {Re $t$};
  \node [left] at (yaxis) {Im $t$};
  \node at (-2.5,0.5) {$t_i$};
  \node at (3,0.5) {$t_f$};
  \node at (1.5,0.5) {$C_1$};
  \node at (1.5,-0.6) {$C_2$};
  \node at (-2.3,-1.6) {$t_i-\ii\beta$};
  \node at (-2.6,-0.8) {$C_\beta$};
\end{tikzpicture}
\begin{tikzpicture}[baseline=-2
  ]
  \draw [>=stealth,->] (-3,0) -- (3,0) coordinate (xaxis);
  \draw [>=stealth,->] (0,-1.9) -- (0,1.9) coordinate (yaxis);
  \path [draw,, line width=0.8pt, postaction={decoration={markings,mark=at position 2cm with {\arrow[line width=1pt,>=stealth]{>}}},decorate}, color=ctcolorblue] (-2.8,0.2) -- (2.7,0.2);
  \path [draw, line width=0.8pt, postaction={decoration={markings,mark=at position 2cm with {\arrow[line width=1pt,>=stealth]{>}}},decorate}, color=ctcolorblue] (2.7,-0.2) -- (-2.8,-0.2) ;
  \node [below] at (xaxis) {Re $t$};
  \node [left] at (yaxis) {Im $t$};
  \node at (-2.8,0.5) {$-\infty$};
  \node at (2.7,0.5) {$+\infty$};
  \node at (1.5,0.5) {$C_1$};
  \node at (1.5,-0.6) {$C_2$};
  \node at (2,1.3) {$C=C_1\cup C_2$};
\end{tikzpicture}
\caption{Left: the Kadanoff-Baym contour on the complex-time plane, composed of a forward branch ($C_1$), a backward branch ($C_2$), and the Matsubara contour ($C_\beta$). Right: the Schwinger-Keldysh contour $C=C_1\cup C_2$, where the forward and backward branches extend in the range $(-\infty,+\infty)$. Note that there is no offset in $C_1$ and $C_2$ from the real axis. The two paths lie on top of each other, infinitely close to the real axis but plotted at a certain distance for illustration.}
\label{fig:hot-contours}
\end{figure}
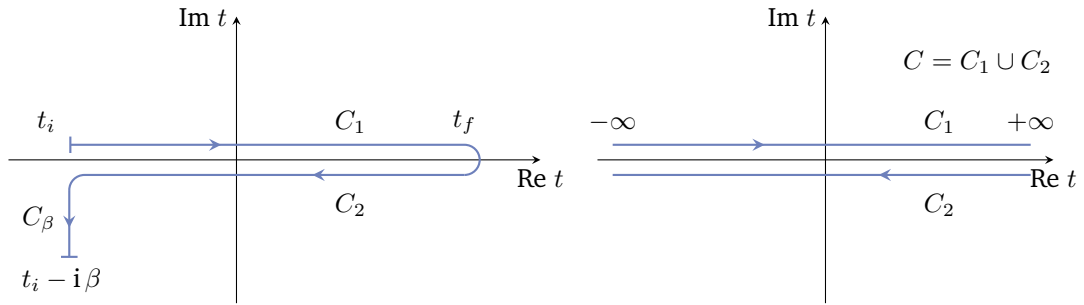

In the \textit{real-time formalism}, meson fields are defined along a contour on the complex-time plane~\cite{kadanoff1962quantum,Blaizot:1999xk,Rammer,Cassing:2008nn}. The so-called Kadanoff-Baym contour, shown in the left panel of Fig.~\ref{fig:hot-contours}, starts at an initial time $t_i$, runs along the real axis up to a larger final time $t_f$, goes back to $t_i$, again along the real axis, and it finally makes a vertical descent down to $t_i-\ii\beta$, where $\beta=1/T$. This vertical path alone, the Matsubara contour $C_\beta$, is the basis of the \gls{itf} (with $t_i=0$), which we exploit in this chapter to compute equilibrium properties. By setting $t_i\rightarrow -\infty$ and $t_f\rightarrow +\infty$, one gets the so-called Schwinger-Keldysh contour $C$, which is composed of a forward branch $C_1$ and a backward branch $C_2$, as shown in the right panel of Fig.~\ref{fig:hot-contours}. 

First, as we are concerned with mesonic fields, for a generic bosonic field $\hat{\phi}(x)=\hat{\phi}(t,\vec{x})$ we define the two-point thermal Green's function ordered in the temporal arguments (along the $C$ contour) as
\begin{equation}
 \ii G(x,x')=\langle\mathcal{T}_C\{\hat{\phi}(x)\hat{\phi}(x')\}\rangle = \left\{ 
 \begin{array}{ccc}
 \langle  \hat{\phi}(x) \hat{\phi}(x') \rangle & \textrm{if} & t \succ t' \\
 \langle  \hat{\phi}(x')  \hat{\phi}(x) \rangle & \textrm{if} & t' \succ t
 \end{array}
 \right. \ , 
\end{equation}
where the contour ordering operator ${\cal T}_C$ orders field operators along the temporal contour $C$. In this context, the notation $t \succ t'$ represents that time $t$ comes later (along $C$) than $t'$.
In the practice, we work with the so-called ``greater'' and ``lesser'' Green's functions, also called Wightman functions,
\begin{align} 
  \ii G^> (x,x') \equiv \langle \hat{\phi}(x) \hat{\phi}(x') \rangle \ , \label{eq:hot-Ggreat}  \\
  \ii G^< (x,x') \equiv \langle \hat{\phi}(x') \hat{\phi}(x) \rangle \ . \label{eq:hot-Gless}
\end{align}
The physical interpretation of the Wightman functions can be found in the classical Ref.~\cite{kadanoff1962quantum}.

Then, one can define the thermal retarded and advanced propagators as follows,
\begin{align} 
  G^\textrm{R}  (x,x') & =   \theta_C(t-t')  \left[ G^> (x,x') - G^< (x,x')\right]   \ , \label{eq:hot-Gret} \\ 
  G^\textrm{A} (x,x') &  = - \theta_C(t'-t) \left[ G^> (x,x') - G^< (x,x')\right] \ , \label{eq:hot-Gadv} 
\end{align}
where the generalized step function $\theta_C(t-t')$ is defined along the Schwinger-Keldysh contour, and it is $0$ if $t \prec t'$, and $1$ if $t \succ t'$. 

These definitions for the Green's functions are valid in equilibrium as well as out of equilibrium, and up to this point we have not explicitly stated whether we have real or imaginary time. To perform actual calculations it is necessary to choose a formalism, keeping in mind that the final results for the physical quantities cannot depend on the choice of the contour $C$.

From now on in this chapter we use the \gls{itf}, which is based on the simplest choice of the integration contour, namely the straight vertical line that connects $t_i$ with $t_i-\ii\beta$ in the left panel of Fig.~\ref{fig:hot-contours}. The time variable may then be parameterized as $t=\ii \tau$ with $\tau\in[0,\,\beta]$. That is, we perform a Wick rotation to Euclidean time. The resulting thermal quantities, for example, the thermal propagator, resemble their counterparts in the zero-temperature theory, with the following differences, in Fourier space:
\begin{itemize}
\item energy variables are replaced by the so-called \textit{Matsubara frequencies}, 
\begin{alignat}{2}\label{eq:hot-matsubarafreqbosons}
 \omega_n&=\displaystyle\frac{2n\pi}{\beta} &&\quad\textrm{for\, bosons} \ , \\
 \omega_n&=\displaystyle\frac{(2n+1)\pi}{\beta}&&\quad \textrm{for\, fermions} \ . \label{eq:hot-matsubarafreqfermions}
\end{alignat}
 \item any integration over internal energies is transformed into a summation over Matsubara frequencies,
 \begin{equation}\label{eq:hot-defsummatsubara}
  \int\frac{d^4q}{(2\pi)^4}\rightarrow \frac{1}{\beta}\sum_n\int\frac{d^3q}{(2\pi)^3} \ .
 \end{equation}
\end{itemize}
See Refs.~\cite{Weldon:1983jn,Das:1997gg,kapustagale,lebellac,fetterwalecka} for details of the formalism.

The practical application to the scattering of heavy mesons off light mesons requires the explicit calculation in the \gls{itf} of the two-meson propagator and the heavy-meson self-energy. The self-energy is the other essential quantity that we need to compute at finite temperature. It is defined as the energy that a particle acquires as a result of its modification through the interactions with the medium. Thus, the self-energy gives a contribution to the particle in-medium mass and decay width. Before computing these two quantities in Sections~\ref{subsec:hot-form-loop} and \ref{subsec:hot-form-selfe}, respectively, using the \gls{itf}, let us discuss the general procedure to perform the sum over Matsubara frequencies.

\subsection{The Matsubara summation}
\label{subsec:hot-form-matsubarasum}
\begin{figure}[b!]\centering 
\begin{tikzpicture}[baseline=-2,
    decoration={%
      markings,
      mark=at position 0.01cm with {\arrow[line width=1pt,>=stealth]{>}},
    }
  ]
  \draw [>=stealth,->] (-1.9,0) -- (1.9,0) coordinate (xaxis);
  \draw [>=stealth,->] (0,-1.9) -- (0,1.9) coordinate (yaxis);
  \node [color=ctcolordarkgreen] at (0,0) {$\times$};
  \node [color=ctcolordarkgreen] at (0,0.5) {$\times$};
  \node [color=ctcolordarkgreen] at (0,1) {$\times$};
  \node [color=ctcolordarkgreen] at (0,1.5) {$\times$};
  \node [color=ctcolordarkgreen] at (0,-0.5) {$\times$};
  \node [color=ctcolordarkgreen] at (0,-1) {$\times$};
  \node [color=ctcolordarkgreen] at (0,-1.5) {$\times$};
  \path [draw, line width=0.8pt, postaction=decorate, color=ctcolorblue] (0.2,0)  arc (0:360:0.2) ;
  \path [draw, line width=0.8pt, postaction=decorate, color=ctcolorblue] (0.2,0.5)  arc (0:360:0.2) ;
  \path [draw, line width=0.8pt, postaction=decorate, color=ctcolorblue] (0.2,1)  arc (0:360:0.2) ;
  \path [draw, line width=0.8pt, postaction=decorate, color=ctcolorblue] (0.2,1.5)  arc (0:360:0.2) ;
  \path [draw, line width=0.8pt, postaction=decorate, color=ctcolorblue] (0.2,-0.5)  arc (0:360:0.2) ;
  \path [draw, line width=0.8pt, postaction=decorate, color=ctcolorblue] (0.2,-1)  arc (0:360:0.2) ;
  \path [draw, line width=0.8pt, postaction=decorate, color=ctcolorblue] (0.2,-1.5)  arc (0:360:0.2) ;
  \node [below] at (xaxis) {Re $z$};
  \node [left] at (yaxis) {Im $z$};
  \node at (0.6,1.) {$C_n$};
  \node at (-0.5,1) {$z_n$};
\end{tikzpicture}
\hspace{1cm}
\begin{tikzpicture}[baseline=-2,
    decoration={%
      markings,
      mark=at position 1cm with {\arrow[line width=1pt,>=stealth]{>}},
    }
  ]
  \draw [>=stealth,->] (-1.9,0) -- (1.9,0) coordinate (xaxis);
  \draw [>=stealth,->] (0,-1.9) -- (0,1.9) coordinate (yaxis);
  \node [color=ctcolordarkgreen] at (0,0) {$\times$};
  \node [color=ctcolordarkgreen] at (0,0.5) {$\times$};
  \node [color=ctcolordarkgreen] at (0,1) {$\times$};
  \node [color=ctcolordarkgreen] at (0,1.5) {$\times$};
  \node [color=ctcolordarkgreen] at (0,-0.5) {$\times$};
  \node [color=ctcolordarkgreen] at (0,-1) {$\times$};
  \node [color=ctcolordarkgreen] at (0,-1.5) {$\times$};
  \path [draw, line width=0.8pt, postaction=decorate, color=ctcolorblue] (0.25,-1.5) -- (0.25,1.5) arc (0:180:0.25) -- (-0.25,-1.5) arc (180:360:0.25) ;
  \node [below] at (xaxis) {Re $z$};
  \node [left] at (yaxis) {Im $z$};
  \node at (-0.5,.4) {$C'$};
\end{tikzpicture}
\caption{Representation of the infinite number of poles of the weighting function $f(z)=(e^{\beta z}-1)^{-1}$ on the imaginary axis, depicted by the green crosses. The contours in the left plot may be deformed to the contour depicted in the right plot.}
\label{fig:hot-form-Matsubara1}
\end{figure}
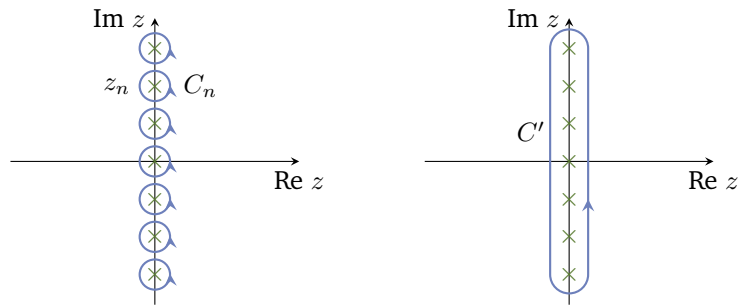

This subsection contains a brief discussion on the general technique of the Matsubara summation, before applying it to the specific calculation of the thermal two-meson propagator and the meson self-energy within the \gls{itf}. The aim is to analytically evaluate a summation of the form
\begin{equation}\label{eq:matsubara_sum_1}
 \frac{1}{\beta}\sum_n F(\ii\,\omega_n) \ ,
\end{equation}
with the bosonic Matsubara frequencies defined as $\ii\omega_n=\ii\,2n\pi/\beta$ (in the case of fermions these would be fermionic frequencies, $\ii\omega_n=\ii (2n+1)\pi/\beta$). The sum over Matsubara frequencies can be replaced by a sum over the poles of $F$ using a standard trick, as described by, for instance, Ref.~\cite{kapustagale}. First of all the purely imaginary Matsubara frequencies are substituted by a general complex variable $z$, so the function $F(\ii\,\omega_n)$ transforms into $F(z)$. After this analytic continuation, the Cauchy residue theorem can be used to replace the sum with a contour integral. This theorem relates a contour integral in the complex plane with the sum over the poles enclosed within the contour,
\begin{equation}\label{eq:matsubara_sum_2}
 \oint_C g(z)=2\pi i\,\sum_n\textrm{Res}\,g(z_n) \ ,
\end{equation}
where $z_n$ are the complex poles of the function $g(z)$ and $\textrm{Res}$ refers to the residues of $g(z)$ around these poles. For simple poles, that is, of order one, one has $\textrm{Res}_{z=z_n}\, g(z)=\lim_{z\rightarrow z_n}\{(z-z_n)g(z)\}$. The essence of the trick is to consider a weighting function $f(z)=(e^{\beta z}-1)^{-1}$ that has simple poles with residue $1/\beta$ at $z_n=\ii\,2n\pi/\beta$, which coincide with the values of the Matsubara frequencies of the sum in Eq.~(\ref{eq:matsubara_sum_1}). The poles of $f(z)$ are illustrated in Fig~\ref{fig:hot-form-Matsubara1} by green crosses on the imaginary axis. The residues at $z_n$ of the product of the functions $f(z)F(z)$ are given by
\begin{equation}\label{eq:matsubara_sum_3}
 \textrm{Res}_{z=z_n}\{f(z)F(z)\}=\lim_{z\rightarrow z_n}\bigg\{(z-z_n)\frac{F(z)}{e^{\beta z}-1} \bigg\}=\frac{1}{\beta}F(z_n) \ .
\end{equation}

Using the result of Eq.~(\ref{eq:matsubara_sum_3}) together with the residue theorem of Eq.~(\ref{eq:matsubara_sum_2}), the Matsubara sum can be written as a contour integral:
\begin{align}\label{eq:matsubara_sum_4}\nonumber
 \frac{1}{\beta}\sum_n F(\ii\,\omega_n)&=\frac{1}{\beta}\sum_nF(z_n) \quad && \textrm{analytic continuation} \\ \nonumber
 &=\sum_n\textrm{Res}_{z=z_n}\frac{F(z)}{e^{\beta z}-1} \quad && \textrm{Eq.~(\ref{eq:matsubara_sum_3})} \\ \nonumber
 &=\sum_n\frac{1}{2\pi  \ii}\oint_{C_n}\frac{F(z)}{e^{\beta z}-1}\phantom{-} \quad && \textrm{residue theorem} \\
 &=\frac{1}{2\pi \ii}\oint_{C'} \frac{F(z)}{e^{\beta z}-1} \ . && 
\end{align}
In the last line, we have combined the $n$ contour integrals along the small contours $C_n$ around each of the purely imaginary poles $z_n$ of the function $f(z)$ (left plot in Fig.~\ref{fig:hot-form-Matsubara1}) in a contour $C'$ enclosing the $n$ poles (right plot in Fig.~\ref{fig:hot-form-Matsubara1}). Yet the function $F(z)$ may have poles itself.

\begin{figure}[b!]\centering 
$0=\,$
\begin{tikzpicture}[baseline=-2,
    decoration={%
      markings,
      mark=at position 1cm with {\arrow[line width=1pt,>=stealth]{>}},
    }
  ]
  \draw [>=stealth,->] (-1.9,0) -- (1.9,0) coordinate (xaxis);
  \draw [>=stealth,->] (0,-1.9) -- (0,1.9) coordinate (yaxis);
  \node [color=ctcolordarkgreen] at (0,0) {$\times$};
  \node [color=ctcolordarkgreen] at (0,0.25) {$\times$};
  \node [color=ctcolordarkgreen] at (0,0.5) {$\times$};
  \node [color=ctcolordarkgreen] at (0,0.75) {$\times$};
  \node [color=ctcolordarkgreen] at (0,1) {$\times$};
  \node [color=ctcolordarkgreen] at (0,1.25) {$\times$};
  \node [color=ctcolordarkgreen] at (0,-0.25) {$\times$};
  \node [color=ctcolordarkgreen] at (0,-0.5) {$\times$};
  \node [color=ctcolordarkgreen] at (0,-0.75) {$\times$};
  \node [color=ctcolordarkgreen] at (0,-1) {$\times$};
  \node [color=ctcolordarkgreen] at (0,-1.25) {$\times$};
  \node [color=ctcolormagenta] at (-1,-0.5) {$\times$};
  \node [color=ctcolormagenta] at (0.75,0.5) {$\times$};
  \path [draw, line width=0.8pt, postaction=decorate, color=ctcolorblue] (1.5,0)  arc (0:360:1.5) ;
  \node [below] at (xaxis) {Re $z$};
  \node [left] at (yaxis) {Im $z$};
  \node at (1,1.5) {$C$};
  \node at (-.65,-0.5) {$z_1$};
  \node at (1.1,.5) {$z_2$};
  \node at (-0.35,1) {$z_n$};
\end{tikzpicture}
$=\,$
\begin{tikzpicture}[baseline=-2,
    decoration={%
      markings,
      mark=at position 1cm with {\arrow[line width=1pt,>=stealth]{>}},
    }
  ]
  \draw [>=stealth,->] (-1.9,0) -- (1.9,0) coordinate (xaxis);
  \draw [>=stealth,->] (0,-1.9) -- (0,1.9) coordinate (yaxis);
  \node [color=ctcolordarkgreen] at (0,0) {$\times$};
  \node [color=ctcolordarkgreen] at (0,0.25) {$\times$};
  \node [color=ctcolordarkgreen] at (0,0.5) {$\times$};
  \node [color=ctcolordarkgreen] at (0,0.75) {$\times$};
  \node [color=ctcolordarkgreen] at (0,1) {$\times$};
  \node [color=ctcolordarkgreen] at (0,1.25) {$\times$};
  \node [color=ctcolordarkgreen] at (0,-0.25) {$\times$};
  \node [color=ctcolordarkgreen] at (0,-0.5) {$\times$};
  \node [color=ctcolordarkgreen] at (0,-0.75) {$\times$};
  \node [color=ctcolordarkgreen] at (0,-1) {$\times$};
  \node [color=ctcolordarkgreen] at (0,-1.25) {$\times$};
  \node [color=ctcolormagenta] at (-1,-0.5) {$\times$};
  \node [color=ctcolormagenta] at (0.75,0.5) {$\times$};
  \path [draw, line width=0.8pt, postaction=decorate, color=ctcolorblue] (0.25,-1.25) -- (0.25,1.25) arc (0:180:0.25) -- (-0.25,-1.25) arc (180:360:0.25) ;
  \node [below] at (xaxis) {Re $z$};
  \node [left] at (yaxis) {Im $z$};
  \node at (-0.5,.4) {$C'$};
\end{tikzpicture}
$+\,$
\begin{tikzpicture}[baseline=-2,
    decoration={%
      markings,
      mark=at position 0.8cm with {\draw[line width=1pt,>=stealth,->] (-0.01,0) -- (0.07,0);},
    }
  ]
  \draw [>=stealth,->] (-1.9,0) -- (1.9,0) coordinate (xaxis);
  \draw [>=stealth,->] (0,-1.9) -- (0,1.9) coordinate (yaxis);
  \node [color=ctcolordarkgreen] at (0,0) {$\times$};
  \node [color=ctcolordarkgreen] at (0,0.25) {$\times$};
  \node [color=ctcolordarkgreen] at (0,0.5) {$\times$};
  \node [color=ctcolordarkgreen] at (0,0.75) {$\times$};
  \node [color=ctcolordarkgreen] at (0,1) {$\times$};
  \node [color=ctcolordarkgreen] at (0,1.25) {$\times$};
  \node [color=ctcolordarkgreen] at (0,-0.25) {$\times$};
  \node [color=ctcolordarkgreen] at (0,-0.5) {$\times$};
  \node [color=ctcolordarkgreen] at (0,-0.75) {$\times$};
  \node [color=ctcolordarkgreen] at (0,-1) {$\times$};
  \node [color=ctcolordarkgreen] at (0,-1.25) {$\times$};
  \node [color=ctcolormagenta] at (-1,-0.5) {$\times$};
  \node [color=ctcolormagenta] at (0.75,0.5) {$\times$};
  \path [draw, line width=0.8pt, postaction=decorate, color=ctcolorblue] (-1.25,-0.5)  arc (-180:180:.25) ;
  \path [draw, line width=0.8pt, postaction=decorate, color=ctcolorblue] (0.5,0.5)  arc (-180:180:.25) ;
  \node [below] at (xaxis) {Re $z$};
  \node [left] at (yaxis) {Im $z$};
  \node at (-1.5,-0.5) {$C_1$};
  \node at (1.35,.5) {$C_2$};
\end{tikzpicture}
\caption{Contour integrals used in the evaluation of the Matsubara summation of Eq.~(\ref{eq:matsubara_sum_1}) for a function $F(z)$ with two poles at $z_1$ and $z_2$. The integral along the contour $C$ enclosing all the poles (left plot) vanishes as the radius goes to infinity. Upon the deformation of the contour $C$, it can be split in the sum of an integral along the contour $C'$ enclosing the poles on the imaginary axis (middle plot) and the integrals along the contours $C_k$ around the poles of $F(z)$ (right plot).}
\label{fig:hot-form-Matsubara}
\end{figure}
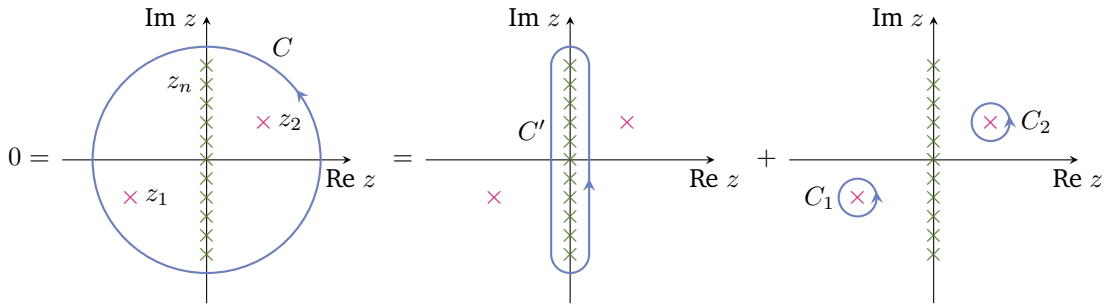

Let us consider an example in which $F(z)$ has $j=2$ poles at $z_1$ and $z_2$ as depicted in Fig.~\ref{fig:hot-form-Matsubara}. Now, if $f(z)F(z)$ goes fast enough to zero as $|z|\rightarrow\infty$, the contour integral along the circular path $C$ enclosing all the poles (left plot in Fig.~\ref{fig:hot-form-Matsubara}) vanishes as the radius of the contour goes to infinity. Moreover, the contour $C$ can be deformed into a contour $C'$ containing the $z_n$ poles of $f(z)$ (middle plot in Fig.~\ref{fig:hot-form-Matsubara}) and small contours around the $z_j$ poles of $F(z)$ (right plot in Fig.~\ref{fig:hot-form-Matsubara}). This allows us to relate the integral around the poles of $f(z)$ with the sum of the integrals around the poles of $F(z)$:
\begin{align}\label{eq:matsubara_sum_5}\nonumber
 \frac{1}{\beta}\sum_n F(\ii\,\omega_n)&=\frac{1}{2\pi \ii}\oint_{C'} \frac{F(z)}{e^{\beta z}-1} \quad && \phantom{blablablablablablablab} \\ \nonumber
 &=-\frac{1}{2\pi \ii}\sum_j\oint_{C_j}\frac{F(z)}{e^{\beta z}-1}  && \\ \nonumber
 &=-\sum_j\textrm{Res}_{z=z_j}\frac{F(z)}{e^{\beta z}-1} \quad && \textrm{residue theorem} \\
 &=-\sum_j\frac{1}{e^{\beta z_j}-1}\lim_{z\rightarrow z_j}\left\{(z-z_j)F(z)\right\} \ . \hspace{-20mm}  &&
\end{align}

In the third line of Eq.~(\ref{eq:matsubara_sum_5}), we have again taken advantage of the Cauchy residue theorem to finally obtain, in the last line, an expression for the summation over the infinite number of Matsubara frequencies involving the sum over just the reduced number of poles $z_j$ of $F(z)$.

\subsection{Thermal two-meson propagator}
\label{subsec:hot-form-loop}
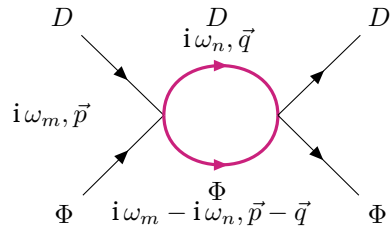
\begin{figure}[b!]%
 \centering
  \begin{tikzpicture}[baseline=(i.base)]
    \begin{feynman}[small]
      \vertex (i);
      \vertex [right = 1.5cm of i] (j);
      \vertex [above left = 1.5cm of i] (a) {\(D\)};
      \vertex [above right = 1.5cm of j] (c) {\(D\)};
      \vertex [below left = 1.5cm of i] (b) {\(\Phi\)};
      \vertex [below right = 1.5cm of j] (d) {\(\Phi\)};
      \diagram* {
        (a) -- [fermion] (i) -- [fermion, half left,very thick,ctcolormagenta, looseness=1.5] (j) -- [fermion] (c), 
        (b) -- [fermion] (i) -- [fermion, half right,very thick,ctcolormagenta, looseness=1.5] (j) -- [fermion] (d),
       };
     \node at (0.7,1.3) {\(D\)};
     \node at (0.7,1) {\(\ii\omega_n,\vec{q}\)};
     \node at (0.7,-1) {\(\Phi\)};
     \node at (0.6,-1.3) {\(\ii\omega_m-\ii\omega_n,\vec{p}-\vec{q}\)};
     \node at (-1.5,0) {\(\ii\omega_m,\vec{p}\)};
    \end{feynman}
  \end{tikzpicture}
\caption{Diagram of the general two-meson loop function with dressed internal meson propagators, denoted with thick magenta lines.}
\label{fig:hot-form-loop}
\end{figure}

We consider two mesons that we denote as $\mathcal{M}\equiv D$ and $\mathcal{M'}\equiv\Phi$. The two-body loop function in the \gls{itf} reads
\begin{align}\label{eq:hot-loop1}\nonumber
 G_{D\Phi}(\ii\omega_m,\vec{p}\,;T)&=-\frac{1}{\beta}\sum_n\int\frac{d^3q}{(2\pi)^3}\mathcal{D}_D(\ii\omega_n,\vec{q}\,)\mathcal{D}_\Phi(\ii\omega_m-\ii\omega_n,\vec{p}-\vec{q}\,) \\ 
 &=-\frac{1}{\beta}\sum_n\int\frac{d^3q}{(2\pi)^3}\frac{1}{\omega_n^2+\vec{q}\,^2+m_D^2}\frac{1}{(\omega_m-\omega_n)^2+(\vec{p}-\vec{q}\,)^2+m_\Phi^2} \ ,
\end{align}
where $p^0\rightarrow\ii\omega_m$ is the total external energy and $q^0\rightarrow\ii\omega_n$ is the internal $D$-meson energy, as shown in Fig.~\ref{fig:hot-form-loop}.

Before performing the Matsubara summation over $\omega_n$ we introduce the Lehmann representation for the propagators in terms of the spectral function,
\begin{equation}\label{eq:hot-propLehmann}
 \mathcal{D}_M(\ii\omega_n,\vec{q}\,)=\int d\omega\frac{S_M(\omega,\vec{q}\,)}{\ii\omega_n-\omega}=\int_0^\infty d\omega\frac{S_M(\omega,\vec{q}\,)}{\ii\omega_n-\omega}-\int_0^\infty d\omega\frac{S_{\bar{M}}(\omega,\vec{q}\,)}{\ii\omega_n+\omega} \ ,
\end{equation}
where the subindex $M$ denotes the meson species ($D$ or $\Phi$) and in the second equality we have separated the particle ($M$) and antiparticle ($\bar{M}$) parts. Using delta-type spectral functions, $S_M(\omega,\vec{q}\,)=\frac{\omega_M}{\omega}\delta(\omega^2-\omega_M^2)$, with $\omega_M=\sqrt{\vec{q}\,^2+m_M^2}$, it is straightforward to see that Eq.~(\ref{eq:hot-loop1}) is recovered. By keeping generic spectral functions $S_D(\omega,\vec{q}\,)$ and $S_\Phi(\omega',\vec{p}-\vec{q}\,)$, that is, by dressing the mesons in the loop,
\begin{align}\label{eq:hot-loop2}\nonumber
 G_{D\Phi}(\ii\omega_m,\vec{p}\,;T)=-\frac{1}{\beta}\sum_n &\int\frac{d^3q}{(2\pi)^3}\int_0^\infty d\omega\left\{\frac{S_D(\omega,\vec{q}\,)}{\ii\omega_n-\omega}-\frac{S_{\bar{D}}(\omega,\vec{q}\,)}{\ii\omega_n+\omega}\right\}  \\
 &\times\int_0^\infty d\omega'\left\{\frac{S_\Phi(\omega',\vec{p}-\vec{q}\,)}{\ii\omega_m-\ii\omega_n-\omega'}-\frac{S_{\bar{\Phi}}(\omega',\vec{p}-\vec{q}\,)}{\ii\omega_m-\ii\omega_n+\omega'} \right\} \ ,
\end{align}
we get four terms of the form $F(\ii\omega_n)\equiv F(z)$, with simple poles at $z_1=\pm\,\omega$ and $z_2=\ii\omega_m\pm\omega'$, respectively, for which we perform the Matsubara summation as described in Section~\ref{subsec:hot-form-matsubarasum}.

Using Eq.~(\ref{eq:matsubara_sum_5}) we obtain the following expression for the loop function:
\begin{align}\label{eq:hot-loop3}\nonumber
 G_{D\Phi}(\ii\omega_m,\vec{p}\,;T)&=\int\frac{d^3q}{(2\pi)^3}\int_0^\infty d\omega\int_0^\infty d\omega'\Bigg\{\left[1+f(\omega,T)+f(\omega',T)\right]  \\ \nonumber
 &\quad\times\left[\frac{S_D(\omega,\vec{q}\,)S_\Phi(\omega',\vec{p}-\vec{q}\,)}{\ii\omega_m-\omega-\omega'}-\frac{S_{\bar{D}}(\omega,\vec{q}\,)S_{\bar{\Phi}}(\omega',\vec{p}-\vec{q}\,)}{\ii\omega_m+\omega+\omega'} \right]  \\ \nonumber
 &+\left[f(\omega,T)-f(\omega',T)\right] \\
 &\quad\times\left[ \frac{S_{\bar{D}}(\omega,\vec{q}\,)S_\Phi(\omega',\vec{p}-\vec{q}\,)}{\ii\omega_m+\omega-\omega'}-\frac{S_{D}(\omega,\vec{q}\,)S_{\bar{\Phi}}(\omega',\vec{p}-\vec{q}\,)}{\ii\omega_m-\omega+\omega'}\right]\Bigg\} \ ,
\end{align}\vspace{-1mm}
where 
\begin{equation}
 f(\omega,T)=\frac{1}{e^{\omega/T}-1}
\end{equation}
 is the equilibrium \gls{be} distribution function at temperature $T$. We have used that the \gls{be} function evaluated at negative energy can be written as \vspace{-1mm}
 \begin{equation}
  f(-\omega,T)=\frac{1}{e^{-\omega/T}-1}=-\left[1+f(\omega,T)\right] \ ,
 \end{equation}
and at the bosonic Matsubara frequency $\omega_m=\ii 2m\pi T$, as \vspace{-1mm}
\begin{equation}
    f(\ii\omega_m\pm\omega',T)=f(\pm\,\omega',T) \ .
\end{equation}
The $1$ preceding the \gls{be} occupancy numbers in the first line of Eq.~(\ref{eq:hot-loop3}) represents the vacuum contribution. To evaluate this equation, one has to place an upper cut-off on the momentum integration, or else analytically continue to $4-2\eta$ dimensions, so as to regularize it (see the discussion on regularization of the loop function in Chapter~\ref{ch:exoticsinfreespace}). 

The above expression for the loop function must be analytically continued from the discrete imaginary frequencies $\ii\omega_m$ to real energies to describe a two-meson system with energy $E$, that is, $\ii\omega_m\rightarrow E+\ii\varepsilon$. Additionally, taking into account that the spectral functions satisfy the relation $S_{\bar{M}}(-\omega,\vec{q}\,;T)=-S_M(\omega,\vec{q}\,;T)$, we can express the thermal loop in the following compact way:
\begin{equation}\label{eq:hot-loop-compact}
 G_{D\Phi}(E,\vec{p}\,;T)=\int\frac{d^3q}{(2\pi)^3}\int d\omega\int d\omega'\frac{S_D(\omega,\vec{q}\,;T)S_\Phi(\omega',\vec{p}-\vec{q}\,;T)}{E-\omega-\omega'+\ii\varepsilon}\left[1+f(\omega,T)+f(\omega',T)\right] \ ,
\end{equation}
where the integrals over energy extend from $-\infty$ to $+\infty$.

At finite temperature the meson masses are dressed by the medium. The effects of finite temperature in the unitarized scattering amplitudes are obtained by solving the \gls{bs} equation in Eq.~(\ref{eq:free-th-BSmatrixinv}) with thermal loops containing dressed mesons. 
In general, the spectral function necessary to dress the meson in the loop is computed from the imaginary part of the retarded meson propagator,
\begin{equation} \label{eq:hot-specfunc}
  S_{M}(\omega,\vec{q}\,;T)=-\frac{1}{\pi}\textrm{Im\,}\mathcal{D}_{M}(\omega,\vec{q}\,;T)=-\frac{1}{\pi}\textrm{ Im\,}\Bigg(\frac{1}{\omega^2-\vec{q}\,^2-m_{M}^2-\Pi_{M}(\omega,\vec{q}\,;T)}\Bigg) \ .
\end{equation}
We take this prescription for the heavy-meson spectral function in our calculations, and the self-energy follows from closing the light-meson line in the $T$ matrix, as discussed later on.

In this chapter, we are interested in analyzing the medium modification of the $D$-meson propagator due to light mesons. However, light mesons also suffer medium modifications and their spectral functions are modified by the interactions among themselves. For a pion gas, we can use previous results in the literature to tell us that the mass modification is small and that the use of a free spectral function is justified. In Appendix~\ref{appendix-modpion} we show this fact, where up to temperatures of $T=150$~MeV the pion mass varies at most $10\%$. At the largest temperature of $T=150$~MeV considered, we have used $m_\pi=120$~MeV mass in our numerical calculation and found that the final charmed-meson masses (both ground and the dynamically generated states) are modified by less than $0.1\%$. Therefore, we use the free pion spectral function. 

In view that taking the vacuum propagator is a good approximation for the light mesons, the expression for the thermal two-meson loop function given in Eq.~(\ref{eq:hot-loop-compact}) can be simplified accordingly.
Furthermore, we have already noticed that the calculation of the heavy-meson propagator dressed with its self-energy depends on some particular matrix elements of the unitarized thermal amplitude. Thus, one realizes that the calculations of the loop function at finite temperature and the heavy-meson self-energy are connected to each other and they have to be solved in an iterative process until the convergence of the results. This point will become clearer below in this chapter. 

The first iteration of the thermal loop function is made in the approximation of two free (undressed) mesons.
For free mesons, the spectral functions become delta distributions:
\begin{equation}\label{eq:hot-deltaSfunc}
 S_M(\omega,\vec{q}\,)=\frac{\omega_M}{\omega}\delta(\omega^2-\omega_M^2)=\frac{1}{2\omega}\left[\delta(\omega-\omega_M)+\delta(\omega+\omega_M)\right] \ ,
\end{equation}
with $\omega_M=\sqrt{\vec{q}\,^2+m_M^2}$ the energy of the $M=\{D,\Phi\}$ meson, and the loop function simplifies to
\begin{align}\label{eq:hot-loopfree1} \nonumber
 G_{D\Phi}(E,\vec{p}\,;T)&=\int\frac{d^3q}{(2\pi)^3}\frac{1}{4\,\omega_D\omega_\Phi}\Bigg\{\left[1+f(\omega_D,T)+f(\omega_\Phi,T)\right] \\ \nonumber &\times\left(\frac{1}{E-\omega_D-\omega_\Phi+\ii\varepsilon}-\frac{1}{E+\omega_D+\omega_\Phi+\ii\varepsilon}\right) \\  \nonumber
 &+\left[f(\omega_D,T)-f(\omega_\Phi,T)\right] \\
 &\times\left(\frac{1}{E+\omega_D-\omega_\Phi+\ii\varepsilon}-\frac{1}{E-\omega_D+\omega_\Phi+\ii\varepsilon} \right)
  \Bigg\} \ .
\end{align}

In the case when only one of the meson spectral functions can be approximated with a delta-type spectral function, which is what we consider for the subsequent iterations of the calculations in this thesis, one gets the following expression for the two-meson loop function: 
\begin{align}\label{eq:hot-loop-oneSE}\nonumber
 G_{D\Phi}(E,\vec{p}\,;T)&=-\int\frac{d^3q}{(2\pi)^3}\int_0^\infty d\omega\frac{S_D(\omega,\vec{q}\,;T)}{2\omega_\Phi}\Bigg\{\left[1+f(\omega,T)+f(\omega_\Phi,T)\right] \\ \nonumber
 &\times\left(\frac{1}{\omega-(E-\omega_\Phi)-\ii\varepsilon}+\frac{1}{\omega+(E+\omega_\Phi)+\ii\varepsilon}\right) \\ \nonumber
 &+\left[f(\omega_\Phi,T)-f(\omega,T)\right] \\
 &\times\left(\frac{1}{\omega-(E+\omega_\Phi)-\ii\varepsilon}+\frac{1}{\omega+(E-\omega_\Phi)+\ii\varepsilon}\right)\Bigg\} \ ,
\end{align}
with $\omega_\Phi=\sqrt{(\vec{p}-\vec{q}\,)^2+m_\Phi^2}$. 

As in the vacuum case, the two-meson propagator needs to be regularized. The most straightforward way to do that at finite temperature is to introduce a hard cut-off in the upper limit in the integral over the modulus of the momentum. The technical details of the nontrivial numerical integration of Eqs.~(\ref{eq:hot-loopfree1}) and (\ref{eq:hot-loop-oneSE}) regularized with a cut-off are given in Appendix~\ref{sec:app-loop}.

\subsection{Physical interpretation and cuts of the thermal propagator}
\label{subsec:hot-form-Landaucut}
The expressions for the two-meson thermal loop function derived above in the \gls{itf} allow a physical interpretation in terms of multiple scattering with the constituents of the thermal mesonic medium. For such an interpretation we follow a similar argument to that of Weldon in Ref.~\cite{Weldon:1983jn}, where he studied and physically interpreted one-loop self-energies.

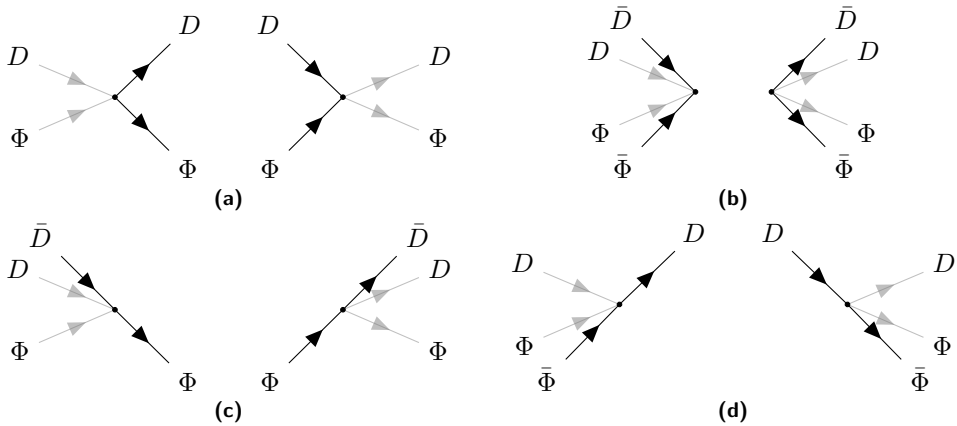
\begin{figure}[b!]
\centering 
\begin{subfigure}[b]{0.45\textwidth}\centering 
\captionsetup{skip=0pt}
\begin{tikzpicture}[baseline=(i1.base)]
    \begin{feynman}
      \vertex (i1) ;
      \vertex [above right = 1cm of i1] (a) {\(D\)};
      \vertex [below right = 1cm of i1] (b) {\(\Phi\)};
      \vertex [right = 3cm of i1] (i2);
      \vertex [above left = 1cm of i2] (c) {\(D\)};
      \vertex [below left = 1cm of i2] (d) {\(\Phi\)};
      \vertex [above left = 0.3cm and 1cm of i1] (a2) {\(D\)};
      \vertex [below left = 0.3cm and 1cm of i1] (b2) {\(\Phi\)};
      \vertex [above right = 0.3cm and 1cm of i2] (c2) {\(D\)};
      \vertex [below right = 0.3cm and 1cm of i2] (d2) {\(\Phi\)};
      \diagram* {
        (i1) -- [fermion] (a), 
        (c) -- [fermion] (i2),
        (i1) -- [fermion] (b), 
        (d) -- [fermion] (i2),
        (a2) -- [fermion, gray, opacity=0.5] (i1),
        (i2) -- [fermion, gray, opacity=0.5] (c2),
        (b2) -- [fermion, gray, opacity=0.5] (i1),
        (i2) -- [fermion, gray, opacity=0.5] (d2),
       } ;   
       \draw[dot] (i1) circle(0.3mm);
       \draw[dot] (i2) circle(0.3mm);
    \end{feynman}
  \end{tikzpicture} 
\caption{}
\label{fig:hot-interp-a}
\end{subfigure}
\begin{subfigure}[b]{0.45\textwidth}\centering 
\captionsetup{skip=0pt}
      \begin{tikzpicture}[baseline=(i1.base)]
    \begin{feynman}
      \vertex (i1) ;
      \vertex [above left = 1cm of i1] (a) {\(\bar{D}\)};
      \vertex [below left = 1cm of i1] (b) {\(\bar{\Phi}\)};
      \vertex [right = 1cm of i1] (i2);
      \vertex [above right = 1cm of i2] (c) {\(\bar{D}\)};
      \vertex [below right = 1cm of i2] (d) {\(\bar{\Phi}\)};
      \vertex [above left = 0.3cm and 1cm of i1] (a2) {\(D\)};
      \vertex [below left = 0.3cm and 1cm of i1] (b2) {\(\Phi\)};
      \vertex [above right = 0.3cm and 1cm of i2] (c2) {\(D\)};
      \vertex [below right = 0.3cm and 1cm of i2] (d2) {\(\Phi\)};
      \diagram* {
        (a) -- [fermion] (i1), 
        (i2) -- [fermion] (c),
        (b) -- [fermion] (i1),
        (i2) -- [fermion] (d),
        (a2) -- [fermion, gray, opacity=0.5] (i1),
        (i2) -- [fermion, gray, opacity=0.5] (c2),
        (b2) -- [fermion, gray, opacity=0.5] (i1),
        (i2) -- [fermion, gray, opacity=0.5] (d2),
       } ;   
       \draw[dot] (i1) circle(0.3mm);
       \draw[dot] (i2) circle(0.3mm);
    \end{feynman}
  \end{tikzpicture} 
\caption{}
\label{fig:hot-interp-b}
\end{subfigure}
 \begin{subfigure}[b]{0.45\textwidth}\centering 
\captionsetup{skip=0pt} 
      \begin{tikzpicture}[baseline=(i1.base)]
    \begin{feynman}
      \vertex (i1) ;
      \vertex [above left = 1cm of i1] (a) {\(\bar{D}\)};
      \vertex [below right = 1cm of i1] (b) {\(\Phi\)};
      \vertex [right = 3cm of i1] (i2) ;
      \vertex [below left = 1cm of i2] (d) {\(\Phi\)};
      \vertex [above right = 1cm of i2] (c) {\(\bar{D}\)};
      \vertex [above left = 0.3cm and 1cm of i1] (a2) {\(D\)};
      \vertex [below left = 0.3cm and 1cm of i1] (b2) {\(\Phi\)};
      \vertex [above right = 0.3cm and 1cm of i2] (c2) {\(D\)};
      \vertex [below right = 0.3cm and 1cm of i2] (d2) {\(\Phi\)};
      \diagram* {
        (a) -- [fermion] (i1), 
        (i2) -- [fermion] (c),
        (i1) -- [fermion] (b), 
        (d) -- [fermion] (i2),
        (a2) -- [fermion, gray, opacity=0.5] (i1),
        (i2) -- [fermion, gray, opacity=0.5] (c2),
        (b2) -- [fermion, gray, opacity=0.5] (i1),
        (i2) -- [fermion, gray, opacity=0.5] (d2),
       } ;   
       \draw[dot] (i1) circle(0.3mm);
       \draw[dot] (i2) circle(0.3mm);
    \end{feynman}
  \end{tikzpicture}
\caption{}
\label{fig:hot-interp-c}
\end{subfigure}
\begin{subfigure}[b]{0.45\textwidth}\centering 
\captionsetup{skip=0pt}
      \begin{tikzpicture}[baseline=(i1.base)]
    \begin{feynman}
      \vertex (i1) ;
      \vertex [above right = 1cm of i1] (a) {\(D\)};
      \vertex [below left = 1cm of i1] (b) {\(\bar{\Phi}\)};
      \vertex [right = 3cm of i1] (i2) ;
      \vertex [above left = 1cm of i2] (c) {\(D\)};
      \vertex [below right = 1cm of i2] (d) {\(\bar{\Phi}\)};
      \vertex [above left = 0.3cm and 1cm of i1] (a2) {\(D\)};
      \vertex [below left = 0.3cm and 1cm of i1] (b2) {\(\Phi\)};
      \vertex [above right = 0.3cm and 1cm of i2] (c2) {\(D\)};
      \vertex [below right = 0.3cm and 1cm of i2] (d2) {\(\Phi\)};
      \diagram* {
        (i1) -- [fermion] (a), 
        (c) -- [fermion] (i2),
        (b) -- [fermion] (i1), 
        (i2) -- [fermion] (d),
        (a2) -- [fermion, gray, opacity=0.5] (i1),
        (i2) -- [fermion, gray, opacity=0.5] (c2),
        (b2) -- [fermion, gray, opacity=0.5] (i1),
        (i2) -- [fermion, gray, opacity=0.5] (d2),
       } ;   
       \draw[dot] (i1) circle(0.3mm);
       \draw[dot] (i2) circle(0.3mm);
    \end{feynman}
  \end{tikzpicture} 
\caption{}
\label{fig:hot-interp-d}
\end{subfigure}
\caption{Absorption and production processes of $D\Phi$ pairs in a thermal bath, obtained from cutting the two-meson loop diagram in Fig.~\ref{fig:hot-form-loop}. Black lines represent the particles that are absorbed/produced, gray lines illustrate the particles that belong to the thermal bath. Processes (a) and (b) require $s\geq(m_D+m_\Phi)^2$ and give rise to the unitary cut; processes (c) and (d) require $s\leq(m_D-m_\Phi)^2$ and give rise to the Landau cut.}
\label{fig:hot-interp}
\end{figure}

We consider the approximation of the loop function with two undressed mesons for simplicity. In this case, after Matsubara summation and analytical continuation to complex energies, the two-meson loop shown in Fig.~\ref{fig:hot-form-loop}, with a heavy meson $D$ and a light meson $\Phi$, is given in Eq.~(\ref{eq:hot-loopfree1}). The separation into explicit partial fractions is mandatory for a straightforward physical interpretation of each of the four terms appearing in this expression as the contribution of the processes sketched in Fig.~\ref{fig:hot-interp}. The imaginary part of the loop function, which is related to the discontinuity across the cuts of the loop function that extend along the real axis, reads
\begin{align}\label{eq:hot-imloop} \nonumber
 \textrm{Im\,}G_{D\Phi}(E,\vec{p}\,;T)&=-\pi\int\frac{d^3q}{(2\pi)^3}\frac{1}{4\omega_D\omega_\Phi} \\ \nonumber
 &\times\Big\{[1+f_D+f_\Phi]\,\delta(E-\omega_D-\omega_\Phi)-[1+f_D+f_\Phi]\,\delta(E+\omega_D+\omega_\Phi)\\ \nonumber 
 &\quad+[f_D-f_\Phi]\,\delta(E+\omega_D-\omega_\Phi)
 -[f_D-f_\Phi]\,\delta(E-\omega_D+\omega_\Phi)\Big\} \\ \nonumber
 &=-\pi\int\frac{d^3q}{(2\pi)^3}\frac{1}{4\omega_D\omega_\Phi} \\ \nonumber &\times\Big\{[(1+f_D)(1+f_\Phi)-f_Df_\Phi]\,\delta(E-\omega_D-\omega_\Phi)\\ \nonumber
 &\quad+[f_{\bar{D}}f_{\bar{\Phi}}-(1+f_{\bar{D}})(1+f_{\bar{\Phi}})]\,\delta(E+\omega_{D}+\omega_{\Phi})\\ \nonumber 
 &\quad+[f_{\bar{D}}(1+f_\Phi)-(1+f_{\bar{D}})f_\Phi]\,\delta(E+\omega_D-\omega_\Phi)\\ 
 &\quad+[(1+f_D)f_{\bar{\Phi}}-f_D(1+f_{\bar{\Phi}})]\,\delta(E-\omega_D+\omega_\Phi)\Big\} \ ,
\end{align}
where we have introduced the short notation $f_i\equiv f(\omega_i,T)$ for the \gls{be} distribution functions. In the second equality, we have added and subtracted the products $f_Df_\Phi$ so as to give a physical interpretation in terms of probabilities for the production and absorption of meson pairs from the bath. The factor $f_i$ weights the probability for absorption of a thermal meson of species $i$ from the bath, while the factor $1+f_i$ gives the probability for the thermal production of a meson $i$. When energy conservation requires $\omega_i<0$, as controlled by the corresponding $\delta$ function, we have identified the meson $i$ with an antimeson $\bar{i}$. Then, the first term in Eq.~(\ref{eq:hot-imloop}) is proportional to the probability for the creation of a pair from the bath, $\textit{bath}\rightarrow\textit{bath}+D\Phi$, with a statistical factor $(1+f_D)(1+f_\Phi)$, \textit{minus} the probability for the inverse process, $\textit{bath}+D\Phi\rightarrow\textit{bath}$, with weight $f_Df_\Phi$ for the absorption. This is depicted by the diagram in Fig.~\ref{fig:hot-interp-a}. Similarly, the second term gives the probability for $\textit{bath}+\bar{D}\bar{\Phi}\rightarrow\textit{bath}$, with weight $f_{\bar{D}}f_{\bar{\Phi}}$, \textit{minus} that for $\textit{bath}\rightarrow\textit{bath}+\bar{D}\bar{\Phi}$, with weight $(1+f_{\bar{D}})(1+f_{\bar{\Phi}})$, as shown in Fig.~\ref{fig:hot-interp-b}. The physical interpretation of the third and fourth terms as probabilities of emission and absorption processes is analogous, and the corresponding diagrams are shown in Figs.~\ref{fig:hot-interp-c} and \ref{fig:hot-interp-d}, respectively.

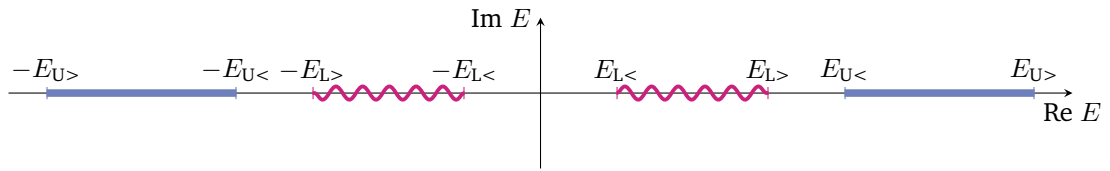
\begin{figure}[b!]\centering 
\begin{tikzpicture}[baseline=-2,
    decoration={%
      markings,
      mark=at position 0.01cm with {\arrow[line width=1pt,>=stealth]{>}},
    }
  ]
  \draw [>=stealth,->] (-7,0) -- (7,0) coordinate (xaxis);
  \draw [>=stealth,->] (0,-1) -- (0,1) coordinate (yaxis);
  \draw [|-|,ctcolorblue] (-4,0) -- (-6.5,0);
  \draw [line width=1mm,ctcolorblue] (-4,0) -- (-6.5,0);
  \draw [|-|,ctcolormagenta,decoration={snake},decorate] (-1,0) -- (-3,0);
  \draw [line width=0.5mm,ctcolormagenta,decoration={snake},decorate] (-1,0) -- (-3,0);
  \draw [|-|,ctcolorblue] (4,0) -- (6.5,0);
  \draw [line width=1mm,ctcolorblue] (4,0) -- (6.5,0);
  \draw [|-|,ctcolormagenta,decoration={snake},decorate] (1,0) -- (3,0);
  \draw [line width=0.5mm,ctcolormagenta,decoration={snake},decorate] (1,0) -- (3,0);
  \node [below] at (xaxis) {Re $E$};
  \node [left] at (yaxis) {Im $E$};
  \node at (-6.5,0.3) {$-E_{\textrm{U>}}$};
  \node at (-4,0.3) {$-E_{\textrm{U<}}$};
  \node at (-3,0.3) {$-E_{\textrm{L>}}$};
  \node at (-1,0.3) {$-E_{\textrm{L<}}$};
  \node at (1,0.3) {$E_{\textrm{L<}}$};
  \node at (3,0.3) {$E_{\textrm{L>}}$};
  \node at (4,0.3) {$E_{\textrm{U<}}$};
  \node at (6.5,0.3) {$E_{\textrm{U>}}$};
\end{tikzpicture}
\caption{Branch cuts of the two-meson thermal loop function in the complex-energy plane: the unitary cut in thick blue lines and the Landau cut in wiggly magenta lines, with the limits $E_{\textrm{L<}}$, $E_{\textrm{L>}}$, $E_{\textrm{U<}}$, and $E_{\textrm{U>}}$ given in Eq.~(\ref{eq:hot-branchlimits}).}
\label{fig:hot-branchcuts}
\end{figure}

The energy conservation imposed by the $\delta$ functions in each of the four terms in Eq.~(\ref{eq:hot-imloop}) implies that the imaginary part of the loop function is nonvanishing only for certain regions of the Mandelstam $s=E^2-\vec{p}\,^2$, giving rise to branch cuts in the loop function. The first two terms are nonzero for $s\geq(m_D+m_\Phi)^2$, giving the usual \textit{unitary cut} that is also present at $T=0$ above the $D\Phi$ threshold, while the third and fourth terms contribute for $s\leq (m_D-m_\Phi)^2$, providing the so-called \textit{Landau cut}. Figure~\ref{fig:hot-branchcuts} shows the location of these cuts in the complex $E$-plane. For positive external energies, the unitary cut arises from the scattering of $D\Phi$ taking place through their conversion into intermediate heavy-light pairs with the same quantum numbers (Fig.~\ref{fig:hot-interp-a}), while the annihilation and subsequent creation of a $\bar{D}\bar{\Phi}$ pair (Fig.~\ref{fig:hot-interp-b}) require negative external energies, $E\leq -(m_D+m_\Phi)$, which is not possible for real particles. The former processes are also allowed in the vacuum but, in the medium, the probabilities of their occurrence are modified by the \gls{be} distributions. On the other hand, the Landau cut arises from scattering processes with the mesons in the medium, in particular those in which the $D\Phi$ pair absorbs a heavy (light) antimeson, the remaining light (heavy) meson propagates in the loop, and finally, the $D\Phi$ pair and the heavy (light) antimeson are produced back, as illustrated in Fig.~\ref{fig:hot-interp-c} (Fig.~\ref{fig:hot-interp-d}). 

The limits of the branch cuts are determined by the constraints imposed by the $\delta$ functions for the conservation of energy and the requirement for the angular variable that $|x|\equiv |\cos\theta|\leq 1$, where $\theta$ is the angle between the external momentum $\vec{p}$ and the internal momentum in the loop $\vec{q}$ \cite{Ghosh:2010hap,Torres-Rincon:2017zbr}. When there is a finite momentum integration cut-off $\Lambda$ in the center-of-mass frame, the limits of the unitary (U) and Landau (L) cuts in Fig.~\ref{fig:hot-branchcuts} can be written as 
\begin{align}\label{eq:hot-branchlimits}\nonumber
   E_{\textrm{L}<}&=\sqrt{\Lambda^2+m_D^2}-\sqrt{\Lambda^2+m_\Phi^2} \ ,\qquad E_{\textrm{L}>}=\sqrt{(m_D-m_\Phi)^2+\vec{p}\,^2} \ , \\
E_{\textrm{U}<}&=\sqrt{(m_D+m_\Phi)^2+\vec{p}\,^2} \ ,\qquad E_{\textrm{U}>}=\sqrt{\Lambda^2+m_D^2}+\sqrt{(\Lambda+|\vec{p}|)^2+m_\Phi^2} \ ,
\end{align}
where $<,>$ denote the lower and upper limits, respectively, and we have defined $m_D>m_\Phi$  to avoid the crossing of the Landau cuts.

\subsection{Meson self-energy}
\label{subsec:hot-form-selfe}
\begin{figure}[b!]%
 \centering
  \begin{tikzpicture}[baseline=(i.base)]
    \begin{feynman}[small]
      \vertex (i);
      \vertex [above left = 1.5cm of i] (a) {\(D\)};
      \vertex [above right = 1.5cm of i] (c) {\(D\)};
      \vertex [below = 0.7cm of i] (d);
      \vertex [below = 1.4cm of i] (b);
      \diagram* {
        (a) -- [fermion] (i) -- [fermion] (c), 
       };
     \draw[dot,minimum size=1mm,thick,fill=black] (i) circle(1mm);
     \draw (d) circle (0.7cm);
     \node at (-1.6,0.8) {\(\ii\omega_n,\vec{q}\)};
     \node at (-1,-0.5) {\(\Phi\)};
     \node at (-2.3,-1.) {\(\ii\omega_m-\ii\omega_n,\vec{p}-\vec{q}\)};
     \node at (1.3,0) {\(T(\ii\omega_m,\vec{p}\,)\)};
    \end{feynman}
  \end{tikzpicture}
\caption{Diagram of the meson self-energy obtained from closing the light-meson line of the $T$ matrix.}
\label{fig:hot-form-selfE}
\end{figure}
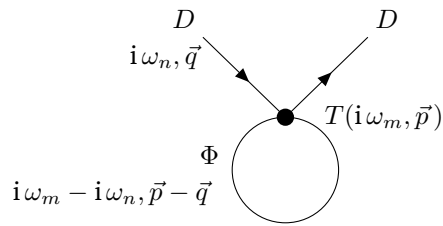

The self-energy of the heavy meson is obtained by closing the light-meson line in the corresponding $T$-matrix element and it is represented by the diagram depicted in Fig.~\ref{fig:hot-form-selfE}. After applying the Feynman rules, one obtains the following expression in the \gls{itf}:
\begin{equation}\label{eq:hot-selfE1}
 \Pi_D(\ii\omega_n,\vec{q}\,;T)=-\frac{1}{\beta}\int\frac{d^3q'}{(2\pi)^3}\sum_m\mathcal{D}_\Phi(\ii\omega_m-\ii\omega_n,\vec{q}\,^\prime)T_{D\Phi}(\ii\omega_m,\vec{p}\,) \ ,
\end{equation}
where $\ii\omega_n$ and $\ii\omega_m$ are the Matsubara frequencies of the external $D$ meson and internal $D\Phi$ system, respectively, and $\vec{q}\,^\prime=\vec{p}-\vec{q}$ is the three-momentum of the light meson.

It is convenient to use the Lehman representation for the light-meson propagator introduced in Eq.~(\ref{eq:hot-propLehmann}), as well as for the $T$ matrix,
\begin{equation}
 T_{D\Phi}(\ii\omega_m,\vec{p}\,)=-\frac{1}{\pi}\int dE \ \frac{\textrm{Im\,} T_{D\Phi}(E,\vec{p}\,)}{\ii\omega_m-E} \ .
\end{equation}
Following the same procedure as for the loop function described above, the expression obtained for the heavy-meson self-energy after the Matsubara summation reads
\begin{align}\label{eq:hot-selfE_Sfuncpi}
    \Pi_{D}(\ii\omega_n,\vec{q}\,;T)&=\frac{1}{\pi} \int\frac{d^3q'}{(2\pi)^3}\int dE \int d\omega
    \frac{S_\Phi(\omega,\vec{p}-\vec{q}\,)\left[f(E,T)-f(\omega,T)\right]}{E-\ii\omega_n-\omega}
    \textrm{Im\,}T_{D\Phi}(E,\vec{p}\,;T) \ .
\end{align}

Just like for the derivation of the thermal loop function, we consider that the interactions of a light meson with a dominantly pionic medium are weak. Given this approximation, the self-energy of the light meson vanishes and its spectral function becomes a delta function like that of Eq.~(\ref{eq:hot-deltaSfunc}). After introducing this in Eq.~(\ref{eq:hot-selfE_Sfuncpi}), the expression for the self-energy of the heavy meson is given by
\begin{align}\label{eq:hot-selfE_4terms}\nonumber
    \Pi_{D}(\ii\omega_n,\vec{q}\,;T)&= \frac{1}{\pi} \int\frac{d^3q'}{(2\pi)^3}\frac{1}{2\omega_\Phi}\int_0^\infty dE
    \Bigg\{\left[1+f(E,T)+f(\omega_\Phi,T)\right] \\ \nonumber
    &\times \left(\frac{1}{E-\ii\omega_n+\omega_\Phi}
    +\frac{1}{E+\ii\omega_n+\omega_\Phi}\right) \\ \nonumber
    &-\left[f(E,T)-f(\omega_\Phi,T)\right] \\
    &\times\left(\frac{1}{E-\ii\omega_n-\omega_\Phi}
    +\frac{1}{E+\ii\omega_n-\omega_\Phi}\right)\Bigg\}
    \textrm{Im\,}T_{D\Phi}(E,\vec{p}\,;T) \ ,
\end{align}
with $\omega_\Phi = \sqrt{q'^2+m_\Phi^2}$. 

Using that $\textrm{Im\,}T_{D\Phi}(-E,\vec{p}\,;T)=-\textrm{Im\,}T_{D\Phi}(E,\vec{p}\,;T)$, and after the analytical continuation $\ii\omega_n\rightarrow \omega+\ii\varepsilon$, the expression can be compactified in the following,
\begin{equation}\label{eq:hot-selfE}
    \Pi_{D}(\omega,\vec{q}\,;T)=\frac{1}{\pi} \int\frac{d^3q'}{(2\pi)^3}\int dE\,\frac{\omega-\omega_\Phi}{\omega_\Phi}\frac{f(E,T)-f(\omega_\Phi,T)}{\omega^2-(\omega_\Phi-E)^2+\textrm{sgn}(\omega)\, \ii\varepsilon } \ \textrm{Im\,}T_{D\Phi}(E,\vec{p}\,;T) \ .
\end{equation}
 
The details of the numerical integration of Eq.~(\ref{eq:hot-selfE}) are given in Appendix~\ref{sec:app-selfe}.

The imaginary part of the self-energy is related to the thermal width acquired by the heavy meson due to the interaction with the light mesons of the medium, while the real part is connected to the thermal modification of the mass. This can be seen easily by considering the expression for the heavy-meson retarded propagator at finite temperature,
\begin{equation}\label{eq:hot-Dmesonprop}
 \mathcal{D}_M(\omega,\vec{q}\,;T)=\frac{1}{\omega^2-\vec{q}^2-m_D^2-\textrm{Re\,}\Pi_D(\omega,\vec{q}\,;T)-\ii\textrm{Im\,}\Pi_D(\omega,\vec{q}\,;T)} \ ,
\end{equation}
where $m_D$ is the vacuum $D$-meson mass in the vacuum, renormalized by the vacuum contribution of the retarded $D$-meson self-energy $\Pi_D$. Therefore, after mass renormalization, the real and imaginary parts of the self-energy can only contain thermal corrections. However, the real part of the self-energy computed with the expressions given above contain both vacuum and thermal corrections. To remove the vacuum part to the real part of $\Pi$ entering the spectral function in Eq.~(\ref{eq:hot-specfunc}), the authors of Ref.~\cite{Cleven:2017fun} pointed out that the real part of the self-energy calculated at $T=0$ has to be subtracted from the corresponding calculation at finite temperature, at each value of $\{\omega,\vec{q}\,\}$. A different approach, which is the one taken here, consists in calculating only the thermal contribution to the real part of the self-energy by dropping the $1$ in the factor $\left[1+f(E,T)+f(\omega_\Phi,T)\right]$ in Eq.~(\ref{eq:hot-selfE_4terms}) (or in the definition of $\mathcal{F}_1$ and $\mathcal{F}_2$ in Eq.~(\ref{eq:hot-selfE_F}) in the appendix). This is the part that survives in the limit $T\rightarrow 0$, in which the \gls{be} factors vanish, $f(\omega,T\rightarrow 0)=0$, and hence this contribution has to be dismissed. 

\vspace{-1mm}
\subsubsection{Self-energy in isospin basis}

The expressions above provide the contribution to the heavy-meson self-energy of a particular species present in the thermal bath of light mesons, $\Phi=\{\pi,K,\bar{K},\eta\}$. The total self-energy is then given by the sum of all the light-meson contributions:
\begin{equation}
 \Pi_D(\omega,\vec{q}\,;T)=\sum_{\Phi=\{\pi,K,\bar{K},\eta\}}\Pi_D^{(\Phi)}(\omega,\vec{q}\,;T) \ ,
\end{equation}
where each of the contributions is computed from the $T$ matrix in the isospin basis for convenience. In the charge basis, to obtain the contribution to the self-energy of a heavy meson with isospin $i_1$ and third component $i_{1_z}$ (for example, a $D^+$ meson with $(i_1,i_{1_z})=(1/2,+1/2)$) coming from a light meson with isospin $i_2$ (for instance, a $\pi$ with $i_2=1$), one has to sum over all the values of $i_{2_z}$ (that is, $\pi^+,\pi^0,\pi^-$),
\begin{equation}
 \Pi_{D(i_1,i_{1_z})}^{(\Phi(i_2))}=\sum_{i_{2_z}}  \Pi_{D(i_1,i_{1_z})}^{(\Phi(i_2,i_{2_z}))} \ .
\end{equation}
Given that the self-energy has to be the same for all the $i_{1_z}$ 
(that is, $\Pi_{D^+}=\Pi_{D^0}$), we take the average over the values of the heavy-meson isospin,
\begin{equation}
 \Pi_{D(i_1)}^{(\Phi(i_2))}=\frac{1}{(2i_1+1)}\sum_{i_{1_z}}\sum_{i_{2_z}}\Pi_{D(i_1,i_{1_z})}^{(\Phi(i_2,i_{2_z}))} \ .
\end{equation}

In the isospin basis, it becomes
\begin{equation}
 \Pi_D^{(\Phi)}=\frac{1}{(2i_1+1)}\sum_I(2I+1)\Pi_D^{(\Phi)}(I) \ ,
\end{equation}
where $I$ is the total isospin of the $D\Phi$ system, and $\Pi_D^{\Phi}(I)$ is calculated from closing the $\Phi$ meson line in the corresponding $T_{D\Phi}^{I}$ matrix element, with isospin $I$.
This provides, for instance, the pionic and kaonic contributions to the $D$- and $D_s$-meson self-energies:
\begin{equation}\label{eq:hot-selfE-isospin}
\def\arraystretch{1.2}
\begin{array}{l}
 \Pi_D^{(\pi)}=\Pi_D^{(\pi)}(I=1/2)+2\Pi_D^{(\pi)}(I=3/2) \ , \\ 
 \Pi_D^{(K)}=\frac12\Pi_D^{(K)}(I=0)+\frac32\Pi_D^{(K)}(I=1) \ , \\ 
 \Pi_D^{(\bar{K})}=\frac12\Pi_D^{(\bar{K})}(I=0)+\frac32\Pi_D^{(\bar{K})}(I=1) \ , \\ 
 \Pi_{D_s}^{(\pi)}=3\Pi_{D_s}^{(\pi)}(I=1) \ , \\ 
 \Pi_{D_s}^{(K)}=2\Pi_{D_s}^{(K)}(I=1/2) \ , \\ 
 \Pi_{D_s}^{(\bar{K})}=2\Pi_{D_s}^{(\bar{K})}(I=1/2) \ .
\end{array}
\end{equation}

For temperatures below the critical temperature for the deconfining transition, $T_c$, the largest contribution to the $D$-meson self-energy comes from pions, as the abundance of heavier light mesons, such as kaons and eta mesons, is suppressed by the \gls{be} factors. We note that the contribution of a kaonic bath can be relevant for temperatures close to $T_c$.  In order to study this contribution, in Section~\ref{sec:hot-results} we analyze the modification of open charm mesons in the presence of a kaonic bath by taking into account the corresponding \gls{be} distributions.

\subsection{Unitarized interactions and self-consistency at finite temperature}
\label{sec:hot-form-unitarization}

At $T < T_c$, and assuming no baryon density, the thermal medium is essentially composed of the lighter mesons of the pseudoscalar meson octet. Their interactions at low energies are governed by \gls{chpt}, based on chiral power counting. The heavy mesons, $D^{(*)}$ and $D_s^{(*)}$ (also the $\bar{B}^{(*)}$ and $\bar{B}_s^{(*)}$), propagate through this medium behaving as Brownian particles, suffering from collisions with any of the light mesons. The interaction of the $D$ and $\bar{B}$ mesons with light particles is described by the effective Lagrangian given in Eqs.~(\ref{eq:free-mm-LagDpiLO}) and (\ref{eq:free-mm-lagrangianNLO}) when describing the interaction of open-heavy flavor mesons with light mesons in free space in Chapter~\ref{ch:exoticsinfreespace}. The tree-level scattering amplitude that follows from this Lagrangian is given in Eq.~(\ref{eq:free-mm-potential}).

This amplitude is used as the kernel of an on-shell \gls{bs} equation within a full coupled-channel basis,  $T=V+VGT$, where $T$ is the unitarized amplitude. 
At finite temperature, we need to account for several modifications, both in the methodology and in the final analysis of the dynamically generated states. 
The two-body loop function $G$ is now modified as compared to the vacuum case since it receives medium corrections due to the light meson gas, and the meson masses are dressed by the medium, as given in Eq.~(\ref{eq:hot-loop-compact}). 
As discussed in Section~\ref{subsec:free-mm-formalism}, the vacuum contribution of the loop function needs regularization, for which we employ a cut-off scheme, a procedure which simplifies the numerical treatment at finite temperature, detailed in Appendix~\ref{sec:app-loop}. The selected value of the UV cut-off, $\Lambda=800$~MeV, has been discussed for the scattering formalism at $T=0$ and is consistent with the regularization scheme in \cite{Guo:2018tjx}. The thermal effects in the unitarized scattering amplitudes are obtained by solving the \gls{bs} equation with thermal loops. In the \gls{itf}, as the thermal corrections enter in loop diagrams~\cite{lebellac,kapustagale}, the tree-level scattering amplitudes $V$ remain the same as in vacuum, with the zeroth component of the four-momentum expressed as a bosonic Matsubara frequency.

In \gls{chpt}, the pion mass and decay constant do not appreciably change with temperature up to two-loops and even in unitary extensions of it~\cite{Schenk:1993ru,Toublan:1997rr}. In addition, the pion damping rate is very much suppressed at the temperatures explored in this dissertation, so we use the pion vacuum spectral function for all temperatures. See the discussion in Appendix~\ref{appendix-modpion}.

The spectral function of the heavy meson is computed from the imaginary part of its retarded propagator as given in Eq.~(\ref{eq:hot-specfunc}). 
The light-meson contribution to the heavy-meson self-energy is computed in the \gls{itf} using Eq.~(\ref{eq:hot-selfE}). As we have discussed in Section~\ref{subsec:hot-form-selfe}, the largest contribution corresponds to that of the thermal pions, as at the temperatures considered, that is, at $T\leq 150$~MeV, these are the most abundant mesons. Unless otherwise stated, in the calculations we only consider the thermal effects due to pions at finite temperature while neglecting other possible medium modifications. 

\begin{figure}[b!]
\centering 
\begin{subfigure}[b]{\textwidth}\centering 
\captionsetup{skip=0pt}
 \begin{tikzpicture}[baseline=(i.base)]
    \begin{feynman}[small]
      \vertex (i) ;
      \vertex [above left = of i] (a) {\(D_i\)};
      \vertex [above right = of i] (b) {\(D_j\)};
      \vertex [below right = of i] (d) {\(\Phi_j\)};
      \vertex [below left = of i] (c) {\(\Phi_i\)};
      \diagram* {
        (a) -- [fermion] (i), 
        (i) -- [fermion] (b),
        (c) -- [charged scalar] (i),
        (i) -- [charged scalar] (d),
       };
     \draw[dot,minimum size=4mm,thick,ctcolormagenta,fill=ctcolorgreen] (i) circle(1.5mm);
    \end{feynman}
  \end{tikzpicture}
  $=$
  \begin{tikzpicture}[baseline=(i.base)]
    \begin{feynman}[small]
      \vertex (i) ;
      \vertex [above left = of i] (a) {\(D_i\)};
      \vertex [above right = of i] (b) {\(D_j\)};
      \vertex [below right = of i] (d) {\(\Phi_j\)};
      \vertex [below left=of i] (c) {\(\Phi_i\)};
      \diagram*{
        (a) -- [fermion] (i), 
        (i) -- [fermion] (b),
        (c) -- [charged scalar] (i),
        (i) -- [charged scalar] (d),
       } ;    
     \draw[dot,ctcolorblue,fill=ctcolorblue] (i) circle(.8mm);
    \end{feynman}
  \end{tikzpicture}
  $+$
  \begin{tikzpicture}[baseline=(i.base)]
    \begin{feynman}[small]
      \vertex (i) ;
      \vertex [above left = of i] (a) {\(D_i\)};
      \vertex [right = of i] (j);
      \vertex [above right = of j] (b) {\(D_j\)};
      \vertex [below right = of j] (d) {\(\Phi_j\)};
      \vertex [below left=of i] (c) {\(\Phi_i\)};
      \vertex [above right =0.5cm of i] (b1) {\(D_k\)};
      \vertex [below right =0.5cm of i] (d1) {\(\Phi_k\)};
      \diagram*{
        (a) -- [fermion] (i), 
        (j) -- [fermion] (b),
        (i) -- [ctcolormagenta, fermion, very thick, half left, looseness=1.2] (j),
        (i) -- [charged scalar, half right, looseness=1.2] (j),
        (c) -- [charged scalar] (i),
        (j) -- [charged scalar] (d),
       } ;    
     \draw[dot,ctcolorblue,fill=ctcolorblue] (i) circle(0.8mm);
     \draw[dot,minimum size=4mm,thick,ctcolormagenta,fill=ctcolorgreen] (j) circle(1.5mm);
    \end{feynman}
  \end{tikzpicture}
\caption{}
\label{fig:hot-selfcons-a}
\end{subfigure}
\\[0.2cm]
\hspace{-1cm}\begin{subfigure}[b]{0.7\textwidth}\centering 
\captionsetup{skip=0pt}
 \begin{tikzpicture}[baseline=(a.base)]
    \begin{feynman}[small]
      \vertex (a);
      \vertex [right = of a] (b);
      \diagram* {
        (a) -- [ctcolormagenta, fermion, very thick,edge label=\(\textcolor{black}{D}\)] (b), 
       };
    \end{feynman}
  \end{tikzpicture}
  $=$
 \begin{tikzpicture}[baseline=(a.base)]
    \begin{feynman}[small]
      \vertex (a);
      \vertex [right = of a] (b);
      \diagram* {
        (a) -- [fermion, edge label=\(D\)] (b), 
       };
    \end{feynman}
  \end{tikzpicture}
  $+$
  \begin{tikzpicture}[baseline=(a.base)]
    \begin{feynman}[small, inline=(a)]
      \vertex (i) ;
      \vertex [left = of i] (a);
      \vertex [right = of i] (b);
      \vertex [below = 0.4cm of i] (d);
      \vertex [below = 0.9cm of i] (e) {\(\pi\)};
      \diagram* {
        (a) -- [fermion,edge label=\(D\)] (i), 
        (i) -- [ctcolormagenta, fermion, very thick,edge label=\(\textcolor{black}{D}\)] (b),
       } ;   
     \draw[dashed] (d) circle(0.3cm);
     \draw[dot,minimum size=4mm,thick,ctcolormagenta,fill=ctcolorgreen] (i) circle(1.5mm);
    \end{feynman}
  \end{tikzpicture}
\caption{}
\label{fig:hot-selfcons-b}
\end{subfigure}\hspace{-1cm}
\begin{subfigure}[b]{0.25\textwidth}\centering
\captionsetup{skip=5pt}
  \begin{tikzpicture}
    \begin{feynman}[small]
      \vertex (i) ;
      \vertex [above left = of i] (a) {\(D\)};
      \vertex [above right = of i] (b) {\(D\)};
      \vertex [below = 0.4cm of i] (d);
      \vertex [below = 1cm of i] (e) {\(\pi\)};
      \diagram* {
        (a) -- [fermion] (i), 
        (i) -- [fermion] (b),
       } ;   
     \draw[dashed] (d) circle(0.4cm);
     \draw[dot,minimum size=4mm,thick,ctcolormagenta,fill=ctcolorgreen] (i) circle(1.5mm);
    \end{feynman}
  \end{tikzpicture}
\caption{}
\label{fig:hot-selfcons-c}
\end{subfigure}
\caption{(a) The \gls{bs} equation in coupled channels. At finite temperature, the $T$ matrix (green big circle) is obtained from the unitarization of the interaction kernel (small blue circle) with dressed internal heavy-meson propagators (thick magenta solid lines).  (b) Dyson equation for the dressed heavy-meson propagator. (c) Heavy-meson self-energy. The heavy meson is dressed by the unitarized interaction with pions.}
\end{figure}
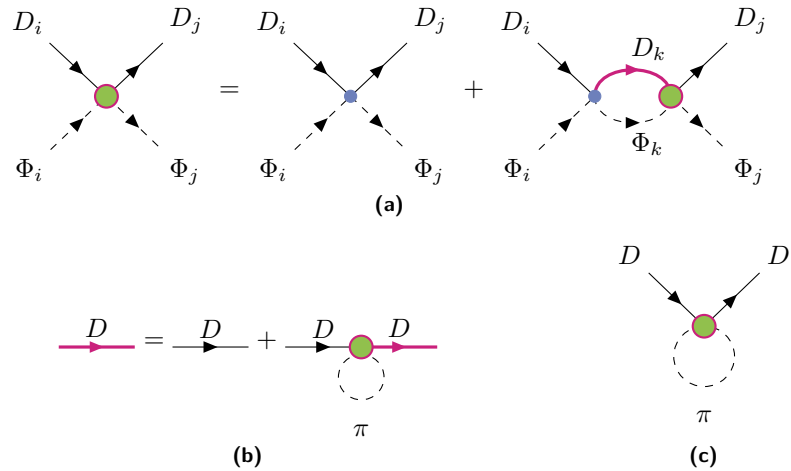

The \gls{bs} equation and the equations for the thermal two-meson loop function, the spectral function, and the self-energy of the heavy meson constitute a set of equations that are coupled to each other. Therefore, this set of equations has to be solved iteratively until self-consistency is reached. The procedure is sketched in Figs.~\ref{fig:hot-selfcons-a}, \ref{fig:hot-selfcons-b} and \ref{fig:hot-selfcons-c}. The $T$-matrix amplitude is represented by a big green circle, whereas the perturbative amplitude $V(s)$ is denoted by a small blue circle. Figure~\ref{fig:hot-selfcons-a} shows the \gls{bs} equation for the two-body scattering. The intermediate propagator of the $D$-meson (thick magenta solid line) is itself dressed by interactions as shown in Fig.~\ref{fig:hot-selfcons-b}, where the $T$ matrix is used in the Dyson equation for the propagator, giving rise to a self-consistent set of equations. As illustrated in Fig.~\ref{fig:hot-selfcons-c}, only the pion contribution, as being the dominant excitation in the thermal bath, is considered in the $D$-meson self-energy.

We recall that for the calculation of the first iteration of the loop function one needs to make use of the approximation with two free (undressed) mesons while, after the calculation of the heavy-meson self-energy, the approximation with a dressed heavy meson can be used for the next iterations, until the convergence of the results is reached. The details of the corresponding approximations in the calculation of the loop function are given in Section~\ref{subsec:hot-form-loop}.

\section{Results}
\label{sec:hot-results}

We conclude this chapter with the presentation and analysis of the results for the thermal effects on the interaction of the open-charm mesons with the light pseudoscalar mesons, together with the consequences for the ground-state spectral functions and the dynamically-generated states at finite temperature. In particular, we describe the thermal dependence of the masses and widths for open-charm ground and excited states for temperatures below both the chiral restoration temperature $T_\chi=156$ MeV~\cite{Aoki:2006we} and the critical temperature for the deconfinement transition $T_c\sim 154$ MeV \cite{Borsanyi:2010bp,Bazavov:2011nk}. The restriction $T<T_\chi\sim T_c$ is evident from the range of validity of an approach based on an effective theory with hadronic degrees of freedom and massive Goldstone bosons. Therefore, we present our results for $T\leq150$~MeV, but one should take the results around the largest value with caution, as the deconfined phase will start to play a role in the system. Furthermore, although the unitarized version of \gls{chpt} extends the validity of this low-energy theory to higher energies, for temperatures above $T=150$~MeV, the thermal energies of the mesons might start lying outside the validity of the theory, as we discussed in Ref.~\cite{Montana:2020lfi}.

\subsection{Results for $D$ mesons}
\label{subsec:hot-results-Dmesons}

\subsubsection{Thermal loops and scattering amplitudes}
Let us start with the discussion of the self-consistent results for the two-meson propagators and the scattering amplitudes in a pionic medium at finite temperature. The in-medium results shown in this section correspond to a medium of thermal pions in the sense that self-consistency is achieved accounting only for the pionic contribution to the heavy-meson self-energies. Nevertheless, the full basis of coupled-channels is considered for the unitarization of the scattering amplitudes.

  \begin{figure}[tb!]
   \centering
   \includegraphics[width=0.88\textwidth]{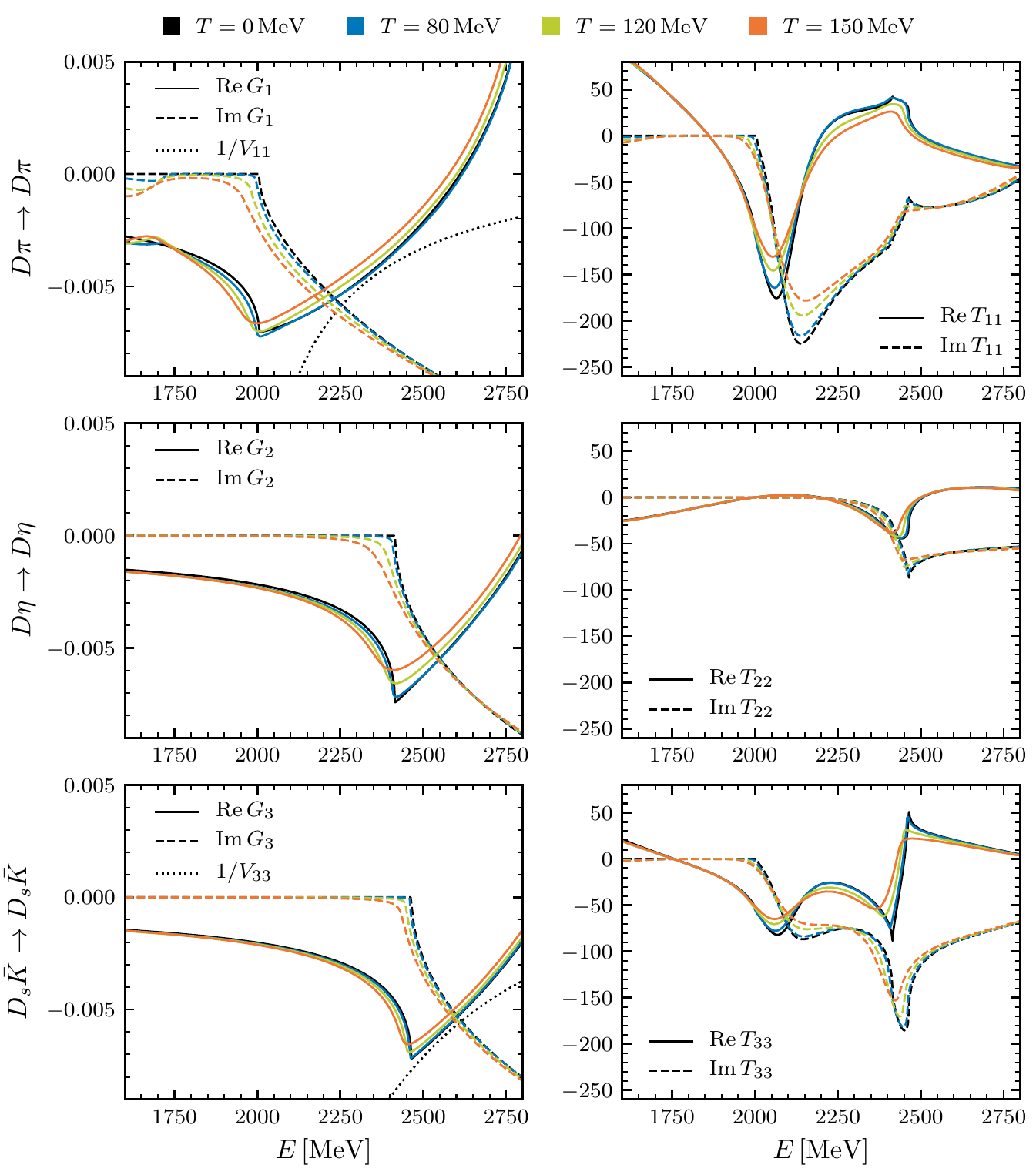}
   \caption{The inverse of the interaction kernel, $1/V_{ii}$, the real and imaginary parts of the loop function, $G_i$, and the real and imaginary parts of the diagonal components of the $T$ matrix, $T_{ii}$, in units of MeV$^0$, in the sector with spin $J=0$ and strangeness and isospin $(S,I)=(0,1/2)$, at various temperatures (colored lines). The subindices $1$, $2$, $3$ refer to the channels $D\pi$, $D\eta$ and $D_s\bar{K}$, respectively.}
   \label{fig:hot-VGT_0_05}
   \end{figure}
   
     \begin{figure}[t!]
    \centering
   \includegraphics[width=0.88\textwidth]{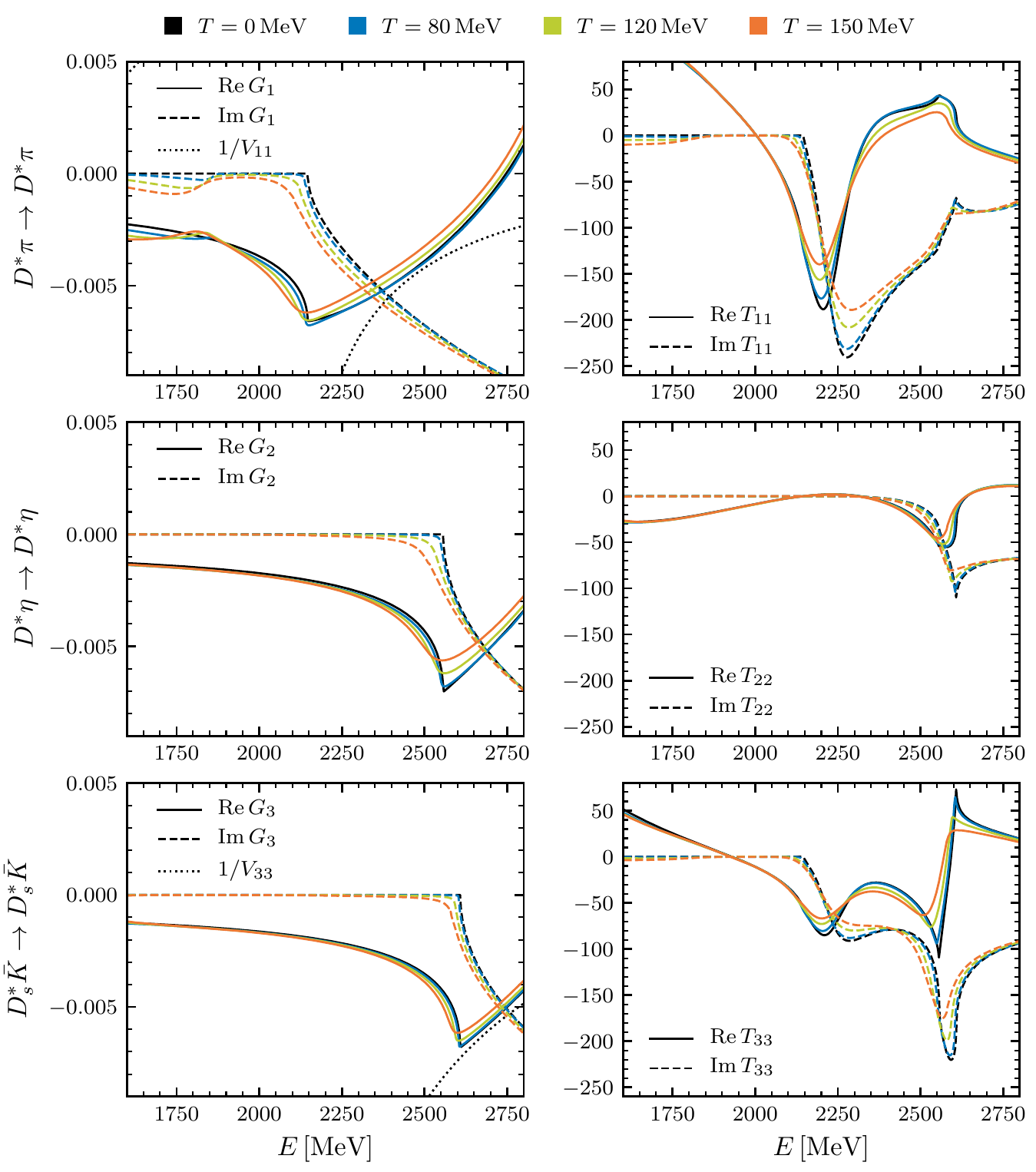}
   \caption{The same as in Fig.~\ref{fig:hot-VGT_0_05} in the sector with $J=1$ and $(S,I)=(0,1/2)$. The subindices $1$, $2$, $3$ refer to the channels $D^*\pi$, $D^*\eta$ and $D_s^*\bar{K}$, respectively.}
   \label{fig:hot-VGT_0_05_vec}
   \end{figure}
   
   \begin{figure}[t!]
   \centering
   \includegraphics[width=0.88\textwidth]{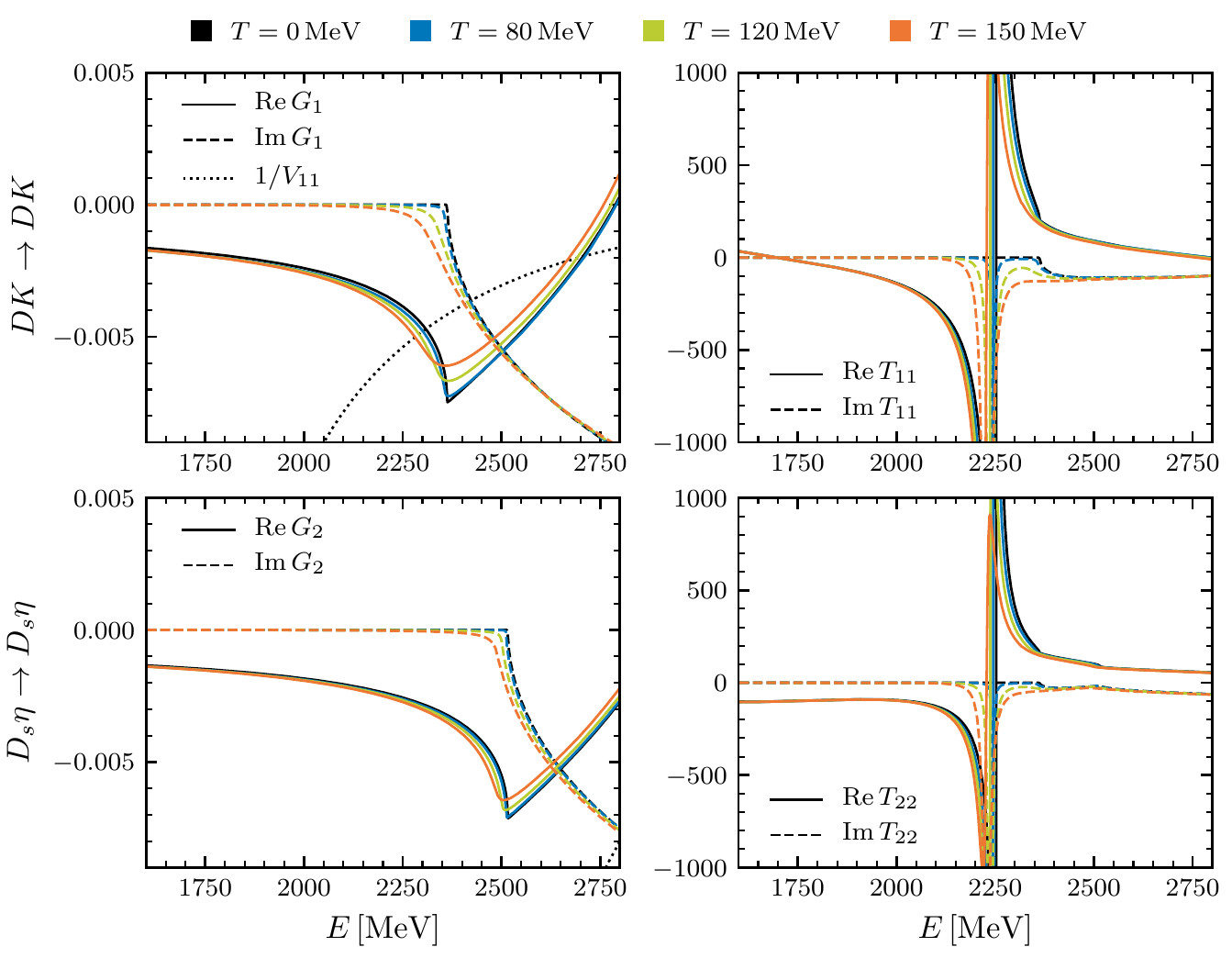}
   \caption{The same as in Fig.~\ref{fig:hot-VGT_0_05} in the sector with $J=0$ and $(S,I)=(1,0)$. The subindices $1$, $2$ refer to the channels $D_s\pi$ and $DK$, respectively.}
   \label{fig:hot-VGT_1_0}
   \end{figure} 
   
     \begin{figure}[b!]
     \centering
   \includegraphics[width=0.88\textwidth]{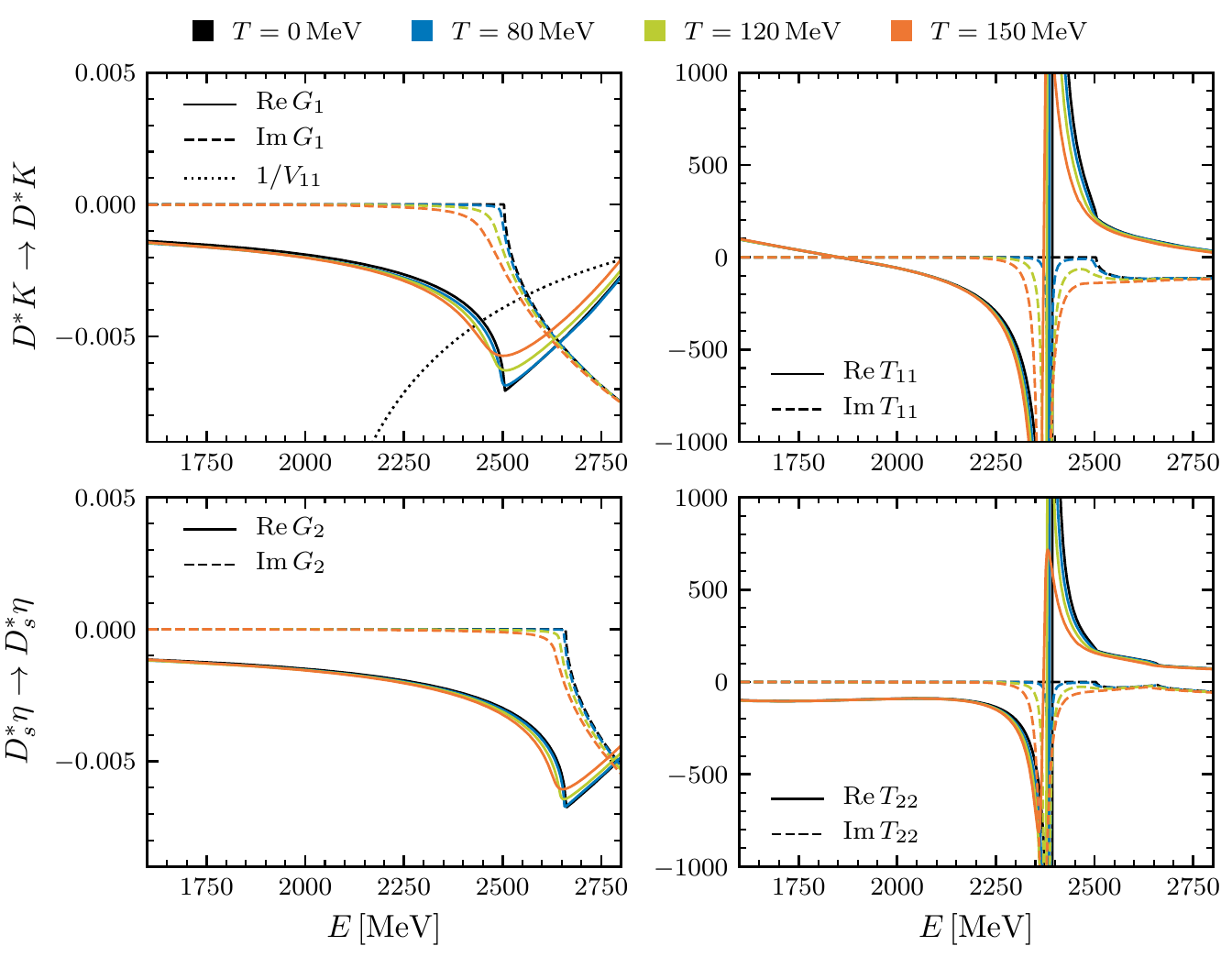}
   \caption{The same as in Fig.~\ref{fig:hot-VGT_0_05} in the sector with $J=1$ and $(S,I)=(1,0)$. The subindices $1$, $2$ refer to the channels $D_s^*\pi$ and $D^*K$, respectively.}
   \label{fig:hot-VGT_1_0_vec}
   \end{figure}

The diagonal amplitudes in the strangeness $S=0$ and isospin $I=1/2$ sector are represented on the right panels of Figs.~\ref{fig:hot-VGT_0_05} and \ref{fig:hot-VGT_0_05_vec} for the interaction of the pseudoscalar meson octet with pseudoscalar and vector charmed mesons, respectively, as functions of the total energy and for a center-of-mass momentum $\vec{P}=0$ and various temperatures (colored lines). 

The corresponding strangeness $S=1$ and isospin $I=0$ scattering amplitudes are shown on the right panels of Figs.~\ref{fig:hot-VGT_1_0} and \ref{fig:hot-VGT_1_0_vec}.

The energy dependence of the loop function $G$ is displayed in the panels on the left of the same figures, together with the inverse of the diagonal element of the interaction kernel, $1/V_{ii}$ (dotted lines), if it falls within the vertical scale employed in the subplots, as its crossing (or proximity, in the case of coupled-channels) with the real part of the loop function (solid lines) gives rise to a pole in the unitarized amplitudes. As mentioned above, in the \gls{itf} only the propagators contain medium corrections, and the interaction potentials remain the same as in the vacuum. In addition to the right-hand unitary cut that starts at $m_D+m_\Phi$ and is present also at $T=0$, a left-hand cut starting at $m_D-m_\Phi$ opens up in the imaginary part of the loop function (dashed lines) at finite temperature. This is the Landau cut arising due to the presence of thermal mesons. Its origin has been discussed in Section~\ref{subsec:hot-form-Landaucut}. For the energy range displayed in the figures above, this cut can only be seen in the top panels, corresponding to the $D\pi$ and $D^*\pi$ channels, and it is more visible as the temperature increases. This is related to the larger abundance of thermal pions compared to the heavier light mesons, that is, kaons and eta mesons, thus confirming our assumption for neglecting the contribution of the latter to the charmed meson self-energy in the self-consistent calculations leading to the results presented in this section.

Furthermore, the structure of the thermal loop function is smoothened in comparison to the vacuum case, both the real and imaginary parts. This smoothening follows from the dressing of the charmed meson in the loop with its spectral function, which acquires a substantial width with increasing temperatures, especially for the nonstrange $D$ and $D^*$ mesons, but also for the strange $D_s$ and $D_s^*$ mesons, as shown below. Besides, as the meson \gls{be} distributions in the intermediate meson-meson propagators extend over higher momenta for larger temperatures, the sharp meson-meson thresholds are diluted and the strength of the loop functions is smoothened out.


Since the interaction kernel is not modified at finite temperature, the changes in the unitarized amplitudes reflect the changes of the meson-meson propagators. Indeed, the inclusion of temperature on the real and imaginary parts of the different meson-meson loop functions results in a smoothening of the real and imaginary parts of the scattering matrices in all the $(S,I)$ sectors, for both pseudoscalar and vectors mesons. As a consequence of the broadening of the charmed-meson spectral functions and the dissolution of the meson-meson thresholds with increasing temperatures, the corresponding scattering amplitudes are smeared out while spreading over a wider energy range. Physically, this can be understood as larger temperatures result in larger available phase space for decay.

We have already discussed in Chapter~\ref{ch:exoticsinfreespace} that the phenomenology at $T=0$ of the interaction of the light mesons with the charmed pseudoscalar and vector mesons is very similar due to \gls{hqss}, and only mild differences are apparent in the structure of the loops and the scattering amplitudes due to the shift towards higher energies related to the mass difference between the $D$ and the $D^*$. Also, the thermal effects on the loop functions and the unitarized amplitudes are comparable for both $J=0$ and $J=1$ sectors, as is clear from the comparison of Figs.~\ref{fig:hot-VGT_0_05} and \ref{fig:hot-VGT_0_05_vec}, as well as Figs.~\ref{fig:hot-VGT_1_0} and \ref{fig:hot-VGT_1_0_vec}.

\subsubsection{Self-energies and open-charm spectral functions}

\begin{figure}[b!]
  \centering
   \includegraphics[width=\textwidth]{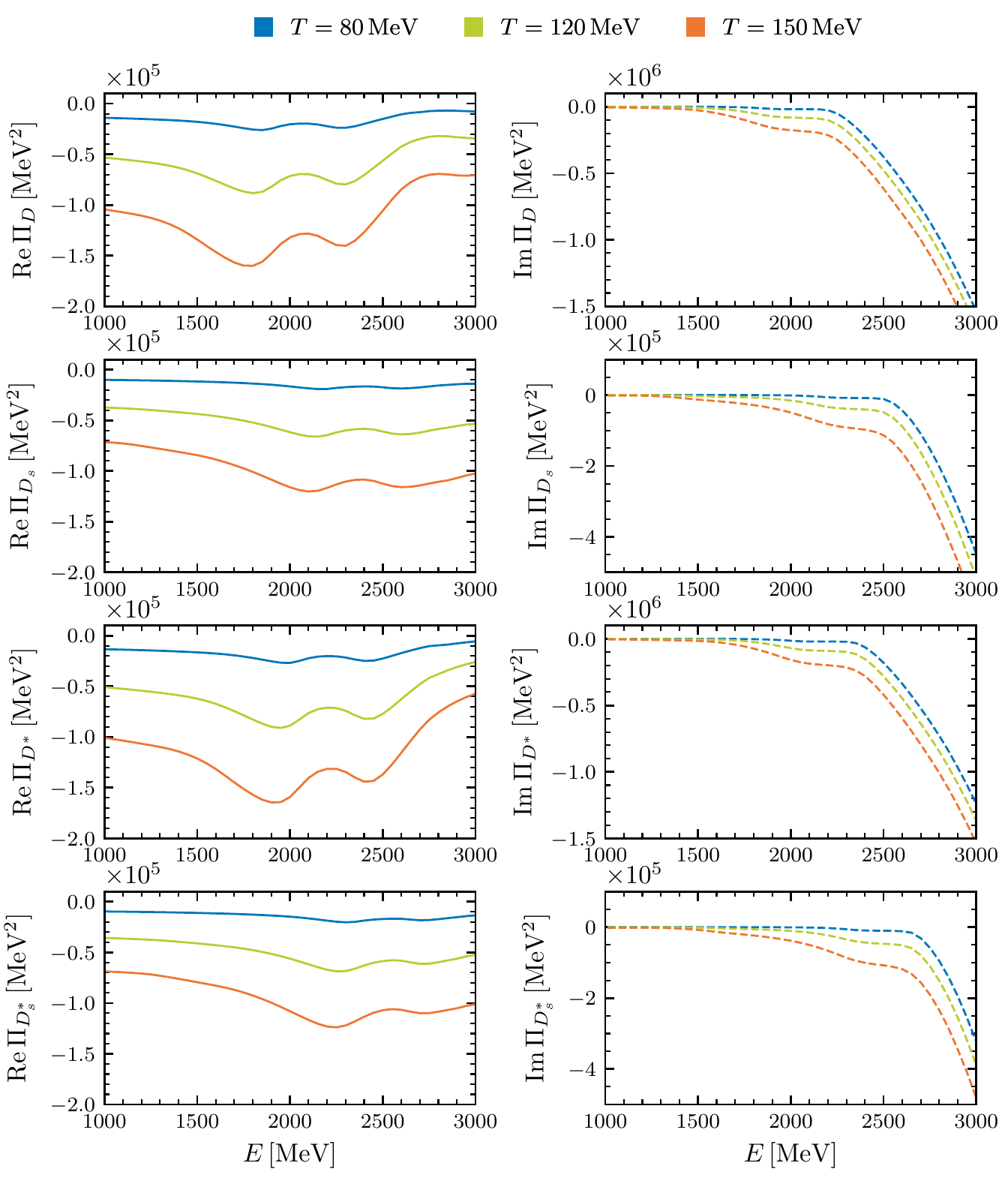}
   \caption{Real and imaginary parts of the pionic contribution to self-energies of the $D$ (first row), the $D_s$ (second row), the $D^*$ (third row), and the $D_s^*$ (fourth row) mesons at various temperatures (colored lines). }
   \label{fig:hot-selfe_charm}
   \end{figure}

In the following, we discuss the results obtained for the self-energies of the pseudoscalar $D$ and $D_s$, and the vector $D^*$ and $D_s^*$ mesons, which are displayed in Fig.~\ref{fig:hot-selfe_charm}, as a function of the energy, for $\vec{q}=\vec{0}\,$, and for different temperatures (colored lines). These results have been obtained considering only the pionic contribution to the heavy-meson self-energy, which is the main one at temperatures below $T_\chi$, but taking into account the interaction of $D^{(*)}$ and $D_s^{(*)}$ mesons with pions in all isospin channels, as given by the first and fourth rows of Eq.~(\ref{eq:hot-selfE-isospin}), respectively. 

The spectral functions of the $D^{(*)}$ and $D_s^{(*)}$ mesons follow the standard definition in terms of the retarded propagator, see Eq.~(\ref{eq:hot-specfunc}). 
In terms of the real and imaginary parts of the heavy-meson self-energy, it reads
\begin{equation}\label{eq:hot-sfunc_reimPi}
 S_D(\omega,\vec{q}\,;T)=-\frac{1}{\pi}\frac{\textrm{Im\,}\Pi_D(\omega,\vec{q}\,;T)}{\left[\omega^2-\vec{q}\,^2-m_D^2-\textrm{Re\,}\Pi_D(\omega,\vec{q}\,;T)\right]^2+\left[\textrm{Im\,}\Pi_D(\omega,\vec{q}\,;T)\right]^2} \ ,
\end{equation}
with the subindex $D$ denoting any of the charmed particles, that is, $D$, $D_s$, $D^*$, or $D_s^*$.

If the spectral function is narrow, namely in the so-called quasiparticle approximation,
the temperature-dependent quasiparticle energy, $\omega_{\textrm{qp}}$, and decay width, $\Gamma_{\textrm{qp}}$, can be defined as
\begin{equation}\label{eq:hot-quasipart-energy}
\omega_{\textrm{qp}}^2-\vec{q}\,^2-m_D^2-\textrm{Re}\,\Pi_D(\omega_{\textrm{qp}},\vec{q}\,;T)=0 \ ,
\end{equation}
and
\begin{equation}\label{eq:hot-quasipart-width}
 \Gamma_{\textrm{qp}}=-\frac{\textrm{Im}\,\Pi_D(\omega_{\textrm{qp}},\vec{q}\,;T)}{\omega_{\textrm{qp}}} \ .
\end{equation}

The real part of the self-energy shown in the left panels of Fig.~\ref{fig:hot-selfe_charm} contains only the thermal contribution, and therefore it is related to the thermal correction to the in-medium mass of the charmed meson. If the pole of the retarded meson propagator is not far from the vacuum one, one can see from the definition of the quasiparticle energy in Eq.~(\ref{eq:hot-quasipart-energy}) that, for $\vec{q}=\vec{0}\,$, the mass shift is approximately given by $\Delta m_D\approx {\textrm{Re}}\,\Pi_D(m_D,\vec{0}\,;T)/(2m_D)$. With this approximation, it is easy to see that the masses of the charmed mesons will move towards lower energies with increasing temperatures, due to the negative character of the real part of the self-energy. In addition, due to the large attractive interaction in the $D^{(*)}\pi$ channel, the mass shift of the nonstrange $D^{(*)}$ meson (first and third rows) will be larger than for the strange $D_s^{(*)}$ (second and fourth rows). The results for $J^P=0^-$ and $J^P=1^-$ are expected to be of comparable size. The authors of Ref.~\cite{Cleven:2017fun} neglected the shift of the in-medium mass of the heavy mesons by setting to zero the real part of the respective self-energies. Although small compared to the vacuum mass, $|\Delta m_D|/m_D\sim 5\%$, we consider that it is still important and keep the full self-energy for the calculation of the spectral function.


 \begin{figure}[t!]
  \centering
   \includegraphics[width=0.9\textwidth]{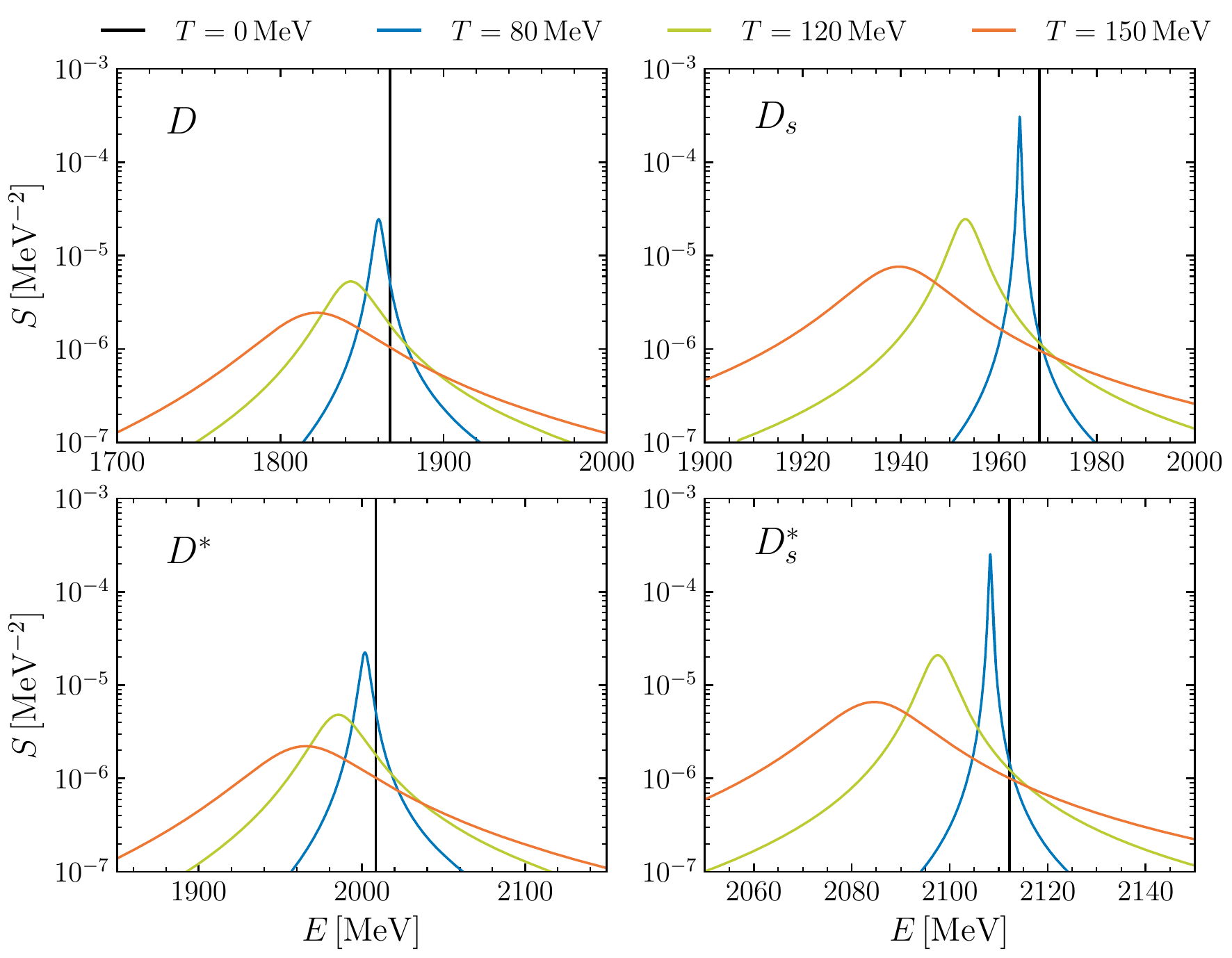}
   \caption{Spectral functions of the $J=0$ ground states ($D$ and $D_s$, top panels) and the $J=1$ ground states ($D^*$ and $D_s^*$, bottom panels) in a pionic bath at different temperatures (colored lines).}
   \label{fig:hot-spectralfunction_charm}
   \end{figure}
   
The right panels of Fig.~\ref{fig:hot-selfe_charm} display the imaginary part of the self-energy of the $D$, $D_s$, $D^*$, and $D_s^*$ mesons, in the successive rows. They are qualitatively very similar for the four charmed mesons. We observe a first increase below $m_D$, followed by a sharp rise at $m_D+2m_\pi$. While the latter corresponds to the energy of the charmed meson at rest from which the decay into two pions becomes possible, the former is related to processes that are exclusively due to the presence of a medium at finite temperature, that is, the absorption of two thermal pions, and thus it becomes more relevant at larger temperature. The imaginary part of the self-energy is related to the heavy-meson decay width through Eq.~(\ref{eq:hot-quasipart-width}). From this relation, we can anticipate a widening of the ground-state spectral functions with increasing temperatures of the thermal medium.
   
In Fig.~\ref{fig:hot-spectralfunction_charm} we show the dependence of the spectral functions, calculated from the corresponding self-energy in a pionic bath using Eq.~(\ref{eq:hot-sfunc_reimPi}), on the meson energy and at zero three-momentum, for different temperatures up to $T=150$ MeV. The top panels display the spectral functions of the pseudoscalar open-charm ground-state mesons, that is, $D$ (left panel) and $D_s$ (right panel), whereas the bottom panels display the case of the vector open-charm ground-state spectral functions, $D^*$ (left panel) and $D_s^*$ (right panel). The vertical lines depict the value of the mass in the vacuum of each of these mesons. We see the increased broadening of all spectral functions with temperature. Moreover, we observe that the maximum of the spectral functions is shifted towards lower energies for higher temperatures, indicating the attractive character of the interaction of open-charm mesons with a pionic bath. 

It is also interesting to note the similar shape of the pseudoscalar and vector open-charm ground-state spectral functions, that is, the parallel behavior with the temperature of the $D$ and $D^*$ spectral functions as well as the $D_s$ and $D^*_s$ ones. As previously mentioned, at \gls{lo} in the heavy-mass expansion pseudoscalar and vector open-charm ground states are related by \gls{hqss}. Therefore, a similar thermal modification of the spectral functions in the pseudoscalar and vector channels is expected.
    
\begin{figure}[t!]
  \centering
   \includegraphics[width=0.9\textwidth]{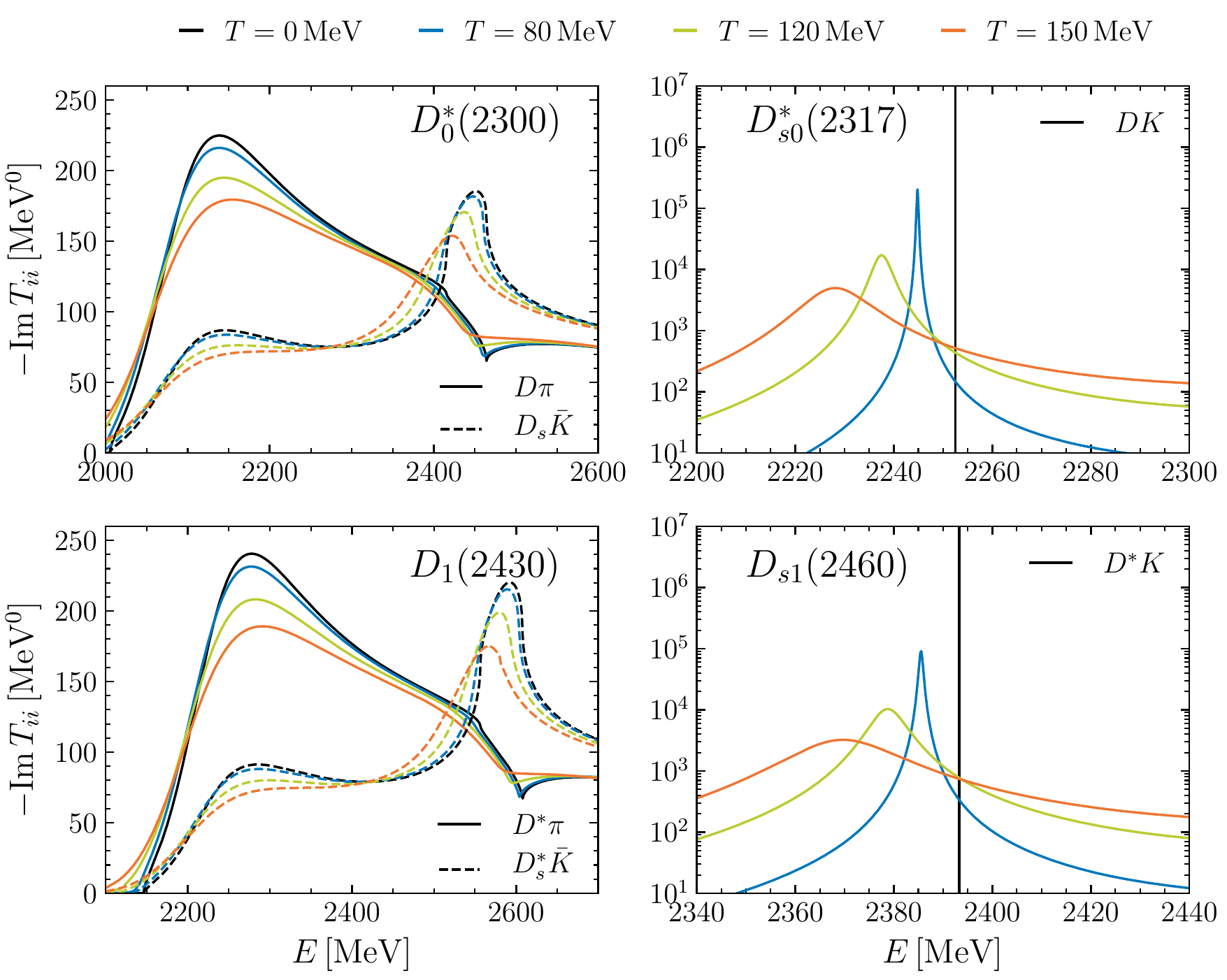}
   \caption{Imaginary part of the scattering amplitudes in the diagonal $D\pi$ and $D_s\bar{K}$ channels in the $J=0$ and $(S,I)=(0,1/2)$ sector (top left), the $DK$ channel in the $J=0$ and $(S,I)=(1,0)$ (top right), the $D^*\pi$ and $D^*_s\bar{K}$ channels in the $J=1$ and $(S,I)=(0,1/2)$ sector (bottom left), and the $D^*K$ channel in the $J=1$ and $(S,I)=(1,0)$ (bottom right), at different temperatures (colored lines).}
   \label{fig:hot-ImT_charm}
   \end{figure}

The properties of the dynamically generated states are directly obtained from the imaginary part of the amplitudes $T_{ii}$, as a proxy for their spectral shape. This is presented in the top panels of Fig.~\ref{fig:hot-ImT_charm} for channels with $J=0$, with $i$ denoting the channel to which the state couples most, that is, $D\pi$ ($D_s \bar{K}$) for the lower (higher) pole of the $D_0^*(2300)$ in the $(S,I)=(0,1/2)$ sector, and $DK$ for the pole of the $D_{s0}^*(2317)$ in the $(S,I)=(1,0)$ sector. 
The bottom panels of the same figure show the corresponding results for the thermal effects on the dynamically generated states with $J=1$, with $i$ indicating $D^*\pi$ ($D^*_s \bar{K}$) for the lower (higher) pole of the $D_1(2430)$ in the sector with $(S,I)=(0,1/2)$, and $D^*K$ for the $D_{s1}^*(2460)$ pole in the $(S,I)=(1,0)$ sector.

In the nonstrange case, peculiar structures appear for both $J=0$ and $J=1$. These are produced by the interplay of the position of the resonance to some nearby channel thresholds. Still, the evolution of the peaks and widths of the amplitudes with $T$ is evident. For the strange sectors, the situation is clearer, but one can observe that, in addition to the typical thermal widening, more strength is visible on the right-hand tail, producing an asymmetric distribution. The reason lies in the fact that the unitary $DK$ threshold is lowered due to the decrease of the $D$ mass and its widening with temperature, hence opening the phase space for decay into this channel at smaller energies.

\subsubsection{Thermal evolution of masses and widths}

Finally, we discuss the evolution with the temperature of the masses and decay widths of both the ground-state mesons and the dynamically generated states. We analyze our results when approaching $T_\chi$, in view that the states that are dynamically generated in our model are the chiral partners of the ground-state $D^{(*)}$ and $D_s^{(*)}$ mesons, and their masses are expected to become degenerate when chiral symmetry is restored ($T>T_\chi$). Differently from the $T=0$ case described in Chapter~\ref{ch:exoticsinfreespace}, the results of which have been presented in Tables~\ref{tab:free-mm-poles0} and \ref{tab:free-mm-poles1}, we find the determination of the position of the poles in the complex-energy plane unfeasible. Apart from complications tied to the analytic continuation of imaginary frequencies to the different \glspl{rs}, a numerical search on the complex plane within self-consistency is computationally challenging.

As a matter of fact, the values of the mass and the width can be extracted from the position and the half-width at half-maximum of the peak of the spectral functions in the real-energy axis. For the ground states, $D^{(*)}$ and $D_s^{(*)}$, this method is totally acceptable as the quasiparticle approximation is entirely justified. However, for the dynamically generated states, at least in the $S=0$ channel, this entails more problems because their poles are located far from the real axis and the width is not a well-defined concept. Because of these problems, we establish the following strategy.

In order to obtain a quantitative description of the thermal dependence of the masses and widths of the open-charm ground states, we analyze the behavior of the corresponding spectral functions with temperature. The mass change with temperature is extracted from the quasiparticle peak, by solving Eq.~(\ref{eq:hot-quasipart-energy}), whereas the variation of the width with temperature can be obtained from Eq.~(\ref{eq:hot-quasipart-width}), which is practically identical to the value that one can obtain from the thermal spectral function at half height. 

The determination of the behavior of the mass and width with the temperature of the dynamically-generated states, such as the $D^*(2300)$ and $D_{s0}^* (2317)$ as well as the $D_1(2430)$ and $D_{s1}(2460)$, is rather delicate. Since the calculation of the poles in the complex plane at finite temperature is unfeasible, we employ the method described in the following to obtain the particle properties on the real axis, through fits of the imaginary part of the unitarized scattering amplitudes at finite temperature shown in Figs.~\ref{fig:hot-ImT_charm}.

For isolated resonances close to the real energy axis and not close to any threshold, one can simply use a Breit-Wigner form. However, in the case of resonances that interact with the background of another resonance due to the coupled-channel case approach, we use a Breit-Wigner-Fano shape~\cite{Fano:1961zz}. This can be used for the lower pole in the double pole structures of the $D_0^*(2300)$ and the $D_1(2430)$. Indeed, we have checked that the values of the mass and the width obtained from the fit at $T=0$ are in very good agreement with the values of the pole mass and the width in Tables~\ref{tab:free-mm-poles0} and \ref{tab:free-mm-poles1}.

The Breit-Wigner-Fano-type distribution provides a simple parametrization to describe the distorted lower resonance at finite temperature:
\begin{equation}\label{eq:hot-BreitWignerFano}
 f^\textrm{BWF}(E;A,m_R,\Gamma_R,q)=A\frac{\Gamma_R/2+ (E-m_R)/q}{(\Gamma_R/2)^2+(E-m_R)^2} \ ,
\end{equation}
where $q$ is the Fano parameter measuring the ratio of resonant-scattering to background-scattering amplitude. In the absence of background, the value of $q$ goes to infinite and Eq.~(\ref{eq:hot-BreitWignerFano}) becomes the usual Breit-Wigner distribution.

For resonances close to a threshold we fit a Flatt\'e-type distribution~\cite{Flatte:1976xu}. In particular, for the higher pole in the double pole structure we first subtract the background and, then, we use a generalized Flatt\'e parametrization with three coupled channels:
\begin{align}\label{eq:hot-Flatte3channels} \nonumber
 \textrm{Im\,} T_{ij}(s;&\,C,m_R,g_1,g_2,g_3) = Cg_ig_j\Bigg[\frac{\rho_1g_1^2}{\big(m_R^2-s+|\rho_2|g_2^2+|\rho_3|g_3^2\big)^2+\big(\rho_1g_1^2\big)^2} \ \theta(m_D+m_\eta-\sqrt{s}) \\ \nonumber
 &+ \frac{\rho_1g_1^2+\rho_2g_2^2}{\big(m_R^2-s+|\rho_3|g_3^2\big)^2+\big(\rho_1g_1^2+\rho_2g_2^2\big)^2} \ \theta(\sqrt{s}-m_D-m_\eta) \ \theta(m_{D_s}+m_{\bar{K}}-\sqrt{s}) \\
 &+ \frac{\rho_1g_1^2+\rho_2g_2^2+\rho_3g_3^2}{\big(m_R^2-s\big)^2+\big(\rho_1g_1^2+\rho_2g_2^2+\rho_3g_3^2\big)^2} \ \theta(\sqrt{s}-m_{D_s}-m_{\bar{K}})\Bigg] \ ,
\end{align}
where $\rho_i$ stands for the phase space of the $i^\textrm{th}$ channel, constituted of two hadrons with masses $m_{ia}$ and $m_{ib}$:
\begin{equation}
 \rho_i(\sqrt{s})=\frac{2p_i(\sqrt{s})}{\sqrt{s}}=\Bigg[\Bigg(1-\frac{(m_{ia}+m_{ib})^2}{s}\Bigg)\Bigg(1-\frac{(m_{ia}-m_{ib})^2}{s}\Bigg)\Bigg]^{1/2} \ .
\end{equation}

The resonance width is obtained from
\begin{equation}\label{eq:hot-widthFlatte}
 m_R\Gamma_R=\rho_1(m_R)g_1^2+\rho_2(m_R)g_2^2+\rho_3(m_R)g_3^2 \ ,
\end{equation}
with the phase spaces evaluated at the resonance mass.
In our case, the subindices correspond to $1\equiv D^{(*)}\pi$, $2\equiv D^{(*)}\eta$, and $3\equiv D_s^{(*)}\bar{K}$. In order to avoid an ill behavior of the fit due to the large number of free parameters, the value of $g_1$ is imposed to vary linearly from its lowest value at $T=0$ to the highest one at $T=150$ MeV.

The Breit-Wigner-Fano distribution is also used for isolated resonances at high temperatures if they become wide enough to be affected by threshold effects. In the case of the $D_{s0}^*(2317)$ and the $D_{s1}(2460)$, the difference between using a Breit-Wigner-Fano distribution instead of a Breit-Wigner one is very small, being $<0.05\%$ for the masses and $<1\%$ for the widths at $T=150$~MeV.

\begin{figure}[t!]
\includegraphics[width=\textwidth]{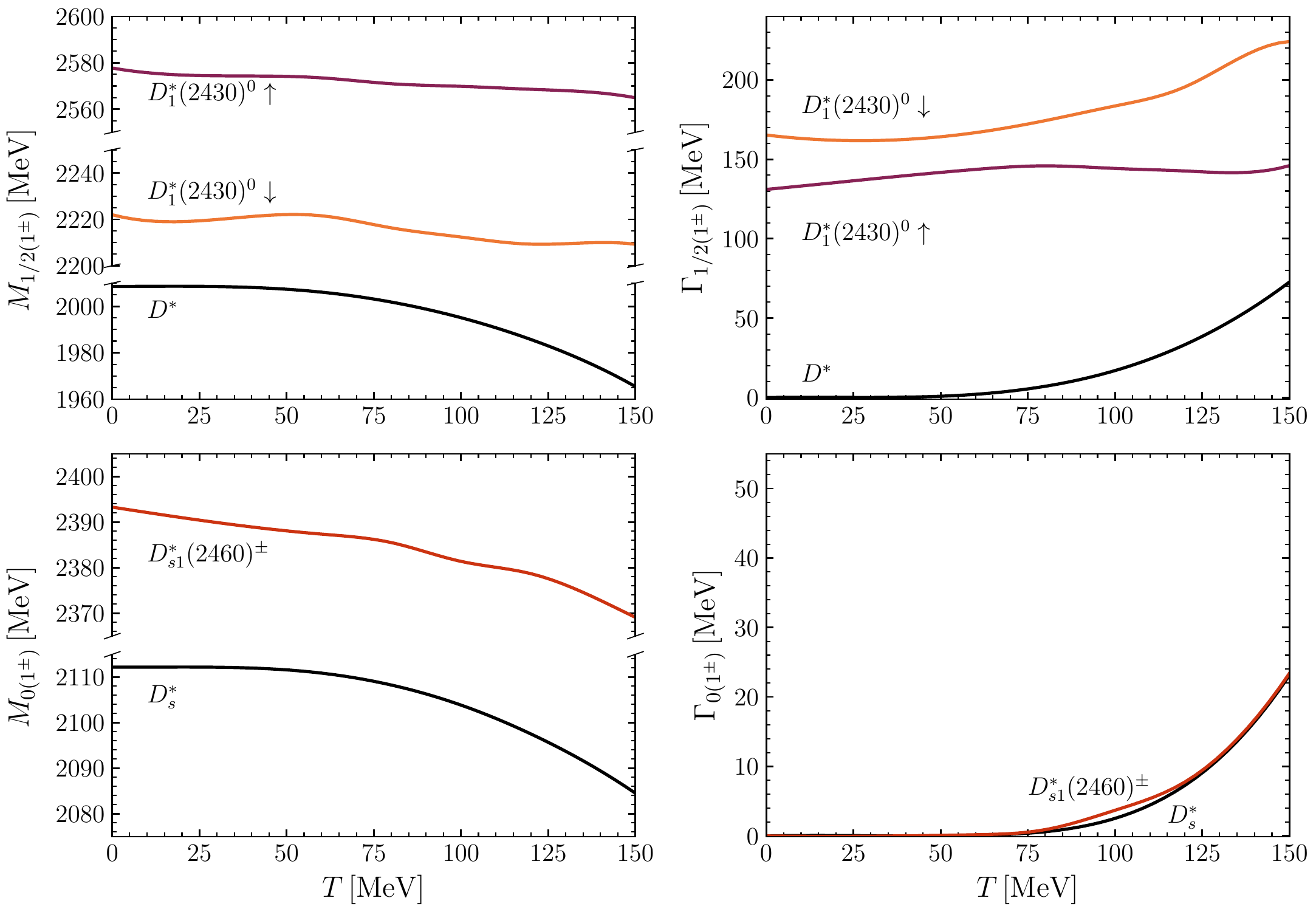}
\caption{Temperature evolution of the mass (left panels) and width (right panels) of the $J=1$ ground-state mesons and dynamically generated states in the $(S, I) = (0, 1/2)$ sector (upper panels) and in the $(S, I) = (1, 0)$ sector (lower panels) in a pionic medium.}
\label{fig:hot-masses_vec}
\end{figure}

The evolution with the temperature of the properties of the $J^P=0^\pm$ open-charm states obtained with these prescriptions is presented in Fig.~\ref{fig:hot-masses_ps}, the masses in the left panels and the widths in the panels on the right, while the corresponding results for the $J^P=1^\pm$ states are given in Fig.~\ref{fig:hot-masses_vec}. We analyze the effect of including the contribution of the kaons and antikaons to the heavy-meson self-energy, in addition to the dominant pion-induced self-energy, in the self-consistent calculations. Therefore, in Figs.~\ref{fig:hot-masses_ps}, the results in a pionic bath (solid lines) are compared with those obtained when the medium is populated by pions, kaons and antikaons (dashed lines). They are summarized as follows:

\begin{figure}[b!]
\includegraphics[width=\textwidth]{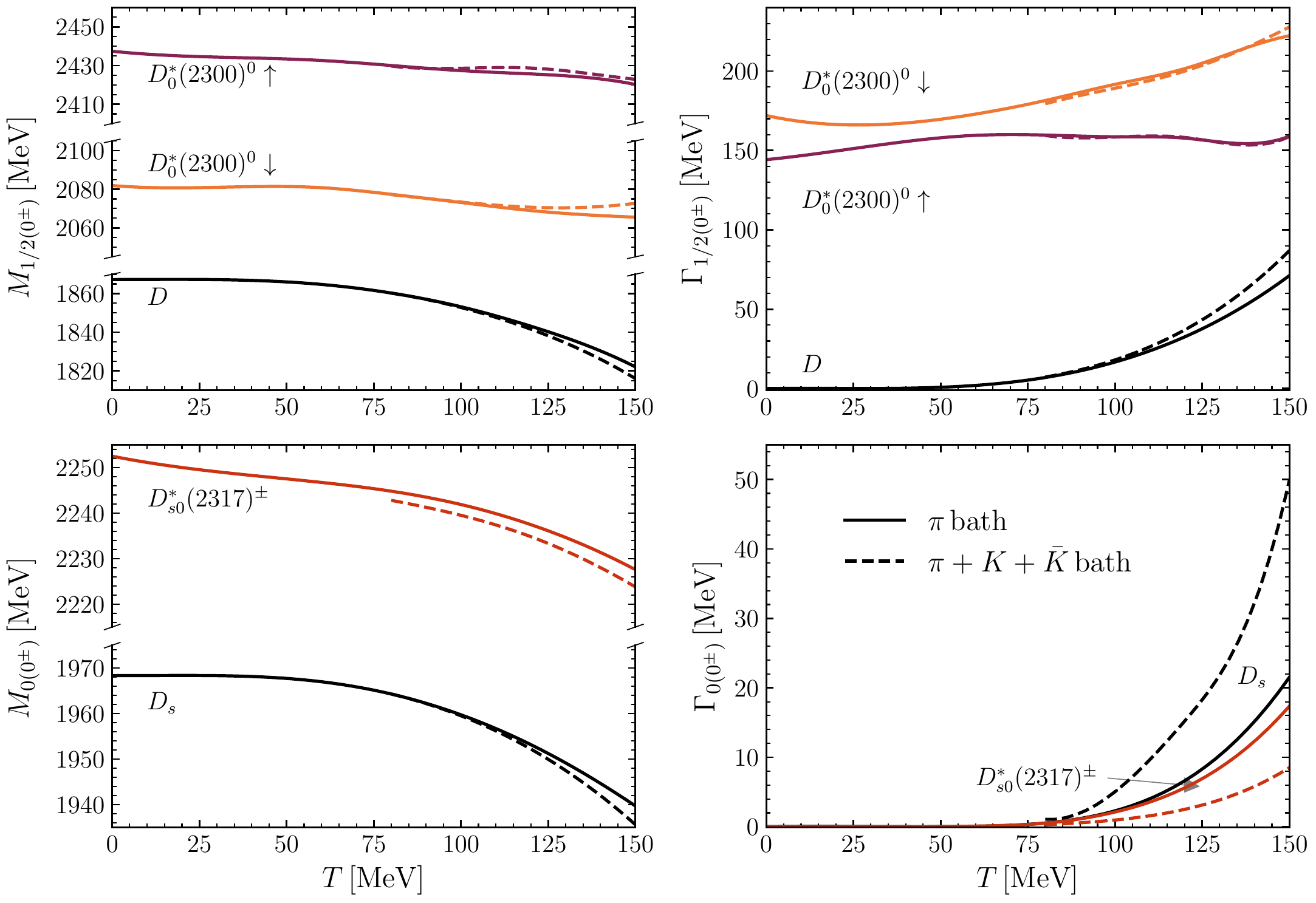}
\caption{Temperature evolution of the mass (left panels) and width (right panels) of the $J=0$ ground-state mesons and dynamically generated states in the $(S,I) = (0, 1/2)$ sector (upper panels) and in the $(S,I) = (1,0)$ sector (lower panels) in a pionic medium (solid lines) and in a medium with $\pi$, $K$ and $\bar{K}$ (dashed lines).}
\label{fig:hot-masses_ps}
\end{figure}

\begin{itemize}
 \item In a pionic medium, the ground-state $D$ mass has a sizable decrease of $\Delta m_D \approx 45$~MeV at the highest temperature $T=150$ MeV. This reduction is consistent, albeit twice larger, with that observed in~\cite{Fuchs2006}, where a more phenomenological approach is used to compute the $D$-meson propagator. Our reduction, on the other hand, is smaller than the one reported in Ref.~\cite{Sasaki:2014asa}, which uses nonunitarized \gls{chpt}. However, in the $\textrm{SU}(4)$ effective approach of \cite{Cleven:2017fun} no significant modification is reported. Regarding the $D^*_0(2300)$, the two poles have a more stable trend compared to the ground state. They slightly move downwards, moderately distancing from each other. As a consequence, in this sector, we cannot conclude that masses of opposite parity states become degenerate close to $T_\chi$, although the temperatures studied might be still low for the chiral symmetry restoration.  In~\cite{Buchheim2018}  a large reduction in the mass of the positive-parity $D$ meson partner, of around $150$~MeV, is found at $T=150$ MeV, but using a constant $D$ mass as an input of the sum-rule analysis. An even larger reduction of close to $200$~MeV is seen in the results of~\cite{Sasaki:2014asa}.
 
 As for the $J=1$ states, the decrease of the $D^*$ mass, which is $\Delta m_{D^*} \approx 43$~MeV at $T=150$ MeV, is similar to the $D$ mass shift, while for the two poles that form the $D_1(2430)$, their masses decrease less rapidly with temperature compared to the ground state, distancing from each other as temperature increases, in an analogous manner as for the two poles of the  $D^*_0(2300)$. As a consequence, also in the $J=1$ case, we cannot conclude that masses of opposite parity states become degenerate with temperature, at least for the range of temperatures studied here.  
\end{itemize}

\begin{itemize}
 \item The width of the nonstrange states increases with temperature, being more relevant the change in width for the ground state. The $D$ and $D^*$ mesons show a similar width of around $\sim70$~MeV at $T =150$ MeV, consistent with~\cite{Cleven:2017fun} and the estimates of Ref.~\cite{Fuchs:2004fh}, and of Ref.~\cite{He:2011yi} for the $D$ meson. The widths of the two poles of the $D_0^*(2300)$ and the $D_1(2430)$ obtained from the fits increase moderately with temperature with respect to their vacuum values, being of the same order for the $J^P=0^+$ and the $J^P=1^+$ states.
 \item In the strangeness sector we observe a clearer picture. The parity partners seem to decrease their mass with temperature, in a similar amount for both states (and for the $J=0$ and the $J=1$ sectors), reaching a reduction of $\approx 28$ MeV for the negative-parity states and $\approx 25$~MeV for the positive-parity partners at $T=150$~MeV. Consequently, they are still far from chiral degeneracy. These behaviors seem to be compatible with the low-temperature trends seen in the linear-sigma model calculation of~\cite{Sasaki:2014asa} for the scalar and pseudoscalar states.
 \item  The decay widths of both strange partners increase from zero at similar rates. The width of the $D_{s0}^* (2317)$ is comparable to that acquired by the $D_s$ ground state at $T=150$ MeV ($\sim 20$~MeV).
 Analogously, the widths of the $D_s^*$ and $D_{s1}(2460)$ increase similarly with temperature, becoming slightly larger compared to the ones of the $0^\pm$ states.
 We note that, whereas the widths of the $D$ and $D^*$ are only due to medium effects, the width of the $D_{s0}^* (2317)$ and the $D_{s1}(2460)$ are affected by the reduction of the mass and the widening of the $D$ and $D^*$ mesons, respectively, due to the dominant contribution of the $DK$ and $D^*K$ channels in their respective dynamical generation.
\item In a bath that includes $K$ and $\bar{K}$ in addition to pions (see dashed lines in Fig.~\ref{fig:hot-masses_ps}),
the masses of the ground states $D$ and $D_s$ decrease an additional amount of around $5$~MeV at $T=150$ MeV.
The modification of the widths is, however, different for the nonstrange and the strange states. 
In the case of the $D$ meson, the width increases around $20\%$ while that for the $D_s$ meson is more than twice larger than the width in a pionic medium at $T=150$ MeV. This follows from the stronger interaction of the $D_s$ meson with kaons than with pions. On the other hand, the effect of the pionic and kaonic bath on the dynamically generated states is rather moderate, with the masses of the two resonances of the $D_0^*(2300)$ increasing a few MeV and no significant modification of the widths, whereas the mass of the $D_{s0}^*(2317)$ drops slightly and the width is reduced by half due to the reduction of the phase space for decay. These results are in agreement with what we found in~\cite{Montana:2020vjg}. In the calculations presented in this thesis, we have increased the cut-offs of the self-consistent integrals, thus the small differences when compared to our previously published results.
\end{itemize}



As we have just discussed, there is a parallelism between the behavior in a thermal medium of the ground-state pseudoscalar and vector mesons, as well as between the dynamically generated scalar and axial-vector states. Again, this is because the interactions of light mesons with pseudoscalar open-charm ground states and those with vector open-charm ones are related by \gls{hqss}. Hence, our conclusions are similar in both sectors.

We note that the results presented in this dissertation have been improved with respect to those that we published in Refs.~\cite{Montana:2020lfi,Montana:2020vjg}. As mentioned, the differences essentially consist of an increase in the values of the cutoffs in the integrals of the \gls{itf}. Despite the slight differences in the quantitative analysis, the discussion at the qualitative level of the thermal modification of the open-charm meson properties is identical.

Apart from the above comparisons with previous models, unfortunately, there is no solid data from first principles to compare to. However, despite the limitations in obtaining reliable information from finite temperature \gls{lqcd} simulations, tied to the difficulties in extracting the spectral function from the lattice Euclidean correlators, we can still aim at a qualitative comparison. We note that a recent \gls{lqcd} calculation~\cite{Kelly:2018hsi} presents the spectral functions of $D$ and $D_s$ channels at different temperatures. The analysis in that paper concludes that no medium modification concerning the $D$ and $D_s$ ground states is seen up to $T_{\chi} $, where  $T_ {\chi} \simeq 185$~MeV in that work. Given the precision of the \gls{lqcd} data this might be in well agreement with our findings here, as our $D$ ($D_s$) mass shift is only $2\%$  ($1\%$) of the mass itself. As a pion mass of $m_\pi \sim 380$ MeV is used in~\cite{Kelly:2018hsi}, one has to re-address the self-consistent calculations with a heavier pion mass and analyze the effects on the charm meson properties for temperatures $T<T_{\chi}$. This issue is further discussed in Chapter~\ref{ch:lattice}, where we compute open-charm Euclidean correlators from the spectral functions obtained with the thermal unitarized effective approach discussed in this chapter and in Ref.~\cite{Montana:2020var}, using the values of the meson masses in~\cite{Kelly:2018hsi}, and compare the results with those extracted from the lattice calculations~\cite{Montana:2020var}.

\subsection{Results for $\bar{B}$ mesons}
\label{subsec:hot-results-Bmesons}
To analyze the modification of open-bottom mesons in a thermal pionic medium, we follow the strategy based on the \gls{itf} that has been explained in Section~\ref{sec:hot-formalism} for open-charm mesons, in which now the heavy meson is identified with a bottomed meson, $\mathcal{M}\equiv\bar{B}$. In the following, we discuss the results obtained within self-consistency for the thermal loops and the unitarized amplitudes of the scattering of the $\bar{B}$ mesons off the light $\Phi$ mesons at finite temperature, as well as for the self-energies and the spectral functions of the bottomed ground-state mesons. We also analyze the changes induced by the medium in the dynamically generated states in this sector.

\subsubsection{Thermal loops and scattering amplitudes}
   
\begin{figure}[h!]
   \centering
   \includegraphics[width=0.88\textwidth]{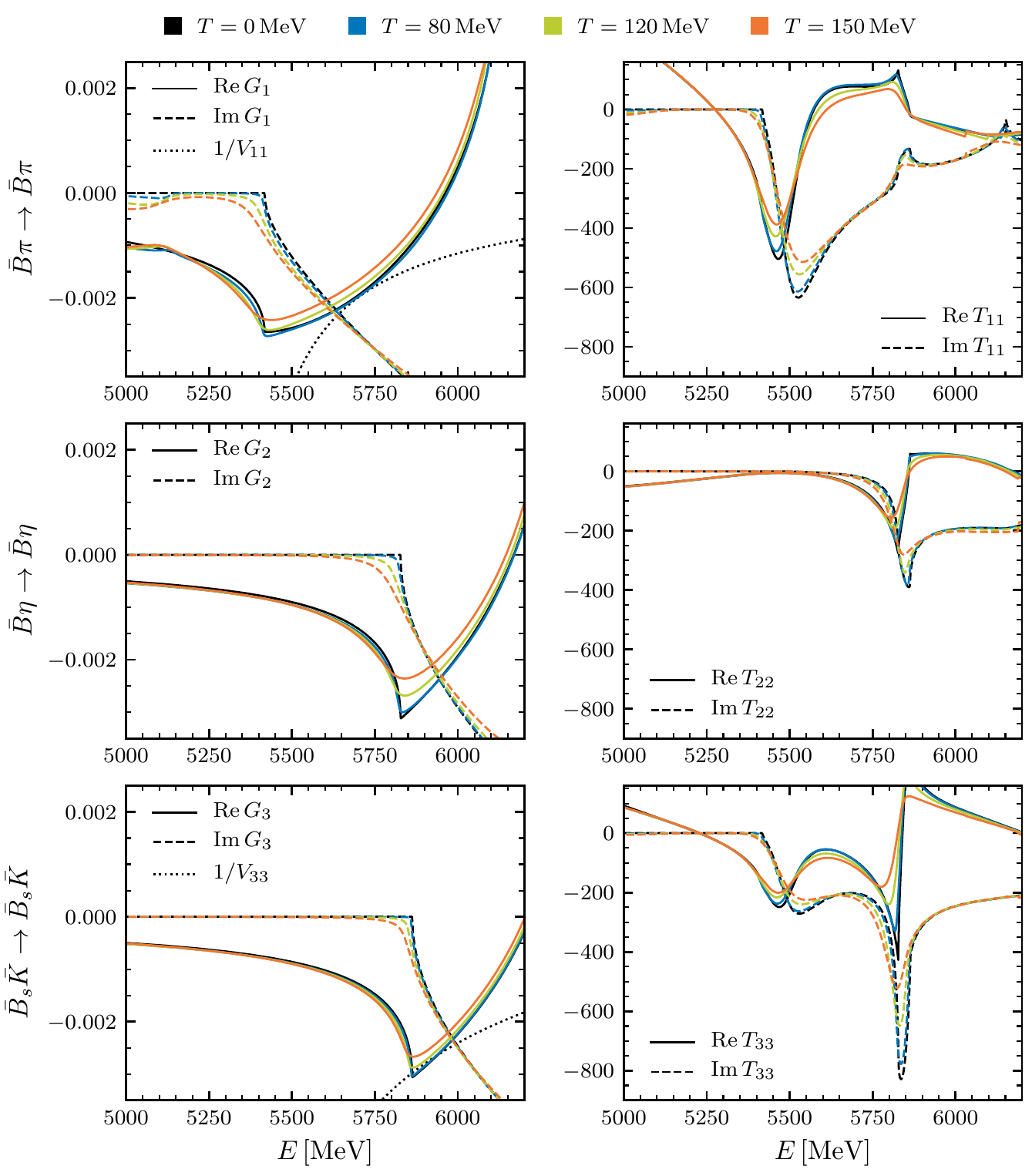}
   \caption{The inverse of the interaction kernel, $1/V_{ii}$, the real and imaginary parts of the loop function, $G_i$, and the real and imaginary parts of the diagonal components of the $T$ matrix, $T_{ii}$, in units of MeV$^0$, in the sector with bottom $-1$, spin $J=0$, and strangeness and isospin $(S,I)=(0,1/2)$, at various temperatures (colored lines). The subindices $1$, $2$, $3$ refer to the channels $\bar{B}\pi$, $\bar{B}\eta$, and $\bar{B}_s\bar{K}$, respectively.}
   \label{fig:hot-VGT_0_05_B}
   \end{figure}

\begin{figure}[p!]
    \centering
   \includegraphics[width=0.88\textwidth]{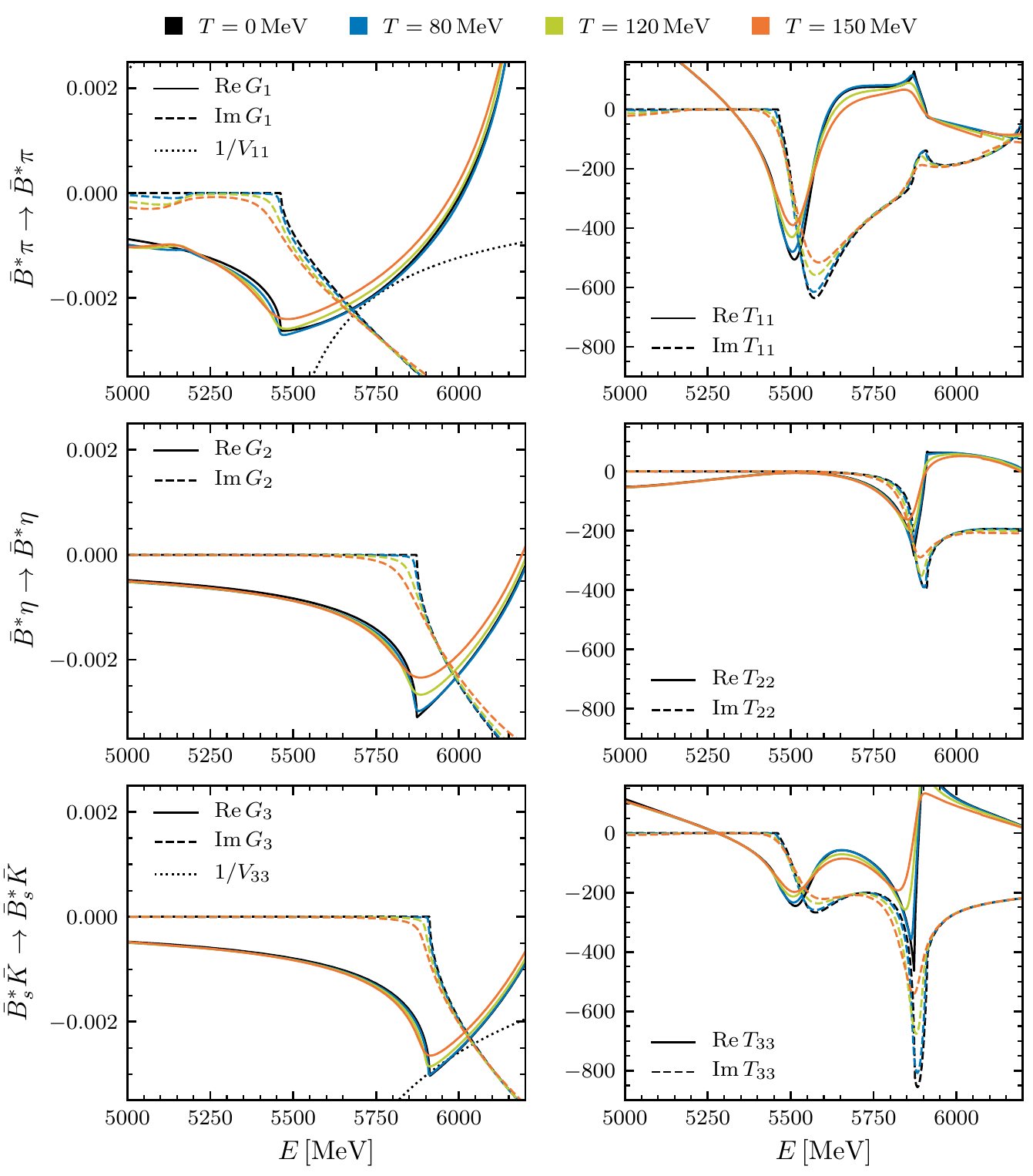}
   \caption{The same as in Fig.~\ref{fig:hot-VGT_0_05_B} in the sector with $J=1$ and $(S,I)=(1,0)$. The subindices $1$, $2$, $3$ refer to the channels $\bar{B}^*\pi$, $\bar{B}^*\eta$, and $\bar{B}_s^*\bar{K}$, respectively.}
   \label{fig:hot-VGT_0_05_vec_B}
   \end{figure}

The thermal heavy-light two-meson loop functions in the sectors with strangeness $S=0$ and isospin $I=1/2$ are shown in the left panels of Fig.~\ref{fig:hot-VGT_0_05_B} for bottomed pseudoscalar mesons, and of Fig.~\ref{fig:hot-VGT_0_05_vec_B} for bottomed vector mesons, as functions of the total energy and for $\vec{P}=0$ and various temperatures (colored lines). We also plot the inverse of the diagonal element of the interaction kernel, $1/V_{ii}$ (dotted lines), if it falls within the vertical scale employed in the subplots, as its proximity to the real part of the loop (solid lines) indicates the appearance of poles in the unitarized scattering amplitudes. 
The corresponding unitarized scattering amplitudes obtained from the solution of the coupled-channel \gls{bs} equation with these thermal loop functions are displayed in the right panels of the respective figures.

\begin{figure}[t!]
   \centering
   \includegraphics[width=0.88\textwidth]{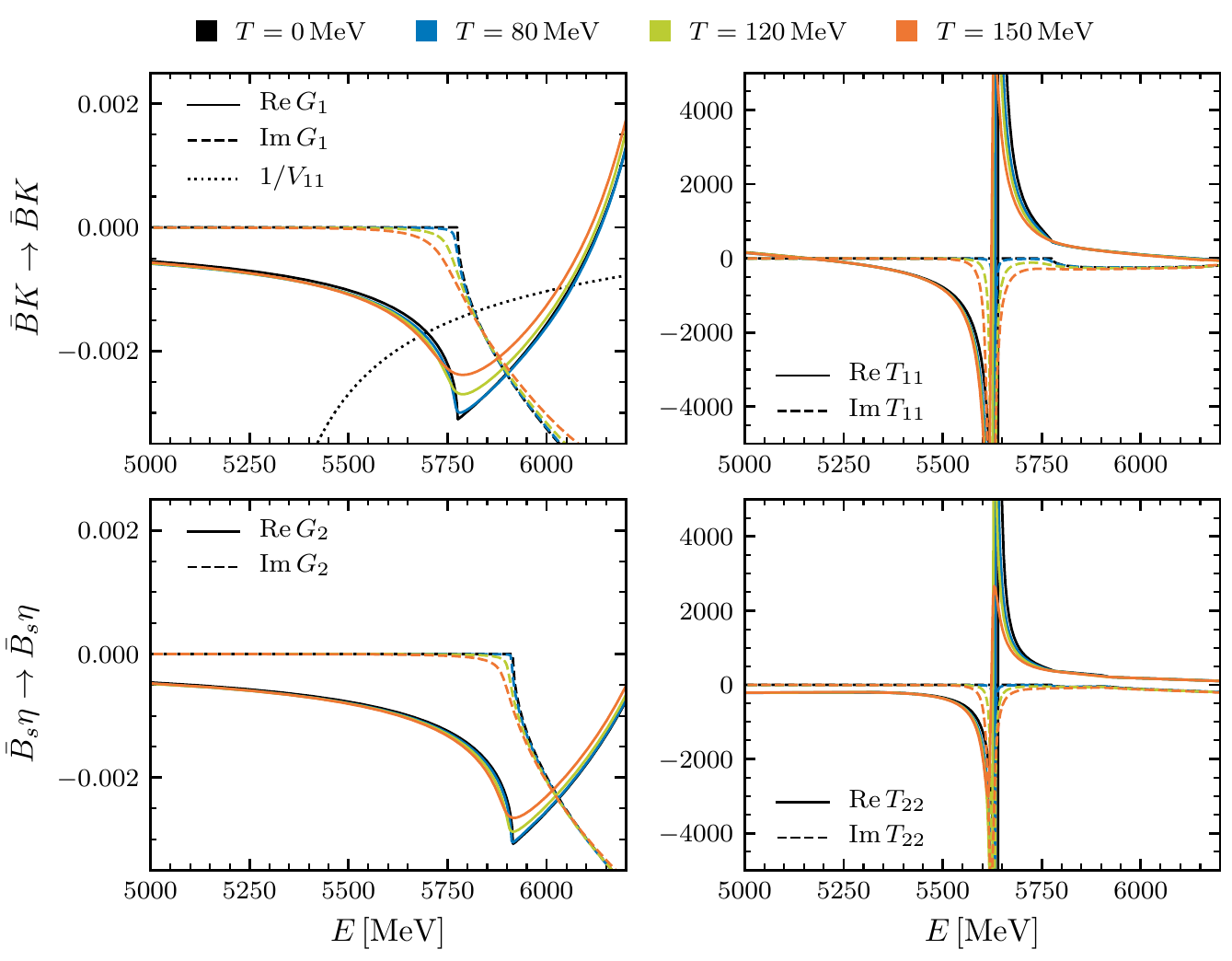}
   \caption{The same as in Fig.~\ref{fig:hot-VGT_0_05_B} in the sector with $J=0$ and $(S,I)=(1,0)$. The subindices $1$, $2$ refer to the channels $\bar{B}_s\pi$ and $\bar{B}K$, respectively.}
   \label{fig:hot-VGT_1_0_B}
   \end{figure}

\begin{figure}[b!]
     \centering
   \includegraphics[width=0.88\textwidth]{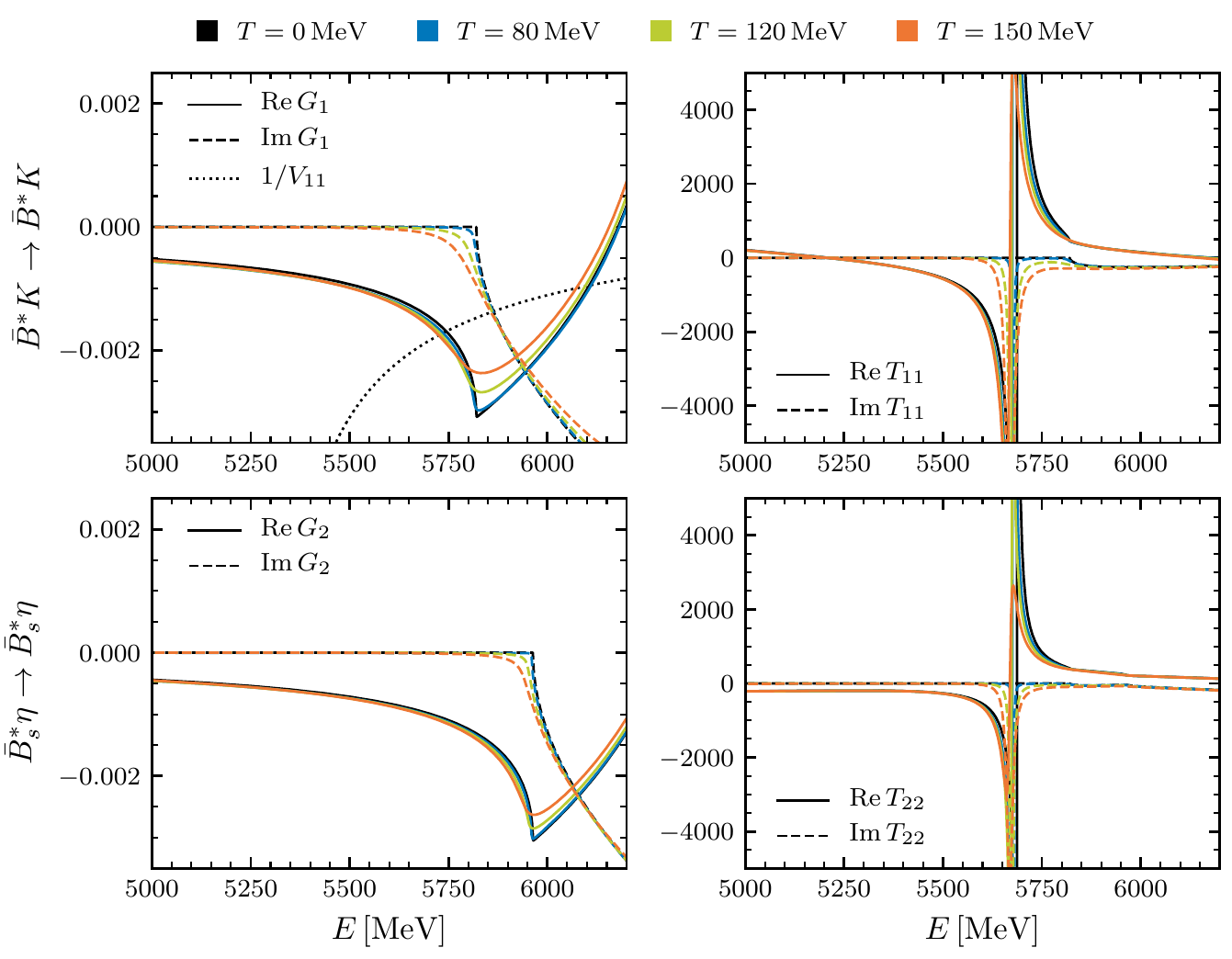}
   \caption{The same as in Fig.~\ref{fig:hot-VGT_0_05_B} in the sector with $J=1$ and $(S,I)=(1,0)$. The subindices $1$, $2$ refer to the channels $\bar{B}_s^*\pi$ and $\bar{B}^*K$, respectively.}
   \label{fig:hot-VGT_1_0_vec_B}
   \end{figure}

Figures~\ref{fig:hot-VGT_1_0_B} and \ref{fig:hot-VGT_1_0_vec_B} show the thermal loop functions (left panels) and the scattering amplitudes (right panels) in the strangeness $S=1$ and isospin $I=0$ sectors, for pseudoscalar $\bar{B}$ mesons and vector $\bar{B}^*$ mesons, respectively.

These results for the thermal loops and the scattering amplitudes in the bottom sector are qualitatively very similar to those shown in Figs.~\ref{fig:hot-VGT_0_05} to \ref{fig:hot-VGT_1_0_vec} in the charm sector.
  
The unitary cut that opens up in the imaginary part of the loop functions (dashed lines) at the $m_{\bar{B}}+m_\Phi$ threshold is smoothened with increasing temperatures as compared to the $T=0$ case due to the dressing of the thermal loop with the spectral function of the bottomed meson. Furthermore, the Landau cut, which is only visible for the $\bar{B}\pi$ loops (upper left panels of Figs.~\ref{fig:hot-VGT_0_05_B} and \ref{fig:hot-VGT_0_05_vec_B}) for the scale used in the plots, arises at finite temperature for energies below $m_{\bar{B}}-m_\Phi$ due to the absorption and production processes of thermal mesons that are only possible in the presence of a thermal medium.

Regarding the unitarized amplitudes, the structures that are visible in the real axis due to the presence of poles in the complex-energy plane (see the discussion in Section~\ref{sec:free-mm} for the vacuum case) reflect the smoothening of the loop functions, and a subsequent widening and moderate melting of the dynamically generated states is observed at large temperatures of the medium.

\clearpage

\subsubsection{Self-energies and open-beauty spectral functions}

  \begin{figure}[b!]
  \centering
   \includegraphics[width=\textwidth]{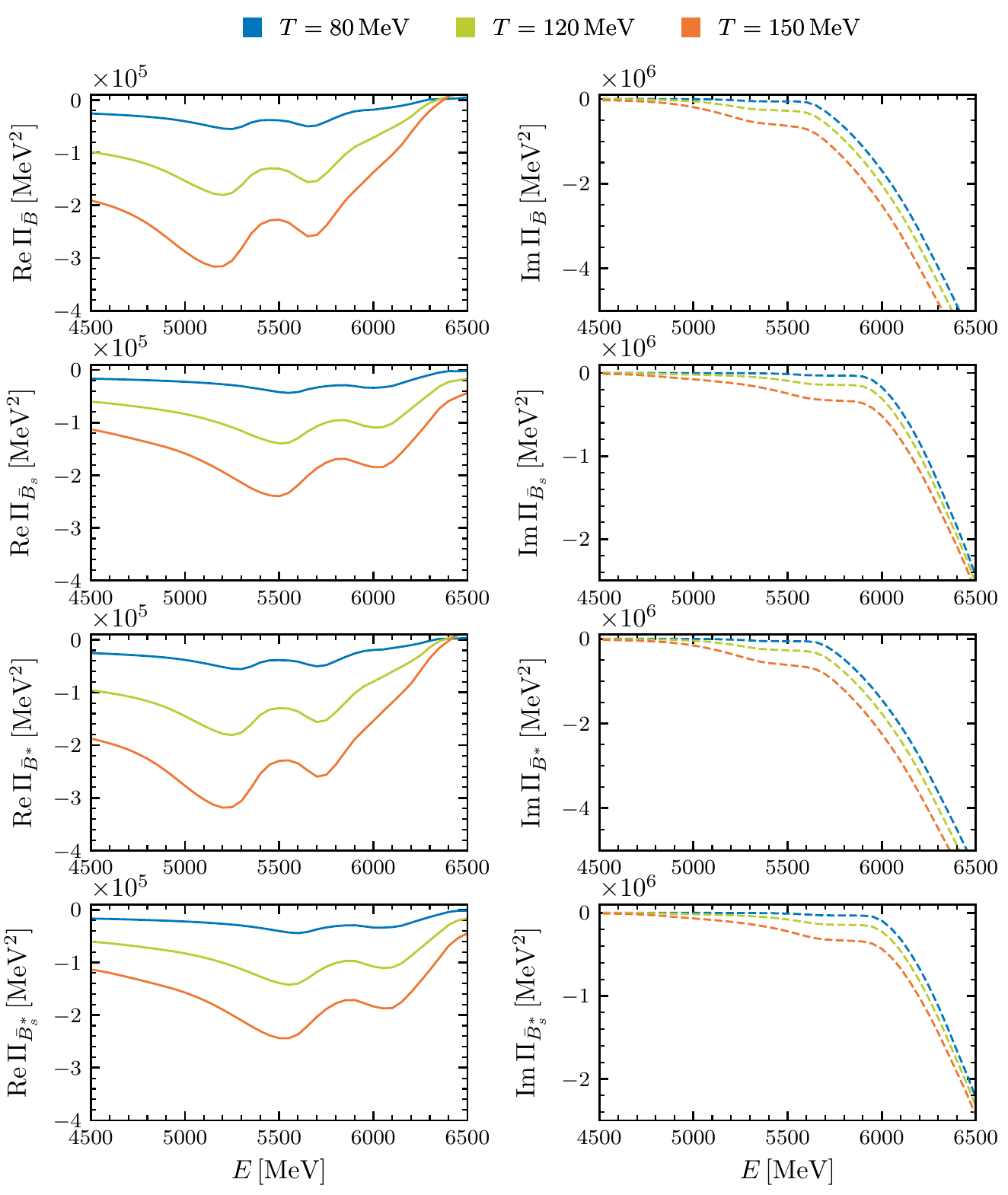}
   \caption{Real and imaginary parts of the pionic contribution to self-energies of the $\bar{B}$ (first row), the $\bar{B}_s$ (second row), the $\bar{B}^*$ (third row), and the $\bar{B}_s^*$ (fourth row) mesons at various temperatures (colored lines). }
   \label{fig:hot-selfe_bottom}
   \end{figure}
   
In Fig.~\ref{fig:hot-selfe_bottom} we show, in the respective rows, the real part (left panels) and the imaginary part (right panels) of the pionic self-energies of the pseudoscalar $\bar{B}$ and $\bar{B}_s$, and the vector $\bar{B}^*$ and $\bar{B}_s^*$ mesons as a function of the energy and for $\vec{q}=\vec{0}\,$, for different temperatures (colored lines).

As we discussed for the charm sector above, the real part of the self-energy is connected to the in-medium mass shift, while the imaginary part is related to the thermal width. For the imaginary part displayed in the right panels of Fig.~\ref{fig:hot-selfe_bottom}, we can see the growth below $m_{\bar{B}}$ that accounts for absorption processes that are only possible in the medium and, thus, it is larger at high temperatures, while the rise at $m_{\bar{B}}+2m_\pi$ takes place at similar rates for all temperatures, since it is linked to the decay of the bottomed meson into two additional pions, and for a large enough energy of an off-shell $\bar{B}$ meson this process is also possible in the vacuum.
   
In comparison with the self-energies of the charmed ground-state mesons shown in Fig.~\ref{fig:hot-selfe_charm}, one can see that the values of the real part of the self-energy over the bottomed meson mass, $\textrm{Re}\,\Pi_{\bar{B}}(m_{\bar{B}},\vec{0}\,;T)/(2m_{\bar{B}})$, at a given temperature, take less negative values in the bottomed sector. This gives a smaller shift towards lower energies with respect to the value of the vacuum mass, as we will show below. On the other hand, the values of the imaginary part of the self-energies in the bottom sector divided by the mass, $\textrm{Im}\,\Pi_{\bar{B}}(m_{\bar{B}},\vec{0}\,;T)/m_{\bar{B}}$, are larger in magnitude than in the charm sector. Therefore, a larger widening of the $\bar{B}$ mesons with temperature is expected.
   
\begin{figure}[b!]
  \centering
   \includegraphics[width=0.9\textwidth]{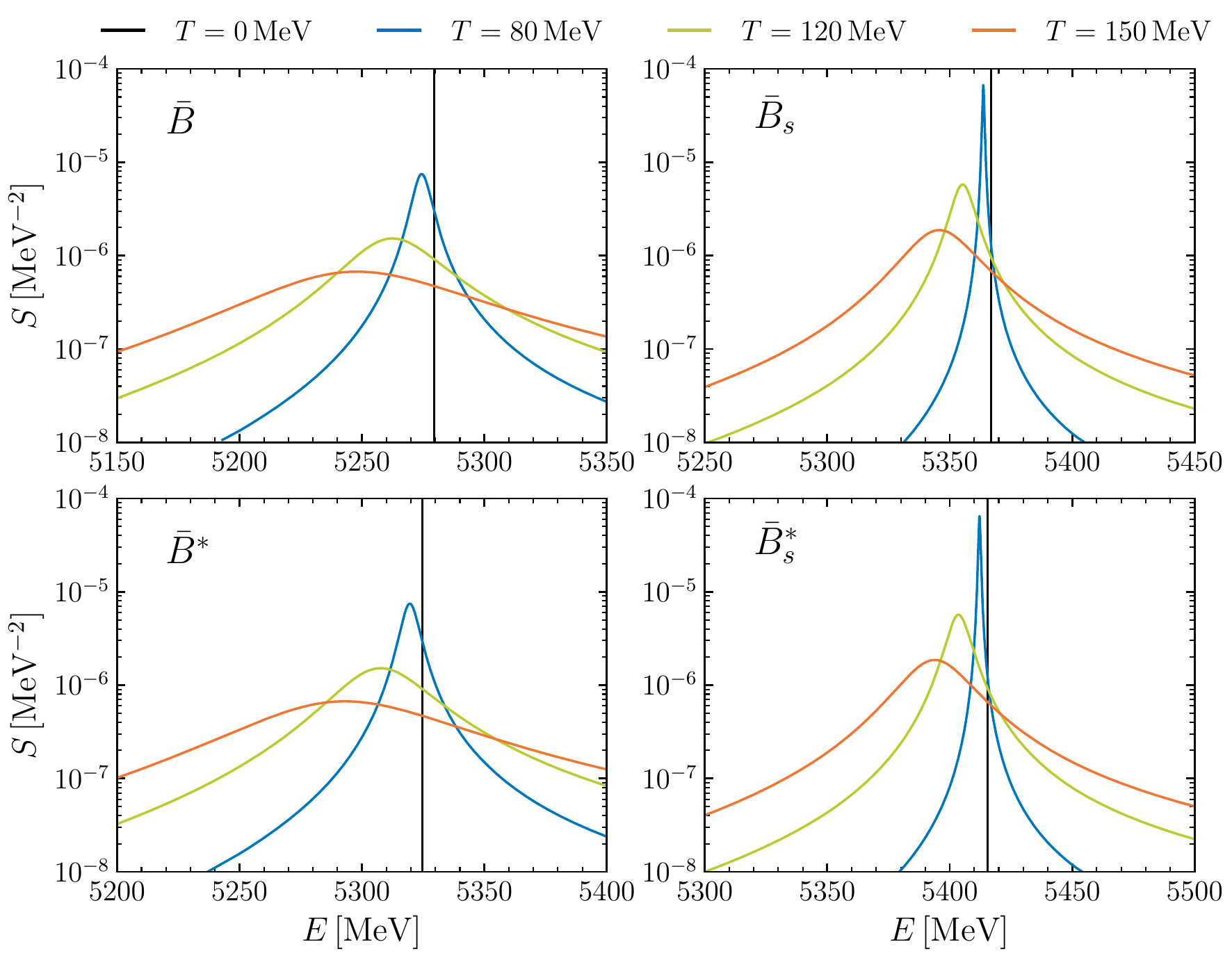}
   \caption{Spectral functions of the $J=0$ ground states ($\bar{B}$ and $\bar{B}_s$, top panels) and the $J=1$ ground states ($\bar{B}^*$ and $\bar{B}_s^*$, bottom panels) in a pionic bath at different temperatures (colored lines).}
   \label{fig:hot-spectralfunction_bottom}
   \end{figure}

The spectral functions of the bottomed ground-state mesons in a pionic bath obtained from Eq.~(\ref{eq:hot-sfunc_reimPi}) are displayed in Fig.~\ref{fig:hot-spectralfunction_bottom}, as functions of the meson energy and $\vec{q}=0$, for different temperatures up to $T=150$ MeV. The pseudoscalar $\bar{B}$ and $\bar{B}_s$ mesons are shown in the top panels, whereas the bottom panels display the vector open-bottom ground-state spectral functions for $\bar{B}^*$ and $\bar{B}_s^*$. The vertical lines represent the value of the vacuum mass of each of these mesons. From these plots, we see the larger broadening with temperature of all the spectral functions in the bottom sector compared to the charm sector (see Fig.~\ref{fig:hot-spectralfunction_charm} for the comparison). Note that the energy range in the subpanels of Fig.~\ref{fig:hot-spectralfunction_bottom} is twice as large as that employed in Fig.~\ref{fig:hot-spectralfunction_charm}. On the other hand, we observe that the shift of the maximum of the spectral functions towards lower energies with increasing temperatures is smaller in the bottom sector, as anticipated from the analysis of the self-energies above. 

  \begin{figure}[t!]
  \centering
   \includegraphics[width=0.9\textwidth]{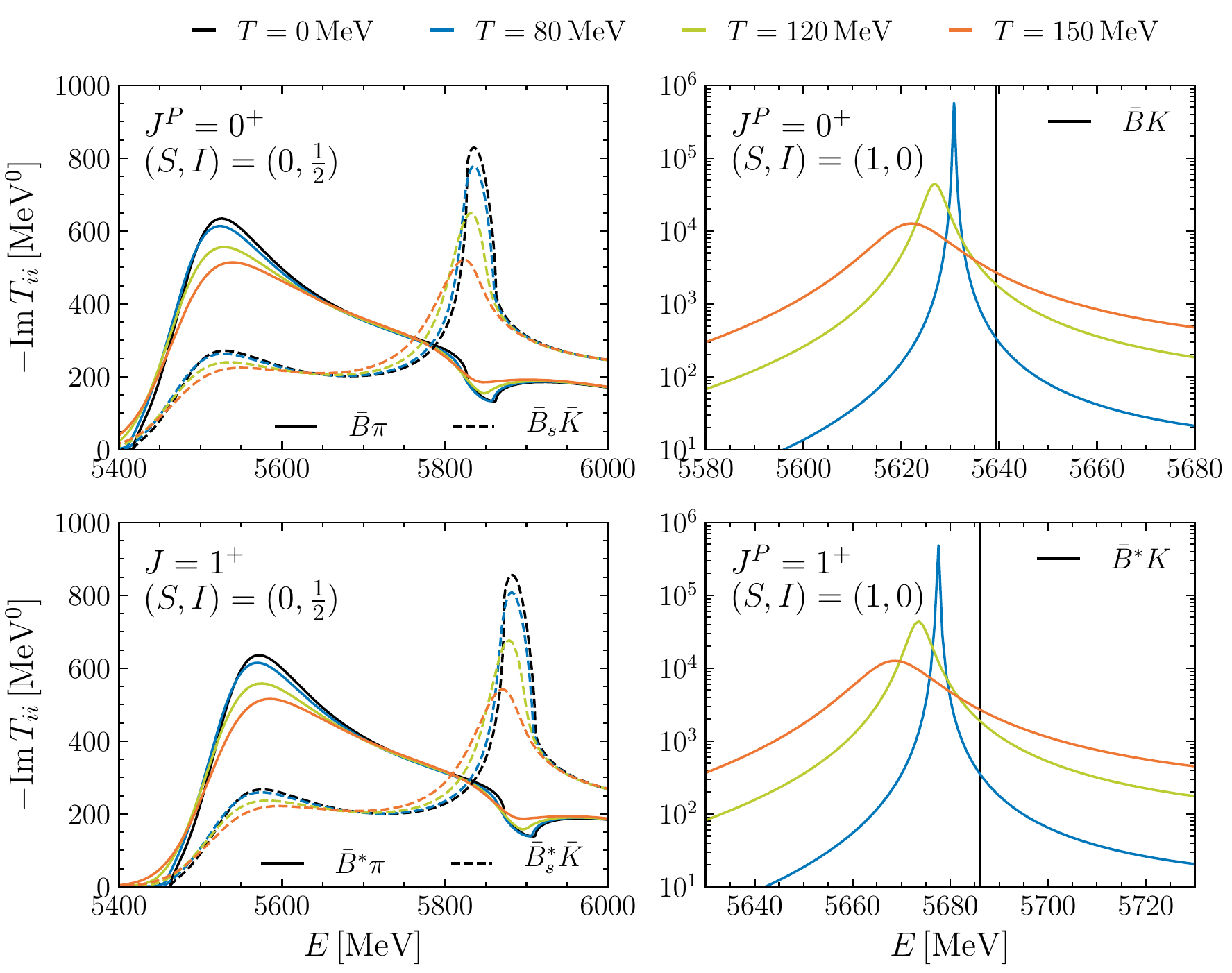}
   \caption{Imaginary part of the scattering amplitudes in the diagonal $\bar{B}\pi$ and $\bar{B}_s\bar{K}$ channels in the $J=0$ and $(S,I)=(0,1/2)$ sector (top left), the $\bar{B}K$ channel in the $J=0$ and $(S,I)=(1,0)$ (top right), the $\bar{B}^*\pi$ and $\bar{B}^*_s\bar{K}$ channels in the $J=1$ and $(S,I)=(0,1/2)$ sector (bottom left), and the $\bar{B}^*K$ channel in the $J=1$ and $(S,I)=(1,0)$ (bottom right), at different temperatures (colored lines).}
   \label{fig:hot-ImT_bottom}
   \end{figure}

Again, there is a parallelism between the thermal effects in the pseudoscalar $\bar{B}$ and in the vector $\bar{B}^*$ spectral functions, as well as between those for the $\bar{B}_s$ and the $\bar{B}^*_s$, due to \gls{hqss}.

In order to analyze the thermal effects on the properties of the dynamically generated states, the imaginary part of the amplitudes $T_{ii}$, which we take as a proxy for their spectral shape, are presented in the top panels of Fig.~\ref{fig:hot-ImT_bottom} for channels with $J=0$, with $i$ denoting the channel to which the state couples most, that is, the $\bar{B}\pi$ ($\bar{B}_s \bar{K}$) channel for the lower (higher) pole in the $(S,I)=(0,1/2)$ sector (left panel), and the $\bar{B}K$ one for the pole in the $(S,I)=(1,0)$ sector (right panel). 
The bottom panels of the same figure show the corresponding results for the thermal effects on the dynamically generated states with $J=1$, with $i$ indicating the $\bar{B}^*\pi$ ($\bar{B}^*_s \bar{K}$) channel for the lower (higher) pole in the sector with $(S,I)=(0,1/2)$ (left panel), and $\bar{B}^*K$ for the $\bar{B}_{s1}^*(2460)$ pole in the $(S,I)=(1,0)$ sector (right panel).

The modification with the temperature of the peaks and the widening of the structures that appear in the nonstrange case are very similar for both $J=0$ and $J=1$ sectors. For the strange sectors, we can see a clear widening of the state, which is generated as a bound state with no width at $T=0$, as well as the shift of its peak maximum towards lower energies. The vertical lines in the right panels show the values of the real energy at which the poles are dynamically generated at $T=0$.

\subsubsection{Thermal evolution of masses and widths}

\begin{figure}[b!]
\includegraphics[width=\textwidth]{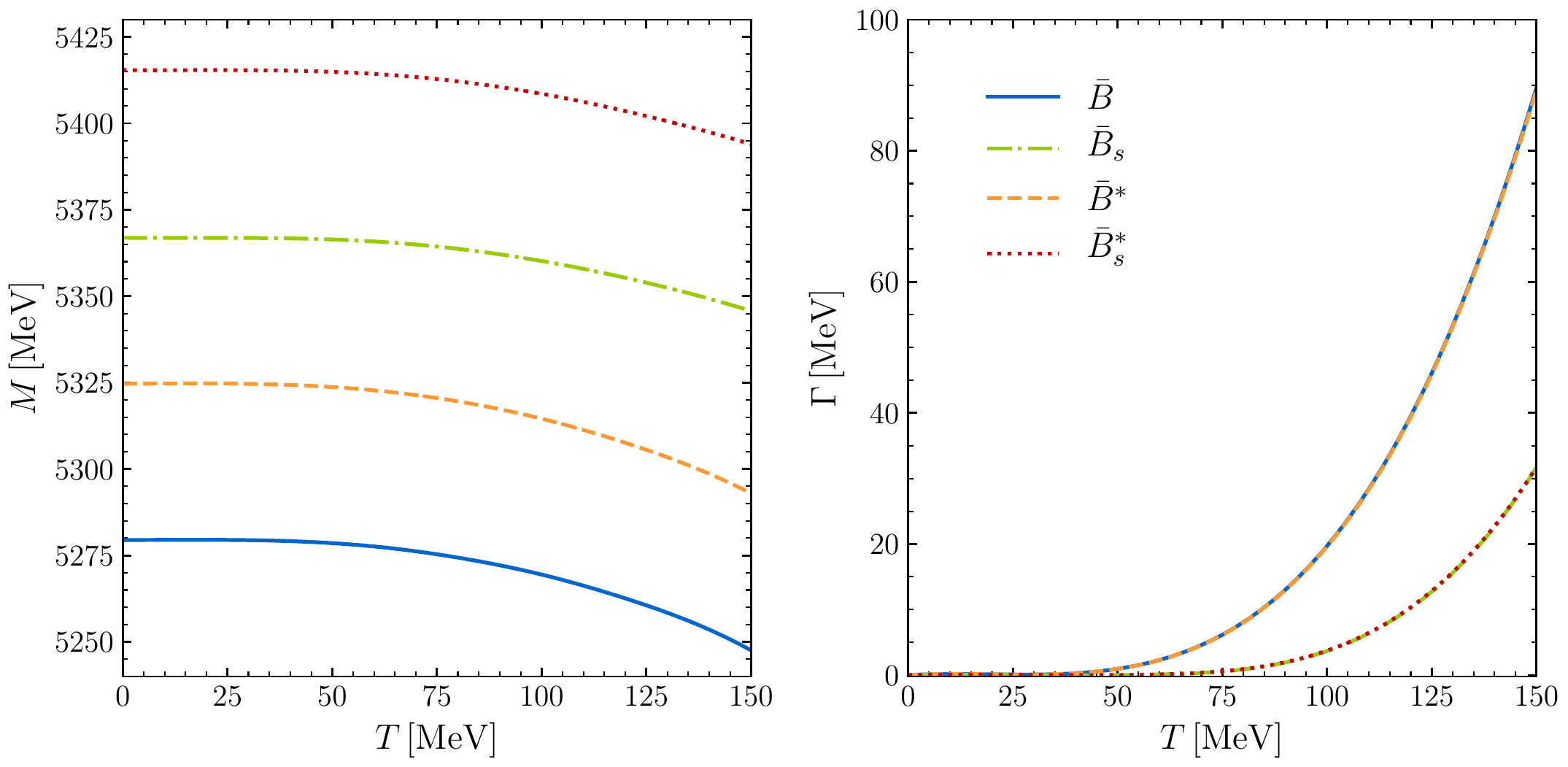}
\caption{Temperature evolution of the mass (left panel) and width (right panel) of the bottomed $J^P=0^-$ and $J^P=1^-$ ground-state mesons.}
\label{fig:hot-masses_bottom}
\end{figure}

The evolution with the temperature of the properties of the bottomed ground-state mesons is shown in Fig.~\ref{fig:hot-masses_bottom} for both the pseudoscalars and the vectors. The values of the masses at each temperature (left panel) have been extracted from the peak of the respective in-medium spectral functions in the real-energy axis, while the thermal widths (right panel) have been calculated using Eq.~(\ref{eq:hot-quasipart-width}).

In a pionic medium, the masses of the ground-state $\bar{B}$ and $\bar{B}^*$ mesons decrease by a similar amount of $\Delta m_{\bar{B}}\approx \Delta m_{\bar{B}^*}\approx 32$~MeV at the highest temperature $T=150$, which is of the same order but slightly smaller than what we found in the case of the nonstrange charmed $D$ and $D^*$ mesons. The width of the bottomed states increases with temperature, up to a value of $\sim 90$~MeV, for both the pseudoscalar and the vector states, which is about $20$~MeV larger than in the case of the charmed states.

In the strangeness sector, the reduction of the mass of $\sim 20$~MeV at the largest temperature is also very similar for both the $\bar{B}_s$ and the $\bar{B}_s^*$ mesons and smaller than in the charm sector, while their thermal width at this temperature ($\sim 30$~MeV) is slightly larger than the value that we have reported in the charm sector.


As we can see from these results, we expect the thermal effects on the masses and widths of the ground-state mesons in the bottom sector to be of comparable size to those in the charm sector. As already mentioned, the somewhat larger widths found for the open-bottom states are certainly related to the greater magnitude of the self-energy (with respect to the heavy-meson mass) in this sector, which is ultimately connected to the strength of the integrated scattering amplitudes, weighted by \gls{be} factors. One can see, for instance, that the $\bar{B}^{(*)}\pi$ amplitudes (with respect to the $\bar{B}^{(*)}$-meson mass) in the left panels of Fig.~\ref{fig:hot-ImT_bottom} are bigger than the $D^{(*)}\pi$ amplitudes (with respect to the $D^{(*)}$-meson mass) shown in the left panels of Fig.~\ref{fig:hot-ImT_charm}. 

Furthermore, it is important to note that the comparison of the results in the two sectors at the quantitive level may be affected by the details of the numerical calculations. For instance, in the \gls{itf} the integrals for the calculation of the thermal loop function and the self-energy should in principle extend to infinity. However, this is not possible to implement in the numerical calculations and one has to introduce a large enough cut-off in the numerical integration, taking also into account that the effective theory breaks down at high energies. Due to the heavier mass of the $\bar{B}^{(*)}$ mesons compared to the $D^{(*)}$ mesons, the energy integrals have to be extended up to larger values in the former case, introducing some uncertainties that are difficult to quantify.

We are not aware of any calculations for open-bottom mesons at finite temperature and vanishing baryon density with which the results presented here can be compared. 
There exist, however, some works that have addressed the properties of open-bottom (and also open-charm) mesons in nuclear matter. 
These works include nuclear mean-field calculations in matter \cite{Pathak:2014vra}, quark-meson coupling models~\cite{Tsushima:2002cc}, models based on $\pi$-exchange implementing heavy-quark symmetries \cite{Yasui:2012rw}, and \gls{qcd} sum-rule computations (see \cite{Gubler:2018ctz} and references therein). These works found that the general trends of the bottomed mesons at finite density are similar to those of the (anti)charmed mesons.

On the other hand, studies investigating the properties of heavy quarkonia in a hot medium are more abundant than in the case of open heavy flavor (for a comprehensive review see \cite{Rothkopf:2019ipj}). Most of them are dedicated to the in-medium properties of charmonia and used approaches that fall mainly into two categories: models based on chiral Lagrangians~\cite{Matinyan:1998cb,Lin:1999ad,Haglin:1999xs,Haglin:2000ar,Bourque:2004av,Abreu:2017cof,Cleven:2017fun}, and quark-model calculations~\cite{Wong:1999zb,Barnes:2003dg,Maiani:2004py,Maiani:2004qj,Zhou:2012vv}; but other approaches such as \gls{qcd} sum rules \cite{Duraes:2002ux,Duraes:2002px}, perturbative \gls{qcd} calculations~\cite{Song:2005yd}, and the comovers interaction model~\cite{Capella:2007jv} have also been explored. In the case of bottomonia, studies of this kind are more scarce \cite{Lin:2000ke,Wong:2001td,Abreu:2018mnc,Abreu:2021mrq}. Finite-temperature correlators and quarkonium spectral functions have also been calculated with \gls{lqcd}~\cite{Umeda:2002vr,Asakawa:2003re,Datta:2003ww,Aarts:2007pk,Jakovac:2006sf,Ding:2012sp,Ohno:2011zc,Larsen:2019bwy,Larsen:2019zqv} and potential models~\cite{Wong:2004zr,Mocsy:2005qw,Cabrera:2006wh,Alberico:2007rg,Mocsy:2007yj}.

\chapter{Open-charm Euclidean correlators}
\label{ch:lattice}

In the present chapter, we determine the Euclidean meson correlators for open-charm mesons and compare them to the \gls{lqcd} simulations of Ref.~\cite{Kelly:2018hsi}. To the best of our knowledge, this work is the only computation of Euclidean correlators of open-charm mesons at the moment of publication of this thesis. To this goal, we adapt our \gls{eft}-based calculations of open-charm spectral functions in a pionic bath described in Chapter~\ref{ch:hot-medium} to the use of the unphysical meson masses determined in Ref.~\cite{Kelly:2018hsi}. 

The chapter is organized as follows. After the introduction of Section~\ref{sec:latt-intro} and the brief overview of \gls{lqcd} methods given in Section~\ref{sec:latt-lqcd}, we introduce in Section~\ref{sec:latt-euclidean} the concept of the meson correlator at finite temperature and its relation to the meson spectral function. Our results are given in Section~\ref{sec:latt-results}, where we summarize our calculation for the open-charm spectral function within the \gls{eft} employing the meson unphysical masses reported in Ref.~\cite{Kelly:2018hsi} and compare our results for the Euclidean correlators with those from \gls{lqcd}. 
The work presented in this chapter was published in Ref.~\cite{Montana:2020var}.

%
%
%

\section{Introduction}
\label{sec:latt-intro}
There exist a few theoretical approaches that can be used to determine the features of the meson spectral function, but none of them is yet conclusive in the full range of energies and temperatures available in the experiments. Perturbative \gls{qcd} can be only applied at very large energies and/or temperatures \cite{Karsch:2000gi,Aarts:2005hg,Mocsy:2007yj}. 
The AdS/CFT duality has also been used to describe some features of the spectral function, but a clear correspondence with \gls{qcd} allowing quantitative studies is still missing \cite{Erdmenger:2007cm,CasalderreySolana:2011us}.

\Gls{lqcd} is a powerful tool to perform calculations from first principles for any energy and temperature, a priori. However, despite the recent progress in the determination of heavy-meson spectral properties in matter from \gls{lqcd} calculations (see Ref.~\cite{Rothkopf:2019ipj} for a recent review and references therein), there are still a few drawbacks that prevent lattice results from being decisive when determining the spectral features of heavy mesons. From \gls{lqcd} one can determine the so-called Euclidean meson correlators and the meson spectral functions are then extracted from them. The reconstruction of the spectral functions from the correlators turns out, however, to be rather complicated \cite{Jarrell:1996rrw}. Furthermore, the simulation of light quarks on the lattice is computationally very demanding, and usually larger (unphysical) masses are used. 


Hadronic models based on \glspl{eft} in matter offer a complementary strategy to \gls{lqcd} to determine the modification of the heavy-meson spectral features in a hot and/or dense medium \cite{Rapp:2011zz,Tolos:2013gta,Hosaka:2016ypm,Aarts:2016hap}. The matter below the deconfinement transition temperature consists of hadrons, essentially light mesons, in the low-density high-temperature regime. In this domain, we have obtained the thermal properties of pseudoscalar and vector charm mesons within a finite-temperature self-consistent unitarized approach based on a chiral effective field theory that implements heavy-quark spin symmetry \cite{Montana:2020lfi,Montana:2020vjg}, as has been described in the previous chapter. Once the spectral features are known, it is then possible to determine the corresponding Euclidean meson correlators and compare them to \gls{lqcd} results. In this way, the ill-posed extraction of the spectral function is avoided while testing directly the results from finite-temperature effective field theories against \gls{lqcd} simulations.

Therefore, only the interplay between these techniques may shed light on this issue. With this aim, we check, in this chapter, the results of the thermal \gls{eft} against \gls{lqcd} calculations. We perform this comparison at the level of Euclidean correlators, rather than the direct comparison of spectral properties, thus avoiding the complications arising from the reconstruction of the spectral functions from the lattice correlators.


\section{Overview of lattice QCD}
\label{sec:latt-lqcd}

In this section, we give a brief overview of the formulation and calculations of \gls{qcd} on the lattice, covering the main aspects necessary for the analysis and discussion of the results presented in Section~\ref{sec:latt-results} in comparison with the \gls{lqcd} data in Ref.~\cite{Kelly:2018hsi}. For a broader introduction to the field, as well as more theoretical and computational details, we refer the reader to textbooks such as Refs.~\cite{DeGrand:2006zz,Gattringer:2010zz} and reviews~\cite{Laermann:2003cv,pdg}. 

Originally proposed by Wilson in the 1970s \cite{Wilson:1974sk}, \gls{lqcd} is nowadays a well-established method to compute the properties, decays, and interactions of hadrons in the nonperturbative regime from the first principles governing the strong interactions between quarks and gluons.

In \gls{lqcd} the Euclidean spacetime is discretized on a four-dimensional grid or lattice of size $N_\sigma^3\times N_\tau$, with lattice spacing $a$. By taking the limit of vanishing $a$, the continuum \gls{qcd} is recovered.  
This formulation of \gls{qcd} on a discrete rather than continuous spacetime introduces a UV cut-off scale, which regularizes the theory in a natural way by restricting the highest momentum to $\Lambda<\frac{\pi}{a}$.

The volume and the temperature of the system are related to the lattice spacing by
\begin{equation}\label{eq:latt-VT}
 V=(aN_\sigma)^3\ , \quad T=\frac{1}{aN_\tau} \ .
\end{equation}

The Lagrangian of \gls{qcd} in Euclidean time follows from the \gls{qcd} Lagrangian in Eq.~(\ref{eq:intro-lagrangianQCD}), defined in Minkowski spacetime, by performing a Wick rotation ($t\rightarrow -\ii\tau$):
\begin{align}\label{eq:latt-ELag} \nonumber
 \mathcal{L}_{\textrm{QCD}}^{\textrm{E}}&=\mathcal{L}_{\textrm{F}}^{\textrm{E}}+\mathcal{L}_{\textrm{G}}^{\textrm{E}}\\
 &=\sum_f\psi_{f,a}(\slashed{D}_{ab}^{\textrm{E}}+m_f\delta_{ab})\psi_{f,b}-\frac14F_{\mu\nu}^AF_A^{\mu\nu} \ ,
\end{align}
with the subindex F denoting the fermionic part of the Lagrangian and G the gluonic one. We recall that the lowercase ($a,b$) and the uppercase ($A$) indices denote the color charge of the quark and gluon fields, respectively, and the subindex $f$ refers to the quark flavor.
The covariant derivative $\slashed{D}^{\textrm{E}}$ and the field-strength tensor are given by
\begin{align}
 \slashed{D}^{\textrm{E}}&=\gamma_\mu^{\textrm{E}}D_\mu^{\textrm{E}}=\left(\partial_\mu+\ii g_st^C\mathcal{A}_\mu^C \right)\gamma_\mu^{\textrm{E}} \  , \\
 F_{\mu\nu}^A&=\partial_\mu\mathcal{A}^A_\nu-\partial_\nu\mathcal{A}^A_\mu-g_sf^{ABC}\mathcal{A}^B_\mu\mathcal{A}^C_\nu \ ,
\end{align}
where $\gamma_\mu^{\textrm{E}}$ are the Euclidean Dirac matrices.

\begin{figure}[t!]
 \centering
\begin{tikzpicture}[scale=1.5,
                    grid/.style={very thin,gray},
                    decoration={markings,mark=at position 0.6 with {\arrow[line width=2pt,>=stealth,ctcolorblue]{>}}}]
    \draw[grid] (0,3,0) -- (3,3,0) -- (3,0,0);
    \draw[grid] (0,3,1) -- (3,3,1) -- (3,0,1);
    \draw[grid] (0,3,2) -- (3,3,2) -- (3,0,2);
    \draw[grid] (0,3,0) -- (0,3,2) -- (0,0,2);
    \draw[grid] (1,3,0) -- (1,3,2) -- (1,0,2);
    \draw[grid] (2,3,0) -- (2,3,2) -- (2,0,2);
    \draw[grid] (3,3,0) -- (3,3,2) -- (3,0,2);
    \draw[grid] (0,0,2) -- (3,0,2) -- (3,0,0);
    \draw[grid] (0,1,2) -- (3,1,2) -- (3,1,0);
    \draw[grid] (0,2,2) -- (3,2,2) -- (3,2,0);
    \foreach \x in {0,1,2}
        \foreach \y in {0,1,2}
            \foreach \z in {0,1}
        {
            \draw[grid] (\x,\y,\z) -- (\x+1,\y,\z);
            \draw[grid] (\x,\y,\z) -- (\x,\y+1,\z);
            \draw[grid] (\x,\y,\z) -- (\x,\y,\z+1);
        }
            \path[draw,line width=2pt,postaction=decorate,color=ctcolorblue] (0,0,2) -- (0,1,2);
            \path[draw,line width=2pt,postaction=decorate,color=ctcolorblue] (1,2,2) -- (1,2,1);
            \path[draw,line width=2pt,postaction=decorate,color=ctcolorblue] (1,2,1) -- (1,2,0);
            \path[draw,line width=2pt,postaction=decorate,color=ctcolorblue] (1,1,1) -- (1,1,2);
            \path[draw,line width=2pt,postaction=decorate,color=ctcolorblue] (1,1,2) -- (1,0,2);
            \path[draw,line width=2pt,postaction=decorate,color=ctcolorblue] (1,0,2) -- (2,0,2);
            \path[draw,line width=2pt,postaction=decorate,color=ctcolorblue] (2,0,2) -- (3,0,2);
            \path[draw,line width=2pt,postaction=decorate,color=ctcolorblue] (3,0,2) -- (3,0,1);
            \path[draw,line width=2pt,postaction=decorate,color=ctcolorblue] (3,0,1) -- (2,0,1);
            \path[draw,line width=2pt,postaction=decorate,color=ctcolorblue] (2,0,1) -- (2,1,1); 
            \path[draw,line width=2pt,postaction=decorate,color=ctcolorblue] (2,3,0) -- (2,2,0);
            \path[draw,line width=2pt,postaction=decorate,color=ctcolorblue] (2,2,0) -- (3,2,0);
            \path[draw,line width=2pt,postaction=decorate,color=ctcolorblue] (3,2,0) -- (3,3,0);
            \path[draw,line width=2pt,postaction=decorate,color=ctcolorblue] (3,3,0) -- (2,3,0);
            \draw[ctcolormagenta,fill=ctcolormagenta] (1,1,1) circle(1mm);
            \draw[ctcolormagenta,fill=ctcolormagenta] (2,1,1) circle(1mm);
            \draw[ctcolormagenta,fill=ctcolormagenta] (1,2,2) circle(1mm);
            \draw[ctcolormagenta,fill=ctcolormagenta] (1,2,0) circle(1mm);
            \draw[ctcolormagenta,fill=ctcolormagenta] (0,1,2) circle(1mm);
            \draw[ctcolormagenta,fill=ctcolormagenta] (0,0,2) circle(1mm);
            \node [left] at (-0.1,1,2) {\(\psi_x\)};
            \node [left] at (-0.1,0,2) {\(\bar{\psi}_x\)};
            \node [left] at (-0.1,0.5,2) {\(U_{x,\mu}\)};
            \node [right] at (3.1,2.5,0) {\(U_{x,\mu\nu}\)};
\end{tikzpicture}
\caption{Sketch of a three-dimensional lattice showing the quark fields lying on the grid sites (magenta circles) and gluon fields represented by the links connecting the grid sites (blue lines). The closed loop in the upper right corner corresponds to a plaquette. }
\label{fig:latt-lattice}
\end{figure}
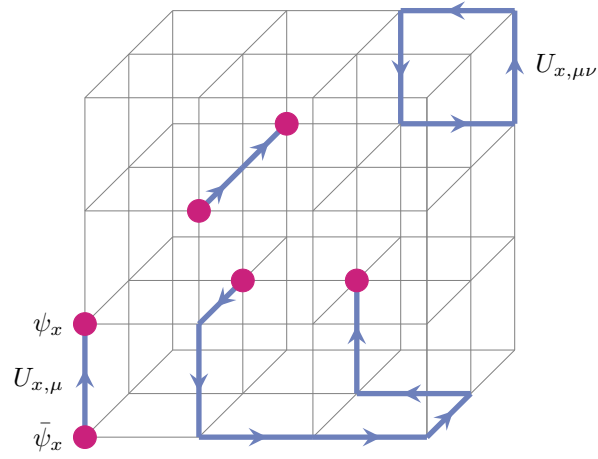

On the lattice, quark fields $\bar\psi_x$, $\psi_x$ live on the sites (labeled with $x\equiv(t,\vec{x})$), whereas the gluons reside on the links connecting neighboring sites, that is, $x$ to $x+a\hat{\mu}$ in the $\mu$ direction in the lattice, and are represented by the gauge links, 
\begin{equation}
 U_{x,\mu}=\exp\left(\ii ag_s\mathcal{A}_\mu(x)\right) \ .
\end{equation}
It is easy to see that $U_{x,\mu}U_{x+a\hat{\mu},-\mu}=\mathbb{1}$. Furthermore $U^\dagger U=\mathbb{1}$, which gives $U_{x,\mu}^\dagger=U_{x+a\hat{\mu},-\mu}$.
The smallest closed loop of gauge links is called \textit{plaquette}:
\begin{align}\nonumber
 U_{x,\mu\nu}&=U_{x,\mu}U_{x+a\hat{\mu},\nu}U_{x+a(\hat{\mu}+\hat{\nu}),-\mu}U_{x+a\hat{\nu},-\nu} \\
 &=U_{x,\mu}U_{x+a\hat{\mu},\nu}U_{x+a\hat{\nu},\mu}^\dagger U_{x,\nu}^\dagger \ .
\end{align}
A schematic illustration of quark and gluon fields on the lattice is given in Fig.~\ref{fig:latt-lattice}, where we have also represented a plaquette.

The lattice regularized grand-canonical partition function $ \mathcal{Z}$ is given in the Euclidean path-integral formalism by
\begin{equation}\label{eq:latt-partfunc}
 \mathcal{Z}=\int\mathcal{D}[U]\mathcal{D}[\bar{\psi},\psi]e^{-S_{\textrm{F}}[U,\bar\psi,\psi]-S_{\textrm{G}}[U]} \ ,
\end{equation}
with the integration measure $\int\mathcal{D}[U]\mathcal{D}[\bar{\psi},\psi]\equiv\int\prod_{x,\mu}dU_{x,\mu}\prod_{x,f}d\psi_{x,f}d\bar{\psi}_{x,f}$ running over all possible configurations of fields.
Each configuration is weighted by the Boltzmann factor $e^{-S}$, where $S_{\textrm{F}}[U,\bar\psi,\psi]$ is the fermion part and $S_{\textrm{G}}[U]$ the gluon part of the \gls{qcd} action, related to the discretized version of the Euclidean Lagrangian in Eq.~(\ref{eq:latt-ELag}) by $S=\int d^4x\,\mathcal{L}^{\textrm{E}}_{\textrm{QCD}}$. In finite temperature studies, due to the relation between the temperature of the system and the temporal lattice extent of Eq.~(\ref{eq:latt-VT}), $x_0$ is taken to run along a temperature direction rather than a time direction, that is, $S=-\int_0^{1/T} dx_0\int_V d^3x\, \mathcal{L}^{\textrm{E}}_{\textrm{QCD}}$.

The expectation value of a physical observable $\mathcal{O}$ is then obtained through
\begin{equation}\label{eq:latt-observable}
\langle\mathcal{O}\rangle=\frac{1}{\mathcal{Z}}\int\mathcal{D}[U]\mathcal{D}[\bar{\psi},\psi]\mathcal{O}e^{-S_{\textrm{F}}[U,\bar\psi,\psi]-S_{\textrm{G}}[U]} \ .
\end{equation}

The simplest gauge-invariant gluonic action, usually called the Wilson gauge action, is given in terms of the plaquette,
\begin{equation}\label{eq:latt-gluonaction}
 S_{\textrm{G}}=\beta\sum_{x,\mu\leq\nu}\left[1-\frac{1}{3}\textrm{Tr\,}\textrm{Re\,}U_{x,\mu\nu} \right] \ ,
\end{equation}
with $\beta=6/g^2$ being the lattice coupling. While, in the limit $a\rightarrow 0$, this simple parametrization reproduces the continuum expression,
discretization effects are of order $\mathcal{O}(a^2)$. The incorporation of larger loops of six links (as opposed to the four links in the plaquette) defines the improved (or Symanzik) gauge action~\cite{Symanzik:1983dc,Symanzik:1983gh} that is commonly used.

As for the fermionic action, one can consider the na\"ive discretization of the Euclidean action in the continuum
 by replacing the covariant derivative 
 with a finite difference, 
\begin{equation}
 D^{\textrm{E}}_\mu\psi(x)\rightarrow\frac{1}{2a}\left(U_{x,\mu}\psi_{x+a\hat{\mu}}-U_{x,\mu}^\dagger\psi_{x-a\hat{\mu}}\right) \ ,
\end{equation}
where the factors $U_{x,\mu}$ ensure that the fermionic action is gauge invariant. However, this fermionic action suffers from the so-called fermion doubling problem, which refers to the presence of unwanted unphysical poles in the fermion propagator, and is closely related to the explicit breaking of chiral symmetry. To deal with the undesired doublers, in the Wilson fermion formulation~\cite{Wilson:1974sk}, for instance, an extra term is added to the fermion action in such a way that the mass of the doublers increases and they decouple in the continuum limit. A variant of this formulation comprises Clover fermions, for which the fermion action contains the Wilson and Clover term in addition to the na\"ive discretization of the Dirac operator~\cite{Sheikholeslami:1985ij}. It is also very extended the use of staggered fermions~\cite{Kogut:1974ag}, with which the doubler problem is not eliminated but reduced, but a remnant of chiral symmetry is preserved. Domain Wall fermions~\cite{Kaplan:1992bt,Shamir:1992im,Shamir:1993zy,Furman:1994ky} and overlap fermions~\cite{Neuberger:1997fp,Neuberger:1998wv}, on the other hand, allow for the preservation of the exact chiral symmetry but are very time consuming.

After a suitable choice of the actions, the Gaussian integral over the quark fields can be performed, leading to the lattice partition function
\begin{equation}\label{eq:latt-partfunc2}
 \mathcal{Z}=\int\mathcal{D}[U]e^{-S_{\textrm{G}}[U]}\textrm{det\,}[M] ,
\end{equation}
where $M=(\slashed{D}^{\textrm{E}} +m_f-\mu\gamma_0)$ is the fermion matrix, in which we have introduced the quark chemical potential, $\mu=\mu_B/3$. Its inverse represents the fermion propagator. The effects of the sea quarks are contained in $\{\textrm{det\,}[M]\}$. 

For $\mu=0$, the expectation value of physical quantities, such as $\mathcal{Z}$, or observables $\langle \mathcal{O}\rangle$, can be calculated by Monte Carlo methods. Simulations at finite baryon density, $\mu\neq 0$, have to deal with the so-called \textit{sign problem}. There are nevertheless some methods to get around this problem, including Taylor expansion around $\mu=0$, imaginary chemical potential simulations, the complex Langevin approach, and reweighting from $\mu=0$, that are, however, only applicable at small $\mu$ (see, for example, Refs.~\cite{Muroya:2003qs,Berger:2019odf} for reviews on finite-density \gls{lqcd}).

Numerical simulations of \gls{qcd} on a lattice are computationally very demanding and, thus, the power of \gls{lqcd} calculations is limited, in practice, by the availability of computational resources and the efficiency of the algorithms. The evaluation of the fermion determinant is the most computationally expensive part and, therefore, simulations with dynamical quarks turn out to be very resource demanding. For this reason, the ``quenched approximation'', which neglects quark-loop contributions by taking $\textrm{det}\,M=1$, is sometimes used in exploratory \gls{lqcd} studies. 

In general, \gls{lqcd} calculations are affected by both statistical and systematic errors. While the former arise from the use of importance sampling in Monte Carlo methods, the latter are related, for example, to the use of nonzero values of $a$ and the finite size of the lattice. The observables have thus to be computed for various lattice spacings and volumes, and extrapolated to the continuum and infinite-volume limits, for a fixed value of $T$. In addition, there are lattice artifacts, that is, cut-off effects, that depend on the particular discretization chosen for the \gls{qcd} action.

The basic input parameters for calculations on the lattice are the strong coupling constant $\alpha_s=g_s^2/(4\pi)$ and the quark masses. The lattice spacing $a$, which depends on the choice of the coupling constant, is then determined from a dimensionful quantity measured by experiments. For instance, from the mass of a hadron $H$ through $a=(am_H)^{\textrm{lat}}/m_H^{\textrm{exp}}$, with $(am_H)^{\textrm{lat}}$ the value of the hadron mass obtained on the lattice in lattice units. A common choice is the mass of the $\Omega^-$ baryon.

As for the quark masses, current \gls{lqcd} simulations often employ $m_u=m_d=<m_s$ ($N_f=2+1$), or even include the $c$ quark mass $m_c$ ($N_f=2+1+1$), and their values are tuned to reproduce the physical values of the ratios between the meson masses ($m_\pi$, $m_K$ and $m_{\eta_c}$) and the quantity used to set the scale (e.g. $m_\Omega$).
Simulations with physical light meson masses are computationally expensive and require large lattices.
In practice, at least until recently, typical \gls{lqcd} simulations are done at some unphysically heavy quark mass, and the results are extrapolated to the physical point, usually using fits inspired in \gls{chpt}. In the last decade, advances in both algorithms and computers, have allowed \gls{lqcd} calculations to approach the physical quark masses.

In finite temperature studies of heavy quarks on the lattice, sometimes anisotropic lattices are used, on which the spacing in the temporal direction, $a_\tau$, is smaller than that in the spatial directions, $a_\sigma$, so as to have a fine enough time discretization, although cut-off effects depend on the spatial lattice spacing. Then, the anisotropy parameter is defined as $\xi=a_\sigma/a_\tau>1$.

\section{Meson Euclidean correlators and spectral functions}
\label{sec:latt-euclidean}

The primary tools in \gls{lqcd} calculations are \textit{Euclidean correlators} of some operators $\widehat{O}$, described by the expectation value:
\begin{align}\nonumber
 \langle \widehat{O}_1(\tau,\vec{x})&\widehat{O}_2(0,\vec{0})\rangle=\frac{1}{\mathcal{Z}}\int\mathcal{D}[U]\mathcal{D}[\bar{\psi},\psi]  \\ &\times
 \widehat{O}_2[U,\bar\psi,\psi]\widehat{O}_1[U,\bar\psi,\psi]
  e^{-S_{\textrm{F}}[U,\bar\psi,\psi]-S_{\textrm{G}}[U]} \ ,
\end{align}
with $ \mathcal{Z}$ being the partition function defined in Eq.~(\ref{eq:latt-partfunc}).

For a meson with quantum numbers $H$, the meson (quark-antiquark pair) operator to consider is the following:
\begin{equation}
 J_H(\tau,\vec{x})=\bar{\psi}_f(\tau,\vec{x})\Gamma_H\psi_f(\tau,\vec{x}) \ ,
\end{equation}
where $\Gamma_H=\mathds{1},\gamma_5,\gamma_\mu,\gamma_5\gamma_\mu$ correspond to the scalar, pseudoscalar, vector and axial vector channels, respectively, and $f$ refers to the flavor of the valence quark. 

After carrying out the explicit integration over fermion fields, the propagation of a meson from time $t=0$ to $\tau$ is given by the Euclidean time correlator
\begin{align}\nonumber
 \langle J(\tau,\vec{x}\,)&J(0,\vec{0}\,)\rangle=-\frac{1}{\mathcal{Z}}\int\mathcal{D}[U]e^{-S_{\textrm{G}}[U]}\,\textrm{det\,}[M] \\ &\times
 \textrm{Tr\,}[\Gamma_HM^{-1}(0,\vec{0}\,;\tau,\vec{x}\,)\Gamma_HM^{-1}(\tau,\vec{x};0,\vec{0}\,)] \ .
\end{align}

In the case of meson correlators at a finite temperature, one has to consider that the temperature of the system on the lattice is related to the temporal extent through Eq.~(\ref{eq:latt-VT}). 

\begin{figure}[t!]
\centering
\includegraphics[scale=0.6]{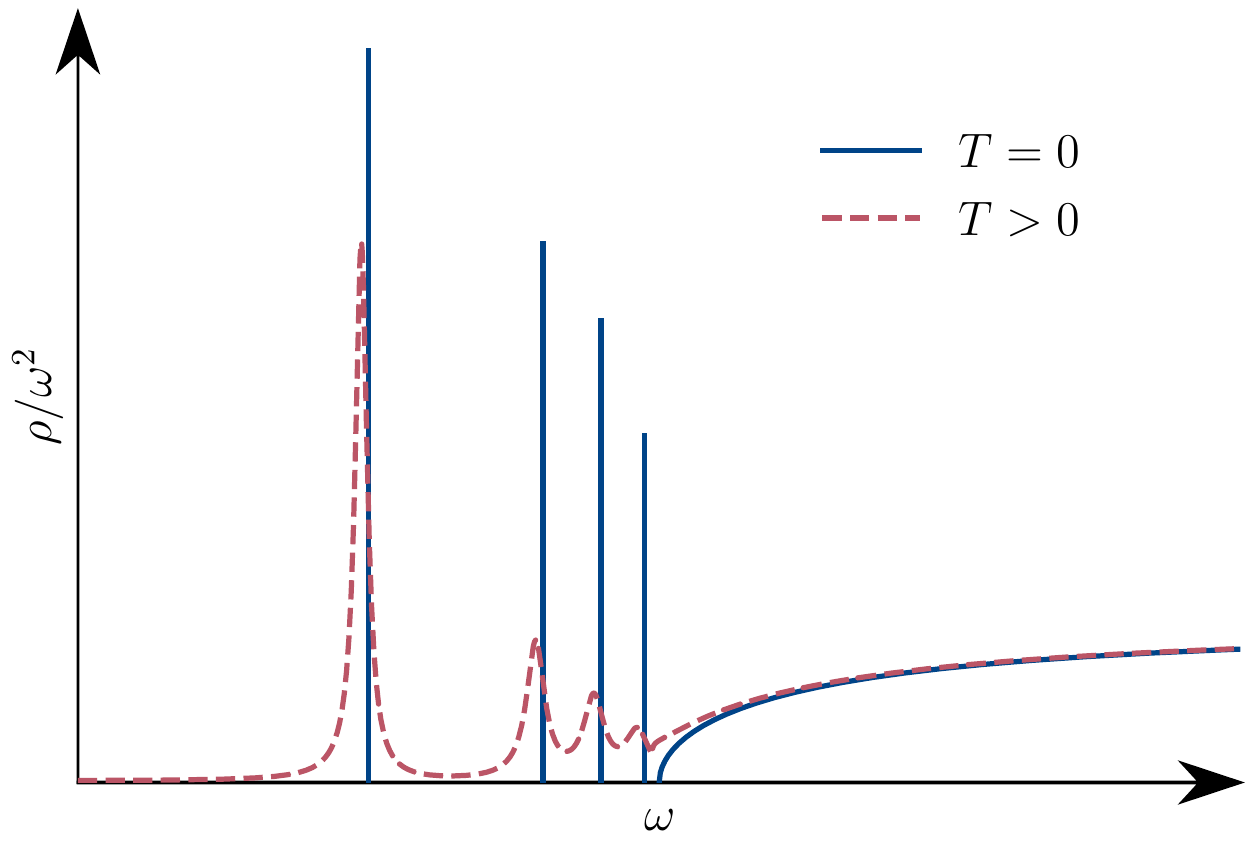}
  \caption{Schematic picture of the meson spectral function at $T=0$ (blue solid line) and at finite temperature (red dashed line).}
  \label{fig:latt-sketch}
\end{figure}

As we have seen in Chapter~\ref{ch:hot-medium}, the in-medium properties of a meson are incorporated into its spectral function, defined as the imaginary part of the retarded propagator (see Eq.~(\ref{eq:hot-specfunc})). This allowed us to obtain the thermal evolution of the ground-state mass and thermal width.
However, the meson spectral function at finite temperature contains information not only on the ground state but also on the masses and widths of the possible excited bound states, as well as the continuum of scattering states. A schematic picture is shown in Fig.~\ref{fig:latt-sketch}. At $T=0$ the spectral function results from the contribution of different delta functions corresponding to the ground state of mass $m$ and the bound excited states, and a continuum distribution starting at $\omega\geq 2m$ for 2-particle states. At finite temperature, one expects the masses to be modified, as well as a broadening of 1-particle states to take place.

The Euclidean temporal correlator in momentum space,
\begin{equation}
 G_E(\tau,\vec{p}\,;T)\equiv\int d^3\vec{x}\,e^{-i\vec{p}\cdot\vec{x}}\langle J(\tau,\vec{x})J(0,\vec{0}\,)\rangle \ ,
\end{equation}
is related to the spectral function $\rho(\omega,\vec{p}\,;T)$ through the convolution with a known kernel:
\begin{equation}\label{eq:latt-corr_to_sfunc}
 G_E(\tau,\vec{p}\,;T)=\int_0^\infty d\omega \,K(\tau,\omega;T)\,\rho(\omega,\vec{p}\,;T) \ ,
\end{equation}
with \vspace{-1mm}
\begin{align}\label{eq:latt-kernel}\nonumber
 K(\tau,\omega;T)&=\frac{\cosh\left[\omega\left(\tau-\frac{1}{2T}\right)\right]}{\sinh \left(\frac{\omega}{2T}\right)}=\\ &=e^{\omega\tau}f(\omega,T)+e^{-\omega\tau}[1+f(\omega,T)] \ ,
\end{align}
where $f(\omega,T)$ is the \gls{be} statistical factor.

In the following, we omit the dependence on the momentum $\vec{p}$, as we focus on spectral functions with $\vec{p}=\vec{0}$ for simplicity, and make the identifications $G_E(\tau;T)\equiv G_E(\tau,\vec{p}=\vec{0}\,;T)$ and $\rho(\omega\,;T)\equiv\rho(\omega,\vec{p}=\vec{0}\,;T)$.

In \gls{lqcd} simulations, the values of the Euclidean correlator are obtained for a set of points in Euclidean time, $\tau=\tau_i$, that is, $\{\tau_i,G_E(\tau_i;T)\}$ for $i=1,...,N_\tau$ and $\tau_i\in[0,1/T]$. In addition, the lattice data $G_E(\tau_i;T)$ have a statistical error due to the fact that only a finite number of gauge configurations can be generated in a Monte Carlo simulation. The inversion of Eq.~(\ref{eq:latt-corr_to_sfunc}) to extract a continuous spectral function $\rho(\omega;T)$ from such a limited number of data points\footnote{Equation~(\ref{eq:latt-kernel}) is symmetric in the Eucliean time at $\tau=1/(2T)$ and thus only half of the data points of the Euclidean correlator, i.e. $N_\tau/2$, provide independent information to extract the spectral function.} with noise is an ill-posed problem. Two methods are usually employed to try to circumvent this problem, both taking specific assumptions on the shape of the spectral function:
\begin{enumerate}
 \item Bayesian methods like the Maximum Entropy Method (MEM)~\cite{Nakahara:1999vy,Asakawa:2000tr} or stochastic reconstruction methods 
\cite{Ding:2017std} perform the kernel inversion by statistically inferring the most probable spectral function.
  \item Fitting the lattice data with suitable \textit{Ans\"atze} for the spectral function, incorporating bound states, a continuum, or perturbative input \cite{Burnier:2017bod}.
\end{enumerate}
The fact that a priori assumptions about the spectral functions are needed makes the determination of their shape and details at finite temperature very challenging, as we do not have much prior information on them.

Besides the commonly used approaches mentioned above, the problem of the reconstruction of the spectral function from the Euclidean correlator has also been recently explored with machine learning techniques~\cite{PhysRevB.98.245101,PhysRevLett.124.056401,Kades:2019wtd,Chen:2021giw}, which have turned out to reach a reconstruction accuracy comparable to the methods based on Bayesian interference, and even potentially superior at large noise levels~\cite{Kades:2019wtd,Chen:2021giw}.

In addition to the statistical errors, the systematic errors tied to the finite lattice spacing, the finite lattice volume, and the large quark masses employed in the Monte Carlo calculations make the results obtained on the lattice differ from the desired physical ones. Although these errors can be minimized by extrapolating to the continuum, to the infinite volume limit, and to the physical mass limit, they are usually tedious and not always performed in lattice data analyses.

By inspecting Eq.~(\ref{eq:latt-corr_to_sfunc}), one can see that the temperature dependence of the correlators does not only come from the spectral function but the integration kernel also carries an inherent dependence on the temperature. When directly comparing correlation functions at different temperatures, one may want to discern the differences due to the modification of the spectral function with temperature alone. To do so, it is useful to define the so-called \textit{reconstructed correlator} at a reference temperature $T_r$,
\begin{equation}\label{eq:latt-corr_reconstructed}
 G_E^r(\tau;T,T_r)=\int_0^\infty d\omega K(\tau,\omega;T)\rho(\omega;T_r) \ .
\end{equation}
The integration kernel is the same as that of $G_E(\tau;T)$ and, therefore, any difference when comparing $G_E(\tau;T)$ and $G_E^r(\tau;T,T_r)$ arises from differences in the spectral functions at $T$ and $T_r$. The value of $T_r$ is usually chosen to correspond to a temperature at which the shape of the spectral function is better known, so one can reliably trust the spectral function obtained from the lattice correlator. Thus, the lowest temperature available is usually chosen. Then, a ratio of unity between the Euclidean and the reconstructed correlators, for instance, indicates that there is no in-medium modification.

\section{Results}
\label{sec:latt-results}

In this chapter, we adopt the lattice setup of Ref.~\cite{Kelly:2018hsi}, where an anisotropic lattice, with spacing in the temporal direction $a_\tau^{-1}=5.63~\textrm{GeV}$ and anisotropy parameter $\xi=3.5$, is used. The ensembles employed contain dynamical light and strange quarks, with unphysical masses for the two mass-degenerate light quarks and roughly physical values for the strange and charm quarks \cite{Aarts:2014nba}. The resulting masses of the light and charm mesons on the lattice are listed in Table~\ref{tab:latt-masses}. While the open-charm mesons have almost the same masses as at the physical point~\cite{pdg}, the light mesons are substantially heavier, especially the pion, which has a mass almost three times larger in the lattice. The pseudocritical temperature determined for this configuration is $T_c=185$~MeV and the correlators have been calculated for temperatures shown in Table~\ref{tab:latt-temp}, which are directly obtained from the number of points in the Euclidean time direction $N_\tau$ through Eq.~(\ref{eq:latt-VT}). For further details, we refer the reader to Ref.~\cite{Kelly:2018hsi} and references therein.

\begin{table}[htbp!]
\setlength{\tabcolsep}{10pt}
\renewcommand{\arraystretch}{1.2}
\begin{tabular}{l c c c c c}
 \hline
 & $m_\pi$ (MeV) & $m_K$  (MeV) & $m_\eta$  (MeV) & $m_D$  (MeV) & $m_{D_s}$  (MeV) \\
 \hline
Lattice & $384$ & $546$ & $589$ & $1880$ & $1943$ \\
Physical & $138$ & $496$ & $548$ & $1867$ & $1968$ \\
\hline
\end{tabular}
\centering
\caption{Values of the masses of the light mesons ($\pi$, $K/\bar{K}$, $\eta$) and the open-charm mesons ($D$, $D_s$) in the lattice setup of Ref.~\cite{Kelly:2018hsi}, and at the physical point~\cite{pdg}.}
\label{tab:latt-masses}
\end{table}

\begin{table}[htbp!]
\setlength{\tabcolsep}{10pt}
\renewcommand{\arraystretch}{1.2}
\begin{tabular}{c c c c}
 \hline
$N_\sigma$ & $N_\tau$ & $T$ (MeV) & $T/T_c$ \\
\hline
$16$ & $128$ & $44$ & $0.24$ \\
$24$ & $40$ & $141$ & $0.76$ \\
$24$ & $36$ & $156$ & $0.84$ \\
$24$ & $32$ & $176$ & $0.95$ \\
$24$ & $28$ & $201$ & $1.09$ \\
$24$ & $24$ & $235$ & $1.27$ \\
$24$ & $20$ & $281$ & $1.52$ \\
$24$ & $16$ & $351$ & $1.90$ \\
\hline
\end{tabular}
\centering
\caption{Lattice volumes $N_\sigma^3\times N_\tau$ and temperatures used in the work of Ref.~\cite{Kelly:2018hsi}, where the lattice spacing in the Euclidean time direction is $a_\tau^{-1}=5.63~\textrm{GeV}$ and the pseudocritical temperature is $T_c=185$~MeV.}
\label{tab:latt-temp}
\end{table}

\subsection{Thermal EFT spectral functions at unphysical meson masses}
\label{subsec:latt-results-sfunc}

The calculation of the spectral functions of the $D$ and $D_s$ mesons in a pionic bath at finite temperature within the \gls{itf} has been discussed in Chapter~\ref{ch:hot-medium}. The unitarized amplitudes of the scattering of heavy mesons off light pseudoscalar mesons ($\pi$, $K$, $\bar{K}$, and $\eta$) have been obtained from solving the coupled-channel \gls{bs} equation with thermal loop functions dressed with the heavy-meson spectral function, within a self-consistent approach.
The effective Lagrangian describing the interactions has been detailed in Section~\ref{sec:free-mm}. See also Refs.~\cite{Montana:2020lfi,Montana:2020vjg}. 

In the hadronic effective approach presented in the previous chapters, the values of the unphysical meson masses of the lattice setup in Ref.~\cite{Kelly:2018hsi} can be implemented straightforwardly, since the meson masses in the vacuum are input parameters of the model. This way, we can directly compare with the \gls{lqcd} calculations.

The unitarized $T$-matrix approach can also be extended to finite volumes and implement the subsequent periodic boundary conditions present in the lattice by replacing the three-momentum  integral in the calculation of the vacuum two-meson propagator by a discrete sum over momenta, $\int d^3q/(2\pi)^3\rightarrow\sum_n1/L^3$~\cite{MartinezTorres:2011pr,Doring:2011vk}. Similar modifications can be also implemented, in principle, in the expressions for the thermal propagator and the meson self-energy given in the previous chapter. While this might reduce the sources for discrepancies between our calculations with \gls{lqcd}, it is out of the scope of the work in this thesis.

Let us reproduce, for completeness, the expression of the spectral function of the charm meson defined in terms of its propagator in a hot medium:
\begin{equation}\label{eq:latt-Sfunc}
 S_D(\omega,\vec{q}\,;T)=-\frac{1}{\pi}\textrm{Im\,}\mathcal{D}_D(\omega,\vec{q}\,;T)=-\frac{1}{\pi}\textrm{Im\,}\left(\frac{1}{\omega^2-\vec{q}\,^2-M_D^2-\Pi_D(\omega,\vec{q}\,;T)}\right) \ ,
\end{equation}
where the self-energy, $\Pi_D$, is obtained from closing the pion line in the $T_{D\pi\rightarrow D\pi}$ matrix element of the unitarized amplitude. Heavier light meson contributions such as the kaons contribution to the dressing of the $D$ meson are neglected, as their abundance in the hot medium is suppressed. Indeed, in the previous chapter, we have quantified the effect of the kaons in the medium to be around $10\%$ of the mass shift of the $D$ mesons induced by the pion-only hot medium at $T=150$~MeV, at the physical point (see also Ref.~\cite{Montana:2020vjg}). On the lattice, the mass difference between the pions and the heavier light mesons is reduced, as the mass of the pion is considerably larger on the lattice and the kaon and $\eta$-meson masses are more similar to their physical values (see Table~\ref{tab:latt-masses}). Despite this, the kaons are still more than $150$~MeV heavier than the pions in the lattice, and, thus, the pionic contribution to the self-energy is certainly the most relevant.

\begin{figure}[b!]
\centering
\includegraphics[width=\textwidth]{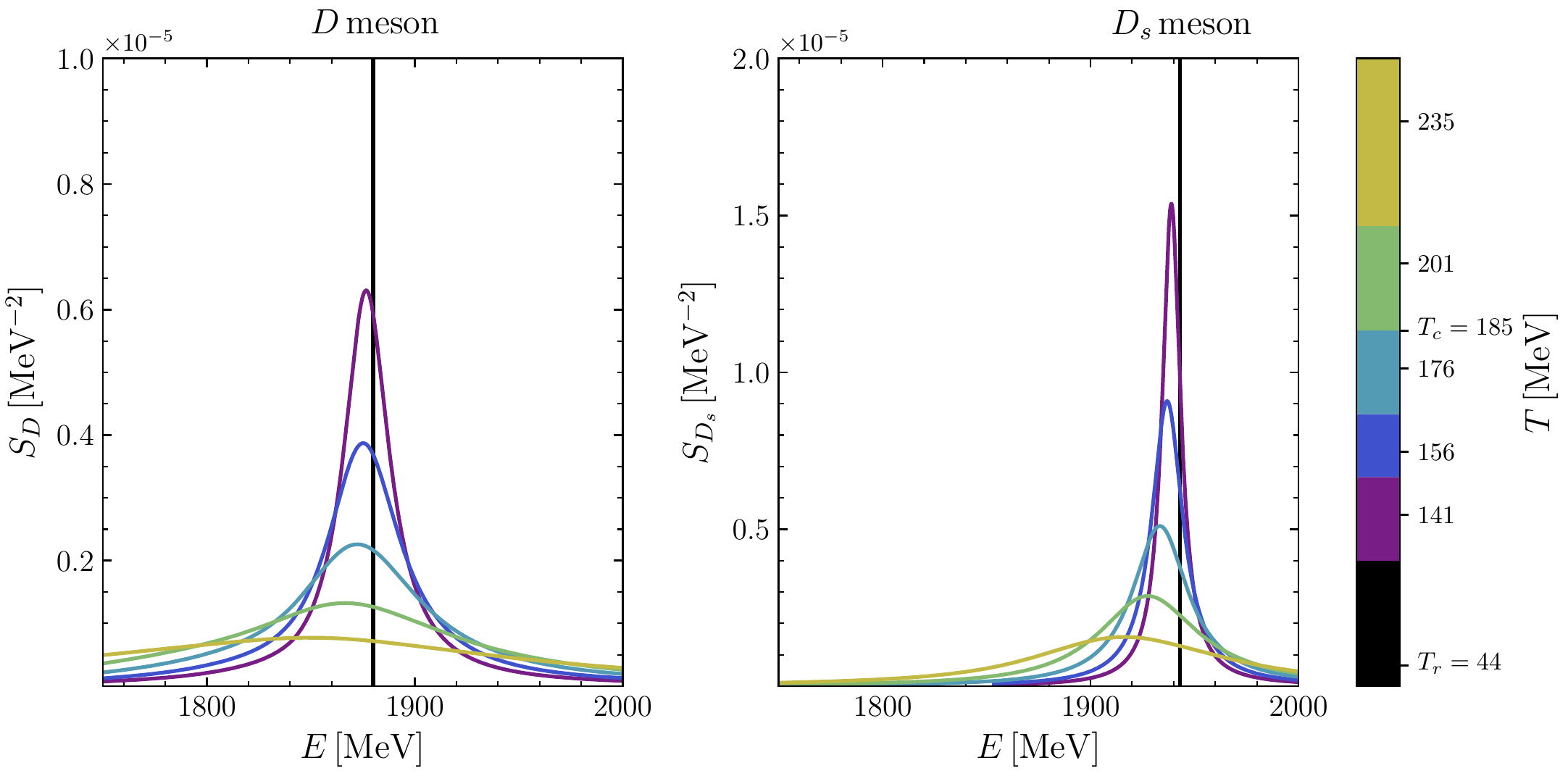}
  \caption{Spectral functions of the $D$ meson (left panel) and $D_s$ meson (right panel) obtained from their effective interaction with unphysically heavy pions at finite temperature.}
  \label{fig:latt-spectral_functions}
\end{figure}

The spectral functions of the $D$ and $D_s$ mesons obtained from their interaction with the unphysically heavy pions in a hot medium are shown in Fig.~\ref{fig:latt-spectral_functions}.  In these figures, we show the spectral functions for $D$ (left panel) and $D_s$ (right panel) as a function of the meson energy for the different temperatures used in Ref.~\cite{Kelly:2018hsi}. As discussed in the previous chapter with physical meson masses, we see a broadening of both spectral functions with increasing temperatures due to the larger available phase space for decay at finite temperatures. Moreover, the maximum of both spectral functions slightly moves to lower energies with temperature due to the attractive character of the heavy-light meson-meson interaction.

A study of net charm fluctuations \cite{Bazavov:2014yba} suggests that open-charm hadrons start to dissolve already close to the chiral crossover. Although the deconfined degrees of freedom necessary to investigate the melting of charm hadrons are absent in the model, we still show the spectral functions coming from our hadronic model for temperatures above the lattice pseudocritical temperature $T_c$ in order to explore the validity of our results for the correlators at those temperatures, as data below $T_c$ is scarce. 

\subsection{Euclidean correlators and comparison with lattice QCD}
\label{subsec:latt-results-corr}
 
Once the $D$ and $D_s$ spectral functions at finite temperature are known, we can obtain the corresponding Euclidean correlators from
Eqs.~(\ref{eq:latt-corr_to_sfunc}) and (\ref{eq:latt-corr_reconstructed}), and compare them to \gls{lqcd} calculations. It is important to notice that the spectral function $S_D(\omega;T)$ of Eq.~(\ref{eq:latt-Sfunc}), displayed in Fig.~\ref{fig:latt-spectral_functions}, differs from $\rho(\omega;T)$, entering in Eq.~(\ref{eq:latt-corr_to_sfunc}). This is due to the fact that the former contains the ground-state peak and an additional continuum, corresponding to $D\pi\pi$ scattering states in the case of the $D$ spectral function and $D\pi K$ states in the case of the spectral function of the $D_s$, while the latter contains all possible quark-antiquark ($c\overline{l}$ or $c\overline{s}$) states. 
 
Furthermore, the dimensions of $S_D(\omega;T)$ are MeV$^{-2}$ while the dimensions of $\rho(\omega;T)$ are MeV$^2$.  
The two quantities are related through the fourth power of the charm meson mass  \cite{Klingl:1996by,Gubler:2016itj}: 
\begin{equation}
 \rho_\textrm{gs}(\omega;T)=M_D^4S_D(\omega;T) \ ,
 \label{eq:latt-rel}
\end{equation}
with $M_D$ the vacuum value in the PDG for the mass of the $D$ (or $D_s$) meson. In Refs.~ \cite{Klingl:1996by,Gubler:2016itj} the authors obtained the relation between the electromagnetic current-current correlation in matter and the vector meson self-energy, and hence the meson spectral function, based on \gls{vmd}. In Eq.~(\ref{eq:latt-rel}) we have similarly connected the Euclidean current-current correlator of the lattice simulations with the open-charm spectral function resulting from the chiral effective theory that implements \gls{hqss}. Also, in \gls{lqcd} studies the reconstructed spectral function is usually identified with the dimensionless quantity $\rho(\omega;T)/\omega^2$.

\begin{figure}[t!]
\centering
\includegraphics[width=0.9\textwidth]{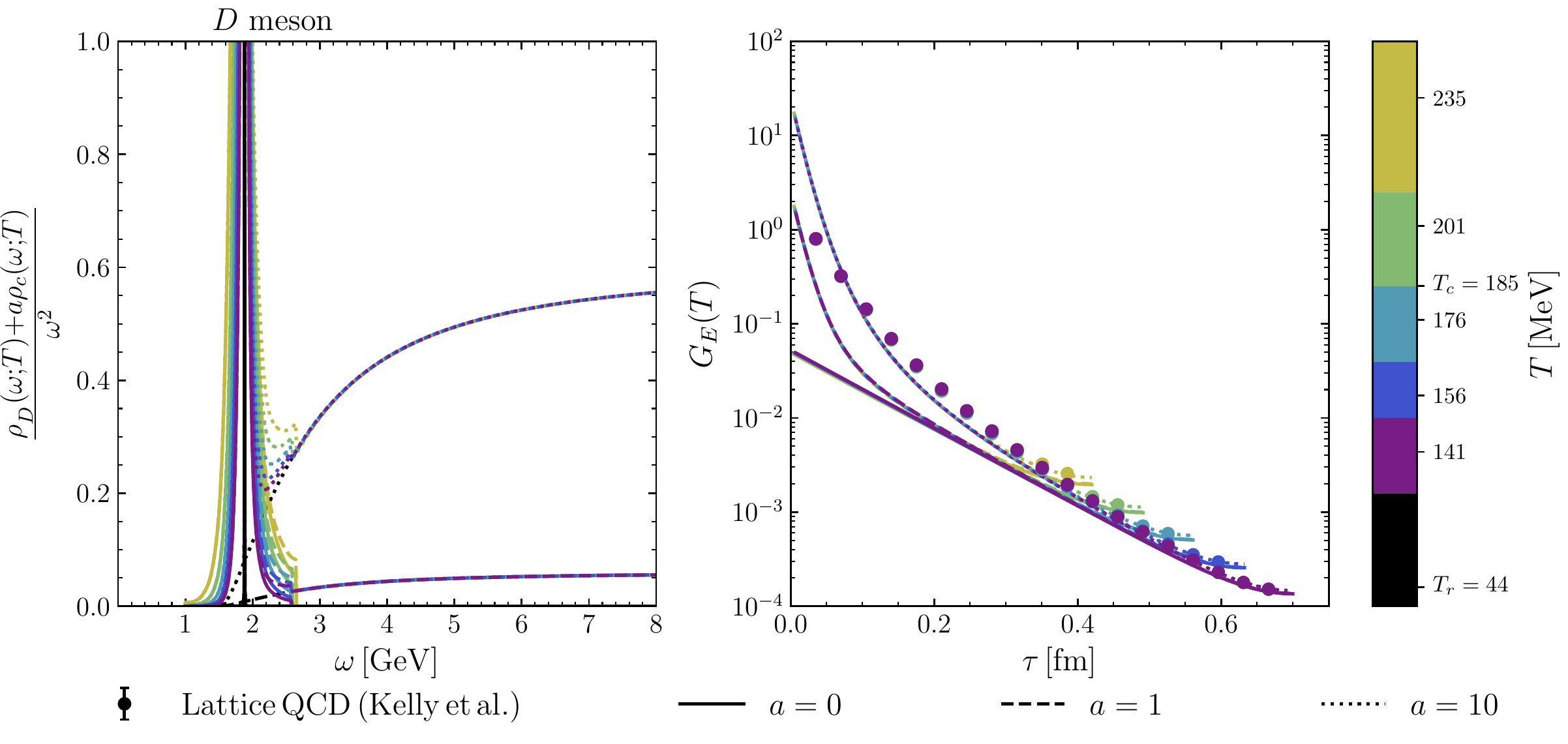}
  \caption{The spectral functions in the $C=1$, $S=0$ sector (left panel) and the Euclidean correlators (right panel) at different temperatures and values of the weight of the continuum spectral function, that is, the value of parameter $a$.}
  \label{fig:latt-corr_D_unphys}
\end{figure}

\begin{figure}[t!]
\centering
\includegraphics[width=0.9\textwidth]{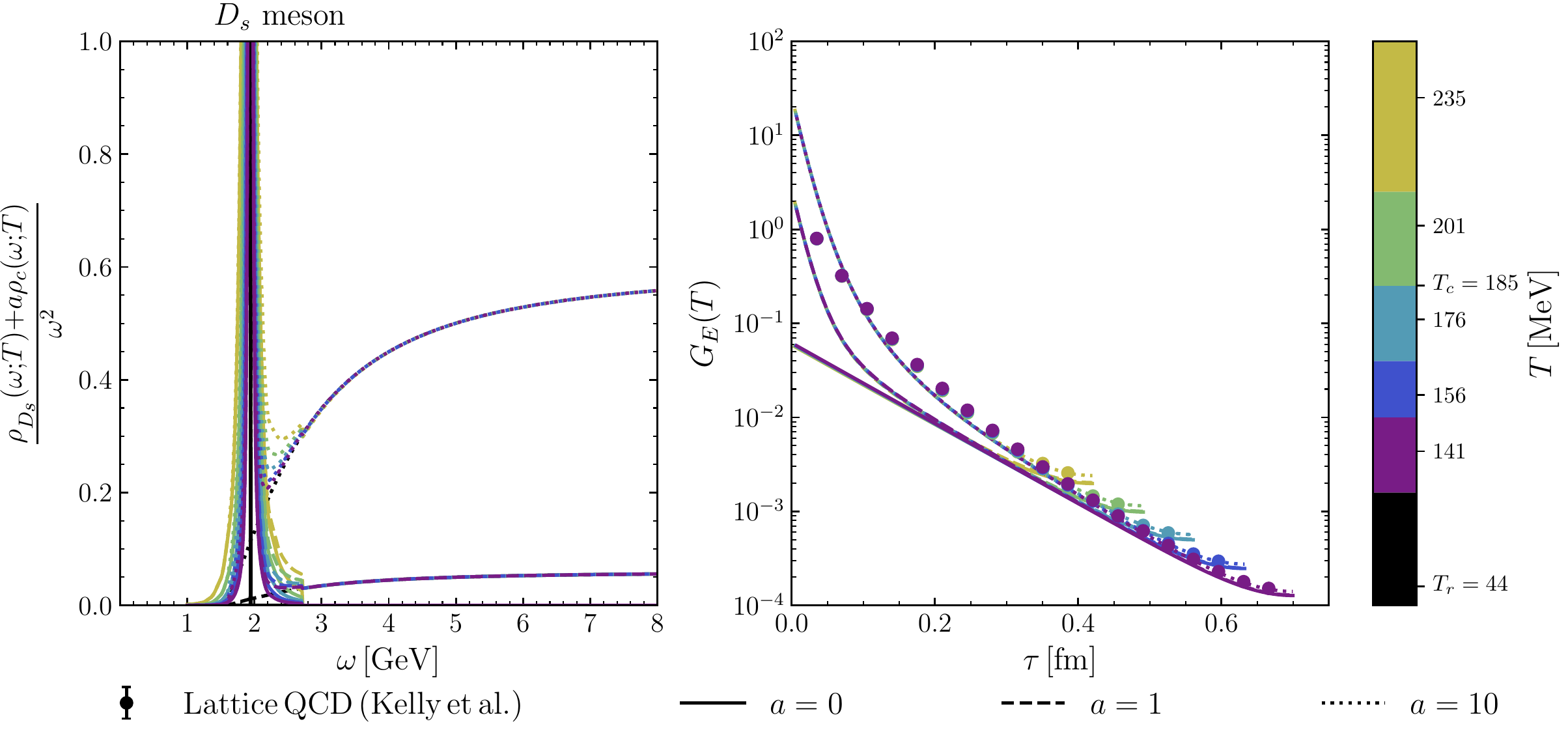}
  \caption{The same as in Fig.~\ref{fig:latt-corr_D_unphys} in the strangeness $S=1$ sector.}
  \label{fig:latt-corr_Ds_unphys}
\end{figure}

\begin{figure}[t!]
\centering
\includegraphics[width=0.9\textwidth]{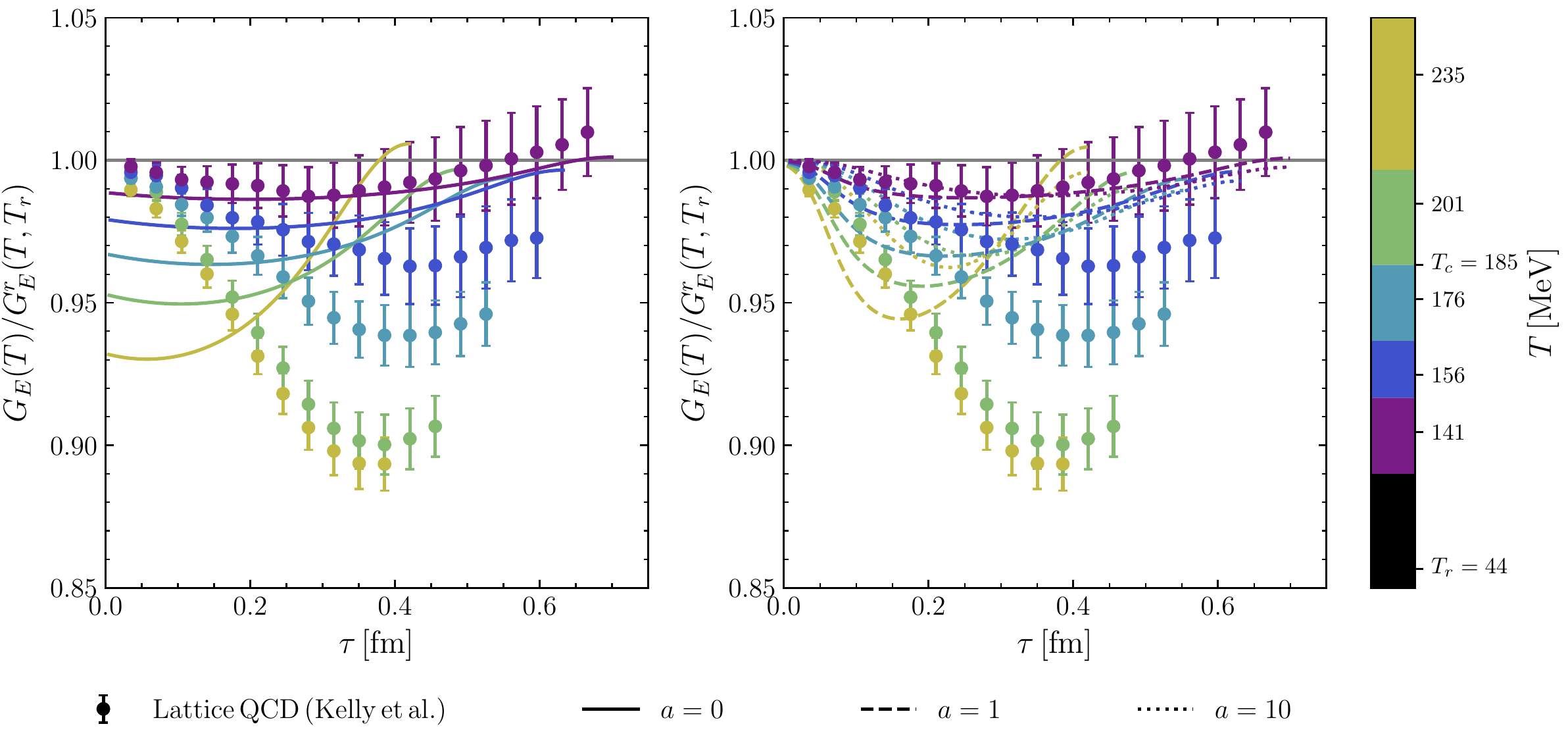}
  \caption{Ratio of the Euclidean correlator, calculated at temperature $T$, to the reconstructed correlator, at $T_r=44\,\textrm{MeV}$, in the $C=1$, $S=0$ sector, considering spectral functions with the ground-state $D$ meson only (left panel) and adding a continuum with a weight $a$ (right panel).}
  \label{fig:latt-ratio_D_unphys}
\end{figure}

\begin{figure}[t!]
\centering
\includegraphics[width=0.9\textwidth]{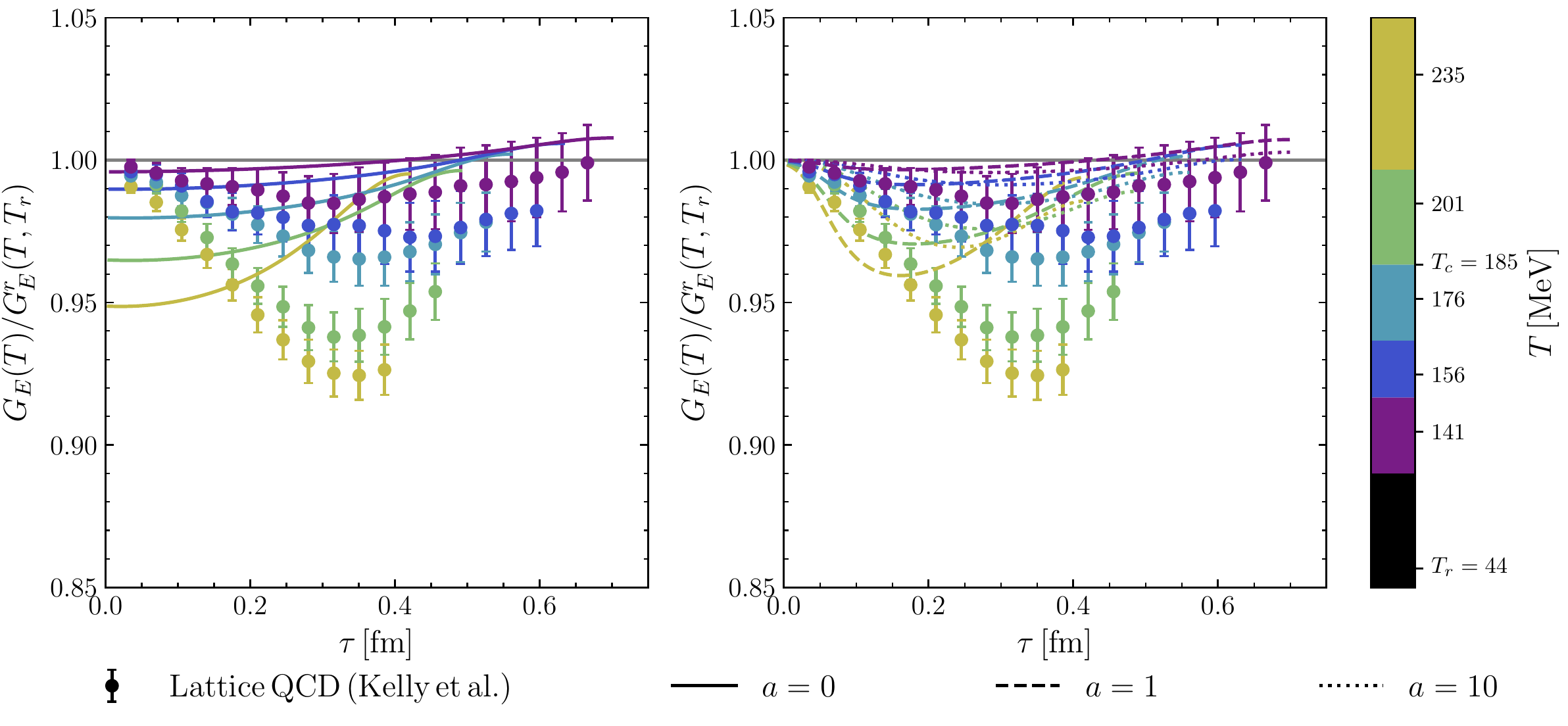}
  \caption{The same as in Fig.~\ref{fig:latt-ratio_D_unphys} in the strangeness $S=1$ sector, where the ground state is the $D_s$ meson.}
  \label{fig:latt-ratio_Ds_unphys}
\end{figure}

 
With $\rho_\textrm{gs}(\omega;T)$ and $\rho_\textrm{gs}(\omega;T_r)$, we can readily calculate Euclidean correlators and the ratios with the reconstructed correlators, $G_E(\tau;T)/G_E^r(\tau;T,T_r)$, for a direct comparison with lattice data. The input spectral function at $T=(141,\,156,\,176,\,201,\,235)$~MeV and $T_r=44$~MeV are shown with solid lines in the left panel of Fig.~\ref{fig:latt-corr_D_unphys} for the $D$ meson and in the left panel of Fig.~\ref{fig:latt-corr_Ds_unphys} for the $D_s$ meson, while the Euclidean correlators are displayed in solid lines in the corresponding right panels. The Euclidean time extends in the range $\tau\in[0,a_\tau N_\tau/2]$ that is smaller at large temperatures. The lattice data of Ref.~\cite{Kelly:2018hsi} for the correlators is displayed with filled circles. In addition, in the left panels of Figs.~\ref{fig:latt-ratio_D_unphys} and \ref{fig:latt-ratio_Ds_unphys}, the ratios of the correlators are shown with solid lines, together with the lattice results of Ref.~\cite{Kelly:2018hsi}, with filled circles with error bars).

The first clear observation is the deviation of the correlators (solid lines in the right panels of Figs.~\ref{fig:latt-corr_D_unphys} and \ref{fig:latt-corr_Ds_unphys}) and the ratio of correlators (left panels of Figs.~\ref{fig:latt-ratio_D_unphys} and \ref{fig:latt-ratio_Ds_unphys}) at small Euclidean times with respect to the lattice data points for all the temperatures and both the $D$ and $D_s$ mesons. However, it is important to note that, for the lowest temperature $T=141$ MeV, the ratio of correlators lies within the error bars of the lattice data. For increasing temperatures, the calculated ratios deviate largely from the lattice calculations. Above or close to the pseudocritical temperature, $T_c=185$~MeV, we do not expect a good matching as the deconfined degrees of freedom are not included in the effective hadronic model, but one would expect a better comparison at lower temperatures.

The discrepancy observed at small $\tau$ for temperatures below $T_c$ might be related to the fact that the spectral functions do not contain the higher-energy states present in the lattice correlators in addition to the ground-state, that is, the possible excited states and the continuum spectrum. As a first approximation, in the following we only add a continuum contribution to the spectral functions. In this way, we aim at understanding the differences with the fewest possible parameters, while trying to improve the comparison of the hadronic and lattice approaches.  

With this goal we define the lattice spectral function as
\begin{equation}
 \rho(\omega;T)= \rho_\textrm{gs}(\omega;T)+a\,\rho_\textrm{cont}(\omega;T) \ ,
\end{equation}
where we add, to the ground-state spectral function obtained from the effective field theory, the contribution of a continuum of scattering states weighted with a factor $a$\footnote{Not to be confused with the lattice spacing in the temporal direction, $a_\tau$.}. 

The continuum contribution to the spectral function is sometimes mimicked with a step function. An analytical expression of the free meson spectral function, obtained from the individual quark spectral functions in the noninteracting limit, is also often used and was first derived for charmonium states in Refs.~\cite{Karsch:2000gi,Karsch:2003wy}. This spectral function describes quark-antiquark pairs with degenerate masses in the limit of infinitely high temperature. An equivalent expression can be derived in the case of nondegenerate quark masses $m_1>m_2$ \cite{FMeyerThesis},
\begin{align}\nonumber
 \rho_M(\omega;T)&=\frac{N_c}{32\pi}\sqrt{\left(\frac{m_1^2-m_2^2}{\omega^2}+1\right)^2-\frac{4m_2^2}{\omega^2}}\,\omega^2 \\ \nonumber &\times \left[(a_M-b_M)+2b_M\frac{m_1^2+m_2^2}{\omega^2}-4c_M\frac{m_1m_2}{\omega^2}-(a_M+b_M)\left(\frac{m_1^2-m_2^2}{\omega^2}\right)^2\right] \\  &\times\left[n(-\omega_0,T)-n(\omega-\omega_0,T)\right]\,\theta\left(\omega-(m_1+m_2)\right) \ ,
\end{align}
where $N_c=3$ is the number of colors, $\omega_0=\frac{1}{2\omega}(\omega^2+m_1^2-m_2^2)$, and $n(\omega,T)=[e^{\omega/T}+1]^{-1}$ is the Fermi-Dirac distribution. The coefficients $(a_M,b_M,c_M)$ are $(1,-1,1)$ for the scalar, $(1,-1,-1)$ for the pseudoscalar, $(2,-2,-4)$ for the vector, and $(2,-2,4)$ for the axial-vector channels. These coefficients result from the traces of the gamma matrices defining the character (scalar, pseudoscalar, vector, axial vector) of the meson. Therefore, for pseudoscalar mesons we have
\begin{align}\nonumber
 \rho_\textrm{cont}(\omega;T)&=\frac{3}{16\pi}\sqrt{\left(\frac{m_1^2-m_2^2}{\omega^2}+1\right)^2-\frac{4m_2^2}{\omega^2}}\,\omega^2  \left(1-\frac{(m_1-m_2)^2}{\omega^2}\right)\\  &\times \left[n(-\omega_0,T)-n(\omega-\omega_0,T)\right]\,\theta\left(\omega-(m_1+m_2)\right) \ .
\end{align} 

In the nonstrange sector, we take $m_1=m_c=1.5$ GeV and $m_2=m_l=0$, whereas in the sector with strangeness $S=1$ we use $m_1=m_c=1.5$ GeV and $m_2=m_s=100$ MeV. The left panels of Figs.~\ref{fig:latt-corr_D_unphys} and \ref{fig:latt-corr_Ds_unphys} display the spectral functions obtained for three different values of the continuum to ground-state contribution: $a=0$ (solid lines, no continuum), $a=1$ (dashed lines), and $a=10$ (dotted lines), for the nonstrange and strange sectors, respectively. The corresponding Euclidean correlators are plotted in the right panels of the same figures, and the ratios with the reconstructed correlators are shown in Figs.~\ref{fig:latt-ratio_D_unphys} and \ref{fig:latt-ratio_Ds_unphys}.

The inclusion of the continuum in the spectral functions improves the behavior of the correlators and the ratio of correlators at small $\tau$ but does not allow for the reproduction of the shape of the lattice correlators. We note that the lattice correlators shown here are not continuum extrapolated and therefore suffer from cut-off effects at small $\tau \lesssim 0.1~\textrm{fm}$, as well as finite-volume effects that have not been introduced in the effective theory. Nevertheless, taking into account the contribution of the continuum permits the ratios to go to one at $\tau\rightarrow 0$ for all temperatures, as the region of very small Euclidean times is essentially governed by the spectral function at very high energies, which has very little dependence on the temperature. 

The modification of the ratios at larger $\tau$ due to the inclusion of the continuum is rather moderate and only the results for the lowest temperature of $T=141$ MeV are compatible with the lattice data within the error bars. 
The region of middle and large values of $\tau$ is rather sensitive to the shape of the spectral functions at low energies (few GeV), where not only variations in the ground-state properties can produce significant modifications of the correlators but also where the free spectral function might be not appropriate to describe the continuum and where excited states are likely to be present. 

In particular, the behavior of the correlators can vary due to a widening of the ground state, which is expected at temperatures close to $T_c$ when considering further contributions to the $D$ and $D_s$ meson self-energies coming from the thermal kaons and antikaons in the medium. In the previous chapter we have shown that, for physical meson masses, the inclusion of $K$ and $\bar{K}$ mesons in the bath in addition to pions starts to have an appreciable effect on the mass shift of the $D$ and $D_s$ mesons at temperatures $T\sim140-150$ MeV. Yet the impact on the $D_s$ width is visible already at $T=80$ MeV and the $D_s$ width in a kaonic and pionic medium is almost a factor two larger than that induced in a pion-only medium at $T=150$ MeV. 
In this chapter, we have used unphysical masses for pions and kaons, with the pions being almost three times heavier than physical pions and the kaons similarly heavy, thus considerably reducing the mass gap between the two lightest pseudoscalar mesons. Consequently, the temperature onset for a sizable thermal modification of the heavy mesons due to pions is larger and closer to that of kaons, in such a way that the kaon-induced width of the charm mesons, especially of the $D_s$, might not be small at temperatures close to $T_c$. Therefore, considering a pionic and kaonic medium in our calculations could improve the comparison of the ratios of Euclidean correlators of the $D_s$ meson at $T=141~\textrm{ MeV}=0.76\,T_c$, and of both $D$ and $D_s$ mesons at larger temperatures below $T_c$.

In regards to the excited states, their properties and thermal modification are not known and their inclusion in the spectral function is not feasible. An alternative way to eliminate this source of discrepancy when comparing the results of the calculated Euclidean correlators with those simulated on a lattice could be to apply techniques for the reduction of excited-state contamination from the correlators in the analysis of the \gls{lqcd} data.

Finally, one might also wonder about the validity of the chiral effective theory employed at an unphysical pion mass of $384$~MeV. However, we consider the Lagrangian to be reliable, as its parameters have been adjusted to finite volume energy levels and scattering lengths, obtained at various unphysical masses but having a smooth extrapolation down to the physical pion mass \cite{Guo:2018tjx}.

\chapter{In-medium kinetic theory of heavy mesons and transport coefficients}
\label{ch:transport}

The way we have dealt with temperature in Chapter~\ref{ch:hot-medium} is by Wick rotating the time dimension from Minkowski to Euclidean spacetime and compactifying it in the range $0\leq \ii t\leq 1/T$, with $T$ being the temperature of the system. In this approach, that is, in the \gls{itf}, the time dependence vanishes, as time is ``converted'' into temperature, and hence we have been able to study equilibrium quantities. In the present chapter, we want to calculate out-of-equilibrium kinetics, as well as transport coefficients, and a real-time formalism that describes the evolution in real time is needed.
The generalization of the imaginary-time and real-time formalisms consists in considering time defined on the complex plane. One can then particularize a specific formalism by taking a certain contour in the complex plane, for instance, the ``Matsubara contour'' for the \gls{itf}, the ``Kadanoff-Baym contour'', or the ``Schwinger-Keldysh contour'' (see Fig.~\ref{fig:hot-contours}).

In this chapter, we analyze the effect on the transport properties of the heavy mesons ($D$ and $\bar{B}$) in a mesonic bath at finite temperature. To this end, we make use of the in-medium unitarized amplitudes in a mesonic environment at finite temperature obtained in Chapter~\ref{ch:hot-medium}, which have been tested in Chapter~\ref{ch:lattice} against \gls{lqcd} calculations of Euclidean correlators below the temperature of the \gls{qcd} phase transition, $T_c$. The final goal is to calculate the $D$- and $\bar{B}$-meson transport coefficients below the transition temperature, paying special attention to the inclusion of off-shell effects coming from the full spectral features of the heavy mesons in a hot mesonic bath, and analyze the matching at $T_c$ of our results with those of \gls{lqcd} and Bayesian analyses of \gls{hic} data, which are available at temperatures close and above $T_c$. 

The chapter is organized as follows. After a brief introduction in Section~\ref{sec:transport-intro}, we present in Section~\ref{sec:transport-nonequilibrium} the derivation of the kinetic equation, implementing the features of the $T$-matrix approximation in the collision terms. In Section~\ref{sec:transport-equilibrium} we particularize it to an equilibrium situation and review the definition of the spectral properties of the heavy mesons at finite temperature, whereas in Section~\ref{sec:transport-kinematic} we show the different kinematic contributions to the heavy-meson thermal width coming from the off-shell treatment of the heavy meson. Finally, in Section~\ref{sec:transport-transport} we obtain, after rederiving the Fokker-Planck equation, the different transport coefficients within the off-shell approach, and compare our findings with the latest results of \gls{lqcd} and Bayesian analyses. The work detailed in this chapter was published in Ref.~\cite{Torres-Rincon:2021yga}.

\section{Introduction}
\label{sec:transport-intro}

Heavy hadrons are considered to be an efficient and unique probe for testing the different \gls{qcd} phases created in \glspl{hic}, in both the \gls{qgp} and hadronic phases (see Refs.~\cite{Aarts:2016hap,Prino:2016cni,Dong:2019unq,Dong:2019byy,Zhao:2020jqu} for recent reviews). Due to the large mass of the heavy (charm and bottom) quarks as compared to the mass of the light-flavor quarks, heavy quarks have large relaxation times and, thus, cannot totally relax to equilibrium during the fireball expansion in \glspl{hic}. For this reason, heavy mesons constitute ideal probes to characterize the \gls{qgp} properties. Determining their in-medium properties in a hadronic medium at extreme conditions is a subject that attracts a lot of interest nowadays. 

The characterization of the different QCD phases can be performed by analyzing experimental observables in \glspl{hic}, such as the nuclear modification ratio as well as the elliptic flow~\cite{Aarts:2016hap,Prino:2016cni,Dong:2019unq,Dong:2019byy,Zhao:2020jqu}. These physical observables are strongly correlated to the behavior of the transport properties of heavy hadrons, and these depend crucially on the interaction of the heavy particles with the surrounding medium.

In particular, the diffusion of open-charm ($D$ mesons) in hadronic matter was initially obtained within an effective theory that incorporated both chiral and heavy-quark symmetries~\cite{Laine:2011is}, and also by using parametrized interactions with light mesons and baryons~\cite{He:2011yi}. 
Following these initial works, unitarized effective interactions of heavy mesons ($D$ and $\bar{B}$) with light mesons and baryons that exploited chiral and heavy-quark symmetries were used to obtain the heavy-meson transport coefficients, as functions of the temperature and the baryochemical potential of the hadronic bath, by means of the Fokker-Planck equation approach (see Refs.~\cite{Tolos:2013kva,Torres-Rincon:2013nfa,Ozvenchuk:2014rpa}). Moreover, the transport coefficients of the low-lying heavy baryons ($\Lambda_c$ and $\Lambda_b$) were examined by employing a similar unitarized framework to account for the interaction of these states with light mesons in Refs.~\cite{Tolos:2016slr,Das:2016llg}. 

Similar approaches using different models or effective descriptions, both below and above the phase transition, have been developed (see \cite{
vanHees:2004gq,Moore:2004tg,Mannarelli:2005pz,vanHees:2005wb,
CasalderreySolana:2006rq,vanHees:2007me,Beraudo:2009pe,
He:2011yi,Ghosh:2011bw,He:2011qa,Das:2012ck,Berrehrah:2013mua,
Ozvenchuk:2014rpa,Das:2015ana,Lang:2016jpe,Liu:2018syc} for some references). These works exploited the Fokker-Planck (or Langevin) equation description for the heavy particles. Some studies pointed out limitations in the \gls{qgp} phase and considered the Boltzmann equation instead~\cite{Das:2013kea,Tolos:2016slr}. While the Fokker-Planck or the Boltzmann kinetic equations seem natural starting points to address the calculation of transport coefficients, the scattering amplitudes used to describe the collisions were computed from a microscopic model, completely independent of the kinetic theory. Therefore, from a purely theoretical perspective, there is a lack of internal consistency in these calculations, as it would be desirable to construct both the interaction rates and the transport equation from the same microscopic theory. 

On the other hand, in most of the previous analyses, the transition amplitudes of the scattering of heavy mesons with light hadrons and, hence, the transport coefficients were calculated without implementing medium corrections for the interactions. Indeed, the off-shell effects cannot be accounted for in the standard Boltzmann or Fokker-Planck equations. For that, an extension using the more general Kadanoff-Baym equations is required. Moreover, we have seen in Chapter~\ref{ch:hot-medium} that the in-medium interactions induce new kinematic domains, namely, the regime of the Landau cut, which affect the $D$-meson properties. These new effects, which would also affect the $D$-meson transport coefficients, can be naturally incorporated upon the derivation of an off-shell kinetic equation. Therefore, to apply our findings in Chapter~\ref{ch:hot-medium}, we are forced to address the derivation of an off-shell kinetic equation and, with this, consistently describe the interaction amplitudes and the transport equation from the same effective theory.

\section{Nonequilibrium description}
\label{sec:transport-nonequilibrium}
\subsection{The off-shell kinetic equation}
In this section we describe the kinetic equation for heavy mesons interacting with light mesons, exploiting the real-time formalism of scalar quantum fields~\cite{kadanoff1962quantum,Danielewicz:1982kk,calzetta1988nonequilibrium,Botermans:1990qi,Davis:1991zc,Blaizot:1999xk,Rammer,Juchem:2003bi,Cassing:2008nn,bonitz2016quantum}.
While the derivation of the kinetic theory based on out-of-equilibrium quantum field theory is rather general, for the practical computation of the physical quantities considered in this chapter, namely, thermal widths and transport coefficients, only equilibrium properties, mostly taken from Chapter~\ref{ch:hot-medium}, will be used. 

Out of equilibrium, the fundamental quantities are the two Wightman functions, defined in Eqs.~(\ref{eq:hot-Ggreat}) and (\ref{eq:hot-Gless}). For the heavy meson, they read   
\begin{align} 
  \ii G_D^> (x,x') \equiv \langle D(x) D(x') \rangle \ , \label{eq:transport-Ggreat}  \\
  \ii G_D^< (x,x') \equiv \langle D(x') D(x) \rangle \ , \label{eq:transport-Gless}
\end{align}
where the subscript in $G^{\lessgtr}(x,x')$ refers to the particular hadron species. They correspond to the time-ordered Green's functions (ordered along the real-time contour, Fig.~\ref{fig:hot-contours}) of the $D$-meson propagator, depending on the relative ordering of the time arguments $t$, $t'$ (see Section~\ref{sec:hot-formalism}).
%
%
%
Such definitions for the heavy meson can be extended to the light-meson sector, for instance for the $\Phi$-meson Wightman functions, $\ii G_\Phi^{\lessgtr} (x,x')$, where $\Phi=\{ \pi,K,\bar{K},\eta\}$.
%

The retarded and advanced heavy-meson propagators are related to Eqs. (\ref{eq:transport-Ggreat}) and (\ref{eq:transport-Gless}) as
\begin{align} 
  G_D^{\textrm{R}}  (x,x') & =   \theta_C(t-t')  \left[ G_D^> (x,x') - G_D^< (x,x')\right]   \ , \label{eq:transport-Gret} \\ 
  G_D^{\textrm{A}} (x,x') &  = - \theta_C(t'-t) \left[ G_D^> (x,x') - G_D^< (x,x')\right] \ , \label{eq:transport-Gadv} 
\end{align}
with $\theta_C(t-t')$ defined along the Schwinger-Keldysh contour $C=C_1\cup C_2 $ (see right panel in Fig.~\ref{fig:hot-contours}). 
 
In coordinate space, the heavy-meson self-energy is also defined along the real-time contour $C$. The time-ordered $D$-meson self-energy~\cite{Blaizot:1999xk} reads
\begin{equation}
  \Pi_D(x,x')= \Pi_D^\delta (x) \delta^{(3)} (\vec{x}-\vec{x}\,') \delta_C(t-t') + \theta_C(t-t') \Pi_D^>(x,x') + \theta_C(t'-t) \Pi_D^<(x,x') \ , \label{eq:transport-selfenergy} 
\end{equation}
where $\Pi^\delta (x)$ is the local (tadpole) contribution~\cite{Blaizot:1999xk}. We have introduced a generalized Dirac delta $\delta_C(t-t')$ along the contour which can be defined as
\begin{align}
\delta_C(t-t') = \left\{ 
\begin{array}{ll}
\delta(t-t') & \textrm{if} \quad t,t' \in C_1 \ , \\
-\delta(t-t') & \textrm{if} \quad t,t' \in C_2 \ , \\
0 & \textrm{otherwise} \ .   \\
\end{array}
\right.
\end{align}

The definitions for $\Pi_D^<(x,x')$ and $\Pi_D^>(x,x')$ are analogous to those for the Green's functions. Similarly, the retarded and advanced self-energies follow,
\begin{align} 
  \Pi^{\textrm{R}}_D (x,x') & = \theta_C(t-t') \left[ \Pi_D^>(x,x')-\Pi_D^<(x,x') \right] \ , \label{eq:transport-Piret} \\ 
  \Pi^{\textrm{A}}_D (x,x') & =-\theta_C(t'-t) \left[ \Pi_D^>(x,x')-\Pi_D^<(x,x') \right] \ . \label{eq:transport-Piadv} 
\end{align}

The equilibrium retarded $D$-meson self-energy can be computed along the Matsubara contour, as has been done in Chapter~\ref{ch:hot-medium} using the \gls{itf}. In the present chapter, we extensively use those results. To simplify the notation, from now on we suppress the subindex in the self-energies, as it is understood that all of them refer to the heavy meson.
 
The Dyson equations for the time-ordered Green's function read~\cite{kadanoff1962quantum,Danielewicz:1982kk,Botermans:1990qi,Davis:1991zc,Blaizot:1999xk},
\begin{align} 
  - \left[\partial_x^2 + m_{D}^2\right]   G_D (x,x') - \int_C d^4z \, \Pi(x,z)  G_D(z,x') & = \delta_C(t-t')\delta^{(3)}(\vec{x}-\vec{x}\,') \label{eq:transport-dyson1} \ , \\
  - \left[\partial_{x'}^2 + m_{D}^2\right]   G_D (x,x') -  \int_C d^4z \, G_D(x,z) \Pi(z,x') & =\delta_C(t-t') \delta^{(3)}(\vec{x}-\vec{x}\,') \ ,
  \label{eq:transport-dyson2}
\end{align}
where the time convolution, $\int_C dz^0$, is taken along the Schwinger-Keldysh path. The bare $D$-meson mass term is denoted as $m_{D}$. Equations~(\ref{eq:transport-dyson1}) and (\ref{eq:transport-dyson2}) are independent Dyson equations, acting on the two possible arguments of the Green's function. While in the absence of interactions the equations of motion are the Klein-Gordon ones, when interactions are considered, the $D$-meson self-energy corrections need to be taken into account. The microscopic theory describing the interactions between heavy and light mesons is the effective theory used in the previous chapters, which incorporates chiral and heavy-quark symmetries. 

One can particularize Eqs.~(\ref{eq:transport-dyson1}) and (\ref{eq:transport-dyson2}) for the ``lesser'' function $G_D^<(x,x')$, for example by fixing $t$ in the forward contour $C_1$ and $t'$ in the backward one $C_2$. One finds that
\begin{align} 
  - \left[\partial_x^2 + m_{D}^2 \right]  G_D^<(x,x') -  \int_C d^4z\, \Pi(x,z) G_D(z,x') & =0 \ , \\
  - \left[\partial_{x'}^2 +m_{D}^2 \right]   G_D^<(x,x') -  \int_C d^4z\, G_D(x,z) \Pi(z,x') & =0 \ ,
\end{align}
where the convolution operation should respect $t \prec t'$. This term can be simplified by applying the Langreth rules~\cite{Rammer} to arrive to
\begin{align} 
&-\left[\partial_x^2 + m_{D}^2 + \Pi^\delta (x)\right]   G_D^< (x,x') \nonumber \\
&\hspace{3cm} =  \int_{-\infty}^\infty d^4z \left[ \Pi^{\textrm{R}} (x,z) G_D^< (z,x') + \Pi^< (x,z) G_D^{\textrm{A}}(z,x') \right] \ , \\
&-\left[\partial_{x'}^2 + m_{D}^2 + \Pi^\delta (x')\right]   G_D^< (x,x') \nonumber \\ 
&\hspace{3cm} =  \int_{-\infty}^\infty d^4z \left[  G_D^< (x,z) \Pi^{\textrm{A}} (z,x') + G_D^{\textrm{R}} (x,z) \Pi^<(z,x') \right] \ , 
\end{align}
where we have introduced the decomposition of the self-energy given in Eq.~(\ref{eq:transport-selfenergy}). Notice that the time integration in $z^0$ is now restricted from $-\infty$ to $+\infty$ only, that is, along $C_1$ in Fig.~\ref{fig:hot-contours}. Equations for $G_D^>(x,x')$ can be obtained by formally replacing ``$<$'' with ``$>$'' in all cases. To lighten the notation we write both equations as
\begin{align} 
&\left[G^{-1}_{0,x\phantom{'}} - \Pi^\delta (x) \right]   G_D^\lessgtr (x,x') = \left(\Pi^{\textrm{R}} \otimes G_D^\lessgtr \right)(x,x') + \left(\Pi^\lessgtr \otimes G_D^{\textrm{A}}\right) (x,x')  \ , \label{eq:transport-Gxeq} \\
&\left[G^{-1}_{0,x'} - \Pi^\delta (x') \right]   G_D^\lessgtr (x,x') =  \left( G_D^\lessgtr \otimes  \Pi^{\textrm{A}}\right) (x,x') + \left( G_D^{\textrm{R}} \otimes \Pi^\lessgtr\right)  (x,x') \ , \label{eq:transport-Gyeq}
 \end{align}
where we have defined $G_{0,x}^{-1} \equiv -\partial_x^2 - m_{D}^2$ and introduced the convolution symbol as
\begin{equation}
  \left(\Pi^{\textrm{R}} \otimes G_D^\lessgtr\right)(x,x') \equiv \int_{-\infty}^\infty d^4z \, \Pi^{\textrm{R}}(x,z) G_D^\lessgtr (z,x') \ . 
\end{equation}

For completeness, the equations of motion for the retarded/advanced (R/A) Green's functions are
\begin{align} 
&\left[G^{-1}_{0,x\phantom{'}} - \Pi^\delta (x) \right] \ G_D^{\textrm{R/A}} (x,x')  -\left(\Pi^{\textrm{R/A}} \otimes G_D^{\textrm{R/A}} \right)(x,x')  = \delta^{(4)}(x- x')    \ , \label{eq:transport-GRAx} \\
&\left[G^{-1}_{0,x'} - \Pi^\delta (x') \right] \ G_D^{\textrm{R/A}} (x,x') -\left( G_D^{\textrm{R/A}} \otimes \Pi^{\textrm{R/A}}\right)  (x,x') = \delta^{(4)}(x-x')   \ , \label{eq:transport-GRAy}
\end{align}
where the Dirac deltas are the standard ones, $\delta^{(4)}(x-x')=\delta(t-t') \delta^{(3)}(\vec{x}-\vec{x}\,') $.
These equations can be obtained using the definitions in Eqs.~(\ref{eq:transport-Piret}) and (\ref{eq:transport-Piadv}). Equations~(\ref{eq:transport-Gxeq}),(\ref{eq:transport-Gyeq}),(\ref{eq:transport-GRAx}), and (\ref{eq:transport-GRAy}) are known as the Kadanoff-Baym equations~\cite{kadanoff1962quantum}.

To construct a kinetic equation from the Kadanoff-Baym equations, we start by taking the difference between Eqs.~(\ref{eq:transport-Gxeq}) and~(\ref{eq:transport-Gyeq}) for $G_D^<(x,x')$,
\begin{align} \nonumber
& \left[ G_{0,x\phantom{'}}^{-1} - \Pi^\delta (x)\right] G_D^< (x,x') -\left[G_{0,x'}^{-1} - \Pi^\delta (x')\right]  G_D^< (x,x') \\ 
& \quad = \left(\Pi^{\textrm{R}} \otimes G_D^<  + \Pi^<  \otimes G_D^{\textrm{A}} -  G_D^< \otimes  \Pi^{\textrm{A}}  - G_D^{\textrm{R}} \otimes \Pi^<\right)  (x,x') \ . \label{eq:transport-diff} 
\end{align}

In order to facilitate the subsequent computations, it is convenient to combine the convolution operator with the commutator and the anticommutator by introducing the following notation: 
\begin{align} 
  \left[A \stackrel{\otimes}{,} B\right]   & \equiv  A\otimes B - B\otimes A \ , \\
\left\{ A \stackrel{\otimes}{,} B \right\} & \equiv  A \otimes B + B \otimes A \ ,
\end{align}
and suppressing the spacetime indices temporarily to lighten the expressions.

With these operators, Eq.~(\ref{eq:transport-diff}) can be written as
\begin{align} 
& \left[ G_{0,x\phantom{'}}^{-1} - \Pi^\delta (x)\right] G_D^< (x,x') -\left[G_{0,x'}^{-1} - \Pi^\delta (x')\right]  G_D^< (x,x')  \nonumber \\ & \hspace{3cm} -  \frac{1}{2} \left[ \Pi^{\textrm{R}} + \Pi^{\textrm{A}} \stackrel{\otimes}{,} G_D^<\right] - \frac{1}{2} \left[ \Pi^< \stackrel{\otimes}{,} G_D^{\textrm{R}}+G_D^{\textrm{A}}\right] \nonumber \\
& \quad =  \frac{1}{2} \left\{ \Pi^{\textrm{R}} \stackrel{\otimes}{,} G_D^< \right\} -\frac{1}{2} \left\{ \Pi^{\textrm{A}} \stackrel{\otimes}{,}  G_D^< \right\}
+ \frac{1}{2} \left\{ \Pi^< \stackrel{\otimes}{,} G_D^{\textrm{A}} \right\} - \frac{1}{2} \left\{ \Pi^< \stackrel{\otimes}{,}  G_D^{\textrm{R}} \right\} \ .  
\end{align}
This equation can be simplified with the help of Eqs.~(\ref{eq:transport-Gret}) and (\ref{eq:transport-Gadv}), as well as Eqs.~(\ref{eq:transport-Piret}) and (\ref{eq:transport-Piadv}), which imply
\begin{align}
G_D^{\textrm{R}} (x,x') - G_D^{\textrm{A}} (x,x')& =   G_D^> (x,x')- G_D^<(x,x') \ , \label{eq:transport-GRALG}\\
\Pi^{\textrm{R}} (x,x')-\Pi^{\textrm{A}} (x,x')& =  \Pi^> (x,x')- \Pi^< (x,x') \ , \label{eq:transport-PiRALG}
\end{align}
leading to the kinetic equation for $G_D^<(x,x')$~\cite{Danielewicz:1982kk,Davis:1991zc,Blaizot:1999xk},
\begin{align} 
& \left[ G_{0,x\phantom{'}}^{-1} - \Pi^\delta (x)\right] G_D^< (x,x') -\left[G_{0,x'}^{-1} - \Pi^\delta (x')\right]  G_D^< (x,x') \nonumber  \\ &\hspace{3cm} -  \frac{1}{2} \left[ \Pi^{\textrm{R}}  + \Pi^{\textrm{A}}  \stackrel{\otimes}{,} G_D^<\right] - \frac{1}{2} \left[ \Pi^< \stackrel{\otimes}{,} G_D^{\textrm{R}} +G_D^{\textrm{A}} \right] \nonumber \\
& \quad =  \frac{1}{2} \left\{ \Pi^> \stackrel{\otimes}{,} G_D^< \right\} - \frac{1}{2} \left\{ \Pi^< \stackrel{\otimes}{,} G_D^>  \right\}  \ .  \label{eq:transport-beforeWT} 
\end{align}
For completeness, the equivalent equation for $G_D^>(x,x')$ is
\begin{align} 
 & \left[ G_{0,x\phantom{'}}^{-1} - \Pi^\delta (x)\right] G_D^> (x,x') -\left[G_{0,x'}^{-1} - \Pi^\delta (x')\right] G_D^> (x,x')  \nonumber \\ &\hspace{3cm} -  \frac{1}{2} \left[ \Pi^{\textrm{R}}  + \Pi^{\textrm{A}}  \stackrel{\otimes}{,} G_D^>\right] - \frac{1}{2} \left[ \Pi^> \stackrel{\otimes}{,} G_D^{\textrm{R}} +G_D^{\textrm{A}} \right] \nonumber \\ 
&\quad  =  \frac{1}{2} \left\{ \Pi^> \stackrel{\otimes}{,} G_D^< \right\} - \frac{1}{2} \left\{ \Pi^< \stackrel{\otimes}{,} G_D^>  \right\}  \ .  \label{eq:transport-beforeWT2} 
\end{align}

These equations are still rather general kinetic equations, and one needs to detail the form of the heavy-meson self-energy in terms of the Green's functions. 

Before doing that, one can  simplify the expressions by implementing a Wigner transformation, together with a gradient expansion~\cite{kadanoff1962quantum}.
For this, we assume that the Green's function depends differently  on the center-of-mass coordinate $X=(x+x')/2$ and the relative coordinate $s=x-x'$. When the dependence on $X$ is much smoother than the one in $s$, it makes sense to Fourier transform the latter and apply a gradient expansion for the former. In that respect, we expand in powers of $\partial^\mu_X$ and perform the so-called Wigner transform to the two-point functions. For a generic operator $A(x,x')$, the Wigner transform is defined as,
\begin{equation}
A(x,x') \xrightarrow{\textrm{WT}} A(X,k) = \int d^4s \ e^{\ii k\cdot s} A \left( X+\frac{s}{2},X-\frac{s}{2} \right) \ . 
\end{equation}
In Appendix~\ref{appendix-Wigner} we provide some details on how to compute the Wigner transform of the product and convolution of operators~\cite{Rammer}. All those rules should be applied for the Wigner transform of Eqs.~(\ref{eq:transport-beforeWT}) and (\ref{eq:transport-beforeWT2}).

After Wigner transform, Eq.~(\ref{eq:transport-beforeWT}) reads 
\begin{align}
  2 k^\mu \frac{\partial \ii G_D^<}{\partial X^\mu} &+   \frac{\partial \Pi^\delta}{\partial X^\mu} \frac{\partial \ii G_D^< }{\partial k_\mu}  - \frac{\partial \textrm{Re } \Pi^{\textrm{R}}}{\partial k_\mu}\frac{\partial \ii G_D^<}{\partial X^\mu} + \frac{\partial \textrm{Re } \Pi^{\textrm{R}}}{\partial X^\mu} \frac{\partial \ii G_D^<}{\partial k_\mu} - \ii \left\{ \Pi^< , \textrm{Re } G_D^{\textrm{R}} \right\}_{\textrm{PB}} \nonumber \\
&=  \ii\Pi^<  \ii G_D^> - \ii \Pi^> \ii G_D^<   \ , \label{eq:transport-afterWT}
\end{align}
where all (unwritten) arguments are now phase-space variables $(X,k)$. To obtain this equation, apart from using the results in Appendix~\ref{appendix-Wigner}, we have applied that
\begin{align} 
&\frac{1}{2} \left[G_D^{\textrm{R}} (X,k)+G_D^{\textrm{A}} (X,k) \right]  = \textrm{Re } G_D^{\textrm{R}} (X,k)  \ , \\
&\frac{1}{2} \left[\Pi^{\textrm{R}}(X,k) +\Pi^{\textrm{A}} (X,k)\right] = \textrm{Re } \Pi^{\textrm{R}} (X,k)\ ,
\end{align}
which follows from 
\begin{align}
&G_D^{\textrm{R}}(X,k)  =\left[G_D^{\textrm{A}} (X,k)\right]^* \ , \label{eq:transport-conjug} \\
&\Pi^{\textrm{R}}(X,k)  =\left[\Pi^{\textrm{A}} (X,k)\right]^* \ . 
\end{align}

The relations in Eqs.~(\ref{eq:transport-GRALG}) and (\ref{eq:transport-PiRALG}) remain the same after the Wigner transform,
\begin{align}
G_D^{\textrm{R}}(X,k)-G_D^{\textrm{A}}(X,k) & = G_D^> (X,k) - G_D^< (X,k) = 2\ii \textrm{Im } G_D^{\textrm{R}} (X,k) \ , \label{eq:transport-greenfuncs} \\ 
\Pi^{\textrm{R}}(X,k)-\Pi^{\textrm{A}}(X,k) & = \Pi^> (X,k) - \Pi^< (X,k) = 2\ii \textrm{Im } \Pi^{\textrm{R}} (X,k) \ . \label{eq:transport-polfuncs} 
\end{align}

Some terms of Eq.~(\ref{eq:transport-afterWT}) can be combined to give
\begin{align} 
& \left(  k^\mu - \frac{1}{2} \frac{\partial \textrm{Re } \Pi^{\textrm{R}} (X,k)}{\partial k_\mu} \right) \frac{\partial \ii G_D^< (X,k)}{\partial X^\mu} + \frac{1}{2}  \frac{\partial \textrm{Re } \Pi^{\textrm{R}} (X,k)}{\partial X^\mu} \frac{\partial \ii G_D^< (X,k)}{\partial k_\mu}  \nonumber \\
& - \frac{\ii}{2}  \{ \Pi^< (X,k), \textrm{Re } G_D^{\textrm{R}} (X,k) \}_{\textrm{PB}}
=\frac{1}{2}  \ii\Pi^< (X,k)  \ii G_D^> (X,k) -\frac{1}{2} \ii\Pi^> (X,k) \ii G_D^< (X,k) \ , 
\label{eq:transport-almostoffGless}
\end{align}
where we have incorporated the local tadpole term $\Pi^\delta(X)$ into the real part of the retarded self-energy, $\textrm{Re } \Pi^{\textrm{R}} (X,k)$. In Eq.~(\ref{eq:transport-almostoffGless}), we have written the explicit dependence on the four-position $X=(\tilde{t},\vec{X})$ and the four-momentum $k=(k^0,\vec{k}\,)$. Note that $k^0$ and $\vec{k}$ are independent variables, although related through the spectral distribution $S_D(X,k)$, to be introduced later. Such a general case where the heavy meson is not on its mass shell, that is, its energy is not determined by its momentum, will be generically denoted as \textit{off-shell}. Already in equilibrium, we have shown in Chapter~\ref{ch:hot-medium} that the heavy meson at $T\neq 0$ is characterized by a continuous spectral function, which represents the distribution of possible energies for a given value of the momentum.

In the present chapter, we will eventually consider an equilibrium situation in a homogeneous background. Therefore, we neglect the mean-field term in Eq.~(\ref{eq:transport-almostoffGless}), proportional to $\partial_{X^\mu} \textrm{Re } \Pi^{\textrm{R}}$~\cite{Blaizot:2001nr}. In addition, we will not consider the Poisson bracket (PB) of the same equation. This term was shown to be unimportant in the quasiparticle limit and can be safely neglected~\cite{Danielewicz:1982kk,Blaizot:2001nr,Cassing:2008nn}. However, one should not forget that in the off-shell case it contributes to the out-of-equilibrium dynamics of $G^>_D (X,k)$~\cite{Botermans:1990qi,Blaizot:2001nr} (see also~\cite{Cassing:2008nn} and references therein).

With these approximations, we arrive at the final form of the ``off-shell'' transport equation,
\begin{align}
&\left(  k^\mu - \frac{1}{2} \frac{\partial \textrm{Re } \Pi^{\textrm{R}} (X,k)}{\partial k_\mu} \right) \frac{\partial \ii G_D^< (X,k)}{\partial X^\mu} \nonumber \\ 
&\hspace{3cm} =\frac{1}{2}  \ii\Pi^< (X,k)  \ii G_D^> (X,k) -\frac{1}{2} \ii\Pi^> (X,k) \ii G_D^< (X,k) \ . \label{eq:transport-offGless} 
\end{align}

For completeness, the transport equation for $G_D^>(X,k)$ is similar to Eq.~(\ref{eq:transport-offGless}), and it shares the same \gls{rhs}, which represents the collision term of the transport equation,
\begin{align}
&\left(  k^\mu - \frac{1}{2} \frac{\partial \textrm{Re } \Pi^{\textrm{R}} (X,k)}{\partial k_\mu} \right) \frac{\partial \ii G_D^> (X,k)}{\partial X^\mu} \nonumber \\ 
& \hspace{3cm} =\frac{1}{2}  \ii\Pi^< (X,k)  \ii G_D^> (X,k) -\frac{1}{2} \ii\Pi^> (X,k) \ii G_D^< (X,k) \ . \label{eq:transport-offGgreat} 
\end{align}

Related to the dispersion relation of the heavy meson, one can consider the equations for the retarded Green's function $G_D^{\textrm{R}}$ in Eqs.~(\ref{eq:transport-GRAx}) and (\ref{eq:transport-GRAy}) and, after performing the Wigner transform along with the gradient expansion, get to
\begin{equation}
\left[ k^2  - m_{D}^2 -  \Pi^{\textrm{R}}(X,k) \right] G_D^{\textrm{R}}(X,k)=1 \ , \label{eq:transport-eqforGR} 
\end{equation}
where the operators of order ${\cal O} (\partial_X^2)$ have been neglected, and, again, we have incorporated the local $\Pi^\delta(X)$ into the retarded self-energy. The pole of the retarded Green's function, that is the zero of the \gls{lhs}, will generically provide the dispersion relation of the $D$ meson modified by the interactions. Notice that the equilibrium and homogeneous $\Pi^{\textrm{R}}(k)$ has been studied in Chapter~\ref{ch:hot-medium}, and it will be exploited here as well.

\subsection{The $T$-matrix approximation}

To close the transport equation for $G_D^<(X,k)$, the remaining step is to detail the heavy-meson self-energies in terms of the Green's functions. To this goal, one needs to apply the microscopic theory to the given system. Here, in consistency with the effective approach described in Chapter~\ref{ch:hot-medium}, we will incorporate exact unitarity constraints to the scattering matrix, implementing an in-medium $T$-matrix resummation of the scattering amplitudes. In the nonequilibrium context, this is called the $T$\textit{-matrix approximation}~\cite{kadanoff1962quantum,Danielewicz:1982kk,Botermans:1990qi}. 

The heavy-meson self-energies can be written in terms of the (retarded) $T$-matrix element as~\cite{kadanoff1962quantum},
\begin{align}
  \ii\Pi^< (X,k) & =  \sum_{ \{ a,b,c \} } \int \frac{d^4 k_1}{(2\pi)^4} \int \frac{d^4 k_2}{(2\pi)^4} \int \frac{d^4 k_3}{(2\pi)^4} (2\pi)^4 \delta^{(4)} (k_1+k_2-k_3-k)  \nonumber \\
  &\times \left|T (k_1^0+k_2^0+\ii\varepsilon, \vec{k}_1 + \vec{k}_2)\right|^2 \, \ii G_{D_a}^< (X,k_1) \ii G_{\Phi_b}^< (X,k_2) \ii G_{\Phi_c}^>(X,k_3) \ , \label{eq:transport-Pil} \\
  \ii\Pi^> (X,k) & =  \sum_{ \{ a,b,c \} } \int \frac{d^4 k_1}{(2\pi)^4} \int \frac{d^4 k_2}{(2\pi)^4} \int \frac{d^4 k_3}{(2\pi)^4} (2\pi)^4 \delta^{(4)} (k_1+k_2-k_3-k)  \nonumber \\
  &\times \left|T (k_1^0+k_2^0+\ii\varepsilon, \vec{k}_1 + \vec{k}_2)\right|^2 \, \ii G_{D_a}^> (X,k_1) \ii G_{\Phi_b}^> (X,k_2) \ii G_{\Phi_c}^<(X,k_3) \ . \label{eq:transport-Pig}
\end{align}

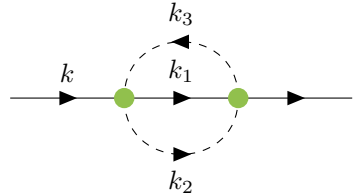
\begin{figure}[b!]
  \centering
\begin{tikzpicture}[baseline=(i.base),inner sep=6pt]
    \begin{feynman}
      \vertex (a);
      \vertex [right = of a] (i);
      \vertex [right = of i] (j);
      \vertex [right = of j] (b);
      \diagram*{
         (a) -- [fermion,edge label=\(k\)] (i);
         (i) -- [fermion,edge label=\(k_1\)] (j);
         (j) -- [fermion] (b);
         (j) -- [charged scalar, half right, looseness=1.7,edge label'=\(k_3\)] (i);
         (i) -- [charged scalar, half right, looseness=1.7,edge label'=\(k_2\)] (j);
     } ;
     \draw[dot,minimum size=4mm,thick,ctcolorgreen,fill=ctcolorgreen] (i) circle(1.2mm);
     \draw[dot,minimum size=4mm,thick,ctcolorgreen,fill=ctcolorgreen] (j) circle(1.2mm);
    \end{feynman}
  \end{tikzpicture} 
\caption{The 2-loop structure of $\Pi^\lessgtr(X,k)$ for the heavy meson. Solid lines represent the heavy-meson propagators, $G_D^\lessgtr(X,k_1)$; dashed lines depict light-meson propagators, $G_\Phi^\lessgtr(X,k_2),G_\Phi^\gtrless(X,k_3)$; green circles correspond to $T$-matrix operators.}
  \label{fig:transport-Pilessgtr}
\end{figure}

Diagrammatically, the self-energies $\Pi^\lessgtr(X,k)$ are given by the 2-loop diagram represented in Fig.~\ref{fig:transport-Pilessgtr}. The solid lines represent heavy mesons, while dashed lines are $\Phi$ propagators, and the vertices stand for the full $T$ matrices.

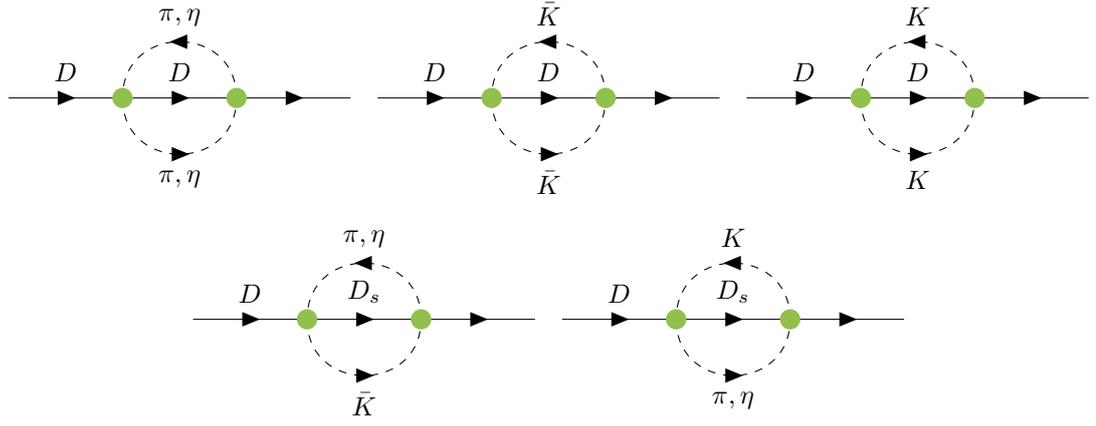
\begin{figure}[t!]
  \centering
   \begin{subfigure}[b]{0.3\textwidth}\centering 
   \begin{tikzpicture}[baseline=(i.base),inner sep=6pt]
    \begin{feynman}
      \vertex (a);
      \vertex [right = of a] (i);
      \vertex [right = of i] (j);
      \vertex [right = of j] (b);
      \diagram*{
         (a) -- [fermion,edge label=\(D\)] (i);
         (i) -- [fermion,edge label=\(D\)] (j);
         (j) -- [fermion] (b);
         (j) -- [charged scalar, half right, looseness=1.7,edge label'={\(\pi,\eta\)}] (i);
         (i) -- [charged scalar, half right, looseness=1.7,edge label'={\(\pi,\eta\)}] (j);
     } ;
     \draw[dot,minimum size=4mm,thick,ctcolorgreen,fill=ctcolorgreen] (i) circle(1.2mm);
     \draw[dot,minimum size=4mm,thick,ctcolorgreen,fill=ctcolorgreen] (j) circle(1.2mm);
    \end{feynman}
  \end{tikzpicture} 
  \end{subfigure}
  \hspace{0.3cm}
   \begin{subfigure}[b]{0.3\textwidth}\centering 
   \begin{tikzpicture}[baseline=(i.base),inner sep=6pt]
    \begin{feynman}
      \vertex (a);
      \vertex [right = of a] (i);
      \vertex [right = of i] (j);
      \vertex [right = of j] (b);
      \diagram*{
         (a) -- [fermion,edge label=\(D\)] (i);
         (i) -- [fermion,edge label=\(D\)] (j);
         (j) -- [fermion] (b);
         (j) -- [charged scalar, half right, looseness=1.7,edge label'={\(\bar{K}\)}] (i);
         (i) -- [charged scalar, half right, looseness=1.7,edge label'={\(\bar{K}\)}] (j);
     } ;
     \draw[dot,minimum size=4mm,thick,ctcolorgreen,fill=ctcolorgreen] (i) circle(1.2mm);
     \draw[dot,minimum size=4mm,thick,ctcolorgreen,fill=ctcolorgreen] (j) circle(1.2mm);
    \end{feynman}
  \end{tikzpicture} 
  \end{subfigure}
  \hspace{0.3cm}
   \begin{subfigure}[b]{0.3\textwidth}\centering 
      \begin{tikzpicture}[baseline=(i.base),inner sep=6pt]
    \begin{feynman}
      \vertex (a);
      \vertex [right = of a] (i);
      \vertex [right = of i] (j);
      \vertex [right = of j] (b);
      \diagram*{
         (a) -- [fermion,edge label=\(D\)] (i);
         (i) -- [fermion,edge label=\(D\)] (j);
         (j) -- [fermion] (b);
         (j) -- [charged scalar, half right, looseness=1.7,edge label'={\(K\)}] (i);
         (i) -- [charged scalar, half right, looseness=1.7,edge label'={\(K\)}] (j);
     } ;
     \draw[dot,minimum size=4mm,thick,ctcolorgreen,fill=ctcolorgreen] (i) circle(1.2mm);
     \draw[dot,minimum size=4mm,thick,ctcolorgreen,fill=ctcolorgreen] (j) circle(1.2mm);
    \end{feynman}
  \end{tikzpicture}  
  \end{subfigure}
   \begin{subfigure}[b]{0.3\textwidth}\centering 
   \begin{tikzpicture}[baseline=(i.base),inner sep=6pt]
    \begin{feynman}
      \vertex (a);
      \vertex [right = of a] (i);
      \vertex [right = of i] (j);
      \vertex [right = of j] (b);
      \diagram*{
         (a) -- [fermion,edge label=\(D\)] (i);
         (i) -- [fermion,edge label=\(D_s\)] (j);
         (j) -- [fermion] (b);
         (j) -- [charged scalar, half right, looseness=1.7,edge label'={\(\pi,\eta\)}] (i);
         (i) -- [charged scalar, half right, looseness=1.7,edge label'={\(\bar{K}\)}] (j);
     } ;
     \draw[dot,minimum size=4mm,thick,ctcolorgreen,fill=ctcolorgreen] (i) circle(1.2mm);
     \draw[dot,minimum size=4mm,thick,ctcolorgreen,fill=ctcolorgreen] (j) circle(1.2mm);
    \end{feynman}
  \end{tikzpicture} 
  \end{subfigure}
  \hspace{0.3cm}
   \begin{subfigure}[b]{0.3\textwidth}\centering 
      \begin{tikzpicture}[baseline=(i.base),inner sep=6pt]
    \begin{feynman}
      \vertex (a);
      \vertex [right = of a] (i);
      \vertex [right = of i] (j);
      \vertex [right = of j] (b);
      \diagram*{
         (a) -- [fermion,edge label=\(D\)] (i);
         (i) -- [fermion,edge label=\(D_s\)] (j);
         (j) -- [fermion] (b);
         (j) -- [charged scalar, half right, looseness=1.7,edge label'={\(K\)}] (i);
         (i) -- [charged scalar, half right, looseness=1.7,edge label'={\(\pi,\eta\)}] (j);
     } ;
     \draw[dot,minimum size=4mm,thick,ctcolorgreen,fill=ctcolorgreen] (i) circle(1.2mm);
     \draw[dot,minimum size=4mm,thick,ctcolorgreen,fill=ctcolorgreen] (j) circle(1.2mm);
    \end{feynman}
  \end{tikzpicture} 
  \end{subfigure}
  \caption{Diagrams contributing to $\Pi_D^\lessgtr(X,k)$ for the $D$ meson in Eqs.~(\ref{eq:transport-Pil}) and (\ref{eq:transport-Pig}). A total of 10 channels are needed due to the coupled-channel problem for the $D$-meson interaction. Green circles are $T$-matrix elements in the appropriate channel.}
  \label{fig:transport-Pilessgtrchannels}
\end{figure}

The sum over $a,b,c$ in Eqs.~(\ref{eq:transport-Pil}) and (\ref{eq:transport-Pig}) encodes the 
different species that can interact and that are fixed by the effective vertices. This sum is restricted to the combinations that respect the conservation of quantum numbers in a full coupled-channel approach. In particular, $D_a$ can describe either $D$ or $D_s$ mesons, and $\Phi_b,\Phi_c$ can represent $\pi,K,\bar{K},\eta$. Figure~\ref{fig:transport-Pilessgtrchannels} shows the 10 allowed diagrams, although not all of them are equally important. We will comment on this when addressing the effect of inelastic processes in our calculations. 

We now discuss the form of the Wightman functions $G_{D,\Phi}^\lessgtr (X,k)$. We have mentioned that light degrees of freedom satisfy their own Kadanoff-Baym equations, which are coupled to those of the heavy mesons. In the context of \glspl{hic}, the standard approach for the heavy-flavor dynamics is to exploit the fact that the light degrees of freedom have reached equilibrium much before the heavy sector, as the latter has a much longer relaxation time, roughly proportional to the mass of the particle. In our goal of accessing transport coefficients of heavy mesons, we also assume this, so there is no need to consider the kinetic equation for light mesons. 

In addition, we apply the thermal local equilibrium solution for $G_\Phi^\lessgtr(X,k)$, which can be expressed as~\cite{kadanoff1962quantum}
\begin{align}
 \ii G_{\Phi}^< (X,k) & = 2\pi S_{\Phi} (X,k) f^{(0)}_{\Phi}(X,k^0)  \ , \label{eq:transport-soleq1} \\
 \ii G_{\Phi}^> (X,k) & = 2\pi S_{\Phi} (X,k) \left[1+f^{(0)}_{\Phi}(X,k^0)\right] \ , \label{eq:transport-soleq2}
 \end{align}
where $S_\Phi(X,k)$ is the equilibrium light-meson spectral function and $f^{(0)}(X,k^0)$ is the equilibrium occupation number, that is, the \gls{be} distribution function. Equation (\ref{eq:transport-soleq2}) incorporates the Bose enhancement factor $1+f^{(0)}(X,k^0)$.

Concerning the heavy mesons, we will assume that they are not far from equilibrium, which is enough to address the calculation of the transport coefficients. Then, we use a similar form as in Eqs.~(\ref{eq:transport-soleq1}) and (\ref{eq:transport-soleq2}) for their Green's function, the so-called Kadanoff-Baym Ansatz,
\begin{align}
 \ii G_{D}^< (X,k) & = 2\pi S_{D} (X,k) f_{D}(X,k^0)  \ , \label{eq:transport-ansatz1} \\
 \ii G_{D}^> (X,k) & = 2\pi S_{D} (X,k) \left[1+f_{D}(X,k^0)\right] \ , \label{eq:transport-ansatz2}
 \end{align}
where the $D$-meson spectral function $S_D(X,k)$ and the distribution function $f_D(X,k^0)$ are out of equilibrium.

Inserting these Ans\"atze into the kinetic equation (see Eq.~(\ref{eq:transport-offGless})), together with the $D$-meson self-energies defined in Eqs.~(\ref{eq:transport-Pil}) and (\ref{eq:transport-Pig}), one obtains 
\begin{align}\nonumber
    \left(  k^\mu - \frac{1}{2} \frac{\partial \textrm{Re } \Pi^{\textrm{R}}}{\partial k_\mu} \right)& \frac{\partial}{\partial X^\mu} [S_D(X,k) f_D(X, k^0)] =  \frac{1}{2}  \int \prod_{i=1}^3 \frac{d^4 k_i}{(2\pi)^3} (2\pi)^4 \delta^{(4)} (k_1+k_2-k_3-k)  \\ \nonumber
  &\times \left|T (k_1^0+k_2^0+\ii\varepsilon, \vec{k}_1 + \vec{k}_2)\right|^2 \ S_D (X,k_1) S_\Phi (X,k_2) S_\Phi (X,k_3) S_D (X,k)  \label{eq:transport-transportG} \\ \nonumber
& \times \left[ f_D (X, k^0_1) f^{(0)}_\Phi (X, k^0_2) \tilde{f}^{(0)}_\Phi (X, k^0_3)  \tilde{f}_D (X, k^0)\right. \\ 
& \left. \qquad-\tilde{f}_D (X, k^0_1) \tilde{f}^{(0)}_\Phi (X, k^0_2)  f^{(0)}_\Phi (X, k^0_3) f_D (X, k^0) \right]  \ , 
  \end{align}
where we have defined $\tilde{f}(X,k^0) \equiv 1+f(X,k^0)$. Notice that we have not written explicitly the sum over $\{a,b,c\}$ on the \gls{rhs}, but it should be understood to account for all possible physical processes.

Focusing on the positive-energy $D$-meson, one can integrate over $dk^0$ along the positive branch in both sides,
\begin{align}\nonumber
   \int_0^{+\infty} dk^0 & \left(  k^\mu - \frac{1}{2} \frac{\partial \textrm{Re } \Pi^{\textrm{R}}}{\partial k_\mu} \right) S_D(X,k) \frac{\partial f_D(X,k^0)}{\partial X^\mu} \label{eq:transport-offshellkinetic} \\ \nonumber
  & = \frac{1}{2} \int_0^{+\infty} dk^0 \int \prod_{i=1}^3 \frac{d^4 k_i}{(2\pi)^3} (2\pi)^4 \delta^{(4)} (k_1+k_2-k_3-k) \left|T (k_1^0+k_2^0+\ii\varepsilon, \vec{k}_1 + \vec{k}_2)\right|^2   \\ \nonumber
  &\quad \times S_D (X,k_1) S_\Phi (X,k_2) S_\Phi (X,k_3) S_D (X,k)  \\ \nonumber
& \quad\times \left[ f_D (X, k^0_1) f^{(0)}_\Phi (X, k^0_2) \tilde{f}^{(0)}_\Phi (X, k^0_3)  \tilde{f}_D (X, k^0) \right. \\ 
& \left. \qquad -   \tilde{f}_D (X, k^0_1) \tilde{f}^{(0)}_\Phi (X, k^0_2)  f^{(0)}_\Phi (X, k^0_3) f_D (X, k^0) \right]  \ , 
  \end{align}
where the transport equation for the spectral function has been used after Eq.~(\ref{eq:transport-transportG})\footnote{From Eqs.~(\ref{eq:transport-ansatz1}) and (\ref{eq:transport-ansatz2}) one has $S_D(X,k)=\ii\left[G_D^>(X,k)-G_D^<(X,k)\right]/(2\pi)$. Then, subtracting the transport equations, Eq.~(\ref{eq:transport-offGgreat}) minus Eq.~(\ref{eq:transport-offGless}), one obtains the transport equation for the spectral density,
$$ \left(  k^\mu - \frac{1}{2} \frac{\partial \textrm{Re } \Pi^{\textrm{R}} (X,k)}{\partial k_\mu} \right) \frac{\partial S_D (X,k)}{\partial X^\mu} = 0 \ . $$}. 

This equation is very similar to the standard Boltzmann equation but the effects of the medium (temperature and density) are now incorporated in the $T$ matrix and the spectral functions of the interacting particles. The reduction of this equation to the classical Boltzmann equation makes an extra assumption for the spectral functions, the \textit{quasiparticle approximation}. In this approximation, the spectral function is only characterized by the quasiparticle energy $E_k$ and the thermal decay width $\gamma_k$, and $S(k)$ admits a Lorentzian shape peaked at $E_k$ and with a spectral width $\gamma_k\ll E_k$. In the limit $\gamma_k/E_k\rightarrow 0$, one can consider the \textit{narrow limit},
\begin{equation} 
S(k) \rightarrow \frac{z_k}{2E_k} \left[\delta(k^0-E_k)-\delta(k^0+E_k)\right] \ . \label{eq:transport-Diracdelta} 
\end{equation}
In this limit, one is effectively treating the quasiparticle as a stable state (no thermal width), but with a medium-modified energy $E_k$ plus a correction due to the $z_k$ factor.

\begin{figure}[t!]
  \centering
   \begin{tikzpicture}[baseline=(i.base)]
    \begin{feynman}
      \vertex (i) ;
      \vertex [above left =1cm of a] (a) {\(k\)};
      \vertex [below left =1cm of i] (b) {\(k_3\)};
      \vertex [above right =1cm of i] (c) {\(k_1\)};
      \vertex [below right =1cm of i] (d) {\(k_2\)};
      \diagram*{
         (a) -- [fermion] (i);
         (i) -- [fermion] (c);
         (b) -- [charged scalar] (i);
         (i) -- [charged scalar] (d);
     } ;
     \draw[dot,minimum size=2mm,thick,ctcolorgreen,fill=ctcolorgreen] (i) circle(1.2mm);
    \end{feynman}
  \end{tikzpicture}
  \caption{Labeling of the incoming and outgoing momenta in a generic binary scattering. Solid lines represent heavy mesons, while dashed lines represent light mesons. The vertex corresponds to a retarded $T$-matrix element.
  \label{fig:transport-naming}}
\end{figure}
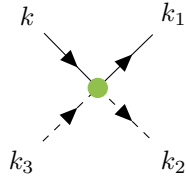

The narrow limit allows to trivially perform the integration over the $k^0_i$ variables in the kinetic equation in Eq.~(\ref{eq:transport-offshellkinetic}) to obtain the on-shell (or Boltzmann) transport equation. When assuming this approximation for all the particles involved, different combinations of the Dirac delta functions can be taken, which describe different scattering processes~\cite{Blaizot:1999xk,Blaizot:2001nr}. However, in this limit, the energy-momentum conservation only allows $2\leftrightarrow 2$ processes~\cite{Blaizot:2001nr,Juchem:2003bi}. Among those, the one with a $D \bar{D}$ pair in the initial state can be neglected due to the Boltzmann suppression factor, and the inverse reaction is suppressed due to the high energy threshold for the two incoming pions). Therefore one only considers one type of collision, namely $D\Phi \rightarrow D\Phi$. We label the momenta of a generic scattering as $k+3 \rightarrow 1+2$ (see Fig.~\ref{fig:transport-naming} for illustration.)

Denoting for simplicity $f_i\equiv f_{D,\Phi} (X,E_i)$, where the species is fixed by the value of $i$, that is $i=\{1,k\}$ for the heavy meson, $i=\{2,3\}$ for the light meson, we obtain
\begin{align}
 & \left[  \frac{\partial}{\partial t} - \frac{\vec{k}}{E_k} \cdot \nabla_X \right] f_k  =  \frac{z_{k}}{2E_k} \int \left( \prod_{i=1}^3 \frac{d^3 k_i }{(2\pi)^3}\frac{z_i}{ 2E_i} \right)  (2\pi)^4 \delta^{(3)} (\vec{k}+ \vec{k}_3-\vec{k}_1-\vec{k}_2) \nonumber \\ 
& \times \left\{ \delta(E_k+E_3-E_1-E_2) \left|T (E_k+E_3, \vec{k} + \vec{k}_3)\right|^2 
 \left( f_1 f_2 \tilde{f}_3 \tilde{f}_k-   \tilde{f}_1 \tilde{f}_2  f_3 f_k \right) \right.  \nonumber \\
& \left. \quad +\ \delta(E_k-E_3-E_1+E_2) \left|T (E_k-E_3, \vec{k} + \vec{k}_3)\right|^2 
\left( f_1 \tilde{f}_2^{(0)} f_3^{(0)}  \tilde{f}_k -   \tilde{f}_1 f_2^{(0)} \tilde{f}_3^{(0)} f_k \right)  \right\} \ ,  \label{eq:transport-onshellkinetic}
\end{align}
where the two different collision terms on the \gls{rhs} of the equality depend on the sign of the energy of the light meson $E_3$. The first one evaluates the scattering amplitude above the mass threshold of the particles $k$ and $3$ and we refer to it as the \textit{unitary contribution}. The second one implies the scattering amplitude below the energy threshold and it is nonzero due to the Landau cut~\cite{Weldon:1983jn,Ghosh:2011bw} arising in the two-meson propagator. This term is referred to as the \textit{Landau contribution}, and it is of key importance in this chapter.

If only elastic collisions are considered, then one can simplify the equation by switching variables $\vec{k}_2$ and $\vec{k}_3$ to obtain
\begin{align}\label{eq:transport-onshellkineticelastic} \nonumber
  \left[  \frac{\partial}{\partial t} - \frac{\vec{k}}{E_k} \cdot \nabla_X \right] f_k  &=  \frac{z_{k}}{2E_k} \int \left( \prod_{i=1}^3 \frac{d^3 k_i }{(2\pi)^3}\frac{z_i}{ 2E_i} \right)  (2\pi)^4 \delta^{(4)} (k+ k_3-k_1- k_2) \\  \nonumber
& \times \left\{ \left|T (E_k+E_3+\ii\varepsilon, \vec{k} + \vec{k}_3)\right|^2 +
\left|T (E_k-E_2+\ii\varepsilon, \vec{k} - \vec{k}_2)\right|^2  \right\} \\ 
& \times \left( f_1 f_2^{(0)} \tilde{f}_3^{(0)} \tilde{f}_k- \tilde{f}_1 \tilde{f}_2^{(0)}  f_3^{(0)} f_k \right) \ .   
\end{align}
This equation looks almost like the Boltzmann equation\footnote{More exactly, the Boltzmann-Uehling-Uhlenbeck equation, as quantum effects are incorporated.} considered previously in the literature, where the effect of the Landau contribution was neglected, vacuum scattering amplitudes were employed, and the factors $z_i$ were set to one. A version of the transport equation where these factors were kept is presented in Ref.~\cite{Davis:1991zc}. In our particular case, the approximation $z_i \simeq 1$ is an excellent one, given the good quasiparticle description of the $D$ mesons. Notice that the quasiparticle energies in Eq.~(\ref{eq:transport-onshellkineticelastic}) and the $T$ matrix do contain medium modifications.

\section{Heavy-meson properties in thermal equilibrium}
\label{sec:transport-equilibrium}
In the previous section, we have derived the kinetic equation satisfied by heavy mesons and commented on the different approximations which can be used to simplify it, from the Kadanoff-Baym equation to the classical Boltzmann equation. These steps were necessary to justify the form of the kinetic equation constructed from the effective theory in Chapters~\ref{ch:exoticsinfreespace} and \ref{ch:hot-medium}. In addition, this derivation will also help to perform the off-shell generalization of the Fokker-Planck equation and the heavy-flavor transport coefficients, which will be pursued in Section~\ref{subsec:transport-fokker-planck}.

In this section, we analyze the different elements appearing in Eq.~(\ref{eq:transport-offshellkinetic}), namely the $T$-matrix elements, the retarded heavy-meson self-energy, and the spectral function, for the particular case of a system in equilibrium.

The $T$ matrix appearing in Eqs.~(\ref{eq:transport-Pil}) and (\ref{eq:transport-Pig}) is the retarded four-point amplitude that follows from solving the \gls{bs} equation in coupled channels, as extensively described in Chapter~\ref{ch:exoticsinfreespace}, in the vacuum, and Chapter~\ref{ch:hot-medium}, at finite temperature in the equilibrium case. 

We have seen that the unitary cut of the $T$ matrix provides a source for the charmed-meson vacuum
decay width through the imaginary part of the corresponding self-energy. As an example, when a $D$ meson interacts with a pion, it can suffer an elastic scattering or, if the energy of
the collision is large enough, then the pair can also convert into a $D\eta$ or $D_s\bar{K}$ pair. At finite temperature, the bath is populated by $\Phi$ mesons, and their relative importance is weighted by the appropriate \gls{be} distribution functions. Consequently, at $T>0$ the contribution of the unitary cut to the decay width is convoluted by statistical weight factors, and additional physical processes appear in the kinematic region of the Landau cut. These new processes include, for example, the absorption of in-medium real light mesons by the $D$ meson and are forbidden in vacuum due to kinematic restrictions. On the other hand, the structure of the $T$ matrix is smeared at finite temperature and the thresholds of the unitary ($\sqrt{s}\geq (m_D + m_\Phi)$) and Landau ($\sqrt{s}\leq |m_D - m_\Phi|$) cuts are smoothened with increasing temperatures as a result of the
widening of the $D$-meson spectral function (see Chapter~\ref{ch:hot-medium} for details).

Regarding the heavy-meson self-energy and spectral function at finite temperature, the equilibrium quantities have been obtained in Chapter~\ref{ch:hot-medium} using the \gls{itf}. Let us reproduce the expressions with the notation used in this chapter, for completeness. The heavy-meson retarded propagator reads
%
%
\begin{equation} 
G^{\textrm{R}}_D (k^0,\vec{k}\,)= \frac{1 }{(k^0)^2-\vec{k}\,^2-m_D^2 -\textrm{Re } \Pi^{\textrm{R}}(k^0,\vec{k}\,;T) - \ii \textrm{Im } \Pi^{\textrm{R}}(k^0,\vec{k}\,;T)} \ , \label{eq:transport-GretT} 
\end{equation}
where $m_{D}$ is the (vacuum) physical $D$-meson mass, renormalized by the vacuum contribution of the retarded $D$-meson self-energy $\Pi^R$. We remind that, after mass renormalization, the real and imaginary parts of the self-energy in Eq.~(\ref{eq:transport-GretT}) only contain thermal corrections. This form can also be obtained from Eq.~(\ref{eq:transport-eqforGR}) when applied to the equilibrium case.

The spectral function is then obtained from the imaginary part of the retarded propagator using Eq.~(\ref{eq:transport-greenfuncs}),
\begin{equation} 
S_D (k^0,\vec{k}\,)=\frac{\ii G_D^> (k^0,\vec{k}\,) - \ii G_D^< (k^0,\vec{k}\,)}{2\pi}  = -\frac{1}{\pi} \textrm{Im } G^{\textrm{R}}_D (k^0,\vec{k}\,) \ .
\end{equation}
This definition of the spectral function in terms of the imaginary part of the retarded Green's function is compatible with the convention in Chapter~\ref{ch:hot-medium} for the equilibrium case.

In terms of the $D$-meson retarded self-energy, the spectral function reads
\begin{equation} 
S_D(k^0,\vec{k}\,) =- \frac{1}{\pi} \frac{ \textrm{Im } \Pi^{\textrm{R}} (k^0,\vec{k}\,;T) }{\left[(k^0)^2-\vec{k}\,^2-m_{D}^2 -\textrm{Re } \Pi^{\textrm{R}}(k^0,\vec{k}\,;T)\right]^2 + \left[ \textrm{Im } \Pi^{\textrm{R}}(k^0,\vec{k}\,;T)\right]^2} \label{eq:transport-spectralfunc} \ . 
\end{equation}

In the quasiparticle approximation, the pole of the retarded Green's function is not far from the vacuum one and the spectral function can be written as
\begin{equation} 
S_D (k^0,\vec{k}\,) \simeq \frac{z_k}{2\pi E_k} \frac{\gamma_k}{(k^0-E_k)^2 +\gamma_k^2} \ , 
\end{equation}
where $E_k$ is the quasiparticle energy, solution of
\begin{equation} 
E^2_k -\vec{k}\,^2-m_{D}^2 -\textrm{Re } \Pi^{\textrm{R}}(E_k,\vec{k}\,;T) =0 \ , \label{eq:transport-quasienergy} 
\end{equation}
and the damping rate $\gamma_k$ is defined as
\begin{equation}
\gamma_k = -\frac{z_k}{2E_k} \textrm{Im } \Pi^{\textrm{R}}(E_k,\vec{k}\,;T) \ , \label{eq:transport-gammak} 
\end{equation}
with the $z_k$ factor,
\begin{equation} 
z_k^{-1} = 1 -\frac{1}{2E_k} \left. \left( \frac{\partial \textrm{Re } \Pi^{\textrm{R}}(k^0,\vec{k}\,;T)}{\partial k^0} \right) \right|_{k^0=E_k} \ . 
\end{equation} 

In particular, this approximation entails that the heavy-meson damping rate should be much smaller than the quasiparticle energy.
We have numerically checked that the $z_k$-factors are very close to one. Rather independent of the quasiparticle momentum $k$, we observe up to 2\% (1\%) deviations from unity for $T=150$ MeV ($T=100$ MeV). For $T=40$ MeV, the factor $z_k$ is fully compatible with one. Therefore the approximation $z_k \simeq 1$ is an excellent one and will be used in what follows.
%

The $D$-meson thermal width, $\Gamma_k$, is defined as twice the damping rate, $\Gamma_k \equiv 2\gamma_k$. This quantity is negligible at low temperatures. The effects of the medium make it sizable at $T=150$~MeV, with thermal widths of the order of $100$~MeV, as seen in Chapter~\ref{ch:hot-medium}. Nevertheless, these values are still small compared to the corresponding $E_k$, which validates the quasiparticle approach.

In the next sections, we explore different approximations to address the $D$-meson thermal width and the transport coefficients. From the results of this section, where the quasiparticle approximation is a very good one, we can anticipate that off-shell effects might not contribute much to these quantities. Nevertheless, we will stay as general as possible and quantify the importance of the different approximations.

\section{Analysis of the heavy-meson thermal width in equilibrium}
\label{sec:transport-kinematic}

In this section, we analyze in detail several effects on the heavy-meson thermal width $\Gamma_k$, defined as twice the damping rate of Eq.~(\ref{eq:transport-gammak}). In particular, we calculate this coefficient for $D$ mesons with two different methodologies. In section~\ref{subsec:transport-kinematic-method1}, the thermal width is obtained from the integration of the imaginary part of the scattering amplitude, whereas in the method described in section~\ref{subsec:transport-kinematic-method2} the integrand is proportional to the square of the amplitude. We also infer the importance of the Landau cut of the scattering amplitude to the thermal width.

Furthermore, the heavy-meson thermal width is given for an on-shell heavy meson, meaning that the heavy-meson energy $E_k$ is automatically fixed by $k$ through Eq.~(\ref{eq:transport-quasienergy}). However, the internal propagators of the retarded self-energy do not need to be on their mass shell. These off-shell effects are addressed in section~\ref{subsec:transport-kinematic-effects}, together with other effects such as the contribution of the inelastic channels or the relevance of the various light mesons of the bath. 

The results for the thermal width of the $\bar{B}$ mesons are summarized in section~\ref{subsec:transport-kinematic-B}.

\subsection{Thermal width from the imaginary part of the scattering amplitude}
\label{subsec:transport-kinematic-method1}
In the definition of the heavy-meson thermal width in terms of the retarded self-energy,
\begin{equation} 
\Gamma_k = - \frac{z_k}{E_k} \textrm{Im } \Pi^{\textrm{R}} (E_k,\vec{k}\,;T) \ , \label{eq:transport-Gammak} 
\end{equation}
only the effect of pions was considered in the self-consistent calculation of the $D$-meson self-energy in the previous chapters, being the contribution of the other light mesons very suppressed. Here we also analyze these contributions because we have access to all the elements of the $T$ matrix.

Let us review the calculation of $\Gamma_k$ in more detail and study the contributions from the different kinematic ranges. 
The calculation of the self-energy in terms of $T$-matrix elements in the \gls{itf} leads, after summing over Matsubara frequencies and performing the analytic continuation to real energies, to the following expression for the imaginary part of the retarded self-energy,
%
%
%
\begin{align}
\textrm{Im }  \Pi^{\textrm{R}}(E_k,\vec{k}\,) & =  \int \frac{d^3p}{(2\pi)^3}   \left[  \frac{f^{(0)}(E_p)}{2E_p}\,   \textrm{Im } T_{D\pi} (E_k+E_p,\vec{p}+\vec{k}\,)  \right. \nonumber \\
& \left. + \frac{1+f^{(0)}(E_p)}{2E_p}\, \textrm{Im } T_{D\pi} (E_k-E_p,\vec{p}+\vec{k}\,)  \right. \nonumber \\ 
& - \left.   \frac{f^{(0)}(E_k+E_p)}{2E_p}\, \textrm{Im } T_{D\pi}(E_k+E_p,\vec{p}+\vec{k}\,) \right. \nonumber \\
& + \left. \frac{f^{(0)}(E_k-E_p)}{2E_p}\, \textrm{Im } T_{D\pi}(E_k- E_p,\vec{p}+\vec{k}\,) \right]  , \label{eq:transport-ImPiR}
\end{align}
where we have already fixed the external $D$-meson energy to the quasiparticle energy $E_k$,  that is a function of $k$, as given by Eq.~(\ref{eq:transport-quasienergy}). We have neglected the $z_k$ factors, because they are very close to one, as previously discussed, but they can be easily incorporated if desired.

Using this result, we can write the $D$-meson thermal width in Eq.~(\ref{eq:transport-Gammak}) as the contribution of four pieces,
\begin{equation}
\Gamma_k = \Gamma_k^{(1)}+\Gamma_k^{(2)}+\Gamma_k^{(3)}+\Gamma_k^{(4)} \ , \label{eq:transport-Gammak4terms} 
\end{equation}
where
\begin{align} 
\Gamma_k^{(1)} & =  -\frac{1}{E_k} \int \frac{d^3p}{(2\pi)^3}  \frac{f^{(0)}(E_p)}{2E_p} \, \textrm{Im } T_{D\pi}(E_k+E_p,\vec{p}+\vec{k}\,) \ , \label{eq:transport-Gammak1} \\ 
\Gamma_k^{(2)} & =  -\frac{1}{E_k} \int \frac{d^3p}{(2\pi)^3}   
\frac{1+f^{(0)}(E_p)}{2E_p} \, \textrm{Im } T_{D\pi} (E_k-E_p,\vec{p}+\vec{k}\,) \ ,  \label{eq:transport-Gammak2} \\ 
\Gamma_k^{(3)} & =  \frac{1}{E_k} \int \frac{d^3p}{(2\pi)^3} \frac{f^{(0)}(E_k+E_p)}{2E_p} \, \textrm{Im } T_{D\pi} (E_k+E_p,\vec{p}+\vec{k}\,) \ , \label{eq:transport-Gammak3} \\ 
\Gamma_k^{(4)} & =  -\frac{1}{E_k} \int \frac{d^3p}{(2\pi)^3}
\frac{f^{(0)}(E_k-E_p)}{2E_p} \, \textrm{Im } T_{D\pi} (E_k- E_p,\vec{p}+\vec{k}\,)  \ . \label{eq:transport-Gammak4}
\end{align}

It is important to realize that both $\Gamma_k^{(1)}$ and $\Gamma_k^{(3)}$ receive a contribution from the scattering amplitude above the two-particle mass threshold, while $\Gamma_k^{(2)}$ and $\Gamma_k^{(4)}$ depend on the values of the $T$ matrix below the threshold. The latter contribution is related to the Landau cut, and only appears at finite temperature when the total momentum of the collision is different from zero or when the masses of the interacting particles are different~\cite{Weldon:1983jn,Das:1997gg,Torres-Rincon:2017zbr}. 

In particular, when vacuum amplitudes are used, the Landau cut disappears, and only $\Gamma_k^{(1)}$  and $\Gamma_k^{(3)}$ contribute. Incidentally, this is the situation in the pion-pion vacuum scattering of Ref.~\cite{Schenk:1993ru}, where only $\Gamma_k^{(1)}$ appears. In our case, $\Gamma_k^{(3)}$ is in fact extremely small, because it is roughly proportional to the product of the pion and $D$-meson densities, and the latter is very scarce~\footnote{This can be shown from the relation $f^{(0)}(E_k+E_p)=f^{(0)} (E_k) f^{(0)} (E_p) / (1+f^{(0)} (E_k)+f^{(0)} (E_p))$.}. However, when considering a medium-dependent interaction, $\Gamma_k^{(2)}$ and $\Gamma_k^{(4)}$ also have a potentially important contribution that we now quantify.

\begin{figure}[t!]
  \centering
  \includegraphics[width=0.333\textwidth]{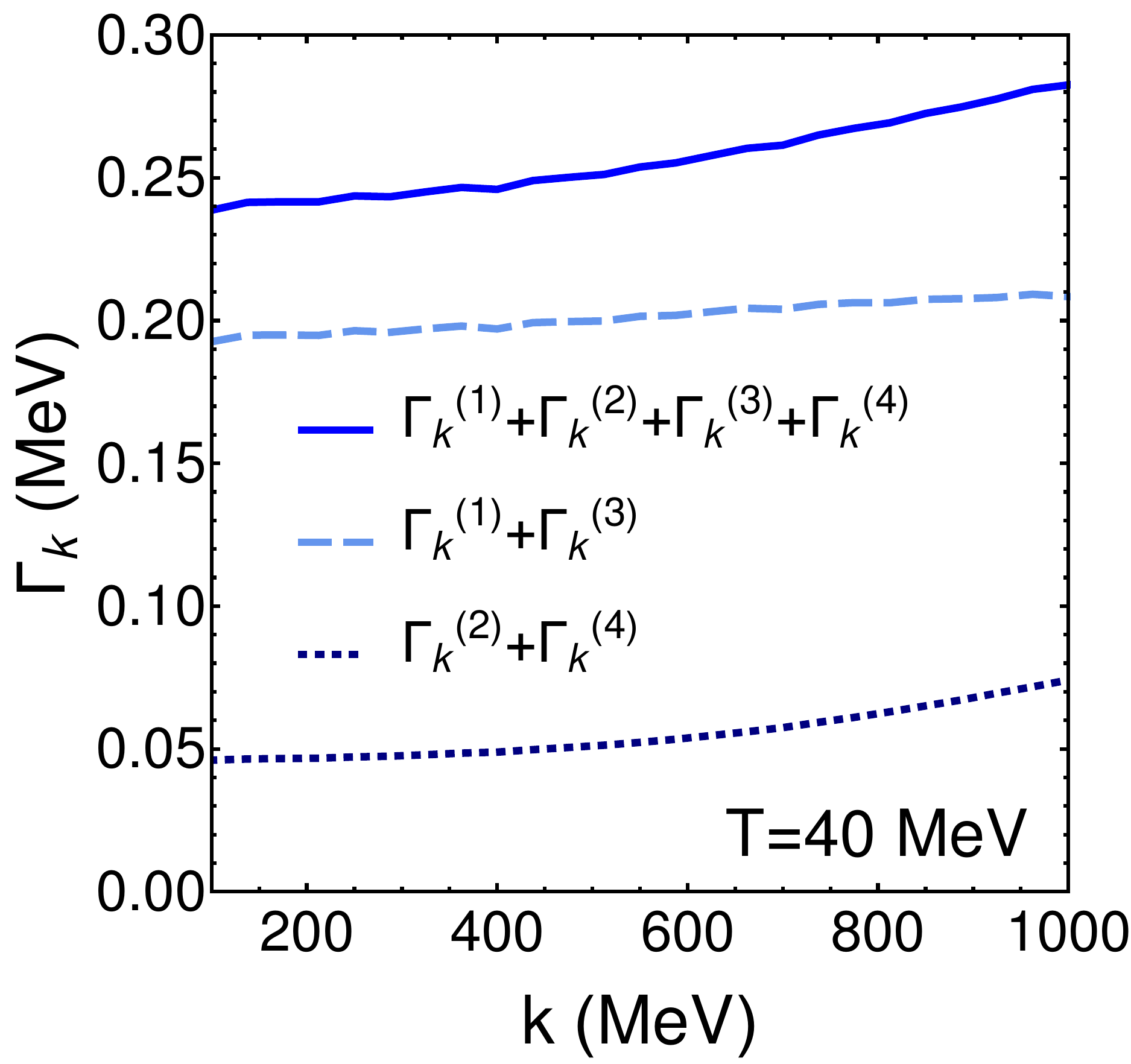}
   \includegraphics[width=0.32\textwidth]{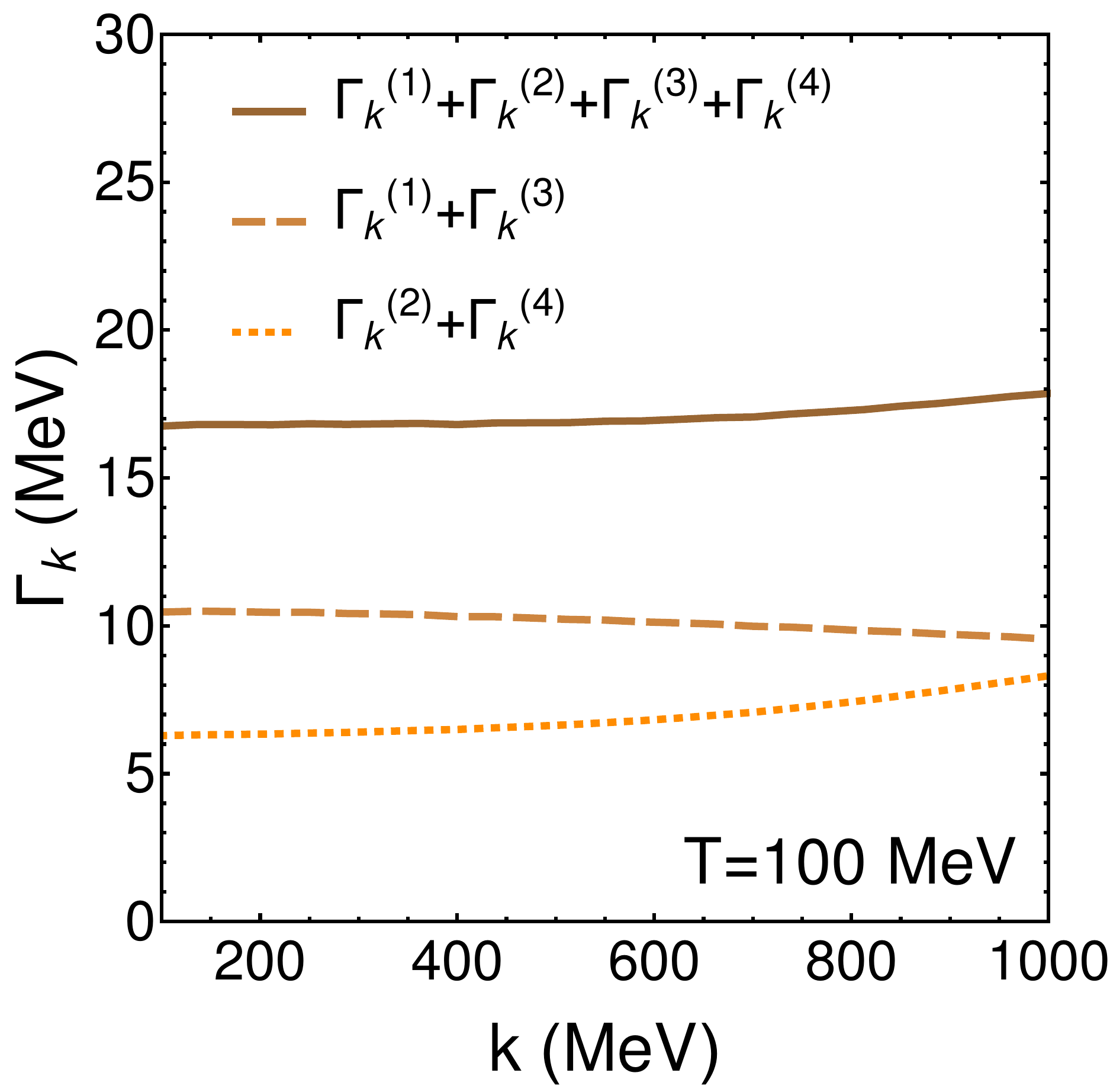}
  \includegraphics[width=0.327\textwidth]{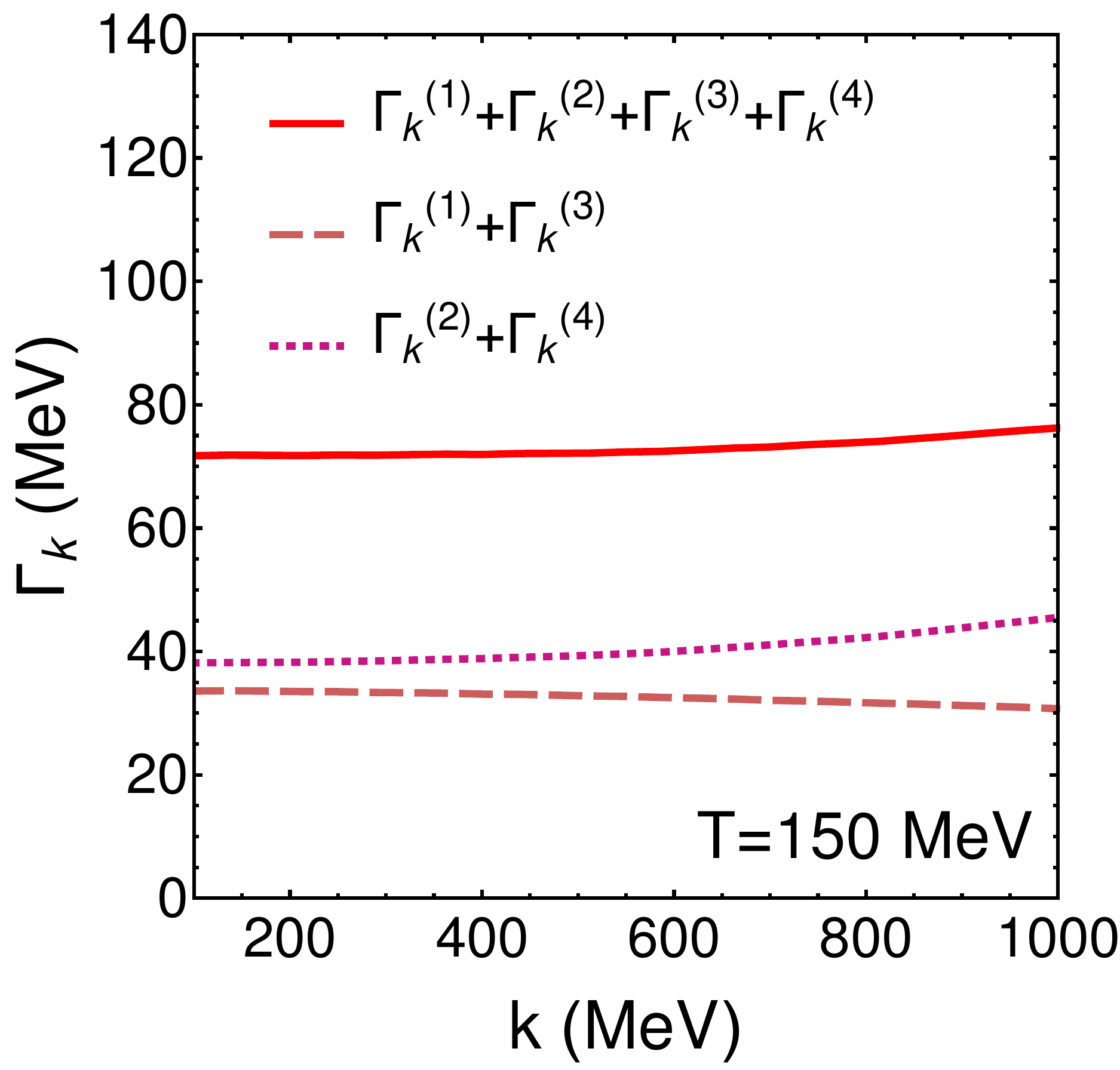}
    \caption{The $D$-meson thermal width as computed from Eq.~(\ref{eq:transport-Gammak4terms}). Dashed lines show the contribution of $\Gamma_k^{(1)}+\Gamma_k^{(3)}$ (Eqs.~(\ref{eq:transport-Gammak1})+(\ref{eq:transport-Gammak3}) above the threshold), whereas dotted lines correspond to the contribution of $\Gamma_k^{(2)}+\Gamma_k^{(4)}$ (Eqs.~(\ref{eq:transport-Gammak2})+(\ref{eq:transport-Gammak4}) below the threshold (Landau cut)).}
  \label{fig:transport-Gammak}
\end{figure}

To gauge the weight of the different terms in Eq.~(\ref{eq:transport-Gammak4terms}) we plot in Fig.~\ref{fig:transport-Gammak} the different contributions at three different temperatures. The input for $T_{D\pi}$ is taken from the results of the scattering amplitudes obtained in Chapter~\ref{ch:hot-medium}.

At small temperatures ($T=40$~MeV) the terms $\Gamma_k^{(1)}+\Gamma_k^{(3)}$ dominate. Nevertheless, the two pieces coming from the Landau cut, $\Gamma_k^{(2)}+\Gamma_k^{(4)}$, give a nonzero contribution resulting in a $20\%$ of the total thermal width. At $T=100$~MeV, the contribution of the unitary cut, $\Gamma_k^{(1)}+\Gamma_k^{(3)}$,  and the Landau cut, $\Gamma_k^{(2)}+\Gamma_k^{(4)}$, are similar. For the higher temperature, $T=150$~MeV, the contribution from the Landau cut already surpasses that of the unitary cut. This means that for temperatures close to $T_c$ there is a dominant contribution to the thermal width coming from the Landau cut, which would be overlooked if vacuum amplitudes were used. We have also checked that $\Gamma_k^{(3)}$ is negligible for all momenta and temperatures. 

%
%
%
\subsection{Thermal width from the scattering amplitude squared}
\label{subsec:transport-kinematic-method2}

The calculation of the thermal width due to thermal pions using Eqs.~(\ref{eq:transport-Gammak4terms}) to (\ref{eq:transport-Gammak4}) allowed us to distinguish the relative weight of the unitary and Landau contributions. However, the effect of the individual collision terms ($D\pi \rightarrow D\pi, D\pi \rightarrow D\eta, D\pi \rightarrow D_s \bar{K}$) cannot be disentangled. On the other hand, the computation of heavy-flavor transport coefficients is performed from the kinetic transport equation, where the collision rates, which are proportional to $|T_{D\pi}|^2$, are used. For these reasons, we derive an alternative expression for $\Gamma_k$, in terms of the scattering amplitude squared, which also serves to cross-check our previous determination of $\Gamma_k$.

We start from the same definition of the thermal width of Eq.~(\ref{eq:transport-Gammak})
and use the relation in Eq.~(\ref{eq:transport-polfuncs}),
\begin{equation} 
\textrm{Im } \Pi^{\textrm{R}} (E_k,\vec{k}\,) = \frac{1}{2 \ii} \left( \Pi^> (E_k,\vec{k}\,) - \Pi^<(E_k,\vec{k}\,)\right) \ . 
\end{equation}
In equilibrium, we can exploit the Kubo-Martin-Schwinger relation for the ``lesser'' and ``greater'' self-energies~\cite{kadanoff1962quantum,Blaizot:1999xk},
\begin{equation} 
\Pi^< (E_k,\vec{k}\,) = e^{-\beta E_k} \Pi^> (E_k,\vec{k}\,) \  , 
\end{equation}
obtaining,
\begin{equation}
  \Gamma_k =  \frac{\ii}{2 E_k } \left[\Pi^> (E_k,\vec{k}\,) - \Pi^< (E_k,\vec{k}\,) \right] =  \frac{\ii}{2   E_k} \frac{1}{\tilde{f}_k^{(0)}} \Pi^>(E_k,\vec{k}\,) \ ,
\end{equation}
where we have denoted $\tilde{f}^{(0)}_k \equiv \tilde{f}^{(0)} (E_k)$, and already simplified $z_k\simeq 1$.

We can now insert the expression in Eq.~(\ref{eq:transport-Pig}) for $\Pi^>(E_k,\vec{k}\,)$, which provides an interpretation of the thermal width in terms of particle collisions. As in the derivation of the off-shell transport theory, we replace the light-meson propagators with those for free particles, but keep the full spectral function of the internal $D$ meson. We can write the thermal width as
\begin{equation}
  \Gamma_k=\Gamma_k^{(\textrm{U})} + \Gamma_k^{(\textrm{L})} \ , \label{eq:transport-widthT2}
\end{equation}
with
\begin{align}
\Gamma_k^{(\textrm{U})} & = \frac{1}{2E_k} \frac{1}{\tilde{f}^{(0)}_k} \sum_{\lambda=\pm} \lambda \int dk_1^0 \int \prod_{i=1}^3 \frac{d^3 k_i }{(2\pi)^3} \frac{1}{2E_2} \frac{1}{2E_3} \left|T (E_k+E_3,\vec{k}+\vec{k}_3)\right|^2 \ S_D(k_1^0,\vec{k}_1) \nonumber \\
&\times (2\pi)^4 \delta^{(3)} ( \vec{k}+\vec{k}_3-\vec{k}_1-\vec{k}_2) \delta(E_k+E_3-k_1^0- \lambda E_2) \tilde{f}^{(0)} (k^0_1) f^{(0)} (E_3) \tilde{f}^{(0)} ( \lambda E_2)  \ ,   \label{eq:transport-widthT2U} 
\end{align}
and
\begin{align}
\Gamma_k^{(\textrm{L})} & = \frac{1}{2E_k} \frac{1}{\tilde{f}^{(0)}_k} \sum_{\lambda=\pm} \lambda \int dk_1^0 \int \prod_{i=1}^3 \frac{d^3 k_i }{(2\pi)^3} \frac{1}{2E_2} \frac{1}{2E_3} \left|T (E_k-E_3,\vec{k}+\vec{k}_3)\right|^2 \ S_D(k_1^0,\vec{k}_1) \nonumber \\
&\times (2\pi)^4 \delta^{(3)} ( \vec{k}+\vec{k}_3-\vec{k}_1-\vec{k}_2) \delta(E_k-E_3-k_1^0- \lambda E_2) \tilde{f}^{(0)} (k^0_1) \tilde{f}^{(0)} (E_3)  \tilde{f}^{(0)} (\lambda E_2)   \ , \label{eq:transport-widthT2L}
 \end{align}
where, like in the expression in Eq.~(\ref{eq:transport-Pig}), there is an implicit summation over particle species, restricted to the allowed scattering channels. In particular, if one focuses on the pion contribution to the $D$-meson thermal width, that is, particle $3$ being a $\pi$, then the remaining sum over species $1$ and $2$ contains three possibilities: $D\pi \rightarrow D\pi$, $D\pi \rightarrow D\eta$, and $D\pi \rightarrow D_s \bar{K}$ scatterings.

The separation made in Eq.~(\ref{eq:transport-widthT2}) makes it clear that $\Gamma_k^{(\textrm{U})}$ evaluates the scattering amplitude above the channel threshold, and it is related to the unitary cut of the scattering amplitude, while $\Gamma_k^{(\textrm{L})}$ evaluates it below the threshold and is therefore related to the Landau cut.

Before showing results, let us mention that, as in the on-shell reduction of the transport theory, if the internal $D$ meson is approximated by a narrow quasiparticle, only the positive branch of the spectral function $S_D(k_1^0,\vec{k}_1)$ and $\lambda=+1$ can hold the energy conservation in Eq.~(\ref{eq:transport-widthT2U}). The same is true for Eq.~(\ref{eq:transport-widthT2L}) but with $\lambda=-1$. Therefore, in the on-shell case (o.s.) the expressions reduce to
\begin{align}
\left. \Gamma_k^{(\textrm{U})} \right|_{\textrm{o.s.}} & = \frac{1}{2E_k} \frac{1}{\tilde{f}^{(0)}_k} \int \prod_{i=1}^3 \frac{d^3 k_i }{(2\pi)^3 2E_i}  \left|T (E_k+E_3,\vec{k}+\vec{k}_3)\right|^2 \nonumber \\
&\times (2\pi)^4 \delta^{(3)} ( \vec{k}+\vec{k}_3-\vec{k}_1-\vec{k}_2) \delta(E_k+E_3-E_1-E_2) \tilde{f}^{(0)} (E_1) f^{(0)} (E_3)  \tilde{f}^{(0)} (E_2)   \ ,   \label{eq:transport-widthT2Uos}  
\end{align}
and
\begin{align}
\left. \Gamma_k^{(\textrm{L})} \right|_{\textrm{o.s.}} & = \frac{1}{2E_k} \frac{1}{\tilde{f}^{(0)}_k}  \int \prod_{i=1}^3 \frac{d^3 k_i }{(2\pi)^3 2E_i} \left|T (E_k-E_3,\vec{k}+\vec{k}_3)\right|^2 \nonumber \\
&\times (2\pi)^4 \delta^{(3)} ( \vec{k}+\vec{k}_3-\vec{k}_1-\vec{k}_2) \delta(E_k-E_3-E_1+E_2) \tilde{f}^{(0)} (E_1) \tilde{f}^{(0)} (E_3)  {f}^{(0)} (E_2)   \ . \label{eq:transport-widthT2Los}
 \end{align}

Given the special kinematics of $\Gamma_k^{(\textrm{L})}$, one cannot express the energy-momentum conservation in terms of a single $\delta^{(4)}$ function. Only in the particular case of elastic (el) scattering, one can make a change of variables $\vec{k}_2 \leftrightarrow -\vec{k}_3$ in $\Gamma_k^{(L)}$ to arrive at
\begin{align}
\left. \Gamma_k^{(\textrm{U})} \right|^{\textrm{el}}_{\textrm{o.s.}}  & = \frac{1}{2E_k} \frac{1}{\tilde{f}^{(0)}_k}  \int \prod_{i=1}^3 \frac{d^3 k_i }{(2\pi)^3 2E_i}
(2\pi)^4 \delta^{(4)} ( k+k_3-k_1-k_2) \left|T (E_k+E_3,\vec{k}+\vec{k}_3)\right|^2 \nonumber \\
& \times \tilde{f}^{(0)} (E_1) \tilde{f}^{(0)} (E_2) f^{(0)} (E_3)  \label{eq:transport-widthT2Uosalt} \ ,
\end{align}
and
\begin{align}
\left. \Gamma_k^{(\textrm{L})} \right|^{\textrm{el}}_{\textrm{o.s.}}  & = \frac{1}{2E_k} \frac{1}{\tilde{f}^{(0)}_k}  \int \prod_{i=1}^3 \frac{d^3 k_i }{(2\pi)^3 2E_i}
(2\pi)^4 \delta^{(4)} ( k+k_3-k_1-k_2) \left|T (E_k-E_2,\vec{k}-\vec{k}_2)\right|^2 \nonumber \\ & \times \tilde{f}^{(0)} (E_1) \tilde{f}^{(0)} (E_2) f^{(0)} (E_3)  \ . \label{eq:transport-widthT2Losalt}
 \end{align}
 
Equations (\ref{eq:transport-widthT2Uosalt}) and (\ref{eq:transport-widthT2Losalt}) can be potentially useful when the $D$ meson is treated as a narrow quasiparticle and inelastic collisions are neglected. Unless otherwise stated, we do not assume this.

Coming back to the general result of Eqs.~(\ref{eq:transport-widthT2U}) and (\ref{eq:transport-widthT2L}), where the full spectral function of the internal $D$ meson is kept, it is possible to analytically check that $\Gamma_k^{(U)}$ in Eq.~(\ref{eq:transport-widthT2U}) is equal to the combination $\Gamma_k^{(1)}+\Gamma_k^{(3)}$ in Eqs.~(\ref{eq:transport-Gammak1}) and (\ref{eq:transport-Gammak3}), while $\Gamma_k^{(L)}$ in Eq.~(\ref{eq:transport-widthT2L}) exactly coincides with $\Gamma_k^{(2)}+\Gamma_k^{(4)}$ in Eqs.~(\ref{eq:transport-Gammak2}) and (\ref{eq:transport-Gammak4}). To do that, one needs to apply the unitarity condition, or optical theorem in the coupled-channel case,
\begin{equation} 
\textrm{Im } T_{D\pi \rightarrow D\pi}(E,\vec{p}\,) = \sum_a T^*_{D\pi \rightarrow a}(E,\vec{p}\,) \, \textrm{Im } G_a^{\textrm{R}} (E,\vec{p}\,)\, T_{a \rightarrow D\pi}(E,\vec{p}\,) \ , \label{eq:transport-optitheo} 
\end{equation}
which follows from the $T$-matrix equation at finite temperature, together with the expression for the retarded two-meson propagator given in Eq.~(\ref{eq:hot-loop-compact}),
where the spectral function of the light meson, $S_\Phi(\omega',\vec{p}-\vec{q}\,)$, is taken in the narrow limit.

Notice that the sum over intermediate states $a$ in the optical theorem in Eq.~(\ref{eq:transport-optitheo}) is to be taken as a sum over species $D_i$ and $\Phi_j$, restricted to the physical states that couple to $D\pi$. If only elastic collisions $D\pi \rightarrow D\pi$ were used, then the optical theorem is necessarily violated. The effect of inelastic processes has been normally ignored in the literature.

\begin{figure}[b!]
  \centering
  \includegraphics[width=0.333\textwidth]{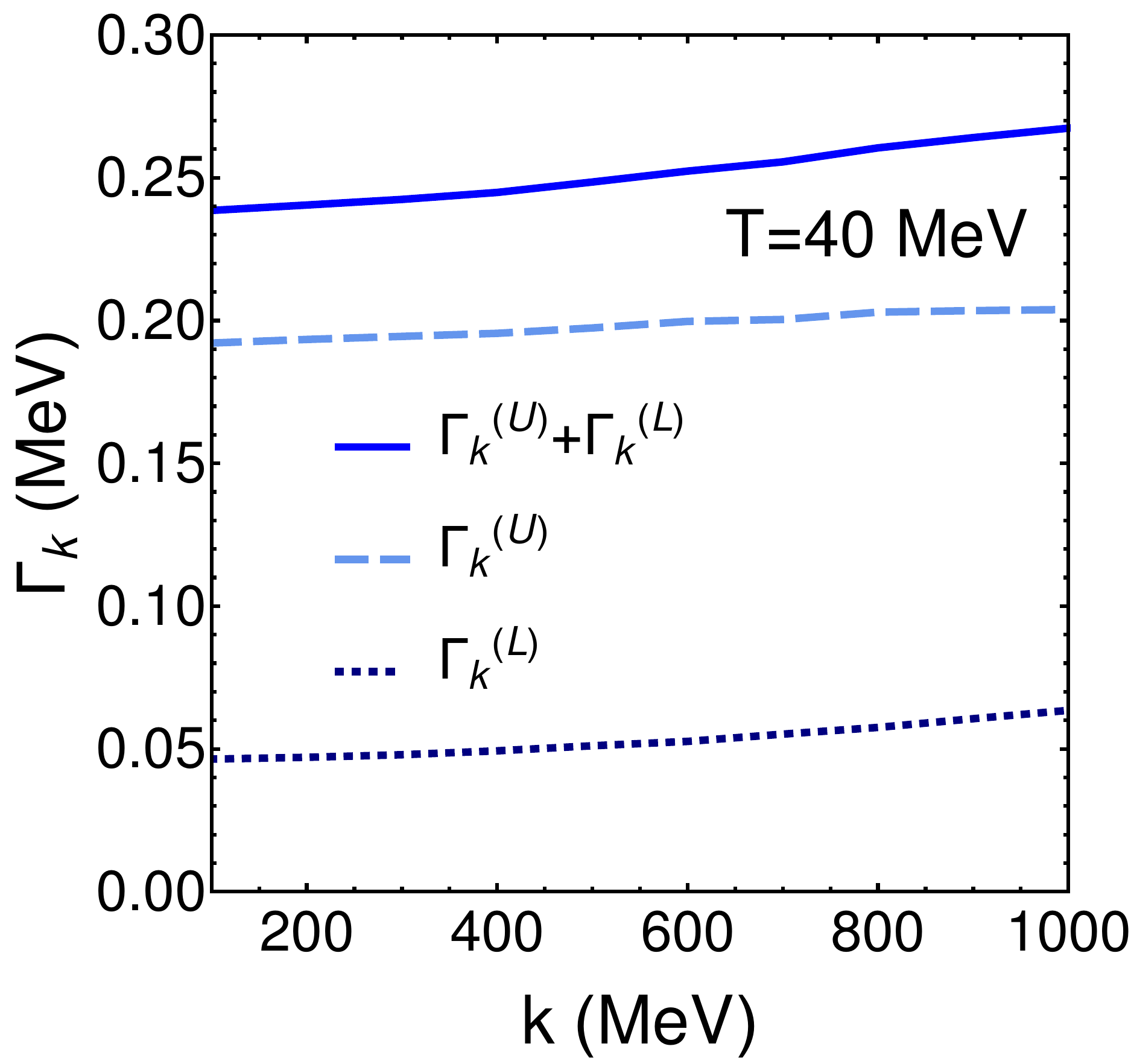}
   \includegraphics[width=0.32\textwidth]{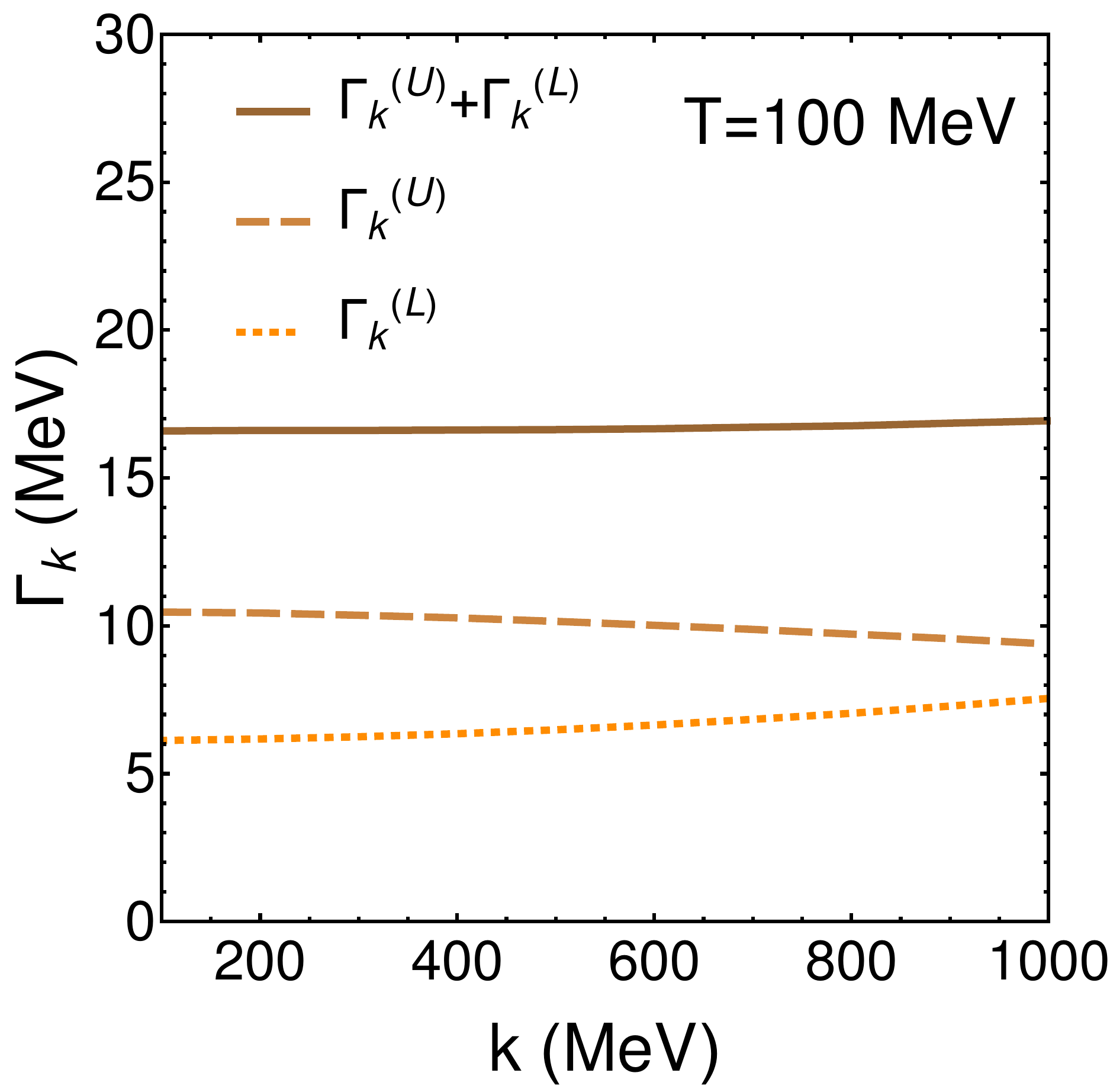}
  \includegraphics[width=0.327\textwidth]{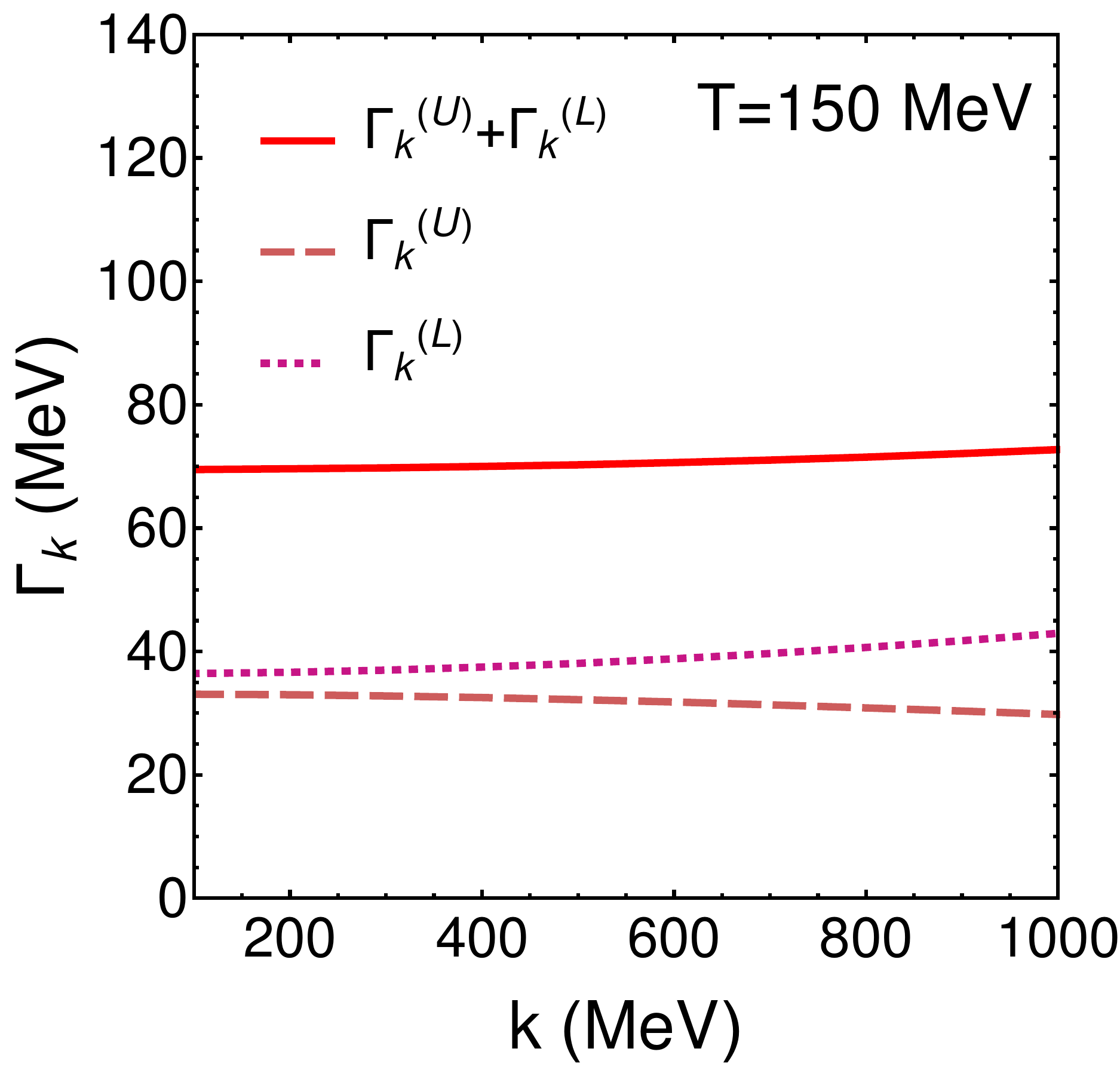}
  \caption{Thermal width of $D$ mesons generated from their interaction with pions as computed  from $\Gamma_k^{(\textrm{U})}$ (Eq.~(\ref{eq:transport-widthT2U})) using dashed lines, and $\Gamma_k^{(\textrm{L})}$ (Eq.~(\ref{eq:transport-widthT2L})) using dotted lines. The full calculation that includes both $\Gamma_k^{(\textrm{U})}$ and $\Gamma_k^{(\textrm{L})}$ is shown with solid lines. Inelastic channels ($D\pi \rightarrow D\eta$ and $D\pi \rightarrow D_s \bar{K}$) are also included.}
    \label{fig:transport-GammakKinetic}
\end{figure}

We present the results of Eqs.~(\ref{eq:transport-widthT2U}) and (\ref{eq:transport-widthT2L}) for temperatures $T=40,100,150$~MeV in Fig.~\ref{fig:transport-GammakKinetic}. We separate the contributions of the unitary and Landau cuts for each temperature, and obtain a similar result to that in Fig.~\ref{fig:transport-Gammak}, where the integration over $\textrm{Im } T_{D\pi \rightarrow D\pi}$ was employed. For consistency with the coupled-channel optical theorem, we have included the three channels $D\pi \rightarrow D\pi, D\pi \rightarrow D\eta$ and $D\pi \rightarrow D_s \bar{K}$.

\subsection{Quantification of different effects}
\label{subsec:transport-kinematic-effects}
The differences between the two approaches, as well as the analysis of several other effects, are summarized in the following.

\subsubsection{Effect of truncation}

As stated, one can analytically prove that the two alternative methods to extract $\Gamma_k$, first via Eq.~(\ref{eq:transport-Gammak4terms}) and second through Eq.~(\ref{eq:transport-widthT2}), are equivalent. In addition, we have stated that this equivalence can also be checked through the direct application of the optical theorem in Eq.~(\ref{eq:transport-optitheo}).

However, the numerical implementation can introduce small differences when a numerical integration cut-off is employed. We name this the effect of truncation. This can be easily understood by looking at Eq.~(\ref{eq:transport-optitheo}). The first method to compute $\Gamma_k$ uses $\textrm{Im } T_{D \pi \rightarrow D\pi}$, and a UV cut-off in $|\vec{p}\,|$ simply truncates the \gls{lhs} of Eq.~(\ref{eq:transport-optitheo}) at that momentum. 
On the other hand, the second method employs the \gls{rhs} of the same equation, where the same cut-off is imposed on $|T_{D\pi \rightarrow a}|^2$, but the term $\textrm{Im } G_a^{\textrm{R}}$ is calculated analytically to perform the integrations. Therefore, the way in which a UV cut-off is imposed in the numerical calculations does not ensure that the truncation effect is the same for the two methods.

\begin{figure}[b!]
  \centering
  \includegraphics[width=0.333\textwidth]{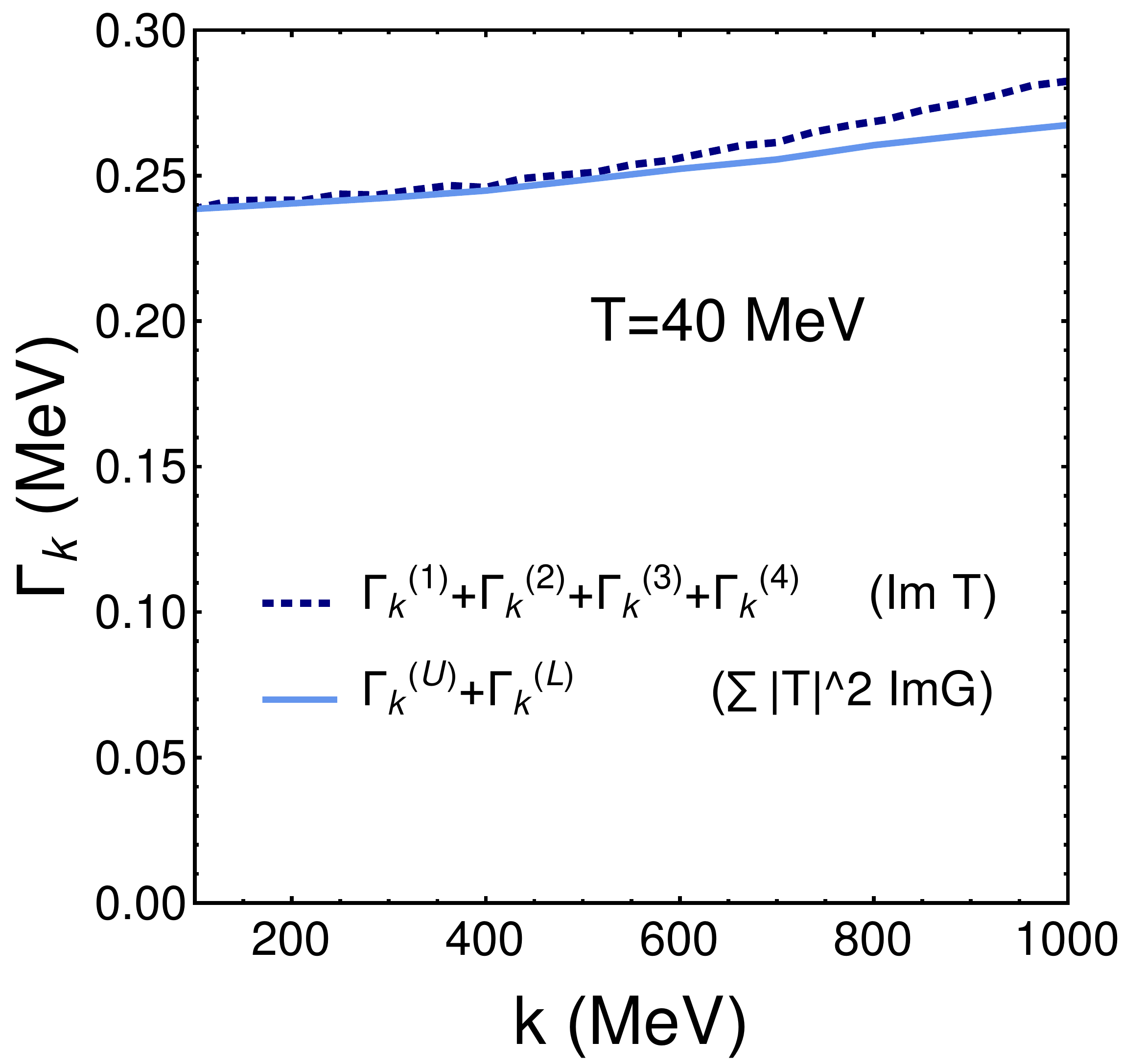}
   \includegraphics[width=0.32\textwidth]{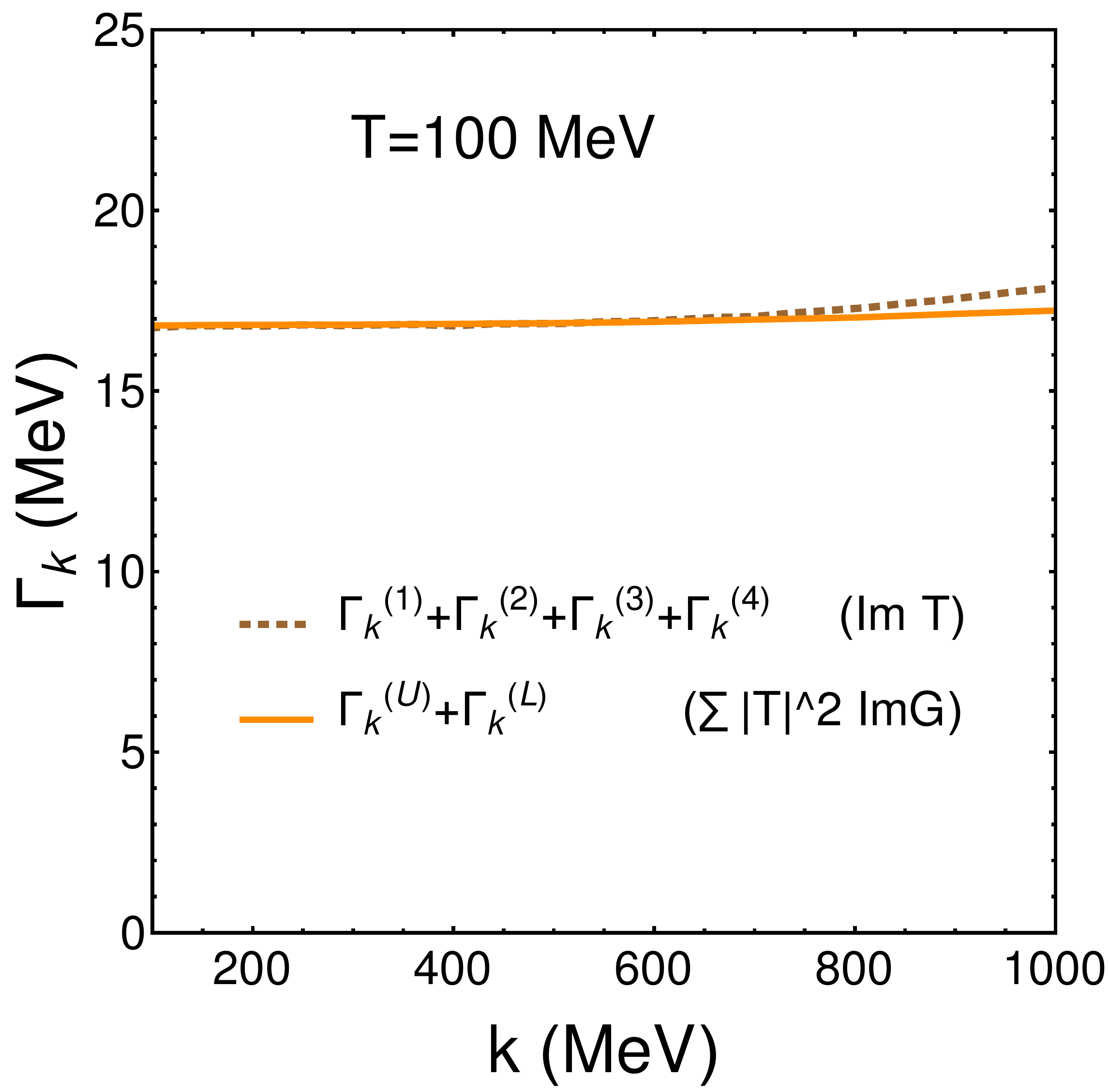}
  \includegraphics[width=0.327\textwidth]{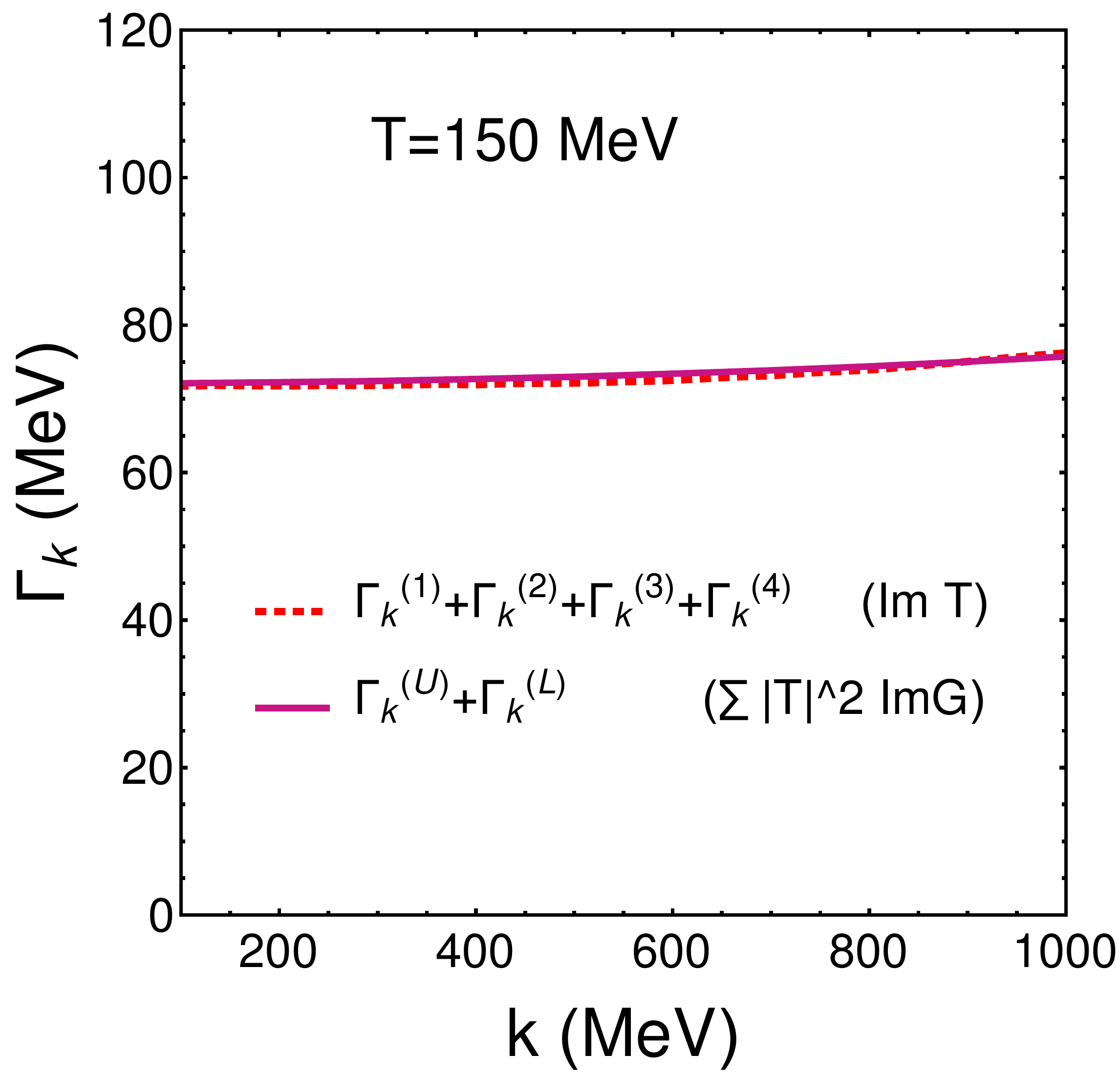}
  \caption{Thermal width of $D$ mesons in a thermal pion gas. Comparison between the two methods described in the text.}
  \label{fig:transport-GammaTall}
\end{figure}

The comparison between Figs.~\ref{fig:transport-Gammak} and~\ref{fig:transport-GammakKinetic} does not show appreciable differences. In Fig.~\ref{fig:transport-GammaTall} we show the total $\Gamma_k$ from the two methods in a single plot for better comparison. Note that the second method includes the three channels (elastic and inelastic) involving pions. Both methods compare very well for low momentum at all temperatures. We have checked that the good comparison persists between unitary and Landau cuts separately. We only obtain deviations at high momentum; hence, cut-off effects. Nevertheless, the differences in $\Gamma_k$ are at most $5\%$, and only for high momenta (which are in any case suppressed when folded with the \gls{be} distribution function).

\subsubsection{Off-shell effects}

We now describe the differences between the use of the on-shell and off-shell approaches. We have extensively described how to implement off-shell effects by keeping the full spectral function of the internal $D$-meson propagator, as opposed to using the narrow limit. To determine the differences, we use the second method to compute $\Gamma_k$, including the three channels involving pions.

\begin{figure}[ht!]
  \centering
  \includegraphics[width=0.333\textwidth]{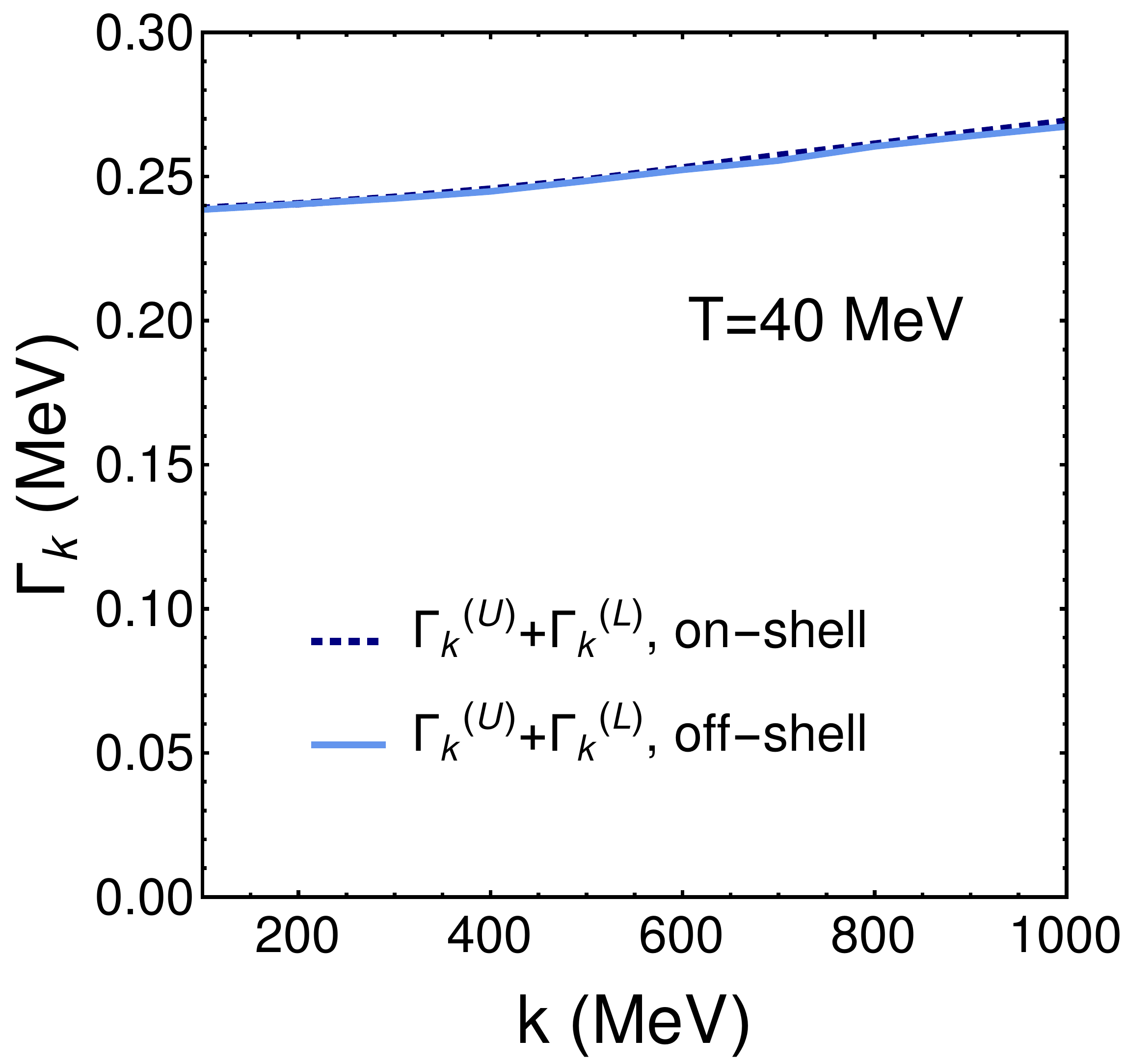}
   \includegraphics[width=0.32\textwidth]{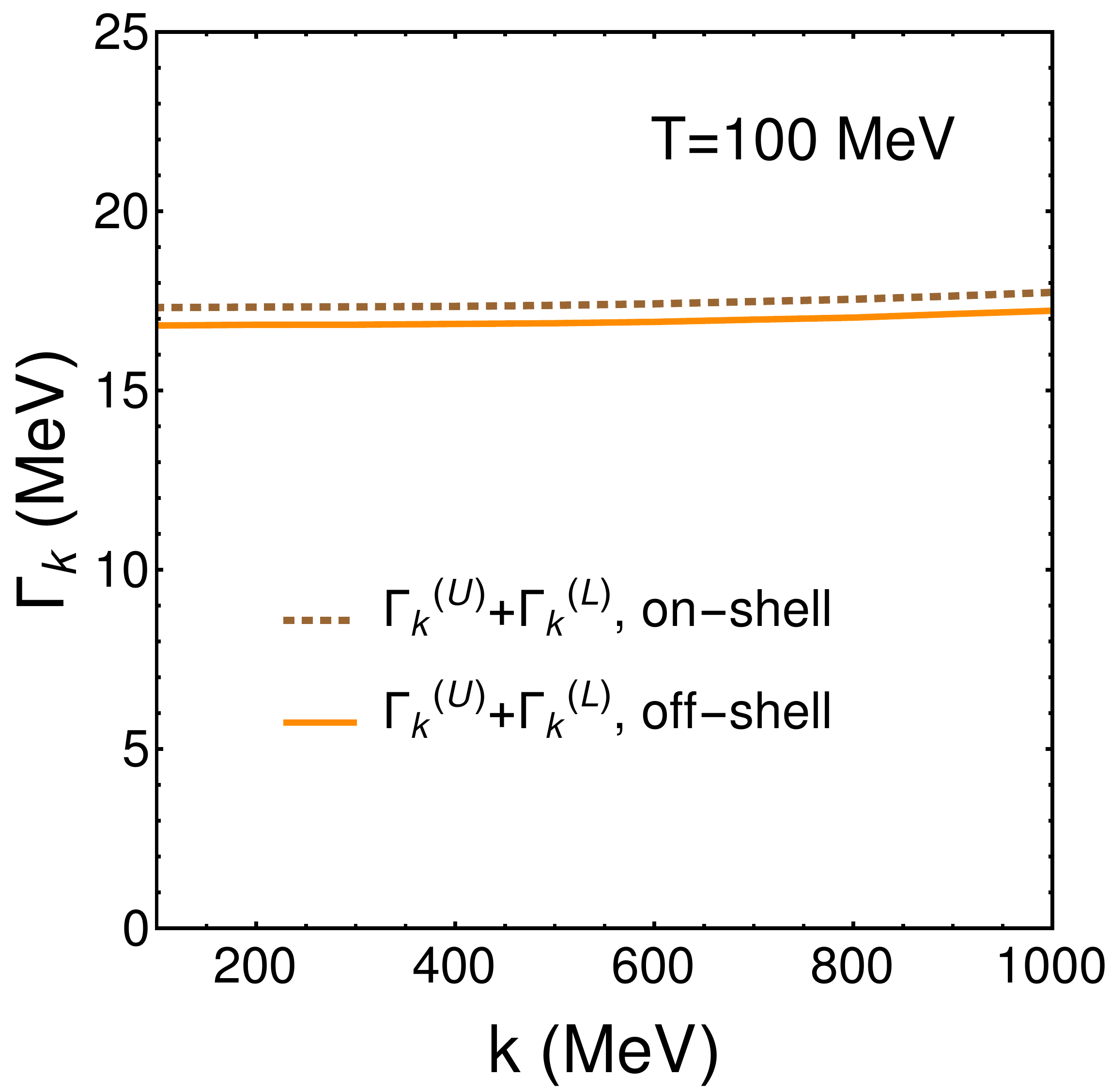}
  \includegraphics[width=0.327\textwidth]{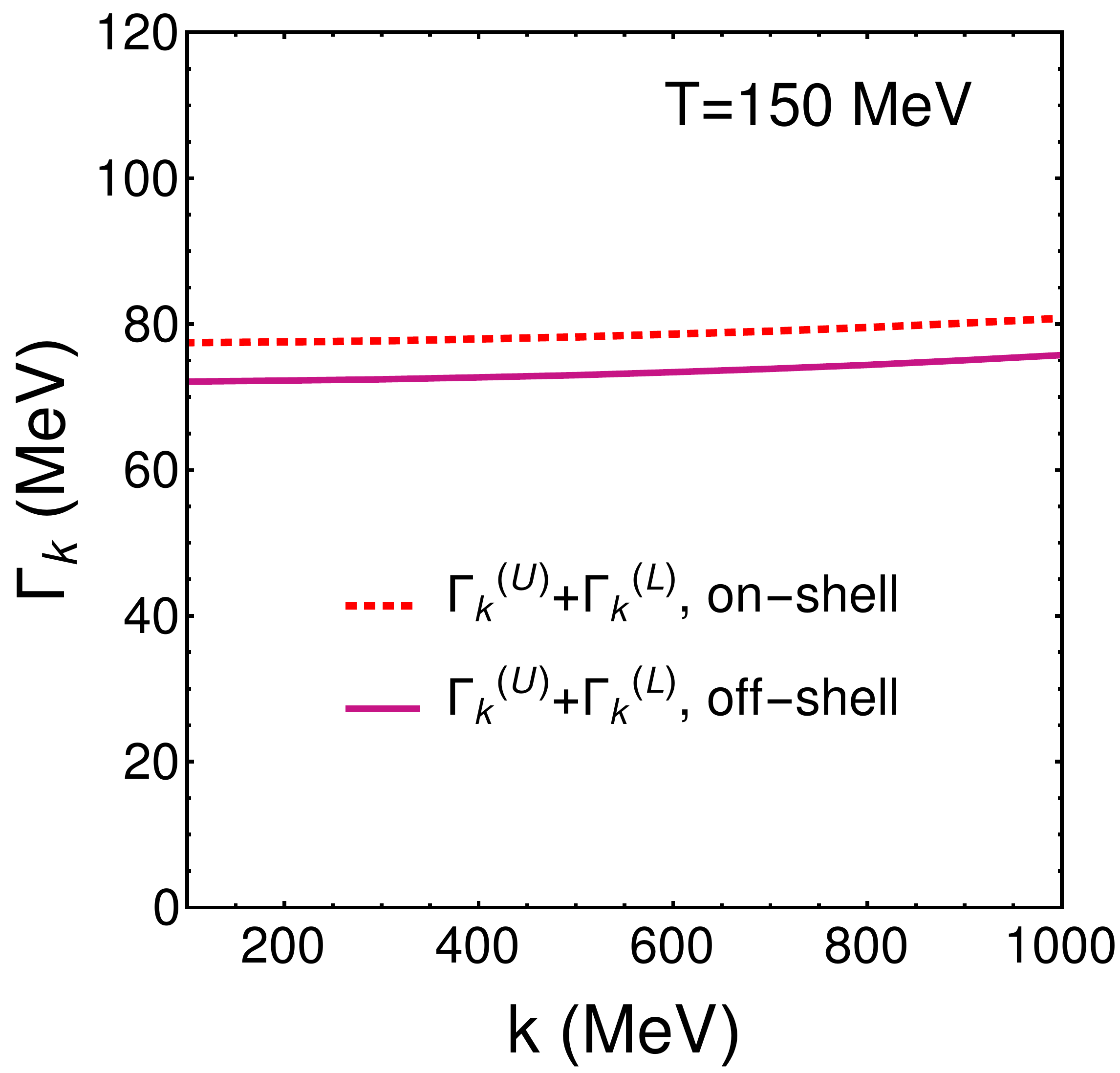}
  \caption{Thermal width of $D$ mesons in a thermal pion gas. Comparison between the off-shell calculation (solid lines, Eqs.~(\ref{eq:transport-widthT2U})+(\ref{eq:transport-widthT2L})) and the on-shell one (dotted lines, Eqs.~(\ref{eq:transport-widthT2Uos})+(\ref{eq:transport-widthT2Los})).}
  \label{fig:transport-Gammaonoff}
\end{figure}

The results are presented in Fig.~\ref{fig:transport-Gammaonoff}. The off-shell calculation is performed through the sum of Eqs.~(\ref{eq:transport-widthT2U}) and (\ref{eq:transport-widthT2L}), and the spectral function $S_D$ is taken from the results in Chapter~\ref{ch:hot-medium}. For the on-shell calculation, we employ the sum of Eqs.~(\ref{eq:transport-widthT2Uos}) and (\ref{eq:transport-widthT2Los}), where the intermediate $D$ meson is taken on shell (narrow limit). In both cases, we use the same temperature-dependent scattering amplitudes. 

As expected, the effects of the spectral function width become more apparent at higher temperatures. When $T \rightarrow 0$ the thermal width goes to zero, and the narrow quasiparticle approximation becomes exact in this limit. In any case, the quasiparticle peak is rather narrow at all temperatures considered, as reported in Chapter~\ref{ch:hot-medium}, and the off-shell effects are generally small. These effects are of the order of $10\%$ for $\Gamma_k$ at the highest temperature, $T=150$~MeV, and rather independent of the external momentum $k$.

\subsubsection{Effect of inelastic channels}

\begin{figure}[t!]
  \centering
  \includegraphics[width=0.333\textwidth]{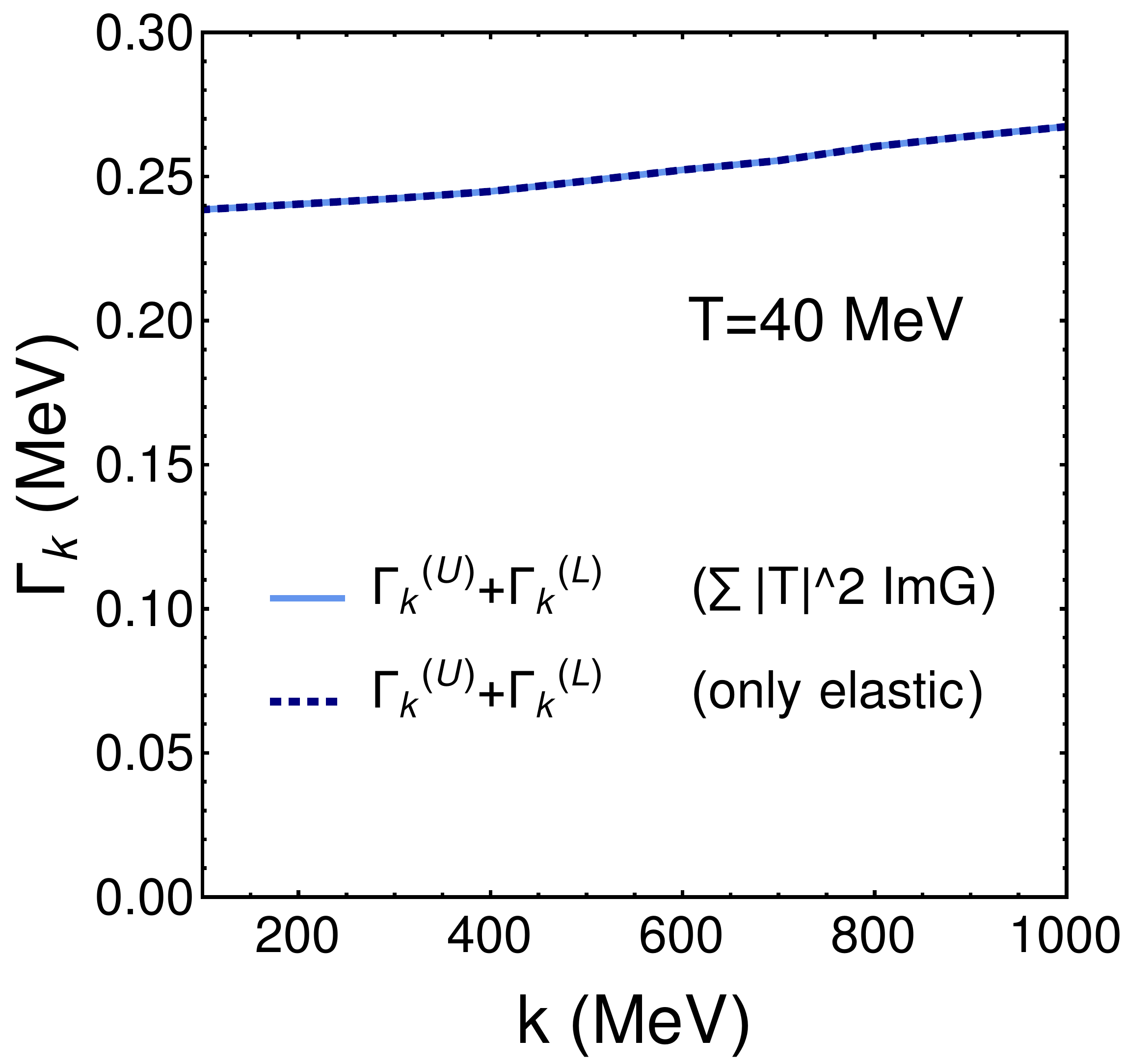}
   \includegraphics[width=0.32\textwidth]{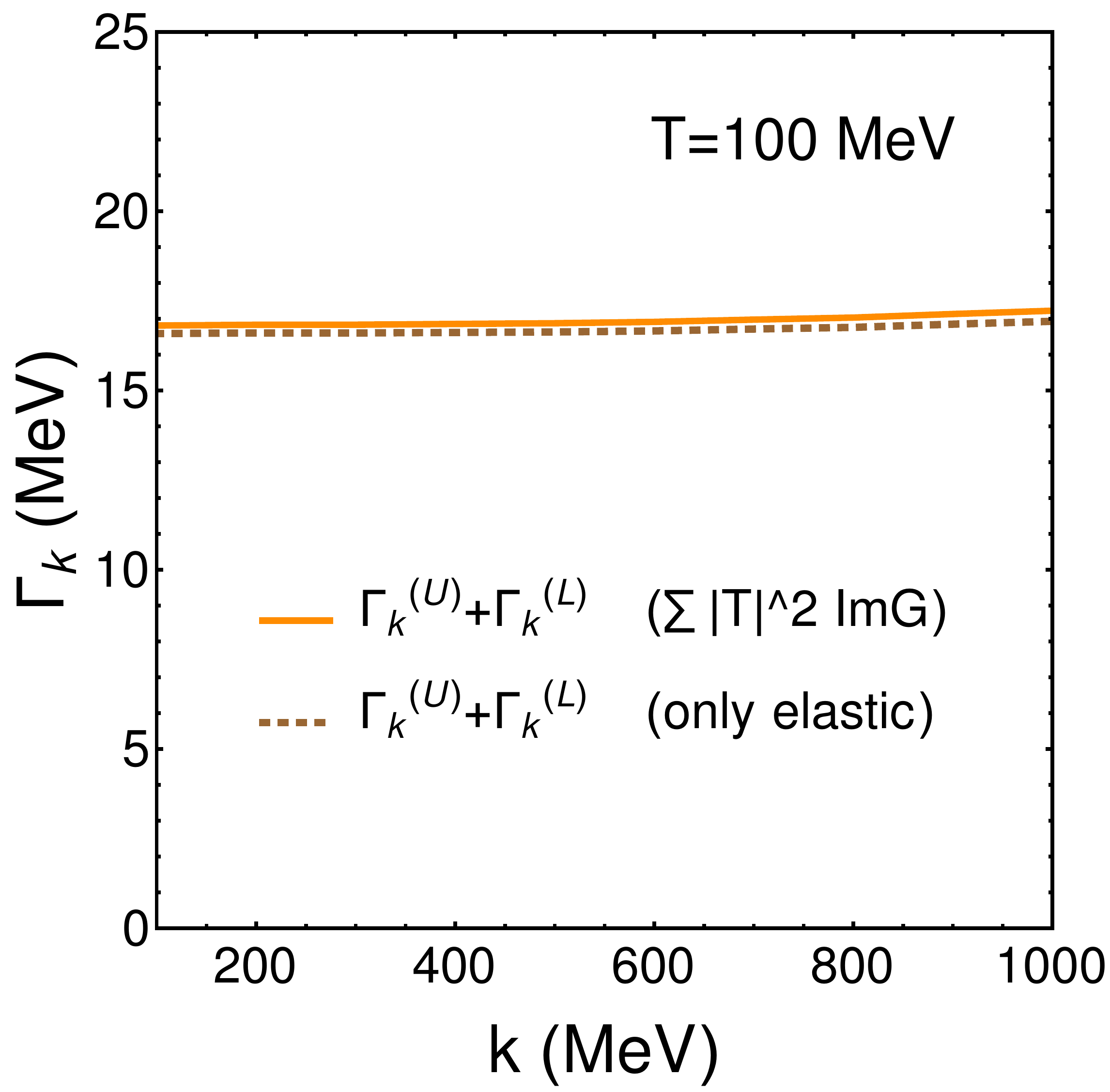}
  \includegraphics[width=0.327\textwidth]{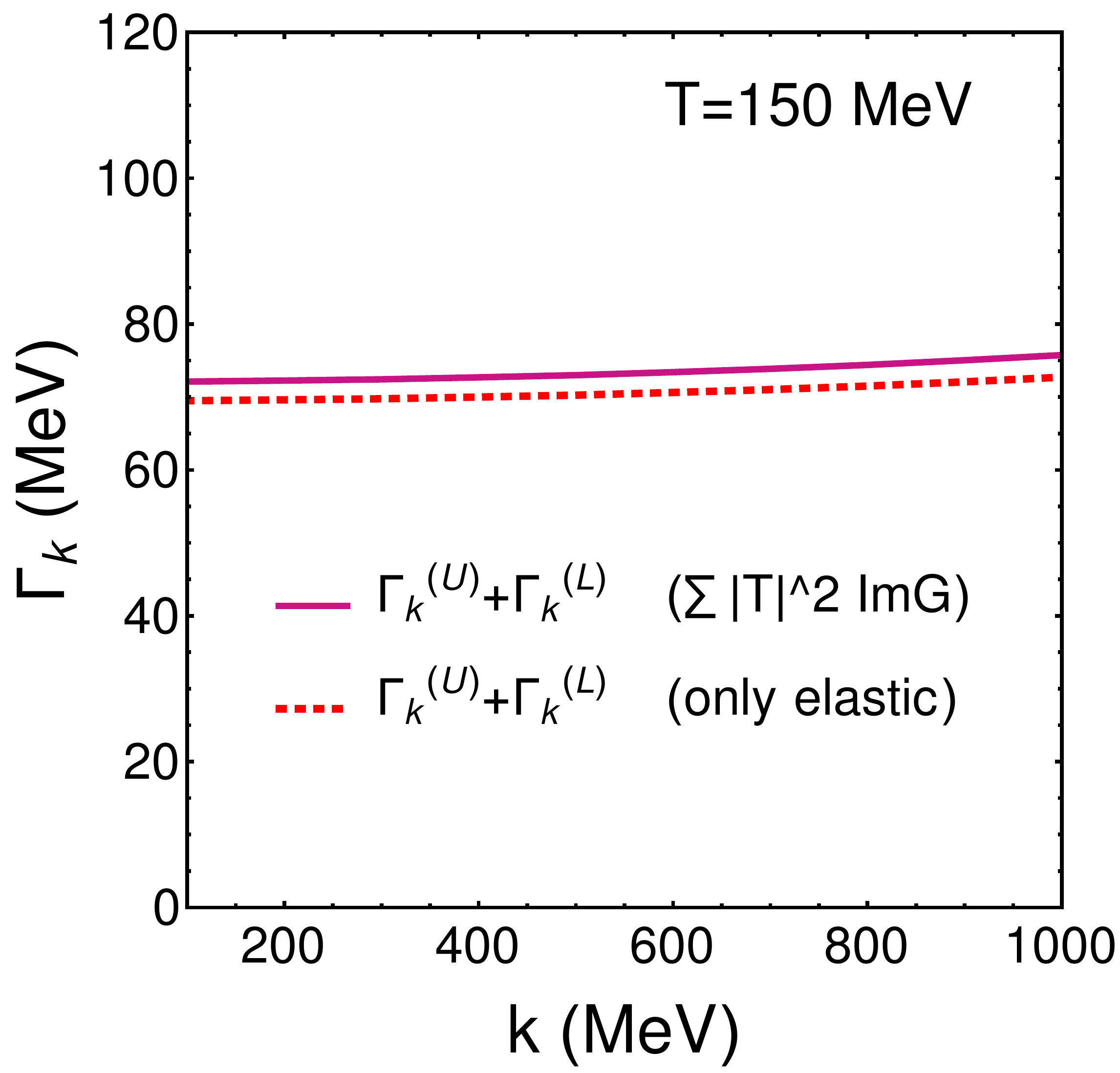}
  \caption{Thermal width of $D$ mesons in a thermal pion gas computed via the off-shell kinetic expressions (Eqs.~(\ref{eq:transport-widthT2U})+(\ref{eq:transport-widthT2L})). The calculations are done using elastic scattering only (dotted lines) and adding also the inelastic channels (solid lines), following the optical theorem in coupled channels.}
  \label{fig:transport-GammaTIne}
\end{figure}

We now discuss the effect of inelastic channels in the second method to compute $\Gamma_k$, given by Eqs.~(\ref{eq:transport-widthT2}), (\ref{eq:transport-widthT2U}), and (\ref{eq:transport-widthT2L}). While their inclusion is strictly required to account for the coupled-channel optical theorem, in the practice, their effect is small. This can be seen in Fig.~\ref{fig:transport-GammaTIne} for the case of the $D$-meson thermal width, only due to the pions of the medium. In this figure, we show in solid lines the complete result with the three inelastic channels and in dashed lines the result with only the elastic channel, $D\pi \rightarrow D\pi$. The effect is rather small for all $T$ and $k$, and for the highest temperature $T=150$~MeV and momentum they are at most $5\%$. Therefore, we will neglect the effect of inelastic channels in the  calculations of transport coefficients.

\subsubsection{Effects of the light mesons in the bath}

While one can safely neglect the inelastic channels, one should not forget that there are four different elastic channels for the interactions of $D$ mesons with light pseudoscalars ($D\pi, DK, D\bar{K}, D\eta$). While the contribution of the most massive mesons is also suppressed, they become increasingly important as the temperature increases. 

To study the effect of the contribution of the different species, we define an averaged thermal width that is only function of temperature as
\begin{equation}
  \Gamma(T)=\frac{1}{n_D} \int d^3k \ f^{(0)} (E_k) \Gamma_k \ , \label{eq:transport-GammaT}
\end{equation}
where $f^{(0)} (E_k)$ is the equilibrium \gls{be} distribution function and $n_D$ is the $D$-meson particle density. 

\begin{figure}[b!]
\centering
\includegraphics[width=0.45\textwidth]{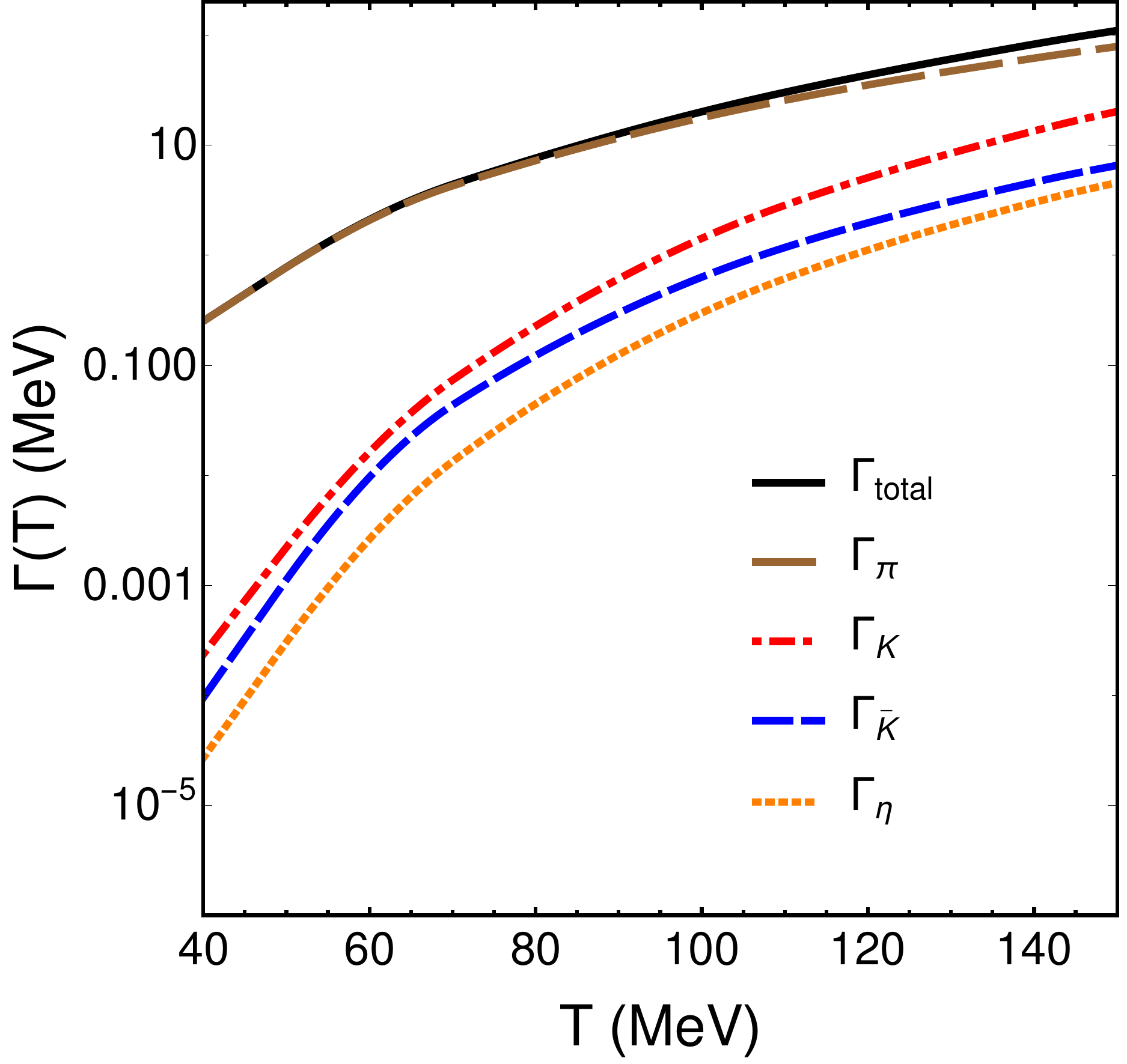}\hspace{0.5cm}
\includegraphics[width=0.42\textwidth]{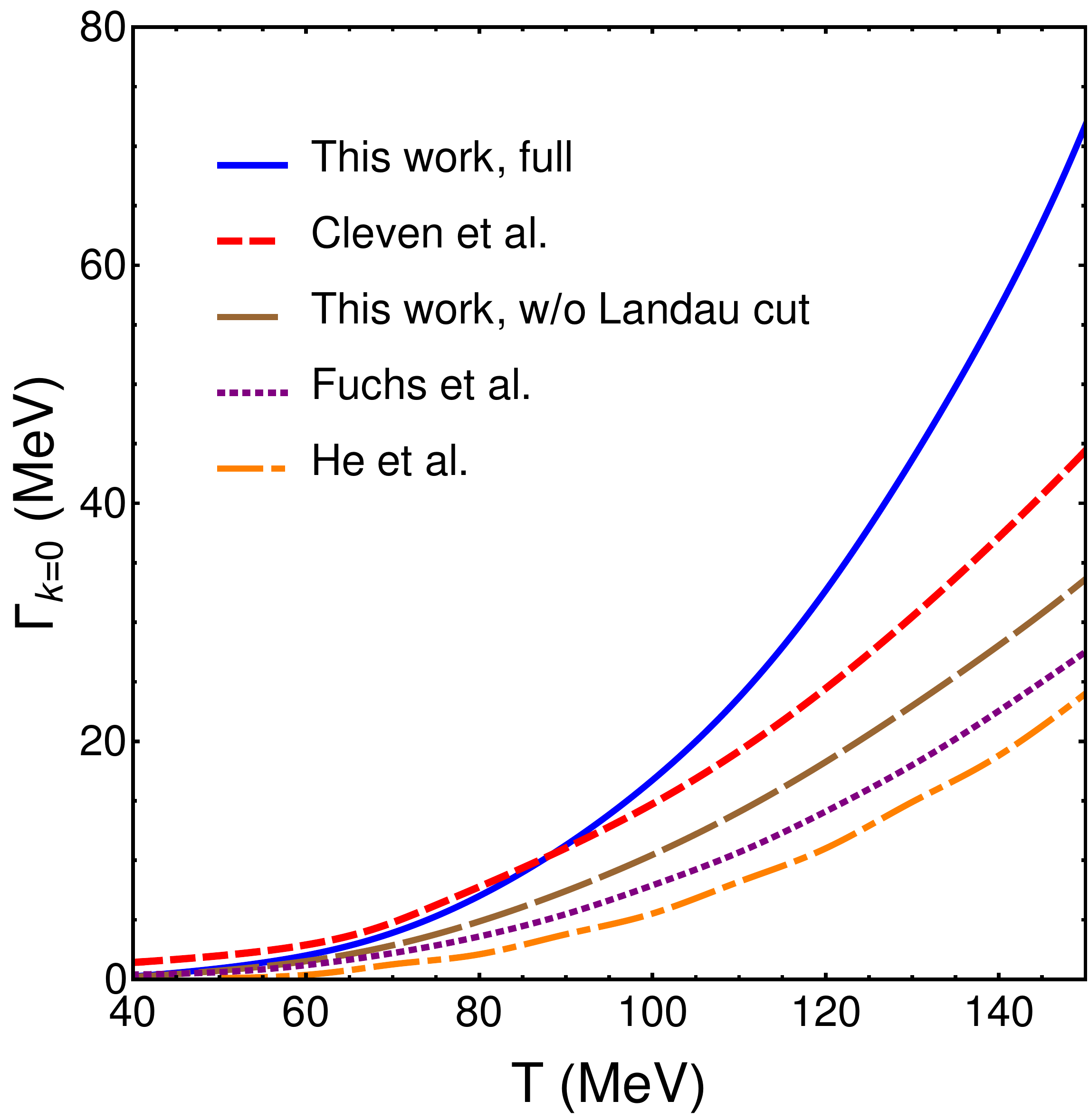}
\caption{Left panel: Contribution to the $D$-meson averaged thermal width from a thermal bath of pions, kaons, antikaons, and $\eta$ mesons. Right panel: Comparison of the $D$-meson thermal width  at $k=0$ in a pion thermal bath for different calculations. See main text for the different sources.
\label{fig:transport-widths}}
\end{figure}

In the left panel of Fig.~\ref{fig:transport-widths} (notice the logarithmic scale), we show the contributions to the $D$-meson width coming from the different meson baths ($\pi,K,\bar{K},\eta$). As expected, the contribution of the more massive mesons is negligible at low temperatures due to the thermal suppression factor, and only the pion term is relevant. At $T=150$~MeV the more massive mesons already contribute with several MeV to the $D$-meson decay width but are still subdominant with respect to the pion one. 

\subsubsection{Comparison with previous approaches}

Finally, we compare our results (labeled as ``full'') \cite{Torres-Rincon:2021yga}, including the unitary and Landau contributions, together with thermal amplitudes, inelastic channels, and off-shell effects, with the calculations of Refs.~\cite{Fuchs:2004fh, He:2011yi,Cleven:2017fun}. These are shown in the right panel of Fig.~\ref{fig:transport-widths}. We focus on the thermal width of $D$ mesons coming from the interaction with a thermal bath of only pions and fix the $D$-meson momentum to $k=0$. Fuchs \textit{et al.} \cite{Fuchs:2004fh} use an effective interaction at \gls{lo} between $D$ mesons and pions and extract the width from the self-energy correction due to pions. He \textit{et al.} \cite{He:2011yi} use a similar interaction based on Ref.~\cite{Fuchs:2004fh} but compute the thermal width using a formula similar to Eq.~(\ref{eq:transport-widthT2Uosalt}). The two calculations provide similar results and they are fairly consistent with our results using the unitary cut alone (label ``w/o Landau cut''). Notice that, apart from the different interaction, we also include inelastic channels while He \textit{et al.} in Ref.~\cite{He:2011yi} do not. This partially
explains why our curve is slightly larger than the other two. Then, Cleven \textit{et al.} \cite{Cleven:2017fun} perform a similar calculation to ours, but the effective approach is based on $\textrm{SU}(4)$ chiral symmetry. Medium effects are also incorporated, including the Landau cut contributions, resulting in a $D$-meson thermal width almost twice larger than the previously discussed two approaches, but still smaller than the present results for temperatures higher than $100$~MeV, the difference reaching around $30\%$ at $T=150$~MeV. This is probably due to the fact that the small mass shift of the $D$-meson, which in our model turns out to be attractive, is ignored in the results of Ref.~\cite{Cleven:2017fun}, thereby making them less affected by the contributions of the sub-threshold Landau cut.

To summarize, in this section we have analyzed several contributions to the $D$-meson width and found that the effect of off-shell dynamics, inelastic channels, and truncation errors are relatively small for the calculation of the $D$-meson thermal width. However the contribution of the Landau cut is essential to describe this coefficient at finite temperatures. We have shown that, thanks to exact unitarity considerations, this contribution appears not only in the imaginary part of the retarded self-energy, but also in the collision term of the kinetic equation. Guided by the results in $\Gamma_k$, we expect that this contribution will be important for the calculation of the $D$-meson transport coefficients as well.

\subsection{Thermal width of bottomed mesons}
\label{subsec:transport-kinematic-B}

In the previous sections we have quantified, in the case of the charmed mesons, the contribution of the unitary and Landau cuts of the unitarized scattering amplitudes to the thermal width at different temperatures of the light-meson bath. In particular, we have seen that the relative importance of the Landau contribution increases with temperature, and it even becomes the dominant one at large temperatures. We have obtained similar results employing either the method based on the imaginary part of the $T$-matrix elements or the one using the scattering amplitudes squared, the small differences at high momentum being related to the effect of truncation.

The off-shell kinetic theory derived in Section~\ref{sec:transport-nonequilibrium} is valid for any heavy species that can be treated as Brownian particles propagating in a mesonic medium. Therefore, one can apply the kinetic equation in Eq.~(\ref{eq:transport-offshellkinetic}) to describe, for instance, the propagation of $\bar{B}$ mesons, by simply replacing $D\rightarrow\bar{B}$, and the approximations that we have made for the charmed mesons are also valid for the bottomed mesons. The narrow limit, for example, is even a better approximation for the $\bar{B}$ mesons, since we have seen in Chapter~\ref{ch:hot-medium} that their thermal width is of the same order as that of the $D$ mesons, but their mass is considerably larger.


\begin{figure}[b!]
  \centering
  \includegraphics[width=0.333\textwidth]{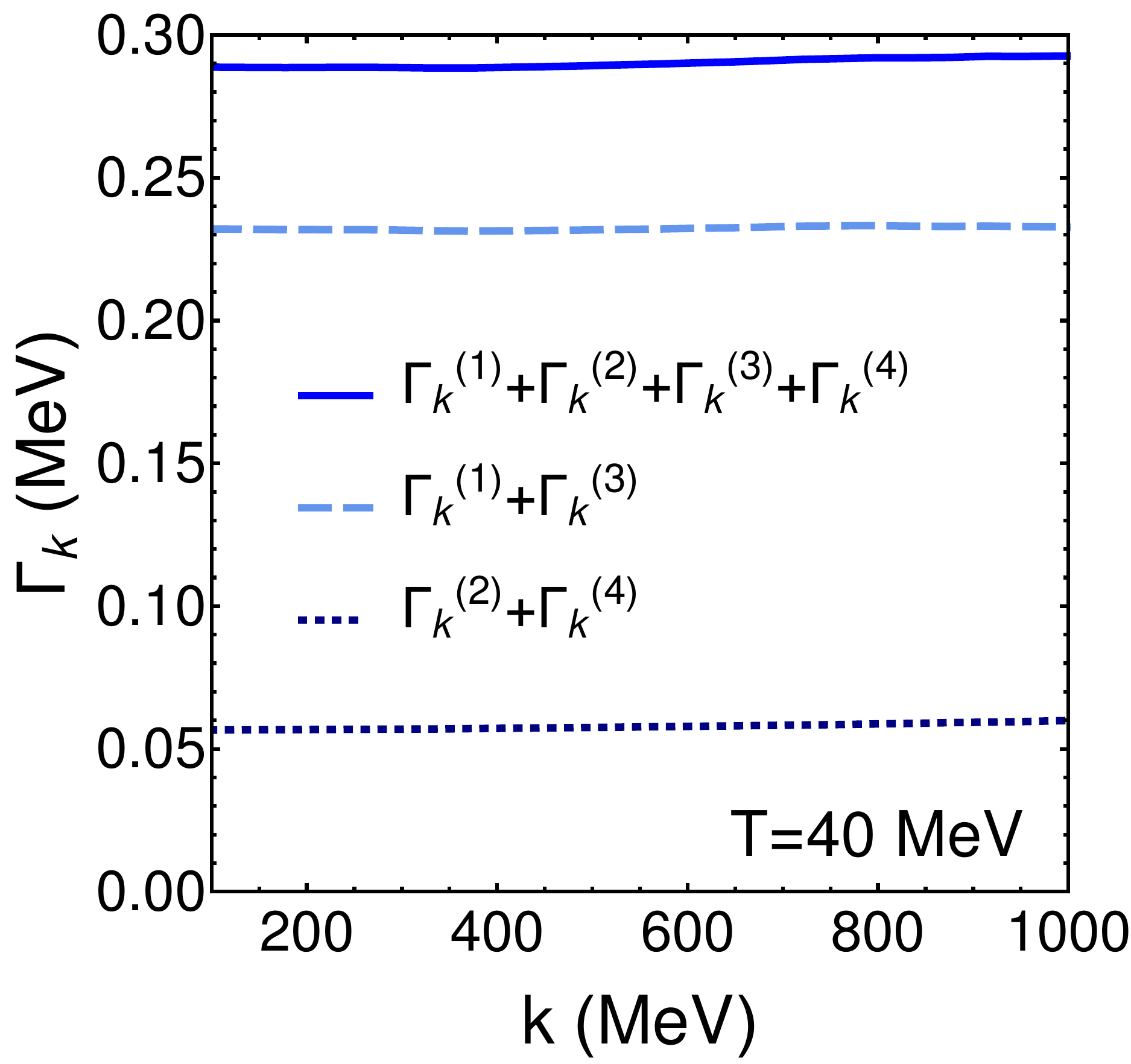}
   \includegraphics[width=0.32\textwidth]{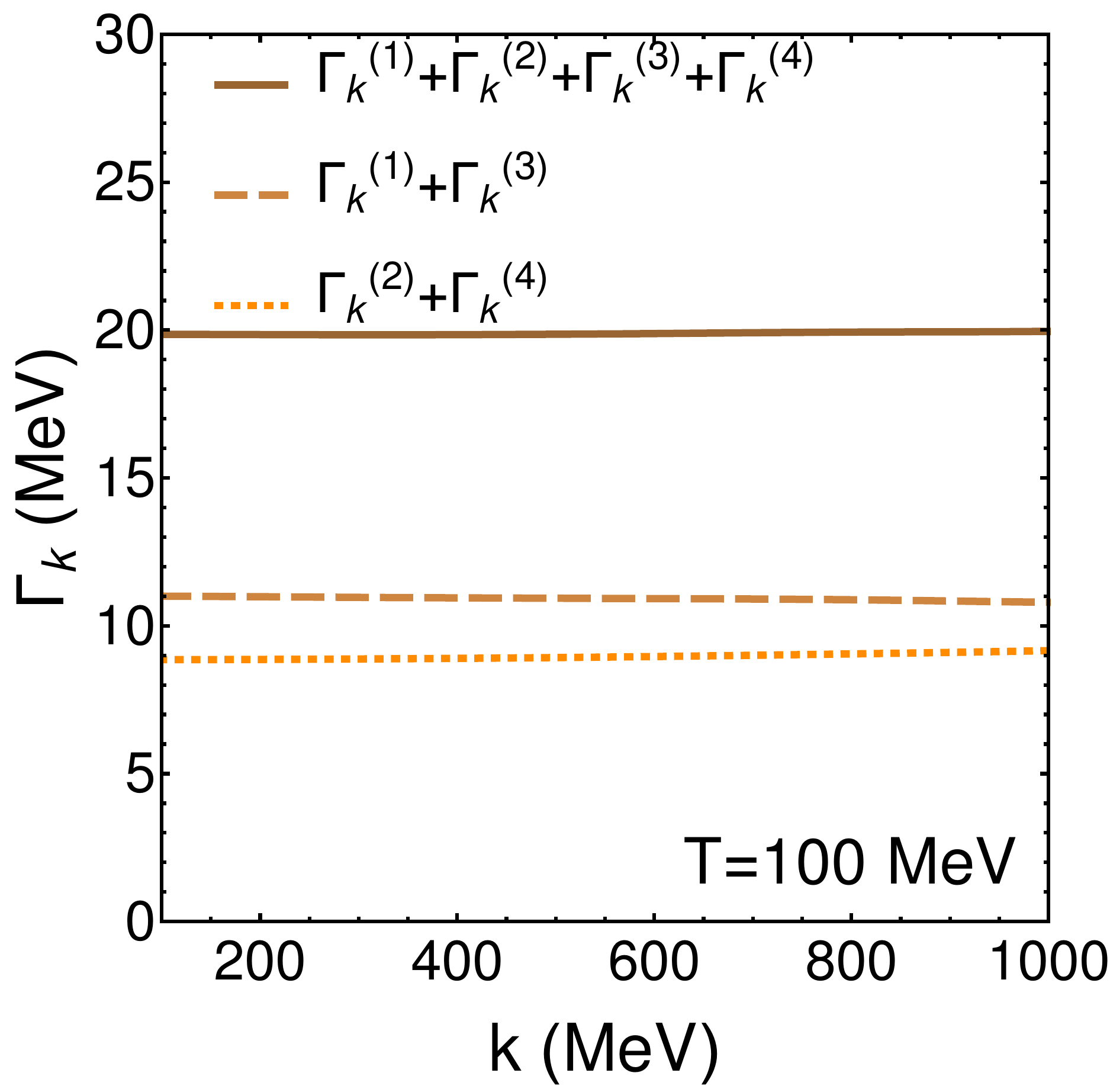}
  \includegraphics[width=0.327\textwidth]{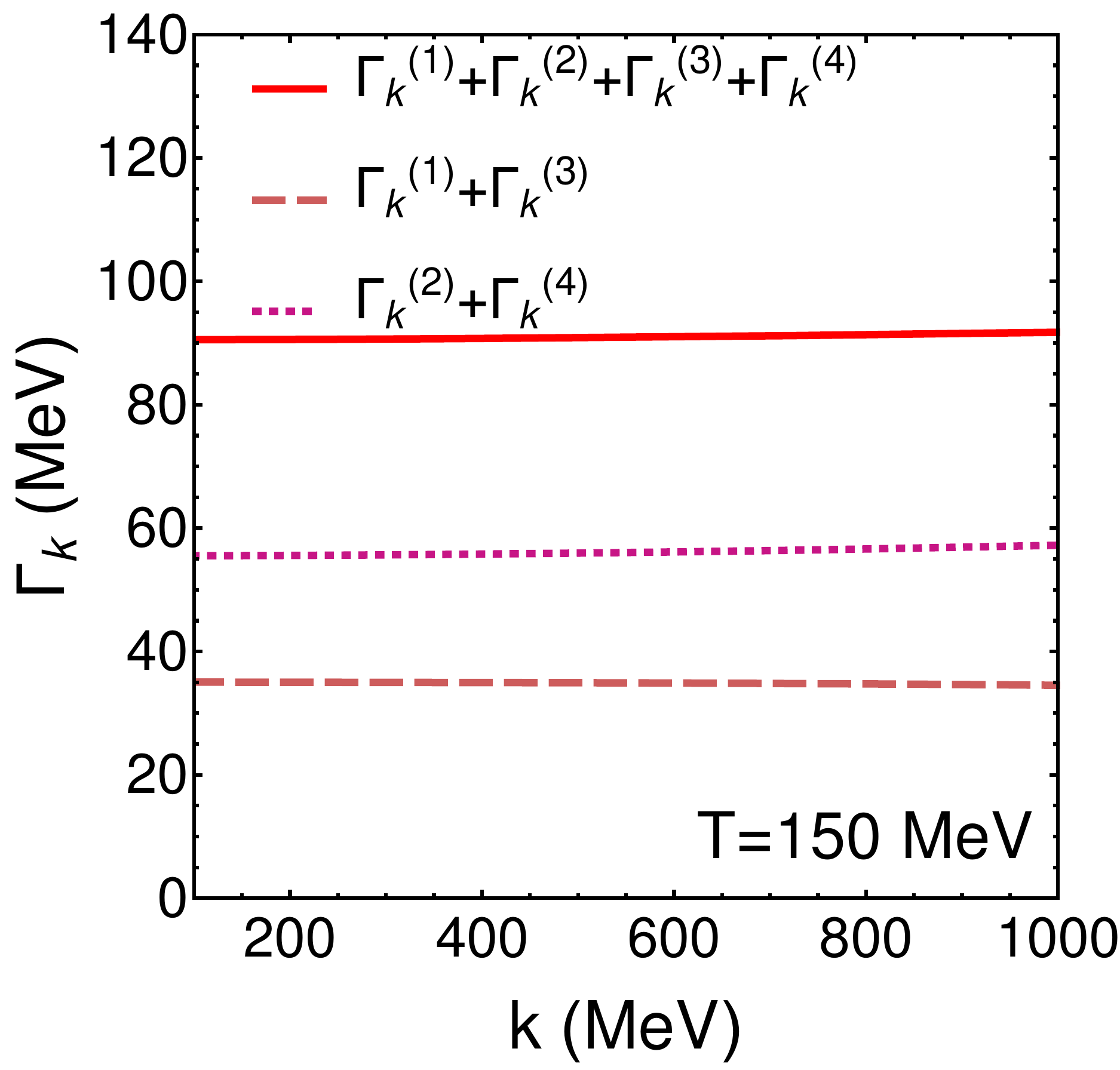}
    \caption{The $\bar{B}$-meson thermal width as computed from Eq.~(\ref{eq:transport-Gammak4terms}). Dashed lines show the contribution of $\Gamma_k^{(1)}+\Gamma_k^{(3)}$ (Eqs.~(\ref{eq:transport-Gammak1})+(\ref{eq:transport-Gammak3}), unitary cut), whereas dotted lines correspond to the contribution of $\Gamma_k^{(2)}+\Gamma_k^{(4)}$ (Eqs.~(\ref{eq:transport-Gammak2})+(\ref{eq:transport-Gammak4}), Landau cut).}
  \label{fig:transport-Gammak-B}
\end{figure}

The contributions of the unitary and Landau cuts to the $\bar{B}$-meson thermal width, as given by $(\Gamma_k^{(1)}+\Gamma_k^{(3)})$ and $(\Gamma_k^{(2)}+\Gamma_k^{(4)})$, respectively, as well as the total thermal width, computed with the set of Eqs.~(\ref{eq:transport-Gammak4terms}) to (\ref{eq:transport-Gammak4}) that use the imaginary part of the scattering amplitude, are shown in Fig.~\ref{fig:transport-Gammak-B} at the same temperatures as for the $D$-meson thermal width (see Fig.~\ref{fig:transport-Gammak}). The results are qualitatively very similar to those obtained for the charm sector: at small temperatures, the contribution from the Landau cut already accounts for $\sim 20\%$ of the total thermal width, while at large temperatures its contribution exceeds that of the unitary cut.

\begin{figure}[b!]
\centering
\includegraphics[width=0.45\textwidth]{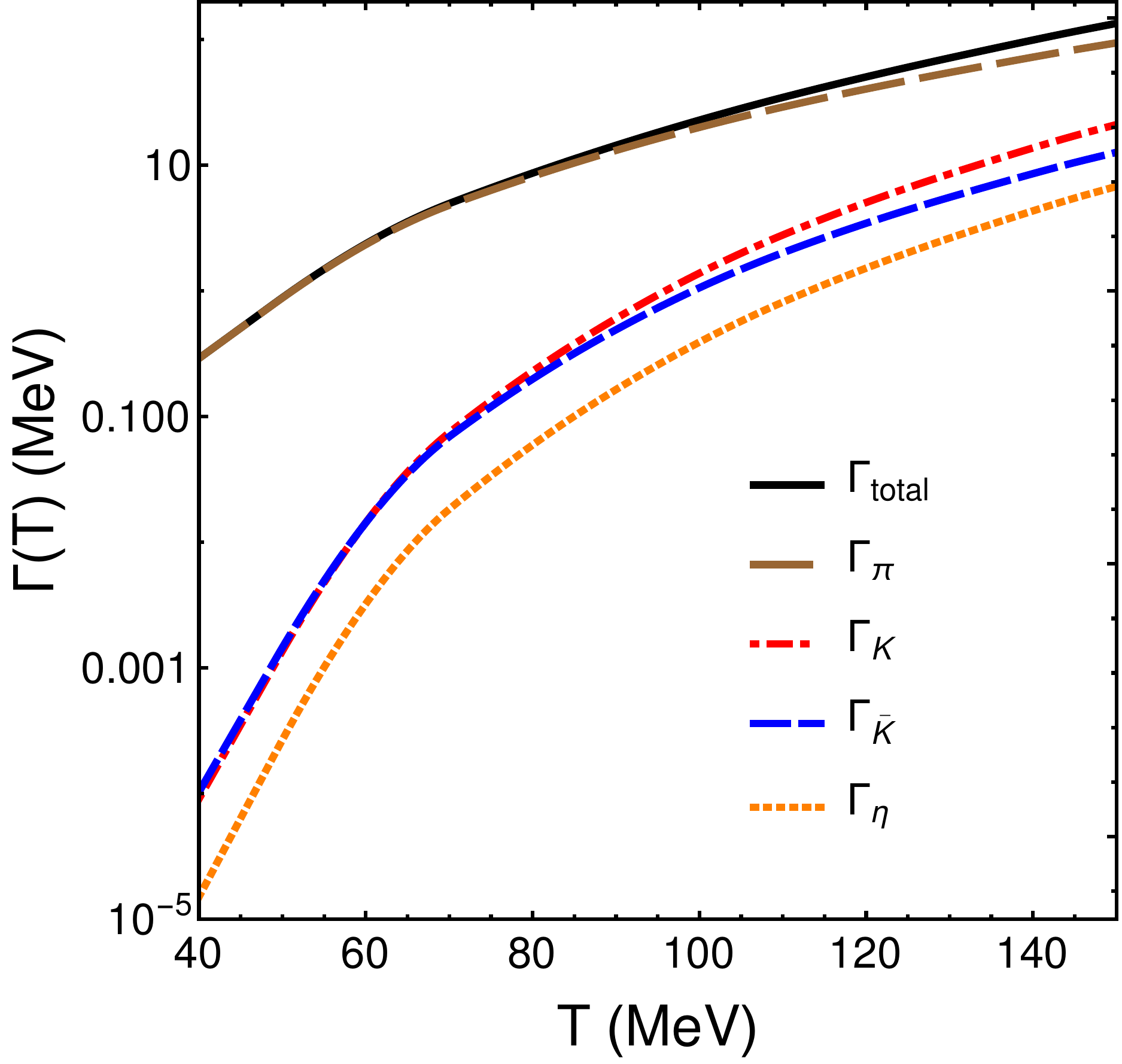}
\caption{Contribution to the $\bar{B}$-meson averaged thermal width from a thermal bath of pions, kaons, antikaons, and $\eta$ mesons.
\label{fig:transport-widths-B}}
\end{figure}

The contribution to the $\bar{B}$-meson width of the heavier light mesons that are present in the thermal bath is displayed in Fig.~\ref{fig:transport-widths-B}. As observed in the charm sector, the dominant contribution at low temperatures is that coming from the pions, as the abundances of heavier mesons, that is, $K$, $\bar{K}$, and $\eta$ mesons, are very small. At temperatures $T\lesssim 150$~MeV, the heavier mesons contribute with a few MeV to the $\bar{B}$-meson width, but pions still account for $\sim 90\%$ of the total thermal width.

In Chapter~\ref{ch:hot-medium} we have found that the thermal effects on the properties of open heavy-flavor mesons in thermal equilibrium with a light-meson bath are comparable for the open-charm and open-bottom sectors. Given the results for the charmed- and bottomed-mesons thermal width presented in this section, we can confirm the similarities that exist between the charm and the bottom sectors. We have also been able to weigh the contribution of the unitary and Landau cuts of the scattering amplitudes to the total thermal width at different temperatures, and found that the separated contributions are of similar size in the two sectors.

\section{Off-shell heavy-meson transport coefficients}
\label{sec:transport-transport}

Let us now turn to the study of the transport coefficients of a heavy meson, when thermal scattering amplitudes are implemented. In order to obtain a sensible definition of the relevant transport coefficients, we need to go back to the kinetic equation described in Section~\ref{sec:transport-nonequilibrium} and incorporate the separation of scales between the heavy-meson mass and the other scales in the system. By doing this, we can convert the off-shell kinetic equation in Eq.~(\ref{eq:transport-transportG}) into a Fokker-Planck equation~\cite{lifshitz1981physical,Svetitsky:1987gq}. From the latter, we can identify the so-called drag force $A$, and the diffusion coefficients $B_0$, $B_1$, and $D_s$, and compute them with thermal effects incorporated.


\subsection{Reduction to an off-shell Fokker-Planck equation}
\label{subsec:transport-fokker-planck}

We start with the off-shell kinetic equation (see Eq.~(\ref{eq:transport-transportG})) where, for simplicity, we keep implicit the sum over the scattering channels,
\begin{align}\label{eq:transport-kinoff}\nonumber
& \left(  k^\mu - \frac{1}{2} \frac{\partial\textrm{Re } \Pi^{\textrm{R}}}{\partial k_\mu} \right) \frac{\partial}{\partial X^\mu}\ii G_D^< (X,k)  \\ 
& \hspace{1.5cm}= \frac{1}{2} \int \prod_{i=1}^3 \frac{d^4k_i}{(2\pi)^4}  (2\pi)^4 \delta^{(4)} (k_1+k_2-k_3-k) \left|T (k_1^0+k_2^0+\ii\varepsilon, \vec{k}_1 + \vec{k}_2)\right|^2  \nonumber \\
&\hspace{2cm} \times  \left[ \ii G_D^<(X,k_1) \ii G_\Phi^<(X,k_2) \ii G_\Phi^>(X,k_3) \ii G_D^>(X,k)\right. \nonumber \\
&\hspace{2.15cm}\left. -\, \ii G_D^>(X,k_1) \ii G_\Phi^>(X,k_2) \ii G_\Phi^<(X,k_3) \ii G_D^<(X,k) \right]  \ . 
\end{align}

Inspired by previous derivations~\cite{lifshitz1981physical,Svetitsky:1987gq,Abreu:2011ic}, we define an off-shell scattering rate
\begin{align}
 & W(k^0,\vec{k},k_1^0,\vec{q}\,) \equiv \int \frac{d^4k_3}{(2\pi)^4} \frac{d^4k_2}{(2\pi)^4}  (2\pi)^4 \delta (k_1^0+k_2^0-k_3^0-k^0) \delta^{(3)} (\vec{k}_2-\vec{k}_3-\vec{q}\,) \nonumber \\
 &\hspace{1cm} \times \left|T (k_1^0+k_2^0+\ii\varepsilon, \vec{k} - \vec{q} + \vec{k}_2)\right|^2  \ii G_\Phi^>(X,k_2)\ii G_\Phi^<(X,k_3) \ii G_D^>(X,k_1^0,\vec{k}-\vec{q}\,) \ , \label{eq:transport-scatrate}
\end{align}
where we have replaced the variable $\vec{k}_1$ with the momentum loss $\vec{q} \equiv \vec{k} - \vec{k}_1$. Equation~(\ref{eq:transport-scatrate}) describes the collision rate of a $D$ meson with energy $k^0$ and momentum $\vec{k}$ to a final $D$ meson with energy $k_1^0$ and momentum $\vec{k}-\vec{q}$. It depends on the spectral weights and the populations of the particles $1$, $2$, and $3$ of the binary collision, encoded in the Wightman functions. It can be interpreted as a collision loss term for a $D$ meson with momentum $\vec{k}$. In fact, the loss term of Eq.~(\ref{eq:transport-kinoff}) can be directly written as,
\begin{equation}
  -\frac12 \int \frac{dk_1^0}{2\pi}\frac{d^3q}{(2\pi)^3} W(k^0,\vec{k},k_1^0,\vec{q}\,) \ii G_D^<(X,k^0,\vec{k}\,) \ .
\end{equation}

The gain term of Eq.~(\ref{eq:transport-kinoff}) can be interpreted as a loss term of an incoming $D$ meson with momentum $\vec{k}_1$, losing the same momentum amount $\vec{q}$ and ending with momentum $\vec{k}$. The energy $k^0$ is an independent free variable. This term reads
\begin{align}
& \frac{1}{2} \int \prod_{i=1}^3 \frac{d^4k_i}{(2\pi)^4}  (2\pi)^4 \delta (k^0+k_3^0-k^0_1-k^0_2) 
\delta^{(3)} (\vec{q}+\vec{k}_3-\vec{k}_2) \left|T (k^0+k_3^0+\ii\varepsilon, \vec{k}+\vec{q} +\vec{k}_3)\right|^2  \nonumber \\
&\qquad \times   \ii G_D^<(X,k^0,\vec{k}+\vec{q}\,) \ii G_\Phi^>(X,k_2) \ii G_\Phi^<(X,k_3) \ii G_D^>(X,k^0_1,\vec{k}\,) \nonumber \\
&= \frac12  \int \frac{dk_1^0}{2\pi}\frac{d^3q}{(2\pi)^3} W(k^0,\vec{k}+\vec{q},k_1^0,\vec{q}\,)\ii  G_D^<(X,k^0,\vec{k}+\vec{q}\,) \ . \label{eq:transport-gainasloss}
\end{align}

Then, Eq.~(\ref{eq:transport-kinoff}) can be written as
\begin{align}
 2 \left(  k^\mu - \frac{1}{2} \frac{\partial \textrm{Re } \Pi^{\textrm{R}}}{\partial k_\mu} \right) \frac{\partial}{\partial X^\mu}\ii G^<_D (X,k)  =& \int \frac{dk_1^0}{2\pi} \frac{d^3q}{(2\pi)^3} \left[  W(k^0,\vec{k}+\vec{q}, k_1^0,\vec{q}\,) \ii G_D^< (X,k^0,\vec{k}+\vec{q}\,)\right. \nonumber \\ 
& \quad \left.  -\, W(k^0,\vec{k},k_1^0,\vec{q}\,) \ii G_D^<(X,k^0,\vec{k}\,)\right] \ .   
\end{align}

This equation is an alternative form of Eq.~(\ref{eq:transport-kinoff}), convenient for the formal reduction to the off-shell Fokker-Planck equation. For this purpose, we exploit the separation of scales between the meson masses, as the mass of the heavy meson is much larger than the temperature and any of the light-meson masses. Such a Brownian picture implies that the typical momentum exchanged in the elastic collision is of the order of $T$, and much smaller than the total momentum of the heavy particle, $|\vec{q}| \ll |\vec{k}|$~\cite{lifshitz1981physical,Svetitsky:1987gq,Abreu:2011ic}. 

We can Taylor expand the combination  $W(k^0,\vec{k}+\vec{q}, k_1^0,\vec{q}\,)\ii G_D^< (X,k^0,\vec{k}+\vec{q}\,)$ around $\vec{k}$ up to second order. In doing so, we consider a homogeneous thermal bath, as the light sector is assumed to be equilibrated in much shorter time scales so that one can employ a space-averaged Green's function~\cite{Svetitsky:1987gq}. In addition, we also set $z_k \simeq 1$ as usual.

After a few steps, one obtains a Fokker-Planck equation for $\ii G_D^< (t,k^0,\vec{k}\,)$
\begin{align} \nonumber
&\frac{\partial}{\partial t} \ii G_D^< (t,k) \\
& \quad = \frac{\partial}{\partial k^i} \left\{ \hat{A} (k;T) k^i\, \ii G_D^< (t,k) + \frac{\partial}{\partial k^j} \left[ \hat{B}_0(k;T) \Delta^{ij} + \hat{B}_1(k;T) \frac{k^i k^j}{\vec{k}\,^2} \right] \ii G_D^< (t,k) \right\} \ , \label{eq:transport-offFP} 
\end{align}
with $\Delta^{ij}=\delta^{ij}-k^ik^j/\vec{k}\,^2$, and we have defined the transport coefficients as~\footnote{There are two notations commonly used in  the literature for the transport coefficients, $F$, $\Gamma_0$, and $\Gamma_1$, or, alternatively, $A$, $B_0$, and $B_1$. The latter one is chosen here, so as to avoid confusion with the thermal width $\Gamma_k$.}
\begin{align}
 \hat{A} (k^0, \vec{k}\,;T) & \equiv \left \langle 1 -\frac{\vec{k} \cdot \vec{k}_1}{\vec{k}\,^2} \right \rangle \ , \label{eq:transport-hatA} \\
 \hat{B}_0 (k^0, \vec{k}\,;T) & \equiv  \frac14 \left \langle \vec{k}_1\,^2 - \frac{(\vec{k} \cdot \vec{k}_1)^2}{\vec{k}\,^2} \right \rangle \ , \label{eq:transport-hatB0} \\
  \hat{B}_1 (k^0, \vec{k}\,;T) & \equiv \frac12 \left \langle  \frac{[ \vec{k} \cdot (\vec{k}-\vec{k}_1)]^2}{\vec{k}\,^2} \right \rangle \ , \label{eq:transport-hatB1}
\end{align}
where $\vec{k}_1$ has been reintroduced, replacing $\vec{q}\,$. The hat is used to denote off-shell transport coefficients, as they depend separately on $k^0$ and $\vec{k}$. The average is defined as
\begin{align}
\left\langle {\cal F}(\vec{k},\vec{k}_1) \right\rangle & = \frac{1}{2k^0} \sum_{\lambda,\lambda'=\pm} \lambda \lambda' \int_{-\infty}^\infty \ dk_1^0 \int \prod_{i=1}^3 \frac{d^3k_i}{(2\pi)^3} \frac{1}{2E_22E_3}  \ S_D(k_1^0,\vec{k}_1)   \nonumber \\
& \times (2\pi)^4 \delta^{(3)} (\vec{k}+\vec{k}_3-\vec{k}_1-\vec{k}_2) \delta (k^0+\lambda' E_3- \lambda E_2-k^0_1) \left|T(k^0+ \lambda' E_3,\vec{k}+\vec{k}_3)\right|^2  \nonumber \\
& \times f^{(0)} (\lambda'E_3) \tilde{f}^{(0)} (\lambda E_2)  \tilde{f}^{(0)} (k_1^0) \ \ {\cal F}(\vec{k},\vec{k}_1)  \ , \label{eq:transport-rateoff}
\end{align}
where the spectral function of the $D$ meson is kept.  

The term for the drag force in Eq.~(\ref{eq:transport-offFP}),
\begin{equation}
 A^i=\hat{A}(k;T)k^i\equiv\left\langle (\vec{k}-\vec{k}_1)^i\right\rangle \ ,
\end{equation}
quantifies the thermal average momentum transfer to the heavy meson due to the collisions with the particles in the medium. On the other hand, the momentum diffusion term, which has been decomposed in the longitudinal and transverse components,
 \begin{equation}
  B^{ij}=\hat{B}_0(k;T)\left(\delta^{ij}-\frac{k^ik^j}{\vec{k}\,^2}\right)+\hat{B}_1(k;T)\frac{k^ik^j}{\vec{k}\,^2}\equiv\frac12\left\langle(\vec{k}-\vec{k}_1)^i(\vec{k}-\vec{k}_1)^j\right\rangle \ ,
 \end{equation}
measures the square of the momentum transfer due to the interactions.

We use the relation of Eq.~(\ref{eq:transport-rateoff}) to compute the off-shell transport coefficients in Eqs. (\ref{eq:transport-hatA}) to (\ref{eq:transport-hatB1}). This approach stands at the same level as Eqs.~(\ref{eq:transport-widthT2U}) and (\ref{eq:transport-widthT2L}), and accounts for thermal modifications, off-shell effects, as well as the Landau cut contributions (case $\lambda'<0$). It is denoted as ``OffShell'' in the following, and it is the most complete calculation of transport coefficients used in this dissertation.

It is important to notice that, in general, it is not possible to derive a Fokker-Planck equation for the heavy-meson distribution, $f_D(t,\vec{k}\,)$, while including off-shell effects in the transport coefficients,because after $k^0$ integration on both sides one cannot factorize the distribution function from the transport coefficients. Only in the particular case of a narrow quasiparticle, it is possible to trivially integrate $k^0$ and obtain the kinetic equation for $f_D(t,\vec{k}\,)$, thus reproducing the previous approaches in the literature~\cite{Berrehrah:2013mua,Liu:2018syc}. For the interested reader, the detailed derivation of the on-shell Fokker-Planck equation is given in Appendix~\ref{appendix-onshell}. 

We remind that, although not explicitly written in Eq.~(\ref{eq:transport-rateoff}), there is a sum over all the allowed channels in these expressions, both the elastic and the inelastic ones. However, we have learned from the thermal width that the contribution of the inelastic processes is very small, and therefore they are neglected in what follows. Nevertheless, all the elastic channels ($D\pi$, $DK$, $D\bar{K}$, and $D\eta$) are considered when computing the coefficients.

The described ``OffShell'' approximation, based on Eq.~(\ref{eq:transport-rateoff}), is rather general. However, we already know that the quasiparticle approximation is excellent for the $D$ mesons. Therefore one can replace the $D$-meson spectral function with the expression in Eq.~(\ref{eq:transport-Diracdelta}), and neglect the $z_k$ factor altogether. 

This brings two consequences. The first one is that the Fokker-Planck equation in Eq.~(\ref{eq:transport-offFP}) for $\ii G^<_D(t,k)$ can be written for $f_D(t,E_k)\equiv f_D(t,\vec{k}\,)$ instead,
\begin{align}\nonumber
&\frac{\partial}{\partial t} f_D(t,E_k) \\
&\quad = \frac{\partial}{\partial k^i} \left\{ A(\vec{k}\,;T) k^i f_D(t,E_k) + \frac{\partial}{\partial k^j} \left[ B_0(\vec{k}\,;T) \Delta^{ij} + B_1(\vec{k}\,;T) \frac{k^i k^j}{k^2} \right] f_D(t,E_k) \right\} \ , \label{eq:transport-onFP} 
\end{align}
where the coefficients only depend on $|\vec{k}|$ as the quasiparticle energy is put on shell,
\begin{align}
 A (\vec{k}\,;T) & \equiv \left \langle 1 -\frac{\vec{k} \cdot \vec{k}_1}{\vec{k}\,^2} \right \rangle_{ \textrm{Thermal U+L}} \ , \label{eq:transport-hatAon} \\
B_0 (\vec{k}\,;T) & \equiv  \frac14 \left \langle \vec{k}_1\,^2 - \frac{(\vec{k} \cdot \vec{k}_1)^2}{\vec{k}\,^2} \right \rangle_{ \textrm{Thermal U+L}}\ , \label{eq:transport-hatB0on} \\
B_1 (\vec{k}\,;T) & \equiv \frac12 \left \langle  \frac{[ \vec{k} \cdot (\vec{k}-\vec{k}_1)]^2}{\vec{k}\,^2} \right \rangle_{ \textrm{Thermal U+L}} \ . \label{eq:transport-hatB1on}
\end{align}

The second consequence is that the scattering rate gets simplified because, as for the Boltzmann equation, only one type of process is able to conserve energy-momentum when all the four particles in the collision are on their mass shell. In this approximation, it is possible to write,
\begin{align}
\left\langle {\cal F} (\vec{k},\vec{k}_1) \right\rangle_{\textrm{Thermal U+L}} & = \frac{1}{2E_k}  \int \frac{d^3k_1}{(2\pi)^4} \frac{d^3k_2}{(2\pi)^3} \frac{d^3k_3}{(2\pi)^3}  (2\pi)^4 \delta^{(4)} (k_1+k_2-k_3-k) \nonumber  \\
  & \times \left[ \left|T(E_k+E_3,\vec{k}+\vec{k}_3)\right|^2+\left|T(E_k-E_2,\vec{k}-\vec{k}_2)\right|^2 \right] \nonumber \\
&  \times \frac{1}{2E_1 2E_2 2E_3} f^{(0)} (E_3) \tilde{f}^{(0)} (E_2) \ {\cal F}(\vec{k},\vec{k}_1) \ , \label{eq:transport-ratetherm}
\end{align}
where we have only considered elastic collisions, as in Eq.~(\ref{eq:transport-onshellkineticelastic}).

This expression looks closer to the previous calculations of the heavy-flavor transport coefficients found in the literature, but the Landau contribution still remains in addition to the unitary one. The scattering amplitudes also include medium effects. We denote this approximation to compute the transport coefficients as ``Thermal U+L''.

One can yet consider another simplification by simply setting the Landau contribution to zero. There is no reason to neglect this term in the medium, but we consider this approximation for the sake of comparison, and denote it as ``Thermal U''. In any case, this approximation should be realistic at low temperatures, where the  Landau cut disappears. The scattering rate to be used in the ``Thermal U'' approximation reads
\begin{align}
\left\langle {\cal F} (\vec{k},\vec{k}_1) \right\rangle_{\textrm{Thermal U}}  & = \frac{1}{2E_k}  \int \frac{d^3k_1}{(2\pi)^4} \frac{d^3k_2}{(2\pi)^3} \frac{d^3k_3}{(2\pi)^3}  (2\pi)^4 \delta^{(4)} (k_1+k_2-k_3-k) \nonumber \\
  & \times \left|T(E_k+E_3,\vec{k}+\vec{k}_3)\right|^2 \frac{1}{2E_22E_3 2E_1} f^{(0)} (E_3) \tilde{f}^{(0)} (E_2) \ {\cal F} (\vec{k},\vec{k}_1)  \ . \label{eq:transport-ratevac}
\end{align}

Finally, to match our results to previous approaches, we simply use Eq.~(\ref{eq:transport-ratevac}) without any thermal effects, neither in the quasiparticle energies nor the scattering amplitudes. We use vacuum amplitudes and standard relativistic expressions for the energies $E_k=\sqrt{\vec{k}\,^2+m_{D}^2}$, $E_1 = \sqrt{\vec{k}_1\,^2 + m_{D}^2}$, where $m_{D}$ is the $D$-meson vacuum mass. This approximation is denoted as ``Vacuum'', as it is the one that most resembles previous calculations.

We summarize in Table~\ref{tab:transport-approxs} the different approximations to compute the heavy-meson transport coefficients. It starts with the simplest one, where vacuum amplitudes without thermal corrections are used, and gets to the most involved one, where thermal and off-shell effects are taken into account.

\begin{table}[t!]
\setlength{\tabcolsep}{9pt}
\renewcommand{\arraystretch}{1.2}
\begin{tabular}{c|cccc}
\hline
 Approximation &  \multirow{2}{*}{Interaction Rate}  & Thermal effects  &  \multirow{2}{*}{Landau cut}  &  \multirow{2}{*}{Off-shell effects}   \\
 name & & on $|T|^2$ and $E_k$ & & \\
\hline
Vacuum & Eq.~(\ref{eq:transport-ratevac}) & \xmark &  \xmark & \xmark \\
Thermal U & Eq.~(\ref{eq:transport-ratevac}) & \cmark & \xmark & \xmark \\
Thermal U+L & Eq.~(\ref{eq:transport-ratetherm}) & \cmark & \cmark & \xmark \\
OffShell & Eq.~(\ref{eq:transport-rateoff}) & \cmark & \cmark & \cmark \\
\hline
\end{tabular}
\centering
\caption{Different approximations for the computation of the heavy-meson transport coefficients used in this thesis. The details are given in the main text.}
\label{tab:transport-approxs}
\end{table}

\subsection{Results for $D$-meson transport coefficients}
\label{subsec:transport-results}
We start the description of the numerical results with the $D$-meson drag coefficient $A(\vec{k}\,;T)$ and the momentum diffusion coefficient $B_0(\vec{k}\,;T)$ (or $\hat{A}(k_0,\vec{k}\,;T)$ and $\hat{B}_0 (k_0,\vec{k}\,;T)$, for the off-shell approximation). The results are presented in the so-called \textit{static limit}, $\vec{k} \rightarrow 0$ ($|\vec{k}|=50$~MeV in the actual computation). In this limit, the equality between the two components of the diffusion coefficients is satisfied, $B_0=B_1$, which we have checked numerically in all the cases.

\begin{figure}[b!]
  \centering
\includegraphics[width=0.445\textwidth]{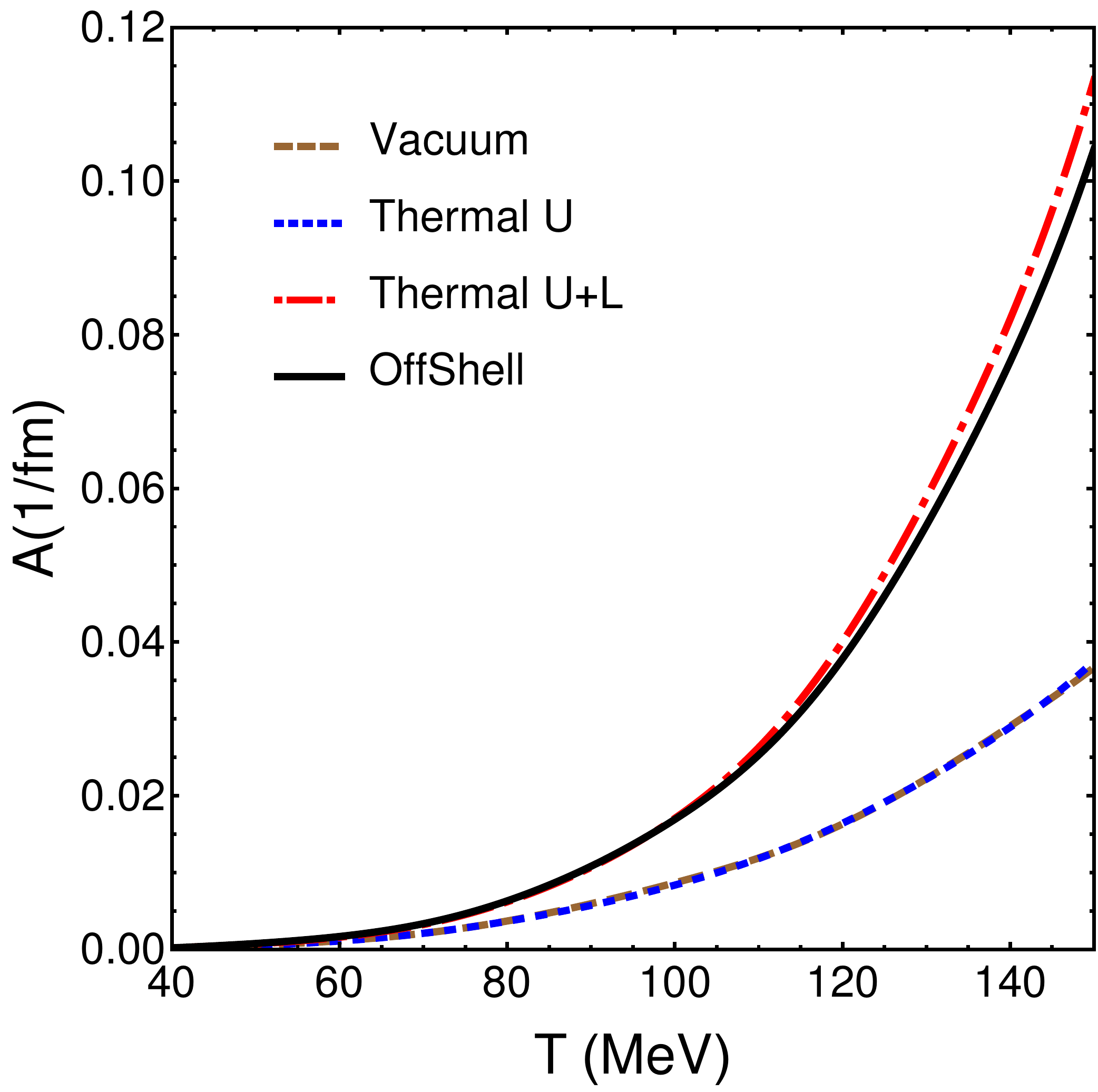}\hspace{0.5cm}
\includegraphics[width=0.46\textwidth]{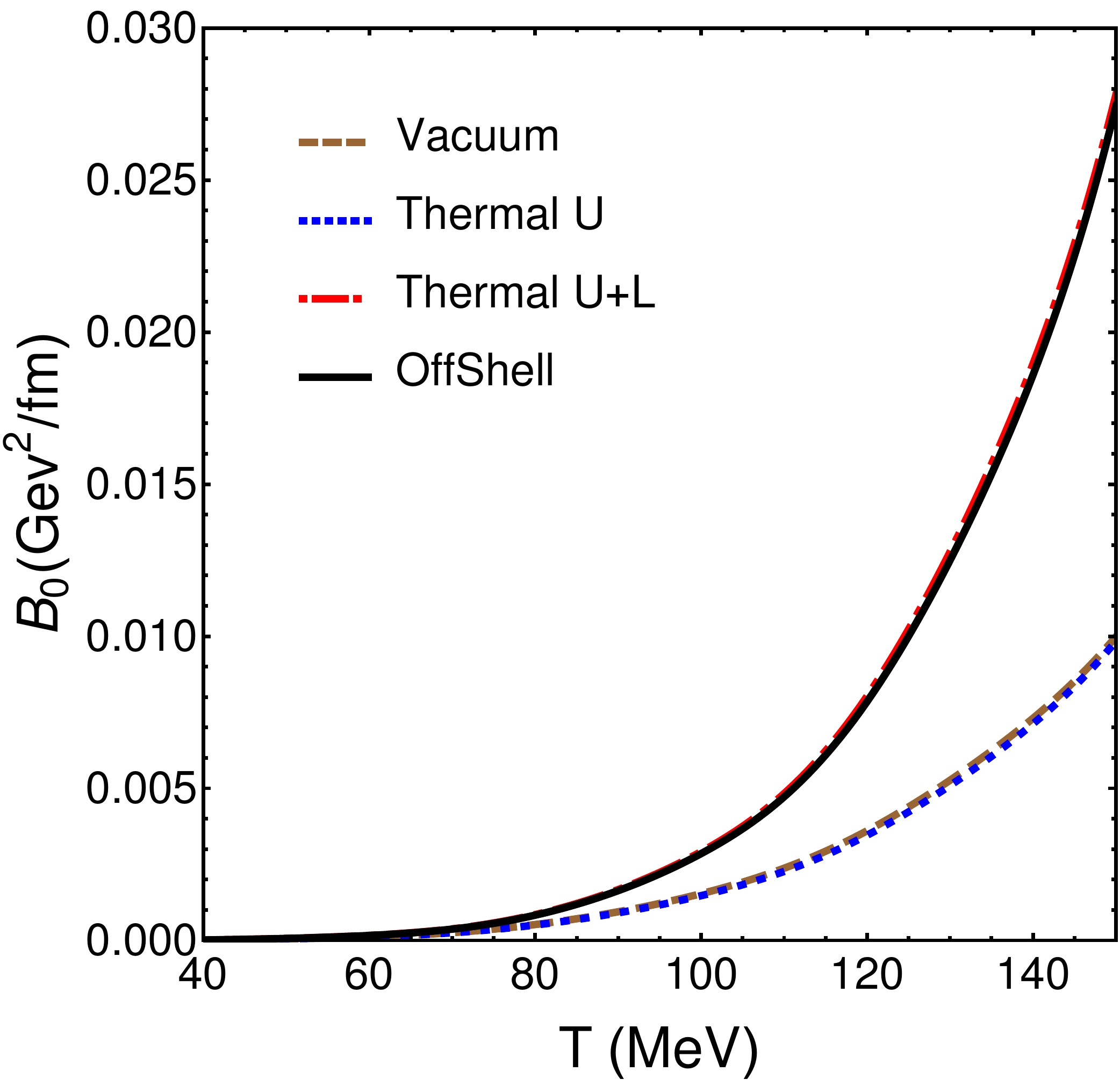}
\caption{Transport coefficients of the $D$ meson in the static limit $\vec{k}\rightarrow 0$ (where $B_1=B_0$), using the different approximations described in Table~\ref{tab:transport-approxs} (see also the main text for details). The curve that incorporates all the thermal and off-shell effects is the one denoted as ``OffShell''.}
  \label{fig:transport-transportcomp}
\end{figure}

In Fig.~\ref{fig:transport-transportcomp} we present the drag coefficient $A$ (left panel) and the diffusion coefficient $B_0$ (right panel) under the different approximations of Table~\ref{tab:transport-approxs}. In particular, in the ``Vacuum'' approximation we employ vacuum scattering amplitudes and masses, which is what was done by the authors of Ref.~\cite{Tolos:2013kva}. 

All the remaining approximations use thermal scattering amplitudes and temperature-dependent quasiparticle energies. ``Thermal U'' only incorporates the unitary cut, and the differences, when compared to ``Vacuum'', are entirely due to thermally dependent interactions. Rather surprisingly, we find no appreciable differences in comparison to ``Vacuum'' even at high temperatures. The main difference comes from the addition of the Landau contribution, which is incorporated in the ``Thermal U+L'' scenario. At our top temperatures, the contribution of this cut is even more important than the one of the unitary cut. This was already anticipated in the analysis of the $D$-meson thermal width in Section~\ref{sec:transport-kinematic}.

In addition, we present our results incorporating off-shell effects, which employ the full spectral distribution of the in-medium $D$ meson. This approximation is denoted ``OffShell'' in Fig.~\ref{fig:transport-transportcomp}. In this case, to fix the external energy dependence we have simply set $k^0=E_k$ with $|\vec{k}|=50$~MeV (static limit).
%
%
Only a small difference can be observed in $A$ at high temperatures, concluding that the genuine off-shell effects are not as important as including the Landau cut contributions. This already happened for the thermal width (see Fig.~\ref{fig:transport-Gammaonoff}). This result is not very surprising as the $D$-meson spectral function is still very narrow for the temperatures considered here, so the quasiparticle approximation works extremely well. 
%
In the off-shell approximation, when the heavy meson carries a finite thermal width, processes $1 \leftrightarrow 3$ are also allowed. In the work presented in this chapter, we have neglected these processes because the required production threshold is higher than the elastic one. Although it is out of the scope of this thesis, it would be very interesting to analyze the Bremsstrahlung processes $D\rightarrow D+\pi+\pi$ and their role in the $D$-meson energy loss.

We finally explore the spatial diffusion coefficient $D_s (T)$~\cite{Abreu:2011ic}\footnote{Not to be confused with the charm-strange mesons, $D_s$, for which the same notation was used in previous chapters.}. This coefficient, which is usually normalized by the thermal wavelength, $1/(2\pi T)$, can be obtained from the static limit of the $B_0(\vec{k}\,; T)$ coefficient,
\begin{equation}
2\pi T D_s (T) = \lim_{\vec{k} \rightarrow 0} \frac{2\pi T^3}{B_0(\vec{k}\,;T)} \ . \label{eq:transport-Dscoeff}
\end{equation}

\begin{figure}[hb!]
  \centering
  \includegraphics[width=0.45\textwidth]{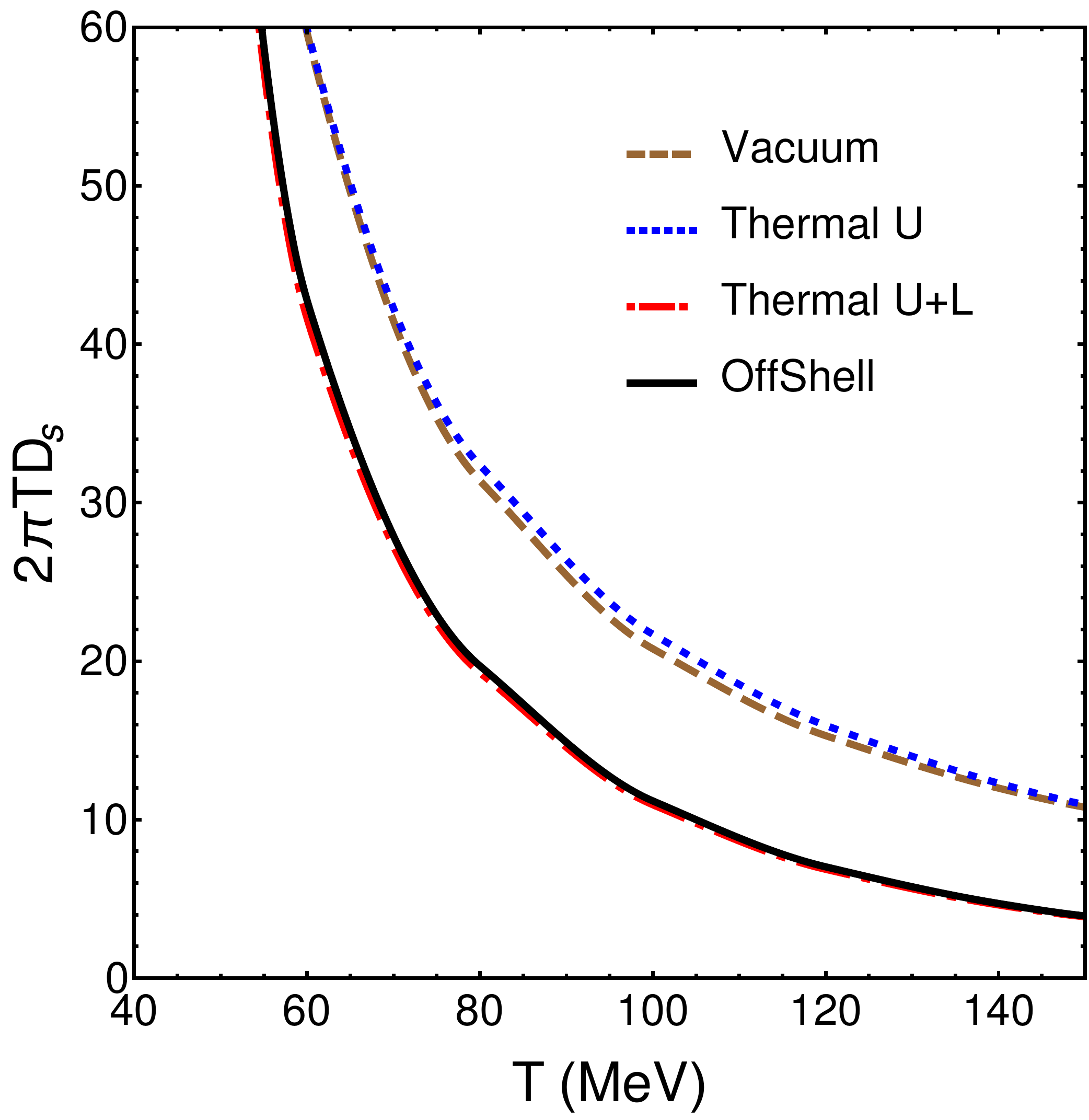}\hspace{0.5cm}
 \includegraphics[width=0.45\textwidth]{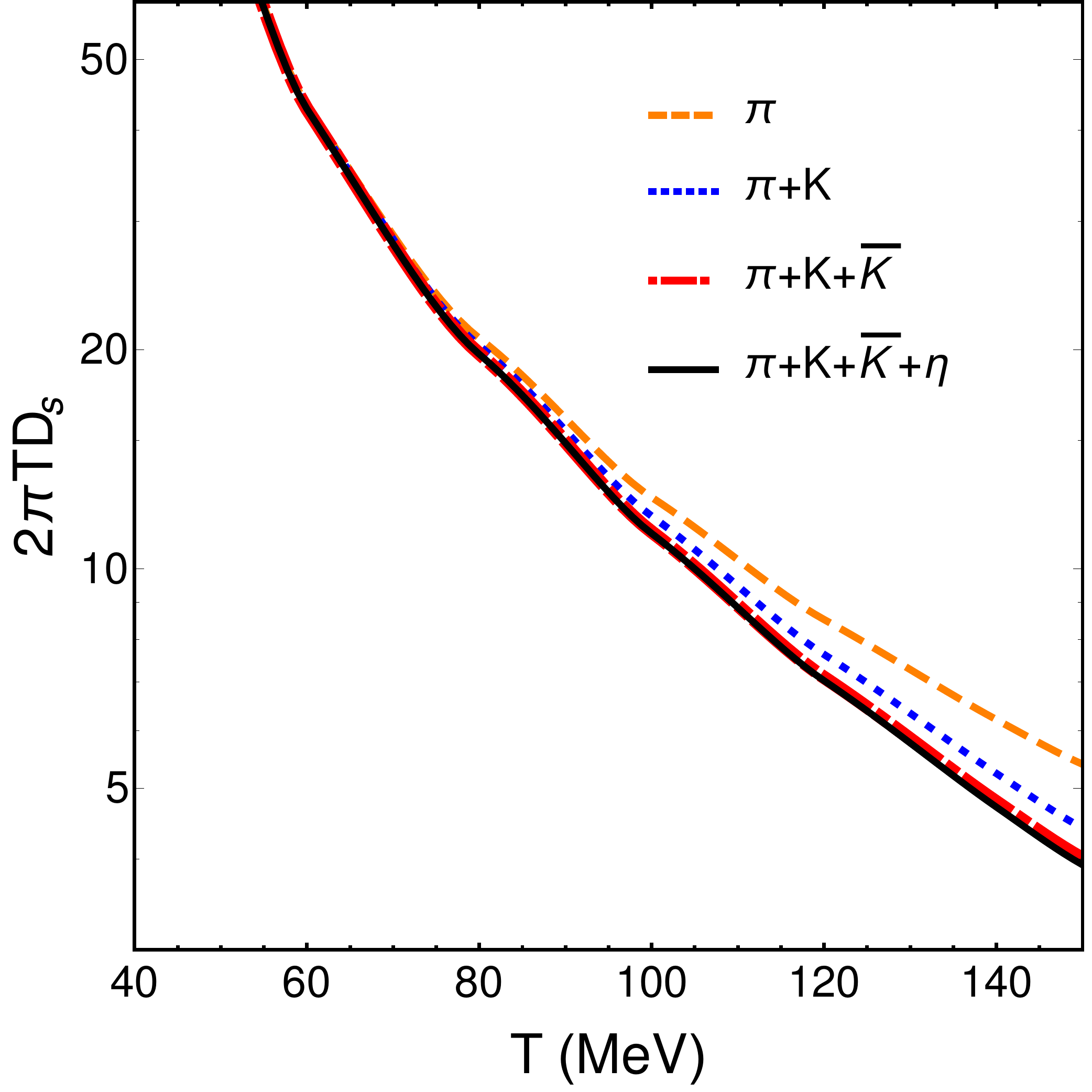}
  \caption{Left panel: Spatial diffusion coefficient divided by the thermal wavelength $(2\pi T)^{-1}$, using the different approximations in Table~\ref{tab:transport-approxs} (see also main text for details). The ``OffShell'' curve is the one incorporating all the thermal and off-shell effects. Right panel: Same coefficient in the ``OffShell'' approximation incorporating sequentially the different light mesons in the calculation (in logarithmic scale).}
  \label{fig:transport-Dscomp}
\end{figure}

This coefficient is shown in Fig.~\ref{fig:transport-Dscomp}. With regards to the past calculations of transport coefficients in the literature, the main difference comes from the Landau contribution, which makes the diffusion coefficient decrease almost by a factor of $3$ close to $T_c$, which is a remarkable effect. The results for the ``Thermal U+L'' and ``OffShell'' are almost identical.

%

Let us comment that an alternative way to fix the $k^0$ dependence of the off-shell transport coefficients is to define an average coefficient weighted by the $D$-meson spectral function. For example, one could define an average $\overline{B}_0$ as,
\begin{equation}
\overline{B}_0 (\vec{k}\,;T) = 2 \int_0^\infty dk^0 k^0 S_D(k^0,\vec{k}\,) \hat{B}_0 (k^0,\vec{k}\,;T) \ . \label{eq:transport-avB0} 
\end{equation}

In the narrow quasiparticle limit, this average coincides with the on-shell evaluation if one uses $z_k \simeq 1$ in addition,
\begin{equation}
\overline{B}_0 (\vec{k}\,;T) = \int_0^\infty dk^0 2k^0 \frac{z_k}{2 E_k} \delta (k^0-E_k) \hat{B}_0 (k^0,\vec{k}\,;T) \simeq \hat{B}_0 (E_k,\vec{k}\,;T)  \ . 
\end{equation} 
Because in our case the quasiparticle approximation works very well, the evaluation of the off-shell coefficient at $k_0 = E_k$ gives similar results to the one using Eq.~(\ref{eq:transport-avB0}).

Finally, we detail the different contributions of adding the light mesons one by one. In Fig.~\ref{fig:transport-GammaTall} we have shown that the pions provide the main contribution to the thermal decay width at temperatures below $T_c$, being the $K$, $\bar{K}$, and $\eta$ mesons subleading even at $T\simeq 150$~MeV. In the right panel of Fig.~\ref{fig:transport-Dscomp}, the spatial diffusion coefficient in the ``OffShell'' approximation is presented, in logarithmic scale, with the light mesons being
sequentially added. One also observes that the contribution of the more massive states is very small due to their thermal suppression. The negligible contribution of baryons at $\mu_B=0$ was studied in Ref.~\cite{Tolos:2013kva}. However, one should keep in mind that close to $T_c$ one could expect the excitation of many states and resonances which can collectively contribute to the transport coefficient in substantially (see Ref.~\cite{Noronha-Hostler:2008kkf} for the shear and bulk viscosities). Therefore, our predictions might not be trustable in the $T\gtrsim T_c$ region, so our results are shown up to $T = 150$~MeV.

\subsubsection{Comparison with other approaches}

To conclude this section on charmed transport coefficients, we compare our results below $T_c$ to recent \gls{lqcd} calculations of the heavy-flavor transport coefficients for temperatures $T \gtrsim T_c$. We also include a recent calculation using Bayesian methods to analyze the \gls{hic} data under a simulation code to obtain a posterior estimation of the spatial diffusion coefficient~\cite{Ke:2018tsh}. In our case, we show our most complete calculation (``OffShell'' approximation) together with the ``Vacuum'' calculation for comparison. From the \gls{lqcd} side, we compile the results presented in Refs.~\cite{Banerjee:2011ra,Kaczmarek:2014jga,Francis:2015daa, Brambilla:2020siz,Altenkort:2020fgs}. All these are given as functions of $T/T_c$. To compare the different results in terms of an absolute temperature, we fix $T_c=156$~MeV~\cite{Bazavov:2018mes}.

\begin{figure}[t]
  \centering
  \includegraphics[width=0.46\textwidth]{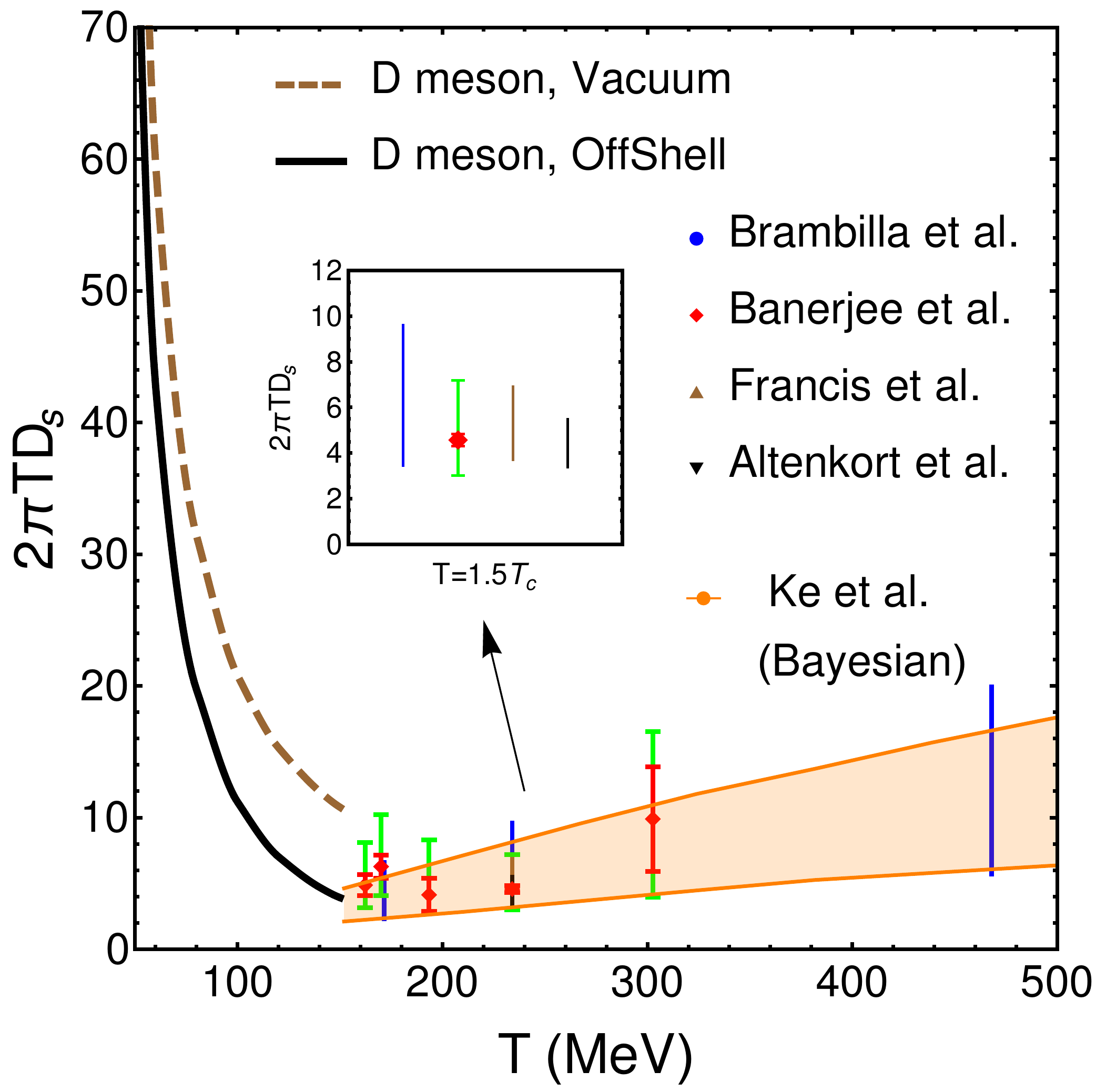}\hspace{0.5cm}
   \includegraphics[width=0.45\textwidth]{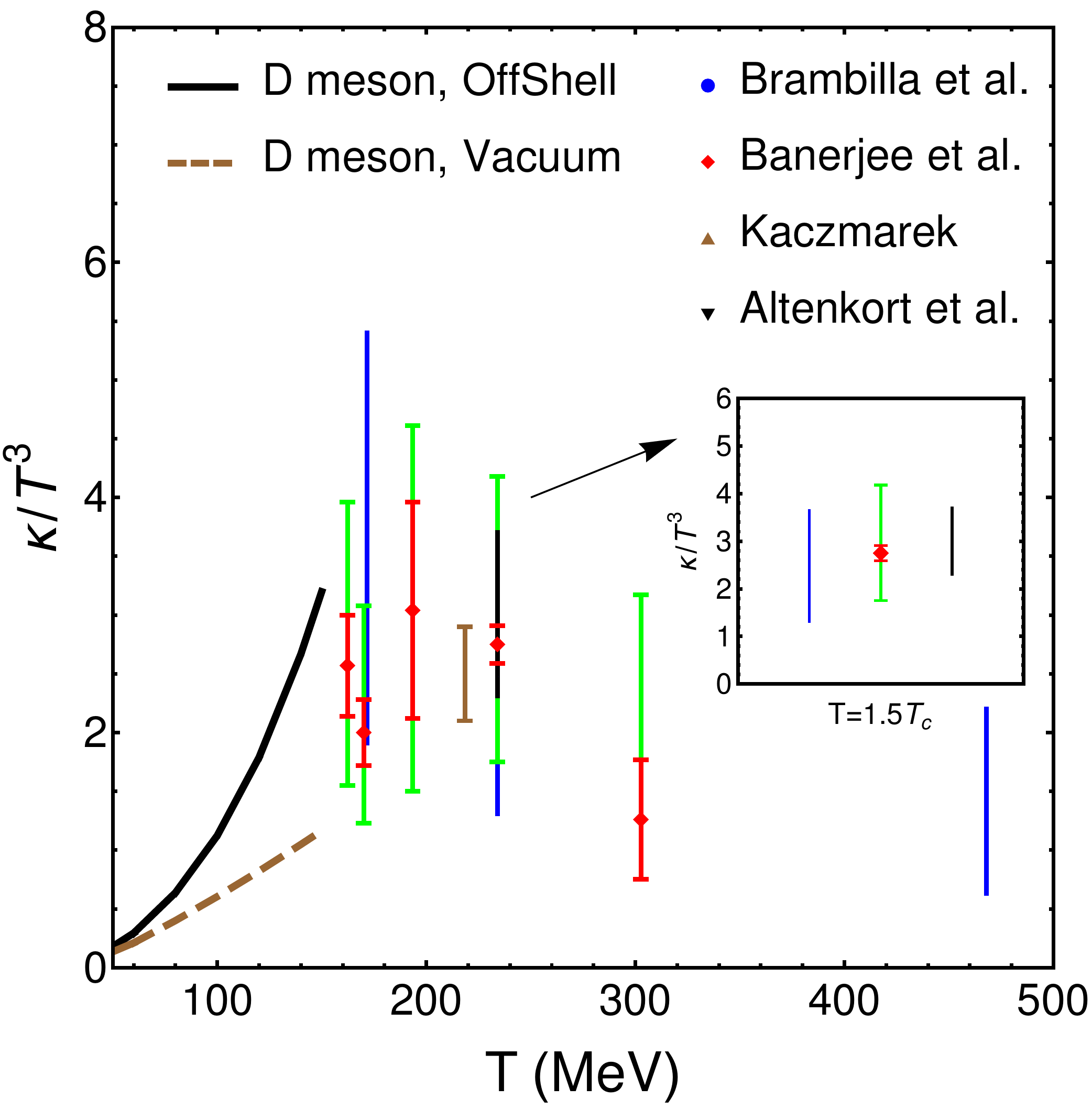}
  \caption{Left panel: Spatial diffusion coefficient normalized by the thermal wavelength around $T_c$. Right panel: Momentum diffusion coefficient $\kappa/T^3=2B_0/T^3$ around $T_c$.}
  \label{fig:transport-latticecomp}
\end{figure}

In the left panel of Fig.~\ref{fig:transport-latticecomp}, we show the spatial diffusion coefficient as defined in Eq.~(\ref{eq:transport-Dscoeff}). In the right panel, we present the momentum diffusion coefficient $\kappa$, as it is usually defined in the \gls{lqcd} community. This coefficient is related to $B_0$ in the static limit as,
\begin{equation}
\kappa(T) = 2 B_0 ( \vec{k}\rightarrow 0;T)\ . 
\end{equation}
This coefficient is not independent of $D_s$ as $\kappa=4\pi T^3/(2\pi T D_s)$. Nevertheless, we provide the results of $\kappa/T^3$ to stress the plausible matching, where a possible maximum happens at the crossover temperature.

Details of the different \gls{lqcd} calculations can be found in their corresponding publications. All of them are characterized by the use of the quenched approximation ($\textrm{SU}(3)$ pure glue plasma) and with different ranges of temperature. References~\cite{Kaczmarek:2014jga,Francis:2015daa,Altenkort:2020fgs} only provide results for a characteristic temperature of $1.5T_c$. Except for the calculation in~\cite{Banerjee:2011ra}, all the results take the lattice continuum limit. The majority of the calculations use a multilevel update to reduce noise~\cite{Banerjee:2011ra,Kaczmarek:2014jga,Francis:2015daa,Brambilla:2020siz}, except for the most recent~\cite{Altenkort:2020fgs}, which employs gradient flow.

We observe a very good matching around $T_c$ between our results, the \gls{lqcd} data, and the result from a Bayesian analysis~\cite{Ke:2018tsh}, especially for the case with thermal and off-shell effects included. This is better seen for the $\kappa/T^3$ coefficient. 

We now comment on the comparison of the ``Vacuum'' approximation to the previous calculation in Ref.~\cite{Tolos:2013kva} using vacuum amplitudes. The authors of that work reported a similar diffusion coefficient with slightly smaller values at high temperatures, for instance at $T=150$~MeV found the value $2\pi TD_s \simeq 8$ , while here we obtain $2\pi TD_s \simeq 12$. In turn, the $A$, $B_0$, and $B_1$ coefficients were systematically larger in~\cite{Tolos:2013kva}. The differences come from several improvements with respect to that work. The \glspl{lec} of the effective Lagrangian are fixed here thanks to recent \gls{lqcd} calculations~\cite{Guo:2018tjx}, while in~\cite{Tolos:2013kva} the authors followed a less rigorous procedure of fixing the \glspl{lec} by matching the mass and width of the $D_0^*(2300)$ resonance (called $D_0(2400)$ in~\cite{Tolos:2013kva}), whose properties reported by the \gls{pdg}, in turn, have changed since then. Furthermore, here we adopt a full consistent coupled-channel approach, while in~\cite{Tolos:2013kva} this was done only partially. For example, the channels involving $D_s$ were not considered. 
 
We should finally mention that for $T>T_c$ we have only shown the results coming from \gls{lqcd} and Bayesian calculations, but there exist many theoretical calculations of these coefficients using different models or effective approaches~\cite{Moore:2004tg,vanHees:2004gq,vanHees:2007me,Das:2010tj,He:2011yi,Mazumder:2011nj,Das:2012ck,Berrehrah:2013mua,Liu:2018syc,Berrehrah:2014kba,Berrehrah:2014tva}. 

\subsection{Results for $\bar{B}$-meson transport coefficients}
\label{subsec:transport-results-b}

Finally, we report our results of the bottom transport coefficients.

\begin{figure}[t!]
  \centering
\includegraphics[width=0.45\textwidth]{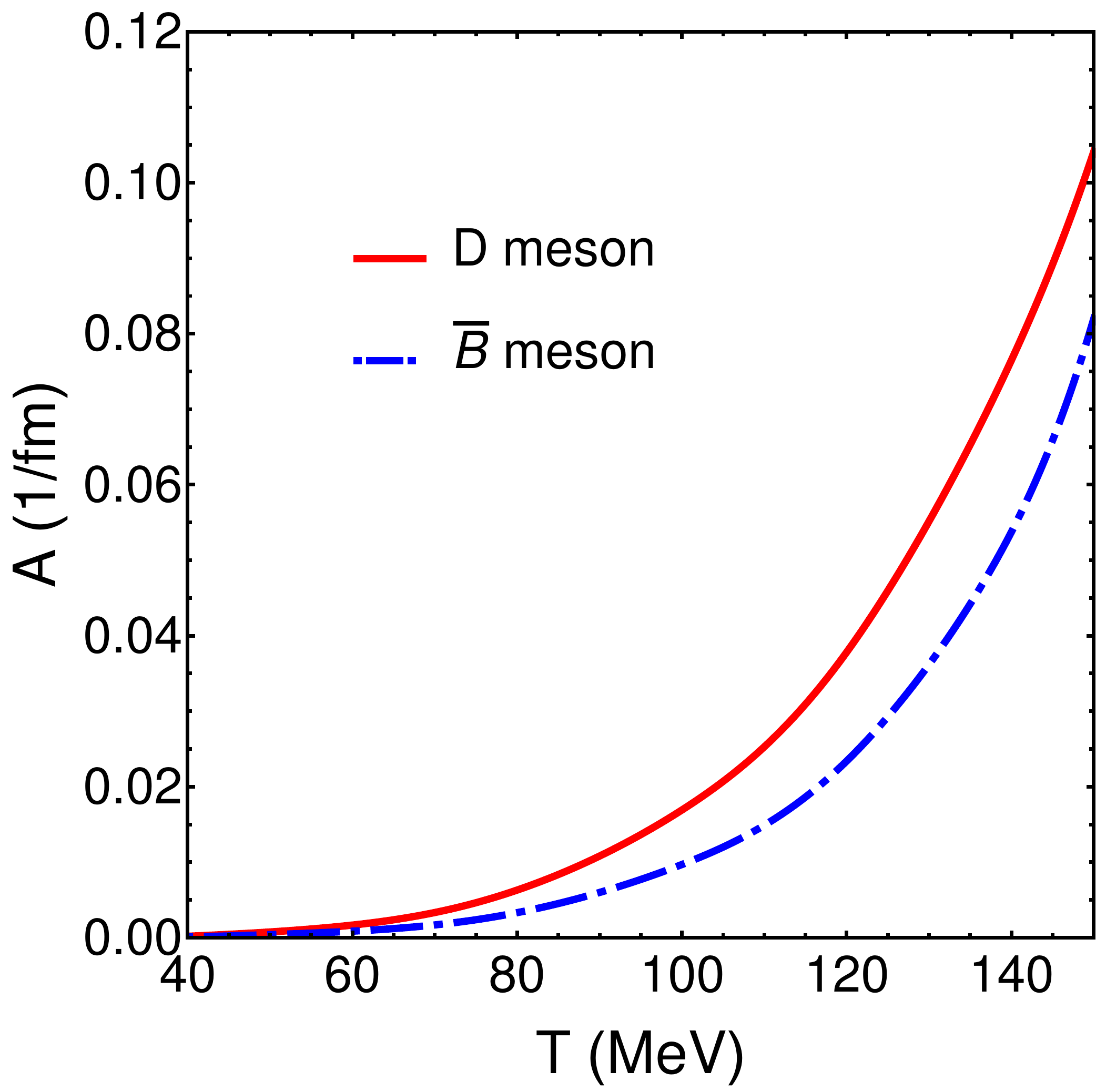}\hspace{0.5cm}
\includegraphics[width=0.45\textwidth]{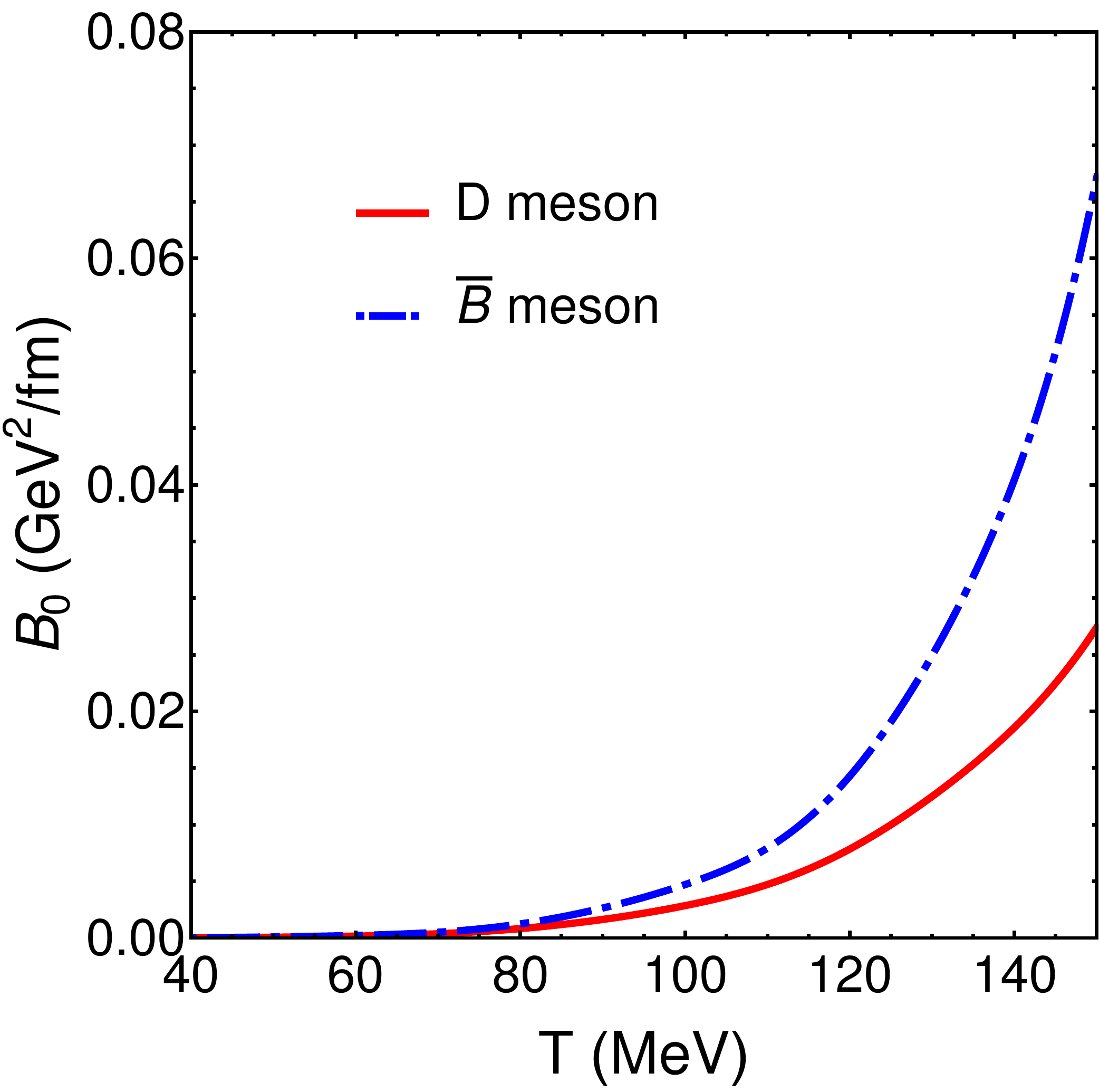}
\caption{Transport coefficients of the $\bar{B}$ meson in the static limit $\vec{k}\rightarrow 0$ (where $B_1=B_0$), compared to the results for the $D$ meson for the calculation that includes all the thermal and off-shell effects (denoted as ``OffShell'' in Fig.~\ref{fig:transport-transportcomp}).}
  \label{fig:transport-transportcomp-b}
\end{figure}

In Fig.~\ref{fig:transport-transportcomp-b} the temperature dependence of the drag coefficient $\hat{A}(k_0,\vec{k}\,;T)$ (left panel) and the momentum diffusion coefficient $\hat{B}_0(k_0,\vec{k}\,;T)$ (right panel) for the $\bar{B}$ meson is compared with that of the corresponding coefficients for the $D$ meson. We have used our most complete derivation for the transport coefficients, incorporating all the thermal and off-shell effects, that is that of Eqs.~(\ref{eq:transport-hatA}) and (\ref{eq:transport-hatB0}), with the average definition in Eq.~(\ref{eq:transport-rateoff}). The $D$ meson coefficients (red solid lines) in Fig.~\ref{fig:transport-transportcomp-b} are thus the same as those denoted as ``OffShell'' (black solid lines) in Fig.~\ref{fig:transport-transportcomp} and are reproduced here to facilitate the comparison between the bottom and charm sectors.  As in Fig.~\ref{fig:transport-transportcomp}, we have taken the static limit, and we have numerically checked that the equality $\hat{B}_0=\hat{B}_1$ is also satisfied in the bottom sector.  


Our results show that the increase of the drag force coefficient (left panel of Fig.~\ref{fig:transport-transportcomp-b}) is less abrupt for the $\bar{B}$ meson than for the $D$ meson, while the momentum diffusion coefficient evolves similarly for both heavy hadrons up to temperatures $\sim 100$~MeV, and then it increases faster for the bottomed meson.
Similar trends for $D$- and $\bar{B}$-meson transport coefficients were found in~\cite{Tolos:2016slr} using vacuum amplitudes, with the similarity between the momentum diffusion coefficients extending to larger temperatures. However, we find here larger values for the drag force and smaller values for the momentum diffusion coefficient, in comparison to those in Ref.~\cite{Tolos:2016slr}, in both heavy sectors.

The normalized off-shell spatial diffusion coefficient $2\pi TD_s$ for the $\bar{B}$ meson as a function of the temperature of the hadronic medium up to $150$~MeV, obtained using Eq.~(\ref{eq:transport-Dscoeff}), is displayed in Fig.~\ref{fig:transport-comp-b}. It is plotted together with the ``OffShell'' result for the $D$ meson, already shown in the left panel of Fig.~\ref{fig:transport-Dscomp} above. We see that the $\bar{B}$-meson spatial diffusion coefficient is about half the value of the corresponding one for the $D$ meson at $T=150$~MeV, in agreement with the results in~\cite{Tolos:2016slr}.

\subsubsection{Comparison with other approaches}

\begin{figure}[t!]
  \centering
  \includegraphics[width=0.45\textwidth]{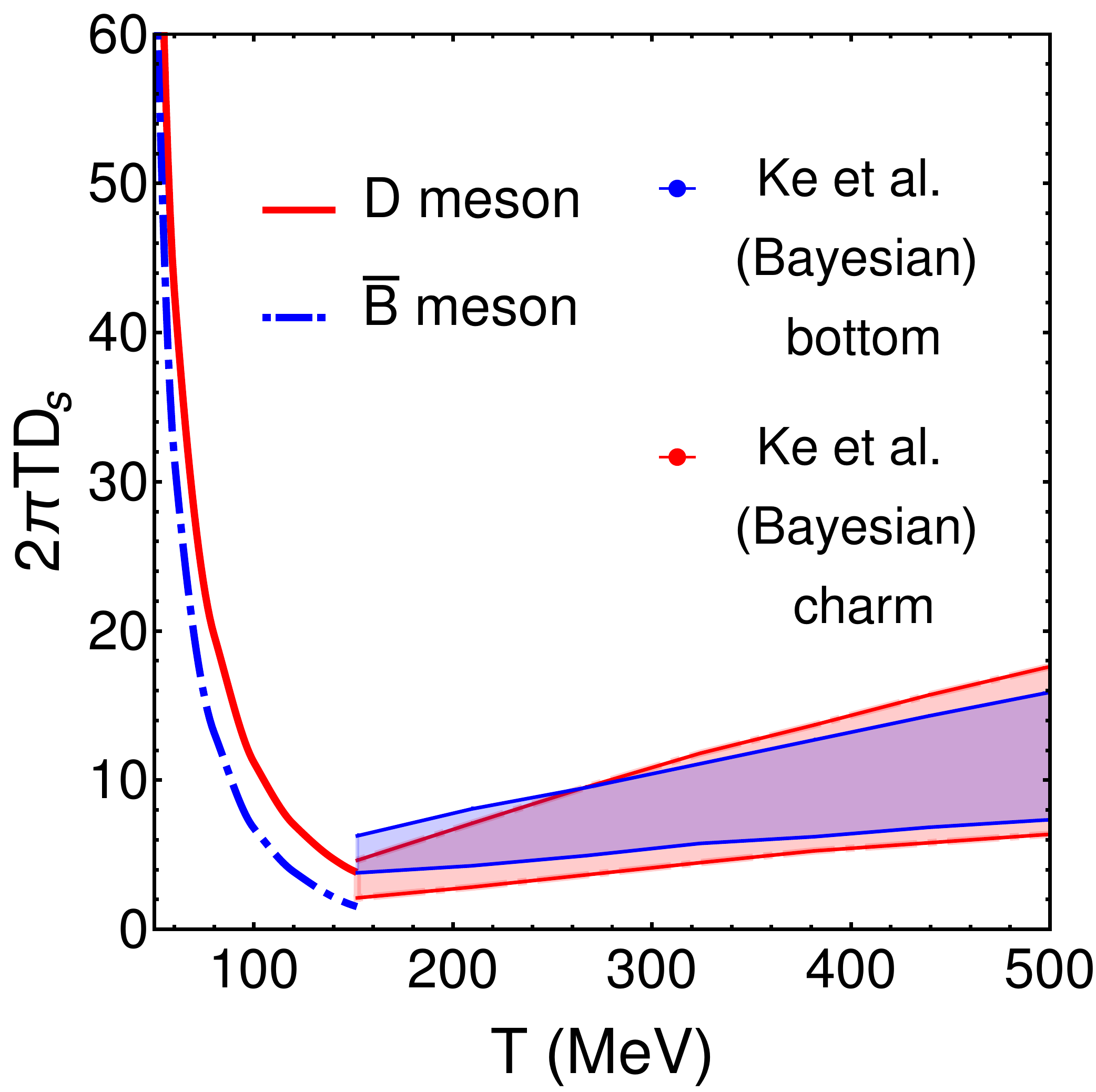}\hspace{0.5cm}
   \includegraphics[width=0.45\textwidth]{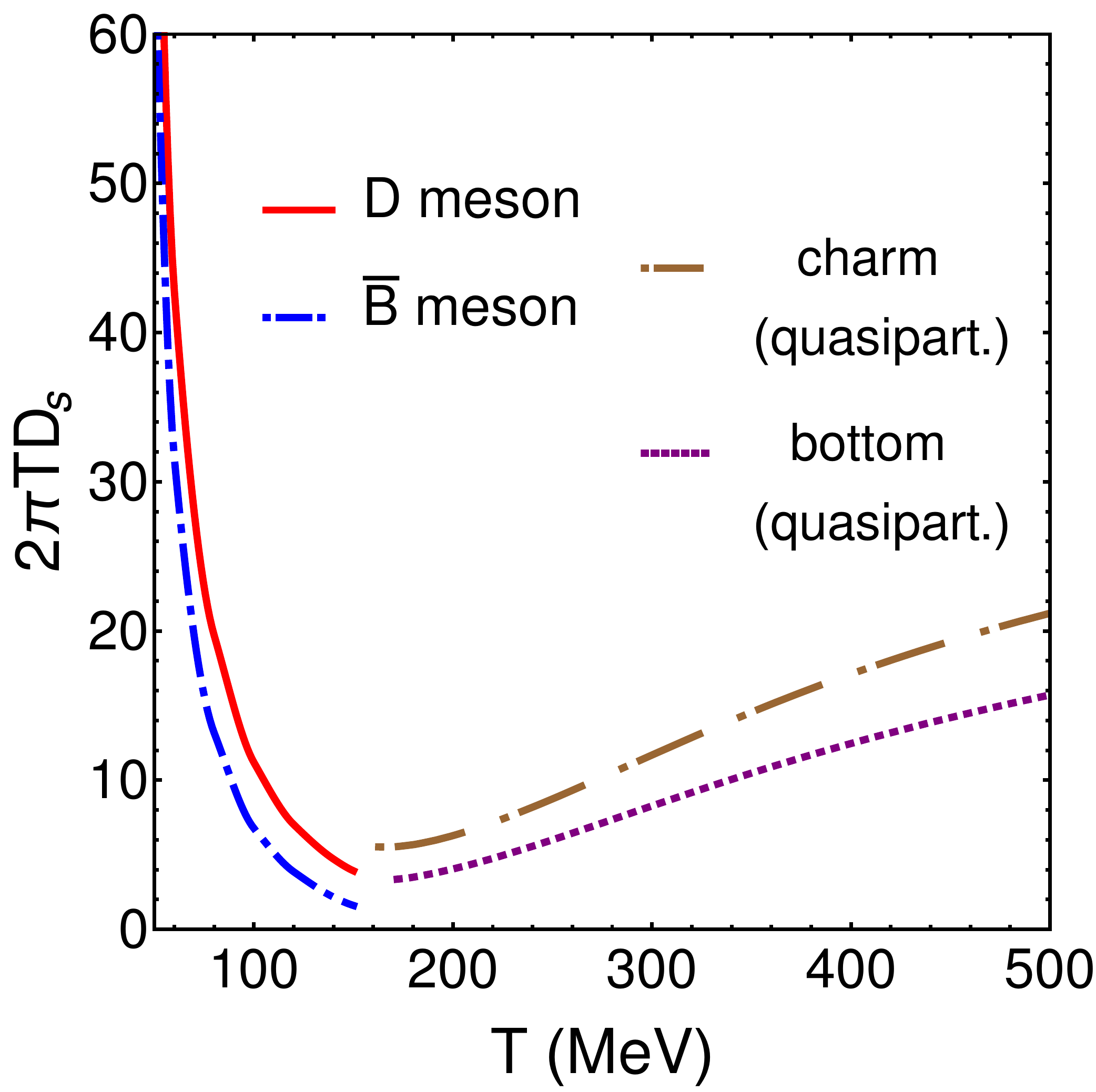}
  \caption{Off-shell spatial diffusion coefficient of the $\bar{B}$ meson (normalized by the thermal wavelength) around $T_c$, together with the results for the $D$ meson, and compared to the calculations above $T_c$ from the Bayesian calculation of~\cite{Ke:2018tsh} (left panel), and from the quasiparticle model of Ref.~\cite{Das:2016llg} (right panel).}
  \label{fig:transport-comp-b}
\end{figure}

In Fig.~\ref{fig:transport-comp-b} we compare our results for the charm and bottom momentum diffusion parameter in the hadronic phase, with two different approaches for the calculation of this transport coefficient in the deconfined phase, above $T\approx 150$~MeV. In the left panel of this figure, we show the $95\%$ credibility region obtained from the Bayesian analysis of \glspl{hic} data of Ref.~\cite{Ke:2018tsh}, for both the charm- (red shaded area) and the bottom-quark (blue shaded area) spatial diffusion coefficient. While in the charm case there is a smooth connection between the results of our thermal effective approach and those of~\cite{Ke:2018tsh}, with a minimum around $T_c$, in the bottom case the matching of these two different ways to obtain $2\pi TD_s$ is less good. Contrary to what we have found in the hadronic phase, the results of the Bayesian analysis lead to larger values of the spatial diffusion coefficient for the $b$ quark than for the $c$ quark just above $T_c$. However, one can see that the width of the shaded areas in the left panel of Fig.~\ref{fig:transport-comp-b} is quite large due to the large uncertainties of the procedure to extract this transport coefficient from \glspl{hic}. At larger temperatures, the results of the Bayesian analysis for charm and bottom basically overlap with each other. 

In the right panel of Fig.~\ref{fig:transport-comp-b} we compare our results for the $D$- and $\bar{B}$-meson diffusion coefficients with the results of the quasiparticle model of Refs.~\cite{Das:2012ck,Berrehrah:2013mua} for the $c$ and $b$ quarks in the \gls{qgp} phase~\cite{Das:2016llg}. This model predicts smaller values for the bottom $D_s$ coefficient compared to the charm result above $T_c$. Although there is not a completely continuous matching with our effective theory results around $T_c$, where one expects further effects coming from a mixed phase of hadronic and deconfined matter, both approaches expect a similar variation of the spatial diffusion coefficient when moving from the charm to the bottom sector.

\chapter{Conclusions}
\label{ch:conclusion}

The purpose of this dissertation has been twofold. On the one hand, the production of heavy hadrons in $B$-factories and heavy-ion colliders has increased exceptionally in the last decades, giving rise to an unprecedented amount of new quarkonia states and heavy flavored hadrons. Many of these states may categorize as exotic composite states, but there is still a lack of clear comprehension of their internal structure (hadronic molecular states, compact multiquark systems, or a mixture of both). The study of exotic hadrons, and of heavy exotics in particular, is currently one of the most active areas of research in hadron physics, both from the experimental and theoretical sides. This thesis has thus been dedicated to the theoretical study of hadrons with charm and bottom quarks that may qualify as molecular states.

On the other hand, the physics of strongly interacting matter under extreme conditions, both at nonzero temperature and baryon density, is also a very challenging field. Understanding the phase diagram of \gls{qcd} requires the combined effort of different communities as diverse as those working on nuclear and particle physics, astrophysics, and general relativity.
In the high-temperature and vanishing baryon density regime, which corresponds to matter filling the entire universe shortly after the Big Bang, heavy-ion collisions at the \gls{rhic} and the \gls{lhc} provide the conditions to create a hot phase of deconfined quarks and gluons. Heavy flavor hadrons are exceptional probes for the properties of the \gls{qgp}, but there are still open questions in this sector that prevent experimentalists and theorists from correctly interpreting the experimental data. The work presented in this thesis has been directed towards understanding the modification of the properties of open heavy-flavor mesons in a hot medium, using an effective theory to describe the interaction of the heavy mesons with the surrounding thermal bath.

In Chapter~\ref{ch:intro} we have reviewed a few aspects of the quark model and the theory of \gls{qcd}, to then motivate the study of exotic hadrons containing heavy quarks. We have also given a brief overview of the present status of strongly interacting matter at extreme conditions of temperature and baryon density and discussed the relevance of heavy-flavor mesons as hard probes of the \gls{qgp} phase formed in \glspl{hic}.

Chapter~\ref{ch:exoticsinfreespace} has been devoted to investigating the properties in vacuum conditions (free space) of exotic hadrons with open heavy flavor in the baryon and meson sectors. Our methodology is based on the use of effective Lagrangians that must be consistent with the symmetries of the strongly interacting system under study.
Moreover, we have unitarized the scattering amplitudes by means of the coupled-channel Bethe-Salpeter equation, paying special attention to the regularization of the two-body propagators. With this strategy, we have been able to dynamically generate states from the interaction between the particles that are the degrees of freedom of our effective theory (baryons and/or mesons). These states have a molecular structure within our approach. In the first section of this chapter, we have described the chiral and heavy-quark symmetries of \gls{qcd}, as well as reviewed some effective theories widely used in the field that are also the basis of the studies that have been performed in the rest of the chapter. Some properties of the scattering amplitudes, in particular unitarity and analyticity in the complex-energy plane, have also been summarized.

The rest of the chapter is divided into two parts.
In the first one (Section~\ref{sec:free-mb}), we have studied the interaction of the low-lying pseudoscalar mesons with the ground-state baryons in the charm $+1$, strangeness $-2$ and isospin $0$ sector, employing a $t$-channel vector meson exchange model. 
The resulting unitarized amplitudes for the scattering of pseudoscalar mesons with baryons show the presence of two resonances, having energies and widths very similar to some of the $\Omega_c^{*0}$ states discovered by the LHCb collaboration in 2017. By exploring the parameter space of our model, we have found several cases that can reproduce the mass and width of the $\Omega_c(3050)^{0}$ and the $\Omega_c(3090)^{0}$ resonances. Our findings allow us to conclude that two of the five $\Omega_c^{*0}$ states observed by the LHCb collaboration could have a meson-baryon molecular origin. 
As our model for the scattering of pseudoscalar mesons with baryons in $s$-wave generates resonances with spin-parity $J^P=1/2^-$, we would anticipate these to be the quantum numbers for these $\Omega_c^{*0}$ states. 
In contrast, some quark model calculations establish either $3/2^-$ or $5/2^-$ for the spin-parity of some of these states. An experimental determination of the spin-parity of the $\Omega_c^{*0}$ states observed at LHCb would be extremely valuable for having a better understanding of their nature. Moreover, further theoretical studies about the molecular interpretation of baryons in this sector, including additional components to the ones considered here, can bring light to this problem. The chapter finalizes with an extension of our study to the bottom sector, which gives rise to several $\Omega_b^{*-}$ resonances that have a molecular meson-baryon structure in the energy region $6400-6800$~MeV, where, despite the low statistics, some structures are visible in the invariant $K^-\Xi_b^0$ mass spectrum from the LHCb experiment. Experimental confirmation of the existence of $\Omega_b^{*-}$ states in this energy region at the future LHC facility, upgraded with increased luminosity, would permit a comparison of their properties with those predicted by quark or hadron molecular models, hence progressing towards establishing their nature.

In the second part (Section~\ref{sec:free-mm}), we have analyzed the interactions of pseudoscalar and vector open heavy-flavor mesons ($D^{(*)}$, $D_s^{(*)}$, $\bar{B}^{(*)}$, $\bar{B}_s^{(*)}$) with light mesons ($\pi$, $K$, $\bar{K}$, $\eta$) using an effective field theory based on chiral and heavy-quark symmetries in vacuum. In the $J^P=0^+$ case, we have described the $D_0^*(2300)$ state as dynamically generated with a double-pole structure, while the $D_{s0}^*(2317)$ has been identified with a molecular bound state. In $J^P=1^+$, we have paid attention to the $D_1(2430)$ resonance (also with a double-pole structure) and the $D_{s1}(2460)$. The parallelism between the $J^P=0^+$ and $J^P=1^+$ sectors is due to \gls{hqss}, which is implemented at \gls{lo} in the Lagrangian, and only broken due to the explicit use of the heavy-meson vacuum masses. We have also extended our calculations to the bottom sector and found the bottomed counterparts of the $J^P=0^+,1^+$ charmed states. In this case, \gls{hqfs} is responsible for the similarities between the charm and bottom sectors.

In Chapter~\ref{ch:hot-medium}, we have extended our analysis of the scattering of open heavy-flavor mesons off the light mesons to finite temperatures and obtained the properties of the former in a thermal medium up to $T=150$~MeV. 
The medium modification of the heavy-meson propagator is calculated in a self-consistent way, in which the heavy-meson self-energy is corrected by the interactions with the medium contained in the unitarized amplitudes, which in turn, are computed by solving the Bethe-Salpeter equation with thermal two-meson propagators.
With this methodology, we have obtained the heavy-meson spectral functions. From these, we have extracted the thermal dependence of the masses and the decay widths of the ground-state $D^{(*)}$, $D_s^{(*)}$ mesons, as well as of the bottomed $\bar{B}^{(*)}$, $\bar{B}_s^{(*)}$, and also of those of the dynamically generated states. We have observed a generic downshift of the thermal masses with the temperature, as large as a few tens of MeV at $T=150$~MeV in a pionic bath, while the decay widths increase with temperature up to values of some tens of MeV at $T=150$~MeV. 
From our results, we have not observed a clear tendency to chiral degeneracy of the $J^P=0^-$ ($1^-$) and the $J^P=0^+$ ($1^+$) partners. However, our calculations were limited by the low-temperature application of the hadron effective theory and such degeneracy might occur at higher temperatures, $T > T_\chi$. One of our main results is that the chiral partner of the $D$ meson, the $D_0^*(2300)$, has a double-pole structure in the complex-energy plane, and it is unclear at this point how the chiral symmetry restoration should be realized for this type of states.
We have also seen that the addition of kaons in the meson bath, along with the pions, results in a slight modification of the heavy-meson masses, and a significant increase of the widths. Nevertheless, the larger contribution to the thermal effects on the properties of the heavy mesons comes from the thermal pions, as they are the most abundant mesons in the bath.


In Chapter~\ref{ch:lattice}, we have computed, for the first time, Euclidean correlators for the charmed $D$ and $D_s$ mesons from their corresponding thermal spectral functions obtained within the finite-temperature effective theory approach developed in this thesis. Our results have been compared with those obtained in \gls{lqcd} simulations at unphysical masses. By considering the full energy-dependent spectral functions, including the continuum of scattering states that is present in the lattice correlators in addition to the ground state, we have found that our calculations of the ratio of the Euclidean and the reconstructed correlators lie within the error bars of the lattice data for temperatures well below the deconfinement transition temperature. At larger temperatures, the ratios deviate significantly from the \gls{lqcd} predictions. Several factors may contribute to these discrepancies. First, we have not considered in our spectral functions the presence of bound excited states that are, however, inherent in the \gls{lqcd} simulations. Second, finite-volume and cut-off effects have not been implemented in our calculations. Moreover, the thermal modification of the charm ground-state meson properties induced by the kaons in the medium may be more relevant in the lattice setup, with unphysically large meson masses, due to the reduction of the mass gap between the kaon and the pion mass.



Finally, in Chapter~\ref{ch:transport}, we have extended the kinetic theory description of heavy mesons at low energy to include the medium (thermal) effects and the spectral properties of the open-charm and open-beauty states. The derivation of the off-shell Boltzmann and Fokker-Planck equations from the heavy-meson effective theory was essential to then calculate the ``off-shell'' heavy-meson transport coefficients, implementing, for the first time, a consistent formulation between the in-medium interactions with light mesons and the kinetic approach. In particular, we have obtained the thermal width, the drag force coefficient, the diffusion coefficient in momentum space, and the spatial diffusion coefficient. Due to their large vacuum mass, the thermal corrections to the mass and spectral broadening of $D$ and $\bar{B}$ mesons are relatively small, so the quasiparticle picture has been found to be a good approximation. 
However, we have found that the use of thermal scattering amplitudes causes the appearance of a new kinematic range in the meson--meson interaction, the so-called Landau contribution, whose contribution to the transport coefficients is rather large at moderate temperatures. In fact, at $T=150$ MeV this new contribution is as large as the standard contribution due to the unitary cut. We have checked that our calculations of the transport coefficients close to $T_c$, including these new effects, are consistent with \gls{lqcd} determinations of the momentum and spatial diffusion coefficients, as well as with Bayesian analyses of \glspl{hic} data. 

To summarize, we have provided, on the one hand, a comprehensive description of $\Omega_c^{*0}/\Omega_b^{*-}$ states and open heavy-flavor excited mesons ($D_0^* (2300)$, $D_{s0}^*(2317)$, $D_1(2430)$ and $D_{s1}(2460)$, and their bottomed counterparts) within the molecular picture, using the appropriate effective Lagrangians to describe the hadron--hadron interactions. In this way, we are confident that we have conveyed the ability of unitarized effective hadronic theories to dynamically generate states that may be identified with experimentally observed hadrons and provide an exotic molecular interpretation for their nature. On the other hand, we have developed a new systematic approach to the in-medium modification of heavy hadrons in a hot mesonic medium below the \gls{qcd} transition temperature, in view of the present and future heavy-ion experiments at low baryon densities as well as the forthcoming results from the \gls{lqcd} simulations at finite temperature. Furthermore, as an application, we have calculated transport coefficients that can be used as an input to perform hydrodynamic simulations and improve our understanding of the \gls{qgp} formation in \glspl{hic}.

%
{%
\setstretch{1.1}
\renewcommand{\bibfont}{\normalfont\small}
\setlength{\biblabelsep}{0pt}
\setlength{\bibitemsep}{0.5\baselineskip plus 0.5\baselineskip}
\printbibliography
}
\cleardoublepage

%

\pdfbookmark[0]{Appendices}{Appendices}
\appendix\cleardoublepage
\noappendicestocpagenum
\cleardoublepage %
\addappheadtotoc
\appendixpage
%
\chapter{Coefficients}
\label{appendix-coef}

In this appendix, we present some tables with useful coefficients for the meson--baryon and meson--meson interactions analyzed in Chapter~\ref{ch:exoticsinfreespace}.

\section{Meson-baryon coefficients}
\label{sec:appendix:mb_coeff}

The coefficients $C_{ij}^v$ of the \gls{tvme} interaction kernel of Eq.~(\ref{eq:free-mb-Vij_1}) denote the strength of the interaction between two \gls{meson-baryon} channels $i$ and $j$ mediated by the exchange of a vector meson $v$, in the sector with isospin $I=0$, strangeness $S=-2$ and charm $C=1$. The coefficients in the isospin basis that are different from zero are listed in Table~\ref{tab:coeff_mb_ps} for the \gls{pseudoscalar-baryon} interaction and in Table~\ref{tab:coeff_mb_v} for the case of \gls{vector-baryon} channels. From these, one gets the value of the summation, $C_{ij}=\sum_vC_{ij}^v$, presented in Tables~\ref{tab:free-mb-coeff} and \ref{tab:free-mb-coeff-v}.
We note that the transitions mediated by a charmed vector meson ($D^*$, $D_s^*$) or a $J/\psi$ meson are suppressed by a factor $\kappa_c$ and $\kappa_{c\bar{c}}$, respectively, as described in Section~\ref{sec:free-mb}.

\begin{table}[htbp!]
\setlength{\tabcolsep}{10pt}
\renewcommand{\arraystretch}{1.2}
\centering
\begin{tabular}{c|c|c|c}
 \hline
 $i$ & $j$ & $v$ & $C_{ij}^v$ \\
 \hline
\multirow{5}{*}{$\bar{K}\Xi_c$} & \multirow{3}{*}{$\bar{K}\Xi_c$} & $\rho$ & $\frac32$ \\
               &                & $\omega$ & $\frac12$ \\
               &                & $\phi$ & $-1$     \\ \cline{2-4}
               & $D\Xi$ & $D_s^*$ & $\sqrt{\frac32}\kappa_c$  \\ \cline{2-4}
               & $\bar{D}_s\Omega_{cc}$ & $D^*$ & $-\sqrt{3}\kappa_c$ \\
 \hline
 \multirow{6}{*}{$\bar{K}\Xi_c'$} & \multirow{3}{*}{$\bar{K}\Xi_c'$} & $\rho$ & $\frac32$  \\
              &                 & $\omega$ & $\frac12$  \\
              &                 & $\phi$ & $-1$  \\ \cline{2-4}
              & $D\Xi$          & $D_s^*$ & $\frac{1}{\sqrt{2}}\kappa_c$ \\ \cline{2-4}
              & $\eta\Omega_c$ & $K^*$ & $-\sqrt{6}$ \\  \cline{2-4}
              & $\bar{D}_s\Omega_{cc}$ & $D^*$ & $\kappa_c$ \\
 \hline
 \multirow{5}{*}{$D\Xi$} & \multirow{2}{*}{$D\Xi$} & $\rho$ & $\frac32$ \\
              &                 & $\omega$ & $\frac12$ \\ \cline{2-4}
              & $\eta\Omega_c$ & $D^*$ & $-\frac{1}{\sqrt{3}}\kappa_c$ \\ \cline{2-4}
              & $\eta'\Omega_c$ & $D^*$ & $-\sqrt{\frac23}\kappa_c$  \\ \cline{2-4}
              & $\eta_c\Omega_{c}$ & $D^*$ & $\sqrt{2}\kappa_c$ \\
 \hline
 $\eta\Omega_c$ & $\bar{D}_s\Omega_{cc}$ & $D_s^*$ & $-\sqrt{\frac23}\kappa_c$ \\
 \hline
 $\eta'\Omega_c$ & $\bar{D}_s\Omega_{cc}$ & $D_s^*$ & $\sqrt{\frac13}\kappa_c$ \\
 \hline
 \multirow{3}{*}{$\bar{D}_s\Omega_{cc}$} & \multirow{2}{*}{$\bar{D}_s\Omega_{cc}$} & $\phi$ & $-1$ \\
             &                  & $J/\psi$ & $2 \kappa_{c\bar{c}}$ \\ \cline{2-4}
             & $\eta_c\Omega_c$ & $D_s^*$ & $-\kappa_c$ \\
 \hline
\end{tabular}
 \caption{Coefficients of the \gls{pseudoscalar-baryon} interaction in the sector with $(I,S,C)=(0,-2,1)$ in the isospin basis.}
 \label{tab:coeff_mb_ps}
\end{table}

\begin{table}[t]
\setlength{\tabcolsep}{10pt}
\renewcommand{\arraystretch}{1.2}
\centering
\begin{tabular}{c|c|c|c}
 \hline
 $i$ & $j$ & $v$ & $C_{ij}^v$ \\
 \hline
 \multirow{5}{*}{$D^*\Xi$} & \multirow{2}{*}{$D^*\Xi$} & $\rho$ & $\frac32$ \\
              &                 & $\omega$ & $\frac12$ \\ \cline{2-4}
                & $\bar{K}^*\Xi_c$ & $D_s^*$ & $\sqrt{\frac32}\kappa_c$  \\ \cline{2-4}
              & $\bar{K}^*\Xi_c$          & $D_s^*$ & $\frac{1}{\sqrt{2}}\kappa_c$ \\ \cline{2-4}
             & $\omega\Omega_c$ & $D^*$ & $-\kappa_c$ \\ \cline{2-4}
              & $J/\psi\Omega_{c}$ & $D^*$ & $\sqrt{2}\kappa_c$ \\
 \hline
 \multirow{5}{*}{$\bar{K}^*\Xi_c$} & \multirow{3}{*}{$\bar{K}^*\Xi_c$} & $\rho$ & $\frac32$ \\
               &                & $\omega$ & $\frac12$ \\
               &                & $\phi$ & $-1$     \\ \cline{2-4}
               & $\bar{D}_s^*\Omega_{cc}$ & $D^*$ & $-\sqrt{3}\kappa_c$ \\
 \hline
 \multirow{6}{*}{$\bar{K}^*\Xi_c'$} & \multirow{3}{*}{$\bar{K}^*\Xi_c'$} & $\rho$ & $\frac32$  \\
              &                 & $\omega$ & $\frac12$  \\
              &                 & $\phi$ & $-1$  \\ \cline{2-4}
              & $\omega\Omega_c$ & $K^*$ & $-\sqrt{2}$ \\  \cline{2-4}
              & $\phi\Omega_c$ & $K^*$ & $2$ \\  \cline{2-4}
              & $\bar{D}_s^*\Omega_{cc}$ & $D^*$ & $\kappa_c$ \\
 \hline
 $\phi\Omega_c$ & $\bar{D}_s^*\Omega_{cc}$ & $D_s^*$ & $\kappa_c$ \\
 \hline
 \multirow{3}{*}{$\bar{D}_s^*\Omega_{cc}$} & \multirow{2}{*}{$\bar{D}_s^*\Omega_{cc}$} & $\phi$ & $-1$ \\
             &                  & $J/\psi$ & $2 \kappa_{c\bar{c}}$ \\ \cline{2-4}
             & $J/\psi\Omega_c$ & $D_s^*$ & $-\kappa_c$ \\
 \hline
\end{tabular}
 \caption{Coefficients of the \gls{vector-baryon} interaction in the sector with $(I,S,C)=(0,-2,1)$ in the isospin basis.}
 \label{tab:coeff_mb_v}
\end{table}

\section{Coefficients of the $D\Phi$ interaction in the charge basis}
\label{sec:appendix:mm_coeff}

Here we provide the coefficients of the $D\Phi$ interaction kernel of Eq.~(\ref{eq:free-mm-potential}) in the charge basis. They are listed in Tables~\ref{tab:coeff_Dphi_Qbasis} and~\ref{tab:coeff_Dphi_Qbasis2}.
  
\begin{table}[b!]
\setlength{\tabcolsep}{10pt}
\renewcommand{\arraystretch}{1.2}
\centering
\begin{tabular}{c l | c c c c c}
\hline
 $(S,Q)$  &  Channel & $C_\textrm{LO}^{jk}$ & $C_0^{jk}$ & $C_1^{jk}$ &  $C_\textrm{24}^{jk}$ & $C_{35}^{jk}$  \\  
\hline
 $(-1,-1)$ & $D^0K^-\rightarrow D^0K^-$ & $1$ & $m_K^2$ & $-m_K^2$ & $1$ & $1$ \\
\hline
 $(-1,0)$  & $D^0\bar{K}^0\rightarrow D^0\bar{K}^0$ & $0$ & $m_K^2$ & $0$ & $1$ & $0$ \\
   & $D^0\bar{K}^0\rightarrow D^+K^-$ & $1$ & $0$ & $-m_K^2$ & $0$ & $1$ \\
   & $D^+K^-\rightarrow D^+K^-$ & $0$ & $m_K^2$ & $0$ & $1$ & $0$ \\
\hline
 $(-1,+1)$  & $D^+\bar{K}^0\rightarrow D^+\bar{K}^0$ & $1$ & $m_K^2$ & $-m_K^2$ & $1$ & $1$ \\
\hline
 $(0,-1)$ & $D^0\pi^-\rightarrow D^0\pi^-$ & $1$ & $m_\pi^2$ & $-m_\pi^2$ & $1$ & $1$ \\
\hline
 $(0,0)$   & $D^0\pi^0\rightarrow D^0\pi^0$ & $0$ & $m_\pi^2$ & $-m_\pi^2$ & $1$ & $1$ \\
   & $D^0\pi^0\rightarrow D^+\pi^-$ & $-\sqrt{2}$ & $0$ & $0$ & $0$ & $0$ \\
   & $D^0\pi^0\rightarrow D_s^+K^-$ & $-\frac{1}{\sqrt{2}}$ & $0$ & $-\frac{1}{2\sqrt{2}}(m_K^2+m_\pi^2)$ & $0$ & $\frac{1}{\sqrt{2}}$ \\
   & $D^0\pi^0\rightarrow D^0\eta$ & $0$ & $0$ & $-\frac{1}{\sqrt{3}}m_\pi^2$ & $0$ & $\frac{1}{\sqrt{3}}$ \\ 
   & $D^+\pi^-\rightarrow D^+\pi^-$ & $-1$ & $m_\pi^2$ & $-m_\pi^2$ & $1$ & $1$ \\
   & $D^+\pi^-\rightarrow D_s^+K^-$ & $-1$ & $0$ & $-\frac{1}{2}(m_K^2+m_\pi^2)$ & $0$ & $1$ \\ 
   & $D^+\pi^-\rightarrow D^0\eta$ & $0$ & $0$ & $-\sqrt{\frac{2}{3}}m_\pi^2$ & $0$ & $\sqrt{\frac{2}{3}}$ \\    
\hline
\end{tabular}
\centering
\caption{Coefficients $C_i^{jk}$ of the LO and NLO terms of the potential for $D\phi\rightarrow D\phi$ in the sectors with charm $C=1$, strangeness $S$ and charge $Q$ in the charge basis. }
\label{tab:coeff_Dphi_Qbasis}
\end{table}
\newpage
\begin{table}[t!]
\setlength{\tabcolsep}{10pt}
\renewcommand{\arraystretch}{1.2}
\centering
\begin{tabular}{c l | c c c c c}
\hline
 $(S,Q)$  &  Channel & $C_\textrm{LO}^{jk}$ & $C_0^{jk}$ & $C_1^{jk}$ &  $C_\textrm{24}^{jk}$ & $C_{35}^{jk}$  \\  
\hline
 $(0,0)$   & $D_s^+K^-\rightarrow D_s^+K^-$ & $-1$ & $m_K^2$ & $-m_K^2$ & $1$ & $1$ \\ 
   & $D_s^+K^-\rightarrow D^0\eta$ & $-\sqrt{\frac{3}{2}}$ & $0$ & $\frac{1}{2\sqrt{6}}(5m_K^2-3m_\pi^2)$ & $0$ & $-\frac{1}{\sqrt{6}}$ \\
   & $D^0\eta\rightarrow D^0\eta$ & $0$ & $m_\eta^2$ & $-\frac{1}{3}m_\pi^2$ & $1$ & $\frac{1}{3}$ \\ 
\hline
$(0,+1)$   & $D^0\pi^+\rightarrow D^0\pi^+$ & $-1$ & $m_\pi^2$& $-m_\pi^2$ & $1$ & $1$ \\
   & $D^0\pi^+\rightarrow D^+\pi^0$ & $\sqrt{2}$ & $0$ & $0$ & $0$ & $0$ \\
   & $D^0\pi^+\rightarrow D_s^+\bar{K}^0$ & $-1$ & $0$ & $-\frac{1}{2}(m_K^2+m_\pi^2)$ & $0$ & $1$ \\
   & $D^0\pi^+\rightarrow D^+\eta$ & $0$ & $0$ & $-\sqrt{\frac{2}{3}}m_\pi^2$ & $0$ & $\sqrt{\frac{2}{3}}$ \\
   & $D^+\pi^0\rightarrow D^+\pi^0$ & $0$ & $m_\pi^2$ & $-m_\pi^2$ & $1$ & $1$ \\
   & $D^+\pi^0\rightarrow D_s^+\bar{K}^0$ & $\frac{1}{\sqrt{2}}$ & $0$ & $\frac{1}{2\sqrt{2}}(m_K^2+m_\pi^2)$ & $0$ & $-\frac{1}{\sqrt{2}}$ \\
   & $D^+\pi^0\rightarrow D^+\eta$ & $0$ & $0$ & $\frac{1}{\sqrt{3}}m_\pi^2$ & $0$ & $-\frac{1}{\sqrt{3}}$ \\
   & $D_s^+\bar{K}^0\rightarrow D_s^+\bar{K}^0$ & $-1$ & $m_K^2$ & $-m_K^2$ & $1$ & $1$ \\
   & $D_s^+\bar{K}^0\rightarrow D^+\eta$ & $-\sqrt{\frac{3}{2}}$ & $0$ & $\frac{1}{2\sqrt{6}}(5m_K^2-3m_\pi^2)$ & $0$ & $-\frac{1}{\sqrt{6}}$ \\
   & $D^+\eta\rightarrow D^+\eta$ & $0$ & $m_\eta^2$ & $-\frac{1}{3}m_\pi^2$ & $1$ & $\frac{1}{3}$ \\
\hline
 $(0,+2)$  & $D^+\pi^+\rightarrow D^+\pi^+$ & $1$ & $m_\pi^2$ & $-m_\pi^2$ & $1$ & $1$ \\
\hline
$(1,0)$   & $D^0K^0\rightarrow D^0K^0$ & $0$ & $m_K^2$ & $0$ & $1$ & $0$ \\
   & $D^0K^0\rightarrow D_s^+\pi^-$ & $1$ & $0$ & $-\frac{1}{2}(m_K^2+m_\pi^2)$ & $0$ & $1$ \\
   & $D_s^+\pi^-\rightarrow D_s^+\pi^-$ & $0$ & $m_\pi^2$ & $0$ & $1$ & $0$ \\
\hline
$(1,+1)$   & $D_s^+\pi^0\rightarrow D_s^+\pi^0$ & $0$ & $m_\pi^2$ & $0$ & $1$ & $0$ \\
   & $D_s^+\pi^0\rightarrow D^0K^+$ & $\frac{1}{\sqrt{2}}$ & $0$ & $-\frac{1}{2\sqrt{2}}(m_K^2+m_\pi^2)$ & $0$ & $\frac{1}{\sqrt{2}}$ \\
   & $D_s^+\pi^0\rightarrow D^+K^0$ & $-\frac{1}{\sqrt{2}}$ & $0$ & $\frac{1}{2\sqrt{2}}(m_K^2+m_\pi^2)$ & $0$ & $-\frac{1}{\sqrt{2}}$ \\
   & $D_s^+\pi^0\rightarrow D_s^+\eta$ & $0$ & $0$ & $0$ & $0$ & $0$ \\
   & $D^0K^+\rightarrow D^0K^+$ & $-1$ & $m_K^2$ & $-m_K^2$ & $1$ & $1$ \\
   & $D^0K^+\rightarrow D^+K^0$ & $-1$ & $0$ & $-m_K^2$ & $0$ & $1$ \\
\hline
 $(1,+1)$  & $D^0K^+\rightarrow D_s^+\eta$ & $\sqrt{\frac{3}{2}}$ & $0$ & $\frac{1}{2\sqrt{6}}(5m_K^2-3m_\pi^2)$ & $0$ & $-\frac{1}{\sqrt{6}}$ \\
   & $D^+K^0\rightarrow D^+K^0$ & $-1$ & $m_K^2$ & $-m_K^2$ & $1$ & $1$ \\
   & $D^+K^0\rightarrow D_s^+\eta$ & $\sqrt{\frac{3}{2}}$ & $0$ & $\frac{1}{2\sqrt{6}}(5m_K^2-3m_\pi^2)$ & $0$ & $-\frac{1}{\sqrt{6}}$ \\
   & $D_s^+\eta\rightarrow D_s^+\eta$ & $0$ & $m_\eta^2$ & $-\frac{4}{3}(2m_K^2-m_\pi^2)$ & $1$ & $\frac{4}{3}$ \\
\hline
 $(1,+2)$   & $D_s^+\pi^+\rightarrow D_s^+\pi^+$ & $0$  & $m_\pi^2$ & $0$ & $1$ & $0$\\
  & $D_s^+\pi^+\rightarrow D^+K^+$ & $1$ & $0$ & $-\frac{1}{2}(m_K^2+m_\pi^2)$ & $0$ & $1$ \\
  & $D^+K^+\rightarrow D^+K^+$ & $0$ & $m_K^2$ & $0$ & $1$ & $0$ \\
\hline
 $(2,+1)$  & $D_s^+K^0\rightarrow D_s^+K^0$ & $1$ & $m_K^2$ & $-m_K^2$ & $1$ & $1$ \\
\hline
 $(2,+2)$  & $D_s^+K^+\rightarrow D_s^+K^+$ & $1$ & $m_K^2$ & $-m_K^2$ & $1$ & $1$ \\
\hline
\end{tabular}
\centering
\caption{Continuation of Table~\ref{tab:coeff_Dphi_Qbasis}. }
\label{tab:coeff_Dphi_Qbasis2}
\end{table}

We also give the expressions relating the coefficients in the isospin basis, $C^{(S,I)}_{i\rightarrow j}$, with those in the charge basis, $\tilde{C}^{(S,Q)}_{i\rightarrow j}$, obtained from the Clebsch-Gordan coefficients coupling the particle isospins $i_1$ and $i_2$ to total isospin $I$,
\begin{equation}   \nonumber
\begin{aligned}[b]
&C_{D\bar{K}\rightarrow D\bar{K}}^{(-1,0)}&=&\,\frac{1}{2}\tilde{C}_{D^+K^-\rightarrow D^+K^-}^{(-1,0)}+\frac{1}{2}\tilde{C}_{D^0\bar{K}^0\rightarrow D^0\bar{K}^0}^{(-1,0)}-\tilde{C}_{D^0\bar{K}^0\rightarrow D^+K^-}^{(-1,0)} \\
&C_{D\bar{K}\rightarrow D\bar{K}}^{(-1,1)}&=&\,\frac{1}{2}\tilde{C}_{D^+K^-\rightarrow D^+K^-}^{(-1,0)}+\frac{1}{2}\tilde{C}_{D^0\bar{K}^0\rightarrow D^0\bar{K}^0}^{(-1,0)}+\tilde{C}_{D^0\bar{K}^0\rightarrow D^+K^-}^{(-1,0)} \\ 
& &=&\,\tilde{C}_{D^0K^-\rightarrow D^0K^-}^{(-1,-1)}=\tilde{C}_{D^+\bar{K}^0\rightarrow D^+\bar{K}^0}^{(-1,+1)} \\
\end{aligned}
\end{equation}
\begin{equation}\nonumber
\begin{aligned}[b]
&C_{D\pi\rightarrow D\pi}^{(0,1/2)}&=&\,\frac{2}{3}\tilde{C}_{D^0\pi^+\rightarrow D^0\pi^+}^{(0,+1)}+\frac{1}{3}C_{D^+\pi^0\rightarrow D^+\pi^0}^{(0,+1)}-\frac{2\sqrt{2}}{3}\tilde{C}_{D^+\pi^0\rightarrow D^0\pi^+}^{(0,+1)}\\ 
& &=&\,\frac{1}{3}\tilde{C}_{D^0\pi^0\rightarrow D^0\pi^0}^{(0,0)}+\frac{2}{3}\tilde{C}_{D^+\pi^-\rightarrow D^+\pi^-}^{(0,0)}+\frac{2\sqrt{2}}{3}\tilde{C}_{D^+\pi^0\rightarrow D^0\pi^+}^{(0,0)} \\
&C_{D\pi\rightarrow D\eta}^{(0,1/2)}&=&\,\sqrt{\frac{2}{3}}\tilde{C}_{D^0\pi^+\rightarrow D^+\eta}^{(0,+1)}-\frac{1}{\sqrt{3}}\tilde{C}_{D^+\pi^0\rightarrow D^+\eta}^{(0,+1)} \\
& &=&\,\frac{1}{\sqrt{3}}\tilde{C}_{D^0\pi^0\rightarrow D^0\eta}^{(0,0)}+\sqrt{\frac{2}{3}}\tilde{C}_{D^+\pi^-\rightarrow D^0\eta}^{(0,0)}\\
&C_{D\pi\rightarrow D_s\bar{K}}^{(0,1/2)}&=&\,\sqrt{\frac{2}{3}}\tilde{C}_{D^0\pi^+\rightarrow D_s^+\bar{K}^0}^{(0,+1)}-\frac{1}{\sqrt{3}}\tilde{C}_{D^+\pi^0\rightarrow D_s^+\bar{K}^0}^{(0,+1)} \\
& &=&\,\frac{1}{\sqrt{3}}\tilde{C}_{D^0\pi^0\rightarrow D_s^+K^-}^{(0,0)}+\sqrt{\frac{2}{3}}\tilde{C}_{D^+\pi^-\rightarrow D_s^+K^-}^{(0,0)}\\
&C_{D\eta\rightarrow D\eta}^{(0,1/2)}&=&\,\tilde{C}_{D^+\eta\rightarrow D^+\eta}^{(0,+1)}=\tilde{C}_{D^0\eta\rightarrow D^0\eta}^{(0,0)}\\ 
&C_{D\eta\rightarrow D_s\bar{K}}^{(0,1/2)}&=&\,\tilde{C}_{D^+\eta\rightarrow D_s^+\bar{K}^0}^{(0,+1)}=\tilde{C}_{D^0\eta\rightarrow D_s^+K^-}^{(0,0)}\\ 
&C_{D_s\bar{K}\rightarrow D_s\bar{K}}^{(0,1/2)}&=&\,\tilde{C}_{D_s^+\bar{K}^0\rightarrow D_s^+\bar{K}^0}^{(0,+1)}=\tilde{C}_{D_s^+K^-\rightarrow D_s^+K^-}^{(0,0)}\\ 
&C_{D\pi\rightarrow D\pi}^{(0,3/2)}&=&\,\frac{2}{3}\tilde{C}_{D^0\pi^0\rightarrow D^0\pi^0}^{(0,0)}+\frac{1}{3}\tilde{C}_{D^+\pi^-\rightarrow D^+\pi^-}^{(0,0)}-\frac{2\sqrt{2}}{3}\tilde{C}_{D^0\pi^0\rightarrow D^+\pi^-}^{(0,0)}\\ 
& &=&\,-\frac{1}{3}\tilde{C}_{D^0\pi^+\rightarrow D^0\pi^+}^{(0,+1)}+\frac{2}{3}\tilde{C}_{D^+\pi^0\rightarrow D^+\pi^0}^{(0,+1)}+\frac{2\sqrt{2}}{3}\tilde{C}_{D^+\pi^0\rightarrow D^0\pi^+}^{(0,+1)} \\ 
& &=&\,\tilde{C}_{D^0\pi^-\rightarrow D^0\pi^-}^{(0,-1)}=\tilde{C}_{D^+\pi^+\rightarrow D^+\pi^+}^{(0,+2)}\\
&C_{DK\rightarrow DK}^{(1,0)}&=&\,\frac{1}{2}\tilde{C}_{D^0K^+\rightarrow D^0K^+}^{(1,+1)}+\frac{1}{2}\tilde{C}_{D^+K^0\rightarrow D^+K^0}^{(1,+1)}+\tilde{C}_{D^+K^0\rightarrow D^0K^+}^{(1,+1)} \\
&C_{DK\rightarrow D_s\eta}^{(1,0)}&=&\,-\frac{1}{\sqrt{2}}\tilde{C}_{D^0K^+\rightarrow D_s^+\eta}^{(1,+1)}-\frac{1}{\sqrt{2}}\tilde{C}_{D^+K^0\rightarrow D_s^+\eta}^{(1,+1)}\\
&C_{D_s\eta\rightarrow D_s\eta}^{(1,0)}&=&\,\tilde{C}_{D_s^+\eta\rightarrow D_s^+\eta}^{(1,+1)}\\
&C_{D_s\pi\rightarrow D_s\pi}^{(1,1)}&=&\,\tilde{C}_{D_s^+\pi^0\rightarrow D_s^+\pi^0}^{(1,+1)}=\tilde{C}_{D_s^+\pi^+\rightarrow D_s^+\pi^+}^{(1,+2)}=\tilde{C}_{D_s^+\pi^-\rightarrow D_s^+\pi^-}^{(1,0)}\\ 
&C_{D_s\pi\rightarrow DK}^{(1,1)}&=&\,\tilde{C}_{D_s^+\pi^-\rightarrow D^0K^0}^{(1,0)}=\tilde{C}_{D_s^+\pi^+\rightarrow D^+K^+}^{(1,+2)}\\  
& &=&\,\frac{1}{\sqrt{2}}\tilde{C}_{D_s^+\pi^0\rightarrow D^0K^+}^{(1,+1)}-\frac{1}{\sqrt{2}}\tilde{C}_{D_s^+\pi^0\rightarrow D^+K^0}^{(1,+1)}\\
&C_{DK\rightarrow DK}^{(1,1)}&=&\,\tilde{C}_{D^0K^0\rightarrow D^0K^0}^{(1,0)}=\tilde{C}_{D^+K^+\rightarrow D^+K^+}^{(1,+2)}\\ 
& &=&\,\frac{1}{2}\tilde{C}_{D^0K^+\rightarrow D^0K^+}^{(1,+1)}+\frac{1}{2}\tilde{C}_{D^+K^0\rightarrow D^+K^0}^{(1,+1)}-\tilde{C}_{D^0K^+\rightarrow D^+K^0}^{(1,+1)}\\
&C_{D_sK\rightarrow D_sK}^{(2,1/2)}&=&\,\tilde{C}_{D_s^+K^0\rightarrow D_s^+K^0}^{(2,+1)}=\tilde{C}_{D_s^+K^+\rightarrow D_s^+K^+}^{(2,+2)} 
\end{aligned}
\end{equation}

\vspace{5mm}
The phase convention for the isospin states, $|I\,I_3\rangle$, is the following:
\begin{equation}\nonumber
 \begin{aligned}[b]
  &|\pi^+\rangle =-|1\,+1\rangle\; &&  |\pi^0\rangle =|1\,0\rangle\; &&  |\pi^-\rangle\; =|1\,-1\rangle && \\
  &|\eta\rangle =|0\, 0\rangle\; && && && \\ 
  &|K^+\rangle =\left|\frac12\, +\frac12\right\rangle\; && |K^0\rangle =\left|\frac12\, -\frac12\right\rangle\; && |K^-\rangle =\left|\frac12\, -\frac12\right\rangle\; && |\bar{K}^0\rangle =-\left|\frac12\, +\frac12\right\rangle \\
  &|D^+\rangle =-\left|\frac12\, +\frac12\right\rangle\; && |D^0\rangle =\left|\frac12\, -\frac12\right\rangle\; && |D^-\rangle =\left|\frac12\, -\frac12\right\rangle\; && |\bar{D}^0\rangle =\left|\frac12\, +\frac12\right\rangle \\  
  &|D_s^+\rangle =|0\, 0\rangle\; && |D_s^-\rangle =|0\, 0\rangle \ . &&  && 
\end{aligned}
\end{equation}

\chapter{Finite-temperature modifications of light mesons}
\label{appendix-modpion}

In Chapter~\ref{ch:hot-medium} we have neglected the medium modifications of the light mesons and used vacuum spectral functions for them, in both the $T$-matrix calculation as well as in the $D$-meson self-energy corrections. This approximation, which should be reasonable at low temperatures, was implemented in Refs.~\cite{Montana:2020lfi,Montana:2020vjg}, where we based our assumption on the pion mass modifications given in Refs.~\cite{Schenk:1993ru,Toublan:1997rr}. In this appendix, we present a validity check using a medium-modified pion mass.

To address the correction of the pion self-energy due to the thermal bath, we have applied the methodology of~\cite{Schenk:1993ru}. As opposed to our calculation for heavy mesons, the method in~\cite{Schenk:1993ru} is not self-consistent but based on the one-loop correction to the meson self-energy in the dilute limit. We have computed the real part of the pole of the pion propagator, whose self-energy is corrected by the thermal medium producing a modified dispersion relation,
\begin{equation}
\omega (p) \simeq \omega_p - \frac{1}{\omega_p}\int \frac{d^3q}{(2\pi)^3 2\omega_q}  f(\omega_q,T) \textrm{Re } T_{\pi \pi} (s) \ , 
\end{equation}
where $\omega_p = \sqrt{p^2+m_\pi(T=0)}$ is the vacuum dispersion relation (with $m_\pi(T=0)=138$~MeV),
$f(\omega_p,T)$ is the \gls{be} distribution function, $T_{\pi\pi} (s)$ is the isospin averaged amplitude of the $\pi\pi \rightarrow \pi\pi$ process, and $s=(p+q)^2$ the Mandelstam variable. 

Here, $T_{\pi\pi}(s)$ is calculated using the unitarized scattering amplitudes coming from the $SU(3)$ \gls{chpt} Lagrangian~\cite{Oller:1997ng,Oller:1998hw}. The unitarization approach used in~\cite{Oller:1997ng,Oller:1998hw} is similar to ours, although not equal. In particular, the scattering amplitudes from~\cite{Oller:1998hw} have no corrections due to the temperature, but this is consistent with the one-loop approximation for the pion self-energy.

\begin{figure}[htp!]\centering
\includegraphics[width=0.55\textwidth]{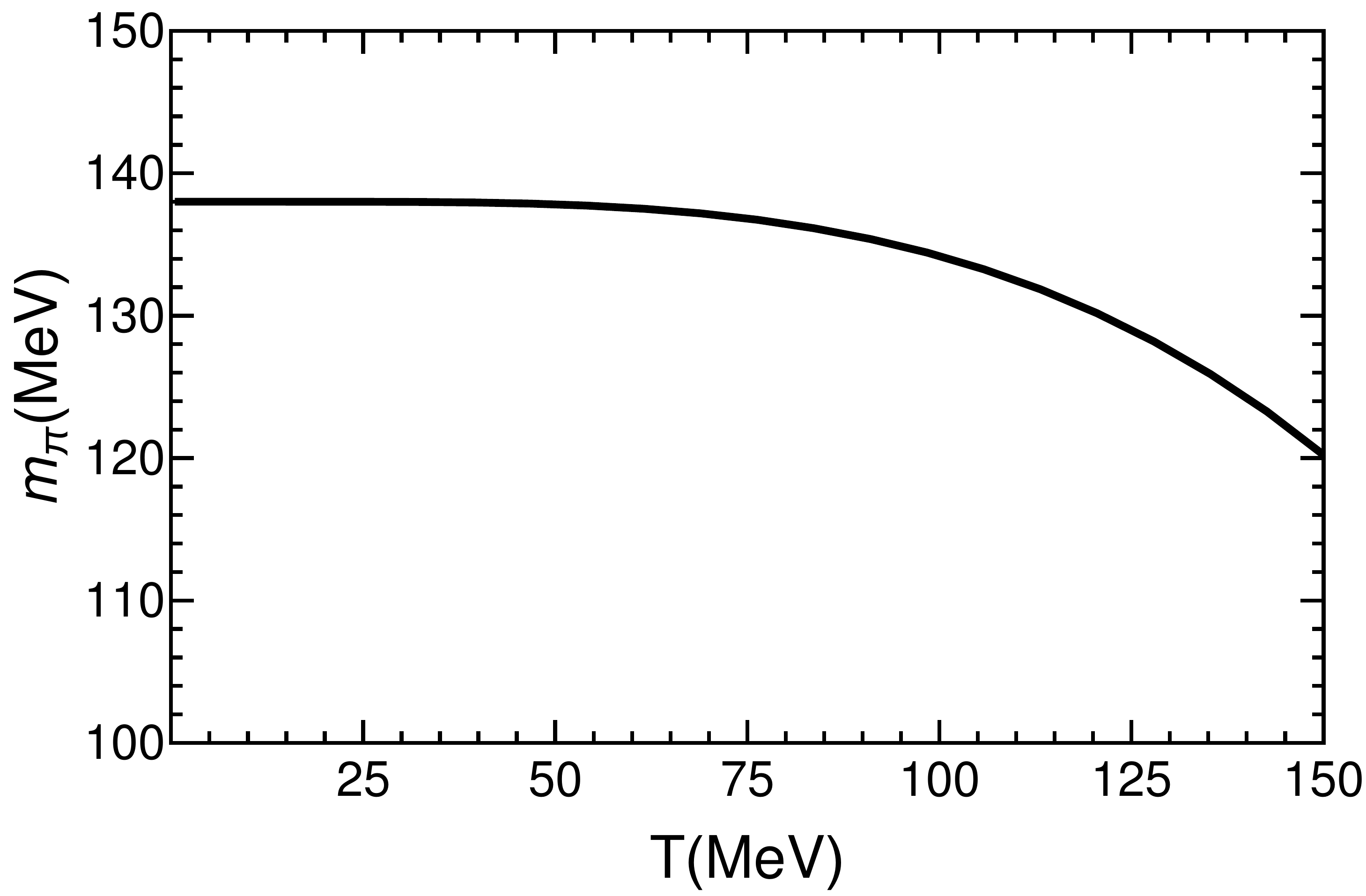}
\caption{Pion mass $m_\pi(T)=\omega(p=0;T)$ as a function of the temperature after incorporating the $SU(3)$ \gls{chpt} amplitudes of Ref.~\cite{Oller:1998hw} into the one-loop pion self-energy correction of Ref.~\cite{Schenk:1993ru}.}
\label{fig:app-pionmass}
\end{figure}

In this appendix we neglect the pion width, which is also generated due to temperature effects, so we can still use Dirac delta spectral functions peaked at $\omega(p)$. We define the thermal pion mass as the value $m_\pi(T) = \omega (p=0;T)$ and plot it in Fig.~\ref{fig:app-pionmass} up to $T=150$~MeV. At this temperature the pion mass is $m_\pi(T=150~\textrm{MeV})=120$~MeV.

We have run our code for the $D$-meson self-energy at $T=150$ MeV using this reduced pion mass. We find that the mass of the ground-state $D^{(*)}$ and $D_s^{(*)}$ mesons are only slightly modified with a decrease of $\Delta m_{D^{(*)}}=4$~MeV and $\Delta m_{D_s^{(*)}}=2$~MeV, with respect to the thermal masses reported in Chapter~\ref{ch:hot-medium}, while the widths do not change appreciably. With regards to the dynamically generated states, the lowest-lying state that corresponds to the $D_0^*(2300)$, as well as the one for $D_1(2430)$, change their masses by $-2$~MeV, being the widths $20$~MeV larger. As for the highest-lying resonances, both change by $-2$~MeV, with a similar change in width. For the bound states $D_{s0}^*(2317)$ and $D_{s1}^*(2460)$ the change in mass is $-2$~MeV, while the width increases by $1$~MeV.

In conclusion, for low temperatures, $T \ll 150$~MeV, it is acceptable to neglect the medium effects on the light mesons. For the largest temperature considered, that is $T=150$~MeV, the effects of a medium-modified pion are noticeable but still small. The incorporation of the medium-modified spectral functions, with both mass and decay width depending on temperature, is imperative to decide whether the widening of the pion can produce a significant change in the properties of heavy-flavor mesons at intermediate temperatures. Also the modification of the other light mesons. However, this is out of the scope of this dissertation.

\chapter{Numerical integration of thermal quantities}
In this appendix, we briefly discuss the details of the procedure followed for the non-trivial numerical integration of the expressions obtained in Chapter~\ref{ch:hot-medium} with the \gls{itf} for quantities such as the two-meson thermal loop and the self-energy of the heavy meson.

\section{Two-meson propagator}
\label{sec:app-loop}
For free mesons, the thermal two-meson loop function is given by Eq.~(\ref{eq:hot-loopfree1}). Using cut-off regularization let us write 
\begin{equation} \label{eq:hot-loop_cutoff}
 G_{D\Phi}(E,\vec{p}\,;T)=\int_0^{\Lambda}\frac{dq}{16\pi^2}\frac{q^2}{\omega_D}\int_{-1}^1 dx\, \frac{1}{\omega_\Phi(x)}\sum_{i=1}^4\frac{\mathcal{F}_i(q,x)}{\mathcal{G}_i(q,x)} \ ,
\end{equation}
with $\omega_D=\sqrt{q^{2}+m_D^2}$ and $\omega_\Phi=\sqrt{(\vec{p}-\vec{q}\,)^2+m_\Phi^2}=\sqrt{p^{2}+q^{2}-2pqx+m_\Phi^2}$. We have introduced $q\equiv|\vec{q}|$ and $p\equiv|\vec{p}|$ to simplify the notation, and $x=\cos\theta$ is the angular variable. The functions $\mathcal{F}_i(q,x)$ and $\mathcal{G}_i(q,x)$ are defined as:
\begin{align}\label{eq:hot-F}\nonumber
  \mathcal{F}_1(q,x)&=\mathcal{F}_2(q,x)=1+f(\omega_D,T)+f(\omega_\Phi,T) \ , \\
  \mathcal{F}_3(q,x)&=\mathcal{F}_4(q,x)=f(\omega_D,T)-f(\omega_\Phi,T) \ , 
\end{align}
and
\begin{align}\label{eq:hot-G}\nonumber
   \mathcal{G}_1(q,x)&=E-(\omega_D+\omega_\Phi)+\ii\varepsilon \ , \\ \nonumber
  \mathcal{G}_2(q,x)&=-E-(\omega_D+\omega_\Phi)-\ii\varepsilon \ , \\ \nonumber
  \mathcal{G}_3(q,x)&=E+(\omega_D-\omega_\Phi)+\ii\varepsilon \ , \\
  \mathcal{G}_4(q,x)&=-E+(\omega_D-\omega_\Phi)-\ii\varepsilon \ .
\end{align}
The $x$-dependence of the integrand is contained in $\omega_\Phi$. When integrating over $x$, we have to be careful in the regions around poles, that is, the roots of $\mathcal{G}_i(q,x)$, where a finer grid is needed.

The result of the direct numerical integration of an integral of the type 
$\int_{-1}^1 dx\, \frac{f(x)}{g(x)} $
with $g(x_0)=0$ is highly unstable and depends on how close we come to $x_0$, as it is ill-defined at this point unless $f(x_0)=0$. The standard way to proceed is to perform a principal value integral.
As $x$ approaches $x_0$ we can do a Taylor expansion of $g(x)$,
\begin{equation}\label{eq:hot-taylor}
 g(x)=g(x_0)+g'(x_0)(x-x_0)+...\approx g'(x_0)(x-x_0) \ ,
\end{equation}
and add and subtract the integral of $f(x_0)/ g'(x_0)(x-x_0)$, which has the same kind of singularity as the function that we want to integrate, 
\begin{align}\label{eq:hot-x-integral}
 \int_{-1}^1 dx\,\frac{f(x)}{g(x)}&= \int_{-1}^1 dx\,\frac{f(x)}{g(x)}+\int_{-1}^1 dx\, \frac{f(x_0)}{g'(x_0)}\frac{1}{(x-x_0)}-\int_{-1}^1 dx\, \frac{f(x_0)}{g'(x_0)}\frac{1}{(x-x_0)} \nonumber \\ 
 & = \int_{-1}^1 dx\,\Bigg[\frac{f(x)}{g(x)}-\frac{f(x_0)}{g'(x_0)}\frac{1}{x-x_0}\Bigg]+\frac{f(x_0)}{g'(x_0)}\int_{-1}^1 dx\,\frac{1}{x-x_0} \ .
\end{align}
The subtracted term allows us to get a numerically stable integrand, while the added term can be solved analytically, using the identity
\begin{equation}
 \int dx \,\frac{1}{x\pm\ii\varepsilon}=\textrm{p.v.\,}\int dx\, \frac{1}{x}\mp\ii\pi\ ,
\end{equation}
where $\textrm{p.v.}$ denotes the principal value integral, defined as
\begin{equation}
 \textrm{p.v.\,}\int_a^b dx\,\frac{1}{x}=\left[\int_a^{-\varepsilon}+\int_{\varepsilon}^b\right]dx\,\frac{1}{x} \ .
\end{equation}
With this, one can write
\begin{equation} \label{eq:hot-x-integral2}
 \int_{-1}^1 dx\,\frac{f(x)}{g(x)}=\int_{-1}^1 dx\Bigg[\frac{f(x)}{g(x)}-\frac{f(x_0)}{g'(x_0)}\frac{1}{x-x_0}\Bigg]+\frac{f(x_0)}{g'(x_0)}\Bigg[\ln\frac{|+1-x_0|}{|-1-x_0|}\mp \ii\pi\Bigg]\ ,
\end{equation}
where the sign of the imaginary part ($+$ or $-$) depends on whether we have $x_0+\ii\varepsilon$ or $x_0-\ii\varepsilon$. 
This way of dealing with the $x$-integration not only allows us to get a numerical stable result for the real part but also provides an imaginary part to the loop function in Eq.~(\ref{eq:hot-loop_cutoff}). We use
\begin{align}\label{eq:hot-loop-x-integral} \nonumber
 \int_0^{\Lambda}\frac{dq}{16\pi^2}\frac{q^2}{\omega_D}&\int_{x_{\textrm{min}}}^1 dx \frac{1}{\omega_\Phi(x)}\frac{\mathcal{F}_i(q,x)}{\mathcal{G}_i(q,x)}=\int_0^{\Lambda}\frac{dq}{16\pi^2}\frac{q^2}{\omega_D}\\ \nonumber
 &\times\Bigg\{ \int_{x_{\textrm{min}}}^1 dx\Bigg[\frac{1}{\omega_\Phi(x)}\frac{\mathcal{F}_i(q,x)}{\mathcal{G}_i(q,x)}-\frac{1}{\omega_\Phi(x_i)}\frac{\mathcal{F}_i(q,x_i)}{\mathcal{G}_i'(q,x_i)}\frac{1}{x-x_i}\Bigg]\\ 
 &\quad+\frac{1}{\omega_\Phi(x_i)}\frac{\mathcal{F}_i(q,x_i)}{\mathcal{G}_i'(q,x_i)}\Bigg[\ln\frac{|+1-x_i|}{|x_{\textrm{min}}-x_i|}+s_i\,\ii\pi\Bigg]\Bigg\}
\end{align}
when $E>\omega_D$ for $i=\{1,3\}$, and $E<\omega_D$ for $i=\{3,4\}$, and for the values of $q$ for which the $i$-term of the integrand in the $x$-integration has a pole at the value $x_i$,
\begin{equation}
 x_i=\frac{p^2+q^2+m_\Phi^2-\omega_i^2}{2pq} \ , \quad \textrm{with} \quad \left\{ 
 \begin{array}{lll}
  \omega_1=\phantom{-}E-\omega_D+\ii\varepsilon \\ 
  \omega_3=\phantom{-}E+\omega_D+\ii\varepsilon \\
  \omega_4=-E+\omega_D -\ii\varepsilon
 \end{array}
 \right. \ ,
\end{equation}
which satisfies $\mathcal{G}_i(q,x_i)=0$ and $|x_i|<1$. It is easy to see that the signs in Eq.~(\ref{eq:hot-loop-x-integral}) are $s_1=s_3=-$ and $s_4=+$. When there is no pole, the numerical integration of Eq.~(\ref{eq:hot-loop_cutoff}) can be done directly.

We note that the lower limit of the $x$-integral in Eq.~(\ref{eq:hot-loop_cutoff}) has been replaced by $x_{\textrm{min}}\geq-1$, as not all the angles are accessible for a certain $q$. Indeed, we are calculating the loop function in the reference frame of the laboratory and we want to apply a certain cut-off, $\Lambda^{\textrm{cm}}=800$~MeV, in the center-of-mass frame, for consistency with the vacuum approach described in Section~\ref{sec:free-mm}. While one possibility would be to boost the meson-meson system to the center-of-mass frame and then calculate the loop, it is easier to calculate the cut-off in the lab frame. In the nonrelativistic limit, the cut-off in the lab frame reads
\begin{equation}\label{eq:hot-cutoff-lab}
 \Lambda^{\textrm{lab}}=\Lambda^{\textrm{cm}}+\xi p \ , \quad \textrm{with}\quad \xi=\frac{m_D}{m_D+m_\Phi} \ ,
\end{equation}
and $p$ the modulus of the external momentum.
Then for $\xi p<\Lambda^{\textrm{cm}}$, we have that
\begin{equation}
 \begin{array}{ll}
  -1< x < +1 \ , & \quad \textrm{for} \quad 0< q< (\Lambda^{\textrm{cm}} -\xi p) \ , \\ 
  x_{\textrm{min}}< x< +1 \ , &\quad \textrm{for} \quad (\Lambda^{\textrm{cm}}-\xi p)< q< (\Lambda^{\textrm{cm}}+\xi p) \ , \\
  \nexists\, x \ , & \quad \textrm{for} \quad q>(\xi p+\Lambda^{\textrm{cm}}) \ ,
 \end{array}
\end{equation}
where $q$ is in the modulus of the internal momenta in the lab frame, and $x_{\textrm{min}}$ is given by
\begin{equation}\label{eq:hot-xmin}
 x_{\textrm{min}}=\frac{q^2+\xi^2 p^2-(\Lambda^{\textrm{cm}})^2}{2q\xi p} \ .
\end{equation}
Similarly, for $\xi p>\Lambda^{\textrm{cm}}$, 
\begin{equation}
 \begin{array}{ll}
  \nexists\, x \ , & \quad \textrm{for} \quad 0< q< (\xi p-\Lambda^{\textrm{cm}}) \ , \\ 
  x_{\textrm{min}}< x< +1 \ , &\quad \textrm{for} \quad (\xi p-\Lambda^{\textrm{cm}})< q< (\xi p+\Lambda^{\textrm{cm}}) \ , \\
  \nexists\, x \ , & \quad \textrm{for} \quad q>(\xi p+\Lambda^{\textrm{cm}}) \ .
 \end{array}
\end{equation}

The considerations above have to be taken into account for the proper numerical calculation of the free two-meson propagator for a generic external center-of-mass frame energy $E$ and momentum $\vec{p}$. The case of $\vec{p}=0$ has to be treated separately because the integrand does not depend on the angular variable. That is, the $x$-integration is trivial, and one has to be careful with the poles in the $q$-integration. One can use, in principle, the principal value integration approach. However, the integral
\begin{equation}
 G_{D\Phi}(E,\vec{p}=0;T)=\int_0^{\Lambda}\frac{dq}{8\pi^2}\frac{q^2}{\omega_D(q)\omega_\Phi(q)}\sum_{i=1}^4\frac{\mathcal{F}_i(q)}{\mathcal{G}_i(q)} \ , \quad \textrm{with}\quad \omega_{D,\Phi}=\sqrt{q^2+m_{D,\Phi}^2}
\end{equation}
needs to be rewritten as an analytically solvable one and, in practice, it is not possible. Nevertheless, the numerical integration in $q$, taking the intervals $\left[0,q_i\right)$, $\left(q_i,\Lambda\right]$ and approaching the pole position $q_i$ symmetrically from both sides, is very stable. The condition $q_i>0$ is only satisfied for the first and fourth terms, with $q_1=q_4=\sqrt{[E-(m_\Phi+m_D)^2][E-(m_\Phi-m_D)^2]}/(2E)$. Then, the imaginary part is given by 
\begin{equation}
 \textrm{ Im\,}G_{D\Phi}(E,\vec{p}=0;T)=-\ii\pi \frac{1}{8\pi^2}\frac{q_i^2}{\omega_D(q_i)\omega_\Phi(q_i)}\frac{\mathcal{F}_i(q_i)}{|\mathcal{G}_i'(q_i)|} \ ,
\end{equation}
with $i=1$ for $E>(m_D+m_\Phi)$ and $i=4$ for $E<|m_D-m_\Phi|$, giving rise to the unitary cut and the Landau cut, respectively (see discussion in Section~\ref{subsec:hot-form-Landaucut}).

For the numerical integration of the loop function in the case where only one of the mesons is dressed with the spectral function, that is, the expression given in Eq.~(\ref{eq:hot-loop-oneSE}), we proceed in a similar way. Using cut-off regularization, we write Eq.~(\ref{eq:hot-loop-oneSE}) as
\begin{equation}\label{eq:hot-loop-oneSE-cutoff}
 G_{D\Phi}(E,\vec{p}\,;T)=-\int_0^{\Lambda}\frac{dq}{8\pi^2}q^2\int_{x_{\textrm{min}}}^1 dx\,\frac{1}{\omega_\Phi(q,x)}\int_0^{\omega_{\textrm{max}}}d\omega\, S_D(\omega,q;T)\sum_{i=1 }^4\frac{\mathcal{F}_i(\omega,q,x)}{\mathcal{G}_i(\omega,q,x)} \ ,
\end{equation}
where we have defined
\begin{align}\label{eq:hot-F_oneSE} \nonumber
  \mathcal{F}_1(\omega,q,x)&=\mathcal{F}_2(\omega,q,x)=1+f(\omega,T)+f(\omega_\Phi,T) \ ,  \\
  \mathcal{F}_3(\omega,q,x)&=\mathcal{F}_4(\omega,q,x)=-f(\omega,T)+f(\omega_\Phi,T)   \ ,
\end{align}
and
\begin{equation}\label{eq:hot-G_oneSE}
 \mathcal{G}_i(\omega,q,x)=\omega-\omega_i \ , \quad i=\{1,2,3,4\} \ ,  \quad \textrm{with} \quad \left\{
 \begin{array}{l}
 \omega_1=\phantom{-}E-\omega_\Phi+\ii\varepsilon  \\ 
 \omega_2=-E-\omega_\Phi-\ii\varepsilon  \\ 
 \omega_3=\phantom{-}E+\omega_\Phi+\ii\varepsilon  \\
 \omega_4=-E+\omega_\Phi-\ii\varepsilon 
\end{array}
\right. \ .
\end{equation}
In this case, we deal with the poles of the integrand, located at $\omega_i>0$, when performing the $\omega$-integration. We use the technique of the principal value integral described above. In addition, one has to make sure to properly capture the strength to the integrand coming from the spectral function, which can be considerably narrow at low temperatures, as shown in Chapter~\ref{ch:hot-medium}.

In Eq.~(\ref{eq:hot-loop-oneSE-cutoff}), the upper limit of the $\omega$-integration, which should in principle extend to infinity, has been replaced by $\omega_{\textrm{max}}\rightarrow\infty$, and in practice we take a large enough value. Furthermore, the cut-off $\Lambda$ in the lab frame is computed using Eq.~(\ref{eq:hot-cutoff-lab}), and the lower limit of the $x$-integration is set by the value of $x_{\textrm{min}}$ in Eq.~(\ref{eq:hot-xmin}).

\section{Heavy-meson self-energy}
\label{sec:app-selfe}
 
For the numerical integration of the self-energy of Eq.~(\ref{eq:hot-selfE}), where a delta-type spectral function has been introduced for the light meson, we use similar tricks to those described for the thermal loop function in the previous section. First, we write Eq.~(\ref{eq:hot-selfE}) as
\begin{equation}\label{eq:hot-selfE2}
 \Pi_{D}(\omega,\vec{q}\,;T)=\frac{1}{\pi}\int_0^{\Lambda}\frac{dq'}{8\pi^2}\frac{q'^2}{\omega_\Phi}\int_{-1}^1 dx\int_0^{E_\textrm{max}} dE\,\textrm{Im\,}T_{D\Phi}(E,\vec{p}\,;T) \sum_{i=1}^4\frac{\mathcal{F}_i(E,q',x)}{\mathcal{G}_i(E,q',x)} \ ,
\end{equation}
with
\begin{align}\label{eq:hot-selfE_F}  \nonumber
 \mathcal{F}_1(E,q,x)&=\mathcal{F}_2(E,q,x)=1+f(E,T)+f(\omega_\Phi,T) \ , \\
 \mathcal{F}_3(E,q,x)&=\mathcal{F}_4(E,q,x)=-f(E,T)+f(\omega_\Phi,T) \ ,
\end{align}
and
\begin{equation}\label{eq:hot-selfE_G}
 \mathcal{G}_i(E,q,x)=E-E_i \ ,   \quad i=\{1,2,3,4\} \ ,  \quad \textrm{with} \quad \left\{
 \begin{array}{l}
 E_1=\phantom{-}\omega-\omega_\Phi+\ii\varepsilon  \\ 
 E_2=-\omega-\omega_\Phi-\ii\varepsilon  \\ 
 E_3=\phantom{-}\omega+\omega_\Phi+\ii\varepsilon  \\
 E_4=-\omega+\omega_\Phi-\ii\varepsilon 
\end{array}
\right. \ .
\end{equation}
Then, we make use of Eqs.~(\ref{eq:hot-taylor}) to (\ref{eq:hot-x-integral2}) for the principal value integral, which gives
\begin{align}\label{eq:hot_selfE_pv}\nonumber
 \frac{1}{\pi}\int_0^{\Lambda}\frac{dq'}{8\pi^2}\frac{q'\,^2}{\omega_\Phi}&\int_{-1}^1 dx\int_0^{E_\textrm{max}} dE\,\textrm{Im\,}T_{D\Phi}(E,\vec{p}\,;T) \frac{\mathcal{F}_i(E,q',x)}{\mathcal{G}_i(E,q',x)} =\frac{1}{\pi}\int_0^{\Lambda}\frac{dq'}{8\pi^2}\frac{q'\,^2}{\omega_\Phi}\int_{-1}^1 dx \\ \nonumber 
 & \times\Bigg\{ \int_0^{E_\textrm{max}}dE\left[\textrm{Im\,}T_{D\Phi}(E,\vec{p}\,;T) \frac{\mathcal{F}_i(E,q',x)}{\mathcal{G}_i(E,q',x)}-\textrm{Im\,}T_{D\Phi}(E_i,\vec{p}\,;T) \frac{\mathcal{F}_i(E_i,q',x)}{E-E_i} \right] \\ 
 &+\textrm{Im\,}T_{D\Phi}(E_i,\vec{p}\,;T) \mathcal{F}_i(E_i,q',x)\left[ \ln\frac{|E_{\textrm{max}}-E_i|}{|E_i|} +s_i \ii\pi \right] \Bigg\} \ ,
\end{align}
with $s_1=s_3=+$ and $s_4=-$, resulting from the $\pm\ii\varepsilon$ term of the analytical continuation in Eq.~(\ref{eq:hot-selfE_G}). Therefore, when there is a pole in the integrand, $E_i>0$, the self-energy is obtained using Eq.~(\ref{eq:hot_selfE_pv}). Otherwise, one can directly perform the numerical integration of Eq.~(\ref{eq:hot-selfE2}).

We note that the upper integration limits in Eqs.~(\ref{eq:hot-selfE_F}) and (\ref{eq:hot_selfE_pv}) do not have the meaning of a physical cut-off; rather they correspond to a truncation of the numerical integrals and their values have to be large enough since, in principle, $\Lambda\rightarrow \infty$ and $E_{\textrm{max}}\rightarrow\infty$, but also having in mind that the effective theory breaks down at high energies.

It is also important to note that the integrand involved in the calculation of the self-energy may vary considerably in certain areas of the integration domain. In addition to the regions around poles that appear due to zeros in the denominators and which are taken into account in the numerical integration with the trick of the principal value as described above, the integrand may vary over several orders of magnitude in the vicinity of bound states and resonances of the $T$ matrix. Because of this, the numerical integration with a fixed number of integration points is not adequate. On the contrary, it is convenient to use a recursive procedure that allows us to automate an adaptive domain and increase the number of integration points around the features of the $T$ matrix. We refer the interested reader to Chapter~5 of the lecture notes in Ref.~\cite{lecturesHjorthJensen} for a detailed discussion on integration methods including adaptive integration methods.

\chapter{Wigner transform}
\label{appendix-Wigner}
In the derivation of the kinetic equation in Chapter~\ref{ch:transport}, we have implemented a Wigner transform to simplify the expressions. In this appendix, we give some details on the Wigner transform of the product and convolution operators following Ref.~\cite{Rammer}.

For any two-point function such as the Wightman function $G_D^<(x,x')$, one can define its Wigner transform  as
\begin{equation}
G_D^< (x,x') \xrightarrow{\ \ \textrm{WT} \ \ }  G_D^< (X,k) \equiv \int d^4s \ e^{\ii k\cdot s} \ G_D^< \left( X+\frac{s}{2},X-\frac{s}{2} \right) \ , 
\end{equation}
where we have introduced the center-of-mass and relative coordinates, $X=(x+x')/2$ and $s=x-x'$, respectively~\cite{Danielewicz:1982kk,Blaizot:1999xk,Rammer}. 

We apply the Wigner transform to all the terms of the off-shell kinetic equation~(See Eq.~(\ref{eq:transport-beforeWT})). The first term includes the following combination,
\begin{equation}
G_{0,x}^{-1} G_D^< (x,x') - G_{0,x'}^{-1} G_D^<(x,x') \ , 
\end{equation}
where $G_0^{-1}(x)=-\partial^2_x-m_D^2$. In terms of $X$ and $s$ it reads
\begin{align}\nonumber
G_{0,x}^{-1} G_D^< (x,x') - G_{0,x'}^{-1} G_D^<(x,x') &=- \left[\partial_x^2  - \partial_{x'}^2\right] G_D^<(x,x') \\ &= - 2 \partial_s \cdot \partial_X G^<_D \left(X+\frac{s}{2},X-\frac{s}{2} \right) \ . 
\end{align}

Then, we multiply this term by $e^{\ii k \cdot s}$ and integrate over $d^4s$ to obtain,
\begin{equation}
G_{0,x}^{-1} G_D^< (x,x') - G_{0,x'}^{-1} G_D^<(x,x') \ \xrightarrow{\textrm{WT}} \ 2\ii k^\mu \frac{\partial G^<_D(X,k)}{\partial X^\mu} \ , 
\end{equation}
where the derivative over $s$ transforms into a four-momentum $k^\mu$.

This technique is applied to all the other terms in the kinetic equation, implementing at the same time a gradient expansion in $X$. The latter allows neglecting higher-order terms in $\partial^\mu_X$, like $\partial_X^2 G_D^>(X,k)$. For example, in the first commutator of Eq.~(\ref{eq:transport-beforeWT}) one finds the factor $\Pi^\delta(x)$, which is expanded as
\begin{equation}
\Pi^\delta (x) =\Pi^\delta \left( X+ \frac{s}{2} \right)=\Pi^\delta(X) + \frac{1}{2} s \cdot \partial_X \Pi^\delta(X) + {\cal O} (\partial_X^2) \ , 
\end{equation}
where we neglect higher-order terms in $\partial_X$. Then, the combination $-\Pi^\delta(x)+\Pi^\delta(x')$ becomes $-s \cdot \partial_X \Pi^\delta(X) +  {\cal O} (\partial_X^2)$. Up to this order, the Wigner transform gives
\begin{equation}
\left[-\Pi^\delta(x)+\Pi^\delta(x')\right]  G_D^< (x,x') \ \xrightarrow{\textrm{WT}} \  \ii \frac{\partial \Pi^\delta(X)}{\partial X} \cdot \frac{\partial G_D^<(X,k)}{\partial k} \ , 
\end{equation}
where the factor $s^\mu$ produces the operator $\partial_k^\mu$.

Let us consider the convolution of two operators,
\begin{equation}
C(x,x') \equiv \left(A \otimes B \right) (x,x') =\int d^4z\, A(x,z) B(z,x') \ . 
\end{equation}
It is possible to show that, after the Wigner transform, it becomes~\cite{Rammer}
\begin{equation}
C(x,x') \  \xrightarrow{\textrm{WT}} \   A(X,k) \exp \left[ - \frac{\ii}{2} \left( \overleftarrow{\partial_X} \cdot \overrightarrow{\partial_k} -  \overleftarrow{\partial_k} \cdot \overrightarrow{\partial_X} \right) \right] B(X,k) \ , 
\end{equation}
where the derivative operators act in the direction marked by the arrows. Applying the gradient expansion and keeping the first order in $\partial_X$, we can simplify it to,
\begin{equation}
C(x,x')  \ \xrightarrow{\textrm{WT}} \  A(X,k) B(X,k) + \frac{\ii}{2} \left\{ A,B \right\}_{\textrm{PB}}  + {\cal O} (\partial_X^2) \ , \label{eq:app-conv} 
\end{equation}
where the Poisson bracket is defined as
\begin{equation}
\left\{ A,B \right\}_{\textrm{PB}} \equiv \frac{\partial A(X,k)}{\partial k_\mu } \frac{\partial B(X,k)}{\partial X^\mu} -\frac{\partial A(X,k)}{\partial X^\mu } \frac{\partial B(X,k)}{\partial k_\mu} \ . 
\end{equation}

Then, using Eq.~(\ref{eq:app-conv}) one finds that the anticommutators appearing in the collision terms of Eqs.~(\ref{eq:transport-beforeWT}) and (\ref{eq:transport-beforeWT2}),
\begin{equation}
\left\{ \Pi^\lessgtr \stackrel{\otimes}{,} G_D^\gtrless \right\} (x,x')= \left(\Pi^\lessgtr \otimes G_D^\gtrless\right) (x,x') + \left(G_D^\gtrless \otimes \Pi^\lessgtr \right) (x,x') \ , 
\end{equation}
transform to
\begin{equation}
\left\{ \Pi^\lessgtr \stackrel{\otimes}{,} G_D^\gtrless \right\} (x,x') \ \xrightarrow{\textrm{WT}} \  2 \Pi^\lessgtr (X,k) G_D^\gtrless (X,k) \ , 
\end{equation}
as the Poisson brackets cancel, $\left\{ A,B \right\}_{\textrm{PB}} = - \left\{ B,A \right\}_{\textrm{PB}}$.

However, for the commutators like
\begin{equation}
-\left[ \Pi^{\textrm{R/A}}  \stackrel{\otimes}{,} G_D^\lessgtr\right] (x,x') =- \left(\Pi^{\textrm{R/A}}  \otimes G_D^\lessgtr\right) (x,x') + \left(G_D^\lessgtr    \otimes  \Pi^{\textrm{R/A}}\right) (x,x') \ ,
\end{equation}
the Poisson bracket is the only remaining piece,
\begin{equation}-\left[  \Pi^{\textrm{R/A}}  \stackrel{\otimes}{,} G_D^\lessgtr\right] (x,x')  \ \xrightarrow{\textrm{WT}} \  -\ii \left\{ \Pi^{\textrm{R/A}} ,G_D^\lessgtr \right\}_{\textrm{PB}} \ . \end{equation}

Similarly,
\begin{equation} 
-\left[ \Pi^\lessgtr \stackrel{\otimes}{,} G_D^{\textrm{R/A}} \right] \ \xrightarrow{\textrm{WT}} \  -\ii \left\{ \Pi^\lessgtr ,  G_D^{\textrm{R/A}}  \right\}_{\textrm{PB}} \ . 
\end{equation}

Applying these rules to the different terms in Eq.~(\ref{eq:transport-beforeWT}), then Eq.~(\ref{eq:transport-afterWT}) follows.

\chapter{On-shell Fokker-Planck equation} 
\label{appendix-onshell}
In Chapter~\ref{ch:transport}, we have obtained the off-shell Fokker-Planck equation for the heavy-meson (homogeneous) Wightman function $\ii G_D^<(t,k^0,\vec{k}\,)$. We reproduce it here again for convenience,
\begin{align} 
\frac{\partial}{\partial t} \ii G_D^< (t,k^0,\vec{k}\,) &= \frac{\partial}{\partial k^i} \Bigg\{ \hat{A}(k^0,\vec{k};T) k^i \ii G_D^< (t,k^0,\vec{k}\,)  \nonumber \\ 
&  \quad + \frac{\partial}{\partial k^j} \left[ \hat{B}_0(k^0,\vec{k};T) \Delta^{ij} + \hat{B}_1(k^0,\vec{k};T) \frac{k^i k^j}{\vec{k}\,^2} \right] \ii G_D^< (t,k^0,\vec{k}\,)\Bigg\} \ , \label{eq:app-FPforG} 
\end{align}
The ``off-shell'' coefficients were defined in Eqs.~(\ref{eq:transport-hatA}), (\ref{eq:transport-hatB0}) and (\ref{eq:transport-hatB1}),
\begin{align}
 \hat{A}(k^0,\vec{k};T) & = \frac{1}{2k^0} \int  \frac{dk_1^0}{2\pi} \frac{d^3q}{(2\pi)^3} W(k^0,\vec{k}, k_1^0,\vec{q}\,) \ \frac{\vec{q} \cdot \vec{k}}{\vec{k}\,^2} , \\
 \hat{B}_0 (k^0,\vec{k};T) & = \frac{1}{2k^0}\frac{1}{4} \int \frac{dk_1^0}{2\pi} \frac{d^3q}{(2\pi)^3} W(k^0,\vec{k}, k_1^0,\vec{q}\,) \ \left[ \vec{q}\,^2 - \frac{(\vec{q} \cdot \vec{k}\,)^2}{\vec{k}\,^2} \right] , \\
  \hat{B}_1 (k^0,\vec{k};T) & = \frac{1}{2k^0}\frac{1}{2} \int \frac{dk_1^0}{2\pi} \frac{d^3q}{(2\pi)^3}  W(k^0,\vec{k}, k_1^0,\vec{q}\,) \ \frac{(\vec{q} \cdot \vec{k}\,)^2}{\vec{k}\,^2} \ ,
\end{align}
where we have expressed the average in terms of the integration of the scattering rate $W(k^0\vec{k},k_1^0,\vec{q}\,)$ integrated over the transferred momentum. These equations follow immediately from the Fokker-Planck reduction of the transport equation.

The scattering rate reads
\begin{align}
  W(k^0,\vec{k},k_1^0,\vec{q}\,) & \equiv \int \frac{d^4k_2}{(2\pi)^4} \frac{d^4k_3}{(2\pi)^4}  (2\pi)^4 \delta (k_1^0+k_2^0+k_3^0-k^0) \delta^{(3)} (\vec{k}_2+\vec{k}_3-\vec{q}\,) \nonumber \\
  & \times \left|T (k_1^0+k_2^0+\ii\varepsilon, \vec{k} - \vec{q} + \vec{k}_2)\right|^2  \ii G_\Phi^>(k_2^0,\vec{k}_2)\ii G_\Phi^<(k_3^0,\vec{k}_3) \ii G_D^>(k_1^0,\vec{k}-\vec{q}\,) \ .
\end{align}

We stress that, upon the integration over $dk^0/(2\pi)$ in Eq.~(\ref{eq:app-FPforG}), it is not possible to obtain the standard Fokker-Planck equation with ``on-shell'' coefficients depending only on $\vec{k}$ due to the presence of a $D$ meson with a generic spectral function. To match the previous results and derive the ``on-shell'' version of the coefficients, one needs to apply the Kadanoff-Baym Ansatz of Eqs.~(\ref{eq:transport-ansatz1}) and (\ref{eq:transport-ansatz2}), and particularize for the narrow quasiparticle limit of the spectral function in Eq.~(\ref{eq:transport-Diracdelta})
with $z_k \simeq 1$,
\begin{equation}
\ii G^<_D (t,k^0,\vec{k}\,) = \frac{2\pi}{2E_k} \left[\delta(k^0-E_k)-\delta(k^0+E_k)\right] f_D(t,k^0) \ , 
\end{equation}
and similarly for particles $1$, $2$, and $3$.

Then, after integrating along the positive branch of $k^0$, one can obtain the Fokker-Planck equation for $f_D(t,\vec{k}\,)$~\footnote{We slightly abuse of notation here, as it should strictly read $f(t,E_k)$.}:
\begin{equation}
 \frac{\partial}{\partial t} f_D (t,\vec{k}\,) =  \frac{\partial}{\partial k^i} \left\{ k^i A(\vec{k};T) f_D (t,\vec{k}\,) + \frac{\partial}{\partial k^j} \left[ B_0(\vec{k};T) \Delta^{ij} + B_1(\vec{k};T) \frac{k^i k^j}{k^2} \right] f_D (t,\vec{k}\,) \right\} \ ,
\end{equation}
where the transport coefficients read
\begin{align}
A( \vec{k};T) & = \int \frac{d^3q}{(2\pi)^3} w(\vec{k}, \vec{q}\,) \ \frac{\vec{q} \cdot \vec{k}}{\vec{k}\,^2} \ , \label{eq:app-onshellA} \\ 
 B_0 (\vec{k};T) & = \frac{1}{4} \int \frac{d^3q}{(2\pi)^3} w(\vec{k},\vec{q}\,) \ \left[ \vec{q}\,^2 - \frac{(\vec{q} \cdot \vec{k}\,)^2}{ \vec{k}\,^2} \right] \ ,  \label{eq:app-onshellB0} \\
  B_1 (\vec{k};T) & = \frac{1}{2} \int \frac{d^3q}{(2\pi)^3}  w(\vec{k}, \vec{q}\,) \ \frac{(\vec{q} \cdot \vec{k}\,)^2}{ \vec{k}\,^2} \ . \label{eq:app-onshellB1}
\end{align}
We have introduced the ``on-shell'' scattering rate
\begin{equation} 
w(\vec{k},\vec{q}\,) \equiv \frac{1}{2E_k} \int \frac{dk_1^0}{2\pi} W(E_k,\vec{k},k_1^0,\vec{q}\,) \ ,  
\end{equation}
which in terms of the scattering amplitude reads
\begin{align} 
  w(\vec{k},\vec{q}\,) & = \int \frac{d^3k_3}{(2\pi)^6} f_\Phi^{(0)}(\vec{k}_3) \tilde{f}_\Phi^{(0)}(\vec{k}_3+\vec{q}\,) \frac{1}{2E_k 2E_{k_3} 2E_{k+q} 2E_{k_3+q}} \nonumber \\
  &\times (2\pi)^4 \delta(E_k+E_{k_3}-E_{k+q}-E_{k_3+q}) \nonumber \\
  &\times \left[  \left|T(E_k+E_3,\vec{k}+\vec{k}_3)\right|^2+\left|T(E_k-E_{k_3+q},\vec{k}-\vec{k}_3-\vec{q}\,)\right|^2 \right] \ . 
  \label{eq:app-onshellw}
\end{align}
The expressions of the coefficients in Eqs.~(\ref{eq:app-onshellA}), (\ref{eq:app-onshellB0}) and (\ref{eq:app-onshellB1}) together with the on-shell scattering rate in Eq.~(\ref{eq:app-onshellw}) coincide with those in Refs.~\cite{Abreu:2011ic,Tolos:2013kva,Tolos:2016slr}, apart from the Landau term arising in Eq.~(\ref{eq:app-onshellw}), which is the new contribution found in this thesis.


\pagestyle{plain}				
%

\addchap{Resum}
\label{sec:abstract-cat}

La física hadrònica tal com l'entenem en l'actualitat es remunta a la concepció del model de quarks, proposat independentment per Gell-Mann \cite{Gell-Mann:1964ewy} i Zweig \cite{Zweig:1964ruk} l'any 1964, en un intent de classificar i entendre les propietats d'un gran nombre de partícules (hadrons) que s'havien anat descobrint al llarg de la dècada anterior. No va ser, però, fins a la formalització de la teoria de la cromodinàmica quàntica (QCD), duta a terme per Fritzsch, Leutwyler i el mateix Gell-Mann \cite{Fritzsch:1973pi} a principis de la dècada del 1970, que la interacció forta es va entendre millor en termes de quarks i gluons.

En el model de quarks original, els hadrons es van classificar en dos grans subgrups: els mesons, formats per una parella quark-antiquark; i els barions, que són estats lligats de tres quarks. Tanmateix, no es va excloure la possibilitat de tenir hadrons amb una composició de quarks de valència diferent de la dels mesons i barions ordinaris, sempre que fos compatible amb les regles de la teoria de la QCD.
Durant dècades no es va trobar cap evidència experimental de l'existència d'aquests hadrons exòtics, però la situació ha canviat en els últims vint anys degut a l'explosió de dades experimentals obtingudes en acceleradors col·lisionadors electró-positró i d'hadrons. La confirmació de l'existència de diversos estats hadrònics multiquark, com per exemple tetraquarks i pentaquarks, ha tingut lloc sobretot en el sector dels hadrons pesants, que són aquells amb almenys un quark \textit{charm} (encant, en català) o \textit{bottom} (fons, en català). L'estudi dels hadrons exòtics, especialment dels hadrons exòtics pesants, és actualment una de les línies de recerca més actives en física hadrònica. Per una banda, hi ha programes dedicats a la cerca de nous mesons i barions exòtics tant en instal·lacions experimentals en curs com futures. D'altra banda, la comunitat teòrica destina un gran esforç a entendre la naturalesa d'aquests estats exòtics i distingir, per exemple, una estructura multiquark compacta d'un estat molecular, és a dir, d'un estat lligat o quasi-lligat de dos o més hadrons.

La teoria de la QCD presenta dues característiques extremadament importants. D'una banda, la constant d'acoblament que descriu la intensitat de la interacció forta entre dues partícules disminueix a mesura que augmenta la seva energia o, el que és el mateix, a mesura que disminueix la seva separació. És el que es coneix com llibertat asimptòtica. Tanmateix, a baixes energies (de l'ordre de la massa dels hadrons, per sota d'uns quants GeV), l'acoblament és tan intens que no permet resoldre la QCD pertorbativament. En aquest règim cal recórrer a mètodes no pertorbatius, com són les teories de camps efectives i la \textit{lattice} QCD (LQCD, o QCD en el reticle, en català). L'altra característica de la QCD és l'anomenat confinament del color, que fa que les partícules amb càrrega de color no es puguin aïllar. Així doncs, els quarks i els gluons no es poden observar lliures experimentalment, sinó confinats dins dels hadrons, que tenen càrrega de color neutra. No obstant això, la llibertat asimptòtica prediu que a altes energies els hadrons perden la seva identitat i donen lloc a una ``sopa'' de quarks i gluons desconfinats. És el que es coneix com a plasma de quarks i gluons (QGP) i només es pot trobar a temperatures i/o densitats bariòniques extremadament altes.

Experimentalment, les condicions d'alta temperatura i baixa densitat bariònica necessàries per crear un QGP calent es poden aconseguir amb col·lisions d'ions pesants a altes energies al Col·lisionador d'Ions Pesants Relativistes (RHIC) del Laboratori Nacional de Brookhaven (BNL, Nova York, Estats Units) i al Gran Col·lisionador d'Hadrons (LHC) del CERN (Organització Europea per a la Recerca Nuclear, Ginebra, Suïssa).
Els hadrons amb charm permeten sondejar la formació de la fase de QGP en les col·lisions d'ions pesants, ja que els quarks i antiquarks charm són creats únicament durant les fases inicials de la col·lisió i, per tant, experimenten tota l'evolució del QGP. Després del procés d'hadronització, aquests quarks i antiquarks pesants donen lloc, predominantment, a mesons amb charm ``obert'', és a dir, a mesons $D$. Per tal de descriure la dades experimentals, cal entendre, des d'un punt de vista teòric, la propagació dels mesons $D$ en la fase hadrònica i la seva interacció amb el medi calent de mesons lleugers que els envolta.  

Aquesta tesi doctoral persegueix un doble objectiu. D'una banda, pretén estudiar hadrons exòtics amb contingut pesant que han estat observats recentment i que es poden descriure com estats hadrònics moleculars. D'altra banda, una part important del treball presentat en aquesta memòria s'orienta a millorar la comprensió de la modificació de les propietats dels mesons amb sabor pesant obert en un medi calent. En tots dos casos, la descripció de les interaccions entre hadrons es basa en l'ús de teories efectives hadròniques.

L'estructura d'aquesta memòria de tesi doctoral és la següent. Al Capítol~\ref{ch:intro} presentem un repàs d'alguns aspectes del model de quarks i de la teoria de la QCD, per tal de motivar, a continuació, l'estudi dels hadrons que contenen quarks pesants. En aquest capítol també donem una visió general de l'estat actual de les fases de matèria QCD que podem trobar en condicions extremes de temperatura i densitat bariònica i, finalment, discutim la rellevància dels mesons amb sabor pesant com a sondes dures de la fase de QGP formada en les col·lisions d'ions pesants.

El Capítol~\ref{ch:exoticsinfreespace} s'ha dedicat a investigar les propietats en el buit d'hadrons exòtics amb sabor pesant obert en els sectors bariònic i mesònic. La metodologia emprada es basa en l'ús de Lagrangians efectius consistents amb les simetries del sistema fortament interaccionant sota estudi. Les amplituds de dispersió s'unitaritzen a través de la resolució de l'equació de Bethe-Salpeter en canals acoblats, prestant especial atenció a la regularitació del propagador de dos cossos. Aquesta estratègia ens permet generar estats a partir de la interacció entre les partícules que són els graus de llibertat de la teoria efectiva, és a dir, entre mesons i/o barions. Els estats generats dinàmicament tenen, per tant, una estructura molecular. En la primera secció d'aquest capítol descrivim les simetries quiral i de quark pesant de la teoria de la QCD, així com algunes de les teories efectives que són la base dels estudis que presentem a la resta del capítol. També comentem breument les propietats d'unitarietat i analiticitat de les amplituds de dispersió.

La resta del capítol es divideix en dues parts. En la primera (Secció~\ref{sec:free-mb}), s'estudia la interacció dels mesons pseudoscalars de més baixa energia amb els barions en estat fonamental, en el sector de charm $+1$, estranyesa $-2$ i isospí $0$, mitjançant un model d'intercanvi de mesons vectorials en canal $t$. Les amplituds unitaritzades obtingudes presenten dues ressonàncies amb propietats molt similars a les d'alguns dels estats $\Omega_c^{*0}$ descoberts per la col·laboració LHCb al 2017. Ajustant els paràmetres del model, podem reproduir els valors experimentals de la massa i l'amplada de decaïment de les ressonàncies $\Omega_c(3050)^{0}$ i $\Omega_c(3090)^{0}$. Així doncs, podem concloure que almenys dos dels estats $\Omega_c^{*0}$ observats experimentalment podrien ser molècules mesó--barió i que, com que són generades dinàmicament a partir de la interacció en ona $s$ entre mesons pseudoscalars i barions, tindrien nombres quàntics d'espín-paritat $J^P=1/2^-$. En canvi, alguns models de quarks estableixen valors de $3/2^-$ o $5/2^-$ per l'espín-paritat d'alguns d'aquests estats. Una futura mesura experimental dels nombres quàntics dels estats $\Omega_c^{*0}$ observats a l'LHCb permetria entendre millor la seva naturalesa. A més, estenem l'estudi d'aquesta secció al sector del bottom, on trobem diversos estats $\Omega_b^{*-}$ amb estructura molecular de tipus mesó--barió en la regió d'energia $6400-6800$ MeV. Malgrat la falta d'estadística, algunes estructures són visibles en aquesta regió de l'espectre experimental de l'LHCb per a les $\Omega_b^{*-}$. La confirmació experimental de l'existència d'estats $\Omega_b^{*-}$ en aquesta regió d'energia a les futures instal·lacions de l'LHC, millorades amb major lluminositat, permetrien comparar les propietats d'aquests estats amb les prediccións dels models de quarks i dels models moleculars, i progressar en la comprensió de la seva naturalesa.

En la segona part (Secció~\ref{sec:free-mm}), analitzem les interaccions en el buit dels mesons pseudoscalars i vectorials amb sabor pesant obert ($D^{(*)}$, $D_s^{(*)}$, $\bar{B}^{(*)}$, $\bar{B}_s^{(*)}$) amb mesons lleugers ($\pi$, $K$, $\bar{K}$, $\eta$) utilitzant una teoria efectiva de camps basada en les teories quiral i de quark pesant. En el cas de $J^P=0^+$, trobem que l'estat $D_0^*(2300)$ és generat dinàmicament amb una estructura de doble pol, mentre que l'estat $D_{s0}^*(2317)$ s'identifica amb un estat lligat de tipus molecular. Amb $J^P=1^+$ tenim la ressonància $D_1(2430)$, que també s'identifica amb dos pols al pla d'energia complexa, i l'estat $D_{s1}(2460)$. El paral·lelisme que s'observa entre els sector amb $J^P=0^+$ i $J^P=1^+$ és degut a la simetria d'espín de quark pesant (HQSS), implementada en el Lagrangià a ordre dominant, i només trencada per l'ús de les masses dels mesons pesants en el buit. També estenem els càlculs al sector del bottom, on el model genera dinàmicament estats excitats homòlegs als dels sector del charm. En aquest cas, és la simetria de sabor de quark pesant (HQFS) la responsable de les similituds entre ambdós sectors.

Al Capítol~\ref{ch:hot-medium} ampliem l'anàlisi de la dispersió de mesons amb sabor pesant obert dels mesons lleugers a temperatura finita. Per tal d'obtenir les propietats dels mesons pesants en un medi tèrmic fins a una temperatura màxima de $T=150$~MeV, calculem la modificació en el medi del propagador del mesó pesant de manera autoconsistent, que inclou correccions en l'autoenergia degut a les interaccions amb el medi a través de les amplituds unitaritzades. Aquestes, al seu torn, s'obtenen a partir de les resolució de l'equació de Bethe-Salpeter amb propagadors de dos mesons modificats, també, pels efectes de la temperatura. Amb aquesta metodologia obtenim les funcions espectrals dels mesons pesants i, a partir d'aquestes, la dependència tèrmica de les masses i les amplades de decaïment dels mesons en estat fonamental, tant en el sector del charm ($D^{(*)}$, $D_s^{(*)}$) com en el sector del bottom ($\bar{B}^{(*)}$, $\bar{B}_s^{(*)}$), i també dels estats generats dinàmicament. Els resultats mostren una reducció generalitzada de les masses tèrmiques amb la temperatura, que és d'unes desenes de MeV en un bany de pions a $T=150$~MeV, mentre que les amplades de decaïment augmenten amb la temperatura fins a valors d'unes desenes de MeV a $T=150$~MeV. No observem una tendència clara vers la degeneració quiral dels companys quirals amb $J^P=0^-$ ($1^-$) i $J^P=0^+$ ($1^+$). Tanmateix, el rang de temperatures dels nostres càlculs està limitat per la validesa de la teoria efectiva hadrònica a baixes temperatures, i la degeneració quiral s'espera que tingui lloc a una temperatura més elevada, $T>T_\chi$. Un dels nostres resultats més rellevants és que el company quiral del mesó $D$, és a dir, el mesó $D_0^*(2300)$, té una estructura de doble pol al pla d'energia complexa, i en aquest cas no està clar com hauria de tenir lloc la restauració de la simetria quiral. En aquest capítol també estudiem l'efecte de considerar els kaons en el bany de mesons, a més dels pions, sobre les propietats dels mesons pesants. Això resulta en una lleugera modificació de les masses dels mesons pesants i en un augment de les amplades, però la major contribució als efectes tèrmics és a causa dels pions, ja que són els mesons més lleugers i, per tant, més abundants en el bany de mesons a les temperatures considerades.

Al Capítol~\ref{ch:lattice} calculem, per primera vegada, els correladors Euclidians dels mesos encantats $D$ i $D_s$ a partir de les seves corresponents funcions espectrals, obtingudes amb la formulació de la teoria efectiva a temperatura finita desenvolupada en aquesta tesi. L'objectiu és comparar els resultats amb simulacions de LQCD i, per tant, adaptem les valors de les masses dels mesons als valors obtinguts amb LQCD, que són majors que els respectius valors físics. Per al càlcul dels correladors considerem la completa dependència en energia de les funcions espectrals, tenint en compte que els correladors de LQCD inclouen tant la contribució de l'estat fonamental com del continu d'estats de dispersió a altes energies. Els resultats dels nostres càlculs del quocient entre el correlador Euclidià i el correlador reconstruït es troben dins de les barres d'error de les dades de LQCD, per a temperaures molt per sota de la temperatura de la transició de QCD, $T_c$. A més altes temperatures aquest quocient es desvia significativament de les prediccions de LQCD degut a diversos factors. En primer lloc, els nostres càlculs no inclouen la presència d'estats lligats excitats en les funcions espectrals i, en canvi, aquests sí que són presents en les simulacions de LQCD. En segon lloc, no hem implementat els efectes de volum finit i de \textit{cut-off} del reticle en el càlculs de la teoria efectiva. A més, la modificació tèrmica de les propietats dels mesons pesants induïdes pel kaons del medi podrien ser més rellevants en aquest cas que en les determinacions del capítol anterior, ja que amb les masses no-físiques de LQCD la diferència de massa entre pions i kaons és més petita.

Finalment, al Capítol~\ref{ch:transport} estenem la descripció de la teoria cinètica dels mesons pesants a baixa energia per tal d'incloure els efectes del medi tèrmic i les propietats espectrals dels estats amb charm i bottom obert. Així doncs, al principi d'aquest capítol derivem les equacions \textit{off-shell} de Boltzmann i de Fokker-Planck a partir de la teoria efectiva utilitzada en els capítols anteriors. Aquest és un pas essencial per tal de poder obtenir, a continuació, els coeficients de transport off-shell dels mesons pesants, implementant, per primer cop, la consistència entre la descripció de les interaccions amb els mesons lleugers del medi i la formulació cinètica. Concretament, calculem l'amplada tèrmica, el coeficient d'arrossegament, el coeficient de difusió en l'espai de moments, i el coeficient de difusió espaial. A causa de la seva gran massa en el buit, les correccions de temperatura finita a la massa i l'amplada espectral dels mesons $D$ i $\bar{B}$ són relativament petites, de manera que, en bona aproximació, els mesons pesants es poden tractar com a quasipartícules. No obstant això, l'ús d'amplituds de dispersió a temperatura finita obre un nou rang cinemàtic per a la interacció mesó--mesó, conegut com a tall de Landau. Aquesta contribució als coeficients de transport és força gran a temperatures moderades. De fet, a $T=150$ MeV aquesta nova contribució s'equipara a la contribució usual deguda al tall unitari. Al final del capítol comparem els càlculs dels coeficients de transport per a temperatures per sota de $T_c$, incloent aquests nous efectes de temperatura finita, amb determinacions de LQCD del coeficient de difusió espaial i en l'espai de moments, així com també amb anàlisis Bayesians de dades de col·lisions d'ions pesats, per sobre de $T_c$. Trobem un bon acord amb els nostres resultats a la temperatura de transició de QCD que fa que hi hagi una continuïtat suau dels càlculs a $T_c$.

Les conclusions es presenten al Capítol~\ref{ch:conclusion}. També s'inclouen un seguit d'apèndixs amb taules, derivacions i informació addicional que s'han omès del text principal per tal de facilitar la seva lectura.

Recapitulant, en aquesta memòria es proporciona, d'una banda, una descripció extensa dels estats $\Omega_c^{*0}/\Omega_b^{*-}$ i dels mesons excitats amb sabor pesant obert ($D_0^* (2300)$, $D_{s0}^*(2317)$, $D_1(2430)$ i $D_{s1}(2460)$, i dels seus homòlegs en el sector del bottom) en termes d'estats hadrònics moleculars, utilitzant els Lagrangians efectius adequats per descriure les interaccions hadró--hadró. Així doncs, tenim la confiança que amb aquest treball demostrem la capacitat de les teories efectives hadròniques unitaritzades per generar estats dinàmicament. Quan aquests estats es poden identificar amb hadrons exòtics observats experimentalment, llavors les teories efectives hadròniques permeten proporcionar una interpretació de molècula hadrònica per a la seva naturalesa. D'altra banda, a la vista dels actuals i futurs experiments de col·lisions d'ions pesants, que permeten generar la fase de QGP a densitat bariònica baixa, i dels prometedors resultats de les simulacions de LQCD a temperatura finita, en aquesta tesi doctoral desenvolupem una nova formulació sistemàtica basada en teories efectives per a l'estudi de la modificació de les propietats dels hadrons pesants en un medi mesònic per sota de la temperatura de transició de QCD. A més, a mode d'aplicació dels càlculs d'aquesta metodologia a temperatura finita, calculem coeficients de transport que poden ser utilitzats en simulacions hidrodinàmiques i millorar, així, la comprensió de la formació del QGP en col·lisions d'ions pesants.
\cleardoublepage


\end{document}